\documentstyle[11pt,epsf,rotate]{article}
\input{psfig}

\newcommand{\RE}{{\rm Re}}
\newcommand{\IM}{{\rm Im}}
\newcommand{\vcb}{|V_{cb}|}
\newcommand{\vtd}{|V_{td}|}
\newcommand{\vub}{|V_{ub}/V_{cb}|}
\newcommand{\vts}{|V_{ts}|}

\newcommand{\svs}{\vbox{\vskip 5mm}}
\newcommand{\mvs}{\vbox{\vskip 8mm}}

\def\R1{\varepsilon_1}
\def\E8{\varepsilon_8}
\def\gat{\tilde{\gamma}}
\def\gh{\hat{g}}
\def\gt{\tilde{g}}
\def\gah{\hat{\gamma}}
\def\ga{\gamma}
\def\gaf{\gamma_{5}}

\def\eps{\varepsilon}
\def\epe{\varepsilon'/\varepsilon}
\def\as{\alpha_s}
\newcommand{\eqn}{\ref}
\def\Heff{{\cal H}_{\rm eff}}
\newcommand{\nn}{\nonumber}
\newcommand{\mt}{m_{\rm t}}
\newcommand{\mtb}{\overline{m}_{\rm t}}
\newcommand{\mcb}{\overline{m}_{\rm c}}
\newcommand{\mc}{m_{\rm c}}
\newcommand{\ms}{m_{\rm s}}
\newcommand{\md}{m_{\rm d}}
\newcommand{\mb}{m_{\rm b}}
\newcommand{\mw}{M_{\rm W}}
\newcommand{\mz}{M_{\rm Z}}
\newcommand{\gev}{\, {\rm GeV}}
\newcommand{\mev}{\, {\rm MeV}}

\newcommand{\Lms}{\Lambda_{\overline{\rm MS}}}

\newcommand{\Bsg}{$B \to X_s \gamma$ }

\newcommand{\bea}{\begin{eqnarray}}
\newcommand{\eea}{\end{eqnarray}}
\newcommand{\bd}{\begin{displaymath}}
\newcommand{\ed}{\end{displaymath}}
\newcommand{\aem}{\alpha}

\newcommand{\beq}{\begin{equation}}
\newcommand{\eeq}{\end{equation}}
\newcommand{\be}{\begin{equation}}
\newcommand{\ee}{\end{equation}}
\newcommand{\bi}{\begin{itemize}}
\newcommand{\ei}{\end{itemize}}
\newcommand{\ord}{{\cal O}}

\newcommand{\f}{\frac}

\def\kpnn{$K^+\rightarrow\pi^+\nu\bar\nu$}
\def\kpn{K^+\rightarrow\pi^+\nu\bar\nu}
\def\klpn{K_{\rm L}\rightarrow\pi^0\nu\bar\nu}
\def\klpnn{$K_{\rm L}\rightarrow\pi^0\nu\bar\nu$}
\def\klm{K_{\rm L} \to \mu^+\mu^-}
\def\aspi{\frac{\as}{4\pi}}
\def\gf{\gamma_5}
\newcommand{\imlt}{\IM\lambda_t}
\newcommand{\relt}{\RE\lambda_t}
\newcommand{\relc}{\RE\lambda_c}

\begin{document}
\thispagestyle{empty}
\begin{flushright}
 TUM-HEP-316/98 \\
 hep-ph/9806471 \\
June 1998
\end{flushright}
\vskip1truecm
\centerline{\Large\bf  Weak Hamiltonian, CP Violation and Rare Decays
   \footnote{\noindent
   Dedicated to my father Bronislaw Buras (1915-1994).\\
To appear in "Probing the Standard Model of Particle Interactions",
F. David and R. Gupta, eds, 1998 Elsevier Science B.V.
   }}
\vskip1truecm
\centerline{\large\bf Andrzej J. Buras}
\bigskip
\centerline{\sl Technische Universit{\"a}t M{\"u}nchen}
\centerline{\sl Physik Department} 
\centerline{\sl D-85748 Garching, Germany}
\vskip1truecm
\centerline{\bf Abstract}
These lectures describe in detail the effective Hamiltonians for weak decays
of mesons constructed by means of the operator product expansion and
the renormalization group method. We calculate Wilson coeffcients of local 
operators,
discuss mixing of operators under renormalization, the anomalous dimensions
of operators and anomalous dimension matrices. We elaborate on the 
renormalzation scheme and renormalization
scale dependences and their cancellations in physical amplitudes. 
In particular
we discuss the issue of $\gamma_5$ in D-dimensions and the role of 
evanescent operators in
the calculation of two-loop anomalous dimensions. We present an explicit
calculation of the $6\times6$ one-loop anomalous dimension matrix involving
current-current and QCD-penguin operators and we give some hints how to
properly calculate two-loop anomalous dimensions of these operators. In the
phenonomenological part of these lectures we discuss in detail: CKM matrix,
the unitarity triangle and its determination, two-body
non-leptonic B-decays and the generalized factorization, the ratio $\epe$,
$B\to X_s\gamma$, $K^+\to\pi^+\nu\bar\nu$, $K_L\to\pi^0\nu\bar\nu$,
$B\to X_s\nu\bar\nu$, $B_s\to\mu\bar\mu$ and 
some aspects of CP violation in B-decays.


\newpage

\thispagestyle{empty}

\mbox{}

\newpage

\pagenumbering{roman}

\tableofcontents

\newpage

\pagenumbering{arabic}

\setcounter{page}{1}

\section{Introduction}
\subsection{General View}
The basic starting point for any serious phenomenology of weak decays of
hadrons is the effective weak Hamiltonian which has the following generic
structure
\be\label{b1}
{\cal H}_{eff}=\frac{G_F}{\sqrt{2}}\sum_i V^i_{\rm CKM}C_i(\mu)Q_i~.
\ee
Here $G_F$ is the Fermi constant and $Q_i$ are the relevant local
operators which govern the decays in question. The Cabibbo-Kobayashi-Maskawa
factors $V^i_{CKM}$ \cite{CAB,KM} 
and the Wilson Coefficients $C_i$ \cite{OPE,ZIMM} describe the 
strength with which a given operator enters the Hamiltonian.

\begin{figure}[hbt]
\vspace{0.10in}
\centerline{
\epsfysize=1.9in
\epsffile{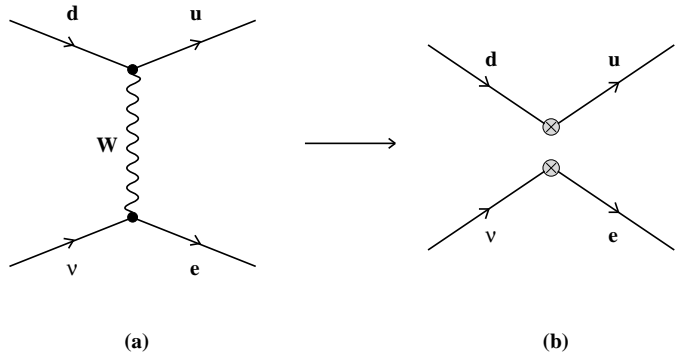}
}
\vspace{0.08in}
\caption[]{
$\beta$-decay at the quark level in the full (a) and effective (b)
theory.
\label{L:1}}
\end{figure}

In the simplest case of the $\beta$-decay, ${\cal H}_{eff}$ takes 
the familiar form
\be\label{beta}
{\cal H}^{(\beta)}_{eff}=\frac{G_F}{\sqrt{2}}
\cos\theta_c[\bar u\gamma_\mu(1-\gamma_5)d \otimes
\bar e \gamma^\mu (1-\gamma_5)\nu_e]~,
\ee
where $V_{ud}$ has been expressed in terms of the Cabibbo angle. In this
particular case the Wilson Coefficient is equal unity and the local
operator, the object between the square brackets, is given by a product 
of two $V-A$ currents. This local operator is represented by the
diagram (b) in fig. \ref{L:1}.
Equation (\ref{beta}) represents the Fermi theory for $\beta$-decays 
as formulated by Sudarshan and
Marshak \cite{SUMA} and Feynman and Gell-Mann \cite{GF} forty years ago, 
except that in (\ref{beta})
the quark language has been used and following Cabibbo a small departure of
$V_{ud}$ from unity has been incorporated. In this context the basic 
formula (\ref{b1})
can be regarded as a generalization of the Fermi Theory to include all known
quarks and leptons as well as their strong and electroweak interactions as
summarized by the Standard Model. It should be stressed that the formulation
of weak decays in terms of effective Hamiltonians is very suitable for the
inclusion of new physics effects. We will discuss this issue briefly in these
lectures.

Now, I am aware of the fact that the formal operator language used here is
hated by experimentalists and frequently disliked by more phenomenological
minded theorists. Consequently the literature on weak decays, in particular
on B-meson decays, is governed by Feynman diagram drawings with W-, Z- and top
quark exchanges, rather than by the operators in (\ref{b1}). 
In the case of the $\beta$-decay we have the diagram (a) in fig.~\ref{L:1}.
Yet such Feynman
diagrams with full W-propagators, Z-propagators and top-quark propagators
really represent the situation at very short distance scales 
$\ord ({\rm M_{W,Z}, m_t})$, whereas the
true picture of a decaying hadron with masses 
$\ord(\mb,\mc,m_K)$ is more properly described by
effective point-like vertices which are represented by the local operators
$Q_i$. The Wilson coefficients $C_i$ can then be regarded as coupling constants
associated with these effective vertices.

Thus ${\cal H}_{eff}$ in (\ref{b1}) is simply a series of effective 
vertices multiplied 
by effective coupling constants $C_i$. This series is known under the name 
of the operator product expansion (OPE) \cite{OPE,ZIMM,WIT}. 
Due to the interplay of electroweak 
and strong interactions the structure of the local operators (vertices) is 
much richer than in the case of the $\beta$-decay. They can be classified 
with respect to the Dirac structure, colour structure and the type of quarks 
and leptons relevant for a given decay. Of particular interest are the 
operators involving quarks only. They govern the non-leptonic decays.

Now what about the couplings $C_i(\mu)$ and the scale $\mu$? The 
important point is that $C_i(\mu)$
summarize the physics contributions from scales higher than $\mu$ and due to
asymptotic freedom of QCD they can be calculated in perturbation theory as
long as $\mu$ is not too small. $C_i$ include the top quark contributions and
contributions from other heavy particles such as W, Z-bosons and charged
Higgs particles or supersymmetric particles in the supersymmetric extensions
of the Standard Model. At higher orders in the electroweak coupling the
neutral Higgs may also contribute. Consequently $C_i(\mu)$ depend generally 
on $m_t$ and also on the masses of new particles if extensions of the 
Standard Model are considered. This dependence can be found by evaluating 
so-called {\it box} and {\it penguin} diagrams with full W-, Z-, top- and 
new particles exchanges and {\it properly} including short distance QCD 
effects. The latter govern the $\mu$-dependence of the couplings $C_i(\mu)$.

The value of $\mu$ can be chosen arbitrarily. As we will see below it serves 
to separate the physics contributions to a given decay amplitude into
short-distance contributions at scales higher than $\mu$ and long-distance
contributions corresponding to scales lower than $\mu$. It is customary 
to choose
$\mu$ to be of the order of the mass of the decaying hadron. 
This is $\ord (\mb)$ and $\ord(\mc)$ for B-decays and
D-decays respectively. In the case of K-decays the typical choice is
 $\mu=\ord(1-2~GeV)$
instead of $\ord(m_K)$, which is much too low for any perturbative 
calculation of the couplings $C_i$.

Now due to the fact that $\mu\ll  M_{W,Z},~ m_t$, large logarithms 
$\ln\mw/\mu$ compensate in the evaluation of
$C_i(\mu)$ the smallness of the QCD coupling constant $\alpha_s$ and 
terms $\alpha^n_s (\ln\mw/\mu)^n$, $\alpha^n_s (\ln\mw/\mu)^{n-1}$ 
etc. have to be resummed to all orders in $\alpha_s$ before a reliable 
result for $C_i$ can be obtained.
This can be done very efficiently by means of the renormalization group
methods \cite{REGM,HV1,Weinberg}. 
The resulting {\it renormalization group improved} perturbative
expansion for $C_i(\mu)$ in terms of the effective coupling constant 
$\alpha_s(\mu)$ does not involve large logarithms and is more reliable.

It should be stressed at this point that the construction of the effective
Hamiltonian ${\cal H}_{eff}$ by means of the operator product expansion and 
the
renormalization group methods can be done fully in the perturbative framework.
The fact that the decaying hadrons are bound states of quarks is irrelevant
for this construction. Consequently the coefficients $C_i(\mu)$ are 
independent of the
particular decay considered in the same manner in which the usual gauge
couplings are universal and process independent.

So far so good. Having constructed the effective Hamiltonian we can proceed
to evaluate the decay amplitudes. An amplitude for a decay of a given meson 
$M= K, B,..$ into a final state $F=\pi\nu\bar\nu,~\pi\pi,~DK$ is simply 
given by
\be\label{amp5}
A(M\to F)=\langle F|{\cal H}_{eff}|M\rangle
=\frac{G_F}{\sqrt{2}}\sum_i V^i_{CKM}C_i(\mu)\langle F|Q_i(\mu)|M\rangle,
\ee
where $\langle F|Q_i(\mu)|M\rangle$ 
are the hadronic matrix elements of $Q_i$ between M and F. As indicated
in (\ref{amp5}) these matrix elements depend similarly to $C_i(\mu)$ 
on $\mu$. They summarize the physics contributions to the amplitude 
$A(M\to F)$ from scales lower than $\mu$.

We realize now the essential virtue of OPE: it allows to separate the problem
of calculating the amplitude
$A(M\to F)$ into two distinct parts: the {\it short distance}
(perturbative) calculation of the couplings $C_i(\mu)$ and 
the {\it long-distance} (generally non-perturbative) calculation of 
the matrix elements $\langle Q_i(\mu)\rangle$. The scale $\mu$, as
advertised above, separates then the physics contributions into short
distance contributions contained in $C_i(\mu)$ and the long distance 
contributions
contained in $\langle Q_i(\mu)\rangle$. By evolving this scale from 
$\mu=\ord(\mw)$ down to lower values one
simply transforms the physics contributions at scales higher than $\mu$ 
from the hadronic matrix elements into $C_i(\mu)$. Since no information 
is lost this way the full amplitude cannot depend on $\mu$. Therefore 
the $\mu$-dependence of the couplings $C_i(\mu)$ has to cancel the 
$\mu$-dependence of $\langle Q_i(\mu)\rangle$. In other words it is a
matter of choice what exactly belongs to $C_i(\mu)$ and what to 
$\langle Q_i(\mu)\rangle$. This cancellation
of $\mu$-dependence involves generally several terms in the expansion 
in (\ref{amp5}).

Clearly, in order to calculate the amplitude $A(M\to F)$, the matrix 
elements $\langle Q_i(\mu)\rangle$ have to be evaluated. 
Since they involve long distance contributions one is forced in
this case to use non-perturbative methods such as lattice calculations, the
1/N expansion (N is the number of colours), QCD sum rules, hadronic sum rules,
chiral perturbation theory and so on. In the case of certain B-meson decays,
the {\it Heavy Quark Effective Theory} (HQET) also turns out to be a 
useful tool.
Needless to say, all these non-perturbative methods have some limitations.
Consequently the dominant theoretical uncertainties in the decay amplitudes
reside in the matrix elements $\langle Q_i(\mu)\rangle$.

The fact that in most cases the matrix elements $\langle Q_i(\mu)\rangle$
 cannot be reliably
calculated at present, is very unfortunate. One of the main goals of the
experimental studies of weak decays is the determination of the CKM factors 
$V_{\rm CKM}$
and the search for the physics beyond the Standard Model. Without a reliable
estimate of $\langle Q_i(\mu)\rangle$ this goal cannot be achieved unless 
these matrix elements can be determined experimentally or removed from the 
final measurable quantities
by taking the ratios or suitable combinations of amplitudes or branching
ratios. However, this can be achieved only in a handful of decays and
generally one has to face directly the calculation of 
$\langle Q_i(\mu)\rangle$.

Now in the case of semi-leptonic decays, in which there is at most one hadron
in the final state, the chiral perturbation theory in the case of K-decays
and HQET in the case of B-decays have already provided useful estimates of
the relevant matrix elements. This way it was possible to achieve
satisfactory determinations of the CKM elements $V_{us}$ and $V_{cb}$ in 
$K\to\pi e\nu$ and $B\to D^*e\nu$ respectively. 
We will also see that some rare decays like $K\to\pi\nu\bar\nu$ and
$B\to\mu\bar\mu$ can be calculated very reliably.

The case of non-leptonic decays in which the final state consists exclusively
out of hadrons is a completely different story. Here even the matrix
elements entering the simplest decays, the two-body decays like 
$K\to\pi\pi$, $D\to K\pi$ or $B\to DK$ cannot be
calculated in QCD reliably at present. For this reason approximative schemes
for these decays can be found in the literature. One of such schemes, the
factorization scheme for matrix elements has been popular for some time among
experimentalists and phenomenologists. The other approach is the diagrammatic
approach \cite{DIAG}, 
in which the decay amplitudes are decomposed into various
contributions corresponding to certain flavour-flow topologies which in the
literature appear unter the names of "trees", "colour-suppressed trees",
"penguins", "annihilation" etc. Supplemented by isospin symmetry, the
approximate SU(3) flavour symmetry and various "plausible" dynamical
assumptions the diagrammatic approach has been used extensively for
non-leptonic B-decays during the nineties.

As we will see in these lectures the factorization approach 
has several limitations and an improved treatment of
non-leptonic B-decays, beyond this approach, is called for. 
We will have no time to discuss the diagrammatic approach, which goes
beyond the factorization approach, but also here improvements
are necessary.
For K-decays
some progress in this direction has been done by means of the 1/N approach,
hadronic sum rules, chiral perturbation theory and lattice calculations.
However, these techniques will not be discussed here as they are the 
subjects of other lectures at this
summer school. We will only collect necessary results obtained in
these approaches.   

Returning to the Wilson coefficients $C_i(\mu)$ it should be stressed that 
similar
to the effective coupling constants they do not depend only on the scale $\mu$
but also on the renormalization scheme used: this time on the 
scheme for the renormalization of local operators. That the local operators 
undergo renormalization is not surprising. After all they represent effective
vertices and as the usual vertices in a field theory they have to be
renormalized when quantum corrections like QCD or QED corrections are taken
into account. As a consequence of this, the hadronic matrix elements 
$\langle Q_i(\mu)\rangle$
are
renormalization scheme dependent and this scheme dependence must be cancelled
by the one of $C_i(\mu)$ so that the physical amplitudes are 
renormalization scheme
independent. Again, as in the case of the $\mu$-dependence, the 
cancellation of
the renormalization scheme dependence involves generally several 
terms in the
expansion (\ref{amp5}).

Now the $\mu$ and the renormalization scheme dependences of the couplings 
$C_i(\mu)$ can
be evaluated efficiently in the renormalization group improved perturbation
theory. Unfortunately the incorporation of these dependences in the
non-perturbative evaluation of the matrix elements  
$\langle Q_i(\mu)\rangle$
remains as an important
challenge and most of the non-perturbative methods on the market are
insensitive to these dependences. The consequence of this unfortunate
situation is obvious: the resulting decay amplitude are $\mu$ and 
renormalization
scheme dependent which introduces potential theoretical uncertainty in the
predictions. On the other hand we will see in the course of these lectures
that in certain decays these dependences are rather mild.

So far I have discussed only  {\it exclusive} decays. It turns out that
in the case of {\it inclusive} decays of heavy mesons, like B-mesons,
things turn out to be easier. In an inclusive decay one sums over all 
(or over
a special class) of accessible final states so that the amplitude for an
inclusive decay takes the form:
\be\label{ampi}
A(B\to X)
=\frac{G_F}{\sqrt{2}}\sum_{f\in X} 
V^i_{\rm CKM}C_i(\mu)\langle f|Q_i(\mu)|B\rangle~.
\ee
At first sight things look as complicated as in the case of exclusive decays.
It turns out, however, that the resulting branching ratio can be calculated
in the expansion in inverse powers of $\mb$ with the leading term 
described by the spectator model
in which the B-meson decay is modelled by the decay of the $b$-quark:
\be\label{hqe}
{\rm Br}(B\to X)={\rm Br}(b\to q) +\ord(\frac{1}{\mb^2})~. 
\ee
This formula is known under the name of the Heavy Quark Expansion (HQE)
\cite{HQE1}-\cite{HQE3}.
Since the leading term in this expansion represents the decay of the quark,
it can be calculated in perturbation theory or more correctly in the
renormalization group improved perturbation theory. It should be realized
that also here the basic starting point is the effective Hamiltonian 
 (\ref{b1})
and that the knowledge of the couplings $C_i(\mu)$ is essential for 
the evaluation of
the leading term in (\ref{hqe}). But there is an important difference 
relative to the
exclusive case: the matrix elements of the operators $Q_i$ can be 
"effectively"
evaluated in perturbation theory. 
This means, in particular, that their $\mu$ and renormalization scheme
dependences can be evaluated and the cancellation of these dependences by
those present in $C_i(\mu)$ can be investigated.

Clearly in order to complete the evaluation of $Br(B\to X)$ also the 
remaining terms in
(\ref{hqe}) have to be considered. These terms are of a non-perturbative 
origin, but
fortunately they are suppressed by at least two powers of $m_b$. 
They have been
studied by several authors in the literature with the result that they affect
various branching ratios by less then $10\%$ and often by only a few percent.
Consequently the inclusive decays give generally more precise theoretical
predictions at present than the exclusive decays. On the other hand their
measurements are harder. There are of course some important theoretical
issues related to the validity of HQE in (\ref{hqe}) which appear in the 
literature under the name of quark-hadron duality. 
Since these matters are discussed in
detail by Mark Wise in his lectures, I will not discuss them here and will
use HQE as God given.

We have learned now that the matrix elements of $Q_i$ are easier to handle in
inclusive decays than in the exclusive ones. On the other hand the evaluation
of the couplings $C_i(\mu)$ is equally  difficult in both cases although 
as stated
above it can be done in a perturbative framework. Still in order to achieve
sufficient precision for the theoretical predictions it is desirable to have
accurate values of these couplings. Indeed it has been realized at the end of
the eighties
that the leading term (LO) in the renormalization group improved perturbation
theory, in which the terms $\alpha^n_s (\ln\mw/\mu)^n$ are summed, is 
generally insufficient and the
inclusion of next-to-leading corrections  (NLO) which correspond to summing
the terms $\alpha^n_s (\ln\mw/\mu)^{n-1}$ is necessary. 
In particular, unphysical left-over $\mu$-dependences
in the decay amplitudes and branching ratios resulting from the truncation of
the perturbative series are considerably reduced by including NLO
corrections. These corrections are known by now for the most important and
interesting decays and will constitute a considerable part of these lectures. 
\subsection{Strategy}
Like in any serious climb we need some strategy for these lectures. We will
see that some parts of our tour will be rather easy, some other parts rather
technical and difficult. 
The tour consists of seven sections (2--8) devoted to the basic formalism
of weak decays and seven sections (9--15), which present in detail some
selected applications of this formalism.
The map of our route is given the contents. Here we go.

As always we begin with {\it First Steps}. They will allow us to collect most
elementary ingredients of the Standard Model including elementary vertices,
propagators and the corresponding Feynman rules. This is also the place to
discuss the CKM matrix, its most convenient parametrizations and the related
unitarity triangle.

Having this general information at hand we will move next to discuss FCNC
processes in rather general terms. The idea here is to collect most important
effective vertices resulting from penguin and box diagrams and give for them
Feynman rules. We will then see that there are seven basic $\mt$-dependent
functions which enter the effective vertices in question and thereby
determine the strength of FCNC transitions. 

We will illustrate  the derivation of effective
rules by calculating explicitly two simplest vertices. 
In this part we will also have a first look at effective weak
Hamiltonians which are the main objectives of these lectures. This will allow
us to discuss briefly GIM mechanism \cite{GIM1} 
and give a description of the so-called
penguin-box expansion \cite{PBE0}: 
version of OPE particularly suitable for the study of
the $\mt$-dependence of FCNC processes. Here, obviously, the seven basic
functions mentioned above will play the crucial role.

The two first parts just described actually have an introductory character.
They are really like a gentle hike to our base  camp. From now on the matters
begin to be more difficult. Particularly difficult are sections 4,5,6 and 8
which are really at the heart of these lectures. Any serious student who
wants to learn the field of weak Hamiltonians at the level needed for
professional applications should study these sections in great detail. The
reason being that it is not sufficient for a good phenomenology to simply copy
from some papers the values of Wilson coefficients and insert them in some
formulae also copied from still other papers. There are so many subtleties in
this field that without a sufficient understanding of section 4,5,6 and 8 it
will be difficult to avoid errors in phenomenological applications of the
formalism presented there. So what can be found in the basic sections 4,5,6
and 8?

Section 4 is devoted to the renormalization and the renormalization group in
QCD. In particular we will discuss the dimensional regularization paying some
tribute to the issue of $\gamma_5$ in $D\not=4$ dimensions. We will discuss 
the MS and ${\rm \overline{MS}}$
renormalization schemes giving the list of the most important renormalization
constants. Some of these constants will be calculated. Subsequently we will
move to discuss renormalization group equations offering several explicit
derivations. This section culminates in the analysis of the running QCD
coupling, the analysis of the running quark mass and the introduction of
the concept of the renormalization group improved perturbation theory. 

The formal discussion of weak hamiltonians is given first in sections 5 and
6. Section 5 introduces the concept of the operator product expansion and
discusses in great detail the implementation of the renormalization group
techniques into OPE in the leading logarithmic approximation (LO). Section 6
can be considered as the generalization of section 5 to include
next-to-leading logarithmic corrections (NLO). The basic actors of sections 5
and 6 are so-called current-current operators $Q_1$ and $Q_2$. 
Only these
operators are discussed in these two sections. My strategy here is to present
in the simplest setting most important concepts of this field. Thus we will
discuss Wilson coefficients of local operators, mixing of operators under
renormalization, anomalous dimensions and anomalous dimension matrices,
matching of the full theory to the effective theory, renormalization scheme
and renormalization scale dependences and their cancellations, the issue of
evanescent operators in the calculation of two-loop anomalous dimensions and
several other things. In particular we will give master formulae for
anomalous dimensions and a procedure for a correct calculation of Wilson
coefficient functions including NLO corrections. We will also present an
explicit calculation of the anomalous dimension matrix for the operators
$(Q_1, Q_2)$ at the one-loop level and we will give some hints 
how to properly calculate two-loop anomalous dimensions of these operators.

After a short break (section 7), in which a numerical study of the results of
sections 5 and 6 will be presented, we will move to section 8 which can be
considered as a generalization of sections 5 and 6 to include other
operators: the penguin operators of various sorts and the operators
originating in the box diagrams. In particular we will derive the proper
matching for penguin operators and we will provide explicit calculations of
the $6\times 6$ one-loop anomalous dimension matrix for the current-current 
and QCD
penguin operators. The material of this section should allow the reader to
follow without difficulties the applications of OPE and renormalization group
to any decay at the NLO level present in the literature.

The remaining sections of these lectures amount simply to the applications of
the formalism developed in sections 4,5,6 and 8. It will also turn out that
our brief numerical analysis of section 7 was not accidental.

I would like to stress that sections 9--15 should not be considered as a
comprehensive review of the phenomenology of the full field of weak decays.
Certainly not! There are several issues which I have omitted completely: one
of them is the $\Delta I=1/2$ issue in non-leptonic decays, 
the other two are the
popular rare decays $K_L\to\pi^0 e^-e^+$ and $B\to X_s\mu\bar\mu$. 
Moreover the presentation of CP-asymmetries in
B-decays is very superficial and the D-meson decays are completely omitted.
Yet there are many excellent reviews of these
topics and I will from time to time give references where this material can
be found. My choice of topics for sections 9-15 was motivated by the wish to
present the techniques and methods developed in the previous sections in some
representative, phenomenologically important settings. 
Here is the choice I have made.

Section 9 deals with two-body non-leptonic B-decays. 
The purpose of this section is to make a
critical look at the existing analyses of these decays in the framework of
{\it factorization} and the so-called {\it generalized factorization.}
 I have decided to
discuss this topic here as it offers an excellent arena for various issues
analyzed already in sections 5,6 and 7.

Section 10 deals with the issues of $K^0-\bar K^0$ and 
$B^0_{d,s}-\bar B^0_{d,s}$ mixings, indirect CP violation in
$K_L\to\pi\pi$ and with the standard construction of the unitarity triangle.

Section 11 deals with a sad story: the efforts to calculate the CP violating
ratio $\epe$. Here I will first summarize the work done in my group at the
Technical University in  Munich. Subsequently I will make a brief review of
other efforts including most recent developments.

Section 12 deals with much more successful efforts: the calculations of the
branching ratio for the inclusive \Bsg decay. Here the matrix elements of the
relevant operators can be effectively calculated and various issues related
to scale dependences which we have discussed formally in sections 6 and 8 can
be analyzed in explicit terms.

Section 13 is my love story: the rare decays $K_L\to\pi^0\nu\bar\nu$
and $K^+\to\pi^+\nu\bar\nu$. Since these decays are
theoretically very clean, also here various formal issues discussed in
previous sections can be analyzed with concrete examples. We will
demonstrate the great potential of $K\to\pi\nu\bar\nu$ 
in the determination of the CKM
parameters: in particular of $V_{td}$, $\IM V^{*}_{ts}V_{td}$ and
$\sin 2\beta$. We will also discuss the rare B-decays 
$B\to X_{s,d}\nu\bar\nu$ and $B_{s,d}\to l^+l^-$.
This section ends with a description of two-loop electroweak contributions to
rare K- and B-decays in the large $\mt$ limit.

Section 14 offers some future visions. First we will discuss briefly
CP-asymmetries in B-decays and their potential in the determination of the
angles of the unitarity triangle. This determination of the CKM parameters 
wil be confronted with the determination by
means of $K\to\pi\nu\bar\nu$ decays. 
Subsequently a number of other strategies for a clean
determination of the CKM matrix will be briefly discussed. 

Section 15 offers a brief outlook of the field of weak decays
 for the next ten years.
Finally in section 16  a few general messages 
 on the Les Houches summer school 1997 will be made.
\subsection{Whatïs New in these Lectures}
In writing these lectures I benefited enormously from a review on NLO
QCD corrections to weak decays written in collaboration with Gerhard 
Buchalla and
Markus Lautenbacher in 1995 \cite{BBL}, from a review on CP violation and 
rare decays
written in collaboration with Robert Fleischer in the spring of 1997 
\cite{BF97} and from
several courses on the renormalization of QCD and the renormalization group
methods in weak decays I have given at the Technical University in Munich
during the nineties. It is unavoidable that there is some overlap between the
present lectures and the reviews in \cite{BBL,BF97}.
On the other hand there are several
differences and many things which are covered here but cannot be found 
there. In particular:
\bi
\item
We discuss the renormalization of QCD and the renormalization group in
more detail offering several derivations.
\item
We cover the issue of $\gamma_5$ in D dimensions.
\item
We analyze the role of evanescent operators in the calculation of
two-loop anomalous dimensions of local operators. 
\item
We present an explicit calculation of the $6\times 6$ one-loop anomalous
dimension matrix involving current-current and QCD penguin operators.
\item
We calculate explicitly a counter-diagram in the evaluation of two-loop
anomalous dimensions in order to exhibit the role of evanescent operators.
\item
We discuss critically the hypothesis of the generalized factorization in
two-body non-leptonic B-decays.
\item
We review the present status of the calculation of the non-perturbative
factors $B^{(1/2)}_{6}$ and $B^{(3/2)}_{8}$ relevant for the calculation 
of $\epe$ and present
an updated analysis of this ratio.
\item
We review in detail the present status of the radiative $B\to X_s\gamma$
decay including most recent developments.
\item
 While our discussion of $K^+\to\pi^+\nu\bar\nu$ and 
$K_L\to\pi^0\nu\bar\nu$ borrows a lot from \cite{BBL,BF97}, it contains a
few derivations absent there as well as new numerical estimates.
Moreover we discuss briefly two--loop electroweak contributions to these
decays.
\item
Our discussion of CP violation in B-decays is very superficial. On the
other hand we make some comments on the recent hot issue: the role of the
final state interactions in the determination of the CP-phases from
CP-asymmetries in B-decays.
\ei
\section{First Steps}
\setcounter{equation}{0}
\subsection{The Basic Lagrangian}
Throughout these lectures we will dominantly work in the context of
the Standard Model with three generations of quarks and leptons and
the interactions described by the gauge group 
$ SU(3)_C\otimes SU(2)_L\otimes U(1)_Y$ which undergoes the spontaneous
breakdown:
\be
SU(3)_C\otimes SU(2)_L\otimes U(1)_Y \to
 SU(3)_C\otimes U(1)_Q
\ee
 Here $Y$ and $Q$ denote the weak hypercharge and  the electric charge
generators respectively. $SU(3)_C$ stands for QCD.

There are excellent text books on the dynamics of the Standard Model
and on the Field Theory. 
My favourites are listed in \cite{MUTA}--\cite{Collins}.
Let us therfore collect here only those ingredients of this model which
are fundamental for the subject of weak decays.

\bi
\item
The strong interactions are mediated by eight gluons $G_a$, the
electroweak interactions by $W^{\pm}$, $Z^0$, $\gamma$ and the
neutral Higgs boson $H^0$. In the non-physical gauges also other
exchanges have to be included. In particular, the contributions
from fictictious Higgs particles $\phi^{\pm}$ have to be taken into
account to obtain gauge independent results.
\item
The dynamics of the theory is described by the fundamental Lagrangian:
\begin{equation}\label{LL}
{\cal L} = {\cal L}({\rm QCD}) + {\cal L}({\rm SU(2)_L\otimes U(1)_Y})+
{\cal L}({\rm Higgs})
\end{equation}
from which - after quantization and spontaneous symmetry breaking - the
Feynman rules can be derived.
\item
Concerning {\it Electroweak Interactions}, the left-handed leptons and
quarks are put into $SU(2)_L$ doublets:
\begin{equation}\label{2.31}
\left(\begin{array}{c}
\nu_e \\
e^-
\end{array}\right)_L\qquad
\left(\begin{array}{c}
\nu_\mu \\
\mu^-
\end{array}\right)_L\qquad
\left(\begin{array}{c}
\nu_\tau \\
\tau^-
\end{array}\right)_L
\end{equation}
\begin{equation}\label{2.66}
\left(\begin{array}{c}
u \\
d^\prime
\end{array}\right)_L\qquad
\left(\begin{array}{c}
c \\
s^\prime
\end{array}\right)_L\qquad
\left(\begin{array}{c}
t \\
b^\prime
\end{array}\right)_L       
\end{equation}
with the corresponding right-handed fields transforming as singlets
under $ SU(2)_L $. The primes in (\ref{2.66}) are discussed a few pages 
below. The electroweak interactions are summarized by the Lagrangian
\begin{equation}\label{3}
{\cal L}_{\rm int}^{\rm EW}={\cal L}_{\rm CC}+{\cal L}_{\rm NC}\,,
\end{equation}
with ${\cal L}_{\rm CC}$ and ${\cal L}_{\rm NC}$ describing 
{\it charged} and {\it neutral} current interactions respectively.
Concentrating on the fermion-gauge-boson electroweak interactions
we have:
\item
{\bf Charged Current Interactions:}
\begin{equation}\label{4}
{\cal L}_{\rm CC}=\frac{g_2}{2 \sqrt{2}}(J^+_\mu W^{+\mu}+J^-_\mu W^{-\mu}),
\end{equation}
where
\begin{equation}\label{6}
J^+_\mu=
(\bar{u} d')_{V-A} +
(\bar{c} s')_{V-A} +
(\bar{t} b')_{V-A} +
(\bar{\nu}_e e)_{V-A} +
(\bar{\nu}_\mu \mu)_{V-A} +
(\bar{\nu}_\tau \tau)_{V-A}
\end{equation}
denotes the charged current and
$g_2$ is the $SU(2)_L$ coupling constant.
\item
{\bf Neutral Current Interactions:}
\begin{equation}\label{5}
{\cal L}_{\rm NC}=
- e J^{\rm em}_\mu A^{\mu}+ \frac{g_2}{2 \cos \Theta_{\rm W}} J^0_\mu Z^\mu,
\end{equation}
where $e$ is the QED coupling constant
 and $\Theta_{\rm W}$  the Weinberg
angle. The neutral electromagnetic and weak currents are given by
\begin{equation}\label{7}
J^{\rm em}_\mu=\sum_f {Q_f \bar f \gamma_\mu f}
\end{equation}
\begin{equation}\label{8}
J^0_\mu=\sum_f \bar f \gamma_\mu (v_f-a_f\gamma_5) f,
\end{equation}
where
\begin{equation}\label{9}
v_f=T^f_3-2 Q_f \sin^2\Theta_{\rm W},
\qquad
a_f=T^f_3.
\end{equation}
Here $Q_f$ and $T^f_3$ denote the charge and the third component of the
weak isospin of the left-handed fermion $f_L$, respectively. These
electroweak charges are collected in table \ref{tab:ewcharges}.
\ei
\begin{table}[htb]
\caption[]{Electroweak Quantum Numbers.
\label{tab:ewcharges}}
\begin{center}
\begin{tabular}{|c||c|c|c|c|c|c|c|}
\hline
 & $\nu^e_L$ & $ e^-_L$ & $ e^-_R$ & $u_L$ & $d_L$ & $u_R$ & $d_R$ \\
\hline
$Q$ & 0 & $-1$ & $-1$ & 2/3 & $-1/3$ & 2/3 & $-1/3$ \\
\hline
$T_3$ &1/2 & $-1/2$ & 0 & 1/2 & $-1/2$ & 0 & 0 \\
\hline
$Y$ & $-1$ & $-1$ & $-2$ & 1/3 & 1/3 & 4/3 & $-2/3$ \\
\hline
\end{tabular}
\end{center}
\end{table}
\subsection{Elementary Vertices }
\begin{figure}[hbt]
\vspace{0.10in}
\centerline{
\epsfysize=3in
\epsffile{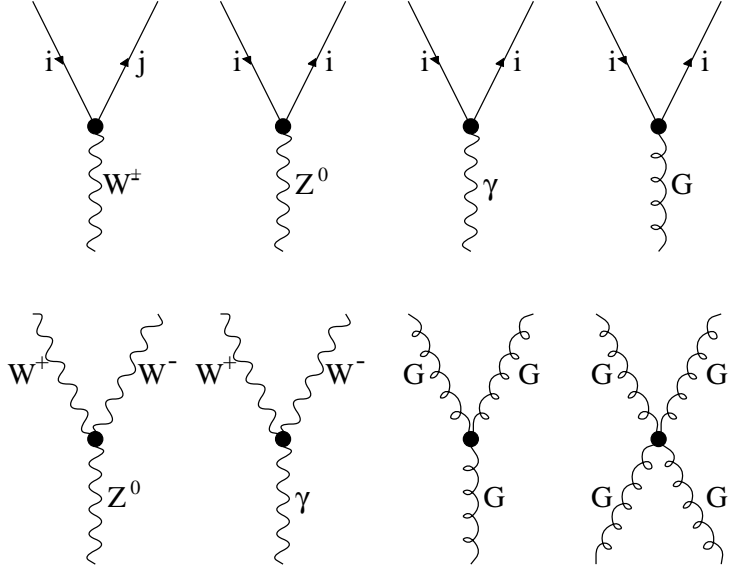}
}
\vspace{0.08in}
\caption[]{
Elementary Vertices
\label{fig:1}}
\end{figure}

Let us next recall those elementary interaction vertices which govern
the physics of weak decays. They are shown in fig. \ref{fig:1}.
The following comments should be made:
\begin{itemize}
\item
The indices $i,j$ denote flavour: $i,j = u,d,c,t,\ldots$
\item
In non--physical gauges also vertices involving fictitious
Higgs particles in place of $W^{\pm}$, $Z^0$ have to be included in
this list.
\item
The quartic
electroweak couplings will not enter these lectures.
\item
The flavour is conserved in vertices involving neutral gauge bosons
$Z^0$, $\gamma$ and $G$. This fact implies the absence of flavour
changing neutral current (FCNC) transitions at the tree level. 
This striking property of neutral interactions in the Standard
Model is quarantieed by the {\rm GIM} mechanism \cite{GIM1}. 
We will return
to this mechanism later on.
\item
The charged current processes mediated by $W^{\pm}$ are
obviously flavour violating with the strength of violation given by
the gauge coupling $g_2$  and effectively at low energies 
by the Fermi constant 
\begin{equation}\label{2.100}
\frac{G_{\rm F}}{\sqrt{2}}=\frac{g^2_2}{8 \mw^2}
\end{equation}
and a {\it unitary} $3\times3$
{\rm CKM} matrix. 
\item
The {\rm CKM} matrix \cite{CAB,KM,GIM1} connects the {\it weak
eigenstates} $(d^\prime,s^\prime,b^\prime)$ and the corresponding {\it mass 
eigenstates} $d,s,b$ through
\begin{equation}\label{2.67}
\left(\begin{array}{c}
d^\prime \\ s^\prime \\ b^\prime
\end{array}\right)=
\left(\begin{array}{ccc}
V_{ud}&V_{us}&V_{ub}\\
V_{cd}&V_{cs}&V_{cb}\\
V_{td}&V_{ts}&V_{tb}
\end{array}\right)
\left(\begin{array}{c}
d \\ s \\ b
\end{array}\right)=\hat V_{\rm CKM}\left(\begin{array}{c}
d \\ s \\ b
\end{array}\right).
\end{equation}
In the leptonic sector the analogous mixing matrix is a unit matrix
due to the masslessness of neutrinos in the Standard Model.
The unitarity of the CKM matrix assures the absence of
elementary FCNC vertices. It
 is consequently at the basis of the GIM mechanism.
On the other hand,
the fact that the $V_{ij}$'s can a priori be complex
numbers allows  CP violation in the Standard Model. The
structure of the CKM matrix is discussed in detail in the
next subsection.
\item
 The most
important Feynman rules are given in figs.\ \ref{fig:2} and \ref{fig:3}.  
It should be noted that
the photonic and gluonic
vertices are vectorlike (V), the $W^{\pm}$ vertices are purely $V-A$
and the $Z^0$ vertices involve both $V-A$ and
$V+A$ structures.
\item
The vertices with 
fictitious Higgs $\phi^\pm$
have not been shown. Of particular interest are the vertices involving 
the top quark. Setting $m_d=0$ we have for instance the following
Feynman rules:
\be\label{HIGG1}
\bar t\phi^+ d: \qquad 
-\frac{g_2}{2\sqrt{2}}V_{td}(1-\gamma_5)\frac{\mt}{\mw}
\ee
\be\label{HIGG2}
\bar d\phi^- t: \qquad 
\frac{g_2}{2\sqrt{2}}V^*_{td}(1+\gamma_5)\frac{\mt}{\mw}
\ee
Due to the proportionality to $\mt$ these vertices play important
role in rare and CP violating decays and transitions.
\item
Finally the vertex involving the neutral Higgs $H^0$ and the fermions is
diagonal in flavour, again due to the {\rm GIM} mechanism. It is given by
\be
\bar f H^0 f: \qquad -i \frac{g_2}{2}\frac{m_f}{\mw}
\ee
The effects of $H^0$ on these lectures are  only at two-loops in electroweak
interactions.
\ei
\begin{figure}[thb]
\centerline{
\epsfysize=9in
\epsffile{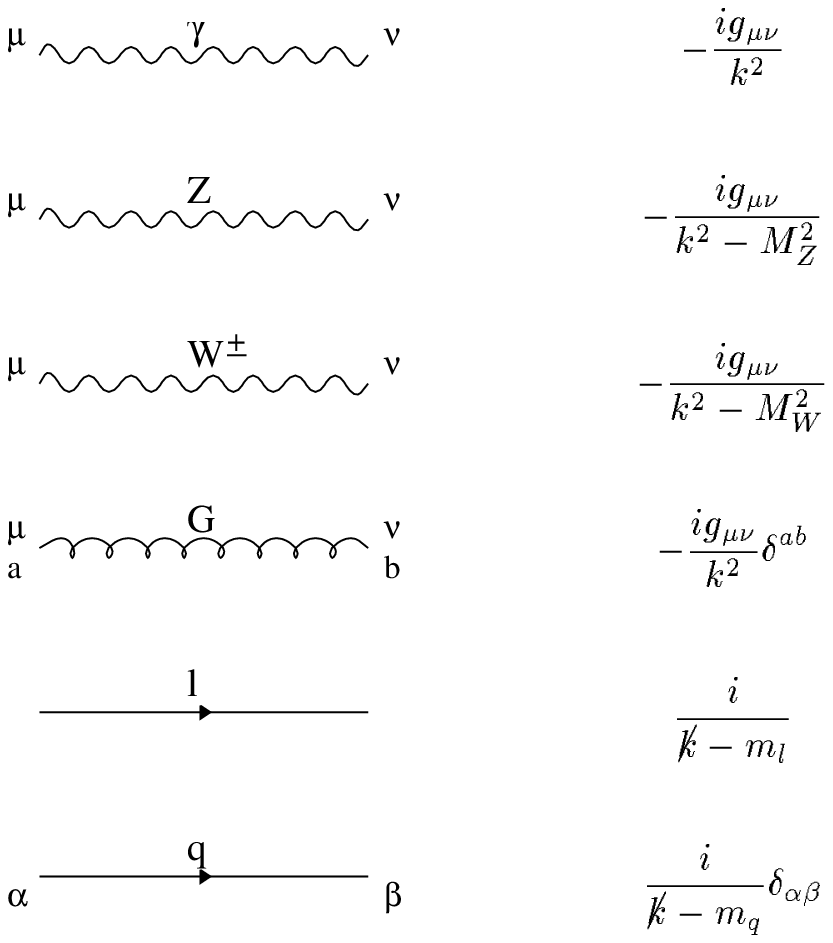}
}
\vspace{-5.25in}
\caption[]{
Feynman Rules (Propagators)
\label{fig:2}}
\end{figure}

\begin{figure}[thb]
\centerline{
\epsfysize=9.5in
\epsffile{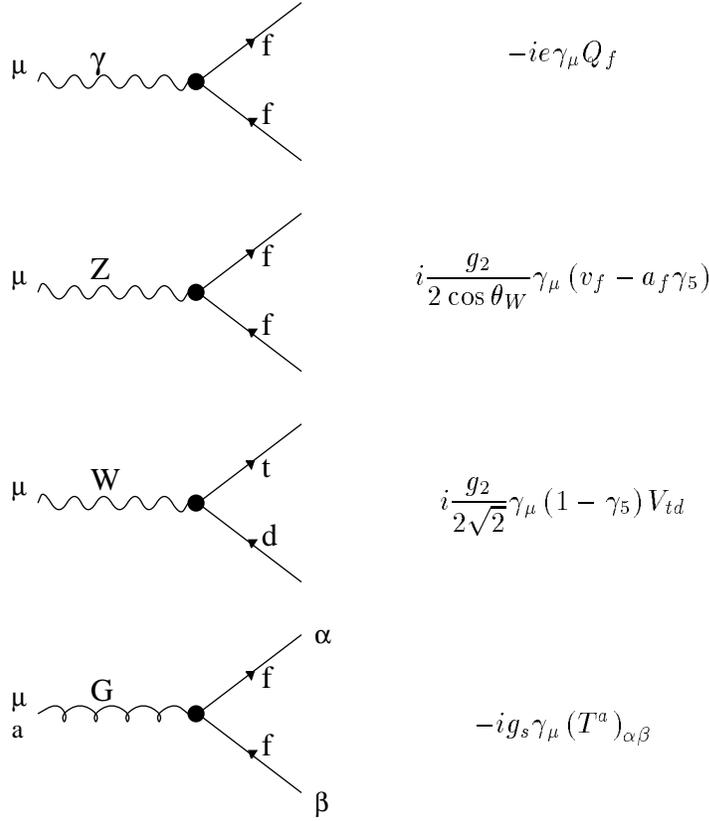}
}
\vspace{-4.95in}
\caption[]{
Feynman Rules (Vertices)
\label{fig:3}}
\end{figure}

\noindent
With the help of the elementary vertices of fig.\ \ref{fig:1},
the propagators and  Feynman rules
at hand, one can build physically interesting
processes and subsequently evaluate them. The simplest of such
processes, which forms the basis for subsequent considerations, is the
$W^{\pm}$ exchange between two fermion lines like the one shown 
in fig. \ref{L:1}a.
Neglecting the momentum of the W-propagator relative to $\mw$ and
multiplying the result by "i",
this process, generalized to arbitrary fermions in the Standard Model,
 gives the following tree level effective Hamiltonian
describing the charged weak interactions of quarks and leptons:
\begin{equation}\label{WE}
{\cal H}^{\rm tree}_{\rm eff} = {{G_{\rm F}}\over{\sqrt2}}\;{\cal J}^+_\mu
{\cal J}^{-\mu}
\end{equation}
with ${\cal J}^+_\mu$  given in (\ref{6}).

\subsection{CKM Matrix}
\subsubsection{General Remarks}
Let us next discuss the stucture of the
quark-mixing-matrix $\hat V_{CKM}$ defined by (\ref{2.67}) in more
detail. We know from the text books that this matrix can be
parametrized by
three angles and a single complex phase.
This phase leading to an
imaginary part of the CKM matrix is a necessary ingredient to describe
CP violation within the framework of the Standard Model.

Many parametrizations of the CKM
matrix have been proposed in the literature.  We will use
two parametrizations in these lectures: the standard parametrization 
\cite{CHAU} recommended by the particle data group  \cite{PDG}  
and the Wolfenstein parametrization \cite{WO}.

\subsubsection{Standard Parametrization}
            \label{sec:sewm:stdparam}
With
$c_{ij}=\cos\theta_{ij}$ and $s_{ij}=\sin\theta_{ij}$ 
($i,j=1,2,3$), the standard parametrization is
given by:
\begin{equation}\label{2.72}
\hat V_{CKM}=
\left(\begin{array}{ccc}
c_{12}c_{13}&s_{12}c_{13}&s_{13}e^{-i\delta}\\ -s_{12}c_{23}
-c_{12}s_{23}s_{13}e^{i\delta}&c_{12}c_{23}-s_{12}s_{23}s_{13}e^{i\delta}&
s_{23}c_{13}\\ s_{12}s_{23}-c_{12}c_{23}s_{13}e^{i\delta}&-s_{23}c_{12}
-s_{12}c_{23}s_{13}e^{i\delta}&c_{23}c_{13}
\end{array}\right)\,,
\end{equation}
where $\delta$ is the phase necessary for {\rm CP} violation.
$c_{ij}$ and
$s_{ij}$ can all be chosen to be positive
and  $\delta$ may vary in the
range $0\le\delta\le 2\pi$. However, the measurements
of CP violation in $K$ decays force $\delta$ to be in the range
 $0<\delta<\pi$. 

From phenomenological applications we know that 
$s_{13}$ and $s_{23}$ are small numbers: $\ord(10^{-3})$ and ${\cal
O}(10^{-2})$,
respectively. Consequently to an excellent accuracy $c_{13}=c_{23}=1$
and the four independent parameters are given as 
\begin{equation}\label{2.73}
s_{12}=| V_{us}|, \quad s_{13}=| V_{ub}|, \quad s_{23}=|
V_{cb}|, \quad \delta.
\end{equation}

The first three can be extracted from tree level decays mediated
by the transitions $s \to u$, $b \to u$ and $b \to c$ respectively.
The phase $\delta$ can be extracted from CP violating transitions or 
loop processes sensitive to $| V_{td}|$. The latter fact is based
on the observation that
 for $0\le\delta\le\pi$, as required by the analysis of CP violation
in the $K$ system,
there is a one--to--one correspondence between $\delta$ and $|V_{td}|$
given by
\begin{equation}\label{10}
| V_{td}|=\sqrt{a^2+b^2-2 a b \cos\delta},
\qquad
a=| V_{cd} V_{cb}|,
\qquad
b=| V_{ud} V_{ub}|\,.
\end{equation} 

The main  phenomenological advantages of (\ref{2.72}) over other
parametrizations proposed in the literature are basically these
two \cite{HALE} : 
\begin{itemize}
\item
$s_{12}$, $s_{13}$ and $s_{23}$ being related in a very simple way
to $| V_{us}|$, $| V_{ub}|$ and $|V_{cb}|$ respectively, can be
measured independently in three decays.
\item
The CP violating phase is always multiplied by the very small
$s_{13}$. This shows clearly the suppression of CP violation
independently of the actual size of $\delta$.
\end{itemize}

For numerical evaluations the use of the standard parametrization
is strongly recommended. However once the four parameters in
(\ref{2.73}) have been determined it is often useful to make
a change of basic parameters in order to see the structure of
the result more transparently. This brings us to the Wolfenstein
parametrization \cite{WO} and its generalization given in 
\cite{BLO}.

\subsubsection{Wolfenstein Parameterization }\label{Wolf-Par}
 The Wolfenstein parametrization 
is an approximate parametrization of the CKM matrix in which
each element is expanded as a power series in the small parameter
$\lambda=| V_{us}|=0.22$,
\begin{equation}\label{2.75} 
\hat V=
\left(\begin{array}{ccc}
1-{\lambda^2\over 2}&\lambda&A\lambda^3(\varrho-i\eta)\\ -\lambda&
1-{\lambda^2\over 2}&A\lambda^2\\ A\lambda^3(1-\varrho-i\eta)&-A\lambda^2&
1\end{array}\right)
+\ord(\lambda^4)\,,
\end{equation}
and the set (\ref{2.73}) is replaced by
\begin{equation}\label{2.76}
\lambda, \qquad A, \qquad \varrho, \qquad \eta \, .
\end{equation}

Because of the
smallness of $\lambda$ and the fact that for each element 
the expansion parameter is actually
$\lambda^2$, it is sufficient to keep only the first few terms
in this expansion. 

The Wolfenstein parametrization is certainly more transparent than
the standard parametrization. However, if one requires sufficient 
level of accuracy, the higher order terms in $\lambda$ have to
be included in phenomenological applications.
This can be done in many ways.
The
point is that since (\ref{2.75}) is only an approximation the {\em exact}
definiton of the parameters in (\ref{2.76}) is not unique by terms of the 
neglected order
${\cal O}(\lambda^4)$. 
This situation is familiar from any perturbative expansion, where
different definitions of expansion parameters (coupling constants) 
are possible.
This is also the reason why in different papers in the
literature different ${\cal O}(\lambda^4)$ terms in (\ref{2.75})
 can be found. They simply
correspond to different definitions of the parameters in (\ref{2.76}).
Since the physics does not depend on a particular definition, it
is useful to make a choice for which the transparency of the original
Wolfenstein parametrization is not lost. Here we present one
way of achieving this.

\subsubsection{Wolfenstein Parametrization beyond LO}
An efficient and systematic way of finding higher order terms in $\lambda$
is to go back to the standard parametrization (\ref{2.72}) and to
 {\it define} the parameters $(\lambda,A,\varrho,\eta)$ through 
\cite{BLO,schubert}
\begin{equation}\label{2.77} 
s_{12}=\lambda\,,
\qquad
s_{23}=A \lambda^2\,,
\qquad
s_{13} e^{-i\delta}=A \lambda^3 (\varrho-i \eta)
\end{equation}
to {\it  all orders} in $\lambda$. 
It follows  that
\begin{equation}\label{2.84} 
\varrho=\frac{s_{13}}{s_{12}s_{23}}\cos\delta,
\qquad
\eta=\frac{s_{13}}{s_{12}s_{23}}\sin\delta.
\end{equation}
(\ref{2.77}) and (\ref{2.84}) represent simply
the change of variables from (\ref{2.73}) to (\ref{2.76}).
Making this change of variables in the standard parametrization 
(\ref{2.72}) we find the CKM matrix as a function of 
$(\lambda,A,\varrho,\eta)$ which satisfies unitarity exactly!
Expanding next each element in powers of $\lambda$ we recover the
matrix in (\ref{2.75}) and in addition find explicit corrections of
$\ord(\lambda^4)$ and higher order terms:. 

\be
V_{ud}=1-\frac{1}{2}\lambda^2-\frac{1}{8}\lambda^4 +\ord(\lambda^6)
\ee
\be
V_{us}=\lambda+\ord(\lambda^7)
\ee
\be
V_{ub}=A \lambda^3 (\varrho-i \eta)
\ee
\be
V_{cd}=-\lambda+\frac{1}{2} A^2\lambda^5 [1-2 (\varrho+i \eta)]+
\ord(\lambda^7)
\ee
\be
V_{cs}= 1-\frac{1}{2}\lambda^2-\frac{1}{8}\lambda^4(1+4 A^2) +\ord(\lambda^6)
\ee
\be
V_{cb}=A\lambda^2+\ord(\lambda^8)
\ee
\be
V_{td}=A\lambda^3 \left[ 1-(\varrho+i \eta)(1-\frac{1}{2}\lambda^2)\right]
+\ord (\lambda^7)
\ee
\begin{equation}\label{2.83d}
 V_{ts}= -A\lambda^2+\frac{1}{2}A(1-2 \varrho)\lambda^4
-i\eta A \lambda^4 +\ord(\lambda^6)
\end{equation}
\be
V_{tb}=1-\frac{1}{2} A^2\lambda^4+\ord(\lambda^6)
\ee

We note that by definition
$V_{ub}$ remains unchanged and the
corrections to $V_{us}$ and $V_{cb}$ appear only at $\ord(\lambda^7)$ and
$\ord(\lambda^8)$, respectively.
Consequently to an 
 an excellent accuracy we have:
\begin{equation}\label{CKM1}
V_{us}=\lambda, \qquad V_{cb}=A\lambda^2,
\end{equation}
\begin{equation}\label{CKM2}
V_{ub}=A\lambda^3(\varrho-i\eta),
\qquad
V_{td}=A\lambda^3(1-\bar\varrho-i\bar\eta)
\end{equation}
with
\begin{equation}\label{2.88d}
\bar\varrho=\varrho (1-\frac{\lambda^2}{2}),
\qquad
\bar\eta=\eta (1-\frac{\lambda^2}{2}).
\end{equation}
The advantage of this generalization of the Wolfenstein parametrization
over other generalizations found in the literature is the absence of
relevant corrections to $V_{us}$, $V_{cb}$ and $V_{ub}$ and an elegant
change in $V_{td}$ which allows a simple generalization of the 
so-called unitarity triangle beyond LO.

Finally let us collect useful analytic expressions
for $\lambda_i=V_{id}V^*_{is}$ with $i=c,t$:
\begin{equation}\label{2.51}
 \IM\lambda_t= -\IM\lambda_c=\eta A^2\lambda^5=
\mid V_{ub}\mid \mid V_{cb} \mid \sin\delta 
\end{equation}
\begin{equation}\label{2.52}
 \RE\lambda_c=-\lambda (1-\frac{\lambda^2}{2})
\end{equation}
\begin{equation}\label{2.53}
 \RE\lambda_t= -(1-\frac{\lambda^2}{2}) A^2\lambda^5 (1-\bar\varrho) \,.
\end{equation}
Expressions (\ref{2.51}) and (\ref{2.52}) represent to an accuracy of
0.2\% the exact formulae obtained using (\ref{2.72}). The expression
(\ref{2.53}) deviates by at most 2\% from the exact formula in the
full range of parameters considered. For $\varrho$ close to zero
this deviation is below 1\%. A careful reader may note that a small
$\ord(\lambda^7)$ has been dropped in deriving (\ref{2.53}). This
has been done both for artistic reasons and in order to increase
the accuracy of this formula.
After inserting the expressions (\ref{2.51})--(\ref{2.53}) in the exact
formulae for quantities of interest, a further expansion in $\lambda$
should not be made. 
\subsubsection{Unitarity Triangle}
The unitarity of the CKM-matrix implies various relations between its
elements. In particular, we have
\begin{equation}\label{2.87h}
V_{ud}^{}V_{ub}^* + V_{cd}^{}V_{cb}^* + V_{td}^{}V_{tb}^* =0.
\end{equation}
Phenomenologically this relation is very interesting as it involves
simultaneously the elements $V_{ub}$, $V_{cb}$ and $V_{td}$ which are
under extensive discussion at present.

The relation (\ref{2.87h})  can be
represented as a ``unitarity'' triangle in the complex 
$(\bar\varrho,\bar\eta)$ plane. 
The invariance of (\ref{2.87h})  under any phase-transformations
implies that the  corresponding triangle
is rotated in the $(\bar\varrho,\bar\eta)$  plane under such transformations. 
Since the angles and the sides
(given by the moduli of the elements of the
mixing matrix)  in these triangles remain unchanged, they
 are phase convention independent and are  physical observables.
Consequently they can be measured directly in suitable experiments.  
The area of the unitarity triangle  is related to the measure of CP violation 
$J_{\rm CP}$ \cite{js}:
\begin{equation}
\mid J_{\rm CP} \mid = 2\cdot A_{\Delta},
\end{equation}
where $A_{\Delta}$ denotes the area of the unitarity triangle.

The construction of the unitarity triangle proceeds as follows:

\bi
\item
We note first that
\begin{equation}\label{2.88a}
V_{cd}^{}V_{cb}^*=-A\lambda^3+\ord(\lambda^7).
\end{equation}
Thus to an excellent accuracy $V_{cd}^{}V_{cb}^*$ is real with
$| V_{cd}^{}V_{cb}^*|=A\lambda^3$.
\item
Keeping $\ord(\lambda^5)$ corrections and rescaling all terms in
(\ref{2.87h})
by $A \lambda^3$ 
we find
\begin{equation}\label{2.88b}
 \frac{1}{A\lambda^3}V_{ud}^{}V_{ub}^*
=\bar\varrho+i\bar\eta,
\qquad
\qquad
 \frac{1}{A\lambda^3}V_{td}^{}V_{tb}^*
=1-(\bar\varrho+i\bar\eta)
\end{equation}
with $\bar\varrho$ and $\bar\eta$ defined in (\ref{2.88d}). 
\item
Thus we can represent (\ref{2.87h}) as the unitarity triangle 
in the complex $(\bar\varrho,\bar\eta)$ plane 
as shown in fig. \ref{fig:utriangle}.
\ei

\begin{figure}[hbt]
\vspace{0.10in}
\centerline{
\epsfysize=2.1in
\epsffile{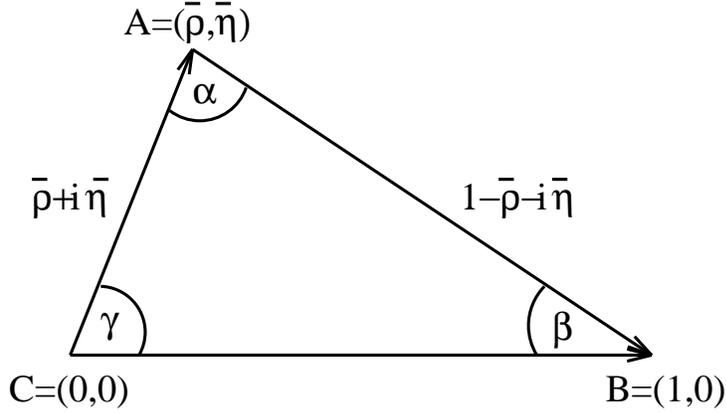}
}
\vspace{0.08in}
\caption{Unitarity Triangle.}\label{fig:utriangle}
\end{figure}

Let us collect useful formulae related to this triangle:
\bi
\item
Using simple trigonometry one can express $\sin(2\phi_i$), $\phi_i=
\alpha, \beta, \gamma$, in terms of $(\bar\varrho,\bar\eta)$ as follows:
\begin{equation}\label{2.89}
\sin(2\alpha)=\frac{2\bar\eta(\bar\eta^2+\bar\varrho^2-\bar\varrho)}
  {(\bar\varrho^2+\bar\eta^2)((1-\bar\varrho)^2
  +\bar\eta^2)}  
\end{equation}
\begin{equation}\label{2.90}
\sin(2\beta)=\frac{2\bar\eta(1-\bar\varrho)}{(1-\bar\varrho)^2 + \bar\eta^2}
\end{equation}
 \begin{equation}\label{2.91}
\sin(2\gamma)=\frac{2\bar\varrho\bar\eta}{\bar\varrho^2+\bar\eta^2}=
\frac{2\varrho\eta}{\varrho^2+\eta^2}.
\end{equation}
\item
The lengths $CA$ and $BA$ in the
rescaled triangle  to be denoted by $R_b$ and $R_t$,
respectively, are given by
\begin{equation}\label{2.94}
R_b \equiv \frac{| V_{ud}^{}V^*_{ub}|}{| V_{cd}^{}V^*_{cb}|}
= \sqrt{\bar\varrho^2 +\bar\eta^2}
= (1-\frac{\lambda^2}{2})\frac{1}{\lambda}
\left| \frac{V_{ub}}{V_{cb}} \right|
\end{equation}
\begin{equation}\label{2.95}
R_t \equiv \frac{| V_{td}^{}V^*_{tb}|}{| V_{cd}^{}V^*_{cb}|} =
 \sqrt{(1-\bar\varrho)^2 +\bar\eta^2}
=\frac{1}{\lambda} \left| \frac{V_{td}}{V_{cb}} \right|.
\end{equation}
\item
The angles $\beta$ and $\gamma$ of the unitarity triangle are related
directly to the complex phases of the CKM-elements $V_{td}$ and
$V_{ub}$, respectively, through
\beq\label{e417}
V_{td}=|V_{td}|e^{-i\beta},\quad V_{ub}=|V_{ub}|e^{-i\gamma}.
\eeq
\item
The angle $\alpha$ can be obtained through the relation
\beq\label{e419}
\alpha+\beta+\gamma=180^\circ
\eeq
expressing the unitarity of the CKM-matrix.
\ei

The triangle depicted in fig. \ref{fig:utriangle} together with $|V_{us}|$ 
and $\vcb$ gives a full description of the CKM matrix. 
Looking at the expressions for $R_b$ and $R_t$, we observe that within
the Standard Model the measurements of four CP
{\it conserving } decays sensitive to $\mid V_{us}\mid$, $\mid V_{ub}\mid$,   
$\mid V_{cb}\mid $ and $\mid V_{td}\mid$ can tell us whether CP violation
($\eta \not= 0$) is predicted in the Standard Model. 
This is a very remarkable property of
the Kobayashi-Maskawa picture of CP violation: quark mixing and CP violation
are closely related to each other. 

\subsubsection{Final Comments}
What do we know about the CKM matrix and the unitarity triangle on the
basis of {\it tree level} decays? A detailed answer to this question
can be found in the reports of
the Particle Data Group \cite{PDG} as well as other reviews.
Here I would like to quote only a few numbers without going into
details how they have been obtained. They are
\begin{equation}\label{vcb}
|V_{us}| = \lambda =  0.2205 \pm 0.0018\,
\quad\quad
\vcb=0.040\pm0.003,
\end{equation}
\begin{equation}\label{v13}
\frac{|V_{ub}|}{\vcb}=0.08\pm0.02, \quad\quad
|V_{ub}|=(3.2\pm0.8)\cdot 10^{-3}.
\end{equation}
The value for $|V_{us}|$ follows from 
$K^+\to \pi^0 e^+ \nu_e$, $K_{\rm L}^0 \to\pi^-e^+ \nu_e$ and
semileptonic hyperon decays. The chiral perturbation theory
\cite{LER1,DHK} plays an important role in these determinations.
$\vcb$ follows from exclusive and inclusive semileptonic B decays
governed by the transition $b \to c$. 
Here the recent improved data \cite{Gibbons} combined with
 HQET in the case of exclusive decays and HQE in the case of
inclusive decays played important role \cite{SUV,Neubert,Braun,CZMI}.
$|V_{ub}|$ follows from from exclusive and inclusive semileptonic B decays
governed by the transition $b \to u$
\cite{Gibbons,CLEOU}.

Setting $\lambda=0.22$, scanning $\vcb$ and $|V_{ub}|$ in
the ranges (\ref{vcb})  and $\cos\delta$ in
the range $-1\leq \cos\delta\leq 1$, we find \cite{BF97}:
\begin{equation}\label{uni1}
4.5\cdot 10^{-3}\leq \vtd \leq 13.7\cdot 10^{-3}\,,
\qquad
0.0353\leq \vts \leq 0.0429
\end{equation}
and
\begin{equation}\label{uni2}
0.9991\leq |V_{tb}| \leq 0.9993,
\qquad
0.9736\leq |V_{cs}| \leq 0.9750.
\end{equation}

From (\ref{uni1}) we observe
that the unitarity of the CKM matrix requires approximate equality of 
$\vts$ and $\vcb$:
$0.954\leq |V_{ts}|/\vcb \leq 0.997$
which is evident if one compares (\ref{CKM1}) with (\ref{2.83d}).
Moreover $|V_{tb}|$ is predicted to be very close to unity.
The experimental value from top-quark decays obtained by CDF is
$|V_{tb}|=0.99\pm 0.15$.

Let us then see what these results imply for the unitarity triangle
of fig. \ref{fig:utriangle}. To this end it is sufficient to insert
the first result in (\ref{v13}) into (\ref{2.94}) to find
\be
 R_b=0.36\pm 0.09
\ee
This tells us only that the apex $A$ of the unitarity triangle lies
in the band shown in fig. \ \ref{L:2}. In order to answer the question where
the apex $A$ lies on this "unitarity clock'' we have to look at different
decays. Most promising in this respect are the so-called "loop induced''
decays and transitions which are the subject of several sections in these
lectures
and CP asymmetries in B-decays which will be briefly discussed in Section 14.
\begin{figure}[hbt]
\vspace{0.001in}
\centerline{
\epsfysize=4in
\rotate[r]{
\epsffile{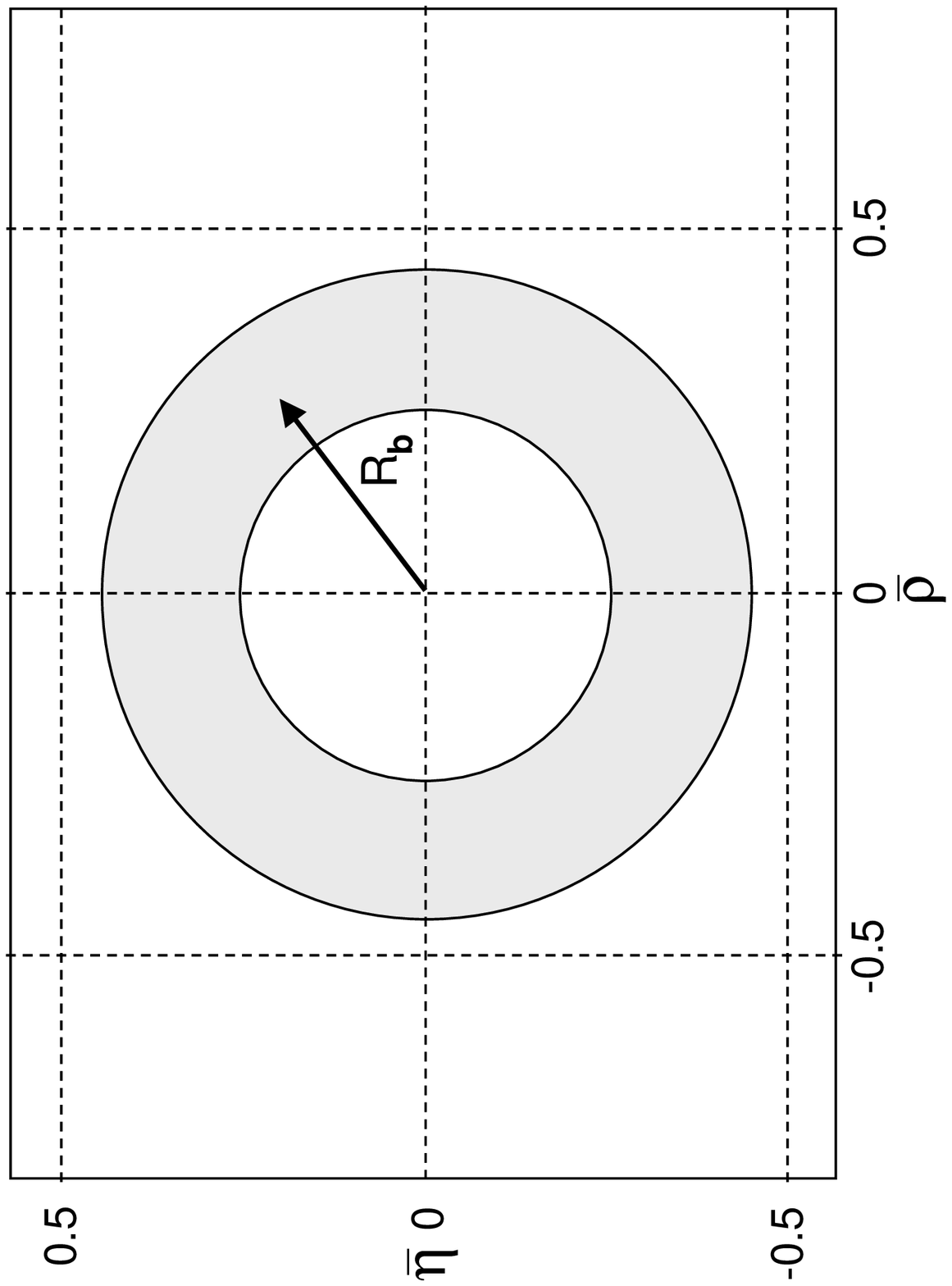}
}}
\vspace{0.005in}
\caption[]{``Unitarity Clock".
\label{L:2}}
\end{figure}
These two different routes for explorations of the CKM matrix
and of the related unitarity triangle may answer the important question, 
whether
the Kobayashi-Maskawa picture
of CP violation is correct and more generally whether the Standard
Model offers a correct description of weak decays of hadrons. Indeed,
in order
to answer these important questions it is essential to calculate as
many branching ratios as possible, measure them experimentally and
check if they all can be described by the same set of the parameters
$(\lambda,A,\varrho,\eta)$. In the language of the unitarity triangle
this means that the various curves in the $(\bar\varrho,\bar\eta)$ plane
extracted from different decays should cross each other at a single point
as shown in fig.~\ref{fig:2011}.
Moreover the angles $(\alpha,\beta,\gamma)$ in the
resulting triangle should agree with those extracted one day from
CP-asymmetries in B-decays. For artistic reasons the value of
$\bar\eta$ in fig.~\ref{fig:2011}
has been chosen to be higher than the fitted central value
$\bar\eta\approx 0.35.$
\begin{figure}[hbt]
\vspace{0.10in}
\centerline{
\epsfysize=4.3in
\rotate[r]{
\epsffile{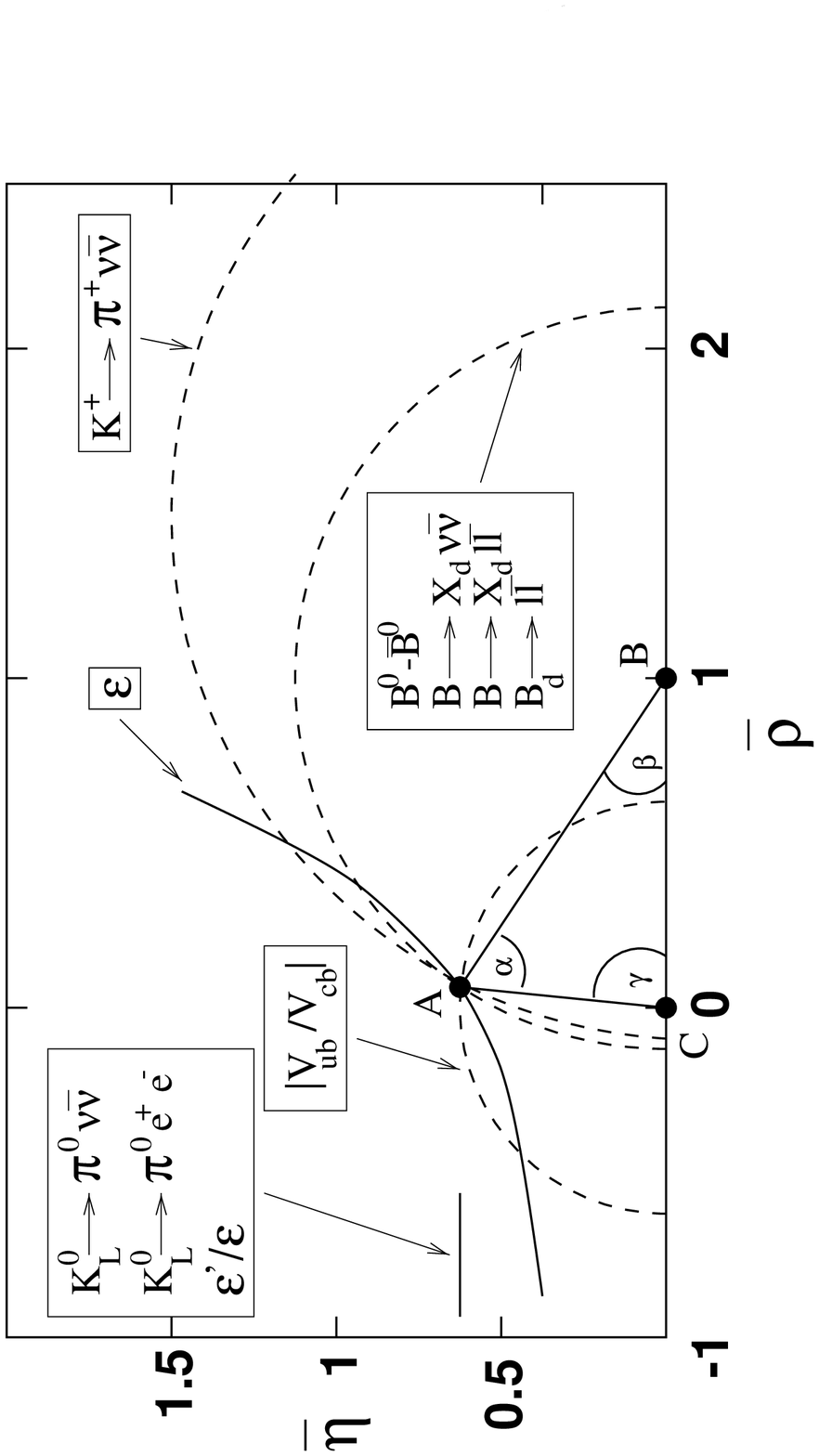}
}}
\vspace{0.08in}
\caption[]{
The ideal Unitarity Triangle. 
\label{fig:2011}}
\end{figure}

Since the CKM matrix is only a parametrization of quark mixing and 
of CP violation and does not offer the explanation of these two
very important phenomena, many physicists hope that a new physics
while providing a dynamical origin of quark mixing and CP violation will
also change the picture given in fig. \ref{fig:2011}. 
That is, the different curves
based on the Standard Model expressions, will not cross each other 
at a single point
and the angles $(\alpha,\beta,\gamma)$ 
extracted one day from
CP-asymmetries in B-decays will disagree with the ones determined from
rare K and B decays.

Clearly the plot in fig. \ref{fig:2011} is highly idealized because in order
to extract such nice curves from various decays one needs perfect
experiments and perfect theory. 
One of the goals of these lectures is to describe the present status
of the theory of weak decays and to identify those decays
for which at least the theory is under control. For such decays,
if they can be measured with a sufficient precision, the curves
in fig. \ref{fig:2011} are  not unrealistic.

The formal discussion of the theory of weak decays is however a real
steep climb and it is advisable to do first a long but gentle hike
by discussing loop induced decays in general terms.
They are known under the name of
{\it Flavour Changing Neutral Current} (FCNC) processes.

\section{FCNC Processes}
\setcounter{equation}{0}
\subsection{General Remarks}
The flavour diagonal structure of the basic vertices involving $\gamma$,
$Z$ and $G$ in fig. \ref{fig:3} 
forbidds the appearance of FCNC processes at the
tree level. With the help of the flavour-changing $W^\pm$-vertex one
can, however, construct one-loop and higher order diagrams which mediate
FCNC processes. The fact that these processes take place only as loop
effects makes them particularly useful for testing the quantum structure
of the theory and in the search of the physics beyond the Standard Model.
At the one--loop level
they can be described by a set
of basic triple and quartic effective vertices. In the literature they
appear under the names of penguin and box diagrams, respectively.
\subsection{Effective Vertices}
\subsubsection{Penguin vertices}
\noindent
These vertices involve only quarks and can be depicted as in 
fig. \ref{fig:4},
where $i$ and $j$ have the same charge but different flavour and $k$
denotes the internal quark whose charge is different from that of $i$
and $j$. 
These effective vertices 
can be calculated by using the elementary vertices and 
propagators of figs.~\ref{fig:2} and \ref{fig:3}.
Important examples are given in fig.~\ref{fig:5}.
The diagrams with fictitious Higgs exchanges in place of $W^\pm$ have 
not been
shown. Strictly
speaking, also self--energy corrections on external lines have to be
included to make the effective vertices finite.
\subsubsection{Box vertices}
These vertices involve in general both quarks and leptons and can be
depicted as in fig. \ref{fig:6},
where again $i,j,m,n$ stand for external quarks or leptons and $k$ and
$l$ denote the internal quarks and leptons. In the vertex (a) the
flavour violation takes place on both sides (left and right) of the
box, whereas in (b) the right--hand side is flavour conserving. These
effective quartic vertices can also be calculated using the elementary
vertices and propagators of figs. \ref{fig:2} and \ref{fig:3}. We have
for instance the vertices in fig.~\ref{fig:7}
which contribute to $B_d^0-\bar B_d^0$ mixing and $K^+\to\pi^+\nu\bar\nu$, 
respectively.
The fictitious Higgs exchanges have not been shown.
Other interesting examples will be discussed in the course of these
lectures. \par
\begin{figure}[hbt]
\centerline{
\epsfysize=1.6in
\epsffile{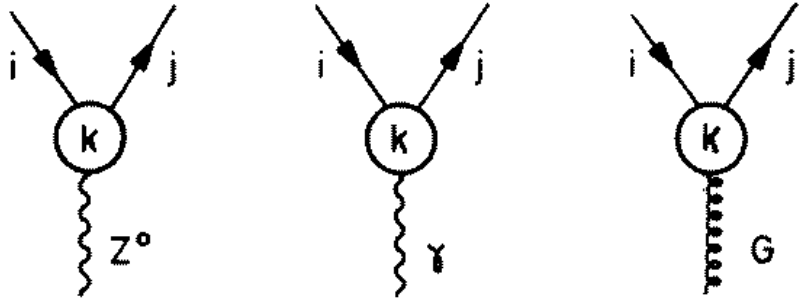}
}
\vspace{0.08in}
\caption[]{
Penguin vertices
\label{fig:4}}
\end{figure}

\begin{figure}[htb]
\centerline{
\epsfysize=3.2in
\epsffile{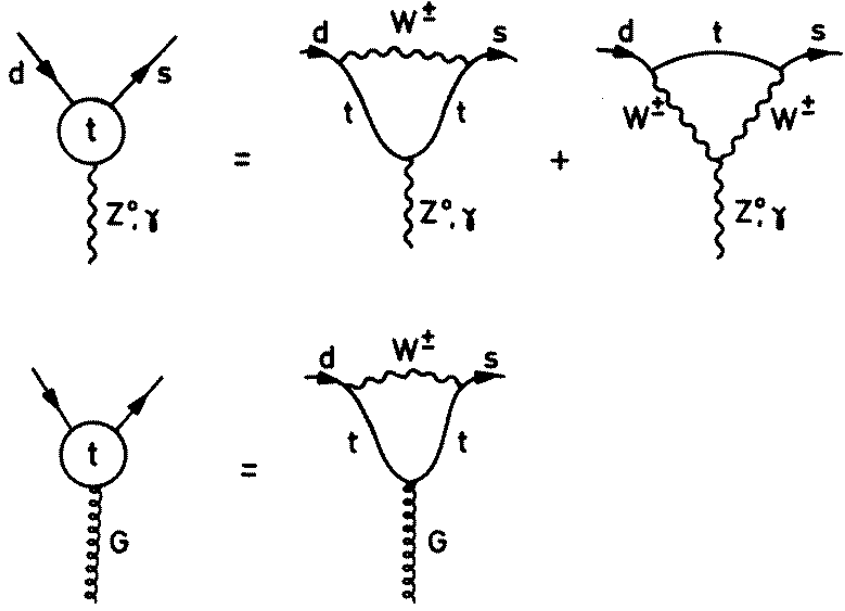}
}
\caption[]{
Penguin vertices resolved in terms of basic vertices
\label{fig:5}}
\end{figure}

\subsubsection{Effective Feynman Rules}
With the help of the elementary vertices and propagators 
shown in figs. \ref{fig:2} and \ref{fig:3}, 
one can  derive
``Feynman rules'' for the effective vertices discussed above by
calculating simply the diagrams on the r.h.s. of the equations in
figs. \ref{fig:5} and \ref{fig:7}. In fig. \ref{fig:5} 
the $Z^0$, $\gamma$ and $gluon$ are
off-shell. In the case of inclusive decays $B \to X\gamma$ and
$B \to X G$ we need also corresponding vertices with on-shell
photons and gluons. For these two cases it is essential to
keep the mass of the external b-quark as otherwise the corresponding 
vertices would vanish.  

The rules for effective vertices are given  
in the 
't Hooft--Feynman gauge for the $W^\pm$ propagator as follows:
\begin{equation}\label{FR}
  {\rm Box} (\Delta S = 2)
~=~ \lambda^2_i {{G^2_{\rm F}}\over{16\pi^2}} \mw^2S_0(x_i) 
   (\bar s d)_{V-A} (\bar s d)_{V-A} 
\end{equation}
\begin{equation}\label{BT12}
 {\rm Box}(T_3= 1/2)~ =~ \lambda_i {{G_{\rm F}}\over{\sqrt 2}}
   {{\alpha}\over{2\pi \sin^2\Theta_{\rm W}}} \lbrack -4 B_0(x_i)\rbrack 
   (\bar s d)_{V-A} (\bar\nu\nu)_{V-A}
\end{equation}
\begin{equation}\label{BTM12}
{\rm Box}(T_3= -1/2)~ =~ \lambda_i {{G_{\rm F}}\over{\sqrt 2}}
   {{\alpha}\over{2\pi \sin^2\Theta_{\rm W}}} B_0(x_i) (\bar s d)_{V-A} 
   (\bar\mu\mu)_{V-A}
\end{equation}
\begin{equation}\label{ZRULE}
 \bar s Z d~ =~i \lambda_i {{G_{\rm F}}\over{\sqrt 2}} {{e}\over{2\pi^2}} 
M^2_Z
   {{\cos\Theta_{\rm W}}\over{\sin\Theta_{\rm W}}} C_0(x_i) \bar s \gamma_\mu 
   (1-\gamma_5)d
\end{equation}
\begin{equation}
 \bar s\gamma d~ =~- i\lambda_i {{G_{\rm F}}\over{\sqrt 2}} {{e}\over{8\pi^2}}
   D_0(x_i) \bar s (q^2\gamma_\mu - q_\mu \not\!q)(1-\gamma_5)d 
\end{equation}
\begin{equation}
 \bar s G^a d~ =~ -i\lambda_i{{G_{\rm F}}\over{\sqrt 
2}} {{g_s}\over{8\pi^2}}
   E_0(x_i) \bar s_{\alpha}(q^2\gamma_\mu - q_\mu \not\!q)
(1-\gamma_5)T^a_{\alpha\beta}d_\beta 
\end{equation}
\begin{equation}\label{MGP}
 \bar s \gamma' b~ =~i\bar\lambda_i {{G_{\rm F}}\over{\sqrt 2}} {{e}\over
   {8\pi^2}} D'_0(x_i) \bar s \lbrack i\sigma_{\mu\lambda} q^\lambda
   \lbrack m_b (1+\gamma_5) \rbrack\rbrack b
\end{equation}
\begin{equation}\label{FRF}
 \bar s G'^a b~ =~ 
i\bar\lambda_i{{G_{\rm F}}\over{\sqrt 2}}{{g_s}\over{8\pi^2}}
   E'_0(x_i)\bar s_{\alpha} \lbrack i\sigma_{\mu\lambda} q^\lambda
   \lbrack m_b (1+\gamma_5) \rbrack\rbrack T^a_{\alpha\beta} b_\beta \,,
\end{equation}
where 
\be
\lambda_i=V^*_{is}V_{id}\quad\quad \bar\lambda_i=V^*_{is}V_{ib}
\ee
In these rules
$q_\mu$ is the {\it outgoing} gluon or photon momentum
and $T_3$ indicates whether $\nu\bar\nu$ or $l^+l^-$
leaves the box diagram. The first rule involves quarks only.
The last two rules involve on-shell photon
and gluon.
We have set $m_s=0$ in these rules.

These rules for effective vertices
together with the rules of figs. \ref{fig:2} and \ref{fig:3}  allow 
the calculation of
the effective Hamiltonians for FCNC processes, albeit without the inclusion
of QCD corrections. The way these rules should be used requires  some care:
\bi
\item
 The penguin vertices should be used in the same manner as the
elementary vertices of fig. \ref{fig:3} which follow from $i {\cal L}$. 
Once a
mathematical expression corresponding to a given diagram has been found,
the contribution of this diagram to the relevant effective Hamiltonian
is obtained by multiplying this mathematical expression by ``i". 
\item
Our conventions for the box vertices are such that they directly 
give the contributions to the effective Hamiltonians. 
\ei
We will give an
example below by calculating the internal top contributions to 
$K^+\to \pi^+\nu\bar\nu$.
First, however, let us make a few
general remarks emphasizing the new features of these effective
vertices as compared to the ones of fig. \ref{fig:1}.
\begin{figure}[hbt]
\centerline{
\epsfysize=1.6in
\epsffile{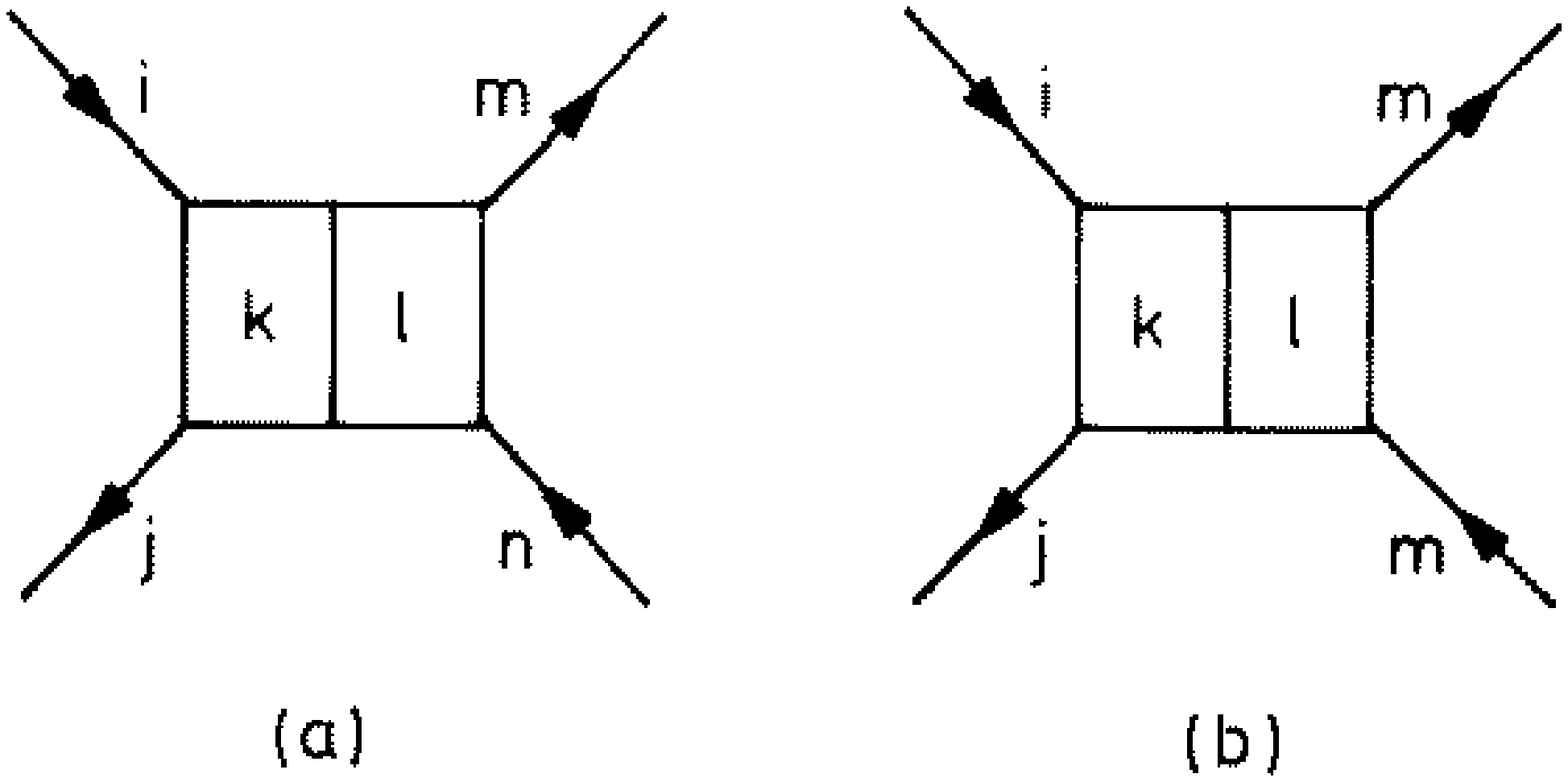}
}
\caption[]{
Box vertices 
\label{fig:6}}
\end{figure}   

\begin{figure}[hbt]
\centerline{
\epsfysize=3.2in
\epsffile{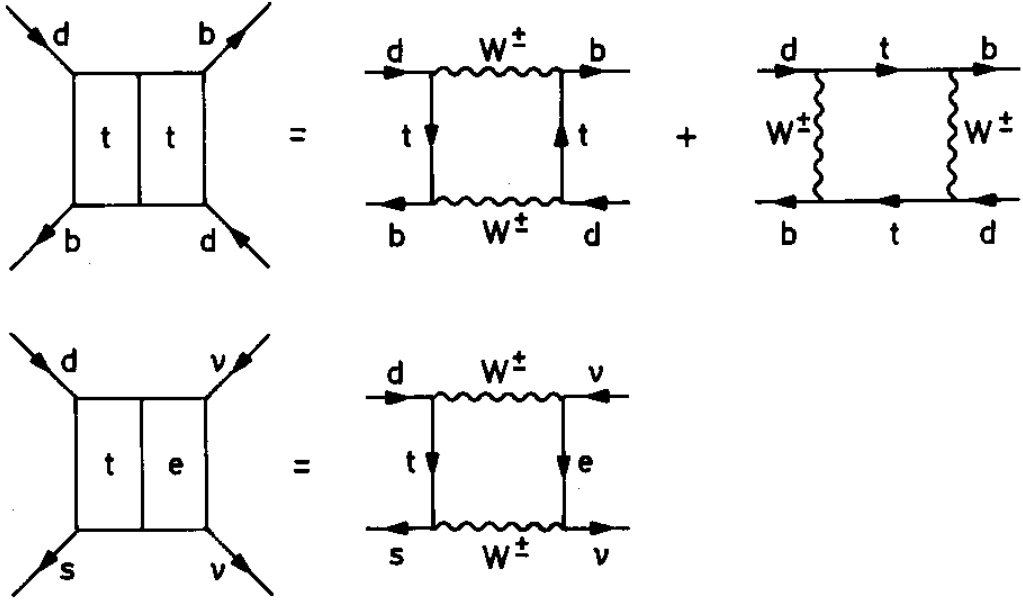}
}
\caption[]{
Box vertices resolved in terms of elementary vertices
\label{fig:7}}
\end{figure}

\begin{itemize}
\item
They are higher order in the gauge couplings and consequently
suppressed relative to elementary transitions. This is consistent
with experimental findings which show very strong suppression
of FCNC transitions relative to tree level processes.
\item
Because of the internal $W^{\pm}$ exchanges all penguin
vertices in fig.\ \ref{fig:5} are purely $V-A$, 
i.e.\ the effective vertices
involving $\gamma$ and $G$ are parity violating as opposed to their
elementary interactions in fig.\ 1! Also the structure of the $Z^0$ coupling
changes since now only $V-A$ couplings are involved. The box vertices
are of the $(V-A)\otimes(V-A)$ type.
\item
The effective vertices depend on the masses of internal
quarks or leptons and consequently are calculable functions of
\begin{equation}
x_i = {{m_i^2}\over{M_W^2}},\qquad i=u,c,t.
\end{equation}
A set of basic universal functions can be found. These functions
govern the physics of all FCNC processes. They are given below.
\item
The effective vertices depend on elements of the CKM
matrix and this dependence can be found directly from the diagrams of
figs. \ref{fig:5} and \ref{fig:7}.
\item
The dependences of a given  vertex on 
 the CKM factors and the masses of internal fermions 
govern  the strength of the vertex in question.
\item
Another new feature of the effective vertices 
 as compared with the elementary vertices is their dependence on
the gauge used for the $W^{\pm}$ propagator. We will return to this
point below.
\end{itemize}

\subsubsection{Basic Functions}
The basic functions present in (\ref{FR})-(\ref{FRF}) 
were calculated by various authors, in
particular by Inami and Lim \cite{IL}.  
They are given explicitly as follows:
\begin{equation}\label{BF}
B_0(x_t)={1\over
4}\left[{x_t\over{1-x_t}}+{{x_t\ln x_t}\over{(x_t-1)^2}}\right]
\end{equation}
\begin{equation}\label{C0xt}
C_0(x_t)={x_t\over 8}\left[{{x_t-6}\over{x_t-1}}+{{3x_t+2}
\over{(x_t-1)^2}}\;\ln x_t\right] 
\end{equation}
\begin{equation}
D_0(x_t)=-{4\over9}\ln x_t+{{-19x_t^3+25x_t^2}\over{36(x_t-1)^3}}
+{{x_t^2(5x_t^2-2x_t-6)} \over{18(x_t-1)^4}}\ln x_t
\end{equation}
\begin{equation}\label{E0}
E_0(x_t)=-{2\over 3}\ln x_t+{{x_t^2(15-16x_t+4x_t^2)}\over{6(1-x_t)^4}}
\ln x_t+{{x_t(18-11x_t-x_t^2)} \over{12(1-x_t)^3}}
\end{equation}
\begin{equation}
D'_0(x_t)= -{{(8x_t^3 + 5x_t^2 - 7x_t)}\over{12(1-x_t)^3}}+ 
          {{x_t^2(2-3x_t)}\over{2(1-x_t)^4}}\ln x_t
\end{equation}
\begin{equation}
E'_0(x_t)=-{{x_t(x_t^2-5x_t-2)}\over{4(1-x_t)^3}} + {3\over2}
{{x_t^2}\over{(1 - x_t)^4}} \ln x_t
\end{equation}
\begin{equation}\label{S0}
S_0(x_t)=\frac{4x_t-11x^2_t+x^3_t}{4(1-x_t)^2}-
 \frac{3x^3_t \ln x_t}{2(1-x_t)^3}
\end{equation}
\begin{equation}\label{BFF1}
S_0(x_c)=x_c
\end{equation}
\begin{equation}\label{BFF}
S_0(x_c, x_t)=x_c\left[\ln\frac{x_t}{x_c}-\frac{3x_t}{4(1-x_t)}-
 \frac{3 x^2_t\ln x_t}{4(1-x_t)^2}\right].
\end{equation}

We would like to make a few comments:
\bi
\item
In the last two expressions we have kept only linear terms in $x_c\ll 1$, 
but of course all orders in $x_t$. The last function generalizes $S_0(x_t)$ 
in (\ref{S0}) to include box diagrams with simultaneous top-quark and
charm-quark exchanges.
\item
The subscript ``$0$'' indicates that 
these functions
do not include QCD corrections to the relevant penguin and box diagrams.
These corrections will be  discussed in detail in subsequent sections. 
\item
In writing the expressions in (\ref{BF})-(\ref{BFF}) we have 
omitted $x_t$--independent 
terms which
do not contribute to decays due to the GIM mechanism. We will discuss this
issue in more detail below.
 Moreover 
\begin{equation}
S_0(x_t)\equiv
F(x_t,x_t) + F(x_u,x_u) - 2 F(x_t,x_u)
\end{equation}
 and 
\begin{equation}
S_0(x_i,x_j)=F(x_i,x_j) + F(x_u,x_u)
- F(x_i,x_u) - F(x_j,x_u),
\end{equation}
where $F(x_i,x_j)$ is the true function  corresponding to a given
 box diagram with $i$ and $j$ quark exchanges. These particular
combinations can be found by drawing all possible box diagrams
(also those with u-quark exchanges), setting $m_u=0$ and using 
unitarity of the
CKM-matrix which implies in particular the relation: 
\begin{equation}
\lambda_u + \lambda_c + \lambda_t =0.
\end{equation}
In this way the effective Hamiltonians for FCNC
 transitions  can be directly obtained 
 by  summing only over $t$ and $c$ quarks.
\item
The expressions
given for  $B_0(x_t)$, $C_0(x_t)$ and $D_0(x_t)$
correspond to the 't Hooft--Feynman gauge ($\xi=1$).
In an arbitrary $R_\xi$ gauge they look differently.
In phenomenological applications it is useful therefore to work
instead with the following gauge independent
combinations \cite{PBE0}: 
\begin{equation}
C_0(x_t,\xi)-4B_0(x_t,\xi,1/2)=C_0(x_t)-4B_0(x_t)=X_0(x_t)
\end{equation}
\begin{equation}
C_0(x_t,\xi)-B_0(x_t,\xi,-1/2)=C_0(x_t)-B_0(x_t)=Y_0(x_t)
\end{equation}
\begin{equation}
C_0(x_t,\xi)+{1\over 4}D_0(x_t,\xi)=C_0(x_t)+{1\over
                                  4}D_0(x_t)=Z_0(x_t).
\end{equation}
\item
 $ X_0(x_t) $ and $ Y_0(x_t) $ are
linear combinations of the $V-A$ components of $ Z^0$--penguin and
box--diagrams with final quarks or leptons having weak isospin $ T_3 $ 
equal to 1/2 and -- 1/2, respectively. 
\item
$ Z_0(x_t) $ is a linear
combination of the vector component of the $ Z^0$--penguin and the
$\gamma$--penguin.
\item
These new functions  are given explicitly as follows:
\begin{equation}\label{X0}
X_0(x_t)={{x_t}\over{8}}\;\left[{{x_t+2}\over{x_t-1}} 
+ {{3 x_t-6}\over{(x_t -1)^2}}\; \ln x_t\right] 
\end{equation}
\begin{equation}\label{Y0}
Y_0(x_t)={{x_t}\over8}\; \left[{{x_t -4}\over{x_t-1}} 
+ {{3 x_t}\over{(x_t -1)^2}} \ln x_t\right]
\end{equation}
\bea\label{Z0}
Z_0(x_t)&=&-{1\over9}\ln x_t + 
{{18x_t^4-163x_t^3 + 259x_t^2-108x_t}\over {144 (x_t-1)^3}}+
\nonumber\\ 
& &+{{32x_t^4-38x_t^3-15x_t^2+18x_t}\over{72(x_t-1)^4}}\ln x_t.
\eea
\ei

Thus 
the set of gauge independet basic functions which govern the
FCNC processes is given by: 
\begin{equation}\label{SXYZ}
S_0(x_t), \quad
X_0(x_t), \quad
Y_0(x_t), \quad
Z_0(x_t), \quad
E_0(x_t), \quad
D_0^{'}(x_t),\quad
E_0^{'}(x_t).
\ee
Finally, we give approximate 
 formulae for the basic functions:
\begin{equation}\label{PBE1}
 S_0(x_t)=0.784~x_t^{0.76},~~~~X_0(x_t)=0.660~x_t^{0.575},  
\end{equation}
\begin{equation}\label{PBE2}
 Y_0(x_t)=0.315~x_t^{0.78},\quad Z_0(x_t)=0.175~x_t^{0.93},  \quad
   E_0(x_t)=0.564~x_t^{-0.51}, 
\end{equation}
\begin{equation}
 D'_0(x_t)=0.244~x_t^{0.30},~~~~~~~~~~E'_0(x_t)=0.145~x_t^{0.19}.
\end{equation}
In the range $150\gev \le \mt \le 200\gev$ 
these approximations reproduce the
exact expressions to an accuracy better than 1\%. We have then 
\begin{equation}
 S_0(x_t)=2.46~\left(\frac{\mt}{170\gev}\right)^{1.52},
\ee
\be
X_0(x_t)=1.57~\left(\frac{\mt}{170\gev}\right)^{1.15},
\quad\quad
Y_0(x_t)=1.02~\left(\frac{\mt}{170\gev}\right)^{1.56},
\end{equation}
\begin{equation}
 Z_0(x_t)=0.71~\left(\frac{\mt}{170\gev}\right)^{1.86},\quad\quad
   E_0(x_t)= 0.26~\left(\frac{\mt}{170\gev}\right)^{-1.02},
\end{equation}
\begin{equation}
 D'_0(x_t)=0.38~\left(\frac{\mt}{170\gev}\right)^{0.60}, \quad\quad 
E'_0(x_t)=0.19~\left(\frac{\mt}{170\gev}\right)^{0.38}.
\end{equation}
These formulae will allow us to exhibit elegantly the $\mt$ dependence
of various branching ratios in the phenomenological sections of
these lectures.
\subsubsection{Explicit Calculation of the Box Diagrams}
Let us derive the rules (\ref{BT12}) and (\ref{BTM12}). 
Beginning with (\ref{BT12})  we consider the relevant
box diagram in fig.~\ref{fig:7}. 
Setting $m_\nu=m_e=0$ the contributions
of the fictitious Higgs exchanges $\phi^\pm$ are also set to zero
and we are left only with the $W^\pm$ exchanges. Concentrating
first on the internal top-quark contribution and using the
Feynman rules of fig. \ref{fig:2} and \ref{fig:3} we have, after
simple manipulations of Dirac matrices,
the following expression for the diagram in fig.~\ref{fig:7}
\be\label{Box1} 
{\cal D}(\nu\bar\nu)
=\left(\frac{g_2}{2\sqrt{2}}\right)^4\lambda_t T_{\sigma\tau}
R^{\sigma\tau}
\ee
where
\be\label{Box2}
T_{\sigma\tau}=-4 \bar s \gamma_\mu\gamma_\sigma\gamma_\nu(1-\gamma_5)d
\otimes \bar\nu\gamma^\mu\gamma_\tau\gamma^\nu(1-\gamma_5)\nu
\ee
and
\be\label{Box3}
R^{\sigma\tau}=\int \frac{d^4 k}{(2\pi)^4}
\frac{k^\sigma k^\tau}
{\lbrack k^2-\mt^2\rbrack k^2 \lbrack k^2-\mw^2\rbrack^2}.
\ee
Note that in view of very massive internal propagators we can set
all external momenta to zero.

The integral $R^{\sigma\tau}$ can be easily evaluated by means of
the standard methods. As the box diagram is finite we do not have
to introduce any regulators and we find
\be\label{Box4}
R^{\sigma\tau}=\frac{g^{\sigma\tau}}{4}\int \frac{d^4 k}{(2\pi)^4}
\frac{1}
{\lbrack k^2-\mt^2\rbrack \lbrack k^2-\mw^2\rbrack^2}
=-\frac{g^{\sigma\tau}}{64\pi^2}\frac{i}{\mw^2}
\lbrack 4 B_0(x_t)+1\rbrack~,
\ee
where $B_0(x_t)$ is given in (\ref{BF}) with $x_t=\mt^2/\mw^2$.

Next using the standard rules for Dirac matrices we find
\be\label{Box5}
T_{\sigma\tau} g^{\sigma\tau}=-64 (\bar s\gamma_\mu(1-\gamma_5)d)
\otimes (\bar\nu \gamma^\mu(1-\gamma_5)\nu)
\equiv -64 (\bar s d)_{V-A} ( \bar\nu \nu)_{V-A}
\ee
We will develop in section 6 a simple method for evaluating in no
time the expressions like (\ref{Box5}).

Collecting all these results we find
\be\label{Box6}
{\cal D}(\nu\bar\nu)
=\lambda_t\frac{g_2^4}{64\pi^2}\frac{i}{\mw^2}
\lbrack 4 B_0(x_t)+1\rbrack
\ee
We can drop ``1" in the square brackets as the inclusion of 
u-quark and c-quark exchanges will cancel it after the unitarity
of the CKM matrix has been used ($\lambda_u+\lambda_c+\lambda_t=0$).

In order to obtain the final result for the effective Hamiltonian
corresponding to the last diagram in fig.~\ref{fig:7} we multiply 
(\ref{Box6}) by ``i" and use
\be\label{Box7}
\frac{g_2^4}{64\pi^2}\frac{1}{\mw^2}=
\frac{G_F}{\sqrt{2}}\frac{\alpha}{2\pi\sin^2\Theta_W}
\ee
to obtain
\be\label{Box8}
H_{eff}(T_3= 1/2) = \lambda_t {{G_{\rm F}}\over{\sqrt 2}}
   {{\alpha}\over{2\pi \sin^2\Theta_{\rm W}}} \lbrack -4 B_0(x_t)\rbrack 
   (\bar s d)_{V-A} (\bar\nu\nu)_{V-A}
\end{equation}
which is simply the rule in (\ref{BT12}).

Having this result it is straightforward to derive the rule (\ref{BTM12}).
In this case the charge flow on the lepton line is opposite to
the one in fig.~\ref{fig:7} and the Dirac structure in (\ref{Box2})
is replaced by
\be\label{Box9}
\tilde T_{\sigma\tau}=
4 \bar s \gamma_\mu\gamma_\sigma\gamma_\nu(1-\gamma_5)d
\otimes \bar\mu\gamma^\nu\gamma_\tau\gamma^\mu(1-\gamma_5)\mu
\ee
with all other expressions unchanged. Consequently
\be\label{Box10}
\tilde T_{\sigma\tau} g^{\sigma\tau}=
16 (\bar s d)_{V-A} (\bar\mu \mu)_{V-A}=
-\frac{1}{4}T_{\sigma\tau} g^{\sigma\tau}
\ee
Thus replacing in (\ref{Box8}) 
$(\bar\nu \nu)_{V-A}$ by $(\bar\mu \mu)_{V-A}$ and multiplying it by
$-1/4$ we recover the rule (\ref{BTM12}).

\subsection{Effective Hamiltonians for FCNC Processes}
\subsubsection{An Example}
With the help of the Feynman rules given in figs. \ref{fig:2} and 
\ref{fig:3} and the effective rules in 
(\ref{BT12}) and (\ref{ZRULE}) 
it is an easy
matter to construct the effective Hamiltonian for 
the decay $K^+\to \pi^+\bar\nu_e\nu_e$ to which
the diagrams in fig.\ \ref{fig:9} contribute. 
\begin{figure}[hbt]
\centerline{
\epsfysize=1.6in
\epsffile{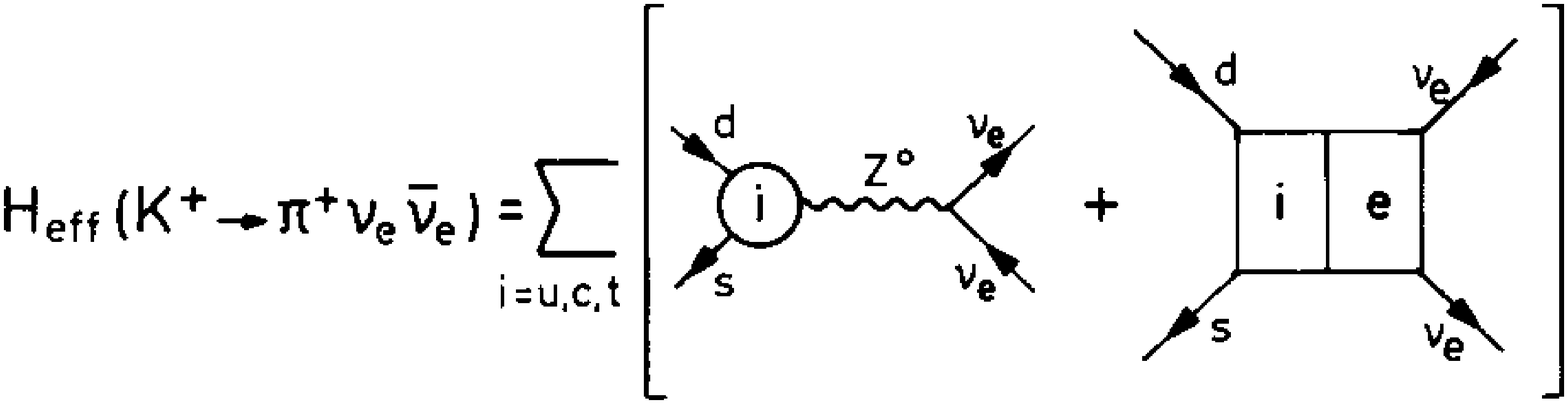}
}
\caption[]{
Calculation of ${\cal H}_{\rm eff}(K^+\to\pi^+\nu_e\bar\nu_e)$
\label{fig:9}}
\end{figure}

Replacing the $Z^0$ propagator by $i g_{\mu\nu}/M_Z^2$
and multiplying the first diagram 
by ``i", we
find the well-known result for the top contribution to this decay:
\begin{equation}\label{kplus}
{\cal H}_{\rm eff}(K^+\to\pi^+\nu_e\bar\nu_e) = {{G_{\rm F}}\over{\sqrt2}}\;
{{\alpha}\over{2\pi\sin^2\Theta_{\rm W}}}\; V_{ts}^* V_{td}\; 
X_0(x_t)\;
(\bar
s d)_{V-A} (\bar\nu_e\nu_e)_{V-A}.
\end{equation}
Here we have expressed the combination $C_0(x_t)-4 B_0(x_t)$ through
the function $X_0(x_t)$.

\subsubsection{Penguin-Box Expansion}
One can generalize such calculations to other processes in which
other basic effective vertices are present. 
For decays involving photonic and/or gluonic penguin vertices, the
$1/q^2$ in the propagator cancels the $q^2$ in the vertex and the
resulting effective Hamiltonian can again be written in terms of local
four--fermion operators. Thus generally an effective Hamiltonian for
any decay considered can be written in the absence of QCD corrections
as
\begin{equation}\label{exp}
{\cal H}_{\rm eff}^{\rm FCNC}=\sum_k C_k O_k,
\end{equation}
where $O_k$ denote local operators such as $(\bar s d)_{V-A}(\bar s
d)_{V-A}$, $(\bar s d)_{V-A}(\bar uu)_{V-A}$ etc. The coefficients
$C_k$ of these operators are simply linear combinations of the
functions of (\ref{SXYZ}) times the corresponding 
CKM factors which can
be read off from our rules. Consequently it is possible to write
the amplitudes for all FCNC decays and transitions 
as linear combinations of the basic,   process independent
$m_t$-dependent functions $F_r(x_t)$ of (\ref{SXYZ}) with
corresponding coefficients $P_r$ characteristic for the decay under
consideration. 
This ``Penguin Box Expansion'' \cite{PBE0} takes
the following general form:
\begin{equation}
A({\rm decay}) = P_0({\rm decay}) + \sum_r P_r({\rm decay}) \, F_r(x_t),
\label{eq:generalPBE}
\end{equation}
where the sum runs over all possible functions contributing to a given
amplitude. 
$P_0$  summarizes contributions stemming from internal quarks
other than the top, in particular the charm quark. As we will
demonstrate in the course of these lectures the general expansion
in (\ref{eq:generalPBE}) can be derived from the Operator Product
Expansion and is valid also in the presence of QCD corrections.
We will encounter many
examples of the expansion (\ref{eq:generalPBE}) in the course of 
these lectures.
We will see that similarly to $K^+\to\pi^+\nu\bar\nu$, there are other
decays which depend only on a single function. However, generally,
several basic functions contribute to a given decay. In particular,
we have the following correspondence between the most interesting FCNC
processes and the basic functions:

\begin{center}
\begin{tabular}{lcl}
$K^0-\bar K^0$-mixing &\qquad\qquad& $S_0(x_t)$, $S_0(x_c,x_t)$ \\
$B^0-\bar B^0$-mixing &\qquad\qquad& $S_0(x_t)$ \\
$K \to \pi \nu \bar\nu$, $B \to X_{d,s} \nu \bar\nu$ 
&\qquad\qquad& $X_0(x_t)$ \\
$K_{\rm L}\to \mu \bar\mu$, $B \to l\bar l$ &\qquad\qquad& $Y_0(x_t)$ \\
$K_{\rm L} \to \pi^0 e^+ e^-$ &\qquad\qquad& $Y_0(x_t)$, $Z_0(x_t)$, 
$E_0(x_t)$ \\
$\varepsilon'$ &\qquad\qquad& $X_0(x_t)$, $Y_0(x_t)$, $Z_0(x_t)$,
$E_0(x_t)$ \\
$B \to X_s \gamma$ &\qquad\qquad& $D'_0(x_t)$, $E'_0(x_t)$ \\
$B \to X_s \mu^+ \mu^-$ &\qquad\qquad&
$Y_0(x_t)$, $Z_0(x_t)$, $E_0(x_t)$, $D'_0(x_t)$, $E'_0(x_t)$
\end{tabular}
\end{center}

\subsection{More about GIM}
At this stage it is useful to return to the GIM mechanism \cite{GIM1}
which did
not allow tree level FCNC transitions. This mechanism is also felt in
the Hamiltonian of (\ref{exp}) and in fact it is fully effective when
the masses of internal quarks of a given charge in loop diagrams
 are set to be equal, e.g.\
$m_u=m_c=m_t$. Indeed the CKM factors in any FCNC process enter in the
combinations
\begin{equation}\label{CK}
C_k \propto \sum_{i=u,c,t}\lambda_i\; F(x_i)~~~~{\rm or}~~~~\sum_{i,j=u,c,t}
\lambda_i\lambda_j\; \tilde F(x_i,x_j),
\end{equation}
where $F,\tilde F$ denote any of the functions of (\ref{SXYZ}), and
the $\lambda_i$ are given in the case of $K$ and $B$ meson decays and 
particle--antiparticle
mixing as follows:
\begin{equation}
\lambda_i=\cases{V_{is}^*V_{id}&~~~$K$--decays, ~~$K^0-\bar K^0$\cr
                   V_{ib}^*V_{id}&~~~$B$--decays, ~~$B^0_d-\bar B^0_d$\cr
                   V_{ib}^*V_{is}&~~~$B$--decays, ~~$B^0_s-\bar B^0_s$\cr} 
\end{equation}
They satisfy the unitarity relation
\begin{equation}
\lambda_u + \lambda_c + \lambda_t =0,
\end{equation}
which implies vanishing  coefficients $C_k$ in (\ref{CK}) if
$x_u=x_c=x_t$. For this reason the mass--independent terms in the calculation
of the basic functions in (\ref{SXYZ})  can always be omitted.
In this limit, FCNC decays and transitions are absent.
Thus beyond tree level the conditions for a complete GIM
cancellation of FCNC processes are:
\begin{itemize}
\item
Unitarity of the CKM matrix
\item
Exact horizontal flavour symmetry which assures the equality
of quark masses of a given charge.
\end{itemize}

\noindent
Now in nature such a horizontal symmetry, even if it exists
at very short distance scales, is certainly broken at low energies by
the disparity of masses of quarks of a given charge. This in fact is
the origin of the breakdown of the GIM mechanism at the one--loop
level and the appearance of FCNC transitions. 
The size of this breakdown, and consequently the size of 
FCNC transitions, depends on the disparity of masses,
on the behaviour of the basic functions of (\ref{SXYZ}), and can be
affected by QCD corrections as we will see in the course of these
lectures. Let us make two
observations: 
\begin{itemize}
\item
For small $x_i\ll 1$, relevant for $i\not= t$, 
the functions (\ref{BF})-(\ref{BFF})
behave as follows:
\begin{equation}
S_0(x_i)\propto x_i, \quad 
  B_0(x_i)\propto x_i \ln x_i, \quad
  C_0(x_i)\propto x_i \ln x_i 
\end{equation}
\begin{equation}
D_0(x_i)\propto \ln x_i, \quad
  E_0(x_i)\propto \ln x_i,  \quad
  D'_0(x_i)\propto x_i, \quad 
  E'_0(x_i)\propto x_i. 
\end{equation}
This implies ``hard'' (quadratic) GIM suppression of FCNC processes 
governed by the
functions $S_0,B_0,C_0,D_0',E_0'$ provided the top quark contributions due 
to small CKM
factors can be neglected. In the case of $D_0(x_i)$ and 
$E_0(x_i)$ only ``soft''
(logarithmic) GIM suppression is present.
\par
\item
For large $x_t$ we have
\begin{equation}
S_0(x_t)\propto x_t, \quad 
  B_0(x_t)\propto {\rm const}, \quad
  C_0(x_t)\propto x_t
\end{equation}
\begin{equation}
D_0(x_t)\propto \ln x_t, \quad
  E_0(x_t)\propto {\rm const},  \quad
  D'_0(x_t)\propto {\rm const}, \quad 
  E'_0(x_t)\propto {\rm const}. 
\end{equation}
\end{itemize}
Thus for FCNC processes governed by top quark contributions, 
the GIM suppression is
not effective at the one loop level and in fact in the case of decays and
transitions receiving contributions from $S_0(x_t)$ and $C_0(x_t)$  
important
enhancements are possible. 
\par

The latter property emphasizes the special role of $K$ and $B$ decays with
regard to FCNC transitions. In these decays the appearance of the
top quark in the internal loop with $m_t>M_W\gg m_c,m_u$ removes the
GIM suppression, making $K$ and $B$ decays a particularly
useful place to test FCNC transitions and to study the physics of
the top quark. Of course the hierarchy of various FCNC transitions is also
determined by the hierarchy of the elements of the CKM matrix allowing
this way to perform sensitive tests of this sector of the Standard
Model. 

The FCNC decays of $D$--mesons are much stronger suppressed because only
$d$, $s$, and $b$ quarks with $m_d,m_s,m_b\ll M_W$ enter internal loops
and the GIM mechanism is much more effective. Also the known structure
of the CKM matrix is less favorable than in $K$ and $B$ decays. For these
reasons we will restrict our presentation to the latter.
In the extensions of the Standard Model, FCNC transitions are possible
at the tree level and the hierarchies discussed here may not apply.
Reviews of FCNC transitions in D-decays can be found in \cite{DDD}.
\subsection{Final Comments}
The discussion presented in this and the previous section left out completely 
the QCD effects in weak decays.
In particular, we have not shown how
to translate the calculations done in terms of quarks into predictions
for the decays of their bound states, the hadrons. Similary
we did not include short distance QCD corrections. 
Still, I hope that this discussion has shown the richness of the field
of weak decays and of FCNC processes and motivated the reader to
learn more about the more technical part of this field. Such
motivation is clearly necessary as from now on our gentle hike
is turning quickly into a real climb. This climb will be rather
steep and in certain parts technically difficult. It will last
with small breaks until we reach section 9. From section 9 on it will
be easy again. 
\section{Renormalization and Renormalization Group}
\setcounter{equation}{0}
\subsection{General Remarks}
This section collects  those basic facts about QCD, its renormalization
and the renormalization group,
which are indispensable for our climb. In particular we discuss
the dimensional regularization, the MS and ${\rm\overline{MS}}$
renormalization schemes and renormalization group equations
for the running QCD coupling and the running quark masses.
We recall solutions of these equations and present numerical
examples. At the end of this section we explain what is meant
by the renormalization group improved perturbation theory.
\subsection{QCD Lagrangian}
The Lagrangian density of QCD, omitting the ghosts and setting the
gauge parameter to $\xi=1$,  reads
\begin{eqnarray}\label{lqcd}
{\cal L}_{QCD} & = &
-{1\over 4}(\partial_\mu A^a_\nu-\partial_\nu A^a_\mu)
(\partial^\mu A^{a\nu}-\partial^\nu A^{a\mu})-{1\over{2}}
(\partial^\mu A^a_\mu)^2   \nonumber \\
&&{}+ \bar q_\alpha (i\not\!\partial-m_q)q_\alpha 
- g_s \bar q_\alpha T^a_{\alpha\beta}\gamma^\mu q_\beta 
A^a_\mu \nonumber \\
&&{}+{g_s\over 2}f^{abc}(\partial_\mu A^a_\nu-\partial_\nu A^a_\mu)
 A^{b\mu}A^{c\nu} - {g_s^2\over 4}f^{abe}f^{cde}A^a_\mu A^b_\nu
 A^{c\mu}A^{d\nu} 
\end{eqnarray}
Here  $A^a_\mu$ are the gluon fields with $(a,b,c=1,..8)$
 and
 $q=(q_1, q_2, q_3)$ is the color triplet of quark flavor $q$,
$q$ = $u$, $d$, $s$, $c$, $b$, $t$. 
$g_s$ is the QCD coupling so that
\be
\as=\frac{g^2_s}{4\pi}.
\ee
Finally $T^a$ and $f^{abc}$ are the generators and
structure constants of $SU(3)_C$, respectively. From this Lagrangian
one can derive the Feynman rules for QCD. Some of these rules have
been given in figs.\ \ref{fig:2} and \ref{fig:3}. 

\subsection{Dimensional Regularization}
In order to deal with divergences that appear in loop
corrections to Green functions we have to regularize the theory
to have an explicit parametrization
of the singularities. In these lectures we will employ
{\it dimensional regularization} (DR). In this regularization
 Feynman diagrams are
evaluated in $D=4-2\,\varepsilon$ space-time dimensions and singularities
are extracted as poles for $\varepsilon \to 0$. Thus the results of
 one-loop or two-loop calculations have the following general structure:
\begin{equation}
{\rm One~ Loop~ Result} = \frac{a_1}{\varepsilon}+b_1~,
\end{equation}
\begin{equation}
{\rm Two~ Loop~ Result} = \frac{a_2}{\varepsilon^2}+
 \frac{b_2}{\varepsilon}+c_2~,
\end{equation}
where $a_i$, $b_i$ and $c_2$ are finite.

Several useful formulae for the evaluation of Feynman diagrams in
$D=4-2\varepsilon$ dimensions 
are collected in the appendix A of my review in \cite{AB80} and
in Muta`s book \cite{MUTA}. Here we only
stress the following important point. Let us consider the second
term in the second line in (\ref{lqcd}). The mass dimensions of
$q_i$, $A^a_\mu$ and ${\cal L}$ are $(D-1)/2$, $(D-2)/2$ and $D$
respectively. Consequently, the dimension of $g_s$ in $D=4-2\varepsilon$
dimensions is simply equal to $\varepsilon$. 
It is more useful to work with a
 dimensionless coupling constant in arbitrary $D$ dimensions. To this
end  we make the replacement in (\ref{lqcd}):
\be
g_s\quad \to \quad g_s \mu^\varepsilon
\ee
where $\mu$ is an arbitrary parameter with the dimensions of mass
and $g_s$ on the r.h.s is dimensionless. The appearence of the
scale $\mu$ has profound impact on these lectures.

\begin{figure}[hbt]
\vspace{0.10in}
\centerline{
\epsfysize=1in
\epsffile{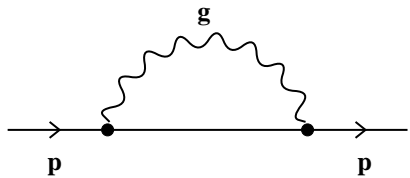}
}
\vspace{0.08in}
\caption[]{Quark-Self-Energy Diagram 
\label{L:3}}
\end{figure}

As an example, consider the calculation of the one-loop self-energy
diagram of fig. \ref{L:3}. Setting
$m_q=0$ and denoting the external quark momentum by $p$ (with $p^2<0$), 
we arrive
by means of standard techniques at
\begin{equation}\label{isigma}
i \Sigma_{\alpha\beta}=i\not\! p~ C_F~ \delta_{\alpha\beta}
g^2_s [2 (1-\varepsilon)] P_{div} B(2-\varepsilon,1-\varepsilon)
\end{equation}
where $C_F=4/3$ is the relevant colour factor. Next
\begin{equation}\label{pdiv}
P_{div}\equiv \frac{\Gamma(\varepsilon)}{(4\pi)^{2-\varepsilon}}
\left(\frac{\mu^2}{-p^2}\right)^\varepsilon
=\frac{1}{16\pi^2}
\lbrack \frac{1}{\varepsilon}+\ln 4\pi-\gamma_E+\ln\frac{\mu^2}{-p^2}
+\ord(\varepsilon)
\rbrack
\end{equation}
and $\Gamma$ and $B$ are the known Euler functions. In expanding
$P_{div}$ around $\varepsilon=0$ we have used
\be
\Gamma(\varepsilon)=\frac{1}{\varepsilon}-\gamma_E+O(\varepsilon)
\qquad
\gamma_E=0.5772...
\ee
where $\gamma_E$ is the Euler constant. Since
\be
B(2-\varepsilon,1-\varepsilon)=\frac{1}{2}(1+2\varepsilon)+
\ord(\varepsilon^2)
\ee 
we arrive at
\begin{equation}\label{se}
i \Sigma_{\alpha\beta}=i\not\! p~ C_F~ \delta_{\alpha\beta}
\frac{\alpha_s}{4\pi}
\lbrack \frac{1}{\varepsilon}+\ln 4\pi-\gamma_E+\ln\frac{\mu^2}{-p^2}+1
\rbrack
\end{equation}
where $\ord(\varepsilon)$ terms have been set to zero.
We have thus extracted the singularity as a $1/\varepsilon$ pole
and have obtained a well-defined finite part. The appearance
of the first four terms
in the square bracket in (\ref{se}) originating from $P_{div}$ in
(\ref{pdiv}) is characteristic for all divergent one-loop calculations.

The dimensional regularization is the favourite regularization in
gauge theories  as it preserves all symmetries of the theory. Possible
problems are connected with the treatment of $\gamma_5$ in $D\not=4$
dimensions, which clearly is of relevance for the study of weak
interactions. Let us discuss this issue now. We follow here
\cite{WEISZ}
\subsection{The Issue of $\gamma_5$ in D Dimensions}
\subsubsection{Preliminaries}
  Let us describe the three distinct sets of computational rules,
 for the manipulation of covariants and Dirac matrices,
 most commonly used in perturbative calculations in the Standard Model.
  These schemes all employ the method
 of dimensional regularization of the Feynman integrals \cite{HV,BM},
 and in each case $D=4-2\varepsilon$ denotes the number of dimensions.
 We will not discuss other regularization schemes such as BPHZ and lattice.
 These work directly in 4-dimensions and hence don't have algebraic
 consistency problems with respect to $\gamma_5$, but their use 
introduces other
 subtleties and two-loop calculations therewith are extremely tedious.
 
\subsubsection{Naive Dimensional Regularization}
  The most commonly used set of rules is one we shall call 'naive dimensional
 regularization' (NDR).
 Only the $D-$ dimensional metric tensor $g$ is introduced satisfying,
\be\label{BW1}
 g_{\mu \nu} = g_{\nu \mu}, \qquad
 g_{\mu \rho} g^{\rho}_{\nu} = g_{\mu \nu}, \qquad
 g_{\mu}^{\mu} = D, 
\ee
 and the Dirac matrices $\gamma_{\mu}$ obey
\be\label{BW2}
 \{ \ga_{\mu} , \ga_{\nu} \} = 2 g_{\mu \nu}.
\ee
 It is standard (but inessential) to set the trace of the unit matrix
 to equal 4; we shall adopt this convention in this and the schemes below.
 When $\gamma_5$ appears in the Feynman vertices
 the manipulation rule adopted in this scheme is that it
 anticommutes with the Dirac matrices,
\be\label{BW3}
 \{ \ga_{\mu} , \gamma_5 \} = 0.
\ee
 It has repeatedly been emphasized in the literature \cite{BM,Bo}
 that this rule leads to obvious algebraic inconsistencies.
 Nevertheless this scheme has been most widely employed for most calculations
 because of its ease to incorporate standard software in computer programs.
 It is known to lead to incorrect results in certain cases, e.g. the axial
 anomaly is not reproduced. On the other hand in many cases
 it does reproduce the correct
 results. A necessary condition for this
 seems to be that the calculated amplitude does not involve the evaluation
 of a closed odd parity
 fermion loop. Indeed, with the NDR rules one does not know
 how to unambiguously handle the expression
$ 
{\rm Tr}(\gamma_5 \gamma_{\mu} \gamma_{\nu} \gamma_{\rho} \gamma_{\lambda}).
$
Beginning with the work of Peter Weisz and myself \cite{WEISZ} it has
been demonstrated in many explicit calculations that the NDR scheme
gives correct results, consistent with the schemes without the $\gamma_5$
problems, provided one can avoid the calculations of traces like the
one given above. In fact all the higher order QCD calculations for weak
decays performed
in the NDR scheme in the last nine years and listed in table 
\ref{TAB1} could
avoid the direct calculation of such traces. To this end 
it is often necessary at the
intermediate stages of the calculation 
to work with a special basis of local operators which differs from
the standard basis discussed in sections 5-8 of these lectures.
Examples of such strategies valid only at two-loop level can be
found in \cite{BJLW1,BJLW2,CURCI}. 
An approach valid apparently to all orders is presented
in \cite{MISTRIK}.

 \subsubsection{Dimensional Reduction}
 A second set of manipulation rules initially introduced by Siegel
 \cite{Si} for the renormalization of supersymmetric theories goes under
 the name of dimensional reduction. Here the Dirac matrices $\gat$ are taken
 to be in 4-dimensions, thus
\be\label{BW4}
 \{ \gat_{\mu} , \gat_{\nu} \} = 2 \gt_{\mu \nu},
\ee
 where $\gt$ is the 4-dimensional metric tensor,
\be\label{BW5}
 \gt_{\mu \nu} = \gt_{\nu \mu}, \qquad
 \gt_{\mu \rho} \gt^{\rho}_{\nu} = \gt_{\mu \nu}, \qquad
 \gt_{\mu}^{\mu} = 4.
\ee
 When evaluating the Feynman integrals the $D-$ dimensional
 $ g_{\mu\nu} $ inevitably makes
 its appearence and it is necessary to supplement the rules with one which
 stipulates the result of contraction of the $4-$ and $D-$ dimensional
 metric tensors. In order to preserve gauge invariance and in apparent concord
 with the reduction to $D<4$ dimensions the rule employed is
\be\label{BW6}
 \gt_{\mu \rho} g^{\rho}_{\nu} = g_{\mu \nu}.
\ee
 
 The advantage of the scheme is that the 4-dimensional Dirac algebra can be
 used to reduce the algebraic complexity of the amplitudes. However there is
 a price to be paid which involves a number of field theoretical subtleties
  some of which are already present in the pure QCD part of the dynamics.
 These are discussed in \cite{ACMP}.
 Again, this scheme has been criticized \cite{Bo,Ma} since it leads
 to similar
 difficulties as the naive dimensional regularization described above.
 In particular it implies that identities homogeneous in the metric tensor in
 $4$-dimensions are also satisfied in generic $D$-dimensions, which is
 manifestly algebraically inconsistent. Although
 the axial anomaly can be reproduced \cite{NT}, and although
 there is to our knowledge as yet no known explicit calculation using
 DRED which gives the wrong result, it
 has not yet been established as a consistent scheme and thus
 maintains at present merely the status of a prescription.

 In the field of weak decays the DRED scheme has been used in \cite{ACMP} 
for the calculation of higher order QCD corrections to $\Delta S=1$
decays.
This result has been confirmed in \cite{WEISZ} and shown to be compatible
with the NDR scheme and the 't Hooft--Veltman scheme discussed below.
Similarly the initial problem of calculating higher order QCD corrections
to the $B\to X_s\gamma$ in the DRED scheme \cite{CAND} has been
resolved by Misiak \cite{MISD}. 
These days the DRED scheme is less popular and the
most calculations of QCD corrections are in the NDR scheme and the
't Hooft--Veltman scheme to which we turn now our attention.
 \subsubsection{The 't Hooft--Veltman Rules}
 The third set of rules is the one originally proposed by 't Hooft and
 Veltman
 \cite{HV} and by Akyeampong and Delbourgo \cite{AD} and systematized by
 Breitenlohner and Maison \cite{BM}. The latter authors showed that this is
 a consistent formulation of dimensional regularization even when 
$\gamma_5$
 couplings are present.
 
 To write down the rules it is convenient to introduce in addition to the
 $D$- and 4- dimensional metric tensors $g$ and $\gt$
 satisfying (\ref{BW1}) and (\ref{BW5}) respectively, 
the $-2\varepsilon$- dimensional tensor $\gh$ satisfying,
\be\label{BW7}
 \gh_{\mu \nu} = \gh_{\nu \mu}, \qquad
 \gh_{\mu \rho} \gh^{\rho}_{\nu} = \gh_{\mu \nu}, \qquad
 \gh_{\mu}^{\mu} = -2\varepsilon.
\ee
 The important difference with respect to dimensional reduction is that
 instead of the rule (\ref{BW6}) for contracting
 the different metric tensors one imposes
\be\label{BW8}
 \gt_{\mu \rho} g^{\rho}_{\nu} = \gt_{\mu \nu},
\ee
 which does not lead to manifest algebraic inconsistencies. In addition to
 (\ref{BW8}) one has,
\be\label{BW9}
 \gh_{\mu \rho} g^{\rho}_{\nu} = \gh_{\mu \nu},  \qquad
 \gh_{\mu \rho} \gt^{\rho}_{\nu} = 0.
\ee
 The D-dimensional Dirac matrix
 is now split into a 4- and $-2\varepsilon$-dimensional parts,
\be\label{BW10}
  \gamma_{\mu} = \gat_{\mu} + \gah_{\mu},
\ee
 with $\gamma$ and $\gat$ obeying the anticommutation relations 
(\ref{BW2})
 and (\ref{BW4}) respectively. $\gah$ on the other hand satisfies
\be\label{BW11}
 \{ \gah_{\mu} , \gah_{\nu} \} = 2 \gh_{\mu \nu},
\ee
 and it anticommutes with $\gat$
\be\label{BW12}
 \{ \gah_{\mu} , \gat_{\nu} \} = 0.
\ee
 Note also by virtue of (\ref{BW9}) it follows
\be\label{BW13}
  \gah_{\mu} \gat^{\mu}  = 0, \qquad
  \gh_{\mu}^{ \nu} \gat_{\nu}  = 0, \qquad
  \gt_{\mu}^{ \nu} \gah_{\nu}  = 0.
\ee
 
 In \cite{BM} it is shown that a $\gamma_5$ can be introduced which
 anticommutes with $\gat$ but commutes with $\gah$,
\be\label{BW14}
 \gamma_5^{2} =1 , \qquad
 \{ \gamma_5 , \gat_{\nu} \} = 0, \qquad
 [ \gamma_5 , \gah_{\nu} ] = 0.
\ee
 Since $\gamma_5$ does not have simple commutation properties
 with $\gamma_{\mu}$ it
 is important to consistently define the coupling to chiral fields in a
 model such as the Standard Model; e.g.
 for coupling to left-handed fields the symmetrically defined vertex
\be\label{BW15}
 {{1}\over{2}} (1 + \gaf) \ga_{\mu} (1 - \gaf)  = \gat_{\mu} (1 - \gaf),
\ee
 should be used \cite{KNS}.
 
 This scheme has admittedly some rather unattractive features.
 In particular it is more
 inconvenient to implement in algebraic computer programs than the
NDR scheme. Nevertheless it
 must
 be stressed again that it is to date the only known scheme (within the
 framework of dimensional regularization) which has been demonstrated to be
 consistent \cite{BM,Bo}, and thus its inconvenience must be tolerated.
For this reason a computer package for Dirac algebra manipulation in 
the HV and NDR
schemes called TRACER has been developed in my group at the Technical
University in Munich by Jamin and Lautenbacher \cite{JaLau}. 
Using this program
one can appreciate the simplicity of the NDR scheme compared with
the HV scheme for which the computer calculations can be sometimes 
really time consuming.

\subsection{Renormalization}
\subsubsection{General Remarks}
In order to eliminate the divergences in Green functions one has to 
renormalize the fields and
parameters in the Lagrangian through
\begin{equation}\label{zqcd}
\begin{array}{lcl}
A^a_{0\mu}=Z^{1/2}_3 A^a_\mu &\qquad& q_0=Z^{1/2}_q q 
 \\
g_{0,s}=Z_g g_s\mu^\varepsilon &\qquad& m_0=Z_m m
\end{array}
\end{equation}
The index ``0'' indicates unrenormalized quantities. $A^a_\mu$ and $q$
are renormalized fields, $g_s$ is the renormalized QCD coupling and $m$
the renormalized quark mass. 
The factors $Z$
are the renormalization constants. They are divergent quantities,
chosen in such a manner that the divergences disappear once the
Greens functions have been expressed in terms of renormalized
quantities only. 

It should be stressed that the unrenormalized parameters
$g_{0,s}$ and $m_0$ are independent of the scale $\mu$. This
implies, in particular, that $g_s$ must be $\mu$-dependent.
Since $Z_i$ have a perturbative expansion in $g_s$ they must
also depend on $\mu$. Consequently also the renormalized mass
$m$ is $\mu$-dependent.
\subsubsection{The Counter--term Method}
A straghtforward way to implement renormalization is provided by the
counter--term method.
Thereby parameters and fields in the
original Lagrangian, considered as unrenormalized
(bare) quantities, are reexpressed through renormalized ones by
means of (\eqn{zqcd}). Thus 

\be\label{COUNT}
{\cal L}^0_{QCD}={\cal L}_{QCD}+{\cal L}_C
\ee
where ${\cal L}_{QCD}$ is given in (\ref{lqcd}). ${\cal L}^0_{QCD}$ is
also given by (\ref{lqcd}) but with $q$ replaced by $q_0$ and similarly
for $A^a_\mu$, $g_s$ and $m$. ${\cal L}_C$ is the {\it counter--term}
Lagrangian. It is simply defined by
(\ref{COUNT}).
For instance:
\begin{equation}\label{ctm}
{\cal L}_q=\bar q_0 i\not\!\partial q_0-m_0\bar q_0 q_0\equiv
\bar q i\not\!\partial q-m\bar q q+(Z_q-1)\bar q i\not\!\partial q-
(Z_q Z_m-1)m\bar q q~.
\end{equation}
${\cal L}_{QCD}$ given entirely in terms of
renormalized quantities leads to the usual Feynman rules of 
figs.\ \ref{fig:2} and \ref{fig:3}.
The counter--terms ($\sim(Z-1)$) can be formally
treated as new interaction terms that contribute
to Green functions calculated in perturbation theory. 
For these new interactions also Feynman rules can be derived.
For instance, the Feynman rule for the counter--terms in (\eqn{ctm}) reads
($p$ is the quark momentum)
\begin{equation}\label{ctex}
i\delta_{\alpha\beta}\lbrack(Z_q-1)\not\! p - (Z_q Z_m-1) m\rbrack~.   
\end{equation}

The constants $Z_i$ are  determined such that the the contributions
from these new interactions cancel the
divergences in the Green functions resulting from the calculations
based on ${\cal L}_{QCD}$ in (\ref{COUNT}) only.
There is some arbitrariness how this can be done because a given
renormalization prescription can in general subtract not only the
divergences but also finite parts. The subtractions of finite
parts is, however, not uniquely defined which results in  the 
{\it renormalization scheme dependence} of $Z_i$ and of the
renormalized fields and parameters. We will elaborate on this
scheme dependence and its cancellation in physical quantities 
at later stages of these lectures.

\subsubsection{MS and $\overline{\bf \rm MS}$ Renormalization Schemes}
The simplest renormalization scheme is the {\it Minimal Subtraction
Scheme} MS  \cite{HV1} 
in which only divergences are subtracted. In this scheme,
the renormalization constants are given by
\begin{equation}\label{zms}
Z_i = \frac{\alpha_s}{4\pi}\frac{a_{1i}}{\varepsilon} 
+\left(\frac{\alpha_s}{4\pi}\right)^2
 \left(\frac{a_{2i}}{\varepsilon^2}+
 \frac{b_{2i}}{\varepsilon}\right)+{\cal O}(\alpha_s^3) 
\end{equation}
where $a_{ji}$ and $b_{ji}$ are $\mu$-independent constants.
The fact that in this scheme the renormalization constants
do not have any explicit $\mu$-dependence and depend on $\mu$
only through $g_s$ is an important virtue of this scheme.
This, in particular, in the context of renormalization group
equations discussed below. Similarly the renormalization constants
$Z_i$ do not depend on masses. Therefore the MS-scheme and the
schemes discussed below belong to the class of mass independent
renormalization schemes \cite{Weinberg}.

Now, starting with the MS scheme, one can construct a whole class
of subtraction schemes which differ from MS by a different continuation
of the renormalized coupling constant to D dimensions. For these
MS-like schemes we have

\begin{equation}
g_{0,s}=Z^k_g g^k_s \mu^\varepsilon_k
\qquad
\mu_k=\mu f_k
\end{equation}
where $f_k$ is an arbitrary number which defines the particular
scheme "k''. Since different schemes in this class differ from the
MS scheme only by a schift in $\mu$, the renormalization constants
for these schemes can be obtained from (\ref{zms}) by replacing
$\alpha_s$ by $\alpha^k_s$ chareacteristic for a given scheme.
The constants $a_{ji}$ and $b_{ji}$, being $\mu$-independent,
remain unchanged.

Of particular interest is the so-called  $\overline{\rm MS}$ scheme
\cite{BBDM} in which
\begin{equation}\label{msms}
\mu_{\overline{MS}}=\mu e^{\gamma_E/2}(4\pi)^{-1/2}
\end{equation}
and $P_{div}$ in (\ref{pdiv}) is replaced by
\begin{equation}\label{pdivb}
\bar P_{div}\equiv \frac{\Gamma(\varepsilon)}{(4\pi)^{2-\varepsilon}}
\left(\frac{\mu^2_{\overline{MS}}}{-p^2}\right)^\varepsilon
=\frac{1}{16\pi^2}
\lbrack \frac{1}{\varepsilon}+\ln\frac{\mu^2}{-p^2}
+\ord(\varepsilon)
\rbrack
\end{equation}
We observe that in this scheme the terms $\ln 4\pi-\gamma_E$,
the artifacts of the dimensional regularization, are absent !

In summary then:
\begin{equation}
\lbrace{\rm MS \to \overline{MS} }\rbrace\quad 
\equiv\quad \lbrace{\mu\to\mu_{\overline{MS}}}\rbrace
\end{equation}
\begin{equation}
\lbrace Z_i^{{MS}} \to Z_i^{\overline{MS}}\rbrace\quad \equiv\quad
\lbrace\alpha_s^{MS} \to \alpha_s^{\overline{MS}}\rbrace.
\end{equation}

In these lectures, we will exclusively work with the $\overline{\rm MS}$
scheme. In order to simplify the notation we will denote
$\mu_{\overline{MS}}$ simply by $\mu$ and simultaneously drop
the $\ln 4 \pi-\gamma_E$ terms in any finite contribution. Similarly
$\alpha_s$ in these lectures will always stand for 
$ \alpha_s^{\overline{\rm MS}}$.

As an example let us find $Z_q$ and $Z_m$. To this end we repeat the
calculation of the self-energy diagram of fig.~\ref{L:3}, this time keeping
the quark mass $m$. Dropping the finite terms, which are of no concern
for finding $Z_i$ in the $\overline{\rm MS}$ scheme, we find

\begin{equation}\label{sigma}
(i \Sigma_{\alpha\beta})_{div}=i C_F \delta_{\alpha\beta}  
\frac{\alpha_s}{4\pi}( \not\! p-4 m)
 \frac{1}{\varepsilon}+{\cal O}(\alpha_s^2) 
\end{equation}
Adding to this result the counter-term (\ref{ctex}) and requiring
the final result to be zero we readly find
\begin{equation}\label{zq0}
Z_q=1- \frac{\alpha_s}{4\pi} C_F \frac{1}{\varepsilon}+ {\cal O}(\alpha_s^2) 
\ee
\be\label{zm}
Z_m=1- \frac{\alpha_s}{4\pi} 3 C_F \frac{1}{\varepsilon}+{\cal O}(\alpha_s^2) 
\end{equation}

Similarly $Z_3$ and $Z_g$ can be found by calculating one-loop corrections
to the gluon propagator and the gluon-$\bar q q$ vertex, respectively.
One finds:
\begin{equation}
Z_3=1- \frac{\alpha_s}{4\pi}\left[\frac{2}{3}f-\frac{5}{3}N \right]
 \frac{1}{\varepsilon}+ {\cal O}(\alpha_s^2) 
\end{equation}

\begin{equation}\label{zg}
Z_g=1- \frac{\alpha_s}{4\pi}\left[\frac{11}{6}N-\frac{2}{6}f \right]
 \frac{1}{\varepsilon}+ {\cal O}(\alpha_s^2) 
\end{equation}
where $N$ denotes the number of colours ($N=3$ in QCD) and $f$ stands
for the number of quark flavours.
\subsubsection{Renormalization of Green Functions}
Let us denote by
\be\label{cgreen}
G^{(n_F,n_G)}(p_j,g_s,m,\mu,\eps)\equiv
\langle 0|T(q_1,...q_{n_F},A^\mu_1,...A^\mu_{n_G})|0\rangle
\ee
a connected renormalized Green function with $n_F$ quark and
$n_G$ gluon external legs carrying momenta $p_j$. Here $m$
indicates general dependence on masses. The corresponding
amputated renormalized one-particle irreducible Green function is 
given by
\be\label{ampgreen}
\Gamma^{(n_F,n_G)}=\frac{G^{(n_F,n_G)}}
{\prod^{n_F}G^{(2,0)}\prod^{n_G}G^{(0,2)}}~.
\ee
Similar expressions exist for the unrenormalized Green functions
$G_0^{(n_F,n_G)}$ and $\Gamma_0^{(n_F,n_G)}$ with all renormalized 
parameters and fields replaced by the corresponding bare quantities.
With (\ref{zqcd}),  $\Gamma^{(n_F,n_G)}$ and  
$\Gamma_0^{(n_F,n_G)}$ are related to each other by
\be\label{grel}
\Gamma^{(n_F,n_G)}(p_j,g_s,m,\mu,\eps)=Z_q^{n_F/2}Z_3^{n_G/2}
\Gamma_0^{(n_F,n_G)}(p_j,g_{0,s},m_0,\eps)~.
\ee
The renormalization then means that when $g_{0,s}$ and $m_0$ on the
r.h.s of (\ref{grel}) are expressed through $g$ and $m$ according to
(\ref{zqcd}), $\Gamma^{(n_F,n_G)}$ are finite and the limit
\be
\lim_{\eps\to 0} \Gamma^{(n_F,n_G)}(p_j,g_s,m,\mu,\eps)=
\Gamma^{(n_F,n_G)}(p_j,g_s,m,\mu)
\ee
exists.

As an example consider the result for the quark self-energy in
(\ref{se}). In the notation of (\ref{grel})
its divergent part added to the ``tree level" propagator is
given by
\begin{equation}
\Gamma_0^{(2,0)}=i C_F \delta_{\alpha\beta}\not\! p (1+ 
\frac{\alpha_s}{4\pi} \frac{1}{\varepsilon})~. 
\end{equation}
The corresponding renormalized two-point function is given by
\be
\Gamma^{(2,0)}=Z_q \Gamma_0^{(2,0)}~
\ee
which with (\ref{zq0}) is indeed finite. In this case at $\ord(\alpha_s)$
only quark field renormalization is needed to obtain finite result.
Coupling renormalization is necessary first at $\ord(\alpha_s^2)$.
\subsection{Renormalization Group Equations}
\subsubsection{The Basic Equations}
In the process of renormalization we have introduced an arbitrary
mass parameter $\mu$. The $\mu$-dependence of the renormalized
coupling constant $g_s$ and of the renormalized quark mass $m$
is governed by the renormalization group equations.
These equations are derived from the
definitions (\eqn{zqcd}) using the fact  that bare quantities are
$\mu$-independent. 
One finds ($g\equiv g_s$):
\begin{equation}\label{rgbe}
{dg(\mu)\over d\ln\mu}=\beta(g(\mu), \eps)  \end{equation}
\begin{equation}\label{rggm}
{d m(\mu)\over d\ln\mu}=-\gamma_m(g(\mu)) m(\mu)  \end{equation}
where
\begin{equation}\label{bete}
\beta(g, \eps)= -\eps g+\beta(g),  \end{equation}

\begin{equation}\label{gamz} 
\beta(g)=-g{1\over Z_g}{d Z_g\over d\ln\mu},\quad\quad
\gamma_m(g)={1\over Z_m}{d Z_m\over d\ln\mu}.
  \end{equation}
(\ref{bete}) is valid in arbitrary
dimensions. In four dimensions $\beta(g, \eps)$ reduces to $\beta(g)$.
Let us prove (\ref{bete}) \cite{Gross}. Using (\eqn{zqcd}) we have 
\bea\label{prove1}
\beta(g,\eps) &=& g_0 \mu {d\over d\mu} [\mu^{-\eps} Z_g^{-1}]
= g_0 
\left[-\eps \mu^{-\eps} Z_g^{-1} + \mu^{-\eps+1}\frac{d Z_g^{-1}}{d\mu}\right]
\nonumber\\ 
&= &-\eps g-g_0 \mu^{-\eps+1}{1\over Z^2_g}{d Z_g \over d\mu}
=-\eps g-g\mu {d Z_g \over d\mu}{1\over Z_g}.
\eea
Similarly one can derive the expression for $\gamma_m$ in (\ref{gamz})
by inserting $m=m_0/Z_m$ into (\ref{rggm}).

$\beta(g)$ and $\gamma(g)$ are called {\it renormalization group functions.}
$\beta(g)$ governs the $\mu$-dependence of $g(\mu)$.
$\gamma_m$, the {\it anomalous dimension} of the mass operator, governs
the $\mu$-dependence of $m(\mu)$.
In the MS $(\overline{\rm MS})$-scheme they depend only on $g$. 
In particular they carry no explicit $\mu$-dependence and are independent of
masses.
Writing
\begin{equation}\label{ziep}
Z_i=1+\sum^\infty_{k=1} {1\over \eps^k} Z_{i, k}(g)  
\end{equation}
and
using (\ref{bete}) and (\ref{gamz}) one finds
\begin{equation}\label{zi1}
\beta(g)=2 g^3{d Z_{g, 1}(g)\over d g^2},
\ee
\be\label{zim}
\gamma_m(g)=-2 g^2{d Z_{m, 1}(g)\over d g^2}.
\end{equation}
Thus $\beta(g)$ and $\gamma_m(g)$ can be
directly obtained from the $1/\eps$-pole parts of the renormalization
constants $Z_g$ and $Z_m$, respectively. This is a very useful
property of the MS-like schemes. Let us demonstrate that (\ref{zi1})
is indeed true. 
We follow Muta \cite{MUTA} and write
\be
\beta(g, \eps)= -\eps g-g f(g),  
\quad\quad
f(g)=\frac{\mu}{Z_g}\frac{d Z_g}{d\mu}.
\end{equation}
Specializing the expansion (\ref{ziep}) to $Z_g$ and inserting
it into formula for $f(g)$ gives
\be\label{mut}
f(g)\left( 1 +\frac{Z_{g,1}}{\eps}+ \frac{Z_{g,2}}{\eps^2}+...\right)
=\frac{1}{\eps} \beta(g,\eps)
\left(\frac{dZ_{g,1}}{dg}+\frac{1}{\eps}\frac{dZ_{g,2}}{dg}+...\right)~.
\ee

Now finitness of $\beta(g)$ implies finitness of $f(g)$. Consequently
the equality (\ref{mut}) should hold for each coefficient of the
power $1/\eps$. In particular the non-singular terms give
\be
f(g)=-g \frac{dZ_{g,1}}{dg}~,
\ee
which with $\beta(g)=-gf(g)$ gives (\ref{zi1}).
The proof of (\ref{zim}) can be done in an analogous
manner using the finitness of $\gamma_m$. It is left as 
a homework problem.

With $Z_g$ and $Z_m$ in (\ref{zg}) and (\ref{zm}) respectively,
the formulae (\ref{zi1}) and (\ref{zim}) give
 immediately the leading terms for $\beta(g)$ and $\gamma_m(g)$:
\be
\beta(g)= -{g^3\over 16\pi^2}\left[\frac{11}{3}N-\frac{2}{3}f \right],
\ee
\be
\gamma_m(g)={g^2\over 16\pi^2} 6 C_F.
\ee
With this technique it is also easy to show that the anomalous
dimensions of the quark field $(\gamma_q)$ and the qluon field
$(\gamma_G)$ defined by
\be
\gamma_q(g)=\frac{1}{2}{1\over Z_q}{d Z_q\over d\ln\mu},\quad\quad
\gamma_G(g)=\frac{1}{2}{1\over Z_3}{d Z_3\over d\ln\mu},
  \end{equation}
are given by
\be
\gamma_i(g)=- g^2{d Z_{i, 1}(g)\over d g^2} \quad\quad (i=q,G).
\end{equation}

\subsubsection{Compendium of Useful Results}
It will be useful to have a collection of results for $\beta(g)$, 
$\gamma(\alpha_s)$ and $Z_{q,1}(\alpha_s)$ including also 
two-loop contributions.
They are:
\begin{equation}\label{bg01}
\beta(g)=-\beta_0{g^3\over 16\pi^2}-\beta_1{g^5\over (16\pi^2)^2}
  \end{equation}

\begin{equation}\label{gama}
\gamma_m(\as)=\gamma^{(0)}_{m}\aspi + \gamma^{(1)}_{m}\left(\aspi\right)^2
\end{equation}

\begin{equation}\label{zq1a} 
Z_{q, 1}(\as)=a_1\aspi + a_2\left(\aspi\right)^2
\end{equation}
where
\begin{equation}\label{b0b1}
\beta_0={{11N-2f}\over 3}\qquad
\beta_1={34\over 3}N^2-{10\over 3}Nf-2C_F f
\ee
\begin{equation}\label{gm01} \gamma^{(0)}_{m}
=6C_F\qquad \gamma^{(1)}_{m}=C_F\left(
     3C_F+{97\over 3}N-{10\over 3}f\right)  \end{equation}
\begin{equation}\label{a1a2} a_1=-C_F\qquad a_2=C_F\left(
     {3\over 4}C_F-{17\over 4}N+{1\over 2}f\right)  
\end{equation}
\be
C_F={{N^2-1}\over{2N}}.
\end{equation}

These results are valid in the MS ($\overline{\rm MS}$) scheme.
$N$ is the number of colours and $f$ the number of quark flavors.
Whereas $\beta_0$, $\beta_1$, $\gamma_m^{(0)}$, $\gamma_m^{(1)}$
are gauge independent, $a_1$ and $a_2$ given here have been obtained
in the $\xi=1$ gauge.

\subsection{Running Coupling Constant}
With the expansion (\ref{bg01}), the renormalization group equation 
(\ref{rgbe}) for $g(\mu)$ can be written as follows:
\begin{equation}\label{rga}
{d\as\over d\ln\mu}=-2\beta_0 {\as^2\over 4\pi}-2\beta_1
  {\as^3\over(4\pi)^2}  \end{equation}
Solving it, one finds \cite{BBDM}:

\begin{equation}\label{QCDC}
{{\alpha_s(\mu)}\over {4\pi}} =
{{1}\over{\beta_0 \ln(\mu^2/\Lambda^2_{\overline{MS}})}}
- {{\beta_1}\over{\beta^3_0}} {{\ln \ln (\mu^2/\Lambda^2_{\overline{MS}})}
\over
{\ln^2(\mu^2/\Lambda^2_{\overline{MS}})}}.
\end{equation} 
Let us make a few comments:
\begin{itemize}
\item
$\Lambda_{\overline{MS}}$ is a QCD scale characteristic for the
$\overline{\rm MS}$ scheme.
It can be determined by
measuring $\alpha_s(\mu)$ at a single value of $\mu$.
To this end the quantity used to determine $\alpha_s(\mu)$ has to
be calculated in the ${\overline{\rm MS}}$ scheme. Strictly speaking
$\alpha_s(\mu)$ should really read $\alpha_{s,\overline{\rm MS}}$ but
we will work exclusively in the ${\overline{\rm MS}}$ scheme and
this complication of the notation is unnecessary. Yet it is useful
to quote the relation to the MS scheme. Using (see (\ref{msms}))
\begin{equation}\label{ms11}
\mu\equiv\mu_{\overline{MS}}=\mu_{MS} e^{\gamma_E/2}(4\pi)^{-1/2}
\end{equation}
in (\ref{QCDC}) one finds the
relation between $\as$ in
the MS and $\overline{\rm MS}$ schemes:
\begin{equation}\label{amsb}
\alpha_{s,MS}=\alpha_{s,\overline{MS}}\left(1+\beta_0(\gamma_E-\ln 4\pi)
 {\alpha_{s,\overline{MS}}\over 4\pi}\right)  \end{equation}
or
\begin{equation}\label{msbl}
\Lambda^2_{\overline{MS}}=4\pi e^{-\gamma_E}\Lambda^2_{MS}
  \end{equation}
\item
$\Lambda_{\overline{MS}}$ and $\alpha_s(\mu)$
depend on $f$, the number of ``effective'' flavours present
in $\beta_0$ and $\beta_1$.
What ``effective'' $f$ really means will be
explained in the next section. For the time being we adopt the following
working procedure:
\begin{equation}\label{feff}
f=\cases{6  &~~~ $\mu\geq\mt$\cr
         5  &~~~ $\mb\leq\mu\leq\mt$ \cr
         4  &~~~ $\mc\leq\mu\leq\mb$ \cr
         3  &~~~ $\mu\leq\mc$.\cr  } 
\end{equation}
Denoting by $\alpha^{(f)}_s$ the effective coupling constant for a
theory with $f$ effective flavours and by $\Lambda^{(f)}_{\overline{MS}}$ 
the corresponding QCD scale parameter, we have the following boundary
conditions which follow from the continuity of $\alpha_s$:
\begin{equation}\label{asc}
\alpha_s^{(6)}(\mt)=\alpha_s^{(5)}(\mt),
\qquad
\alpha_s^{(5)}(\mb)=\alpha_s^{(4)}(\mb),
\qquad
\alpha_s^{(4)}(\mc)=\alpha_s^{(3)}(\mc).
\end{equation}
\ei

The above continuity conditions allow to find values of 
$\Lambda^{(f)}_{\overline{MS}}$
for different $f$ once one particular $\Lambda^{(f)}_{\overline{MS}}$
is known. In table \ref{tab:alphas} we show different 
$\alpha^{(f)}_s(\mu)$ and
$\Lambda^{(f)}_{\overline{MS}}$ corresponding to
\begin{equation}\label{asexp}
\alpha_s^{(5)}(\mz)=0.118 \pm 0.005,
\end{equation}
which is in the ball park of the present world average extracted from
different processes \cite{Schmelling}.
To this end we have set
$\mc=1.3~\gev$, $\mb=4.4~\gev$ and $\mt=170~\gev$.
We observe that for $\mu\geq\mc $ the values of
$\alpha_s(\mu)$ are sufficiently small that
the effects of strong interactions can be
treated in perturbation theory. When one
moves to low energy scales, $\alpha_s$ increases and at
$\mu\approx {\cal O}(1~\gev)$ and high values of 
$\Lambda^{(3)}_{\overline{MS}}$  one finds
$\alpha_s^{(3)}(\mu)>0.5$. This signals breakdown of
perturbation theory for scales lower than $1~\gev$. Yet it is
gratifying that strong interaction contributions to weak decays coming
from scales higher than $1~\gev$ can be treated by perturbative
methods.

\begin{table}[thb]
\caption[]{Values of $\alpha^{(f)}_s(\mu)$ and
$\Lambda^{(f)}_{\overline{MS}}$ corresponding to given values of
$\alpha_s^{(5)}(\mz)$.
\label{tab:alphas}}
\begin{center}
\begin{tabular}{|c|c|c|c|c|c|}\hline
 $\alpha_s^{(6)}(\mt)$& $0.1037$& $0.1054$& $0.1079$ & $0.1104$ &  $0.1120$ 
\\ \hline
 $\Lms^{(6)}[\mev]$& $66$& $76$& $92$ & $110$ &  $123$ 
\\ \hline\hline
 $\alpha_s^{(5)}(\mz)$& $0.113$& $0.115$& $0.118$ & $0.121$ &  $0.123$ 
\\ \hline
 $\Lms^{(5)}[\mev]$& $169$& $190$& $226$ & $267$ &  $296$ 
\\ \hline
 $\alpha_s^{(5)}(\mb)$& $0.204$& $0.211$& $0.222$ & $0.233$ &  $0.241$ 
\\ \hline\hline
 $\Lms^{(4)}[\mev]$& $251$& $278$& $325$ & $376$ &  $413$ 
\\ \hline
 $\alpha_s^{(4)}(\mc)$& $0.336$& $0.357$& $0.396$ & $0.443$ &  $0.482$ 
\\ \hline\hline
 $\Lms^{(3)}[\mev]$& $297$& $325$& $372$ & $421$ &  $457$ 
\\ \hline
 $\alpha_s^{(3)}(1\gev)$& $0.409$& $0.444$& $0.514$ & $0.605$ &  $0.690$ 
\\ \hline
 \end{tabular}
\end{center}
\end{table}
Finally we would like to give an equivalent expression for $\alpha_s$,
which allows to calculate $\alpha_{s}(\mu)$ directly from
the experimental value given in (\ref{asexp}):
\be\label{alphaNLL}
\as(\mu) = \frac{\as(M_Z)}{v(\mu)} \left[1 - \f{\beta_1}{\beta_0} 
           \frac{\as(M_Z)}{4 \pi}    \f{\ln v(\mu)}{v(\mu)} \right],
\ee
where 
\be\label{v(mu)}
v(\mu) = 1 - \beta_0 \frac{\as(M_Z)}{2 \pi} 
\ln \left( \frac{M_Z}{\mu} \right),
\ee
Strictly speaking (\ref{alphaNLL}) is valid for the $f=5$ theory.
In order to find $\as(\mu)$ for $f\not=5$ one has to proceed as in
(\ref{feff}) and (\ref{asc}).
\subsection{Running Quark Mass}
Let us next find the $\mu$-dependence of $m(\mu)$. With
$dg/d\ln\mu=\beta(g)$ the solution of
\begin{equation}\label{rggm2}
{d m(\mu)\over d\ln\mu}=-\gamma_m(g) m(\mu)  
\end{equation}
is obviously
\begin{equation}\label{UMM}
 m(\mu) = m(\mu_0) \exp \left[ 
  - \int_{g(\mu_0)}^{g(\mu)}{dg' \frac{\gamma_m(g')}{\beta(g')}}\right]. 
\end{equation}
Here
$m(\mu_0)$ is the
value of the running mass at the scale $\mu_0$. For instance:
$\ms(2~\gev)$. 
Inserting the expansions for $\gamma_m(g)$ and $\beta(g)$ into (\ref{UMM}) 
and expanding in $\as$ gives:
\begin{equation}\label{mmu}
m(\mu)=
m(\mu_0)
\left[{\as(\mu)\over\as(\mu_0)}\right]^{\gamma^{(0)}_m\over 2\beta_0}
\left[1+\left({\gamma^{(1)}_{m}\over 2\beta_0}-{\beta_1\gamma^{(0)}_{m}\over
  2\beta^2_0}\right){\as(\mu)-\as(\mu_0)\over 4\pi}\right].
\end{equation}
In the literature the running quark mass is often denoted by
$\overline{m}(\mu)$. In these lectures we will use both notations:
$\overline{m}(\mu)\equiv {m}(\mu)$.

Since formulae similar to (\ref{rggm2})--(\ref{mmu}) will
often appear in these lectures, it is useful to derive at least the
leading term in (\ref{mmu}).
Keeping the leading terms in $\gamma_m(g)$ and $\beta(g)$ we have
\be
  - \int_{g(\mu_0)}^{g(\mu)} dg' \frac{\gamma_m(g')}{\beta(g')}
  = \int_{g(\mu_0)}^{g(\mu)} dg' \frac{\gamma_m^{(0)}}{\beta_0}
    \frac{1}{g'}=\frac{1}{2} \frac{\gamma_m^{(0)}}{\beta_0}
    \ln\frac{g^2(\mu)}{g^2(\mu_0)}
\ee
which inserted in (\ref{UMM}) gives the leading term in (\ref{mmu}).
Keeping also the NLO terms in $\gamma_m(g)$ and $\beta(g)$ and
proceeding in a similar manner one readily finds the NLO term
in (\ref{mmu}).

Let us practice a bit the formula (\ref{mmu}). Since the power
$\gamma^{(0)}_m/2\beta_0$ is positve, $m(\mu)$ similarly to $\alpha_s$
decreases with increasing $\mu$. 
Using  $\Lms^{(4)}=325~\mev$ and
$\mc=1.3~\gev$ we find for instance a dictionary between the values
of the strange quark mass $m_s$ evaluated at different scales
using as the input the values $m_s(\mc)$ with $\mc=1.3~\gev$. 
We note a rather
sizable dependence of $m_s$ on $\mu$.
\begin{table}[thb]
\caption[]{The dictionary between the values of $m_s(\mu)$ 
in units of $\mev$.
\label{tab:ms}}
\begin{center}
\begin{tabular}{|c|c|c|c|c|c|}\hline
  $\ms(\mc)$& $ ~75$& $100$& $125$ & $150$ &  $175$ \\ \hline
 $\ms(2~\gev)$& $ ~64$& $~86$& $107$ & $129$ &  $150$ \\ \hline
 $\ms(1~\gev)$& $ ~86$& $115$& $144$ & $173$ &  $202$ \\ \hline
 \end{tabular}
\end{center}
\end{table}

On the other hand the $\mu$ dependence of the top quark mass 
$\mt(\mu_t)$ is much weaker. Taking
$\mt(170\gev) = 170\gev$,  $\alpha_s^{(5)}(\mz)=0.118$ and
scanning $\mu_t$ in the range  $100\gev\le \mu_t \le 300\gev$
we find 
\be
163.0\gev\le \mt(\mu_t)\le 177.4\gev~.
\ee

\subsection{RG Improved Perturbation Theory}
The structure of (\ref{alphaNLL}) and (\ref{mmu})
makes it clear that RG approach goes
 beyond the usual perturbation theory. In order to see what is going on,
let us consider the leading term in (\ref{alphaNLL}):
\be \label{alphaLO}
\as(\mu) = \frac{\as(M_Z)}{1 - \beta_0 \frac{\as(M_Z)}{2 \pi} 
\ln \left( \frac{M_Z}{\mu} \right)}~.
\ee
Expanding it  in $\as(M_Z)$ we find:
\be \label{EXPLO}
\as(\mu) = \as(M_Z)\left[1 +\sum_{n=1}^\infty
\left( \beta_0 \frac{\as(M_Z)}{2 \pi} 
\ln \frac{M_Z}{\mu}\right)^n\right]  
\ee
We conclude that the
solution of the renormalization group equations sums automatically 
large logarithms $ \log (\mz/\mu) $ which appear for $ \mu<<\mz $.
More generally
\be
 {\rm LO}:\quad {\rm Summation~of}~~
 \left(\as(M_Z)\ln \frac{M_Z}{\mu}\right)^n~,
\ee
\be
{\rm  NLO}:\quad {\rm Summation~ of}~~
 \as(M_Z)^n\left(\ln \frac{M_Z}{\mu}\right)^{n-1}~.
\ee
In particular we note that
the expansion (\ref{mmu}) in terms of 
$\as(\mu)$ does not
involve large logarithms and a few terms suffice to obtain reliable
result. 
(\ref{mmu}) is an example of a
{\it Renormalization Group Improved Perturbative Expansion}.
We will encounter similar expansions for other quantities in the
course of these lectures
\subsection{Final Comments}
We have collected certain information about QCD and tools
like  renormalization group methods
which allow to sum large logarithms. We have also discussed
the $\mu$ dependences of the running QCD coupling and the running
quark masses. Yet all these nice and powerful tools are still
insufficient to attack the question of Weak Decays. Yes, what we
still need is the operator product expansion.

\section{Operator Product Expansion in Weak Decays}
            \label{sec:basicform:prelim}
\setcounter{equation}{0}
\subsection{Preliminaries}
Weak Decays of Hadrons are mediated through weak interactions of quarks,
whose strong interactions, binding the quarks into hadrons, are
characterized by typical hadronic energy scale of $\ord (1~\gev)$,
much lower than the
scale of weak interactions: $\ord (M_{W,Z})$.
Our goal is therfore to  derive
an effective low energy theory describing the weak interactions of
quarks. 
The formal framework to achieve this is precisely the 
Operator Product Expansion (OPE) \cite{OPE,ZIMM,WIT}. 
\begin{figure}[hbt]
\vspace{0.10in}
\centerline{
\epsfysize=2in
\epsffile{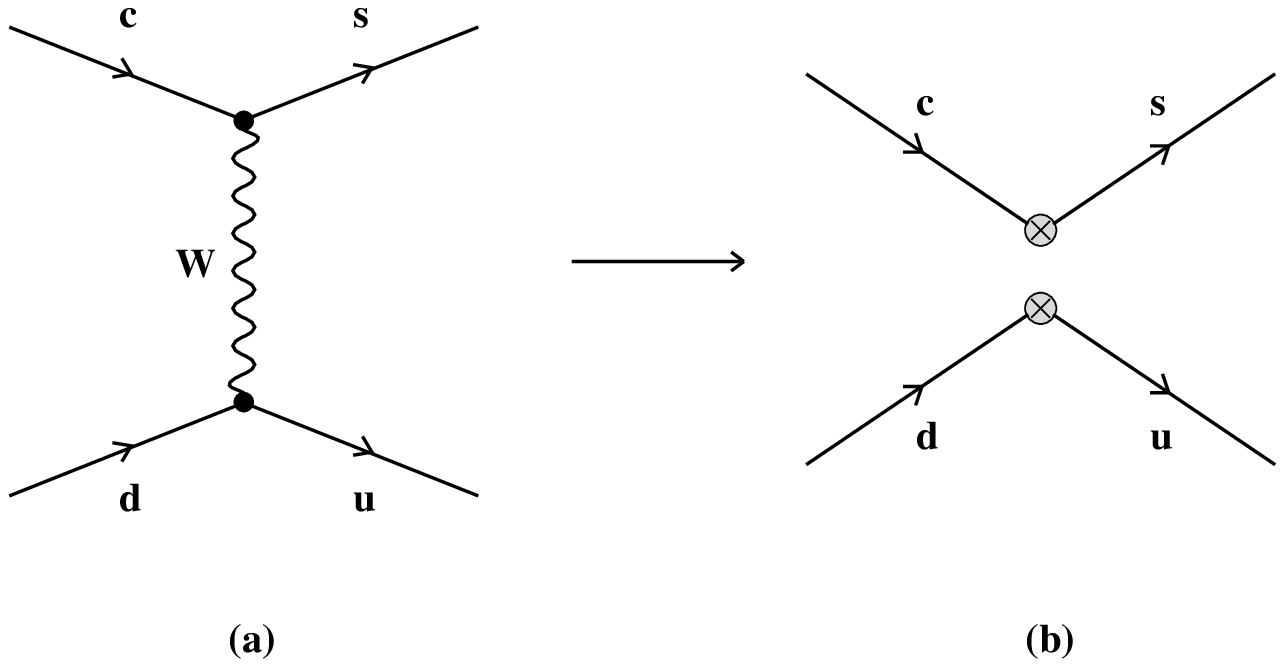}
}
\vspace{0.08in}
\caption[]{$c\to su\bar d$ at the Tree-Level.
\label{L:4}}
\end{figure}

\subsection{Basic Idea}
Consider 
the quark level transition  $c\to su\bar d$.
Disregarding QCD effects for the moment,  the corresponding
tree-level W-exchange
amplitude (fig.~\ref{L:4}a multiplied by ``i") is given by
\begin{eqnarray}\label{aope}
A&=&-{G_F\over\sqrt{2}}V^*_{cs}V_{ud}^{}{M^2_W\over k^2-M^2_W}
  (\bar sc)_{V-A}(\bar ud)_{V-A} \nonumber\\
 &=& {G_F\over\sqrt{2}}V^*_{cs}V_{ud}^{}
  (\bar sc)_{V-A}(\bar ud)_{V-A} + \ord({k^2\over M^2_W})
\end{eqnarray}
where 
\be
(\bar sc)_{V-A}\equiv
\bar s\gamma_{\mu} (1-\gf)c~.
\ee
Since $k$, the momentum transfer through the $W$ propagator, is very
small as compared to $\mw$, terms of the order $\ord({k^2/M^2_W})$
can safely be neglected and the full amplitude $A$ can be approximated
by the first term on the r.h.s of (\ref{aope}).
Now the result in (\ref{aope}) may  also be obtained from 
\begin{equation}\label{hc0}
{\cal H}_{eff}={G_F\over\sqrt{2}}V^*_{cs}V_{ud}^{}
  (\bar sc)_{V-A}(\bar ud)_{V-A} + {\rm High~ D~ Operators},
  \end{equation}
 where the higher dimension operators, typically involving derivative
terms, correspond to the terms $\ord({k^2/M^2_W})$ in (\ref{aope}).
Neglecting the latter terms corresponds to the neglect of higher
dimensional operators. In what follows we will always neglect 
the higher dimensional operators keeping only the operators with
dimensions five
and six.
This
 simple example illustrates the basic idea of  OPE:
the product of two charged current
operators is expanded into a series of local operators,
whose contributions are weighted by effective coupling constants,
the Wilson coefficients. In this particular example the leading
operator has the dimension 6 and its Wilson coefficient in the
normalization of the pilot formula (\ref{b1}) is simply equal
unity. This value will be changed by QCD corrections as we will
see few pages below. Moreover QCD corrections to the diagrams in
fig. \ref{L:4} will generate another operator.

\subsection{Formal Approach}
Let us be a bit more formal for a moment and investigate whether
the same result can be obtained using the path integral formalism.
We will see that this is indeed the case. This discussion will
on the one hand provide a formal basis for the simple procedure
given above and on the other hand will give us more insight
in the virtues of OPE. Simultaneously we will discover that
there is no need to be very formal for the rest of the lectures
and we can proceed by simply generalizing our simple procedure
of section 5.2 to more complicated situations in which also QCD 
effects and more complicated diagrams are
present. 

Our formal discussion follows \cite{BBL} and consists of four steps.

{\bf Step 1}

Consider
the generating functional for Green functions in the path integral
formalism. The relevant part for our discussion is
\begin{equation}\label{zw1}
Z_W\sim\int [dW^+][dW^-] \exp(i\int d^4x {\cal L}_W)  \end{equation}
where 
\begin{eqnarray}\label{lwjw}
\lefteqn{{\cal L}_W=
-{1\over 2}(\partial_\mu W^+_\nu-\partial_\nu W^+_\mu)
 (\partial^\mu W^{-\nu}-\partial^\nu W^{-\mu})+M^2_W W^+_\mu W^{-\mu}}
\hspace{3cm} \nonumber\\
& & +{g_2\over 2\sqrt{2}}(J^+_\mu W^{+\mu}+J^-_\mu W^{-\mu}),
\end{eqnarray}
\begin{equation}\label{jpn}
J^+_\mu=V_{pn} \bar p\gamma_\mu(1-\gf)n\qquad p=(u, c, t)
\quad n=(d, s, b)\qquad J^-_\mu=(J^+_\mu)^\dagger.  \end{equation}

{\bf Step 2:}

We use the unitary gauge for
the W field. 
Introducing the operator:
\begin{equation}\label{kxy}
K_{\mu\nu}(x, y)=\delta^{(4)}(x-y)\left[g_{\mu\nu}(\partial^2+
  M^2_W)-\partial_\mu\partial_\nu\right]   \end{equation}
we have, after discarding a total derivative in the W kinetic term,

\begin{eqnarray}\label{zw2}    \lefteqn{
Z_W\sim\int [dW^+][dW^-] \exp\biggl[ i\int d^4x d^4y W^+_\mu(x)
K^{\mu\nu}(x, y) W^-_\nu(y)  } \hspace{3cm}\nonumber\\
& & {}+i{g_2\over 2\sqrt{2}}\int d^4x
 J^+_\mu W^{+\mu}+J^-_\mu W^{-\mu} \biggr].
\end{eqnarray}
The inverse of $K_{\mu\nu}$, denoted by $\Delta_{\mu\nu}$, and
defined through
\begin{equation}\label{kde1}
\int d^4y K_{\mu\nu}(x, y) \Delta^{\nu\lambda}(y, z)=
 g^{\ \lambda}_\mu \delta^{(4)}(x-z)  \end{equation}
is the W propagator in the unitary gauge
\begin{equation}\label{dexy}
\Delta_{\mu\nu}(x, y)=\int{d^4k\over (2\pi)^4}\Delta_{\mu\nu}(k)
  e^{-i k(x-y)}  \end{equation}
\begin{equation}\label{dek}
\Delta_{\mu\nu}(k)={-1\over k^2-M^2_W}\left(g_{\mu\nu}-
  {k_\mu k_\nu\over M^2_W}\right)~.   \end{equation}

{\bf Step 3:}

Performing the gaussian functional integration over $W^\pm(x)$ in
(\eqn{zw2}) explicitly, we arrive at
\begin{equation}\label{zetw}
Z_W\sim\exp\left[ -i\int{g^2_2\over 8}J^-_\mu(x)\Delta^{\mu\nu}(x, y)
J^+_\nu(y) d^4x d^4y \right]   \end{equation}
This  result implies a nonlocal action functional for the quarks:
\begin{equation}\label{snl}
{\cal S}_{nl}=\int d^4x {\cal L}_{kin}-
{g^2_2\over 8}\int d^4x d^4y J^-_\mu(x)\Delta^{\mu\nu}(x, y)
J^+_\nu(y)    \end{equation}
where the second term represents
 charged current interactions of quarks.

{\bf Step 4:}

Finally, we expand this second, nonlocal term in powers of
$1/M^2_W$ to obtain a series of local interaction operators of
dimensions that increase with the order in $1/M^2_W$. To lowest order
\begin{equation}\label{dloc}
\Delta^{\mu\nu}(x, y)\approx{g^{\mu\nu}\over M^2_W}\delta^{(4)}(x-y) 
\end{equation}
and the second term in (\eqn{snl}) becomes
\begin{equation}\label{jjx}
-{g^2_2\over 8M^2_W}\int d^4x J^-_\mu(x) J^{+\mu}(x)  \end{equation}
corresponding to the usual effective charged current interaction
Lagrangian
\begin{equation}\label{leff}
{\cal L}_{int,eff}=-{G_F\over\sqrt{2}} J^-_\mu(x) J^{+\mu}(x)=-
{G_F\over\sqrt{2}}V^*_{pn}V_{p'n'}^{}(\bar np)_{V-A}
  (\bar p'n')_{V-A}
\end{equation}
which contains, among other terms, the leading contribution to 
(\ref{hc0}).

Let us note several  basic aspects of this approach:

\begin{itemize}
\item  
Formally, the procedure to approximate the interaction term in
(\ref{snl}) by (\ref{jjx}) is an example of short distance OPE.
The product
of the local operators $J^-_\mu(x)$ and $J^+_\nu(y)$, to be taken at
short-distances due to the convolution with the massive, short-range
W propagator $\Delta^{\mu\nu}(x, y)$, is expanded
into a series of composite local operators. The leading term
is shown in (\eqn{jjx}).
\item The dominant contributions in the short-distance expansion come
from the operators of lowest dimension (six in the present example).
The operators of higher
dimensions can usually be neglected in weak decays.
\item 
 OPE series is
equivalent to the original theory, when considered to all orders in
$1/M^2_W$. 
The truncation of the operator series  yields a systematic
approximation scheme for low energy processes, neglecting contributions
suppressed by powers of $k^2/M^2_W$. 
\item 
In going from the full to the effective theory the W boson is
removed as an explicit, dynamical degree of freedom: it is
 ``integrated out'' in step 3 of our procedure.
Alternatively in the canonical operator
formalism the W field  gets
``contracted out'' through the application of Wick's theorem.
From the point of view of low energy dynamics, the effects
of a short-range exchange force mediated by a heavy boson
approximately corresponds to a point interaction familiar from
the Fermi Theory.
\item
Similarly one can ``integrate out'' or ``contract out'' heavy quarks.
This gives {\it Effective f-quark theories} where f denotes the
"light" quarks which have not been integrated out. We now understand
what the effective number of flavours introduced in connection
with the formula (\ref{feff}) really means. By going from higher to
lower $\mu$ scales one integrates out systematically flavours with
masses higher than the actual value of $\mu$. However, as we will
stress below, in connection with renormalization group ideas, 
there is some freedom at which $\mu$ a given flavour is integrated out.
For instance one can extend the five flavour theory down to $\mu=\mb/2$. 
\ei

All this was a bit formal but fortunately we make still another
observation. 
The approach of evaluating the relevant Green functions (or amplitudes)
directly in order to construct the OPE, as in (\eqn{aope}), gives
the same result as the more formal technique employing path integrals.
Consequently we can return, putting aside path integrals, to our
Feynman diagram calculations. Our first task is to investigate
how (\ref{aope}) or (\ref{hc0}) changes when QCD effects are included. 

\subsection{OPE and Short Distance QCD Effects}
            \label{sec:basicform:ope}
\subsubsection{Preliminaries}
Due to the asymptotic freedom of QCD, the short distance
QCD corrections to weak decays, that is the contribution
of hard gluons at energies of the order $\ord(M_W)$ down to hadronic
scales $\ord(1\gev)$, can be treated  in 
the renormalization group (RG) improved perturbation theory. 
We will illustrate this on a simple example of the
$c\to su\bar d$ transition beginning with the ordinary
perturbation theory, subsequently summing leading logarithms
by the RG method and finally generalizing the result to include
next-to-leading logarithms. We will do this in some detail
emphasizing certain characteristic features of this approach.
In particular we will discuss at length the scale and
renormalization scheme dependences advertised in the pilot
section of these lectures. Once all these features are well
understood it will be straightforward to proceed to other
transitions and to generalize the approach to more exciting
situations involving penguins and boxes.

For the $c\to su\bar d$ transition we had without QCD effects
\begin{equation}\label{amp0}
{\cal H}^{(0)}_{eff}={G_F\over\sqrt{2}}V^*_{cs}V_{ud}^{}
  (\bar s_\alpha c_\alpha)_{V-A}(\bar u_\beta d_\beta)_{V-A}
\end{equation}
where the summation over repeated color indices is understood.  

With QCD effects ${\cal H}^{(0)}_{eff}$ is generalized to
\begin{equation}\label{hq12}
{\cal H}_{eff}={G_F\over\sqrt{2}}V^\ast_{cs}V_{ud}
(C_1(\mu) Q_1+C_2(\mu) Q_2) 
\end{equation}
where
\begin{equation}\label{q1c} Q_1=
(\bar s_\alpha c_\beta)_{V-A}(\bar u_\beta d_\alpha)_{V-A}  
 \end{equation}
\begin{equation}\label{q2c} Q_2
=(\bar s_\alpha c_\alpha)_{V-A}(\bar u_\beta d_\beta)_{V-A}
  \end{equation}

The essential features of this Hamiltonian are:
\begin{itemize}
\item In addition to the original operator $Q_2$ (with index 2 for 
historical reasons) 
 a new operator $Q_1$ with the {\it same flavour} form
but {\it different colour structure} is generated. 
That a new operator has to be introduced is evident if we
inspect the colour structure of the diagrams (b) and (c) in 
fig.~\ref{L:13}. They contain the
product of the color charges $T^a_{\alpha\beta}$ and 
$T^a_{\gamma\delta}$ which
using the colour algebra can be rewritten as follows
\begin{equation}\label{tata}
T^a_{\alpha\beta}T^a_{\gamma\rho}
=-{1\over 2N}\delta_{\alpha\beta}\delta_{\gamma\delta}+{1\over 2}
\delta_{\alpha\delta}\delta_{\gamma\beta}   \end{equation}
The first term on the r.h.s gives a correction to the coefficient
of the operator $Q_2$ and the second term gives life to the new
operator $Q_1$.
\item The Wilson coefficients $C_1$ and $C_2$, the coupling constants
for the interaction terms $Q_1$ and $Q_2$, become calculable nontrivial
functions of $\as$, $M_W$ and the renormalization scale $\mu$.
\item
If QCD is neglected, $C_1=0$, $C_2=1$ and
(\eqn{hq12}) reduces to (\eqn{amp0}).
\end{itemize}

\subsubsection{Calculation of Wilson Coefficients}
Our first task is the calculation of the coefficients $C_{1, 2}$
in the ordinary perturbation theory.
$C_{1, 2}$ can be  determined
by the requirement that the amplitude $A_{full}$ in the full theory be
reproduced by the corresponding amplitude in the effective theory
(\eqn{hq12}):
\begin{equation}\label{acq}
A_{full}=A_{eff}=
{G_F\over\sqrt{2}}V^*_{cs}V_{ud}^{}(C_1\langle Q_1\rangle +
C_2\langle Q_2\rangle)   \end{equation}

This procedure is called ``the {\it matching} of the 
full theory onto the
effective theory''. We recall that the full theory is the one in
which all particles appear as dynamical degrees of freedom.
In the case at hand the effective theory is constructed by
integrating out the $W$ field only. The matching procedure which
gives the values of $C_1$ and $C_2$ proceeds in three steps \cite{BBDM}.
The explicit three steps presented below are sufficient for
the subsequent summation of the leading logarithms or equvalently
for the leading term of the RG improved perturbation theory. 
We will generalize these steps in the next section in order to be able
to include also the NLO term in this expansion.

Here we go:

{\bf Step 1: Calculation of $A_{full}$}

The current-current diagrams of fig.~\ref{L:13}\,(a)--(c) and their
symmetric counterparts,
give for the full amplitude $A_{full}$ to $\ord(\as)$ ($m_i=0$, $p^2<0$):

\begin{eqnarray}\label{amp}
A_{full}
&=&
{G_F\over\sqrt{2}}V^\ast_{cs}V_{ud}
\Bigl[ \left( 1+2C_F \aspi (\frac{1}{\varepsilon}+\ln{\mu^2\over -p^2}) \right)
S_2 +{3\over N}\aspi\ln{M^2_W\over -p^2} S_2
\nonumber\\
& &
-3\aspi\ln{M^2_W\over -p^2} S_1 \Bigr] 
\end{eqnarray}
Here:
\begin{equation}\label{s1c} 
S_1\equiv \langle Q_1\rangle_{tree}=
(\bar s_\alpha c_\beta)_{V-A}(\bar u_\beta d_\alpha)_{V-A}
\end{equation}
\begin{equation}\label{s2c} 
S_2\equiv \langle Q_2\rangle_{tree}=
(\bar s_\alpha c_\alpha)_{V-A}(\bar u_\beta d_\beta)_{V-A}
\end{equation}
are just the tree level matrix elements of $Q_1$ and $Q_2$. A few comments
should be made.
\begin{itemize}
\item
We use the term ``amplitude'' in the meaning of an ``amputated Green
function'' (multiplied by "i"). 
Correspondingly operator matrix elements are
amputated Green functions with operator
insertion. Thus gluonic self energy
corrections on external legs are not included.
\begin{figure}[hbt]
\vspace{0.10in}
\centerline{
\epsfysize=1.25in
\epsffile{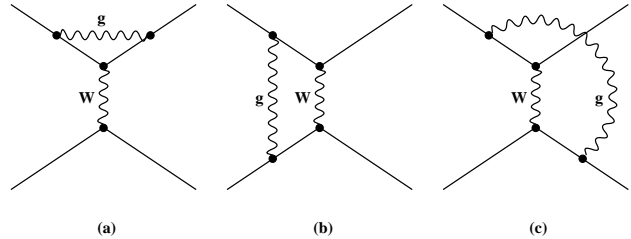}
}
\vspace{0.08in}
\caption[]{One-loop current-current 
diagrams in the full theory.
\label{L:13}}
\end{figure}
\item
For simplicity we have chosen all external momenta $p$ to be equal
and set all quark masses to zero. As we will see below this choice
has no impact on the coefficients $C_i$.
\item
We have kept only logarithmic corrections
$\sim\as\cdot\log$ and discarded constant contributions of order
$\ord(\as)$, which corresponds to the leading log approximation (LO).
\item
The singularity $1/\eps$ can be removed by the quark field renormalization.
This is, however, not necessary for finding $C_i$ as we will see soon.
\end{itemize}

\begin{figure}[hbt]
\vspace{0.10in}
\centerline{
\epsfysize=1.25in
\epsffile{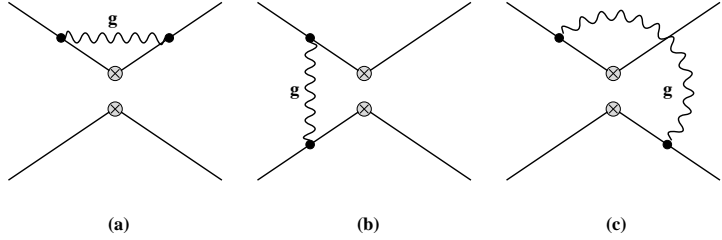}
}
\vspace{0.08in}
\caption[]{One loop current-current  diagrams
in
the effective theory. The 4-vertex ``$\otimes\,\,\otimes$'' denotes the
insertion of a 4-fermion operator $Q_i$.
\label{L:15}}
\end{figure}

{\bf Step 2: Calculation of Matrix Elements $\langle Q_i\rangle$}

The unrenormalized current-current matrix elements
of $Q_1$ and $Q_2$ are found at $\ord(\as)$ by calculating 
the diagrams in  fig.~\ref{L:15}\,(a)-(c) and their symmetric counter-parts.
Adding the contributions without QCD corrections 
($S_1$ and $S_2$ respectively) and using
the same assumptions about the external legs as in step 1, 
we have
\begin{eqnarray}\label{q10}
\langle Q_1\rangle^{(0)} &= & 
\left(1+2C_F \aspi\left({1\over\eps}+\ln{\mu^2\over -p^2}
\right)\right)S_1+{3\over N}\aspi\left({1\over\eps}+\ln{\mu^2\over -p^2}
\right)S_1
\nonumber\\
& &-3\aspi\left({1\over\eps}+\ln{\mu^2\over -p^2}\right) S_2  
\end{eqnarray}
\begin{eqnarray}\label{q20}
\langle Q_2\rangle^{(0)} &=& 
\left(1+2C_F \aspi\left({1\over\eps}+\ln{\mu^2\over -p^2}
\right)\right)S_2+{3\over N}\aspi\left({1\over\eps}+\ln{\mu^2\over -p^2}
\right)S_2
\nonumber\\
& & -3\aspi\left({1\over\eps}+\ln{\mu^2\over -p^2}\right) S_1  
\end{eqnarray}

The divergences in the first terms can again be  eliminated through 
the quark field
renormalization. However, in contrast to the full amplitude
in (\ref{amp}), the
resulting expressions are still divergent after this renormalization.
To remove these additional divergences
multiplicative renormalization, refered to as {\it operator renormalization},
is necessary:
\begin{equation}
Q_i^{(0)} = Z_{ij} Q_j~.
\label{AL}
\end{equation}
We observe that the renormalization constant is in this case a $2\times 2$
matrix $\hat Z$.
Using (\ref{grel})  with $(n_F,n_G)=(4,0)$, we find 
the relation between the unrenormalized
($\langle Q_i\rangle^{(0)}$) and the renormalized amputated Green
functions ($\langle Q_i\rangle$):
\begin{equation}\label{q0zq}
\langle Q_i\rangle^{(0)}=Z^{-2}_q Z_{ij}\langle Q_j\rangle~.
\end{equation}

$Z^{-2}_q$ removes the $1/\varepsilon$ divergences in the
first terms in (\eqn{q10}) and (\eqn{q20}). $Z_{ij}$ remove
the remaining divergences.
From (\eqn{q10}), (\eqn{q20})  we read off 
($\overline{\rm MS}$-scheme)
\begin{equation}\label{zll} \hat Z = 1+ \aspi {1\over\eps}
 \left(\begin{array}{cc}  3/N & -3 \\
                          -3 & 3/N
    \end{array}\right)   \end{equation}

Thus the renormalized matrix elements $\langle Q_i\rangle$
are given by
\begin{equation}\label{q1re}
\langle Q_1\rangle=\left(1+2C_F \aspi\ln{\mu^2\over -p^2}
\right)S_1+{3\over N}\aspi\ln{\mu^2\over -p^2}S_1-
3\aspi\ln{\mu^2\over -p^2} S_2~,   \end{equation}
\begin{equation}\label{q2re}
\langle Q_2\rangle=\left(1+2C_F \aspi\ln{\mu^2\over -p^2}
\right)S_2+{3\over N}\aspi\ln{\mu^2\over -p^2}S_2-
3\aspi\ln{\mu^2\over -p^2} S_1~.   \end{equation}

{\bf Step 3: Extraction of $C_i$}

Inserting $\langle Q_i\rangle$ into (\eqn{acq}) and comparing 
with (\eqn{amp})
we can now extract the coefficients $C_1$ and $C_2$. Yet, we have to be
a bit careful. In the full theory we did not perform any quark field 
renormalization whereas we did this renormalization in the effective
theory as seen in (\ref{q0zq}). This is clearly inconsistent and this
inconsistency is signalled by the divergent Wilson coefficient which is 
clearly wrong. To proceed correctly we have to either remove the
divergence in (\eqn{amp}) by performing quark field renormalization 
as in (\ref{q0zq}) or to leave (\eqn{amp}) as it is and remove
the quark field renormalization from (\ref{q0zq}). In both cases
the matching (\eqn{acq}) gives the same result
\begin{equation}\label{cc12}
C_1(\mu)=-3\aspi\ln{M^2_W\over\mu^2}~,   \qquad
C_2(\mu)=1+{3\over N}\aspi\ln{M^2_W\over\mu^2}~.   
\end{equation}
This simple example shows that it is essential in the process of
matching to treat the external
states in the full and the effective theory in the same manner in order
to obtain the correct result for the Wilson coefficients. In this example 
we were lucky. The inconsistency, which I made for pedagogical reasons,
was signalled by a leftover divergence. In the case of NLO calculations
were also finite non-logarithmic corrections have to be kept, a possible
inconsistency in matching is much harder to see and it is crucial that
at all stages of the matching the treatment of the external legs
on both sides of (\eqn{acq}) is the same. For this reason we are free
to decide whether we perform external field renormalization or not.
In the latter case the left-over divergences in the full and the effective
theory will simply cancel each other in the process of matching.
I discussed here the issue of the cancellation of the {\it ultraviolet} 
divergences
related to  external fields. 
The same comments apply to the {\it infrared}
divergences. For strategic reasons I will now discuss something else
and will return to the issue of infrared divergences in the context of 
matching a few pages below. 

\subsubsection{A Different Look}
The renormalization of
the interaction terms $C_i Q_i$ in the effective theory can also be achieved
in a different, but equivalent, way by using
the standard counter-term method. Here $C_i$ are treated as coupling
constants, which have to be renormalized. We follow here \cite{BBL}.

To this end let us
consider ${\cal H}_{eff}$ as the
starting point with fields and "coupling constants" $C_i$ regardes 
as bare quantities.
They are renormalized according to ($q$=$s$, $c$, $u$, $d$)
\begin{equation}\label{c0z2}
q^{(0)}=Z^{1/2}_q q  \qquad
 C^{(0)}_i=Z^c_{ij} C_j \end{equation}
where $\hat Z^c$ denotes the renormalization matrix for the couplings
$C_j$. It is evident that $\hat Z^c$ must be somehow related to the
renormalization matrix $\hat Z$ in (\ref{zll}). Let us find this relation.

Omitting the factor
${G_F\over\sqrt{2}}V^\ast_{cs}V_{ud}$ we have
\begin{equation}\label{ctcq}
{\cal H}_{eff}=C^{(0)}_iQ_i(q^{(0)})\equiv Z^2_qZ^c_{ij} C_jQ_i\equiv
C_iQ_i+(Z^2_q Z^c_{ij}-\delta_{ij})C_jQ_i  \end{equation}
where the first term on the r.h.s is written in terms of renormalized
couplings and fields $(C_i Q_i)$ and the second term is a counter--term.
The argument $q^{(0)}$ on the l.h.s of
(\eqn{ctcq}) indicates that the interaction vertices $Q_i$ are
composed of bare fields. 
Using
(\eqn{ctcq}) we get the
finite renormalized result
\begin{equation}\label{zqzc}
A_{eff}= Z^2_qZ^c_{ij} C_j\langle Q_i\rangle^{(0)} \end{equation}
On the other hand using (\ref{q0zq}), we have
\begin{equation}\label{zqa}
A_{eff}=C_j\langle Q_j\rangle=C_j Z^{-1}_{ji} Z^2_q \langle Q_i\rangle^{(0)}
\ee
Hence comparing the last two equations we finally find the relation
\begin{equation}\label{zcz}
Z^c_{ij}=Z^{-1}_{ji}.  \end{equation}
This relation will turn out to be very useful in deriving 
the renormalization
group equations for the couplings $C_i$.

\subsubsection{Operator Mixing and Diagonalization}
We have just seen, that gluonic corrections to the matrix element of 
the original operator $Q_2$ are not just proportional to $Q_2$ itself, 
but involve the additional structure $Q_1$.
Therefore, besides a $Q_2$-counter--term, a counter--term $\sim Q_1$ 
is needed to renormalize this matrix element. Similarly
the renormalization of $Q_1$ requires both $Q_1$ and $Q_2$
counter--terms. We say that the operators  
$Q_1$ and $Q_2$ {\it mix under renormalization}. 

For the study of the renormalization group properties of the
system $(Q_1,Q_2)$ it is useful to diagonalize it 
by going to a different operator basis defined by 
\be\label{dbasis}
Q_\pm=\frac{Q_2\pm Q_1}{2} \qquad
C_\pm=C_2\pm C_1~.
\ee
The new operators  $Q_+$ and $Q_-$ are renormalized
independently of each other:
\be\label{dren}
Q_\pm^{(0)} = Z_\pm Q_\pm
\ee
where
\be\label{z+-}
Z_\pm=1+\aspi {1\over\eps} \left(\mp 3\f{N\mp1}{N}\right)~.
\ee
In this new basis the OPE reads
\begin{equation}\label{apam}  A\equiv A_++A_- =
{G_F\over\sqrt{2}}V^\ast_{cs}V_{ud}(C_+(\mu)\langle Q_+(\mu)\rangle +
C_-(\mu)\langle Q_-(\mu)\rangle)~,   \end{equation}
where ($S_\pm=(S_2\pm S_1)/2$)
\begin{equation}\label{aspm}
A_\pm={G_F\over\sqrt{2}}V^\ast_{cs}V_{ud}\left[\left(1+2C_F \aspi
\ln{\mu^2\over -p^2}\right)S_\pm+({3\over N}\mp 3)\aspi\ln{M^2_W\over -p^2} 
S_\pm
\right]   \end{equation}
and
\begin{equation}\label{qmpm}
\langle Q_\pm(\mu)\rangle=\left(1+2C_F \aspi\ln{\mu^2\over -p^2}
\right)S_\pm+({3\over N}\mp 3)\aspi\ln{\mu^2\over -p^2} S_\pm~,  
\end{equation}
\begin{equation}\label{cpm}
C_\pm(\mu)=1+({3\over N}\mp 3)\aspi \ln{M^2_W\over\mu^2}~.
\end{equation}
 
\subsubsection{Factorization of SD and LD}
We have just witnessed in explicit terms
 the most important feature of the OPE, advertised already at the beginning
of these lectures:
 factorization of short-distance (coefficients) and 
long-distance
(operator matrix elements) contributions.
 Schematically, this factorization has the following structure:
\begin{equation}\label{fact}
(1+\as G \ln{M^2_W\over -p^2})\doteq
(1+\as G \ln{M^2_W\over\mu^2})\cdot
(1+\as G \ln{\mu^2\over -p^2})        \end{equation}
which is achieved by the following splitting of the logarithm
\begin{equation}\label{splt}
\ln{M^2_W\over -p^2}=\ln{M^2_W\over\mu^2}+ \ln{\mu^2\over -p^2}  
\end{equation}
or from the point of view of the integration over some virtual momenta
through the splitting
\begin{equation}\label{pmuw}
\int^{M^2_W}_{-p^2}{d k^2\over k^2}=
\int^{M^2_W}_{\mu^2}{d k^2\over k^2} +
\int^{\mu^2}_{-p^2}{d k^2\over k^2}.   \end{equation}

In particular the last formula makes it clear that the Wilson coefficients 
contain the contributions
from large virtual momenta of the loop correction from scales
$\mu=\ord(1\gev)$ to $\mw$, whereas the low energy contributions are
separated into the matrix elements.
The renormalization scale 
 $\mu$ acts as the scale at which the full
contribution to the amplitude 
is separated into a low energy and a high energy part.
\subsubsection{Independence of $C_i$ from External States}
Let us next return to the issue of the infrared divergences in the
process of matching. In the matching discussed explicitly above
they are regulated by taking $p^2\not=0$. They appear both in
 $A_{full}$ and $A_{eff}$. Yet as we have shown above
 the dependence of $A_{full}$ on $p^2$,
representing the long-distance structure of $A$ is, from the
point of view of the effective theory, fully contained
in $\langle Q_i\rangle$ and 
the Wilson coefficients $C_i$  are free from this
dependence. 

Since the coefficient functions do not depend on the
external states, any external state can be used for their
extraction, the only requirement being that the infrared
(and mass) singularities are properly regularized.
In our example an off-shell momentum $p$ for massless external
quarks has been used, but such a choice is clearly one of
several possibilities.
In general one could work with any other arbitrary momentum
configuration,  on-shell or off-shell, with or without external
quark mass, with infrared divergences regulated by off-shell
momenta, quark masses, a fictitious gluon mass or by dimensional
regularization. All these methods would give the same results for
$C_i$. 

In particular the dimensional regularization of 
infrared divergences is very convenient as many integrals
simplify considerably. Older discussions of dimensional
infrared regularization can be found in Muta`s book \cite{MUTA} and
also in a paper by Marciano \cite{MAR80}. Recently this method has
been used in calculating NLO corrections to $K\to\pi\nu\bar\nu$
\cite{BB1} and also for the matching conditions in $B\to X_s\gamma$
\cite{GH97,BKP2}. In particular as we stressed in \cite{BKP2}, 
the distinction of
$1/\varepsilon$ ultraviolet divergences from the infrared
ones is not necessary because after proper renormalization
of ultraviolet singularities, the left-over divergences are
of infrared origin only. These singularities cancel then automatically
in the process of matching. To this end, however, it is essential
to perform the matching at all stages in $D=4-2\varepsilon$
dimensions. This implies that already at the NLO level,
$\ord(\varepsilon)$ terms in Wilson coefficients have to be kept
at the intermediate stages of the calculation. More details on this
efficient technique can be found in the papers quoted above.
\subsection{OPE and the Renormalization Group}
            \label{sec:basicform:rg}
\subsubsection{Preliminaries}
               \label{sec:basicform:rg:basic}
So far we have computed 
\begin{equation}\label{cpm1}
C_\pm(\mu)=1+({3\over N}\mp 3)\aspi \ln{M^2_W\over\mu^2} 
 \end{equation}
 in ordinary
perturbation theory. 
Unfortunately for $\mu=1\gev$
the first order correction term amounts  to
65 -- 130\% although $\as/4\pi\approx 4\%$. 
This finding illustrates explicitly
the breakdown of the naive perturbative expansion 
caused by the appearance of large logarithms originating in the
presence of largely disparate scales $M_W$ and
$\mu$.

Clearly, the result in (\ref{cpm1}) can only be used for $\mu=\ord(\mw)$.
For $\mu\ll \mw$ we have to sum the large logarithms to all orders
of perturbation theory before we can trust our result for $C_\pm$.
Fortunately we have developed in section 4 a very powerful technique to
sum such logarithms and we know exactly what we have to do. 
Yes, in order to sum these large logs we have to find renormalization
group equations for $C_\pm$ and  solve them.
\subsubsection{Renormalization Group Equations for $C_\pm$}
The renormalization group equations for $C_\pm$
follow
 from the fact, that the unrenormalized Wilson coefficients
$C_\pm^{(0)}$ do not depend on $\mu$.
Using the relation (\ref{zcz}), properly adapted to the diagonal
basis, we have first
\be\label{drc}
C_\pm = Z_\pm C^{(0)}_\pm \qquad  Q_\pm^{(0)} = Z_\pm Q_\pm
\ee
and subsequently
\begin{equation}\label{rc+-}
{d C_\pm(\mu)\over d\ln\mu}=\gamma_\pm(g) C_\pm(\mu).
\end{equation}
Here $\gamma_\pm$ is the anomalous dimension of the operator $Q_\pm$
and given by
\begin{equation}\label{gam+-} 
\gamma_\pm(g)={1\over Z_\pm}{d Z_\pm\over d\ln\mu}.
  \end{equation}
Comparing (\ref{rc+-}) and (\ref{gam+-}) with (\ref{rggm}) and (\ref{gamz}),
respectively we see great similarities with the case of the running
quark mass. The only modification is the opposite sign in (\ref{rc+-}).
Consequently many relevant formulae of section 4 can be
immediately employed. Here we go:

\bi
\item
In the MS $(\overline{\rm MS})$-scheme
\begin{equation}\label{zie+-}
Z_\pm=1+\sum^\infty_{k=1} {1\over \eps^k} Z_{\pm, k}(g)  
\end{equation}
and consequently
\begin{equation}\label{zi1+-}
\gamma_\pm(g)=-2 g^2{\partial Z_{\pm, 1}(g)\over\partial g^2}~.
\end{equation}
\item
Using then 
\be\label{z1+-}
Z_\pm=1+\aspi {1\over\eps} \left(\mp 3\f{N\mp1}{N}\right)
\ee
as obtained in (\ref{z+-}) gives the one-loop anomalous
dimensions of $Q_\pm$:
\begin{equation}\label{gpm0}
\gamma_\pm(\as)=\aspi\gamma^{(0)}_\pm \qquad  \gamma^{(0)}_\pm=
   \pm 6{N\mp1\over N}.  \end{equation}
\item
The solution of (\ref{rc+-}) is given as for $m(\mu)$ in (\ref{UMM}):
\begin{equation}\label{C+-}
 C_\pm(\mu) = U_\pm(\mu,\mu_W) C_\pm(\mu_W) 
  \end{equation}
where $\mu_W=\ord(\mw)$ and  $U_\pm(\mu,\mu_W)$ is the evolution function:

\begin{equation}\label{U+-}
U_\pm(\mu,\mu_W)= \exp \left[ 
  \int_{g(\mu_W)}^{g(\mu)}{dg' \frac{\gamma_\pm(g')}{\beta(g')}}\right]~. 
\end{equation}
\item
Using (\ref{gpm0}) and $\beta(g)=-\beta_0 g^3/16\pi^2$ we can now find
$C_\pm(\mu)$ by using the leading term in the formula (\ref{mmu}) for
$m(\mu)$. Setting $\mu_0=\mw$ and taking into account the relative
sign between (\ref{rggm2}) and (\ref{rc+-}) we have

\begin{equation}\label{cpmrg}
C_\pm(\mu)=\left[{\as(M_W)\over\as(\mu)}\right]^{\gamma^{(0)}_\pm\over
  2\beta_0} C_\pm(M_W)  \end{equation}
\item
In order to complete the calculation we use the fact
that at  $\mu=\mw$ no large logarithms are present and $C_\pm(\mw)$
can be calculated in ordinary perturbation theory.
From (\ref{cpm}) we have in LO
\begin{equation}\label{cmw1}  C_\pm(M_W)=1  
\end{equation}
and consequently for $\mu=\mu_b=\ord(\mb)$
\begin{equation}\label{cpmrg1}
C_\pm(\mu_b)=\left[{\as(M_W)\over\as(\mu_b)}\right]^{\gamma^{(0)}_\pm\over
  2\beta_0}~.   \end{equation}
\ei

 We have now summed all leading logarithms and the important
formula (\ref{cpmrg1}) 
gives the  coefficients $C_\pm$ 
in the leading log approximation or in other words the leading term of the
RG improved perturbation theory. For instance, specializing to the case of 
$f=5$ and $\mu_b=\ord(\mb)$ we obtain
\begin{equation}\label{cpmrg12}
C_+(\mu_b)=\left[{\as(M_W)\over\as(\mu_b)}\right]^{6\over 23}   
\quad\quad
C_-(\mu_b)=\left[{\as(M_W)\over\as(\mu_b)}\right]^{-12\over 23}   
\end{equation}
with $\as$ given by the leading expression (\ref{alphaLO}).
For $\mu_b=5.0~\gev$ and $\Lms^{(5)}=225~\gev$ one finds
$C_+(\mu_b)=0.847$ and $C_-(\mu_b)=1.395$, i.e. suppression of
$C_+$ and an enhancement of $C_-$ relative to $C_-=C_+=1$
without QCD corrections. The corresponding enhancements and suppressions
for scales $\ord(1~\gev)$ reflect to some extent the dominance of
the $\Delta I=1/2$ transitions over $\Delta 3/2$ transitions in
$K\to\pi\pi$ decays (the $\Delta I=1/2$ rule) first analyzed in QCD
in \cite{MAIANI}. These short distance effects are insufficient, however,
to explain the dominance of $\Delta I=1/2$ transitions observed
experimentally. We will return briefly to this issue in section 11.

\subsubsection{Choice of the Matching Scale}
In calculating (\ref{cpmrg}) we have set the high energy matching
scale to $\mw$.
The choice of the high energy
matching scale, to be denoted by $\mu_W$, is of course not unique.
 The only
requirement is that  
$\mu_W=\ord(\mw)$
in order to avoid large logarithms $\ln(M_W/\mu_W)$. However, we
know from
(\ref{cpm1}) that in the LO approximation, in which 
$\ord(\alpha_s)$ terms are dropped in $C_\pm(\mu_W)$, we have 
using (\ref{cmw1})
\be
C_\pm(\mu_W)=C_\pm(\mw)+\ord(\alpha_s)=1.
\ee
Consequently in this approximation we also have
\begin{equation}\label{cpmr2}
C_\pm(\mu_b)=
\left[{\as(\mu_W)\over\as(\mu_b)}\right]^{\gamma^{(0)}_\pm\over
  2\beta_0} =
\left[{\as(M_W)\over \as(\mu_b)}\right]^{\gamma^{(0)}_\pm\over
  2\beta_0} (1+\ord(\as)) 
\end{equation}
which differs from (\ref{cpmrg1}) by $\ord(\alpha_s)$ corrections. 

We observe that
a change of $\mu_W$ around the value of $\mw$ causes an ambiguity
of $\ord(\as)$ in the coefficient. 
This ambiguity represents a
theoretical uncertainty in the determination of $C_\pm(\mu_b)$. 
In order to reduce it, it is necessary to go beyond the leading order.
We will do this in the following section. Similar ambiguity exists
in the choice of the low energy scale $\mu_b$ as we will discuss
at various places in these lectures, in particular in connection
with $B\to X_s\gamma$ decay.

\subsubsection{Threshold Effects in LO}
The evolution function $U$ depends on $f$
 through $\alpha_s^{(f)}$ and $\beta_0$ in the exponent.
One can generalize the renormalization group evolution from
$\mw$ down to say $\mu_c=\ord(\mc)$ to include the threshold effect of
the b-quark as follows
\begin{equation}\label{cmub}
 C_\pm(\mu_c)=U^{(f=4)}_\pm(\mu_c,\mu_b)U^{(f=5)}_\pm(\mu_b,\mw) 
C_\pm(\mw)
\end{equation}
which is valid in LO. Here $\mu_b=\ord(\mb)$. Thus
$(\ref{cpmrg12})$ generalizes to
\begin{equation}\label{cpmrg12+}
C_+(\mu_c)=\left[{\as^{(4)}(\mu_b)\over\as^{(4)}(\mu_c)}\right]^{6\over 25}
\left[{\as^{(5)}(M_W)\over\as^{(5)}(\mu_b)}\right]^{6\over 23}~,
\quad\quad
C_-(\mu_c)
=\left[{\as^{(4)}(\mu_b)\over\as^{(4)}(\mu_c)}\right]^{-12\over 25}
\left[{\as^{(5)}(M_W)\over\as^{(5)}(\mu_b)}\right]^{-12\over 23}.
\end{equation}
Again also here there is an ambiguity in $\mu_c$ which 
can only be reduced by going to NLO.

\subsubsection{RGE for $C_i$: Case of Operator Mixing}
The coefficients $C_i(\mu)$ can be now calculated by inverting
(\ref{dbasis}) with the result
\be\label{c12}
C_1(\mu) =\frac{C_+(\mu)-C_-(\mu)}{2}~,    
\qquad   C_2(\mu)=\frac{C_+(\mu)+C_-(\mu)}{2}~,    
\ee
where $C_{\pm}(\mu)$ is given in (\ref{cpmrg1}) or (\ref{cpmrg12+}).

Yet, it is instructive to derive (\ref{c12}) by using a procedure which
one can also apply to more complicated situations in which several
operators mix under renormalization.
To this end we write 
\be\label{mix1}
\vec C^T=(C_1, C_2)~, \qquad  \vec Q^T=(Q_1, Q_2).
\ee
Then
\be\label{mix2}
\vec C^{(0)}=\hat Z_c \vec C \qquad \vec Q^{(0)}=\hat Z \vec Q
\ee
with $\hat Z^T_c=\hat Z^{-1}$.
Defining next the 
 anomalous dimension matrix $\hat\gamma$ by
\begin{equation}\label{gazz} 
\hat\gamma=\hat Z^{-1}{d\hat Z\over d\ln\mu},
\end{equation}
the $\mu$-independence of $\vec C^{(0)}$ implies 
\begin{equation}\label{rgc}
{d \vec C(\mu)\over d\ln\mu}=\hat\gamma^T(\as) \vec C(\mu).  
\end{equation}
The solution of this equation is 
\begin{equation}\label{rgcu}
\vec C(\mu)=\hat U(\mu, M_W) \vec C(M_W)  
\end{equation}
where 
\be\label{Umatrix}
\hat U(\mu,\mw)= \exp \left[ 
  \int_{g(\mw)}^{g(\mu)}{dg' \frac{\hat\gamma^T(g')}{\beta(g')}}\right] 
\end{equation}
is the $\mu$-evolution matrix.

In the MS $(\overline{\rm MS})$-scheme we have
\begin{equation}\label{ziem}
\hat Z=\hat 1+\sum^\infty_{k=1} {1\over \eps^k} \hat Z_{k}(g)  
\end{equation}
and 
\begin{equation}\label{zi1m}
\hat\gamma(g)=-2 g^2{\partial \hat Z_{1}(g)\over\partial g^2}
\end{equation}
Consequently using
\begin{equation}\label{zll0} \hat Z = 1+ \aspi {1\over\eps}
 \left(\begin{array}{cc}  3/N & -3 \\
                          -3 & 3/N
    \end{array}\right)   \end{equation}
we have to first order in $\as$ \cite{MAIANI}
\begin{equation}\label{g120} \hat\gamma(\as)=\aspi \hat\gamma^{(0)}=\aspi
 \left(\begin{array}{cc} -6/N & 6 \\
                          6 & -6/N
    \end{array}\right).   \end{equation}

In order to find $C_i(\mu)$  let us write the LO evolution matrix
as
\begin{equation}\label{u0vd0} 
\hat U^{(0)}(\mu,\mw)= \hat V
\left({\left[{\as(\mw)\over\as(\mu)}
\right]}^{{\vec\gamma^{(0)}\over 2\beta_0}}
   \right)_D \hat V^{-1}   \end{equation}
where $\hat V$ diagonalizes ${\hat\gamma^{(0)T}}$
\begin{equation}\label{ga0d} 
\hat\gamma^{(0)}_D=\hat V^{-1} {\hat\gamma^{(0)T}} \hat V
  \end{equation}
and $\vec\gamma^{(0)}$ is the vector containing the diagonal elements of
the diagonal matrix :
\begin{equation}\label{g120d} \hat\gamma^{(0)}_D=
 \left(\begin{array}{cc} \gamma^{(0)}_+ & 0 \\
                          0 & \gamma^{(0)}_-
    \end{array}\right)   \end{equation}

with $\gamma^{(0)}_\pm$ given in (\ref{gpm0}). Using
\be
\hat V = \hat V^{-1} =\frac{1}{\sqrt{2}}
\left(\begin{array}{cc} 1  & 1 \\
                          1 & -1
    \end{array}\right)   \end{equation}
and
\be
\vec C^T(\mw)=(C_1(\mw), C_2(\mw))=(0,1)
\ee
we reproduce (\ref{c12}) with $C_\pm(\mu)$ given by (\ref{cpmrg12}).

The threshold effects can be incorporated as in (\ref{cmub})
\begin{equation}\label{cmub1}
 \vec C(\mu_c)=\hat U^{(f=4)}(\mu_c,\mu_b)\hat U^{(f=5)}(\mu_b, \mu_W) 
\vec C(\mu_W).
\end{equation}

It is evident that this procedure is valid for arbitrary number
of operators mixing under renormalization. However for more
complicated situations one has to use computer programs like 
{\it Mathematica}
to obtain analytic formulae like (\ref{c12}). We will give
some examples later on.

\subsection{Summary of Basic Formalism }
            \label{sec:basicform:summary}
It is a good moment to make a break and to summarize what we have
achieved in our climb so far. This will also allow us to make
a strategy for the next steps, which as we will see are technically
more advanced.

Ultimately our goal is the evaluation of weak decay amplitudes
involving hadrons in the framework of a low energy effective theory,
of the form
\begin{displaymath}
\langle {\cal H}_{eff}\rangle={G_F\over\sqrt{2}}V_{CKM}
\langle \vec Q^T(\mu)\rangle \vec C(\mu),
\end{displaymath}
where $\mu$ denotes a scale of the order of the mass of the decaying
hadron.
The procedure for this calculation can be divided into the
following three steps.

\noindent
{\bf Step 1: Matching in Perturbation Theory}
\\
Calculation of Wilson coefficients $\vec C(\mu_W)$ at
$\mu_W=\ord(\mw)$ to the desired order in $\as$.  
Since
logarithms of the form $\ln(\mu_W/\mw)$ are not large, this can be
performed in ordinary perturbation theory. 
In the case
of the operators $Q_{1,2}$ and in the LO approximation we
simply have $C_1(\mu_W)=0$, $C_2(\mu_W)=1$ or $C_\pm(\mu_W)=1$.

This step
amounts to matching
the full theory onto a five quark effective theory.
In this process $W^\pm$, $Z^0$, the top-quark and generally
all heavy particles with masses higher than $\mw$ are integrated
out. In the case of $Q_{1,2}$ analyzed so far, the effect of 
integrating out the top-quark has only been seen in that for
$\mu\le\mw$ we have used $\alpha^{(5)}(\mu)$ instead of
 $\alpha^{(6)}(\mu)$. Later when we move to other decays,
the effect of integrating out the top-quark will be more profound.

The matching in question is achieved using the following procedure:

\begin{itemize}
\item
Calculation of the amplitude in the full theory,
\item
Calculation of the operator matrix elements,
\item
Extraction of $C_i(\mu_W)$ from $A_{full}=A_{eff}$.
\end{itemize}
The resulting $C_i(\mu_W)$ depend generally on the masses of the
heavy particles which have been integrated out. Again in the
special case of $Q_{1,2}$ this dependence is absent.

\noindent
{\bf Step 2: RG Improved Perturbation Theory}
\begin{itemize}
\item Calculation of the anomalous dimensions of the operators,
\item Solution of the renormalization group equation for $\vec{C}(\mu)$,
\item Evolution of the coefficients from $\mu_W$ down to the
appropriate low energy scale $\mu$
\begin{displaymath}
\vec C(\mu)=\hat U(\mu, \mu_W)\vec C(\mu_W)~.
\end{displaymath}
\end{itemize}

\medskip
\noindent
{\bf Step 3: Non-Perturbative Regime}
\\
Calculation of hadronic matrix elements $\langle\vec Q(\mu)\rangle$,
normalized at the appropriate low energy scale $\mu$, by means of
some non-perturbative method.

\bigskip
\noindent
Important issues in this procedure are:
\begin{itemize}
\item {\bf Factorization\/} of short- and long
distance contributions:
\begin{itemize}
\item
 $\vec C(\mu)$: contributions from scales {\it higher} than $\mu$
\item 
$\langle \vec Q(\mu)\rangle$: contributions from scales {\it lower}
than $\mu$
\item 
Cancellation of  the $\mu$-dependence
between $C_i(\mu)$ and
$\langle Q_i(\mu)\rangle$.
\end{itemize}
\item  {\bf Summation of large logs\/} by means of the RG method.
 More specifically, in the n--the  order of
RG improved perturbation theory the terms 
\begin{displaymath}
\as^n(\mu)
\left(\as(\mu)\ln{M_W\over\mu}\right)^k
\end{displaymath}
are summed to all orders in $k$ ($k$=0, 1, 2,$\ldots$). This approach
is justified as long as $\as(\mu)$ is small enough. The leading order
corresponds in most cases to $n=0$, the NLO to $n=1$. In certain
processes these canonical values of $n$ may change.
\end{itemize}

\subsection{Future Generalizations}

Until now, our application of the basic formalism summarized above,
concentrated on the current-current operators $Q_1$ and $Q_2$ in the
LO approximation. In the following sections we will generalize this
discussion in several aspects:

\begin{itemize}
\item
We will generalize the calculation of the couplings $C_i(\mu)$ beyond
the LO approximation,
\item
We will include new operators originating in penguin diagrams 
of various sort (Gluon-penguins, Photon-penguins, $Z^0$-penguins).
These operators are generally called {\it Penguin Operators}. This
generalization will bring the $\mt$ dependence into $C_i(\mu_W)$
of these new operators.
\item
We will also include new operators originating in {\it Box Diagrams}. 
This generalization will also bring the $\mt$ dependence into $C_i(\mu_W)$
of these new operators.
\item
In the process of including operators it will turn out to be necessary
to consider also renormalization group equations involving simultaneously
$C_i$ of order  $\ord (1)$,  $\ord (\as)$ and $\ord(\alpha)$ with
$\alpha=\alpha_{QED}$.
\item
Finally we will develop efficient methods for the calculation of the
anomalous dimensions of the operators $Q_i$.
\end{itemize}

We begin these generalizations by including
NLO QCD corrections to $C_\pm (\mu)$. We will do this in such a manner
that the generalization of the formulae listed below to more complicated
processes will be straightforward.

\subsection{Motivations for NLO}
Going beyond the LO approximation is certainly an important but a 
non-trivial step. For this reason we need some motivations to perform
this step. Here are the main reasons for going beyond LO:
\begin{itemize}
\item The NLO is first of all necessary to test the validity of
the renormalization group improved perturbation theory.
\item Without going to NLO the QCD scale $\Lambda_{\overline{MS}}$
extracted from various high energy processes cannot be used 
meaningfully in weak decays.
\item 
Due to renormalization group invariance the physical
amplitudes do not depend on the scales $\mu$ present in $\alpha_s$
or in the running quark masses, in particular $\mt(\mu)$, 
$\mb(\mu)$ and $\mc(\mu)$. However,
in perturbation theory this property is broken through the truncation
of the perturbative series. Consequently one finds sizable scale
ambiguities in the leading order, which can be reduced considerably
by going to NLO. An example of such an ambiguity is the choice of the
high energy matching scale $\mu_W$ discussed above.
\item
The Wilson Coefficients are renormalization scheme dependent quantities.
This scheme dependence appears first at NLO. For a proper matching of
the short distance contributions to the long distance matrix elements
obtained from lattice calculations it is essential to calculate NLO.
The same is true for inclusive heavy quark decays in which the hadron
decay can be modeled by a decay of a heavy quark and the matrix elements
of $Q_i$ can be effectively calculated in an expansion in $1/\mb$.
\item 
In several cases the central issue of the top quark mass dependence
is strictly a NLO effect.
\end{itemize}

\section{Wilson Coefficients Beyond Leading Order}
\setcounter{equation}{0}
\subsection{Preliminaries}
We will now generalize the formulae of the previous section beyond
the LO approximation concentrating on the Wilson coefficients $C_{\pm}$
and $C_{1,2}$. We will begin with the case without operator mixing.
Subsequently we will generalize our discussion 
 to the case of the operator mixing. Next we will
develop methods for the calculation of anomalous dimensions
generalizing our previous discussion 
to the mixing of operators with different canonical
dimensions. In particular the mixing between six and five dimensional
operators.
While we do not have space to present an explicit two-loop calculation
of anomalous dimensions, we will derive explicitly the one-loop anomalous
dimension matrix (\ref{g120}). In section 8.5 we will generalize this
calculation to include the penguin operators.
A detailed discussion of renormalization scheme
and renormalization scale dependences and of their cancellations in
physical amplitudes is an important part of this section. Finally,
we will discuss the issue of the so-called {\it evanescent}
operators which have to be taken into account in a proper calculation
of the anomalous dimensions at the two-loop level. 
\subsection{The Case without Operator Mixing}
Let us consider the coefficients $C_\pm(\mu)$
for which we have the general expression:
\begin{equation}\label{C1+-}
 C_\pm(\mu) = U_\pm(\mu,\mw) C_\pm(\mw) 
  \end{equation}
where 
\begin{equation}\label{U1+-}
U_\pm(\mu,\mw)= \exp \left[ 
  \int_{g(\mw)}^{g(\mu)}{dg' \frac{\gamma_\pm(g')}{\beta(g')}}\right] 
\end{equation}
and we have set $\mu_W=\mw$ in order to simplify the formulae below.
This restriction will be relaxed whenever it will turn out to be 
appropriate.

At NLO we use:
\begin{equation}\label{B8}
C_\pm(M_W)=1+\frac{\as(M_W)}{4\pi}B_\pm
\end{equation}

\begin{equation}\label{gg01P}
\gamma_\pm(\as)=\gamma_\pm^{(0)}\aspi + \gamma_\pm^{(1)}\left(\aspi\right)^2
\end{equation}

\begin{equation}\label{bg01P}
\beta(g)=-\beta_0{g^3\over 16\pi^2}-\beta_1{g^5\over (16\pi^2)^2}
  \end{equation}

Inserting the last two formulae into (\ref{U1+-}) and expanding in
$\alpha_s$ we find
\begin{equation}\label{B9P}
U_\pm(\mu,\mw)=\left[1+\frac{\as(\mu)}{4\pi}J_\pm\right]
      \left[\frac{\as(M_W)}{\as(\mu)}\right]^{d_\pm}
\left[1-\frac{\as(M_W)}{4\pi} J_\pm\right]
\end{equation}
with
\begin{equation}\label{B10P}
J_\pm=\frac{d_\pm}{\beta_0}\beta_1-\frac{\gamma^{(1)}_\pm}{2\beta_0}
\qquad\qquad
d_\pm=\frac{\gamma^{(0)}_\pm}{2\beta_0}.
\end{equation}
This is similar to the $\mu$-dependence of the quark mass discussed
in section 4.8 and given in (\ref{mmu}) except that we have 
written the evolution function
in a particular way: the couplings $\as(\mu)$ increase by going from
right to left. This is clearly not necessary but is useful for the
future generalization to the case of operator mixing.

Inserting (\ref{B9P}) and (\ref{B8}) into (\ref{C1+-}) we find
an important formula for $C_\pm(\mu)$ in the NLO approximation:
\begin{equation}\label{B9PP}
C_\pm(\mu)=\left[1+\frac{\as(\mu)}{4\pi}J_\pm\right]
      \left[\frac{\as(M_W)}{\as(\mu)}\right]^{d_\pm}
\left[1+\frac{\as(M_W)}{4\pi}(B_\pm-J_\pm)\right]~.
\end{equation}

Let us next outline the procedure for finding $B_\pm$.
Since the operators $Q_+$ and $Q_-$ do not mix under renormalization,
$B_+$ and $B_-$ can be found separately. 
The procedure for finding $B_\pm$ amounts to the generalization of
the matching procedure in
LO to include in addition to logarithms also
constant $\ord (\as)$ terms.

{\bf Step 1:}

\begin{equation}\label{amp2}
A^\pm_{full} =
{G_F\over\sqrt{2}}(1+{\as(\mu_W)\over 4\pi}\left[-{{\gamma^{(0)}_\pm}\over 2}
\ln{M^2_W\over -p^2}+\tilde A^{(1)}_\pm\right]) S_\pm~,  \end{equation}
where $S_\pm$ are the tree matrix elements.

{\bf Step 2:}

\vspace{-0.41truecm}
\begin{eqnarray}\label{aef2}
A^\pm_{eff}&=&{G_F\over\sqrt{2}}C_\pm (\mu_W)\langle Q_\pm(\mu_W)\rangle \\
&=&{G_F\over\sqrt{2}}C_\pm(\mu_W)
(1+{\as(\mu_W)\over 4\pi}\left[{{\gamma^{(0)}_\pm}\over 2}
 \ln{-p^2\over\mu^2_W}+\tilde r_\pm
 \right])S_\pm~.   \nonumber
\end{eqnarray}

{\bf Step 3:}

Comparison of (\ref{amp2}) and (\ref{aef2}) yields 
\begin{equation}\label{cmw2}
 C_\pm(\mu_W)=1+{\as(\mu_W)\over 4\pi}\left[-{{\gamma^{(0)}_\pm}\over 2}
 \ln{M^2_W\over\mu^2_W}+B_\pm\right]~, \end{equation}
where
\begin{equation}\label{ba1r} B_\pm=\tilde A_\pm^{(1)}-\tilde r_\pm~.
 \end{equation}

Setting $\mu_W=\mw$ we reproduce (\ref{B8}).
Any infrared dependence like $\ln{-p^2}$ or any special properties
of the external quark states present in $\tilde A_\pm^{(1)}$ and 
$\tilde r_\pm$
cancel in the difference (\ref{ba1r}) so that $B_\pm$ are just  numerical 
constants independent of external states. We will give the numerical values of
$B_\pm$ in the next section.

\subsection{The Case of Operator Mixing}
\subsubsection{Preliminaries}
Let us generalize the preceeding discussion to the case of operator
mixing. Now
\begin{equation}\label{hqtc}
{\cal H}_{eff}={G_F\over\sqrt{2}}\sum_i C_i(\mu)Q_i(\mu)\equiv
  {G_F\over\sqrt{2}}\vec Q^T(\mu) \vec C(\mu)~,   \end{equation}
where the index $i$ runs over all contributing operators, in our
example $Q_1$ and $Q_2$. 

The Wilson coefficient functions are given then by
\begin{equation}\label{cucw}
\vec C(\mu)=\hat U(\mu, \mu_W)\vec C(\mu_W)~.
\end{equation}
Our goal is to find $\vec C(\mu_W)$ and the evolution matrix
$\hat U(\mu, \mu_W)$ keeping NLO corrections.   
\subsubsection{Determination of $\vec C(\mu_W)$}
The procedure for finding $\vec C(\mu_W)$ proceeds again in three
steps:

{\bf Step 1:}

The  amplitude in the full theory after
field renormalization is given by:
\begin{equation}\label{aa01} A_{full}=
{G_F\over\sqrt{2}}\vec S^T(\vec A^{(0)}
+{\as(\mu_W)\over 4\pi}\vec A^{(1)}).
\end{equation}
Here $\vec S$ denotes the tree level matrix elements of the
operators $\vec Q$. In order to simplify the presentation we
have absorbed the logarithms in the $\ord(\alpha_s)$ term
$\vec A^{(1)}$ which also contains non-logarithmic terms as in
(\ref{amp2}). Later on we will discuss a specific example which
will exhibit the detail structure of (\ref{aa01}) more transparently.

{\bf Step 2:}

In the effective theory,
after quark field renormalization and the
renormalization of the operators through
\begin{equation}\label{q0z3}
\vec Q^{(0)}=\hat Z\vec Q,  
\end{equation}
the renormalized matrix elements of the operators
are
\begin{equation}\label{qars}
\langle\vec Q(\mu_W)\rangle=(\hat 1
+{\as(\mu_W)\over 4\pi} \hat r)\vec S  \end{equation}
and consequently
\begin{equation}\label{aeff}  A_{eff}=
{G_F\over\sqrt{2}}\vec S^T(1+{\as(\mu_W)\over 4\pi} \hat r^T) 
\vec C(\mu_W).\end{equation}
Again $\hat r$ contains the relevant logarithms together with
the non-logarithmic terms as in (\ref{aef2}).

{\bf Step 3:}

Equating (\ref{aa01}) and (\ref{aeff}) we obtain
\begin{equation}\label{cmuw} 
\vec C(\mu_W)=
\vec A^{(0)}+{\as(\mu_W)\over 4\pi}
(\vec A^{(1)}-r^T\vec A^{(0)}).
\end{equation}

\subsubsection{Renormalization Group Evolution}
The renormalization group equation for $\vec C$
\begin{equation}\label{rgcv}
{d \vec C(\mu)\over d\ln\mu}=\hat\gamma^T(g)\vec C(\mu)   
\end{equation}
has to be solved now with the boundary condition (\ref{cmuw}). 

The general solution can be written down iteratively
\begin{equation}\label{uit}
\hat U(\mu,\mu_W)=1+\int^{g(\mu)}_{g(\mu_W)}
dg_1{\hat\gamma^T(g_1)\over\beta(g_1)}+
\int^{g(\mu)}_{g(\mu_W)}dg_1\int^{g_1}_{g(\mu_W)}dg_2
{\hat\gamma^T(g_1)\over\beta(g_1)}{\hat\gamma^T(g_2)\over\beta(g_2)}+\ldots 
\end{equation}
which using $dg/d\ln\mu=\beta(g)$  solves
\begin{equation}\label{rgu}
{d\over d\ln\mu}\hat U(\mu, \mu_W)=\hat\gamma^T(g)\hat U(\mu, \mu_W)~.
   \end{equation}

The series in (\ref{uit}) can be written more compactly:
\begin{equation}\label{utge}
\hat U(\mu, \mu_W)=
T_g \exp\int^{g(\mu)}_{g(\mu_W)}dg'{\hat\gamma^T(g')\over\beta(g')}~,
\end{equation}
where in the case $g(\mu)>g(\mu_W)$ the $g$-ordering operator $T_g$ is
defined through
\begin{equation}\label{tgdf}
T_g f(g_1)\ldots f(g_n)=\sum_{perm}
\Theta(g_{i_1}-g_{i_2})\ldots
\Theta(g_{i_{n-1}}-g_{i_n})f(g_{i_1})\ldots f(g_{i_n}). \end{equation}
It brings ordering of the functions $f(g_i)$ such that the
coupling constants increase from right to left. The sum in (\eqn{tgdf}) runs
over all permutations $\{i_1,\ldots, i_n\}$ of $\{1, 2,\ldots, n\}$.
The $T_g$ ordering is necessary because at NLO
$[\hat\gamma(g_1), \hat\gamma(g_2)]\not= 0$. Indeed the matrices
$\hat\gamma^{(0)}$ and $\hat\gamma^{(1)}$ in the perturbative expansion
of the anomalous dimension matrix
\begin{equation}\label{gg01}
\hat\gamma(\as)=\hat\gamma^{(0)}\aspi + \hat\gamma^{(1)}\left(\aspi\right)^2~.
\end{equation}
do not commute with each other.

Inserting (\ref{gg01}) and the expansion (\ref{bg01P}) for $\beta(g)$
into (\ref{utge}) we can write the evolution matrix in analogy to
(\ref{B9P}) as
\begin{equation}\label{u0jj}
\hat U(\mu,\mu_W)=
\left[1+{\as(\mu)\over 4\pi} \hat J\right] \hat U^{(0)}(\mu,\mu_W) 
\left[1-{\as(\mu_W)\over 4\pi} \hat J\right]
\end{equation}
Now it is clear why we have written (\ref{B9P}) in a special manner.
It can be nicely generalized to the mixing case where the ordering
of matrices matters.

$\hat U^{(0)}$ in (\ref{u0jj}) is the leading evolution matrix which
we already discussed in the LO section:
\begin{equation}\label{u0vdP} 
\hat U^{(0)}(\mu,\mu_W)= \hat V
\left(
{\left[{\as(\mu_W)\over\as(\mu)}\right]}^{{\vec\gamma^{(0)}\over 2\beta_0}}
   \right)_D \hat V^{-1}   \end{equation}
where $\hat V$ diagonalizes ${\hat\gamma^{(0)T}}$
\begin{equation}\label{ga0dP} \hat\gamma^{(0)}_D=\hat V^{-1} {\gamma^{(0)T}} 
\hat V
\end{equation}
and $\vec\gamma^{(0)}$ is the vector containing the diagonal elements of
the diagonal matrix $\hat\gamma^{(0)}_D$.

The derivation of the analytic expression for the matrix $\hat J$ follows
\cite{AB80} and is also given in the appendix of the first paper in
\cite{BJLW1}. Here we give
only the final result.
In order to write down the expression for the matrix $\hat J$,
 we define the matrix
\begin{equation}\label{gvg1} 
\hat G=\hat V^{-1} {\hat\gamma^{(1)T}} \hat V   \end{equation}
and a matrix $\hat H$ whose elements are
\begin{equation}\label{sij} 
H_{ij}=\delta_{ij}\gamma^{(0)}_i{\beta_1\over 2\beta^2_0}-
    {G_{ij}\over 2\beta_0+\gamma^{(0)}_i-\gamma^{(0)}_j}.  \end{equation}
Then
\begin{equation}\label{jvs} \hat J=\hat V \hat H \hat V^{-1}.
 \end{equation}

\subsubsection{Final Result for $\vec C(\mu)$}
Putting all things together we obtain the final result
\begin{equation}\label{cjua} 
\vec C(\mu)=(1+{\as(\mu)\over 4\pi} \hat J)\hat U^{(0)}(\mu,\mu_W)
(\vec A^{(0)}+{\as(\mu_W)\over 4\pi}[\vec A^{(1)}
-(\hat r^T+\hat J)\vec A^{(0)}])\end{equation}
which we will discuss in more detail below.

In the case of $(Q_1,Q_2)$ the inclusion of the flavour thresholds in 
(\ref{cucw}) is very similar to the  LO case:
\begin{equation}\label{cthrP}
\vec C(\mu)=\hat U_3(\mu,\mu_c)\hat U_4(\mu_c,\mu_b)
\hat U_5(\mu_b,\mu_W)\vec C(\mu_W)  \end{equation}
where $\hat U_f$ is the evolution matrix for $f$ effective flavors
given in (\ref{u0jj}). This formula has to be slightly modified
if the penguin operators are present. We will return to this point
at a suitable moment of these lectures.

\subsection{The Calculation of the Anomalous Dimensions}
               \label{sec:basicform:wc:adm}
\subsubsection{Master Formulae}
In the previous section we have calculated the anomalous dimension
matrix in the process of the matching of the full and effective
theories. In fact looking back one can see that the anomalous dimensions
can be read off from the coefficients of the logarithms in the matrix
elements $\langle Q_{1,2}\rangle$ in (\ref{q1re}) and (\ref{q2re}).
In more complicated situations, where many operators are present,
such a method is not very useful and it is important
to develop an efficient method for the calculation of 
anomalous dimensions. Here it comes:

\bi
\item
The evaluation of the amputated Green functions with
insertion of the operators $\vec Q$ as in fig. \ref{L:15} gives the relation
\begin{equation}\label{qzgf}
\langle\vec Q\rangle^{(0)}=Z^{-2}_q \hat Z\langle\vec Q\rangle
\equiv \hat Z_{GF}\langle\vec Q\rangle   \end{equation}
 where $\hat Z$ is the
renormalization constant matrix of the operators $\vec Q$ and
$\hat Z_{GF}$ is just defined above.
\item
Next, the anomalous dimension matrix is given by
\begin{equation}\label{gaz2}
\hat\gamma(g)=\hat Z^{-1}{d \hat Z\over d\ln\mu}
\end{equation}
\item
In the MS (or $\overline{\rm MS}$) scheme we have 
\begin{equation}\label{zeps}
\hat Z=1+\sum^\infty_{k=1}{1\over\eps^k} \hat Z_k(g)  
\end{equation}
and consequently as derived in the previous section
\begin{equation}\label{ggz1}
\hat\gamma(g)=-2g^2{\partial \hat Z_1(g)\over\partial g^2}
=-2\as{\partial \hat Z_1(\as)\over\partial \as}.
\end{equation}
\item
For $Z_q$ and $\hat Z_{GF}$ we have  
\begin{equation}\label{zqep}
Z_q=1+\sum^\infty_{k=1}{1\over\eps^k}Z_{q,k}(g)  \end{equation}
\begin{equation}\label{zgfe}
\hat Z_{GF}=1+\sum^\infty_{k=1}{1\over\eps^k} \hat Z_{GF,k}(g)  
\end{equation}
As the matrix elements $\langle\vec Q\rangle$ are finite, the singularities 
in $\hat Z_{GF}$ are  found directly from the calculation of the 
unrenormalized Green functions (\ref{qzgf}).
\item
From (\ref{qzgf}), (\ref{zeps}), (\ref{zqep}), (\ref{zgfe})
we find
\begin{equation}\label{zqgf}
\hat Z_1=2Z_{q,1}\hat 1+\hat Z_{GF,1}.  \end{equation}
\item
With
\begin{equation}\label{zq1b} 
Z_{q, 1}=a_1\aspi + a_2\left(\aspi\right)^2
\end{equation}
and
\begin{equation}\label{zgf1}
\hat Z_{GF,1}=\hat b_1 \aspi + \hat b_2 \left(\aspi\right)^2  
\end{equation}
we  obtain by means of (\ref{ggz1}) {\it Master Formulae} 
for the one- and two-loop
anomalous dimension matrices:

\begin{equation}\label{a1b1}
\gamma^{(0)}_{ij}=-2[2a_1 \delta_{ij}+(b_1)_{ij}],
\quad\quad
\gamma^{(1)}_{ij}=-4[2a_2 \delta_{ij}+(b_2)_{ij}]~.  \end{equation}
\item
In the case without mixing between operators these expressions
reduce to:
\begin{equation}\label{a1b1s}
\gamma^{(0)}=-2[2a_1 +b_1],  
\quad\quad
\gamma^{(1)}=-4[2a_2 +b_2]~.
\end{equation}
\ei
\subsubsection{How to Use One-Loop Master Formulae}
Let us illustrate how the first formula in (\ref{a1b1}) can be
used to obtain the one-loop anomalous dimension matrix
(\ref{g120}). From (\ref{q10}) we extract the coefficients of the
$1/\varepsilon$ singularities to be
\be\label{b11}
(b_1)_{11}=2 C_F+\frac{3}{N},
\quad\quad
(b_1)_{12}=-3~.
\ee
Similarly from (\ref{q20}) we find
\be\label{b22}
(b_1)_{21}=-3,
\quad\quad
(b_1)_{22}=2 C_F+\frac{3}{N}~.
\ee
Now $a_1=-C_F$. Consequently the term $2a_1$ in (\ref{a1b1})
cancels precisely the $2C_F$ term present in $(b_1)_{11}$ and 
$(b_1)_{22}$.
The leftover entries give the one-loop anomalous dimension matrix
(\ref{g120}).

The fact that the renormalization of the external quark fields cancels
the terms $2C_F$ in $(b_1)_{11}$ and $(b_1)_{22}$ is by no means accidental. 
It
is a consequence of the vanishing of the anomalous dimension of
the conserved weak current. Indeed the master formula for the anomalous
dimension of a current can be obtained by considering a two-point Green
function insted of four-point functions considered in deriving the
master formulae (\ref{a1b1s}). One finds this time
\be\label{curr}
\gamma_c^{(0)}=-2[a_1 +b^c_1],  
\quad\quad
\gamma_c^{(1)}=-4[a_2 +b^c_2]  
\end{equation}
where $b^c_1$ and $b^c_2$ are obtained by calculating the relevant
one-loop and two-loop diagrams, respectively. 
$b^c_1$ is simply obtained by calculating the one-loop upper vertex of
the diagram (a) in fig.~\ref{L:15}.
 $b^c_2$ is found by calculating the corresponding
two-loop generalization of this vertex.
One finds $b_1^c=C_F$. The factor
of two in $(b_1)_{11}$ in front of $C_F$ in (\ref{b11}) 
comes from a symmetric diagram to the
the diagram (a) in fig.\ \ref{L:15} with the gluon exchanged 
between the lower quark legs. With $a_1=-C_F$ we find $\gamma_c^{(0)}=0$ 
as it should be.

We get the following useful message from this discussion. The only diagrams
responsible for the non-vanishing anomalous dimensions of current-current 
operators $Q_1$ and $Q_2$ are the diagrams in which the gluons connect the
quark legs belonging to different weak currents. At the one-loop level these
are the diagrams (b) and (c) in fig.\ \ref{L:15} and the 
corresponding symmetric diagrams. One should stress that this simple
rule is not valid for the insertion of penguin operators into
current-current diagrams. We will see this explicitly in section 8.
\subsubsection{How to Use Two-Loop Master Formulae}
The calculation of $(b_2)_{ij}$ in (\ref{a1b1}) is a bit tricker
and technically more difficult. To this end one has to calculate first
two-loop diagrams with $Q_i$ insertions. Examples are given 
in fig. \ref{L:5}.
Next the corresponding two-loop counter-diagrams have to be 
{\it subtracted}. 
In the MS-like schemes the latter are obtained by retaining
only the $1/\eps$ parts in the subdiagrams. The counter-diagrams
corresponding to the two-loop diagrams in fig.~\ref{L:5} are shown 
in fig.~\ref{L:6}
where the small boxes stand for the singular parts of 
the corresponding subdiagrams in fig.~\ref{L:5}. 
For instance the Feynman rule
for the box in fig.~\ref{L:6}b is, in accordance with (\ref{isigma})  
given by
\begin{equation}\label{sigma1}
i C_F \delta_{\alpha\beta}  
\frac{\alpha_s}{4\pi} \not\! p
 \frac{1}{\varepsilon} 
\end{equation} 
where $p$ is the momentum of the quark. Since MS-like schemes are
the so-called mass-independent renormalization schemes the quark masses
can be set to zero in evaluating anomalous dimensions. 

Dropping colour factors and Dirac tensors the result for each diagram
including the counter-diagram has the structure
\be\label{IMIC}
I-I_C=\left(\frac{\alpha}{4\pi}\right)^2
\left[\frac{\mu^2}{-p^2}\right]^{2\eps}
\left[\frac{F}{\eps^2}+\frac{G}{\eps}+...\right]
-\left(\frac{\alpha}{4\pi}\right)^2
\left[\frac{\mu^2}{-p^2}\right]^{\eps}
\left[\frac{F_C}{\eps^2}+\frac{G_C}{\eps}+...\right]
\ee
where the second term represents the counter-diagram. 
Note that the power of $\mu^2$ in the counter-term is $\varepsilon$
and not $2 \varepsilon$ as in I.
It turns out
that for diagrams with non-vanishing $F$ one has diagram by diagram
the relation $F_C=2 F$ and consequently the pole part does not depend on
$\mu$ as it should be:
\be\label{IMICA}
I-I_C=\left(\frac{\alpha}{4\pi}\right)^2
\left[-\frac{F}{\eps^2}+\frac{G-G_C}{\eps}+...\right]
\ee
The coefficient $G-G_C$ can then be identified with the contribution
of a given diagram (after the inclusion of colour factors) to the
coefficient $(b_2)_{ij}$ entering the master formula (\ref{a1b1}).

\begin{figure}[hbt]
\vspace{0.10in}
\centerline{
\epsfysize=2in
\epsffile{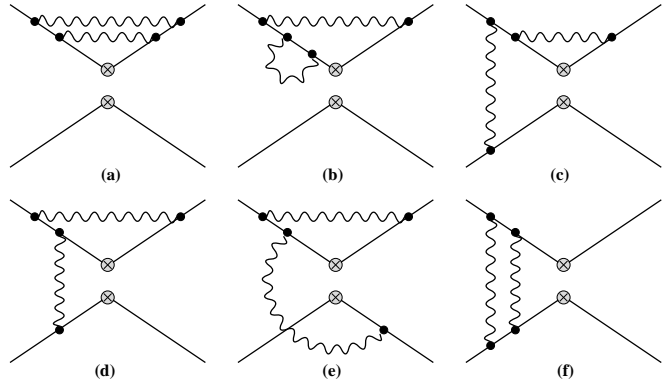}
}
\vspace{0.08in}
\caption[]{Examples of two loop current-current diagrams
contributing to the NLO anomalous dimensions of the operators $Q_1$ and 
$Q_2$.
\label{L:5}}
\end{figure}

\begin{figure}[hbt]
\vspace{0.10in}
\centerline{
\epsfysize=2in
\epsffile{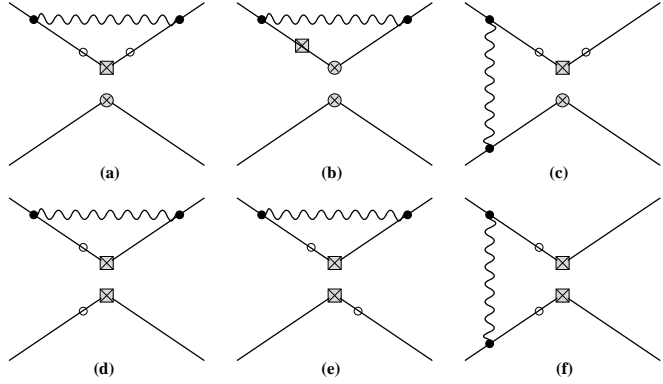}
}
\vspace{0.08in}
\caption[]{Counter-diagrams to the diagrams of fig.~\ref{L:5}.
\label{L:6}}
\end{figure}

Again, as in the one-loop case, the diagrams with gluons exchanged only
between quark legs belonging to the same weak current can be omitted in
the evaluation of two-loop anomalous dimensions of the operators
$Q_1$ and $Q_2$, provided the anomalous dimension of the weak current
vanishes at two-loop level. In this case their contributions to
$(b_2)_{ij}$ are canceled by the $a_2$ term in the master formula
(\ref{a1b1}). It should be stressed that this
feature might not be preserved by some regularization schemes. In particular,
it depends on the treatment of $\gamma_5$ in $D\not=4$ dimensions.
In the NDR scheme, in which $\gamma_5$ anticommutes with $\gamma_\mu$ in
$D\not=4$ dimensions, one has indeed $\gamma^{(1)}_c=0$. But this is
not true in the HV scheme, where $\gamma_5$ has more complicated
properties. Indeed one finds \cite{WEISZ}
\be\label{anc}
[\gamma_c^{(1)}]_{HV}=4 C_F \beta_0.
\ee 
We will return to this issue in a moment. For the time being we 
give the two-loop generalizations of the one-loop anomalous
dimension matrix (\ref{g120}) in the NDR scheme \cite{WEISZ}:
\begin{equation}\label{gNDR} 
 \hat\gamma^{(1)}_{NDR}=
 \left(\begin{array}{cc}
 -\frac{22}{3}-\frac{57}{2 N^2}-\frac{2f}{3N} & 
  \frac{39}{N}-\frac{19N}{6}+\frac{2f}{3} \\
 \frac{39}{N}-\frac{19N}{6}+\frac{2f}{3} & 
 -\frac{22}{3}-\frac{57}{2 N^2}-\frac{2f}{3N}
    \end{array}\right)   
\end{equation}
and in the HV scheme \cite{WEISZ}:
\begin{equation}\label{gHV} 
 \hat\gamma^{(1)}_{HV}=
 \left(\begin{array}{cc}
 -\frac{110}{3}-\frac{57}{2 N^2}+\frac{44 N^2}{3}+
\left(\frac{14}{3N}-\frac{8N}{3}\right) f & 
  \frac{39}{N}+\frac{23N}{2}-2f \\
  \frac{39}{N}+\frac{23N}{2}-2f  & 
 -\frac{110}{3}-\frac{57}{2 N^2}+\frac{44 N^2}{3}+
\left(\frac{14}{3N}-\frac{8N}{3}\right) f
    \end{array}\right)   
\end{equation}
The corresponding result in the DRED scheme has been first
calculated in \cite{ACMP} and confirmed in \cite{WEISZ}.
We observe substantial renormalization scheme dependence. 
In particular the diagonal elements in (\ref{gHV}) contain
terms $\ord(N^2)$ whereas such terms are absent in (\ref{gNDR}).
The origin of these terms can be traced back to the non-vanishing
of $[\gamma_c^{(1)}]_{HV}$. Indeed these terms cancel in the
difference $\hat\gamma^{(1)}_{HV}-2 [\gamma_c^{(1)}]_{HV} \hat 1 
\equiv[\hat\gamma^{(1)}_{HV}]_{\rm eff}$.

As we will discuss in subsection 6.7,
such a shift of two-loop anomalous dimensions is always possible
provided also the matching conditions for the Wilson coefficients
at $\mu_W$ are appropriately changed. Yet eventually this shift
modifies the Wilson coefficients and this modification will be
compensated by the corresponding change in the matrix elements
of the operators so that physical quantities are independent
of these manipulations. All this should be clearer after
subsection 6.7 where the
cancellation of this scheme dependences in physical quantities are
discussed in explicit terms.
\subsubsection{A Warning on the HV Scheme}
In this context we should warn the reader that the numerical
values of the Wilson coefficients in the HV scheme presented here
and in \cite{BBL,BJLW}
 correspond to the choice $[\hat\gamma^{(1)}_{HV}]_{\rm eff}$.
This differs from the treatment of my Italian friends 
\cite{ROMA1,ROMA2} who use
$\hat\gamma^{(1)}_{HV}$ of (\ref{gHV}) instead. For this reason the
NLO corrections to Wilson coefficients in the HV scheme presented 
here are generally smaller
than the ones found by the Rome group. The final physical results
are, however, the same.
\subsection{Explicit Calculation of $2\times 2$ 
Anomalous Dimension Matrix}
\subsubsection{Current-Current Insertions: Generalities}
The set of six diagrams contributing to one--loop anomalous dimension 
matrix through operator
insertions into current-current topologies is given by the diagrams in 
fig.~\ref{L:15} and their symmetric counterparts.
We begin by developing the technology for the calculation of insertions
of any operator with arbitrary colour and Dirac structure into the
diagrams of fig.~\ref{L:15}. 
This will allow us to calculate later also insertions
of penguin operators into the current-current topologies of 
fig.~\ref{L:15}.

Let us then denote the colour and Dirac structure of any operator
by
\be\label{CD}
\hat V_1\otimes\hat V_2,\quad\quad\quad \Gamma_1\otimes\Gamma_2~,
\ee
respectively, so that an operator can be generally written as follows:
\be\label{OG}
O=(\bar s_\alpha \Gamma_1 \hat V_1^{\alpha\beta} c_\beta)\otimes 
  (\bar u_\gamma \Gamma_2 \hat V_2^{\gamma\delta} d_\delta)
\ee
Here we have made  specific choice of quark flavours adapted to the
operators discussed in section 5, but it is trivial to generalize
the following discussion to any other choice of flavours.

Let us consider a few examples:
\be\label{V1}
\hat V_1^{\alpha\beta}\otimes\hat V_2^{\gamma\delta}=
\delta_{\alpha\beta}\otimes\delta_{\gamma\delta}
\equiv{\bf 1}_{\alpha\beta,\gamma\delta}~,
\ee
\be\label{V2}
\hat V_1^{\alpha\beta}\otimes\hat V_2^{\gamma\delta}=
\delta_{\alpha\delta}\otimes\delta_{\gamma\beta}
\equiv{\bf\tilde 1}_{\alpha\beta,\gamma\delta}~,
\ee
\be\label{V3}
\hat V_1^{\alpha\beta}\otimes\hat V_2^{\gamma\delta}=
(T^a)_{\alpha\beta}\otimes(T^a)_{\gamma\delta}
\equiv{\bf \Pi}_{\alpha\beta,\gamma\delta}~.
\ee
Then the colour identity (\ref{tata}) is simply given by
\be\label{CI}
{\bf \Pi}=\frac{1}{2}\left({\bf \tilde 1}-\frac{1}{N}{\bf 1}\right)~.
\ee
In this notation the operators $Q_1$ and $Q_2$ have both the
structure $\Gamma_1=\Gamma_2=\gamma_\mu(1-\gamma_5)$. With the
ordering of flavours as in (\ref{OG}), the colour structure of
$Q_1$ is ${\bf\tilde 1}$. The one of $Q_2$ is ${\bf 1}$.

In order to gain some insight into the calculation of the diagrams
in fig.~\ref{L:15}, let us consider the diagram (a) with the insertion
of the operator $Q_2$. The flavour labels are as in the diagram of
fig.~\ref{L:4}. 
For this diagram we have then
\be\label{DIAG}  
{\cal D}_a^{(1)}=-i g^2\mu^{2\varepsilon} C_F
\int \frac{d^D k}{(2\pi)^D}
\frac{[\bar s T_\nu c]\otimes[\bar u \Gamma^\nu d]}
{k^2[(k+p)^2]^2}
\ee
where
\be\label{TNU}
T_\nu=\gamma_\mu(\not\! k+\not\! p)\Gamma_\nu (\not\! k+\not\! p)\gamma^\mu
=\gamma_\mu\gamma_\rho\Gamma_\nu\gamma_\sigma\gamma^\mu
(k+p)^\rho (k+p)^\sigma
\ee
and $\Gamma_\nu=\gamma_\nu(1-\gamma_5)$. We have used $T^a T^a=C_F$.

Keeping only the divergent part in the relevant D-dimensional integral
we find
\be\label{INTR}
\int \frac{d^D k}{(2\pi)^D}
\frac{(k+p)^\rho (k+p)^\sigma}
{k^2[(k+p)^2]^2}=i g^{\rho\sigma} \frac{1}{16\pi^2}\frac{1}{4}
\frac{1}{\varepsilon}+{\rm finite}
\ee
Thus the divergent part of ${\cal D}_a^{(1)}$ is given by
\be\label{DADIV}
{\cal D}_a^{(1)}=C_F\frac{\as}{4\pi}
\left[\frac{1}{4}\frac{1}{\varepsilon}\right]
\bar s\gamma_\mu\gamma_\rho\Gamma_\nu\gamma^\rho\gamma^\mu c\otimes
\bar u \Gamma^\nu d
\ee

It is straightforward to extend this calculation to other diagrams
in fig.~\ref{L:15} and to the arbitrary operator given in (\ref{OG}).
To this end we fix the ordering of the four flavours as in (\ref{OG})
and drop the external spinors. We find then
\be\label{DA}
{\cal D}_a=\frac{\as}{4\pi}
\left[\frac{1}{4}\frac{1}{\varepsilon}\right]
\left({\cal C}_a^{(1)}
\gamma_\mu\gamma_\rho\Gamma_1\gamma^\rho\gamma^\mu \otimes \Gamma_2
+{\cal C}_a^{(2)}
\Gamma_1\otimes\gamma_\mu\gamma_\rho\Gamma_2\gamma^\rho\gamma^\mu 
\right)
\ee
\be\label{DB}
{\cal D}_b=-\frac{\as}{4\pi}
\left[\frac{1}{4}\frac{1}{\varepsilon}\right]
\left({\cal C}_b^{(1)}
\Gamma_1\gamma_\rho\gamma_\mu\otimes\Gamma_2\gamma^\rho\gamma^\mu 
+{\cal C}_b^{(2)}
\gamma_\mu\gamma_\rho\Gamma_1\otimes\gamma^\mu\gamma^\rho\Gamma_2 
\right)
\ee
\be\label{DC}
{\cal D}_c=\frac{\as}{4\pi}
\left[\frac{1}{4}\frac{1}{\varepsilon}\right]
\left({\cal C}_c^{(1)}
\Gamma_1\gamma_\rho\gamma_\mu\otimes\gamma^\mu\gamma^\rho\Gamma_2 
+{\cal C}_c^{(2)}
\gamma_\mu\gamma_\rho\Gamma_1\otimes\Gamma_2\gamma^\rho\gamma^\mu 
\right)
\ee
where the index (1) stands for the diagrams shown in fig.~\ref{L:15}
and the index (2) for their symmetric counter-parts.
The colour factors are given by 
\be\label{CA}
{\cal C}_a^{(1)}=T^a\hat V_1 T^a\otimes\hat V_2
\qquad 
{\cal C}_a^{(2)}=\hat V_1\otimes T^a\hat V_2 T^a
\ee
\be\label{CB}
{\cal C}_b^{(1)}=\hat V_1 T^a\otimes\hat V_2 T^a
\qquad 
{\cal C}_b^{(2)}=T^a\hat V_1\otimes T^a\hat V_2 
\ee
\be\label{CC}
{\cal C}_c^{(1)}=\hat V_1 T^a\otimes\hat T^a V_2 
\qquad 
{\cal C}_c^{(2)}=T^a\hat V_1\otimes \hat V_2 T^a 
\ee
\subsubsection{Anomalous Dimensions of $Q_1$ and $Q_2$}
Let us apply these general formulae to the case of the operator
$Q_2$ for which we have
\be\label{Q2O}
\hat V_1\otimes\hat V_2 ={\bf 1},\quad\quad
\Gamma_1=\Gamma_2=\gamma_\nu(1-\gamma_5)\equiv\Gamma~.
\ee
Since we are interested only in the $1/\varepsilon$ singularity
we can use in one-loop formulae the rules for $\gamma$-algebra
valid in four dimensions.
Then
\be\label{GR1}
\gamma_\mu\gamma_\rho\Gamma\gamma^\rho\gamma^\mu \otimes \Gamma
=\Gamma\otimes\gamma_\mu\gamma_\rho\Gamma\gamma^\rho\gamma^\mu 
=4 \Gamma\otimes\Gamma
\ee
\be\label{GR2}
\Gamma\gamma_\rho\gamma_\mu\otimes\Gamma\gamma^\rho\gamma^\mu 
=\gamma_\mu\gamma_\rho\Gamma\otimes\gamma^\mu\gamma^\rho\Gamma 
=16 \Gamma\otimes\Gamma
\ee
\be\label{GR3}
\Gamma\gamma_\rho\gamma_\mu\otimes\gamma^\mu\gamma^\rho\Gamma 
=\gamma_\mu\gamma_\rho\Gamma\otimes\Gamma\gamma^\rho\gamma^\mu 
=4 \Gamma\otimes\Gamma
\ee
These results can be most efficiently found by using a trick which
I will call the {\it Greek Method} \cite{GREEK} from now on.
Let me illustrate this method by 
deriving (\ref{GR2}). Following \cite{GREEK} let us write
\be\label{GR4}
\gamma_\mu\gamma_\rho\Gamma\otimes\gamma^\mu\gamma^\rho\Gamma =
{\rm A}~\Gamma\otimes\Gamma~,
\ee
where A is the coefficient we are looking for.
In order to find it we replace $\otimes$ in (\ref{GR4})
by a matrix $\gamma_\tau$ to obtain
\be\label{GR5}
\gamma_\mu\gamma_\rho\Gamma\gamma_\tau\gamma^\mu\gamma^\rho\Gamma =
A~\Gamma\gamma_\tau\Gamma.
\ee 
Inserting $\Gamma$ of (\ref{Q2O}) into this equality and contracting
indices we determine A to be 16.

Using (\ref{GR1})--(\ref{GR3}) in our master formulae 
(\ref{DA})--(\ref{DC}) and summing all diagrams we find
\be\label{SUMD}
\sum_i {\cal D}_i=\frac{\as}{4\pi} \frac{1}{\varepsilon}
\Gamma\otimes\Gamma
\left[{\cal C}_a^{(1)}+{\cal C}_a^{(2)}-
4 ({\cal C}_b^{(1)}+{\cal C}_b^{(2)})+
{\cal C}_c^{(1)}+{\cal C}_c^{(2)}\right].
\ee
For the colour structure in (\ref{Q2O}), the colour structures in this formula
can be easily found to be 
\be\label{colour1}
{\cal C}_a^{(1)}={\cal C}_a^{(2)}=C_F {\bf 1},
\ee
\be\label{colour2}
{\cal C}_b^{(1)}={\cal C}_b^{(2)}={\cal C}_c^{(1)}={\cal C}_c^{(2)}
=\frac{1}{2}\left({\bf \tilde 1}-\frac{1}{N}{\bf 1}\right).
\ee

Now ${\bf\tilde 1}$ stands for the operator $Q_1$. Consequently
inserting (\ref{colour1}) and (\ref{colour2}) into (\ref{SUMD})
and comparing the coefficient of $1/\varepsilon$ with (\ref{zgf1})
we extract
\be\label{b22a}
(b_1)_{21}=-3,
\quad\quad
(b_1)_{22}=2 C_F+\frac{3}{N}.
\ee

The insertion of the operator $Q_1$ represented by ${\bf \tilde 1}$
into diagrams of fig.~\ref{L:15} can be evaluated in an analogous manner
by using the master formulae (\ref{DA})--(\ref{DC}). Because the
colour structure is more complicated the calculation is now a bit more
involved. In order to avoid this complication it is useful to
make a Fierz reordering in $Q_1$ and $Q_2$ so that
\be\label{fierz}
Q_1=(\bar s_\alpha d_\alpha)_{V-A}(\bar u_\beta c_\beta)_{V-A}\quad\quad
Q_2=(\bar s_\alpha d_\beta)_{V-A}(\bar u_\beta c_\alpha)_{V-A}~.
\ee
Now the roles of $Q_1$ and $Q_2$ are interchanged: $Q_1$ is ${\bf 1}$
and $Q_2$ is ${\bf\tilde 1}$. Since gluons are flavour blind we
find immidiately
\be\label{b11a}
(b_1)_{11}=2 C_F+\frac{3}{N},
\quad\quad
(b_1)_{12}=-3~.
\ee
(\ref{b22a}) and (\ref{b11a}) are precisely the values given in 
(\ref{b22}) and (\ref{b11}) respectively.
Upon
inserting them into the one-loop master formula (\ref{a1b1}) and using
$a_1=-C_F$ we finally
reproduce the anomalous dimension matrix 
(\ref{g120}). We will extend this calculation to penguin operators in
subsection 8.5. 

\subsection{Mixing of Operators with different Dimensions}
It is useful to know the following properties of mixing of operators
with different canonical dimensions
\bi
\item
The operators of a given dimension mix only into operators of the
same or lower dimension. In a more formal terminology: to
renormalize an operator of a given dimension one needs only 
operators as counter-terms of the same or lower dimension.
\item
This means in particular that the operators of dimension six,
as $Q_1$ and $Q_2$, can mix into other six dimensional operators
and five dimensional magnetic penguin operators of sections 8.7 and 12.
On the other hand the magnetic penguin operators cannot mix into
dimension six operators.
\item
Consequently whereas the $Q_1$ and $Q_2$ operators influence the
Wilson coefficients of the magnetic penguin operators, the
latter operators have no impact on $C_1$ and $C_2$.
\ei

The proof of these properties is based essentially on dimensional
analysis. It can be found on page 149 of the book by Collins 
\cite{Collins}.

Here comes another useful remark. As we will discuss in section 12
the mixing between the operators $(Q_1,Q_2)$ and the magnetic
penguin operators appears first at the two-loop level. That is
the leading anomalous dimension is obtained by calculating two-loop
diagrams and not one-loop diagrams as discussed sofar.
The next-to-leading anomalous dimensions are then obtained from
three-loop calculations. In this particular case our master
formulae in (\ref{a1b1})  change to ($i\not=j$)
\begin{equation}\label{a1b1m}
\gamma^{(0)}_{ij}=-4[(b_2)_{ij}],
\quad\quad
\gamma^{(1)}_{ij}=-6[(b_3)_{ij}]~.  
\end{equation}
with $(b_2)_{ij}$ and $(b_3)_{ij}$ obtained from $1/\varepsilon$
singularities in two-loop and three-loop diagrams respectively.

\subsection{Renormalization Scheme Dependence}
At NLO various quantities like 
the Wilson coefficients
and the anomalous dimensions depend on the renormalization scheme for
operators.
This dependence  arises because the renormalization
prescription involves an arbitrariness in the finite parts to be
subtracted along with the ultraviolet singularities.  
Two different schemes are then related by a finite renormalization.  

A particular example of the RS  dependence is the dependence on the
treatment of $\gf$ in $D$ dimensions. We have seen in (\ref{gNDR})
and (\ref{gHV}) that the two-loop anomalous dimension matrix for
the operators $(Q_1,Q_2)$ in the NDR scheme differs from the one
in the HV scheme. 

Returning back to our discussion of the NLO corrections to the
Wilson coefficients of subsection 6.3, one finds that 
\begin{equation}\label{rsi}
\beta_0,\quad\beta_1,\quad\gamma^{(0)},\quad\vec A^{(0)},\quad
\vec A^{(1)},\quad \hat r^T+\hat J,\quad\langle\vec Q\rangle^T\vec C  
\end{equation}
are {\it scheme independent}, whereas
 \begin{equation}\label{rsd}
\hat r,\quad\gamma^{(1)},\quad \hat J,\quad\vec C,\quad\langle\vec Q\rangle 
\end{equation}
are {\it scheme dependent.} Let us demonstrate this.

First of all, it is clear that the product
\begin{equation}\label{qtc}
\langle\vec Q(\mu)\rangle^T\vec C(\mu)  \end{equation}
representing the full amplitude is independent of RS. 
The factorization of the amplitude into $\vec C$ and
$\langle\vec Q\rangle$
makes them, however, scheme dependent.
Explicitly, for two different schemes
(primed and unprimed) we have
\begin{equation}\label{rsqc}
\langle\vec Q\rangle'
=(1+\aspi \hat s)\langle\vec Q\rangle~,\qquad
 \vec C'=(1-\aspi \hat s^T)\vec C~,  \end{equation}
 where $\hat s$ is a constant matrix representing a finite
renormalization of $\vec C$ and $\langle\vec Q\rangle$.

Having the relations (\ref{rsqc}) at hand it is straightforward
to find relations between various quantities in the primed and
unprimed schemes. From
\begin{equation}\label{qarsP}
\langle\vec Q(\mu_W)\rangle=(\hat 1
+{\as(\mu_W)\over 4\pi} \hat r)\vec S,  \end{equation}
where $\vec S$ is a vector of tree level matrix elements, 
we immediately obtain
\begin{equation}\label{rprs}  
\hat r'=\hat r+ \hat s~.  \end{equation}

Next from
\begin{equation}\label{qtuc} \langle\vec Q(\mu)\rangle^T\vec C(\mu)\equiv
  \langle\vec Q(\mu)\rangle^T \hat U(\mu,M_W) \vec C(M_W)  \end{equation}
we have
\begin{equation}\label{upus} \hat U'(\mu,M_W)=
(1-{\as(\mu)\over 4\pi}\hat s^T)
\hat U(\mu,M_W)(1+{\as(M_W)\over 4\pi}\hat s^T).
\end{equation}
A comparison with 
\begin{equation}\label{u1jjP}
\hat U(\mu,\mu_W)=
(1+{\as(\mu)\over 4\pi} \hat J) \hat U^{(0)}(\mu,\mu_W) 
(1-{\as(\mu_W)\over 4\pi} \hat J)
\end{equation}
 yields then
\begin{equation}\label{jpjs} \hat J'=\hat J-\hat s^T \end{equation}

Next from (\ref{q0zq}) and (\ref{rsqc}) we clearly have
\begin{equation}\label{zpzs}  \hat Z'=\hat Z~ (\hat 1- \aspi \hat s)  
\end{equation}
Using next the defintion of the anomalous dimension matrix
(\ref{gaz2}) and the expansion (\ref{gg01}) we find 
\begin{equation}\label{gpgs}
\hat \gamma^{(0)\prime}=\hat\gamma^{(0)} \qquad
 \hat\gamma^{(1)\prime}=\hat\gamma^{(1)}+[\hat s,\hat\gamma^{(0)}]+
2\beta_0 \hat s \end{equation}

Let us make a few observations:
\bi
\item
From (\ref{rprs}) and (\ref{jpjs}) follows the
scheme independence of $\hat r^T+\hat J$. 
Next, $\vec A^{(0)}$ and $\vec A^{(1)}$, obtained from the calculation
in the full theory, are clearly independent of the renormalization
of operators. Consequently,
the factor on the right hand
side of $\hat U^{(0)}$
in $\vec C(\mu)$ in (\eqn{cjua}), related to the ``upper end'' 
of the evolution,
is independent of RS.
\item
 The same is true for $\hat U^{(0)}$ as $\hat\gamma^{(0)}$ and
$\beta_0$ are scheme independent. 
\item
$\vec C$ 
depends on RS through  $\hat J$ to the
left of $\hat U^{(0)}$. 
This dependence is compensated for by the
corresponding scheme dependence of $ \langle\vec Q\rangle$ in (\ref{rsqc}).
\end{itemize}

In the absence of operator mixing the relations between various
quantities in two different schemes simplify. Going back to
$C_{\pm}(\mu)$ in (\ref{B9PP}) we have
\begin{equation}\label{gbjs}
{\gamma_\pm^{(1)}}'=\gamma_\pm^{(1)} + 2\beta_0 s_\pm~,
\qquad B_\pm'=B_\pm - s_\pm~,
\qquad J_\pm'=J_\pm - s_\pm~,
\end{equation}
where $s_\pm$ are constant numbers analogous to $\hat s$ in (\ref{rsqc}).

Recalling
\begin{equation}\label{jg01}
J_\pm={1\over 2\beta_0}\left({\beta_1\over\beta_0}\gamma_\pm^{(0)}
  -\gamma_\pm^{(1)}\right)   \end{equation}
we verify  the scheme independence  of
$B_\pm-J_\pm$ in (\eqn{B9PP}).
Again the scheme dependence of $C_\pm(\mu)$ originates in the scheme
dependence of $J_\pm$ present in the first factor in (\eqn{B9PP}).

We should emphasize that the renormalization scheme dependence discussed
here refers to the renormalization of operators and should be
distinguished from the renormalization scheme dependence of $\alpha_s$
discussed in section 4.7. The issue of the latter scheme dependence
in the context of OPE is discussed in detail at the end of section
III in \cite{BBL} and will not be repeated here.
\subsection{Renormalization Scale Dependence}
A physical amplitude cannot depend on the arbitrary renormalization
scale $\mu$. The $\mu$-dependence of the Wilson coefficients has to
be canceled by the $\mu$-dependence of the matrix elements
$\langle Q_i(\mu)\rangle$. Due to the mixing under renormalization
this cancelation may involve simultaneously several operators.
Now, whereas $C_i(\mu)$ can be calculated in perturbation theory,
this is not the case for the matrix elements $\langle Q_i(\mu)\rangle$.
Unfortunately, the existing non-perturbative methods are still
insufficient to study the $\mu$-dependence of $\langle Q_i(\mu)\rangle$
and to verify the $\mu$-independence of physical amplitudes in explicit
terms. Notable exceptions are inclusive decays of heavy mesons
like $B\to X_s \gamma$ and $B \to X_s e^+e^-$, where one can analyze
the cancellation of the $\mu$-dependence using perturbative calculations
of the relevant matrix elements $\langle Q_i(\mu)\rangle$. We refer
to \cite{BuMu:94,BKP1} for the full exposition of this issue in
$B\to X_s \gamma$ and $B \to X_s e^+e^-$. Here it sufficies to illustrate
the cancellation of the $\mu$ dependence by considering a toy model
in which only a single operator $Q$ is present and its matrix element
$\langle Q(\mu)\rangle$ is calculated in perturbation theory.

Let us consider then the amplitude
\begin{equation}\label{hqtcP}
A=\langle {\cal H}_{eff}\rangle=
{G_F\over\sqrt{2}}
  \langle Q(\mu_b)\rangle C(\mu_b)   
\end{equation}
with
\be\label{coff}
C(\mu_b)=U(\mu_b,\mu_W) C(\mu_W)
\ee
where $\mu_b=\ord(m_b)$ and $\mu_W=\ord(\mw)$.
We want to discuss the cancelation of the $\mu_b$ and $\mu_W$ dependences
in (\ref{hqtcP}) and (\ref{coff}) in explicit terms.

Beginning with the leading logarithmic approximation we have
\begin{equation}\label{B9PP0}
U(\mu_b,\mu_W)=  U^{(0)}(\mu_b,\mu_W) =
\left[\frac{\as(\mu_W)}{\as(\mu_b)}\right]^{\frac{\gamma^{(0)}}{2\beta_0}}, 
\qquad C(\mu_W)=1~.
\end{equation}
Moreover $ \langle Q(\mu_b)\rangle=\langle Q \rangle_{tree} $ carries no
$\mu_b$ dependence.
Consequently in LO the amplitude depends on $\mu_b$ and $\mu_W$.
Since $\as(\mu_b)\gg\as(\mu_W)$, the $\mu_b$-dependence is stronger
than the $\mu_W$-dependence. 
If $\gamma^{(0)}/2\beta_0$ is $\ord(1)$
the $\mu_b$-dependence of the resulting amplitudes and branching ratios 
may be rather
disturbing. Known example of such a situation is the strong 
$\mu_b$-dependence
of the branching ratio $Br(B\to X_s\gamma)$ at LO.
We will discuss this in detail in section 12.

Let us next include NLO corrections. Now  the various entries in
(\ref{B9PP0}) are generalized as follows:
\begin{equation}\label{u0jjP}
 U(\mu_b,\mu_W)=
(1+{\as(\mu_b)\over 4\pi} J )  U^{(0)}(\mu_b,\mu_W) 
(1-{\as(\mu_W)\over 4\pi} J)
\end{equation}

\be\label{cp}
C(\mu_W)=1+ {\as(\mu_W)\over 4\pi}\left ( \frac{\gamma^{(0)}}{2}
\ln\frac{\mu_W^2}{\mw^2}+B\right)
\ee
\be\label{MAT5}
\langle Q(\mu_b)\rangle=\langle Q \rangle_{tree}\left[1+
{\as(\mu_b)\over 4\pi}\left( \frac{\gamma^{(0)}}{2}
\ln\frac{m_b^2}{\mu_b^2}+\tilde r\right)\right]
\ee

We can now show that the amplitude $A$ in (\ref{hqtcP})  
is independent of $\mu_b$ and $\mu_W$ 
in $\ord(\as)$.
Using the following useful formula 
\be
\frac{\as(m_1)}{\as(m_2)}=1+\frac{\as}{4\pi}\beta_0\ln\frac{m_2^2}{m_1^2}
\ee
and keeping only logarithmic terms
we can rewrite (\ref{u0jjP}) as
\be\label{UNEW}
U(\mu_b,\mu_W)=\left (1+
{\as(\mu_b)\over 4\pi} \frac{\gamma^{(0)}}{2}
\ln\frac{\mu_b^2}{m_b^2}\right)
\left[\frac{\as(\mw)}{\as(\mb)}\right]^{\frac{\gamma^{(0)}}{2\beta_0}} 
\left(1+ {\as(\mu_W)\over 4\pi} \frac{\gamma^{(0)}}{2}
\ln\frac{\mw^2}{\mu_W^2}\right)
\ee
Inserting (\ref{cp}), (\ref{MAT5}) and (\ref{UNEW}) into
(\ref{hqtcP}) we find that $\mu_b$ and $\mu_W$ dependences cancel
at $\ord(\as)$.

This simple example illustrates very clearly the virtue of NLO
corrections. They reduce considerably various $\mu$-dependences
present in the LO approximation. On the other hand we recover the
well known fact that at fixed order of perturbation theory there
remain unphysical $\mu$-dependences which are of the order of the
neglected higher order contributions. In our simple example the
leftover $\mu_b$ and $\mu_W$ dependences can be investigated
numerically by inserting expressions (\ref{u0jjP})--(\ref{MAT5}) into 
(\ref{hqtcP}) and varying $\mu_b$ and $\mu_W$ say in the ranges
$\mb/2\le \mu_b \le 2 \mb$ and $\mw/2\le \mu_W\le 2 \mw$, respectively.
By comparing the result of this exercise with an analogous exercise
in LO one can on the one hand appreciate the importance of NLO
calculations. On the other hand the leftover $\mu_W$ and $\mu_b$
dependences
at NLO give a rough estimate of the theoretical uncertainty due
to the truncation of the perturbative series. We will illustrate
all this with numerical examples at later stages in these
lectures.

The $\mu$--dependences discussed here are related to the renormalization
group evolution of the Wilson coefficients from high to low energy
scales. This evolution originates in the non-vanishing of the
anomalous dimensions of the corresponding operators $Q_i$.
On the other hand, as we have seen in section 4, the nonvanishing of 
the anomalous dimension
$\gamma_m$ of the mass operator implies the $\mu$--dependence of
the quark masses, in particular $\mt(\mu_t)$, 
$\mb(\mu_b)$ and $\mc(\mu_c)$. These $\mu$--dependences and their 
cancellation in decay amplitudes
will be briefly discussed in section 8 and in the phenomenological
sections of these lectures.

\subsection{Evanescent Operators}
\subsubsection{Origin of Evanescent Operators}
In evaluating the anomalous dimensions of $Q_1$ and $Q_2$ we have used
the Greek Method to reduce the complicated Dirac structures given in
(\ref{GR1})--(\ref{GR3}) to $\Gamma\otimes\Gamma$.
 Since we were only
interested in the $1/\varepsilon$ singularity in a one-loop diagram,
this reduction has been performed in $D=4$ dimensions. In the case of
two-loop calculations, in which the diagrams of fig. \ref{L:15} are
subdiagrams of the diagrams in fig. \ref{L:5}, this reduction has to 
be performed in arbitrary
D-dimensions. Indeed, in a two-loop diagram, the leading singularity
is $1/\varepsilon^2$. The $\ord(\varepsilon)$ terms arising from
reductions like (\ref{GR1})--(\ref{GR3}) in arbitrary D-dimensions, 
when multiplied
by $1/\varepsilon^2$, will contribute to the $1/\varepsilon$
singularity relevant for the calculation of the two-loop
anomalous dimensions.

The question then is how to find $\ord(\varepsilon)$ corrections
to (\ref{GR1})--(\ref{GR3}). I will follow here the work done almost 
ten years ago in collaboration with Peter Weisz \cite{WEISZ}. 
In this paper we have
proposed a simple method for finding these terms. Although more
general methods have been developed subsequently, I still think
that our method is most useful for practical purposes. Yet 
other methods \cite{DuGr,HNE,UN95,SH94}, 
in particular the one of Herrlich and Nierste \cite{HNE,UN95,SH94},
give a deeper inside into these matters and I will briefly
discuss them at the end of this subsection.

The simplest method to find the $\ord(\varepsilon)$ terms in question
would be to apply the Greek Method in D--dimensions. That is, evaluate
(\ref{GR5}) in D-dimensions in order to determine the coefficient $A$.
For instance in the case of (\ref{GR2}) we would find $16-4\varepsilon$
 instead of 16 when using the NDR scheme for $\gamma_5$.
This is what has been done in \cite{GREEK}. Yet as pointed out in
\cite{WEISZ}, the mere replacement of 16 in (\ref{GR2}) 
by $16-4\varepsilon$ with
analogous replacements in (\ref{GR1}) and (\ref{GR3}) would eventually 
give incorrect
two-loop anomalous dimensions. As demonstrated in \cite{WEISZ} the
correct procedure is to supplement
the Greek Method in D dimensions  by the
addition of other operators to the r.h.s of (\ref{GR1})--(\ref{GR3})
which vanish in $D=4$ dimensions. Such operators are called
{\it evanescent} operators.

We will explain the role of evanescent operators in the calculation
of two-loop anomalous dimensions below. First, however, we want to
give the generalizations of (\ref{GR1})--(\ref{GR3}) to $D\not=4$ 
dimensions. In the case of the NDR scheme for $\gamma_5$ they are 
given as follows \cite{WEISZ}
\be\label{GRD1}
\gamma_\mu\gamma_\rho\Gamma\gamma^\rho\gamma^\mu \otimes \Gamma
=4 (1-2\varepsilon)\Gamma\otimes\Gamma
\ee
\be\label{GRD2}
\Gamma\gamma_\rho\gamma_\mu\otimes\Gamma\gamma^\rho\gamma^\mu 
=4(4-\varepsilon) \Gamma\otimes\Gamma+E^{\rm NDR}
\ee
\be\label{GRD3}
\Gamma\gamma_\rho\gamma_\mu\otimes\gamma^\mu\gamma^\rho\Gamma 
=4 (1-2\varepsilon) \Gamma\otimes\Gamma-E^{\rm NDR}~,
\ee
where $\ord(\varepsilon^2)$ terms have been dropped.
Identical results are found for the structures in the second column
of the set (\ref{GR1})--(\ref{GR3}). 
$E^{\rm NDR}$ stands for the evanescent operator
given explicitly by
\be\label{EVAN}
E^{\rm NDR}=\frac{1}{2}
[\gamma_\mu\gamma_\rho\Gamma\gamma^\rho\gamma^\mu\otimes\Gamma
+\Gamma\otimes\gamma_\mu\gamma_\rho\Gamma\gamma^\rho\gamma^\mu
-\Gamma\gamma_\rho\gamma_\mu\otimes\gamma^\mu\gamma^\rho\Gamma
-\gamma_\mu\gamma_\rho\Gamma\otimes\Gamma\gamma^\rho\gamma^\mu]
\ee
As one can verify using the Greek Method, $E^{\rm NDR}$ vanishes in
$D=4$. In the case of HV and DRED schemes the formulae 
(\ref{GRD1})--(\ref{GRD3}) 
are modified and the evanescent operators have more
complicated structures. They can be found in \cite{WEISZ}.
It should be noted that there is no contribution from evanescent
operators to (\ref{GRD1}) in the NDR scheme. 

From calculational point of view the insertions of evanescent operators
into the relevant diagrams are most efficiently evaluated by defining
$E^{\rm NDR}$ simply as the difference between the structures on the
l.h.s of (\ref{GRD2}) and (\ref{GRD3}) and the respective terms on the
r.h.s involving $\Gamma\otimes\Gamma$. We will demonstrate this 
explicitly below.  

\subsubsection{Including Evanescent Operators in the Master
Formulae}
In subsection 6.4 we have derived the master formulae (\ref{a1b1}) and 
(\ref{a1b1s}) for
the computation of two loop anomalous dimensions. This derivation
did not take into account the presence of evanescent operators.
Therefore in cases in which the contributions of these operators
matter, our formulae are strictly speaking incomplete. It is the
purpose of the next few pages to correct for it and to derive
a procedure for the calculation of two-loop anomalous dimensions
which takes into account the evanescent operators. I follow here
again my work in collaboration with Peter Weisz \cite{WEISZ}.

Let us go back to the equation (\ref{IMIC}) which invoves the coefficients
$(F,G)$ and $(F_C,G_C)$ in the singularities of the two-loop
diagrams and the corresponding counter-diagrams respectively.
The evaluation of F and G  is still 
straightforward. Having the final result for a two-loop
diagram with complicated Dirac  structure one 
can simply project on the space of physical operators, denoted
generically by $\Gamma\otimes\Gamma$,
by using the Greek method. 
In this way one can easily deduce the coefficients of the terms 
proportional to  
$\Gamma\otimes\Gamma$ . As an example let us consider
the Dirac structure resulting from the diagram (f) in
fig.~\ref{L:5}. Then
the projection by means of the Greek Method gives:
\be\label{GRTW}
\Gamma\gamma_\mu\gamma_\nu\gamma_\rho\gamma_\tau\otimes\Gamma
\gamma^\mu\gamma^\nu\gamma^\rho\gamma^\tau=
16~(16-14\varepsilon)~\Gamma\otimes\Gamma~.
\ee
The treatment of counter-diagrams needs more care. After the evaluation
 of the subdiagrams the $1/\varepsilon$ contributions are multiplied
by the structures in (\ref{GRD1})--(\ref{GRD3}) 
i.e. they 
include E operators. Making projection onto $\Gamma\otimes\Gamma$ 
 already at this stage would be 
incorrect. Indeed inserting E into  counter-diagrams of
 fig.~\ref{L:6}, generates  back the 
original operator $\Gamma\otimes\Gamma$ and introduces a correction 
to $G_C$ and consequently a correction
to two-loop  anomalous dimension 
of the original operator. It is precisely this correction which
 we have neglected in our master formulae. We will now find how
 our method has to be modified in order to include the effects
 of E-operators.

 For two-loop computation it is sufficient to consider the effects
 of mixing with the evanescent operators   specified
 in the previous subsubsection.
  However, higher loop computations would require,
 in the NDR and HV schemes, consideration of
 an ever increasing number of independent
 operators. Thus generally
 the renormalized operators in generic
 $D$- dimensions are given by,
\be
 O_{i} = (\hat Z^{-1})_{i j} O^{(0)}_{j}~.
\ee
Let us consider the case where the set $O^{(0)}_{i}$ includes the initial 
bare operators $Q_+$ and $Q_-$ introduced in section 5 which are expected
not to mix under renormalization. In what follows we will denote them by
$O^{(0)}_{j}$ with  $j=1,2,$ respectively. All operators $O_{j}$  
with $j>2$  correspond to evanescent operators. It is convenient
 to choose the basis such that the operators
 $O^{(0)}_{j}$ with $j=3,4$ respectively are the
 evanescent operators $E^{+},E^{-}$. As mentioned above it is, for
 our purposes, not necessary to specify the basis further nor to
give explicit formulae of $E^\pm$.
 
 The renormalization matrix $\hat Z$ has a perturbative expansion of the 
form,
\be\label{Zhat}
   \hat Z = \hat 1 + \frac{\as}{4\pi}\hat Z^{(1)} 
+\frac{\as^2}{4\pi} \hat Z^{(2)} + ...
\ee
 Only the first four columns and first four rows of these
 (a priori infinite
 dimensional) matrices are of interest here. The understanding of
 the form of the matrices $\hat Z^{(1)},\hat Z^{(2)}$ is crucial.
 First we have
\be\label{Zmatrix}
\hat Z^{(1)}=\left(\matrix{* & 0 & * &0 \cr 
                       0 & * & 0 & * \cr
                       * & * & - & - \cr
                       * & * & - & - \cr }\right) 
\ee
 where a * denotes non-zero entries and the elements 
"--" are of no interest to us.
 In particular we have,
 $Z^{(1)}_{1 2} = Z^{(1)}_{2 1} = 0$. This situation need
 not, however, continue at higher loops, since in generic
 $D$-dimensions the bare operators $Q^{(0)}_{\pm}$ do not have definite
 Fierz transformation properties in the NDR and HV schemes. Hence it can,
 and in fact does in the NDR and HV schemes, happen that,
\be
   Z^{(2)}_{1 2} \not= 0, \quad\quad  Z^{(2)}_{2 1} \not= 0~.
\ee
 At the same time, at the one-loop level not only do we have
$Z^{(1)}_{3 1} \not=0$,  $Z^{(1)}_{4 2} \not= 0$
 but to define renormalized evanescent operators which can really be
 neglected in precisely 4-dimensions one must, in general, take into
 account the mixing with operators of differing 'naive' Fierz
 symmetry i.e. it can happen that
 \be
   \hat Z^{(1)}_{3 2} \not= 0, \quad\quad \hat Z^{(1)}_{4 1} \not= 0~.
\ee
 We will soon see that this is necessary so that the renormalized operators
 when restricted to precisely 4-dimensions have the correct Fierz symmetry.

  Consider now
 the renormalization group equations for Green functions containing
 one renormalized operator $O_{j}$ insertion, in the regularized
 $D$-dimensional theory. They take the standard form 
 but  due to the mixing with the evanescent
 bare operators, an anomalous dimension matrix occurs
 \be\label{ANOEV}
  \hat\gamma = \hat Z^{-1} \mu {{\partial}\over{\partial \mu}} \hat Z
= \hat Z^{-1}(-\varepsilon g + \beta (g) ){{\partial}\over{\partial g}} 
\hat Z.
\ee
 Expanding $\beta(g)$, $\hat\gamma$ and $\hat Z$ 
 in powers of the renormalized coupling $g$ as in the previous
sections we obtain from (\ref{ANOEV})

\be\label{ga0new} 
  \hat\gamma^{(0)} = -2 \varepsilon \hat Z^{(1)},
\ee
 and at two loops,
\be\label{g1new}
  \hat\gamma^{(1)} = -4 \varepsilon \hat Z^{(2)} 
   - 2\beta_{0} \hat Z^{(1)} +
2\varepsilon \hat Z^{(1)} \hat Z^{(1)}.
\ee

 Expanding the $\hat Z^{(r)}$ in inverse powers of $\varepsilon$,
\be\label{ZEXP}
\hat Z^{(1)}= \hat Z_0^{(1)}+\frac{1}{\varepsilon} \hat Z_1^{(1)}
\quad\quad
\hat Z^{(2)}= \hat Z_0^{(2)}+\frac{1}{\varepsilon} \hat Z_1^{(2)}
+\frac{1}{\varepsilon^2} \hat Z_2^{(2)}
\ee
 and using the fact that the anomalous dimension
 matrix has a finite limit for
 $\varepsilon \rightarrow 0$
 we must have the relation,
\be\label{COND}
4 \hat Z^{(2)}_{2} + 2\beta_{0} \hat Z^{(1)}_{1} - 
2 \hat Z^{(1)}_{1} \hat Z^{(1)}_{1}=0,
\ee
 which has been explicitly checked for the physical $\pm$ submatrix
in \cite{WEISZ}.
We also get
\be\label{GAMMA1}
\gamma^{(1)}=-4 \hat Z^{(2)}_1-2\beta_0 \hat Z^{(1)}_0
   + 2 (\hat Z^{(1)}_1 \hat Z^{(1)}_0+\hat Z^{(1)}_0 \hat Z^{(1)}_1)~.
\ee
Note that we have introduced in (\ref{ZEXP}) the nonsingular terms 
$\hat Z_0^{(1)}$ and $\hat Z_0^{(2)}$ in order to be able to incorporate
the effects of evanescent operators.
In particular the presence of the finite renormalization $\hat Z_0^{(1)}$
allows in the approach of \cite{WEISZ} to remove the finite contributions
from evenescent operators to the matrix elements of physical operators.
On the other hand, as we will see in a moment, this finite renormalization
has an impact on the two-loop anomalous dimensions of the physical operators
and consequently on their Wilson coefficients.
In this  context we note that
\be\label{COND2}
 (\hat Z^{(1)}_0)_{ij} = 0~~~(i,j=1,2)~,
\quad\quad
   (\hat Z^{(1)}_{1})_{3 1} = (\hat Z^{(1)}_{1})_{4 2} = 0.
\ee
The latter property assures that $1/\varepsilon^2$ terms are not
affected by the evanescent operators at the two-loop level.
Finally using the properties (\ref{Zmatrix}) and (\ref{COND2}) in 
(\ref{GAMMA1})
we find
\be\label{R1}
 \gamma^{(1)}_{+}=\gamma^{(1)}_{11}=
   -4(\hat Z^{(2)}_1)_{11}+2(\hat Z^{(1)}_1)_{13}(\hat Z^{(1)}_0)_{31}
\ee
\be\label{R2}
 \gamma^{(1)}_{-}=\gamma^{(1)}_{22}=
   -4(\hat Z^{(2)}_1)_{22}+2(\hat Z^{(1)}_1)_{24}(\hat Z^{(1)}_0)_{42}
\ee
\be\label{R3}
\gamma^{(1)}_{+-}=\gamma^{(1)}_{12}=
   -4(\hat Z^{(2)}_1)_{12}+2(\hat Z^{(1)}_1)_{13}(\hat Z^{(1)}_0)_{32}
\ee
\be\label{R4}
\gamma^{(1)}_{-+}=\gamma^{(1)}_{21}=
   -4(\hat Z^{(2)}_1)_{21}+2(\hat Z^{(1)}_1)_{24}(\hat Z^{(1)}_0)_{41}
\ee
 The first term in  (\ref{R1}) and (\ref{R2}) 
represents (after addition of wave function renormalization)
 our master formula (\ref{a1b1}) 
which was obtained neglecting the mixing with the E-operators.
 The remaining 
terms reflecting the mixing in question are the corrections we
 were looking for.

Without these corrections and corresponding corrections
 in eqs. (\ref{R3}) and (\ref{R4}) the renormalized operators $Q_\pm$
would not have  the correct Fierz symmetry 
and they would mix under renormalization at the two-loop level
 i.e. $\gamma^{(1)}_{+-}$ and $\gamma^{(1)}_{-+}$ 
would be non-zero. The inclusion of E-operators restores the
 Fierz symmetry and removes this mixing i.e.
 $ \gamma^{(1)}_{+-}=\gamma^{(1)}_{-+}=0$. This is explicitly
demonstrated in \cite{WEISZ}.

 Looking at the 'extra contribution'
 from the evanescent operators, one realizes that it is proportional
 to the contribution that the counter-terms involving an
 evanescent operator
 insertion yield to the
 computation of $\hat Z^{(2)}$. 
 This 
is precisely what we stated at the beginning of this subsubsection.
 Note however, 
that the correction terms in eq. (\ref{R1}) and (\ref{R2}) 
are by factor 2
 smaller than 
the corresponding counter-terms (involving $Q_\pm$ operators)
 present in the main 
terms. In the language of diagrams the result just means that
in calculating $\gamma^{(1)}$ the 
contributions to counter-term
 diagrams involving an evanescent
 operator should be multiplied by a factor 1/2.

\subsubsection{How to Use the Improved Master Formulae}
We can now summarize the improved procedure for the calculation
of the two-loop anomalous dimensions.

{\bf Step 1:}

Calculate the full two-loop diagrams and project the 
Dirac structures onto the physical operators by means of the
Greek Method. This gives in particular the coefficient G in
(\ref{IMIC}).

{\bf Step 2:}

Calculate the usual contribution to the counter-term by taking
the relevant subdiagram of a given two-loop diagram, 
projecting it on to the physical
operators $\Gamma\otimes\Gamma$ by means of the Greek Method, 
inserting the result
of this projection into the remaining subdiagram  of a given
 two-loop diagram
and projecting the resulting expression again on to the physical
operators by means of the Greek Method.
 This step gives the first part of $G_C$ in (\ref{IMIC}).
We will denote it by $G^a_C$.

\newpage
{\bf Step 3:} 

Calculate the contribution of the evanescent operator to the
counter diagram by simply inserting
the difference of the two structures in (\ref{GRD2}) or
(\ref{GRD3}), which define E,
into the remaining one-loop subdiagram of a given two-loop
diagram and project the result onto the physical operators
by means of the Greek Method. This step gives the correction
to the counter-term. We will denote the coefficient of $1/\varepsilon$
from this part by $G^b_C$.

Then the two loop anomalous dimension matrix is found by calculating
\be\label{Masternew}
\gamma^{(1)}_{ij}= -4 \lbrack 2a_2 \delta_{ij} 
+(G-G^a_C-\frac{1}{2}G^b_C)_{ij}
\rbrack.
\ee 
Note the factor $1/2$ in the evanescent contribution. Formula
(\ref{Masternew}) generalizes the master formula (\ref{a1b1})
to include the contributions of evanescent operators.

Let us illustrate this procedure by calculating the contribution 
of the diagram (f) in fig.~\ref{L:5} and of its counter-diagram (f)
in fig.~\ref{L:6} to the two--loop anomalous dimension of the
operator with the Dirac structure 
$\gamma_\mu(1-\gamma_5)\otimes\gamma_\mu(1-\gamma_5)$.
We drop the colour factors in what follows.

{\bf Step 1:}

Calculating the diagram \ref{L:5}f, using the projection
(\ref{GRTW}) and multiplying by two (inclusion of the symmetric
counterpart) one finds \cite{WEISZ}
\be\label{66}
F=16~, \quad\quad G=66
\ee
with (F,G) defined in (\ref{IMIC})

{\bf Step 2:}

We first calculate the diagram \ref{L:15}b, as in (\ref{DB}).
We find
\be\label{DS1}
{\cal D}^{(1)}_b=-\frac{\as}{4\pi}
\left[\frac{1}{4}\frac{1}{\varepsilon}\right]
(1+2\varepsilon)
\left[\Gamma\gamma_\rho\gamma_\mu\otimes\Gamma\gamma^\rho\gamma^\mu\right],
\ee
where $(1+2\varepsilon)$ is an additional correction to the integral
(\ref{INTR}), which has to be kept now.

In order to find $G^a_C$ we first project ${\cal D}^{(1)}_b$
on $\Gamma\otimes\Gamma$ by using the Greek Method and keeping
only the divergent part:
\be\label{DS2}
[{\cal D}^{(1)}_b]_{div}=-\frac{\as}{4\pi}
\left[\frac{1}{4}\frac{1}{\varepsilon}\right] 16~\Gamma\otimes\Gamma~.
\ee
Inserting this into diagram \ref{L:6}f gives by means of (\ref{DS1})
\be\label{DS3}
I^{(a)}_C=\left(\frac{\as}{4\pi}\right)^2
\left[\frac{1}{4}\frac{1}{\varepsilon}\right]^2
(1+2\varepsilon)
\left[\Gamma\gamma_\rho\gamma_\mu\otimes\Gamma\gamma^\rho\gamma^\mu\right].
\ee
Using next the projection
\be\label{DS4}
\Gamma\gamma_\rho\gamma_\mu\otimes\Gamma\gamma^\rho\gamma^\mu=
4(4-\varepsilon) \Gamma\otimes\Gamma
\ee
and including the symmetric counterpart of \ref{L:6}f gives
\be\label{DS5}
I^{(a)}_C=\left(\frac{\as}{4\pi}\right)^2
\left[\frac{32}{\varepsilon^2}+\frac{56}{\varepsilon}\right]
\Gamma\otimes\Gamma
\ee
where finite terms have been dropped. Consequently
\be\label{DS6}
F_C^a=32~, \quad\quad G_C^a=56
\ee

{\bf Step 3:}

In order to calculate $G^b_C$ we take first the evanescent part
of ${\cal D}^{(1)}_b$. Dropping the $\ord(\varepsilon)$ from
the integral (it contributes only to finite parts of $I^{(b)}_C$) we have
\be\label{DS7}
[{\cal D}^{(1)}_b]_{ev}=-\frac{\as}{4\pi}
\left[\frac{1}{4}\frac{1}{\varepsilon}\right]
\left[\Gamma\gamma_\rho\gamma_\mu\otimes\Gamma\gamma^\rho\gamma^\mu
-4 (4-\varepsilon)\Gamma\otimes\Gamma\right]~,
\ee
where we have used (\ref{GRD2}). Inserting $[{\cal D}^{(1)}_b]_{ev}$
into the diagram \ref{L:6}f and multiplying by two for the symmetric
counterpart we get
\be\label{DS8}
I^{(b)}_C=2
\left(\frac{\as}{4\pi}\right)^2
\left[\frac{1}{4}\frac{1}{\varepsilon}\right]^2
(1+2\varepsilon)
\left[
\Gamma\gamma_\rho\gamma_\mu\gamma_\nu\gamma_\tau
\otimes\Gamma\gamma^\rho\gamma^\mu\gamma^\nu\gamma^\tau
-4 (4-\varepsilon)\Gamma\gamma_\nu\gamma_\tau\otimes
\Gamma\gamma^\nu\gamma^\tau\right]~.
\ee
Projecting on $\Gamma\otimes\Gamma$ by means of (\ref{GRTW}) and
(\ref{GRD2}) we obtain
\be\label{DS9}
I^{(b)}_C=\left(\frac{\as}{4\pi}\right)^2
\left[-\frac{12}{\varepsilon}\right]
\Gamma\otimes\Gamma
\ee
and
\be
F^b_C=0~, \quad\quad G^b_C=-12.
\ee
We observe that the $1/\varepsilon^2$ singularity is unaffected by
the evanescent contribution. We can now calculate the relevant
combination in (\ref{Masternew}) to be
\be\label{MaNe}
G-G^a_C-\frac{1}{2}G^b_C=16~,
\ee
which is precisely the $1/\varepsilon$ singularity given in table 3
(diagram 5) of \cite{WEISZ}. Also the $1/\varepsilon^2$ singularity
$F-F^a_C=-16$ agrees with \cite{WEISZ} and $F_C=2F$ as promised
after (\ref{IMIC}). Great! Everything works! I hope you are now
motivated to calculate the remaining 26 two--loop diagrams and
corresponding counter-diagrams necessary to reproduce the
matrix (\ref{gNDR}). Actually the calculation of counter-diagrams
is rather straightforward. The difficult part is the calculation
of the two--loop diagrams, like the ones in fig.~\ref{L:5}. 
\subsubsection{Evanescent Scheme Dependences}
The definition of evanescent operators is not unique as stressed
by Dugan and Grinstein \cite{DuGr} 
and in particular by Herrlich and Nierste \cite{HNE}. As an example
consider the Dirac structure  on the l.h.s of (\ref{GRD2}).
Following \cite{HNE} we can generalize this formula to
\be\label{GEVE}
\Gamma\gamma_\rho\gamma_\mu\otimes\Gamma\gamma^\rho\gamma^\mu 
=(16+a\varepsilon) \Gamma\otimes\Gamma+E^{\rm NDR}(a)
\ee
where $ ``a"$ is an arbitrary parameter, which defines the evanescent
operator $E^{\rm NDR}(a)$. For $a=-4$ the definition in (\ref{GRD2})
is chosen.

Now as the preceeding discussion has shown, the presence of evanescent
operators influences the two-loop anomalous dimensions of physical
operators $\Gamma\otimes\Gamma$. Consequently as emphasized in
\cite{HNE}, the arbitrariness in the definition of the evanescent 
operators translates into an additional scheme dependence of two-loop
anomalous dimensions which can be effectively parametrized by  
$``a"$ in (\ref{GEVE}). Therefore, when giving the results for two-loop
anomalous dimensions, it is not sufficient to state simply that
they correspond to NDR, HV or any other renormalization scheme.
One has to specify in addition the definition of evanescent operators.
This is essential as this scheme dependence of two-loop anomalous
dimensions can only be cancelled in physical amplitudes by the 
corresponding scheme dependences present in the matching conditions
(for instance $B_\pm$) at scales $\ord(\mw)$ and by the one present
in the finite matrix elements of operators at scales $\ord(\mu)$.

This means that the treatment of evanescent operators in the
process of matching and in the calculation of
matrix elements of operators at scales $\ord(\mu)$ must be
consistent with the one used in the calculation of two-loop
anomalous dimensions. This issue is elaborated at length in
\cite{HNE,SH94,UN95} and in the appendix B of \cite{BLMM}.
See also the appendix of \cite{Potte}.

There are two virtues of the definition of evanescent operators
proposed by Weisz and myself and discussed in detail above:

\bi
\item
The evanescent operators defined in \cite{WEISZ} influence only
two-loop anomalous dimensions. By definition they do not contribute
to the matching and to the finite corrections to matrix elements
at scales $\ord(\mu)$. They are simply subtracted away in the process
of renormalization.
\item
As a consequence of this, the Fierz symmetry is preserved separately
in two-loop anomalous dimensions, matching conditions and
matrix elements at scales $\ord(\mu)$.
\ei
The second property assures that the operators $Q_+$ and $Q_-$
do not mix under renormalization separately in two-loop anomalous
dimensions, in the matching conditions and in the matrix elements
so that objects like $B_{+-}$, $B_{-+}$, $\gamma^{(1)}_{+-}$,
$\gamma^{(1)}_{-+}$ are assured to vanish in this scheme.
In other schemes (see \cite{DuGr}) this is not the case and
the Fierz symmetry is only recovered after the two-loop anomalous
dimensions are combined with the matching conditions which 
makes the calculations unnecessarily rather involved.

Now comes the most important message of this subsection: 
most of the existing NLO calculations adopt the definition of
evanescent operators in \cite{WEISZ} and all the two-loop
anomalous dimensions and matching conditions given in these
lectures and in the review \cite{BBL} correspond to this definition.

\section{The Effective $\Delta F=1$ Hamiltonian: 
        Current-Current \\ Operators}
   \label{sec:HeffdF1:22}
\setcounter{equation}{0}
\subsection{Basic Formalism}
Let us summarize the results for the coefficients $C_{1,2}(\mu)$ of
the current-current operators $Q_{1,2}$ discussed extensively in the
previous two sections and let us evaluate them for the cases of
$\Delta B=1$, $\Delta C=1$ and $\Delta S=1$ decays.

To be specific let us consider
\begin{equation}\label{B1}
Q_1=(\bar b_\alpha c_\beta)_{V-A} (\bar u_\beta d_\alpha)_{V-A}
\qquad 
Q_2=(\bar b_\alpha c_\alpha)_{V-A} (\bar u_\beta d_\beta)_{V-A}
\end{equation}
\begin{equation}\label{B2}
Q_1=(\bar s_\alpha c_\beta)_{V-A} (\bar u_\beta d_\alpha)_{V-A}
\qquad 
Q_2=(\bar s_\alpha c_\alpha)_{V-A} (\bar u_\beta d_\beta)_{V-A}
\end{equation}
\begin{equation}\label{B3}
Q_1=(\bar s_\alpha u_\beta)_{V-A} (\bar u_\beta d_\alpha)_{V-A}
\qquad 
Q_2=(\bar s_\alpha u_\alpha)_{V-A} (\bar u_\beta d_\beta)_{V-A}
\end{equation}
for $\Delta B=1$, $\Delta C=1$ and $\Delta S=1$ decays respectively.

The corresponding  effective Hamiltonians are given by
\begin{equation}\label{B4}
H_{eff}(\Delta B=1)=\frac{G_F}{\sqrt{2}}V_{cb}^{*}V_{ud}
\lbrack C_1(\mu) Q_1+C_2(\mu)Q_2 \rbrack
\qquad
(\mu=O(m_b))
\end{equation}
\begin{equation}\label{B5}
H_{eff}(\Delta C=1)=\frac{G_F}{\sqrt{2}}V_{cs}^{*}V_{ud}
\lbrack C_1(\mu) Q_1+C_2(\mu)Q_2 \rbrack
\qquad
(\mu=O(m_c))
\end{equation}
\begin{equation}\label{B6}
H_{eff}(\Delta S=1)=\frac{G_F}{\sqrt{2}}V_{us}^{*}V_{ud}
\lbrack C_1(\mu) Q_1+C_2(\mu)Q_2 \rbrack
\qquad
(\mu=O(1\gev))
\end{equation}
In subsequent sections the Hamiltonian (\ref{B6}) will be
generalized to include also penguin operators. 
The Hamiltonians (\ref{B4}) and (\ref{B5}) having operators
built out of four different flavours are unaffected by
penguin contributions. On the other hand there are other
$\Delta B=1$ and $\Delta C=1$ Hamiltonians which contain
important penguin contributions. A well known example is
the Hamiltonian for $B\to X_s\gamma$ decay which will be
analyzed in detail in section 12.
However, the inclusion of penguin operators does not
change the Wilson coefficients $C_{1,2}(\mu)$.

In a numerical analysis it is convenient to work with the operators
 $Q_{\pm}$ and their coefficients $z_{\pm}$ defined by
\begin{equation}\label{B7}
Q_{\pm}=\frac{1}{2} (Q_2\pm Q_1)
\qquad
\qquad
z_\pm=C_\pm=C_2\pm C_1 \, .
\end{equation}
$Q_+$ and $Q_-$ do not mix under renormalization and 
\begin{equation}\label{B9}
z_\pm(\mu)=\left[1+\frac{\as(\mu)}{4\pi}J_\pm\right]
      \left[\frac{\as(M_W)}{\as(\mu)}\right]^{d_\pm}
\left[1+\frac{\as(M_W)}{4\pi}(B_\pm-J_\pm)\right]
\end{equation}
with
\begin{equation}\label{B10}
J_\pm=\frac{d_\pm}{\beta_0}\beta_1-\frac{\gamma^{(1)}_\pm}{2\beta_0},
\qquad\qquad
d_\pm=\frac{\gamma^{(0)}_\pm}{2\beta_0} \, ,
\end{equation}
\begin{equation}\label{B11}
\gamma^{(0)}_\pm=\pm 12 \frac{N\mp 1}{2N}~,
\end{equation}
\begin{equation}\label{B12}
\gamma^{(1)}_{\pm}=\frac{N\mp 1}{2N}
\left[-21\pm\frac{57}{N}\mp\frac{19}{3}N \pm
\frac{4}{3}f-2\beta_0\kappa_\pm\right]~,
\end{equation}
\begin{equation}\label{B13}
B_\pm=\frac{N\mp 1}{2N}\left[\pm 11+\kappa_\pm\right]~.
\end{equation}
The parameters
 $\kappa_\pm$, introduced in \cite{AJB94a}, distinguish between various
renormalization schemes. In particular
\begin{equation}
\kappa_\pm = \left\{ \begin{array}{rl}
    0 & \qquad {\rm NDR}  \\
\mp 4 & \qquad {\rm HV} \\
\mp 6-3 & \qquad {\rm DRED}
\end{array}\right. \, .
\label{B14}
\end{equation}
We recall that
 $B_\pm-J_\pm$ is scheme independent
and the scheme
dependence of  $z_\pm(\mu)$ originates 
entirely from the scheme dependence of $J_\pm$ at the lower end of the
evolution in (\ref{B9}). Using $N=3$ one has
\begin{equation}\label{B15}
J_\pm=(J_\pm)_{\rm NDR}+\frac{3\mp 1}{6}\kappa_\pm
=(J_\pm)_{\rm NDR}\pm\frac{\gamma^{(0)}_\pm}{12}\kappa_\pm~.
\end{equation}

In order to exhibit the $\mu$ dependence at the same footing as the
scheme dependence, it is useful to rewrite (\ref{B9}) in the case
of B-decays as follows:
\begin{equation}\label{20}
z_\pm(\mu)=\left[1+\frac{\alpha_s(m_b)}{4\pi} \tilde J_\pm(\mu)\right]
      \left[\frac{\alpha_s(M_W)}{\alpha_s(m_b)}\right]^{d_\pm}
\left[1+\frac{\alpha_s(M_W)}{4\pi}(B_\pm-J_\pm)\right]
\end{equation}
with 
\begin{equation}\label{21}
\tilde J_\pm(\mu)=(J_\pm)_{NDR}\pm 
\frac{\gamma^{(0)}_\pm}{12}\kappa_\pm
+\frac{\gamma^{(0)}_\pm}{2}\ln(\frac{\mu^2}{m^2_b})
\end{equation}
summarizing both the renormalization scheme dependence and the 
$\mu$--dependence. Note that in the first parenthesis in (\ref{20}) 
we have
set $\alpha_s(\mu)=\alpha_s(m_b)$ as the difference in the
scales in this correction is still of a higher order.
We also note that the scheme and the $\mu$--dependent terms
are both proportional to $\gamma^{(0)}_\pm$. This implies that the
change of renormalization scheme can be compensated by the change
in $\mu$. From (\ref{21}) we find generally
\begin{equation}\label{21a}
\mu_i^\pm=\mu_{NDR}\exp\left(\mp\frac{\kappa_\pm^{(i)}}{12}\right)
\end{equation}
where $i$ denotes a given scheme. From (\ref{B14}) we have then
\begin{equation}\label{22}
\mu_{HV}=\mu_{NDR}\exp\left(\frac{1}{3}\right)
\qquad
\mu_{DRED}^{\pm}=
\mu_{NDR}\exp\left(\frac{2\pm 1}{4}\right)
\end{equation}
Evidently whereas the change in $\mu$ relating HV and NDR is the
same for $z_+$ and $z_-$ and consequently for 
$C_i(\mu)$, the relation between NDR and DRED is more involved. In any
case $\mu_{HV}$ and $\mu_{DRED}^\pm$ are larger than $\mu_{NDR}$. 
This discussion shows that a meaningful analysis of the $\mu$
dependence of $C_i(\mu)$ can only be made simultaneously with the
analysis of the scheme dependence. This observation will be important
for the analysis of two-body B-decays in section 9.

\subsection{Numerical Results for B-Decays}
For B-decays we have simply:
\begin{equation}\label{B20}
C_1(\mu)=\frac{z_+(\mu)-z_-(\mu)}{2}
\qquad\qquad
C_2(\mu)=\frac{z_+(\mu)+z_-(\mu)}{2}
\end{equation}
We set $f=5$ in the formulae above and use the two-loop
$\as(\mu)$ of (\ref{QCDC}) with $\Lms^{(5)}$. 
The results for LO and NLO in NDR and HV schemes are shown in
the table \ref{tab:c2B}.
\begin{table}[htb]
\caption{The coefficients $C_{1,2}(\mu)$ for B-decays at 
$\mu=\overline{m}_{\rm b}(\mb)=
4.40\gev$ 
\label{tab:c2B}}
\begin{center}
\begin{tabular}{|c|c|c|c||c|c|c||c|c|c|}
\hline
& \multicolumn{3}{c||}{$\Lms^{(5)}=160\mev$} &
  \multicolumn{3}{c||}{$\Lms^{(5)}=225\mev$} &
  \multicolumn{3}{c| }{$\Lms^{(5)}=290\mev$} \\
\hline
Scheme & LO & NDR & HV & LO & 
NDR & HV & LO & NDR & HV \\
\hline
$C_1$ & --0.270 & --0.169 & --0.206 & --0.295 & 
--0.184 & --0.226 & --0.317 & --0.198 & --0.243 \\
$C_2$ & 1.119 & 1.071 & 1.089 & 1.132 & 
1.078 & 1.100 & 1.144 & 1.085 & 1.109 \\
\hline
\end{tabular}
\end{center}
\end{table}
\subsection{Numerical Results for D- and K-Decays}
In the case of D-decays and K-decays the relevant scales are
$\mu=\ord(m_c)$ and $\mu=\ord(1\gev)$, respectively
and consequently a more complicated formula with thresholds
should in principle be used. Yet it is possible to use
the following trick which avoids these complications.
We can simply 
use the
master formulae given above with $\Lms^{(5)}$ replaced by $\Lms^{(4)}$
and an ``{effective}'' number of active flavours $f=4.15$. The latter
effective value for $f$ allows to obtain an agreement with the exact
results to better than $1.5\%$. The results are shown in
the table \ref{tab:c2KD} for $\Lms^{(4)}=325\mev$. The calculation for
different values of $\Lms^{(4)}$ is left as a homework
problem.

\begin{table}[htb]
\caption[]{$C_{1,2}(\mu)$ for K-decays and D-decays and 
$\Lms^{(4)}=325\mev$.}
\label{tab:c2KD}
\begin{center}
\begin{tabular}{|c|c|c|c||c|c|c|}
\hline
& \multicolumn{3}{c||}{$C_1(\mu)$} &
  \multicolumn{3}{c| }{$C_2(\mu)$} \\
\hline
$\mu [{\rm GeV}]$ &LO & NDR & HV & LO & NDR & HV  \\
\hline
\hline
1.00 & --0.742 & --0.510 & --0.631 & 1.422 & 1.275 & 1.358 \\
\hline
1.25 & --0.636 & --0.430 & --0.523 & 1.346 & 1.221 & 1.282 \\
\hline
1.50 & --0.565 & --0.378 & --0.457 & 1.298 & 1.188 & 1.237 \\
\hline
1.75 & --0.514 & --0.340 & --0.410 & 1.264 & 1.165 & 1.207 \\
\hline
2.00 & --0.475 & --0.311 & --0.375 & 1.239 & 1.148 & 1.185 \\
\hline
\end{tabular}
\end{center}
\end{table}

\subsection{Discussion}
Let us make a few remarks:

\begin{itemize}
\item
The scheme dependence of the Wilson coefficients is sizable.  
In particular for $C_1$ which vanishes in the absence of QCD
corrections.
\item
The differences between LO and NLO results in the case of $C_1$ are
large showing the importance of NLO corrections. 
Roughly half of this difference is due to the fact that
for the chosen values of $\Lms$ one has $\as^{(LO)}(M_Z)=0.135$ 
 to be compared with $\as(M_Z)=0.118 \pm 0.005$ .
\item
It is  important to keep in mind
that these features are  specific to the schemes chosen.
\end{itemize}

We will use the results of this section in the analysis of two-body
non-leptonic  B-decays in section 9.  
\section{Generalizations}
\setcounter{equation}{0}
\subsection{Preliminaries}
In section 5.7 we have made a strategy for the generalizations of the
simple LO effective Hamiltonian involving only the current-current
operators $Q_1$ and $Q_2$. In the previous two sections we have extended
the analysis of section 5 by calculating NLO corrections to the
Wilson coefficients $C_{1,2}(\mu)$. The goal of the present section
is a description of further generalizations of weak effective
Hamiltonians which will include other operators mentioned
in Section 3. 
Except for an explicit calculation of one-loop anomalous
dimensions of penguin operators,
we will mainly discuss new features skeeping often
derivations. Indeed, this section should be considered as a guide
to the weak effective Hamiltonians. Further details can be found in
the cited literature and in the phenomenological sections of these
lectures.
\subsection{\kpnn}
Let us begin with the rare decay \kpnn. It proceeds through penguin
and box diagrams with internal charm and top exchanges. The internal
u-quark contribution is needed only for the GIM mechanism but can  
otherwise be neglected. Let us concentrate here on the internal top
contribution. We will briefly discuss the charm contribution in
sections 8.9 and 13.

The relevant effective Hamiltonian without QCD corrections has been
constructed already in Section 3. It is given in (\ref{kplus}).
Since the relevant operator is a product of a quark and a leptonic
current, its anomalous dimension vanishes. This simplifies the
QCD analysis considerably. There is no renormalization group
evolution from high-energy scales to low energy scales and the
calculation of the relevant Wilson coefficient amounts to the
matching of the full to the effective theory at scales $\ord(\mw,\mt)$.
The most difficult task here is the calculation of gluon corrections
to the $Z^0$-penguin and the relevant box diagram. It is a two-loop
calculation with massive $W^\pm$ and top quark propagators. In
order to keep gauge invariance, fictitious $\phi^\pm$ Goldstone
bosons, in place of $W^\pm$ propagators have to be also included.
Moreover, looking back at the Feynman rules (\ref{HIGG1}) and
(\ref{HIGG2}), it is evident that for a heavy top quark these are
precisely the dominant contributions. 
Examples of contributing two-loop diagrams are given in fig. \ref{L:7}.
This calculation, involving
roughly 40 two-loop diagrams, is rather tedious, but can be done
analytically due to the fact that the presence of heavy internal
propagators allows to set the external quark momenta
and masses to zero. The infrared divergences can then be regulated
dimensionally. 

The effect of the inclusion of QCD corrections
to the effective Hamiltonian  (\ref{kplus}) amounts to 
the replacement of the function $X_0(x_t)$ by a corrected
function $X(x_t)$ given below. We have then
\begin{equation}\label{hkpn0} 
{\cal H}_{\rm eff}={G_{\rm F} \over{\sqrt 2}}{\alpha\over 2\pi 
\sin^2\Theta_{\rm W}}
 \sum_{l=e,\mu,\tau}
V^{\ast}_{ts}V_{td} X(x_t)~Q(\nu\bar\nu)
 \end{equation}
where
\be
Q(\nu\bar\nu)=(\bar sd)_{V-A}(\bar\nu_l\nu_l)_{V-A} 
\ee
and
\begin{equation}\label{xx} 
X(x_t)=X_0(x_t)+\aspi X_1(x_t) 
\end{equation}
with
\be
x_t=\f{\mtb^2(\mu_t)}{\mw^2},
\qquad
\mu_t={\cal O}(\mt).
\ee
Here $\mtb^2(\mu_t)$ is the running top quark mass defined at the
scale $\mu_t$. Next \cite{IL}
\begin{equation}\label{X01}
X_0(x_t)={{x_t}\over{8}}\;\left[{{x_t+2}\over{x_t-1}} 
+ {{3 x_t-6}\over{(x_t -1)^2}}\; \ln x_t\right] 
\end{equation}
is the leading contribution considered before and \cite{BB1,BB2} 
\begin{eqnarray}\label{xx2}
X_1(x_t)=&-&{23x_t+5x_t^2-4x_t^3\over 3(1-x_t)^2}
+{x_t-11x_t^2+x_t^3+x_t^4\over (1-x_t)^3}\ln x_t
\nonumber\\
&+&{8x_t+4x_t^2+x_t^3-x_t^4\over 2(1-x_t)^3}\ln^2 x_t
-{4x_t-x_t^3\over (1-x_t)^2}L_2(1-x_t)
\nonumber\\
&+&\gamma^{(0)}_m~
 x_t{\partial X_0(x_t)\over\partial x_t}\ln \f{\mu_t^2}{\mw^2}
\end{eqnarray}
with
\begin{equation}\label{l2} 
L_2(1-x)=\int^x_1 dt {\ln t\over 1-t}   \,.
\end{equation}
results from the two-loop calculation discussed briefly above.

\begin{figure}[hbt]
\vspace{0.10in}
\centerline{
\epsfysize=3in
\epsffile{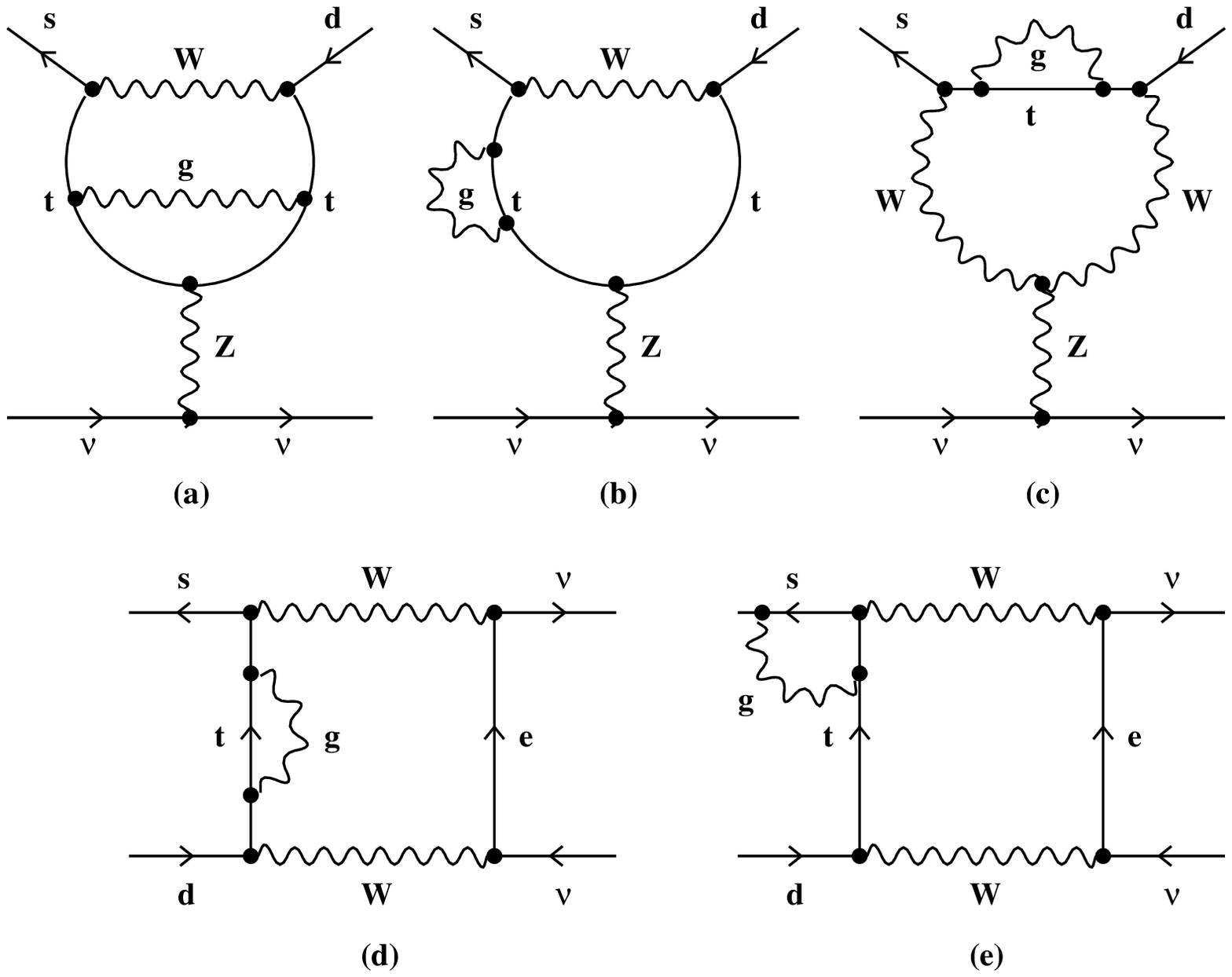}
}
\vspace{0.08in}
\caption[]{Examples of two-loop diagrams contributing to \kpnn .
\label{L:7}}
\end{figure}
 
The $\mu_t$--dependence of the last term in (\ref{xx2}) cancels at
$\ord(\as)$ the $\mu_t$--dependence of the leading term 
$X_0(x_t(\mu_t))$.
Varying $\mu_t$ in the range  $100\gev\le\mu_t\le 300\gev$ gives
a theoretical uncertainty of $\pm 10 \% $ in Br(\kpnn) at LO which
is reduced to $\pm 1 \% $ when the QCD correction in (\ref{xx}) is
included. 
For $\mu_t=m_t$, the complete function $X(x_t)$ can  be written as
\begin{equation}\label{xeta}
X(x_t)=\eta_X\cdot X_0(x_t), \qquad\quad \eta_X=0.985,
\end{equation}
with the QCD factor $\eta_X$
practically independent of $\mt$ and $\Lambda_{\overline{MS}}$.
Thus for this choice of $\mu_t$ the QCD corrections turn out to
be small. They are larger for $\mu_t=\ord(\mw)$ without changing
the final result. That is, the contributions of $X_0$ and $X_1$
to the full function $X$ depend on the particular value of $\mu_t$
but the full result is practically independent of $\mu_t$ after
the NLO corrections have been included. We will return to the
phenomenological aspects of this decay in section 13, where we
will derive the expression for $Br(K^+\to\pi^+\nu\bar\nu)$.
\subsection{$B^0_d-\bar B^0_d$ Mixing}
\subsubsection{Preliminaries}
The next generalization on our list is the $B^0_d-\bar B^0_d$ mixing,
which proceeds to an excellent approximation only through box diagrams
with internal top quark exchanges. The contributions of the internal
$u$ and $c$ quarks are only needed for GIM mechanism. Otherwise they
can be set to zero due to the smallness of $m_u$ and $m_c$ 
relative to $m_t$. The effective Hamiltonian
$\Heff(\Delta B=2)$ for $B_d^0-\bar B_d^0$
mixing, relevant for scales $\mu_b=\ord(m_b)$ is
given by
\begin{equation}\label{hdb2x}
{\cal H}^{\Delta B=2}_{\rm eff}=\frac{G^2_{\rm F}}{16\pi^2}M^2_W
 \left(V^\ast_{tb}V_{td}\right)^2 
 C_Q(\mu_b) Q(\Delta B=2) + h. c.
\end{equation}
where
\begin{equation}\label{qbdbdx}
Q(\Delta B=2)=(\bar b_\alpha d_\alpha)_{V-A}(\bar b_\beta d_\beta)_{V-A}~.
\end{equation}
(\ref{hdb2x}) can be easily derived by using the rules of Section 3.
In the absence of QCD corrections one finds 
\be
C_Q=S_0(x_t)
\ee
where
\begin{equation}\label{S0a}
S_0(x_t)=\frac{4x_t-11x^2_t+x^3_t}{4(1-x_t)^2}-
 \frac{3x^3_t \ln x_t}{2(1-x_t)^3}
\end{equation}
is a function analogous to $X_0(x_t)$ in (\ref{X01}).
\subsubsection{LO Analysis}
The study of QCD corrections to 
$B^0_d-\bar B^0_d$ mixing  is more involved 
than in the case of the top contribution to $K^+\to\pi^+\nu\bar\nu$
as
the operator $Q(\Delta B=2)$, in contrast to $Q(\nu\bar\nu)$, carries
a non-vanishing anomalous dimension $\gamma_Q$. This anomalous dimension
can be shown to be equal to the anomalous dimension of the operator
$Q_+$ considered in the previous sections. Indeed, in the case at hand
the $Q_1$ and $Q_2$ operators are given by
\be
Q_1(\Delta B=2)=(\bar b_\alpha d_\beta)_{V-A}(\bar b_\beta d_\alpha)_{V-A},
\quad\quad
Q_2(\Delta B=2)=Q(\Delta B=2)~.
\ee
and using Fierz symmetry we have
\be
Q_1(\Delta B=2)=Q(\Delta B=2).
\ee
Consequently
\be
Q_+=\frac{Q_2+Q_1}{2}=Q(\Delta B=2)~,
\qquad
Q_-=\frac{Q_2-Q_1}{2}=0  
\ee 
and $\gamma_Q=\gamma_+$. In particular, in LO
\begin{equation}\label{gamq0}
\gamma_Q=\gamma_Q^{(0)}\frac{\as}{4\pi}~,
\qquad
\gamma_Q^{(0)}=
\gamma_+^{(0)}=4~.
\end{equation}
This in turn implies
\be
C_Q(\mu_b)=U^{(0)}(\mu_b,\mu_W) C_Q(\mu_W),
\ee
\be
U^{(0)}(\mu_b,\mu_W)=
\left[\f{\as(\mu_W)}{\as(\mu_b)}\right]^{\f{\gamma_Q^{(0)}}{2\beta_0}},
\qquad
C_Q(\mu_W)=S_0(x_t)
\ee
where $\beta_0=23/3$.
Thus in LO the Wilson coefficient $C_Q(\mu_b)$ is given by
\be
C_Q(\mu_b)=\left[\f{\as(\mu_W)}{\as(\mu_b)}\right]^{6/23} S_0(x_t).
\ee

Before going to the NLO case, let us calculate the matrix element
\begin{equation}\label{mhdb2}
\langle \bar B^0|{\cal H}^{\Delta B=2}_{\rm eff}|B^0\rangle
=\frac{G^2_{\rm F}}{16\pi^2}M^2_W
 \left(V^\ast_{tb}V_{td}\right)^2 
 C_Q(\mu_b) \langle \bar B^0|Q(\Delta B=2)(\mu_b) |B^0\rangle
\end{equation}
where
\begin{equation}
\langle \bar B^0|Q(\Delta B=2)(\mu_b)|B^0\rangle
\equiv \frac{8}{3} B_{B}(\mu_b) F_{B}^2 m_{B}^2
\label{eq:BbarB0}
\end{equation}
and $F_{B}$ is the $B$-meson decay constant. (\ref{mhdb2}) will be an 
important quantity in Section 10.

The $\mu_b$--dependent parameter $B_B(\mu_b)$ parametrizes 
the non-perturbative effects in the hadronic matrix element of the
operator $Q(\Delta B)$. Its value is $\ord(1)$. In phenomanological
applications it is useful to define two
$\mu_b$--independent quantities:
\be
\eta^{(0)}_B=\left[\as(\mu_W)\right]^{6/23},
\qquad
\hat B^{(0)}_{B} = B_{B}(\mu_b) \left[ \as(\mu_b) \right]^{-6/23}~.
\ee
Then:
\begin{equation}\label{mhdb3}
\langle \bar B^0|{\cal H}^{\Delta B=2}_{\rm eff}|B^0\rangle
=\frac{G^2_{\rm F}}{6\pi^2}M^2_W
 \left(V^\ast_{tb}V_{td}\right)^2 
 \hat B^{(0)}_B F^2_B m_B^2\eta^{(0)}_B S_0(x_t)
\end{equation}
We note that there is a left over $\mu_W$--dependence in $\eta_B$ and
$\mu_t$ dependence in $S_0(x_t(\mu_t)$. In order to reduce these dependences
we have to include NLO corrections.

\subsubsection{NLO Analysis}
Applying the standard procedure of matching one finds \cite{BJW90}
\be\label{NLOC}
C_Q(\mu_W)=S_0(x_t)+
\frac{\as(\mu_W)}{4\pi} \left[
S_1(x_t)+F(\mu_W,\mu_t)S_0(x_t)+B_t S_0(x_t)\right]
\ee
\be\label{FUNC}
F(\mu_W,\mu_t)=\frac{\gamma_Q^{(0)}}{2}
 \ln\frac{\mu^2_W}{M^2_W}+\gamma^{(0)}_{m} x_t
 \frac{\partial S_0(x_t)}{\partial x_t}\ln\frac{\mu^2_t}{M^2_W}
\ee

\begin{equation}\label{btndr}
B_t=5\frac{N-1}{2N}+3\frac{N^2-1}{2N}\qquad ({\rm NDR})
\end{equation}  
\be
S_1(x_t)= {\rm Complicated~Function}
\ee
The function $S_1(x_t)$ given in (XII.12) of ref. \cite{BBL}
is a result of a two-loop calculation \cite{BJW90} involving
gluon corrections to the box diagrams. Typical diagrams are shown in 
fig. \ref{L:8}.
The interested reader should consult the detailed analysis in \cite{BJW90}
where a spectacular cancellation of infrared divergences and gauge 
dependences present in the diagrams of the full theory is achieved by 
the corresponding diagrams in the effective theory.

\begin{figure}[hbt]
\vspace{0.10in}
\centerline{
\epsfysize=1.5in
\epsffile{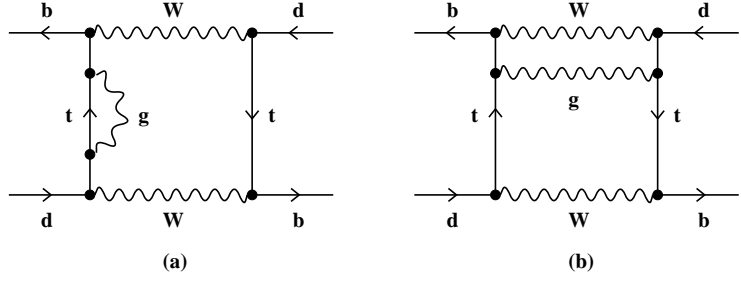}
}
\vspace{0.08in}
\caption[]{Examples of two-loop diagrams contributing to 
$B_d^0-\bar B_d^0$ mixing.
\label{L:8}}
\end{figure}

The second log in $F(\mu_W,\mu_t)$ cancels the $\mu_t$
dependence in $S_0(x_t)$ in analogy to a similar logarithm in
(\ref{xx2}). The first logarithm in (\ref{FUNC}) cancels the
$\mu_W$ dependence present in $U^{(0)}(\mu_b,\mu_W)$.
For $\mu_W=\mu_t$ the formulae given above reduce to the ones given
in \cite{BJW90} and \cite{BBL}. But as discussed already there,
$\mu_W$ and $\mu_t$ can differ from each other and for
pedagogical reasons we do not put them equal here.

The NLO evolution function is given simply by
\begin{equation}\label{u02}
 U(\mu_b,\mu_W)=
\left[1+{\as(\mu_b)\over 4\pi} J_5 \right]  U^{(0)}(\mu_b,\mu_W) 
\left[1-{\as(\mu_W)\over 4\pi} J_5\right]
\end{equation}
with
\be
J_5=J_+=1.627
\qquad (\rm{NDR},~f=5).
\ee

We can now define $\mu_b$ and $\mu_W$ independent quantities at
the NLO level, which moreover are renormalization scheme independent:
\be\label{ETANLO}
\eta_B=\left[\as(\mu_W)\right]^{6/23}\left[1+
\frac{\as(\mu_W)}{4\pi} \left(
 \frac{S_1(x_t)}{S_0(x_t)}+F(\mu_W,\mu_t)
+B_t - J_5\right)\right],
\ee
\begin{equation}\label{Def-Bpar0}
\hat B_{B} = B_{B}(\mu_b) \left[ \as^{(5)}(\mu_b) \right]^{-6/23} \,
\left[ 1 + \frac{\as^{(5)}(\mu_b)}{4\pi} J_5 \right]~.
\end{equation}
Then:
\begin{equation}\label{mhdb4}
\langle \bar B^0|{\cal H}^{\Delta B=2}_{\rm eff}|B^0\rangle
=\frac{G^2_{\rm F}}{6\pi^2}M^2_W
 \left(V^\ast_{tb}V_{td}\right)^2 
 \hat B_B F^2_B m^2_B \eta_B S_0(x_t)~.
\end{equation}

It should be noted that both $\eta_B$ and $S_0(x_t)$ depend on
$\mu_t$ but the product
$\eta_B \cdot S_0(x_t)$ is $\mu_t$-independent in $\ord(\as)$
as the second logarithm in (\ref{FUNC}) cancels the
$\mu_t$ dependence in $S_0(x_t(\mu_t))$. If one varies $\mu_t$ in the
range $100~\gev\le\mu_t\le 300~\gev$, the $\mu_t$ dependence of
$\langle \bar B^0|{\cal H}^{\Delta B=2}_{\rm eff}|B^0\rangle$ amounts in
LO to $\pm 9\%$ and is reduced to $\pm 1\%$ in NLO.

It is customary to evaluate $\eta_B$ at  $\mu_t=\mu_W=m_t$, then
practically $\eta_B$ is independent of $\mt$ and the full $\mt$
dependence of $B^0_d-\bar B^0_d$ mixing resides in $S_0(x_t)$ with
$\mtb(\mt)$. Then for $\as(\mz)=0.118\pm0.03$ one has
\be\label{etab}
\eta_B=0.55\pm 0.01
\ee
where the error includes also the leftover scale uncertainties,
which can only be reduced by calculating $\ord(\as^2)$ corrections.

\begin{figure}[hbt]
\vspace{0.10in}
\centerline{
\epsfysize=1.5in
\epsffile{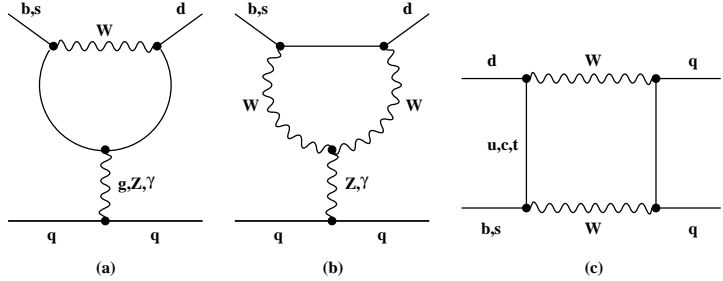}
}
\vspace{0.08in}
\caption[]{One loop penguin and box diagrams in the full theory.
\label{L:14}}
\end{figure}

\subsection{QCD Penguins}
         \label{sec:HeffdF1:66}
\subsubsection{Operators}
The next generalization involves the inclusion of penguin operators
in $\Delta F=1$ transitions. They originate in the QCD penguin diagram 
(a) of fig. \ref{L:14}. Evaluating this diagram one can clearly see 
that there are two
colour structures as in the case of $Q_1$ and $Q_2$. They follow simply
from the decomposition
\begin{equation}\label{tata1}
T^a_{\alpha\beta}T^a_{\gamma\rho}
=-{1\over 2N}\delta_{\alpha\beta}\delta_{\gamma\delta}+{1\over 2}
\delta_{\alpha\delta}\delta_{\gamma\beta}
\end{equation}
The upper effective FCNC vertex in the penguin diagram has $V-A$
structure as seen in the rules of Section 3. The lower vertex is
vectorial ($V$). It turns out however that the renormalization
of the operators with the Dirac structure $(V-A)\otimes V$ requires
the introduction of two new operators with the Dirac structure 
$(V-A)\otimes A$ and the colour structures as the $(V-A)\otimes V$
operators. Indeed, when the $(V-A)\otimes V$ operators are inserted
into the Green functions of fig. \ref{L:16} two new operators in question
are generated. The full set of operators which closes under
renormalization consists then of two current-current operators
$(Q_1,Q_2)$ and the four penguin operators. It is customery to
work with the  $(V-A)\otimes(V-A)$ and  $(V-A)\otimes(V+A)$
penguin operators rather than with the
$(V-A)\otimes V$ and  $(V-A)\otimes A$ structures. Then the
basis of the operators  necessary for the description of
$\Delta B=1$ decays with $\Delta S=1$ 
is given (in the limit $\alpha\equiv\alpha_{QED}=0$) 
as follows:

{\bf Current--Current: }
\begin{equation}\label{O1} 
Q_1 = (\bar c_{\alpha} b_{\beta})_{V-A}\;(\bar s_{\beta} c_{\alpha})_{V-A}
~~~~~~Q_2 = (\bar c b)_{V-A}\;(\bar s c)_{V-A} 
\end{equation}

{\bf QCD--Penguins :}
\begin{equation}\label{O2}
Q_3 = (\bar s b)_{V-A} \sum_{q=u,d,s,c,b}(\bar qq)_{V-A}~~~~~~   
 Q_4 = (\bar s_{\alpha} b_{\beta})_{V-A}\sum_{q=u,d,s,c,b}(\bar q_{\beta} 
       q_{\alpha})_{V-A} 
\end{equation}
\begin{equation}\label{O3}
 Q_5 = (\bar s b)_{V-A} \sum_{q=u,d,s,c,b}(\bar qq)_{V+A}~~~~~  
 Q_6 = (\bar s_{\alpha} b_{\beta})_{V-A}\sum_{q=u,d,s,c,b}
       (\bar q_{\beta} q_{\alpha})_{V+A} 
\end{equation}
The corresponding operators for other B--decays and the D- and
K-decays can be obtained from this basis by an appropriate
change of flavours.

\subsubsection{Effective Hamiltonian}
The effective Hamiltonian for $\Delta B=1$ decays with
$\Delta S=1$ is given then by
\begin{eqnarray}\label{HB4} 
H_{eff}(\Delta B=1) &=&
\frac{G_F}{\sqrt{2}} \Bigg[ \lambda_u(C_1(\mu_b) 
Q^u_1+C_2(\mu_b)Q^u_2) 
+\lambda_c( C_1(\mu_b) Q^c_1+C_2(\mu_b)Q^c_2)
\nonumber\\  
&& -\lambda_t \sum_{i=3}^6 C_i(\mu)Q_i \Bigg] ,
\end{eqnarray}~ 
where
\be\label{lb}
\lambda_q=V_{qs}^{*}V_{qb}
\ee
and
\be\label{Qq}
Q_1^q=(\bar q_\alpha b_\beta)_{V-A}(\bar s_\beta q_\alpha)_{V-A}~,
\quad\quad
Q_2^q=(\bar q_\alpha b_\alpha)_{V-A}(\bar s_\beta q_\beta)_{V-A}~.
\ee
In particular $Q_i^c=Q_i$ in (\ref{O1}).
\subsubsection{Wilson Coefficients}
The calculation of the Wilson coefficients $C_i(\mu_b)$ of
the QCD penguin operators proceeds as outlined in Section 6.3.

The matching at $\mu_W=\mw$ gives, in the presence of the penguin 
diagrams, the values of $\vec C(\mw)$. In the NDR scheme they 
are given by:
\begin{eqnarray}
C_1(\mw) &=&     \frac{11}{2} \; \frac{\as(\mw)}{4\pi} \, ,
\label{eq:CMw1QCD} \\
C_2(\mw) &=& 1 - \frac{11}{6} \; \frac{\as(\mw)}{4\pi} \, ,
\label{eq:CMw2QCD} \\
C_3(\mw) &=& -\frac{\as(\mw)}{24\pi} \widetilde{E}_0(x_t) \, ,
\label{eq:CMw3QCD} \\
C_4(\mw) &=& \frac{\as(\mw)}{8\pi} \widetilde{E}_0(x_t) \, ,
\label{eq:CMw4QCD} \\
C_5(\mw) &=& -\frac{\as(\mw)}{24\pi} \widetilde{E}_0(x_t) \, ,
\label{eq:CMw5QCD} \\
C_6(\mw) &=& \frac{\as(\mw)}{8\pi} \widetilde{E}_0(x_t) \, ,
\label{eq:CMw6QCD}
\end{eqnarray}
where
\begin{eqnarray}
E_0(x_t) &=& 
-\frac{2}{3} \ln x_t + \frac{x_t (18 -11 x_t - x_t^2)}{12 (1-x_t)^3} +
          \frac{x_t^2 (15 - 16 x_t  + 4 x_t^2)}{6 (1-x_t)^4} \ln x_t \, ,
\label{eq:Ext1} \\
\widetilde{E}_0(x_t) &=& E_0(x_t) - \frac{2}{3}
\label{eq:Exttilde1}
\end{eqnarray}
with
\begin{equation}
x_t = \frac{m_t^2}{\mw^2} \, .
\label{eq:xt}
\end{equation}
$C_{1,2}(\mw)$ are simply obtained using $(\ref{B8})$ and $(\ref{B13})$.
We will derive the QCD penguin coefficients $C_i(\mw)~(i=4-6)$
two pages below. The constant $-2/3$ in (\ref{eq:Exttilde1})
is characteristic for the NDR scheme. It is absent in the HV scheme.
In LO $C_2(\mw)=1$ with all remaining coefficients set to zero.
We observe that the $\mt$-dependence in the case at hand enters
first at the NLO level.

The anomalous dimension matrix is $6\times6$:
\begin{equation}
\hat\gamma_s(\as)=\hat\gamma_s^{(0)}\aspi + 
\hat\gamma_s^{(1)}\left(\aspi\right)^2.
\label{eq:gsexpKpp}
\end{equation}
The one loop coefficient $\hat \gamma_s^{(0)}$ is given for $N=3$  by
\cite{PENGUIN}
\begin{equation}
\hat \gamma^{(0)}_s = 
\left(
\begin{array}{cccccc}
{{-2}} & 6 & 0 & 0 & 0 & 0 \\ \svs
6 & {{-2}} & {{-2}\over {9}} & {2\over 3} & {{-2}\over {9}} &\
  {2\over 3} \\ \svs
0 & 0 & {{-22}\over {9}} & {{22}\over 3} & {{-4}\over {9}} & {4\over 3}
\\ \svs
0 & 0 & 6 - {{2 f}\over {9}} & {{-2}} + {{2 f}\over 3} & {{-2\
  f}\over {9}} & {{2 f}\over 3} \\ \svs
0 & 0 & 0 & 0 & {2} & -6  \\ \svs
0 & 0 & {{-2 f}\over {9}} & {{2 f}\over 3} & {{-2 f}\over {9}} & -16
   + {{2 f}\over 3}
\end{array}
\right)
\label{eq:gs0Kpp}
\end{equation}
The explicit calculation of this matrix is given in subsection 8.5.

The two-loop anomalous dimension matrix $\hat \gamma^{(1)}_s$ in the
NDR scheme looks truly  horrible:
\begin{equation}
\left(
\begin{array}{cccccc}
-{{21}\over 2} - {{2\,f}\over 9} & {7\over 2} + {{2\,f}\over 3} & {{79}\over\
  9} & -{7\over 3} & -{{65}\over 9} & -{{7}\over{3}} \\ \mvs
{7\over 2} + {{2\,f}\over 3} & -{{21}\over 2} - {{2\,f}\over 9} &\
  -{{202}\over {243}} & {{1354}\over {81}} & -{{1192}\over {243}} &
{904 \over 81} \\ \mvs
0 & 0 & -{{5911}\over {486}} + {{71\,f}\over 9} & {{5983}\over {162}} +\
  {f\over 3} & -{{2384}\over {243}} - {{71\,f}\over 9} &
{1808 \over 81} - {f \over 3} \\ \mvs
0 & 0 & {{379}\over {18}} + {{56\,f}\over {243}} & -{{91}\over 6} +\
  {{808\,f}\over {81}} & -{{130}\over 9} - {{502\,f}\over {243}} &
-{14 \over 3} + {{646\,f} \over 81} \\ \mvs
0 & 0 & {{-61\,f}\over 9} & {{-11\,f}\over 3} & {{71}\over 3} + {{61\,f}\over\
  9} & -99 + {{11\,f} \over 3} \\ \mvs
0 & 0 & {{-682\,f}\over {243}} & {{106\,f}\over {81}} & -{{225}\over 2} +\
  {{1676\,f}\over {243}} & -{1343 \over 6} + {{1348\,f} \over 81}
\end{array}
\right)
\label{eq:gs1ndrN3Kpp}
\end{equation}
The corresponding matrix in the HV scheme can be found in \cite{BJLW1}.
These two loop matrices have been first calculated in
\cite{BJLW1,ROMA1,ROMA2}. The result in the NDR scheme has been confirmed
subsequently in \cite{CZMM}.

With all these results at hand one can now evaluate $C_i(\mu_b)$
by using
\be\label{EVOLC}
\vec C(\mu_b)=\hat U_5(\mu_b,\mw) \vec C(\mw)
\ee
with $\hat U_5(\mu_b,\mw)$ given in (\ref{u0jj}). With the help
of Mathematica we find
\be
C_j(\mu_b)=C_j^{(0)}(\mu_b)+\frac{\as(\mu_b)}{4\pi}C_j^{(1)}(\mu_b)
\ee
where
\be
C_j^{(0)}(\mu_b)=\sum_{i=3}^8 k_{ji}\eta^{a_i}
\ee

\be
C_j^{(1)}(\mu_b)=\sum_{i=3}^8 \lbrack e_{ji}\eta E_0(x_t)+
f_{ji}+g_{ji}\eta\rbrack \eta^{a_i}
\ee
with
\be
\eta=\left[\frac{\as(\mw)}{\as(\mu_b)}\right]~.
\ee
The magic numbers $a_i$,
$k_{ij}$, $e_{ij}$, $f_{ij}$ and $g_{ij}$ are  collected in tables
\ref{tab:akhLO} and \ref{tab:akhNLO}. The indices $i=1,2$ in these
tables are reserved for magnetic penguin operators discussed in
sections 8.7 and 12. These tables have been calculated by means
of the methods developed in sections 6 and 7.

\begin{table}[htb]
\caption[]{Magic Numbers.
\label{tab:akhLO}}
\begin{center}
\begin{tabular}{|r|r|r|r|r|r|r|}
\hline
$i$ & 3 & 4 & 5 & 6 & 7 & 8 \\
\hline
$a_i $& $ \frac{6}{23} $&$
-\frac{12}{23} $&$
0.4086 $&$ -0.4230 $&$ -0.8994 $&$ 0.1456 $\\
$k_{1i} $&$ \frac{1}{2} $&$ - \frac{1}{2} $&$
0 $&$ 0 $&$ 0 $&$ 0 $\\
$k_{2i} $& $ \frac{1}{2} $&$  \frac{1}{2} $&$
0 $&$ 0 $&$ 0 $&$ 0 $\\
$k_{3i} $& $ - \frac{1}{14} $&$  \frac{1}{6} $&$
0.0510 $&$ - 0.1403 $&$ - 0.0113 $&$ 0.0054 $\\
$k_{4i} $& $ - \frac{1}{14} $&$  - \frac{1}{6} $&$
0.0984 $&$ 0.1214 $&$ 0.0156 $&$ 0.0026 $\\
$k_{5i} $& $ 0 $&$  0 $&$
- 0.0397 $&$ 0.0117 $&$ - 0.0025 $&$ 0.0304 $\\
$k_{6i} $&$ 0 $&$  0 $&$
0.0335 $&$ 0.0239 $&$ - 0.0462 $&$ -0.0112 $\\
\hline
\end{tabular}
\end{center}
\end{table}

\begin{table}[htb]
\caption[]{More Magic Numbers.
\label{tab:akhNLO}}
\begin{center}
\begin{tabular}{|r|r|r|r|r|r|r|}
\hline
$i$ & 3 & 4 & 5 & 6 & 7 & 8 \\
\hline
$a_i $& $ \frac{6}{23} $&$
-\frac{12}{23} $&$
0.4086 $&$ -0.4230 $&$ -0.8994 $&$ 0.1456 $\\
$e_{1i} $&$ 0 $&$ 0 $&$
0 $&$ 0 $&$ 0 $&$ 0 $\\
$f_{1i} $&$ 0.8136 $&$ 0.7142 $&$
0 $&$ 0 $&$ 0 $&$ 0 $\\
$g_{1i} $&$ 1.0197 $&$ 2.9524 $&$
0 $&$ 0 $&$ 0 $&$ 0 $\\
\hline
$e_{2i} $&$ 0 $&$ 0 $&$
0 $&$ 0 $&$ 0 $&$ 0 $\\
$f_{2i} $&$ 0.8136 $&$ - 0.7142 $&$
0 $&$ 0 $&$ 0 $&$ 0 $\\
$g_{2i} $&$ 1.0197 $&$ - 2.9524 $&$
0 $&$ 0 $&$ 0 $&$ 0 $\\
\hline
$e_{3i} $&$ 0 $&$ 0 $&$
0.1494 $&$ -0.3726 $&$ 0.0738 $&$ -0.0173 $\\
$f_{3i} $&$ -0.0766 $&$ - 0.1455 $&$
-0.8848 $&$ 0.4137 $&$ -0.0114 $&$ 0.1722 $\\
$g_{3i} $&$ -0.1457 $&$ - 0.9841 $&$
0.2303 $&$ 1.4672 $&$ 0.0971 $&$ -0.0213 $\\
\hline
$e_{4i} $&$ 0 $&$ 0 $&$
0.2885 $&$ 0.3224 $&$ -0.1025 $&$ -0.0084 $\\
$f_{4i} $&$ -0.2353 $&$ - 0.0397 $&$
0.4920 $&$ -0.2758 $&$ 0.0019 $&$-0.1449 $\\
$g_{4i} $&$ -0.1457 $&$ 0.9841 $&$
0.4447 $&$ -1.2696 $&$ -0.1349 $&$ -0.0104 $\\
\hline
$e_{5i} $&$ 0 $&$ 0 $&$
-0.1163 $&$ 0.0310 $&$ 0.0162 $&$ -0.0975 $\\
$f_{5i} $&$ 0.0397 $&$  0.0926 $&$
0.7342 $&$ -0.1262 $&$ -0.1209 $&$ -0.1085 $\\
$g_{5i} $&$ 0 $&$ 0 $&$
-0.1792 $&$ -0.1221 $&$ 0.0213 $&$ -0.1197 $\\
\hline
$e_{6i} $&$ 0 $&$ 0 $&$
0.0982 $&$ 0.0634 $&$ 0.3026 $&$ 0.0358 $\\
$f_{6i} $&$ -0.1191 $&$ - 0.2778 $&$
-0.5544 $&$ 0.1915 $&$ -0.2744 $&$ 0.3568 $\\
$g_{6i} $&$ 0 $&$ 0 $&$
0.1513 $&$ -0.2497 $&$ 0.3983 $&$ 0.0440 $\\
\hline
\end{tabular}
\end{center}
\end{table}

\subsubsection{Matching Conditions for QCD Penguins}
It is instructive to derive the matching conditions for 
QCD penguin operators in (\ref{eq:CMw3QCD})--(\ref{eq:CMw6QCD}). 
In particular it is useful to
see how the scheme dependent constant $-2/3$ is generated.
Afterall we stated in section 3 that all mass independent
constants in the evaluation of penguin vertices involving
$\mt$-dependent functions like $E_0(x_t)$ can be dropped
because of GIM mechanism. Yet as we will see in a moment
such statements are valid only  in the full theory. In the effective
theory the top quark is absent as a dynamical degree of freedom,
GIM is no longer true and constants like $-2/3$ remain. 

In order to demonstrate this explicitly let us consider first
the tree level Hamiltonian for $\Delta B=1$ decays:

\begin{equation}\label{HB5}
H^{(0)}_{eff}(\Delta B=1)=\frac{G_F}{\sqrt{2}}
\left[\lambda_u Q^u_2
+\lambda_c Q^c_2
+\lambda_t Q^t_2 \right]
\end{equation}
where $\lambda_q$ and $Q^q_2$  are defined in (\ref{lb}) and (\ref{Qq})
respectively. Note the appearance of the operator $Q^t_2$.

Next let us include QCD corrections and perform matching of the
full theory to an effective five quark theory in which the top
quark is no longer a dynamical degree of freedom. Since we are
only interested in the penguin coefficients we can leave out
the QCD corrections to $Q^q_2$ operators and also drop the
$Q^q_1$ operators. Calculating then the usual QCD penguin diagram
\ref{L:14}a
with full $W^\pm$ and internal $u,c,t$ quarks and adding the result
to the tree level matrix element of the Hamiltonian (\ref{HB5})
we find the amplitude in the full theory:

\begin{eqnarray}\label{AMPFULL} 
{\cal A}_{full} & = &
\frac{G_F}{\sqrt{2}} 
 \left( \lambda_u \left[\langle Q^u_2 \rangle^0  
 -\frac{\as(\mw)}{8\pi} G_u(m_u) \langle Q_P\rangle ^0\right] \right.
\nonumber\\ 
&& ~~~~~~~+ \lambda_c \left[\langle Q^c_2 \rangle^0  
 -\frac{\as(\mw)}{8\pi} G_c(m_c) \langle Q_P\rangle ^0\right] 
\nonumber\\ 
&& ~~~~~~~+ \left.\lambda_t 
\left[~~~~~~~-\frac{\as(\mw)}{8\pi} E_0(x_t) 
\langle Q_P\rangle ^0\right] \right).
\end{eqnarray} 

Note that as a preparation for the matching we have already
removed the tree level matrix element of $Q_2^t$ in which
the top quark field is a dynamical degree of freedom.
Next
\be\label{PING}
Q_P=Q_4+Q_6-\frac{1}{3} (Q_3+Q_5)
\ee
where $Q_i$ with $i=3-6$ are the penguin operators defined
in (\ref{O2}) and (\ref{O3}).

The functions $G_i(m_i)$ result from calculating penguin diagrams
with internal $u$ and $c$ quarks. They are given explicitly in
the appendix of the first paper in \cite{BJLW1}. 
As we will see in a moment they will
cancel out in the process of matching and their analytic expression
is not needed here.

Now the effective theory involves only $Q_2^u$, $Q_2^c$ and $Q_P$.
Calculating the insertions of $Q_2^u$ and $Q_2^c$ into QCD penguin
diagrams of fig.~\ref{L:16} and adding the tree level contributions 
of $Q_2^q$ operators
as in the full theory, we find
 
\begin{eqnarray}\label{AMPEFF} 
{\cal A}_{eff} &=& 
\frac{G_F}{\sqrt{2}} 
\Bigg(\lambda_u \Bigg[\langle Q^u_2 \rangle^0  
 -\frac{\as(\mw)}{8\pi} (G_u(m_u)-r) \langle Q_P\rangle
^0\Bigg]
\nonumber\\ 
&& ~~~~~+\lambda_c \Bigg[\langle Q^c_2 \rangle^0  
 -\frac{\as(\mw)}{8\pi} (G_c(m_c)-r) \langle Q_P\rangle ^0\Bigg] 
\nonumber \\
&& ~~~~-\lambda_t C_P \langle Q_P\rangle ^0 \Bigg]\Bigg),
\end{eqnarray} 
where $C_P$ is the coefficient we are looking for. The minus sign
in front of $\lambda_t$ is a convention which has no impact on
physics. Next $r$ is a scheme dependent constant
equal to $2/3$ and $0$ for NDR and HV schemes respectively.
Finally it should be remarked that the insertions of penguin
operators into penguin diagrams contribute only at $\ord(\as^2)$
to (\ref{AMPEFF}) and do not contribute at this order.

Comparing (\ref{AMPFULL}) and (\ref{AMPEFF}) and using the unitarity
relation $\lambda_u+\lambda_c=-\lambda_t$, we determine $C_P$ to
be
\be\label{COFFCP}
C_P=\frac{\as(\mw)}{8\pi}[E_0(x_t)-r]
\ee
Inserting this result into (\ref{AMPEFF}), using the expression
for $Q_P$ in (\ref{PING}) and comparing the coefficient of $\lambda_t$
with the one of (\ref{HB4}) we derive the matching conditions 
(\ref{eq:CMw3QCD})--(\ref{eq:CMw6QCD}).  
\subsubsection{Numerical Values for $C_i(\mu_b)$}
In table \ref{tab:wc6b} we give numerical values of the coefficients
$C_i(\mu_b)$ in LO and the two NLO schemes in question.

\begin{table}[htb]
\caption{$\Delta B=1$ Wilson coefficients at 
$\mu=\overline{m}_{\rm b}(\mb)=
4.40\gev$ for $\mt=170\gev$.
\label{tab:wc6b}}
\begin{center}
\begin{tabular}{|c|c|c|c||c|c|c||c|c|c|}
\hline
& \multicolumn{3}{c||}{$\Lms^{(5)}=160\mev$} &
  \multicolumn{3}{c||}{$\Lms^{(5)}=225\mev$} &
  \multicolumn{3}{c| }{$\Lms^{(5)}=290\mev$} \\
\hline
Scheme & LO & NDR & HV & LO & 
NDR & HV & LO & NDR & HV \\
\hline
$C_1$ & -0.270 & -0.169 & -0.206 & -0.295 & 
-0.184 & -0.226 & -0.317 & -0.198 & -0.243 \\
$C_2$ & 1.119 & 1.071 & 1.089 & 1.132 & 
1.078 & 1.100 & 1.144 & 1.085 & 1.109 \\
\hline
$C_3$ & 0.012 & 0.012 & 0.011 & 0.013 & 
0.013 & 0.012 & 0.015 & 0.014 & 0.013 \\
$C_4$ & -0.028 & -0.032 & -0.026 & -0.030 & 
-0.035 & -0.029 & -0.032 & -0.038 & -0.031 \\
$C_5$ & 0.008 & 0.008 & 0.008 & 0.008 & 
0.009 & 0.009 & 0.009 & 0.009 & 0.010 \\
$C_6$ & -0.034 & -0.037 & -0.029 & -0.038 & 
-0.041 & -0.033 & -0.041 & -0.044 & -0.035 \\
\hline
\end{tabular}
\end{center}
\end{table}

Let us make just a few observations:
\bi
\item
Penguin coefficients are much smaller than $C_1$ and $C_2$.
\item
The largest penguin coefficients are
$C_4$ and $C_6$. In the NDR and HV schemes their values are by a factor
of 6-7 smaller than the coefficient $C_1$.
\item
A numerical analysis shows that in the range $\mt = (170\pm 15)\gev$
the $\mt$ dependence of the QCD penguin coefficients can be neglected. 
\ei
\subsubsection{Threshold Effects in the Presence of Penguins}
In (\ref{cthrP}) we have given a formula for $\vec C(\mu)$ in
the presence of flavour thresholds. This formula implies in particular
that $\vec C_{f-1}(\mu_f)=\vec C_{f}(\mu_f)$ where $\mu_f$ is the
threshold between an effective $f$-flavour theory and an effective theory
with $f-1$ flavours. In the presence of penguin operators the matching
is more involved. One finds now
\be\label{MAT}
\vec C_{f-1}(\mu_f)=\hat M(\mu_f)\vec C_{f}(\mu_f)
\ee
where $\hat M(\mu_f)$ is a matching matrix given by
\be
\hat M(\mu_f)=\hat 1+\frac{\as(\mu_f)}{4\pi}\delta\hat r^T
\ee
The matrix $\delta\hat r^T$ can be found in section VID of \cite{BBL}.
With (\ref{MAT}) the formula (\ref{cthrP}) generalizes to
\begin{equation}\label{cthrPP}
\vec C(\mu)=\hat U_3(\mu,\mu_c)\hat M(\mu_c)\hat U_4(\mu_c,\mu_b)
\hat M(\mu_b) \hat U_5(\mu_b,\mu_W)\vec C(\mu_W).  \end{equation}
\subsection{Explicit Calculation of $6\times 6$ Anomalous Dimension
            Matrix}
\subsubsection{Preliminaries}
It is time to do a real climb by calculating the 
matrix (\ref{eq:gs0Kpp}). This involves the operator insertions
into the penguin diagrams and into the current-current diagrams. Since
master formulae for the latter insertions have already been derived
and applied for the case of $(Q_1,Q_2)$ in section 6.5, we begin our
climb
by discussing the penguin insertions.
\subsubsection{Penguin Insertions: Generalities}
The two diagrams contributing to the anomalous dimension matrix of
$(Q_1,....Q_6)$ through the penguin insertions are given in fig.
\ref{L:16}.
We observe that two types of insersions of a given operator into
a penguin diagram are possible. Type A insertions represented by the
diagram (a) are constructed by joining two quarks belonging to 
two different disconnected parts of an operator 
and attaching the gluon to the resulting internal quark line.
For instance in the case of $(\bar c b)_{V-A}(\bar s c)_{V-A}$
one can join $\bar c$ and $c$ into one line. Type B insertions
represented by the diagram (b) are constructed by joining two quarks
belonging to the same part of a given operator and attaching the
gluon to the resulting quark loop. For instance in the case of
$(\bar c c)_{V-A}(\bar s b)_{V-A}$ we have a c-quark loop. Since
qluons conserve flavour, penguin insertions are only possible
 if a given operator contains at least two quarks with the same
flavour. Note that in the case of $(\bar sb)_{V-A}(\bar s s)_{V-A}$
both types of insertions are possible and have to be taken into
account. Finally the bottom quark line attached to the lower end
of the gluon represents any quark present in the effective theory.
In calculating the contribution of a given diagram one has to sum
over all quark flavours. In this manner the penguin operators
are generated from insertions into penguin diagrams.
\begin{figure}[hbt]
\vspace{0.10in}
\centerline{
\epsfysize=1.5in
\epsffile{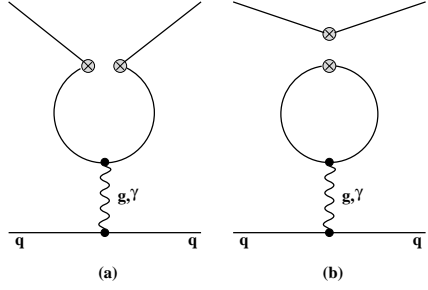}
}
\vspace{0.08in}
\caption[]{One loop penguin diagrams in the effective theory.
\label{L:16}}
\end{figure}
It is clear from the last statement that the insertion of any
operator $(Q_1,....Q_6)$ into the penguin diagrams of fig.~\ref{L:16}
always results in a linear combination of penguin operators.
That is penguin operators  mix under renormalization among
themselves and the current-current operators  mix into
penguin operators but the mixing of penguin operators into
$Q_1$ and $Q_2$ does not take place. This last feature is
not affected by the insertions of penguin operators into
the current-current diagrams of fig.~\ref{L:15} as we will see
explicitly few pages below. Consequently without any calculation
we can state that
\be
(\hat\gamma)_{i1}=(\hat\gamma)_{i2}=0\quad\quad i=3,..6
\ee
and this is also true for the electroweak penguins discussed
in subsection 8.6. 

After these general remarks let us derive two master formulae
for penguin insertions which are analogous to the three master
formulae for current-current insertions given in 
(\ref{DA})--(\ref{DC}).

As in the case of current-current insertions, we consider an 
arbitrary operator with the colour structure $\hat V_1\otimes \hat V_2$
and the Dirac structure $\Gamma_1\otimes\Gamma_2$. Dropping the external
spinors, the insertion of this operator into the penguin vertex of
fig.~\ref{L:16}a  gives
\be\label{WL}
W_\lambda=-i g \mu^\varepsilon \hat V_1 T^a \hat V_2 I^{\mu\nu}
T^\lambda_{\mu\nu}
\ee
where
\be
T^\lambda_{\mu\nu}=\Gamma_1 \gamma_\nu \gamma_\lambda\gamma_\mu \Gamma_2
\ee
and
\be
I^{\mu\nu}=\int\frac{d^Dk}{(2\pi)^D}
\frac{k^\nu (k-q)^\mu}{k^2 (k-q)^2}=
-\frac{i}{16\pi^2} \left[\frac{1}{\varepsilon}\right]
\left[\frac{1}{6}q^\mu q^\nu+q^2\frac{g^{\mu\nu}}{12}\right]
\ee
with $q$ being the gluon momentum. In evaluating $I^{\mu\nu}$
we have kept only the divergent part.

The master formula for type A insertions is then obtained by including
the gluon propagator and the lower vertex in the penguin diagram (a)
of fig. \ref{L:16}. We find
\be\label{PA}
{\cal P}_A= -{\cal C}_A \frac{\as}{4\pi} 
\left[\frac{1}{\varepsilon}\right]
\left[\frac{1}{6}\frac{q^\mu q^\nu}{q^2}+\frac{g^{\mu\nu}}{12}\right]
\Gamma_1\gamma_\nu\gamma_\lambda\gamma_\mu\Gamma_2\otimes \gamma^\lambda~,
\ee
where the colour factor is given by
\be\label{PAC}
{\cal C}_A=\hat V_1 T^a\hat V_2\otimes T^a~.
\ee
It is understood that the Dirac and colour structures on the l.h.s of 
$\otimes$ are sandwiched between free spinors belonging to the
inserted operator and $\gamma^\lambda T^a$ standing on the r.h.s of 
$\otimes$ between the spinors representing the bottom line of the
penguin diagram. Formula (\ref{PA}) serves to calculate the
coefficients $(b_1)_{ij}$ in the master formula (\ref{a1b1}). We will
demonstrate this explicitly below.

The master formula for type B insertions can be derived in an
analogous manner. We find 
\be\label{PB}
{\cal P}_B= {\cal C}_B \frac{\as}{4\pi} 
\left[\frac{1}{\varepsilon}\right]
\left[\frac{1}{6}\frac{q^\mu q^\nu}{q^2}+\frac{g^{\mu\nu}}{12}\right]
{\rm Tr}
(\Gamma_1\gamma_\mu\gamma_\lambda\gamma_\nu)\Gamma_2\otimes \gamma^\lambda
\ee
where
\be\label{PBC}
{\cal C}_B={\rm Tr}(\hat V_1 T^a)\hat V_2\otimes T^a
\ee
with ``Tr" in (\ref{PBC}) standing for the trace in the colour space.
Note that in this formula we have closed the part $\hat V_1\Gamma_1$
of the inserted operator in the loop. If the part $\hat V_2\Gamma_2$
is closed instead, the indices ``1" and ``2" in (\ref{PB}) and
(\ref{PBC}) should be interchanged. The rules for the incorporation
of the external spinors into (\ref{PB}) should be evident in view of the
comments made after (\ref{PAC}). The difference in the overall sign
compared to (\ref{PA}) is a consequence of ``$-1$" for the fermion loop.

\subsubsection{Explicit Calculation of Penguin Insertions}
Let us apply our master formulae to the case of the operator
\be\label{Q2B}
Q_2=(\bar c_\alpha b_\alpha)_{V-A} (\bar s_\beta c_\beta)_{V-A}
\ee
for which we have
\be\label{DICO}
\hat V_1\otimes \hat V_2={\bf 1} \quad\quad 
\Gamma_1=\Gamma_2=\gamma_\tau(1-\gamma_5)
\ee
The flavour structure in (\ref{Q2B}) tells us that only type A insertions
are possible. We use therefore the master formula (\ref{PA}).
As we are only interested in the coefficient of $1/\varepsilon$,
we calculate all Dirac structures in D=4 dimensions. This gives
\be\label{DIRA}
\left[\frac{1}{6}\frac{q^\mu q^\nu}{q^2}+\frac{g^{\mu\nu}}{12}\right]
\Gamma_1\gamma_\nu\gamma_\lambda\gamma_\mu\Gamma_2=
\frac{4}{3}\left[\gamma_\lambda-\frac{q_\lambda\not\! q}{q^2}\right]
\ee
where we have used the identity
\be
\not\! q \gamma_\lambda \not\! q=2 q_\lambda \not\! q-q^2\gamma_\lambda.
\ee

The term $q_\lambda\not\! q/q^2$ does not contribute here as using
the Dirac equation one has $\bar s\not\! q b=0$ for massless quarks. 
More
care is needed when magnetic penguins of sections 8.7 and 12 are considered
and $m_b$ has to be kept.

Dropping then the second term on the r.h.s of (\ref{DIRA}), using
\be
{\cal C}_A=T^a\otimes T^a =\frac{1}{2}
\left[{\bf \tilde 1}-\frac{1}{N} {\bf 1}\right]
\ee
with ${\bf \tilde 1}$ defined in (\ref{V2}) and inserting 
the relevant spinors
we arrive at
\be\label{PAQ20}
{\cal P}_A(Q_2)=-\frac{\as}{4\pi} 
\left[\frac{1}{\varepsilon}\right]\left[\frac{2}{3}\right]
\left[{\bf \tilde 1}-\frac{1}{N} {\bf 1}\right]
[\bar s_\alpha \gamma_\lambda (1-\gamma_5) b_\beta]\otimes
\sum_q \bar q_\gamma \gamma^\lambda q_\delta
\ee
Next we decompose $\gamma^\lambda$ on the r.h.s of $\otimes$
into $V-A$ and $V+A$ parts
\be
\gamma^\lambda=\frac{1}{2}\gamma^\lambda(1-\gamma_5)+
\frac{1}{2}\gamma^\lambda(1+\gamma_5)
\ee 
which allows to express (\ref{PAQ20}) in terms of the penguin
operators 
\be\label{PAQ2}
{\cal P}_A(Q_2)=-\frac{\as}{4\pi} 
\left[\frac{1}{\varepsilon}\right]\left[\frac{1}{3}\right]
\left[Q_4+Q_6-\frac{1}{N} (Q_3+Q_5)\right]
\ee

The contribution of penguin insertions to the coefficients $(b_1)_{2j}$ 
relevant for the master formula (\ref{a1b1})
are consequently given by
\be\label{Q2ROW}
(b_1)_{23}^P=(b_1)_{25}^P=\frac{1}{3N}, \quad\quad
(b_1)_{24}^P=(b_1)_{26}^P=-\frac{1}{3}.
\ee

We next consider $Q_1$ and rewrite it using Fierz reordering
as
\be
Q_1=(\bar c_\alpha c_\alpha)_{V-A} (\bar s_\beta b_\beta)_{V-A}
\ee
so that the colour and Dirac structures are again given by
(\ref{DICO}). This time only type B insertions are possible.
However, ${\rm Tr}(T^a)=0$ and consequently the colour factor in
(\ref{PBC}) vanishes. Thus
\be\label{PBQ1}
{\cal P}_B(Q_1)=0
\ee
implying
\be\label{Q1ROW}
(b_1)_{13}^P=(b_1)_{14}^P=
(b_1)_{15}^P=(b_1)_{16}^P=0.
\ee

In the case of $Q_3$, the type B insertion vanishes as in the case of
$Q_1$ but now two type-A insertions are possible. One involves the
internal b-quark, the other the s-quark. 
Since gluons are flavour-blind and $Q_3$ has the same colour and 
Dirac structures as $Q_2$ we can find immediately 
\be\label{PAQ3}
{\cal P}_A(Q_3)= 2 {\cal P}_A(Q_2)
\ee
Consequently using (\ref{Q2ROW}) we find
\be\label{Q3ROW}
(b_1)_{33}^P=(b_1)_{35}^P=\frac{2}{3N}, \quad\quad
(b_1)_{34}^P=(b_1)_{36}^P=-\frac{2}{3}.
\ee

Next comes $Q_4$. Performing Fierz reordering
we have
\be
Q_4=\sum_q(\bar s_\alpha q_\alpha)_{V-A} (\bar q_\beta b_\beta)_{V-A}.
\ee
The type B insertions involving s and b quarks vanish as in 
the case of $Q_1$. On the other hand we have $f$ type A insertions
involving all quark flavours. Thus
\be\label{PAQ4}
{\cal P}_A(Q_4)=f {\cal P}_A(Q_2) 
\ee
and
\be\label{Q4ROW}
(b_1)_{43}^P=(b_1)_{45}^P=\frac{f}{3N}, \quad\quad
(b_1)_{44}^P=(b_1)_{46}^P=-\frac{f}{3}.
\ee

The penguin insertions of $Q_5$ vanish. The type B insertions vanish because
of ${\rm Tr}(T^a)=0$. The type A insertions vanish because now
\be\label{DI5}
\Gamma_1=\gamma_\tau(1-\gamma_5)~, \quad\quad
\Gamma_2=\gamma_\tau(1+\gamma_5)
\ee
and the Dirac structure in the master formula (\ref{PA}) vanishes. Thus
\be\label{Q5ROW}
(b_1)_{53}^P=(b_1)_{54}^P=
(b_1)_{55}^P=(b_1)_{56}^P=0~.
\ee
Finally the insertions of $Q_6$ have to be considered. Performing
Fierz reordering we have
\be\label{FQ6}
Q_6=-2 \sum_q (\bar s_\alpha(1+\gamma_5) q_\alpha)
 (\bar q_\beta (1-\gamma_5) b_\beta),
\ee
implying
\be\label{DICO6}
\hat V_1\otimes \hat V_2={\bf 1} \quad\quad 
\Gamma_1=(1+\gamma_5) \quad\quad \Gamma_2=(1-\gamma_5)
\ee
Again as in the case of $Q_4$ the type B insertions vanish. The
type A insertion of $Q_6$ is expected at first sight to give different
result than the
one of $Q_4$ because of the different Dirac structure.
However, application of the master formula (\ref{PA}) gives

\be\label{PAQ6}
{\cal P}_A(Q_6)= {\cal P}_A(Q_4) 
\ee
and consequently
\be\label{Q6ROW}
(b_1)_{63}^P=(b_1)_{65}^P=\frac{f}{3N} \quad\quad
(b_1)_{64}^P=(b_1)_{66}^P=-\frac{f}{3}.
\ee
\subsubsection{Explicit Calculation of Current-Current Insertions}
In section 6.5 we have calculated the $2\times 2$ anomalous dimension
matrix for the pair $(Q_1,Q_2)$ by inserting these operators into
the current-current diagrams of fig.~\ref{L:15}. Using the master formulae
(\ref{DA})--(\ref{DC}) for these diagrams together with the basic formula
(\ref{a1b1}) we have found the matrix (\ref{g120}) 
which as seen  in the left
upper corner of (\ref{eq:gs0Kpp}) constitutes a  part of the 
$6\times 6$ matrix
we are trying to reproduce.

What remains to be done are the insertions of the penguin operators into
the current-current diagrams. The case of the pair $(Q_3,Q_4)$ is
simple. From the point of view of current-current insertions the
pair $(Q_3,Q_4)$ behaves as $(Q_1,Q_2)$ and we can write immediately

\be\label{Q3CROW}
(b_1)_{33}^{cc}=2 C_F+\frac{3}{N}, \quad\quad
(b_1)_{34}^{cc}=-3,
\ee
\be\label{Q4CROW}
(b_1)_{43}^{cc}=-3,\quad\quad
(b_1)_{44}^{cc}=2 C_F+\frac{3}{N}. 
\ee

The case of $Q_5$ and $Q_6$ operators is different as they have the
$(V-A)\otimes (V+A)$ structure. Let us consider $Q_5$ first.
Using master formulae (\ref{DA})--(\ref{DC}) for
\be\label{DICO5}
\hat V_1\otimes \hat V_2={\bf 1}, \quad\quad 
\Gamma_1=\gamma_\tau(1-\gamma_5), \quad\quad
\Gamma_2=\gamma_\tau(1+\gamma_5)~,
\ee
we arrive at
\be\label{Q5SUMD}
\sum_i {\cal D}_i(Q_5)=\frac{\as}{4\pi} \frac{1}{\varepsilon}
\Gamma_1\otimes\Gamma_2
\left[{\cal C}_a^{(1)}+{\cal C}_a^{(2)}-
 ({\cal C}_b^{(1)}+{\cal C}_b^{(2)})+
4({\cal C}_c^{(1)}+{\cal C}_c^{(2)})\right]
\ee
with colour factors ${\cal C}_i^{(j)}$ given in (\ref{colour1}) and
(\ref{colour2}).

In order to perform the reduction of Dirac structures in the
master formulae (\ref{DA})--(\ref{DC}) we had to generalize the Greek
Method to the $(V-A)\otimes(V+A)$ operators. In this case $\otimes$
should be replaced by 1 as otherwise the Dirac structures would
identically vanish. Now the coefficients (4,16,4) in 
(\ref{GR1})--(\ref{GR2}) are
replaced by (4,4,16) respectively, which implies a different
weighting of the colour factors in (\ref{Q5SUMD}) 
relative to (\ref{SUMD}). Noting
that $Q_5$ is represented by ${\bf 1}$ and $Q_6$ by ${\bf \tilde 1}$
we obtain from (\ref{Q5SUMD})
\be\label{Q5CROW}
(b_1)_{55}^{cc}=2 C_F-\frac{3}{N},\quad\quad
(b_1)_{56}^{cc}=3.
\ee

Finally we consider $Q_6$. Here it is useful to use the form
(\ref{FQ6}). As $\Gamma_i$ are now given by (\ref{DICO6}) one 
easily finds that the usual Greek Method with $\otimes=\gamma_\tau$
applies. The master formulae (\ref{DA})--(\ref{DC}) then give
\be\label{Q6SUMD}
\sum_i {\cal D}_i(Q_6)=\frac{\as}{4\pi} \frac{1}{\varepsilon}
[-2\Gamma_1\otimes\Gamma_2]
\left[4({\cal C}_a^{(1)}+{\cal C}_a^{(2)})-
 ({\cal C}_b^{(1)}+{\cal C}_b^{(2)})+
({\cal C}_c^{(1)}+{\cal C}_c^{(2)})\right]~.
\ee
The weighting of colour factors differs from the cases $Q_2$ and $Q_5$.
In particular using (\ref{colour2}) we find that the two 
last terms in the square
bracket cancel each other. Effectively then only the insertions of
$Q_6$ into the diagrams (a) of fig.~\ref{L:15}
and its symmetric counterpart contribute. However, contrary to
the case of the operators $Q_{1-5}$ this contribution will not be canceled
by the $\delta_{ij}$ term in (\ref{a1b1}) as now ${\cal C}_a^{(j)}$ 
are multiplied
by 4 instead of 1. Consequently noting that in this case
$Q_6$ is represented by ${\bf 1}$ and $Q_5$ by ${\bf \tilde 1}$
we find
\be\label{Q6CROW}
(b_1)_{65}^{cc}=0,\quad\quad
(b_1)_{66}^{cc}=8 C_F. 
\ee

\subsubsection{Putting Things together}
Let us add the results for penguin and current-current insertions 
obtained above. Setting $N=3$ we find the matrix $\hat b_1$ in
(\ref{a1b1}):
\begin{equation} 
\hat b_1 =  
\left( 
\begin{array}{cccccc} 
2 C_F +1 & -3 & 0 & 0 & 0 & 0 \\ \svs 
 -3 & 2C_F +1 & {{1}\over {9}} & -{1\over 3} & {{1}\over {9}} & 
  -{1\over 3} \\ \svs 
0 & 0 &  2C_F+\frac{11}{9} & -{{11}\over 3} & {{2}\over {9}}  
& -{2\over 3} \\ \svs 
0 & 0 & {-3 +{f\over 9}} & 2C_F+1 -\frac{f}{3}  
& {{f}\over {9}} & -{{f}\over 3} \\ \svs 
0 & 0 & 0 & 0 & 2C_F-1 & 3  \\ \svs 
0 & 0 & {{f}\over {9}} & -{{f}\over 3} & {{f}\over {9}} &  
8C_F - \frac{f}{3} 
\end{array} 
\right) 
\label{bhat} 
\end{equation}
Inserting this matrix into the one-loop master formula (\ref{a1b1}) 
we reproduce
the full $6\times 6$ matrix in (\ref{eq:gs0Kpp}). 
Fantastic! We have reproduced all the magic numbers 
in this matrix. This is almost like reaching the top of
Mont Blanc.

\subsubsection{An Advice}
I hope that this long exercise and the exercise in subsection 6.5 were
useful for those students who have never calculated anomalous dimension
matrices.
But there is another lesson from these exercises. The corresponding
two-loop calculations of current-current and penguin insertions 
involving many more diagrams, more complicated colour factors and
evanescent operators are truly horrible. They are not like climbing
Mont Blanc but rather Mount Everest. They take several months rather
than a day or two. Consequently it is advisable for beginners to
take an experienced guide in order to climb these Himalayas.
Fortunately the guides in physics, as opposed to those in the real
Himalayas, are doing it for free. 

Yet one useful advice
is mandatory here. In our field there are unfortunately sponsors
(thesis supervisors) who send  their students to climb ``Mount Everest"
without having the slightest idea how difficult this climb is. Moreover
they are of little help once the student gets stuck in the middle of the
climb. Here is my advice. If you are not experienced in such Mount
Everest calculations and your sponsor is as described above, there
are only two solutions: either your sponsor has to provide you with
a strong sherpa who has climbed  Everest at least once, or you have 
to find another sponsor before it is too late \cite{Krakauer}!

\subsection{Electroweak Penguins}
\subsubsection{Operators}
The inclusion of the electroweak penguins and box diagrams of 
fig. \ref{L:14}
generates two additional operators $Q_7$ and $Q_9$. With respect to
the colour structure they are analogous to $Q_5$ and $Q_3$ operators,
respectively. When QCD effects are also taken into account, two
additional operators $Q_8$ and $Q_{10}$ are needed to close the system
under renormalization. They are analogous to $Q_6$ and $Q_4$, respectively.
The full set of operators necessary for the description of
$\Delta F=1$ decays including electroweak effects consists 
then of 10 operators.
The 4 electroweak penguin operators relevant for $\Delta B=1$ decays
with $\Delta S=1$ are given by
\begin{equation}\label{O4} 
Q_7 = {3\over 2}\;(\bar s b)_{V-A}\sum_{q=u,d,s,c,b}e_q\;(\bar qq)_{V+A} 
\ee
\be\label{O4a}
 Q_8 = {3\over2}\;(\bar s_{\alpha} b_{\beta})_{V-A}\sum_{q=u,d,s,c,b}e_q
        (\bar q_{\beta} q_{\alpha})_{V+A}
\end{equation}
\begin{equation}\label{O5} 
 Q_9 = {3\over 2}\;(\bar s b)_{V-A}\sum_{q=u,d,s,c,b}e_q(\bar q q)_{V-A}
\ee
\be\label{O6}
Q_{10} ={3\over 2}\;
(\bar s_{\alpha} b_{\beta})_{V-A}\sum_{q=u,d,s,c,b}e_q\;
       (\bar q_{\beta}q_{\alpha})_{V-A} 
\end{equation}
The overall factor $3/2$ is introduced for convenience. The charge
$e_q$ is the charge of the quark coupled to the lower vertex of the
photon or $Z^0$-propagator.
\subsubsection{Wilson Coefficients}
In order to generate the electroweak penguin operators $Q_7-Q_{10}$
it is sufficient to include the photon penguin together with the
relevant QCD renormalization. However, in order to keep the gauge
invariance also $Z^0$-penguins and box-diagrams have to be included
at the NLO level. The latter two sets of diagrams involving only
heavy fields $(W^\pm,Z^0,t)$ contribute only to the Wilson coefficients
at $\mu=\ord(\mw)$ and have no impact on the renormalization group
evolution down to low energy scales. On the other hand the inclusion
of $Z^0$-penguins introduces a strong $m_t$-dependence into Wilson
coefficients of the electroweak penguin operators, which in several
cases has important phenomenological implications. We will discuss
several of them in the phenomenological sections of these lectures. 
Here we
give some information on the Wilson coefficients of electroweak
penguin operators.

The matching at $\mu_W=\mw$ gives in the presence of the electroweak
penguin and box diagrams the values of $C_i(\mw)$ with
 $i=1,..10$. In the NDR scheme they are given by:
\begin{eqnarray}
C_1(\mw) &=&     \frac{11}{2} \; \frac{\as(\mw)}{4\pi} \, ,
\label{eq:CMw1} \\
C_2(\mw) &=& 1 - \frac{11}{6} \; \frac{\as(\mw)}{4\pi}
               - \frac{35}{18} \; \frac{\aem}{4\pi} \, ,
\label{eq:CMw2} \\
C_3(\mw) &=& -\frac{\as(\mw)}{24\pi} \widetilde{E}_0(x_t)
             +\frac{\aem}{6\pi} \frac{1}{\sin^2\theta_W}
             \left[ 2 B_0(x_t) + C_0(x_t) \right] \, , 
\label{eq:CMw3} \\
C_4(\mw) &=& \frac{\as(\mw)}{8\pi} \widetilde{E}_0(x_t) \, ,
\label{eq:CMw4} \\
C_5(\mw) &=& -\frac{\as(\mw)}{24\pi} \widetilde{E}_0(x_t) \, ,
\label{eq:CMw5} \\
C_6(\mw) &=& \frac{\as(\mw)}{8\pi} \widetilde{E}_0(x_t) \, ,
\label{eq:CMw6} \\
C_7(\mw) &=& \frac{\aem}{6\pi} \left[ 4 C_0(x_t) + \widetilde{D}_0(x_t)
\right]\, ,
\label{eq:CMw7} \\
C_8(\mw) &=& 0 \, ,
\label{eq:CMw8} \\
C_9(\mw) &=& \frac{\aem}{6\pi} \left[ 4 C_0(x_t) + \widetilde{D}_0(x_t) +
             \frac{1}{\sin^2\theta_W} (10 B_0(x_t) - 4 C_0(x_t)) \right] \, ,
\label{eq:CMw9} \\
C_{10}(\mw) &=& 0 \, ,
\label{eq:CMw10}
\end{eqnarray}
We recall (see Section 3) that
\begin{eqnarray}
B_0(x_t) &=& \frac{1}{4} \left[ \frac{x_t}{1-x_t} 
+ \frac{x_t \ln x_t}{(x_t-1)^2}
\right]\, , \label{eq:Bxt} \\
C_0(x_t) &=& \frac{x_t}{8} \left[\frac{x_t-6}{x_t-1} 
+ \frac{3 x_t + 2}{(x_t-1)^2}
\ln x_t \right]\, ,
\label{eq:Cxt} \\
D_0(x_t) &=& -\frac{4}{9} \ln x_t + 
\frac{-19 x_t^3 + 25 x_t^2}{36 (x_t-1)^3} +
\frac{x_t^2 (5 x_t^2 - 2 x_t - 6)}{18 (x_t-1)^4} \ln x_t \, ,
\label{eq:Dxt} \\
\widetilde{D}_0(x_t) &=& D_0(x_t) - \frac{4}{9} \, .
\label{eq:Dxttilde} 
\end{eqnarray}
$B_0(x_t)$
results from the evaluation of the box diagrams, $C_0(x_t)$ from the
$Z^0$-penguin, $D_0(x_t)$ from the photon penguin and $E_0(x_t)$ 
discussed already in the previous subsection 
from the gluon penguin diagram.
The constant $-4/9$ in (\ref{eq:Dxttilde})
is characteristic for the NDR scheme. It is absent in the HV scheme.
We note that the presence of electroweak effects modifies the
values of $C_2(\mw)$ and $C_3(\mw)$ by small $\ord(\alpha)$ corrections.
We also note that $C_8(\mw)=C_{10}(\mw)=0$. For $\mu\not=\mw$ non-vanishing
$C_8$ and $C_{10}$ are generated through QCD effects.

The anomalous dimension matrices are $10\times 10$:
\begin{equation}\label{ggew}
\hat\gamma(\as,\alpha)=\hat\gamma_s^{(0)}\aspi + 
\hat\gamma_e^{(0)}\frac{\alpha}{4\pi}+ 
\hat\gamma_s^{(1)}\left(\aspi\right)^2+
\hat\gamma_{se}^{(1)}\aspi\frac{\alpha}{4\pi}
\end{equation}
with $\gamma_s^{(0)}$ and $\gamma_s^{(1)}$ being $10\times 10$
generalizations of the corresponding $6\times 6$ matrices considered
previously. Since now $\ord(\alpha)$ effects are included in the
coefficients at scales $\ord(\mw)$, the anomalous dimension matrix
must also include $\ord(\alpha)$ contributions which are represented
by $\hat\gamma_e^{(0)}$ and $\hat\gamma_{se}^{(1)}$ at LO and NLO
respectively. The four matrices in (\ref{ggew}) can be found in 
\cite{BBL}, where the references to the original literature is
given. See also table \ref{TAB1}.

The calculation of the $6\times 6$ submatrix of
$\hat\gamma_s^{(0)}$ has been presented in detail in section 8.5.
The evaluation of $\hat\gamma_e^{(0)}$ proceeds in an analogous
manner except that the colour factors have to be properly replaced
by electric charges and the closed fermion loops coupled to
the photon have to be multiplied by N=3. Any reader, who succeeded
in calculating
$\hat\gamma_s^{(0)}$ should have no difficulties in calculating
within two hours the matrix
$\hat\gamma_e^{(0)}$. This is a very nice exercise
indeed. 
The calculations of $\hat\gamma_s^{(1)}$ and $\hat\gamma_{se}^{(1)}$ 
are even nicer but take more time. Typically six months for
$\hat\gamma_s^{(1)}$ and then a month for
$\hat\gamma_{se}^{(1)}$~.

Due to the simultaneous apperance of $\alpha$ and $\alpha_s$, the
RG analysis is more involved than the one discussed until now. In
particular the evolution matrix takes now the general form
\be
\hat U(m_1,m_2,\alpha)=\hat U(m_1,m_2)+\frac{\alpha}{4\pi} \hat R(m_1,m_2).
\ee
Here $\hat U(m_1,m_2)$ represents the pure QCD evolution matrix given
in (\ref{u0jj}). $\hat R(m_1,m_2)$ describes the additional evolution
in the presence of electromagnetic interactions. 
It includes both LO and NLO corrections.
Let us recall that
$\hat U(m_1,m_2)$ sums the logarithms $(\as t)^n$ and $\as(\as t)^n$
with $t=\ln(m_2^2/m_1^2)$. On the other hand $\hat R(m_1,m_2)$ sums
the logarithms $t(\as t)^n$ and $(\as t)^n$. The expresion for
$\hat R(m_1,m_2)$ is rather complicated. It can be found in 
\cite{BBL,BJLW}.
The Wilson coefficients are then found by using
\be
\vec C(\mu_b)=\hat U_5(\mu_b,\mw,\alpha)\vec C(\mw)
\ee
with $\vec C(\mw)$ given in (\ref{eq:CMw1})--(\ref{eq:CMw1}).
\subsubsection{Numerical Values}
In table \ref{tab:wc10b} we give numerical values of the coefficients
$C_i(\mu_b)$ in LO and the two NLO schemes in question.
\begin{table}[htb]
\caption{$\Delta B=1$ Wilson coefficients at 
$\mu=\overline{m}_{\rm b}(\mb)=
4.40\gev$ for $\mt=170\gev$.
\label{tab:wc10b}}
\begin{center}
\begin{tabular}{|c|c|c|c||c|c|c||c|c|c|}
\hline
& \multicolumn{3}{c||}{$\Lms^{(5)}=160\mev$} &
  \multicolumn{3}{c||}{$\Lms^{(5)}=225\mev$} &
  \multicolumn{3}{c| }{$\Lms^{(5)}=290\mev$} \\
\hline
Scheme & LO & NDR & HV & LO & 
NDR & HV & LO & NDR & HV \\
\hline
$C_1$ & -0.283 & -0.171 & -0.209 & -0.308 & 
-0.185 & -0.228 & -0.331 & -0.198 & -0.245 \\
$C_2$ & 1.131 & 1.075 & 1.095 & 1.144 & 
1.082 & 1.105 & 1.156 & 1.089 & 1.114 \\
\hline
$C_3$ & 0.013 & 0.013 & 0.012 & 0.014 & 
0.014 & 0.013 & 0.016 & 0.016 & 0.014 \\
$C_4$ & -0.028 & -0.033 & -0.027 & -0.030 & 
-0.035 & -0.029 & -0.032 & -0.038 & -0.032 \\
$C_5$ & 0.008 & 0.008 & 0.008 & 0.009 & 
0.009 & 0.009 & 0.009 & 0.009 & 0.010 \\
$C_6$ & -0.035 & -0.037 & -0.030 & -0.038 & 
-0.041 & -0.033 & -0.041 & -0.045 & -0.036 \\
\hline
$C_7/\aem$ & 0.043 & -0.003 & 0.006 & 0.045 & 
-0.002 & 0.005 & 0.047 & -0.002 & 0.005 \\
$C_8/\aem$ & 0.043 & 0.049 & 0.055 & 0.048 & 
0.054 & 0.060 & 0.053 & 0.059 & 0.065 \\
$C_9/\aem$ & -1.268 & -1.283 & -1.273 & -1.280 & 
-1.292 & -1.283 & -1.290 & -1.300 & -1.293 \\
$C_{10}/\aem$ & 0.302 & 0.243 & 0.245 & 0.328 & 
0.263 & 0.266 & 0.352 & 0.281 & 0.284 \\
\hline
\end{tabular}
\end{center}
\end{table}
Let us make just a few observations:
\bi
\item
Electroweak penguin coefficients being $\ord(\alpha)$ are smaller than 
$C_{1-6}$ coefficients. Notable exception is the coefficient $C_9$
which is in the ball park of the smallest QCD penguin coefficients
$C_3$ and $C_5$. It is the operator $Q_9$ which is the dominant
electroweak penguin in B-decays \cite{FLEISCHP}. 
With decreasing $\mu$ the coefficient $C_8$ increases
considerably. While its role in B-decays can be fully neglected,
it plays considerable role in the CP violation in $K\to\pi\pi$ decays
where also its hadronix matrix element is large.
\item
A numerical analysis shows that 
in contrast to $C_1,\ldots,C_6$, the additional coefficients
$C_7,\ldots,C_{10}$ increase strongly with $\mt$.
This strong $\mt$ dependence originates in the $Z^0$-penguin represented
by the function $C_0(x_t)$ in (\ref{eq:Cxt}).
Even in the range
$\mt = (170 \pm 15)\gev$ with in/decreasing $\mt$ there is 
 a relative variation of $\ord(\pm 19\%)$ and $\ord(\pm
10\%)$ for the absolute values of $C_8$ and $C_{9,10}$, respectively.
\ei
\subsection{Magnetic Penguins}
The inclusive decay $B \to X_s\gamma$ with an on-shell $\gamma$ 
is governed by
the operator $Q_{7\gamma}$ which originates in the photon-penguin
vertex with $q^2=0$, where $q_\mu$ is the momentum of the emitted
photon. In order to obtain a non-vanishing result one has to
keep external b-quark mass as well as external momenta. 
This {\it mass insertion} together with the expansion to second order in
external momenta generates  $Q_{7\gamma}$, which due to the appearance
of $\sigma^{\mu\nu}$ is known under the name of a magnetic photon 
penguin.
The corresponding gluon--penguin vertex with $q^2=0$ results in a 
magnetic gluon penguin operator
$Q_{8G}$ which plays the dominant role in the inclusive 
$B \to X_s~{\rm gluon}$ decay. The magnetic penguins are given by
\begin{equation}\label{O6U}
Q_{7\gamma}  =  \frac{e}{8\pi^2} m_b \bar{s}_\alpha \sigma^{\mu\nu}
          (1+\gamma_5) b_\alpha F_{\mu\nu}\qquad            
Q_{8G}     =  \frac{g}{8\pi^2} m_b \bar{s}_\alpha \sigma^{\mu\nu}
   (1+\gamma_5)T^a_{\alpha\beta} b_\beta G^a_{\mu\nu}.
\end{equation}
The renormalization group analysis of $B\to X_s\gamma$ involves in
addition to $Q_{7\gamma}$ and $Q_{8G}$ also the operators $Q_1...Q_6$
discussed previously. The peculiar feature of this analysis is the
vanishing of the mixing under renormalization between the sets
$(Q_{7\gamma},Q_{8G})$ and $(Q_1...Q_6)$ at the one loop level.
That is in order to calculate the leading entry (LO), representing
this mixing, in the relevant anomalous dimension matrix one is
forced to perform two-loop calculations. At NLO the corresponding
three loop calculations are necessary. Because this mixing has
a very important impact on the resulting decay rate, these
calculations are mandatory before a meaningful theoretical
prediction for $B\to X_s\gamma$ can be obtained.

The decay $B\to X_s\gamma$ is one of the central decays in the
rare decays phenomenology.
Therefore, we will devote to it  section 12 where both
technical and phenomenological aspects of $B\to X_s\gamma$ will be
reviewed.
\subsection{Semi-Leptonic Operators}
In the case of $\kpn$ we have encountered the operator
$Q(\nu\bar\nu)=(\bar sd)_{V-A}(\bar\nu\nu)_{V-A}$. This operator
governs also the decay $K_L\to\pi^0\nu\bar\nu$. An analogous operator 
\be
Q^B(\nu\bar\nu)=(\bar sb)_{V-A}(\bar\nu\nu)_{V-A}
\ee
governs the inclusive decay $B\to X_s\nu\bar\nu$. We will briefly discuss
this decay in section 13.
$Q(\nu\bar\nu)$ and $Q^B(\nu\bar\nu)$ have no anomalous dimensions and
the RG analysis of their Wilson coefficients in the case of internal top
contributions is very simple. We have
discussed this in the case of $\kpn$ at the beginning of this section.
This simplification is caused by the fact that neutrinos do not couple
neither to gluons nor photons. 

Now, in the case of $B \to X_s \mu^+\mu^-$ and $K_L\to \pi^0 e^+e^-$
the following operators play the dominant role:

\begin{equation}\label{9V}
Q_{9V}  = (\bar s b  )_{V-A} (\bar \mu\mu)_{V}~~~~~
Q_{10A}  = (\bar s b )_{V-A} (\bar \mu\mu)_{A}
\end{equation}
and
\be\label{9VS}
Q_{7V}  = (\bar s d  )_{V-A} (e^+e^-)_{V}~~~~~
Q_{7A}  = (\bar s d )_{V-A} (e^+e^-)_{A}
\end{equation}
respectively. We will discuss here briefly $Q_{9V}$ and $Q_{10A}$. The
analysis of $Q_{7V}$ and $Q_{7A}$ is analogous, but more involved
because of lower
renormalization scales involved and related threshold effects.
Detailed expositions of NLO analyses of 
$B \to X_s \mu^+\mu^-$ and $K_L\to \pi^0 e^+e^-$ can be found in
\cite{Mis:94,BuMu:94}  and \cite{BLMM} respectively.

\begin{figure}[hbt]
\vspace{0.10in}
\centerline{
\epsfysize=1.5in
\epsffile{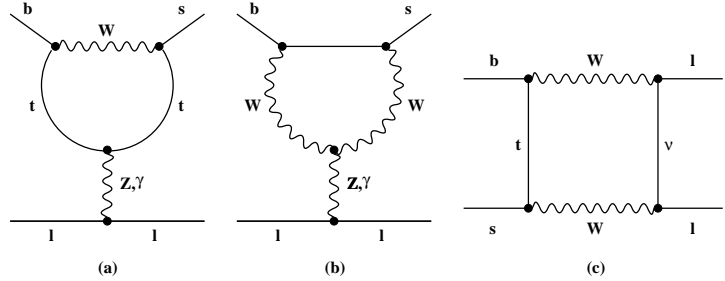}
}
\vspace{0.08in}
\caption[]{One loop diagrams in the full theory contributing to rare decays
with charged leptons in the final state.
\label{L:17}}
\end{figure}

\begin{figure}[hbt]
\vspace{0.10in}
\centerline{
\epsfysize=1.5in
\epsffile{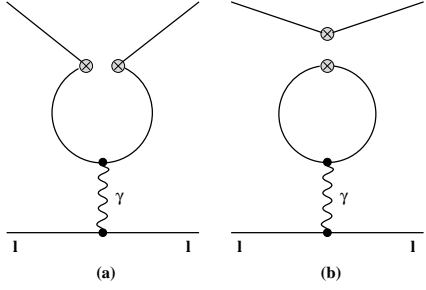}
}
\vspace{0.08in}
\caption[]{One loop diagrams in the effective theory contributing to 
rare decays with charged leptons in the final state.
\label{L:18}}
\end{figure}

As in the case of $Q(\nu\bar\nu)$ and $Q^B(\nu\bar\nu)$, the semi-leptonic
operators $Q_{9V}$ and $Q_{10A}$ have vanishing anomalous dimensions.
However, the fact that charged leptons couple to photons makes the
RG analysis of their coefficients more involved. Indeed these operators
originate in the diagrams of fig. \ref{L:17}. 
Moreover, in the effective theory
the diagrams in fig. \ref{L:18} have to be considered, 
where the inserted operators
are our good friends $(Q_1,....Q_6)$. The operator $Q_{10A}$ involving the
$\gamma_\mu\gamma_5$ current is uneffected by the diagrams involving
photons and its Wilson coefficient can be calculated in the same manner as
for the case of $Q(\nu\bar\nu)$. Including gluon corrections to the
one-loop diagrams in fig. \ref{L:17} one finds:
\be
C_{10A}(\mw)=-\frac{\alpha}{2\pi}\frac{Y(x_t)}{\sin^2\Theta_W}
\ee
where
\begin{equation}\label{yy}
Y(x_t) = Y_0(x_t) + \aspi Y_1(x_t)\end{equation}
with the one-loop function given by \cite{IL}
\begin{equation}\label{yy0}
Y_0(x_t) = {x_t\over 8}
\left[{4-x_t\over 1-x_t}+{3x_t\over (1-x_t)^2}\ln x_t\right]
\end{equation}
and 
\begin{eqnarray}\label{yy1}
Y_1(x_t) = &&{4x_t + 16 x_t^2 + 4x_t^3 \over 3(1-x_t)^2} -
     {4x_t - 10x_t^2-x_t^3-x_t^4\over (1-x_t)^3} \ln x_t\nonumber\\
     &+&{2x_t - 14x_t^2 + x_t^3 - x_t^4\over 2(1-x_t)^3} \ln^2 x_t
           + {2x_t + x_t^3\over (1-x_t)^2} L_2(1-x_t)\nonumber\\
         &+&8x {\partial Y_0(x) \over \partial x} 
          \ln\frac{\mu_t^2}{\mw^2} 
\end{eqnarray}
resulting from two-loop calculations \cite{BB2}.
 
The $\mu_t$-dependence of the last term in (\ref{yy1}) cancels to the
considered order the $\mu_t$-dependence of the leading term 
$Y_0(x_t(\mu_t))$.
The leftover $\mu_t$ dependence in $Y(x_t)$ is below $1\%$.
For $\mu_t=\mt$, the complete function $Y(x_t)$ can  be written as
\begin{equation}\label{yeta}
Y(x_t)=\eta_Y\cdot Y_0(x_t), \qquad\quad \eta_Y=1.026\pm0.006,
\end{equation}
with the QCD factor $\eta_Y$
practically independent of $\mt$. The range in (\ref{yeta}) corresponds
to $150~\gev\le\mt\le 190~\gev$. The dependence on 
$\Lambda_{\overline{MS}}$ can be neglected.

The fate of the operator $Q_{9V}$ is different. Now the diagrams with
photon exchanges contribute in an important way. The presence of the 
photon penguin diagrams in the full theory brings in the function
$D_0(x_t)$ of $(\ref{eq:Cxt})$. On the other hand, the presence of
$Q_1-Q_6$ insertions into the photon penguin diagrams of fig. 
\ref{L:18}
introduces the mixing between $Q_1,...Q_6$ operators and $Q_{9V}$
under renormalization. That is the insertion of any of the four-quark 
operators into the diagrams in fig. \ref{L:18} results in 
$Q_{9V}$ multiplied
by a certain coefficient. From these coefficients the entries
$\gamma^{(0)}_{i9}$ with $i=1,...6$ in
a $7\times 7$ anomalous dimension matrix involving 
$(Q_1,....Q_6,Q_{9V})$ can be found. 
Including gluon corrections to fig.~\ref{L:18} gives
$\gamma^{(1)}_{i9}$.
Now, 
$\gamma^{(0)}_{i9}$ and $\gamma^{(1)}_{i9}$ are coefficients of
$\alpha$ and $\alpha\alpha_s$ respectively. On the other hand
the entries $\gamma^{(0)}_{ij}$ and $\gamma^{(1)}_{ij}$ with
$i,j=1,...6$ are the coefficients of $\alpha_s$ and $\alpha_s^2$
respectively. In order to work with an anomalous dimension
matrix which has a usual expansion in $\alpha_s$, it is convenient
to introduce a new operator
\be\label{newop}
Q^\prime_{9V}=\frac{\alpha}{\alpha_s(\mu)} Q_{9V},
\quad\quad
C^\prime_{9V}(\mu)=\frac{\alpha_s(\mu)}{\alpha} C_{9V}(\mu)
\ee
and perform the RG evolution for the set $(C_1,...C_6,C^\prime_{9V})$.
Once $C^\prime_{9V}(\mu)$ has been calculated by means of the standard
method developed in sections 5 and 6, the coefficient $C_{9V}(\mu)$
can be obtained by using (\ref{newop}).

That the rescaling trick in (\ref{newop}) works at all, is related to
the fact that $Q_{9V}$ cannot mix back into the set $(Q_1,....Q_6)$.
That is $\gamma^{(0)}_{9i}=\gamma^{(1)}_{9i}=0$ for $i=1,...6.$
This trick cannot be used in the case of electroweak penguin contributions
to $\Delta F=1$ decays discussed in subsection 8.6. 
There the mixing between
the QCD penguin and electroweak penguin operators takes place in both
directions and one does not gain anything by making a rescaling of
electroweak four-quark operators. Fortunately the electroweak penguin
operators in (\ref{O4}) and (\ref{O5}) contribute to
$B \to X_s \mu^+\mu^-$ and $K_L\to \pi^0 e^+e^-$ first at $\ord(\alpha^2)$
and consequently they can be fully neglected in these decays.

The anomalous dimensions $\gamma^{(0)}_{i9}$ and $\gamma^{(1)}_{i9}$
can be found in the formulae (VIII.11) and (VIII.12) of ref. 
\cite{BBL}. One
should note that, in contrast to $Q_{9V}$, the rescaled operator
$Q^\prime_{9V}$ has effectively a non-vanishing anomalous dimension resulting
from the presence of $\alpha_s(\mu)$ in (\ref{newop}):
\be
\gamma^{(0)}_{99}=-2\beta_0, \quad\quad \gamma^{(1)}_{99}=-2 \beta_1~.
\ee

For completeness we give the result for $C_{9V}(\mu_b)$ including
NLO corrections in the NDR scheme:
\begin{equation}\label{C9tilde}
C_9^{NDR}(\mu_b)  = \frac{\aem}{2\pi}\left[  
P_0^{NDR} + \frac{Y_0(x_t)}{\sin^2\Theta_W} -4 Z_0(x_t)\right],
\end{equation}
where a negligible contribution proportional to $E_0(x_t)$ has been 
omitted.
$P_0^{NDR}$ is a $\mt$-independent constant which for $\mu_b=5.0 \gev$
and $\as(\mz)=0.118$ equals 2.59. 
The renormalization group improved perturbative expansion for
$P_0^{NDR}$ and $C_9^{NDR}(\mu_b)$ has the structure
$1/\as+\ord(1)+\ord(\as)...$, as seen explicitly in the analytic
formula (X.6) in \cite{BBL}. For this reason in an NLO analysis of
$B\to X_s\mu\bar\mu$ and also $K_L \to\pi^0 e^-e^+$ 
only the leading terms in $Y(x_t)$ and $Z(x_t)$,
i.e. $Y_0(x_t)$ and $Z_0(x_t)$, contribute. 
\subsection{Charm Quarks in Electroweak Loops}
Our discussion of QCD effects in penguin and box diagram contributions
to rare decays concentrated on diagrams with internal top quark
propagators. In the process of matching of the full theory onto
effective five quark theory the top quark is integrated out together
with $W^\pm$ and $Z^0$ bosons and the resulting operators 
are local. The
evolution of their coefficients down to low energy scales proceeds
in the standard manner as discussed in the preceeding sections.

In the case of penguin and box diagrams with internal charm quarks
the situation is more complicated. After the matching at scales
$\ord(\mw)$ charm quarks remain as dynamical degrees of freedom
and after $W^\pm$ and $Z^0$ bosons have been integrated out one has to
deal with bi-local structures rather than with local operators.
An example is shown for the case of $\kpn$ in fig. \ref{L:11}a. 
It results from $Z^0$-penguin contributing to $K^+\to\pi^+\nu\bar\nu$.
These
structures have, in contrast to $Q(\nu\bar\nu)$, anomalous
dimensions which makes the renormalization group analysis
non-trivial. 
Another important example are box diagram contributions to
$K^0-\bar K^0$ mixing where two internal charm propagators 
(see fig.~\ref{L:11}b) or
one charm propagator and one top propagator may appear simultaneously.

\begin{figure}[hbt]
\vspace{0.10in}
\centerline{
\epsfysize=2in
\epsffile{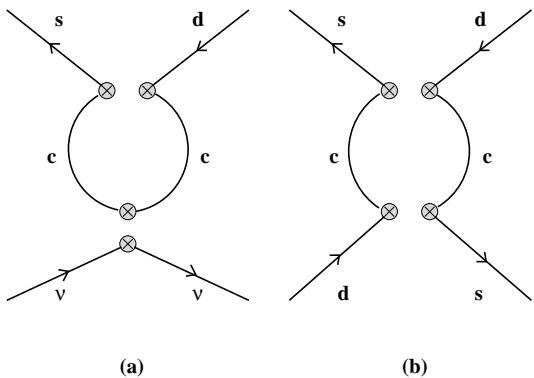}
}
\vspace{0.08in}
\caption[]{Bilocal Structures. 
\label{L:11}}
\end{figure}

In all these cases the RG evolution from scales $\mu_W=\ord(M_W)$
down to $\mu_c=\ord(\mc)$ is more involved than the one presented
sofar. In the process of matching of four-quark theory onto
three-quark theory charm is integrated out and the effective theory
below $\mu_c$ involves only local operators which can be analized
in the standard manner. The renormalization group analysis of
bi-local structures is beyond the scope of these lectures.
On the other hand the tools collected in this and preceeding sections
are sufficient for following detailed expositions of this subject
without great difficulties. The internal charm contributions to
$\kpn$ are calculated in detail in \cite{BB3} where the full two-loop
renormalization group evolution is performed. A brief account of this
analysis can be found in chapter XIB of the review \cite{BBL}. We will use
the numerical results obtained there in section 13. Similarly
   the contributions of charm to $K^0-\bar K^0$ mixing are analyzed
in detail in \cite{HNa,HNb}. A brief account of these papers can be found
in chapters XIIC and XIID of the review \cite{BBL}.
The issue of internal charm contributions to non-leptonic two-body 
B-decays
in the form of the so-called ``charming penguins" 
is discussed in \cite{BF95,ITAL}.
\subsection{Penguin--Box Expansion from OPE}
In section 3 prior to the discussion of QCD effects we have formulated
the FCNC decays in terms of effective vertices corresponding to
various penguin and box diagrams. These effective vertices depend on
a set of basic universal (process independent) 
$\mt$-dependent functions $F_r(x_t)$ listed in 
(\ref{SXYZ}). We have also stated that any decay amplitude can be
written as
\begin{equation}
A({\rm decay}) = P_0({\rm decay}) + \sum_r P_r({\rm decay}) \, F_r(x_t),
\label{generalPBE1}
\end{equation}
where the coefficients $P_r$ are process dependent. It is
straightforward to derive these {\it Penguin-Box Expansion} \cite{PBE0}
from OPE.
To this end we use OPE and 
and rewrite a given decay amplitude $A(M\to F)$ as follows
\begin{equation}\label{USE}
A(M\to F)=\frac{G_{\rm F}}{\sqrt 2} V_{\rm CKM}
\sum_{i,k} \langle F\mid O_k(\mu)\mid M\rangle \;\hat U_{ki}\;(\mu,\mw) 
          \; C_i(\mw),
\end{equation}
where $\hat U_{kj}(\mu,M_W)$ is the renormalization group
transformation from $\mw$ down to $\mu$ given already at several
places in these lectures.

Now, as we have seen in several examples in this section,
$C_i(\mw)$ are linear combinations of the basic functions
$F_r(x_t)$ so that we can write
\be\label{CIA}
C_i(\mw)=c_i+\sum_r h_{ir} F_r(x_t)
\ee
where $c_i$ and $h_{ir}$ are $\mt$-independent constants. 
Inserting (\ref{CIA})
into (\ref{USE}) and summing over $i$ and $k$ we recover
(\ref{generalPBE1}) with
\be\label{PBE8}
P_0({\rm decay})= \sum_{i,k} \langle F\mid O_k(\mu)\mid M\rangle 
\;\hat U_{ki}\;(\mu,\mw)c_i~,
\ee 
\be\label{PBE9}
P_r({\rm decay})= \sum_{i,k} \langle F\mid O_k(\mu)\mid M\rangle 
\;\hat U_{ki}\;(\mu,\mw)h_{ir}~,
\ee 
where we have suppressed the overall factor $(G_F/\sqrt{2})V_{CKM}$.

The process dependence of $P_0$ and $P_r$ enters through
$\langle F\mid O_k(\mu)\mid M\rangle$. In certain cases like
$K\to\pi\nu\bar\nu$ these matrix elements are very simple implying
simple formulae for the coefficients $P_0$ and $P_r$. In other
situations, like $\epe$ discussed in section 11, this is not the
case.

From the perspective of the formula (\ref{USE}) the relation between
the usual OPE and its PBE-version is clear. OPE puts the last two
factors together by summing over "i" to obtain $C_k(\mu)$. The
PBE is realized on the other hand by putting the first two factors
together through the summation over "k" and subsequent rewriting
of $C_i(\mw)$ in terms of $F_r(x_t)$ as explicitly shown in
(\ref{CI}). Equivalently PBE is obtained by setting $\mu=\mw$
in (\ref{USE}) as $\hat U_{ki}\;(\mu,\mw)=\delta_{ki}$.

PBE is very well suited for the study of the extentions of the
Standard Model in which new particles are exchanged in the loops.
We know already that these particles are heavier than W-bosons
and consequently they can be integrated out together with
the weak bosons and the top quark. If there are no new local operators
the mere change is to modify the functions $F_r(x_t)$ which now
acquire the dependence on the masses of new particles such as
charged Higgs particles and supersymmetric particles. The process
dependent coefficients $P_0$ and $P_r$ remain unchanged unless
new effective operators with different Dirac and colour structures
have to be introduced. Examples of the applications of PBE to physics
beyond the Standard Model can be found in \cite{BBHLS,MW96,AAA}.

The universality of the functions $F_r(x_t)$ listed in (\ref{SXYZ})
can be violated partly when QCD corrections to one loop penguin
and box diagrams are included. For instance in the case of
semi-leptonic FCNC transitions there is no gluon exchange in
a $Z^0$-penguin diagram parallel to the $Z^0$-propagator but
such an exchange takes place in non-leptonic decays in which the
bottom line is a quark-line. Thus the general universality of $F_r(x_t)$
present at one loop level is reduced to two universality classes
relevant for semi-leptonic and non-leptonic transitions.
However, as we have seen in the case of $K\to\pi\nu\bar\nu$,
the $\ord(\as)$ corrections to the function $X_0(x_t)$ could
be absorbed into an overall QCD factor $\eta_X$ which moreover
with a proper definition of $\mt$ turned out to be essentially
$\mt$-independent. Similar situations take place in other decays
so that to a very good approximation the top mass dependence is
governed even at the two-loop level through the one-loop functions
and the inclusion of QCD effects plays mainly the role in reducing 
the $\mu_t$-dependences.

\subsection{Status of NLO Calculations}
We end this section by  listing all 
existing NLO calculations for weak decays in table \ref{TAB1}.
Further details on these calculations can be found in the orignal
papers and in the review \cite{BBL}. Some of the implications
of these calculations will be analyzed in detail in subsequent
sections.
\begin{table}[thb]
\caption{References to NLO Calculations}
\label{TAB1}
\begin{center}
\begin{tabular}{|l|l|}
\hline
\bf \phantom{XXXXXXXX} Decay & \bf Reference \\
\hline
\hline
\multicolumn{2}{|c|}{$\Delta F=1$ Decays} \\
\hline
current-current operators     & \cite{ACMP,WEISZ} \\
QCD penguin operators         & \cite{BJLW1,BJLW,ROMA1,ROMA2} \\
electroweak penguin operators & \cite{BJLW2,BJLW,ROMA1,ROMA2} \\
magnetic penguin operators    & \cite{MisMu:94,CZMM} \\
$Br(B)_{SL}$                  & \cite{ACMP,Buch:93,Bagan} \\
inclusive $\Delta S=1$ decays       & \cite{JP} \\
\hline
\multicolumn{2}{|c|}{Particle-Antiparticle Mixing} \\
\hline
$\eta_1$                   & \cite{HNa} \\
$\eta_2,~\eta_B$           & \cite{BJW90} \\
$\eta_3$                   & \cite{HNb} \\
\hline
\multicolumn{2}{|c|}{Rare $K$- and $B$-Meson Decays} \\
\hline
$K^0_L \rightarrow \pi^0\nu\bar{\nu}$, $B \rightarrow l^+l^-$,
$B \rightarrow X_{\rm s}\nu\bar{\nu}$ & \cite{BB1,BB2} \\
$K^+   \rightarrow \pi^+\nu\bar{\nu}$, $K_{\rm L} \rightarrow \mu^+\mu^-$
                                      & \cite{BB3} \\
$K^+\to\pi^+\mu\bar\mu$               & \cite{BB5} \\
$K_{\rm L} \rightarrow \pi^0e^+e^-$         & \cite{BLMM} \\
$B\rightarrow X_s \mu^+\mu^-$           & \cite{Mis:94,BuMu:94} \\
$B\rightarrow X_s \gamma$      & 
\cite{AG2}-\cite{strum}, \cite{GH97,BKP2} \\
\hline
\end{tabular}
\end{center}
\end{table}
\subsection{Final Remarks}
We are roughly half way through these lectures. The last seven
sections dealt with the basic formalism of weak decays. The
next seven sections will present some phenomenological applications
of this formalism.

\section{Non-Leptonic Two-body Decays and Factorization}
\setcounter{equation}{0}
\subsection{Preliminaries}
We will begin the applications of the formalism developed in the
previous seven sections by discussing two-body non-leptonic decays.
Although our discussion will concentrate on two-body B-decays, it
can be generalized in a straightforward manner to D-decays.

I should state from the beginning that it is not my intention
to give here a review of two-body decays and present detailed
comparision with the available data. My intention is rather to 
reanalyze critically the concepts of the {\it factorization}
hypothesis and in particular of the {\it generalized factorization}
hypothesis discussed in the literature. As we will see soon, it
is an excellent battle field for the formalism developed in
previous sections.

Now comes a rather unfortunate move. In this and only this section
we have to modify slightly our notation by interchanging the 
indices 1 and 2 in current-current operators. This we have to do
 in order to conform to the notation used in the literature
on two--body non--leptonic decays. Thus we introduce the operators
\be\label{T1}
O_1=Q_2, \quad\quad O_2=Q_1,
\ee
and their respective coefficients
\be\label{T2}
\bar C_1(\mu)=C_2(\mu), \quad \quad  \bar C_2(\mu)=C_1(\mu).
\ee
Correspondingly
\be\label{T3}
O_\pm=\frac{O_1\pm O_2}{2}, \quad\quad 
z_\pm= \bar C_1(\mu)\pm \bar C_2(\mu),
\ee
and
\begin{equation}\label{10N}
\bar C_1(\mu)=\frac{z_+(\mu)+z_-(\mu)}{2},
\qquad\qquad
\bar C_2(\mu)=\frac{z_+(\mu)-z_-(\mu)}{2},
\end{equation}
with all formulae (\ref{B9})--(\ref{B15}) unchanged.
We have introduced a ``bar", omitted in the literature, in order
to avoid possible confusion. 

This section is based on \cite{AJB94a} and the recent collaboration
with Luca Silvestrini \cite{BUSI}. We do not cover here more dynamical
approaches to non-leptonic decays like QCD sum reles. 
A very nice review of the 
applications of QCD sum rules to non-leptonic decays has been presented
this year by Khodjamirian and R\"uckl \cite{KR98} and is strongly 
recommended.
\subsection{Factorization}
In the factorization approach to non-leptonic meson decays
\cite{FEYNMAN,STECHF} one can
distinguish three classes of decays for which the amplitudes have the
following general structure \cite{BAUER,NEUBERT}:
\begin{equation}\label{1}
A_{\rm I}=\frac{G_F}{\sqrt{2}} V_{CKM}a_1(\mu)\langle O_1\rangle_F 
\qquad {\rm (Class~I)},
\end{equation}
\begin{equation}\label{2}
A_{\rm II}=\frac{G_F}{\sqrt{2}} V_{CKM}a_2(\mu)\langle O_2\rangle_F 
\qquad {\rm (Class~II)},
\end{equation}
\begin{equation}\label{3N}
A_{\rm III}=
\frac{G_F}{\sqrt{2}} V_{CKM}[a_1(\mu)+x a_2(\mu)]\langle O_1\rangle_F
 \qquad {\rm (Class~III)}.
\end{equation}
Here $V_{CKM}$ denotes symbolically the CKM factor characteristic for a
given decay. 
$\langle O_i\rangle_F$ are factorized 
hadronic matrix
elements of the operators $O_i$ given as products of matrix elements of
quark currents and $x$ is a non-perturbative factor equal to unity in
the flavour symmetry limit. Finally $a_i(\mu)$ are QCD factors which
are given as follows
\begin{equation}\label{BS23}
a_1(\mu)=\bar C_1(\mu)+\frac{1}{N} \bar C_2(\mu), \qquad
a_2(\mu)=\bar C_2(\mu)+\frac{1}{N} \bar C_1(\mu).
\end{equation}
We will soon give explicit examples and we will rederive these
formulae as limiting cases of the generalized factorization 
hypothesis. First, however, we would like to make a few general
comments on the weak points of this approach.

At first sight the simplicity of this approach is very appealing.
Once the matrix elements $\langle O_i\rangle_F $
have been expressed in terms of various meson
decay constants and generally model dependent formfactors, predictions
for non-leptonic heavy meson decays can be made. 
Moreover relations between non-leptonic and semi-leptonic decays can
be found which allow to test factorization in a model independent
manner. An incomplete list of analyses of this type is given in 
\cite{LNF,NEUBERT} and will be extended below.
 
On the other hand,
it is well known that
non-factorizable contributions must be present in the hadronic matrix
elements of the current--current operators $O_1$ and $O_2$ in order
to cancel the $\mu$ dependence of $\bar C_i(\mu)$ or $a_i(\mu)$ so that
the physical amplitudes do not depend on the arbitrary renormalization
scale $\mu$. 
$\langle O_i\rangle_F$ being products of matrix elements of
conserved currents
are $\mu$--independent and the cancellation of the $\mu$ dependence
in (\ref{1})--(\ref{3N}) does not take place.
Consequently from the point of view of QCD
the factorization approach can be at best correct at a single value
of $\mu$, the so-called factorization scale $\mu_f$. Although the
approach itself does not provide the value of $\mu_f$, the  
proponents
of factorization expect $\mu_f=O(m_b)$ and $\mu_f=O(m_c)$ for
B-decays and D-decays respectively. 

Here we would like to point out that beyond the leading logarithmic
approximation for $\bar C_i(\mu)$ a new complication arises. 
As we have discussed in previous sections,
 at next to leading level in the renormalization
group improved perturbation theory the coefficients $\bar C_i(\mu)$
depend on the renormalization scheme for operators. Again only
the presence of non-factorizable contributions
 in $\langle O_i\rangle$ can
remove this scheme dependence in the physical amplitudes. 
However $\langle O_i\rangle_F$ are renormalization scheme
independent and the factorization approach is of course unable
to tell us whether it works better with an anti-commuting $\gamma_5$
in $D\not=4$ dimensions (NDR scheme) or with another definition 
of $\gamma_5$ such as used in HV or DRED schemes. Moreover there
are other renormalization schemes parametrized by $\kappa_\pm$
in (\ref{B12})--(\ref{B15}).
The renormalization scheme dependence emphasized here is rather
annoying from the factorization point of view as it precludes
a unique phenomenological determination of $\mu_f$ as we will
show explicitly below. 

On the other hand, arguments have been given
\cite{BJORKEN,DUGAN,NEUBERT}, that 
 factorization approach could be
approximately true in the case of two-body decays with high
energy release \cite{BJORKEN}, or in certain kinematic regions
\cite{DUGAN,ISGUR,ITALY}. We will not repeat here these arguments, which
can be found in the original papers.
Needless to say the issue of factorization does not only 
involve the short distance gluon corrections discussed here
but also final state interactions as stressed in particular
in \cite{ITALY}.

It is difficult to imagine that factorization
can hold even approximately in all circumstances. 
In spite of this, it
became fashonable these days to
test this idea, 
to some extent, by using certain set of formfactors to calculate
$ \langle O_i\rangle_F  $ and by making global fits of the 
formulae (\ref{1})--(\ref{3N})
to the data treating
$ a_1 $ and $ a_2 $ as free independent parameters. As an example we
give the result of a recent
analysis of this type for non-leptonic two-body B-decays
\cite{NS97}  
\begin{equation}\label{8N}
a_1\approx 1.08\pm0.04
\qquad
a_2\approx 0.21\pm0.05
\end{equation}
which is compatible with other analyses 
\cite{Cheng,Soares,LNF,GNF,AKL98}.
At the level of accuracy of the existing  experimental data and 
because of strong model
dependence in the relevant formfactors it is not yet possible  
to conclude on the basis of these analyses whether the
factorization approach is a useful approximation in general or not. 
It is
certainly  conceivable that factorization may apply better to some
non-leptonic decays than to others 
\cite{NEUBERT}-\cite{ISGUR}
and using all decays in a global fit may misrepresent the true
situation. 

The fact that $\langle O_i\rangle_F$ are $\mu$-independent but
$a_i(\mu)$ are $\mu$-dependent, which is clearly inconsistent,
inspired a number of authors \cite{Cheng,Soares, NS97,GNF,AKL98} to
generalize the concept of factorization. The presentation given
in the next subsection, done in collaboration with Silvestrini 
\cite{BUSI},
follows closely the generalization due to 
Neubert and Stech
\cite{NS97} which deals exclusively with the operators $O_1$ and
$O_2$. The generalization presented in \cite{Cheng,GNF,AKL98} are
similar in spirit but includes also the penguin contributions. I will
discuss it briefly at the end of this section. In particular
the very recent analysis of Ali, Kramer and L\"u \cite{AKL98} 
is very informative.
\subsection{Generalized Factorization}
In the generalized factorization framework the formulae 
(\ref{1})--(\ref{3N}) are simply replaced by
\begin{equation}\label{1a}
A_{\rm I}=\frac{G_F}{\sqrt{2}} V_{CKM}
a^{\rm eff}_1\langle O_1\rangle_F 
\qquad {\rm (Class~I)},
\end{equation}
\begin{equation}\label{2a}
A_{\rm II}=\frac{G_F}{\sqrt{2}} V_{CKM}
a^{\rm eff}_2\langle O_2\rangle_F 
\qquad {\rm (Class~II)},
\end{equation}
\begin{equation}\label{3a}
A_{\rm III}=
\frac{G_F}{\sqrt{2}} V_{CKM}
[a^{\rm eff}_1+x a^{\rm eff}_2]\langle O_1\rangle_F
 \qquad {\rm (Class~III)},
\end{equation}
where $a^{\rm eff}_i$ are $\mu$-independent and renormalization
scheme independent parameters to be extracted from experimental data.
From phenomenological point of view there is no change here
relative to the standard factorization as only $a_i(\mu)$ have
been replaced by $a^{\rm eff}_i$. On the other hand, as stressed 
in particular in \cite{NS97}, the new formulation should allow in 
principle some insight into the importance of non-factorizable
contributions.  

In this context I should remark that in the recent literature
mainly the $\mu$-dependence of the non-factorizable contributions
has been emphasized. Their scheme dependence has  been only discussed
in \cite{AJB94a}. It is the latter issue which will be important in
the discussion below. Let us then derive the formulae for
$a_i^{\rm eff}$ including NLO corrections. 

In order to describe generalized factorization in explicit terms
let us consider the decay
$\bar B^0\to D^+\pi^-$. Then the
relevant effective Hamiltonian is given by
\begin{equation}\label{BS1}
H_{eff}=\frac{G_F}{\sqrt{2}}V_{cb}V_{ud}^{*}
\lbrack \bar C_1(\mu) O_1+\bar C_2(\mu)O_2 \rbrack~,
\end{equation}
where
\begin{equation}\label{BS2}
O_1=(\bar d_\alpha u_\alpha)_{V-A} (\bar c_\beta b_\beta)_{V-A}
\qquad 
O_2=(\bar d_\alpha u_\beta)_{V-A} (\bar c_\beta b_\alpha)_{V-A}~.
\end{equation}
$\bar C_1(\mu)$ and $\bar C_2(\mu)$ are 
computed at the renormalization scale $\mu=O(m_b)$.
Since all four quark flavours entering the operators in (\ref{BS2})
are different from each other, no penguin operators contribute to
this decay. 

Using Fierz reordering and colour identities one can rewrite the
amplitude for $\bar B^0\to D^+\pi^-$ as
\begin{equation}\label{BS3}
A(\bar B^0\to D^+\pi^-)=\frac{G_F}{\sqrt{2}}V_{cb}V_{ud}^{*}
a^{\rm eff}_1 \langle O_1\rangle_F
\end{equation}
where
\be
\langle O_1\rangle_F=
\langle\pi^-\mid(\bar d u)_{V-A}\mid 0\rangle
\langle D^+\mid (\bar c b)_{V-A}\mid \bar B^0\rangle
\ee
is the factorized matrix element of the operator $O_1$ and
summation over colour indices in each current is understood.

The effective parameter $a^{\rm eff}_1$ is then given by \cite{NS97}
\be\label{BS4}
a^{\rm eff}_1=\left(\bar C_1(\mu)+\frac{1}{N} \bar C_2(\mu)\right)
[1+\varepsilon_1^{(BD,\pi)}(\mu)]
+\bar C_2(\mu)\varepsilon_8^{(BD,\pi)}(\mu).
\ee
$\varepsilon_1^{(BD,\pi)}(\mu)$ and $\varepsilon_8^{(BD,\pi)}(\mu)$
are two hadronic parameters defined by
\be\label{BS5}
\varepsilon_1^{(BD,\pi)}(\mu)\equiv
\frac{\langle \pi^-D^+|(\bar d u)_{V-A}(\bar c b)_{V-A}|\bar B^0\rangle}
{\langle O_1 \rangle_F}-1
\ee
and
\be\label{BS6}
\varepsilon_8^{(BD,\pi)}(\mu)\equiv 2
\frac{\langle \pi^-D^+|(\bar dT^a u)_{V-A}
(\bar cT^a b)_{V-A}|\bar B^0\rangle}
{\langle O_1\rangle_F}
\ee
with $T^a$ denoting the colour matrices in the standard Feynman
rules. $\varepsilon_i(\mu)$ parametrize the non-factorizable 
contributions to
the hadronic matrix elements of operators. In the case of strict
factorization $\varepsilon_i$ vanish and $a_1^{\rm eff}$ reduces
to $a_1(\mu)$.

It should be emphasized that no approximation has been made
in (\ref{BS3}). Since the matrix element $\langle O_1 \rangle_F$
is scale and renormalization scheme independent this must also
be the case for the effective coefficient $a^{\rm eff}_1$.
Indeed the scale and scheme dependences of the coefficients
$\bar C_1(\mu)$ and $\bar C_2(\mu)$ are cancelled by those present in
the hadronic parameters $\varepsilon_i(\mu)$. We will give
explicit formulae for the latter dependences below.

A similar exercise with the
amplitude for $\bar B^0\to D^0\pi^0$ gives
\begin{equation}\label{BS7}
A(\bar B^0\to D^0\pi^0)=\frac{G_F}{\sqrt{2}}V_{cb}V_{ud}^{*}
a^{\rm eff}_2 \langle O_2 \rangle_F,
\end{equation}
where
\be\label{fact2}
\langle O_2 \rangle_F=
\langle D^0\mid(\bar c u)_{V-A}\mid 0\rangle
\langle \pi^0\mid (\bar d b)_{V-A}\mid \bar B^0\rangle
\ee
is the factorized matrix element of the operator $O_2$.

The effective parameter $a^{\rm eff}_2$ is given by \cite{NS97}
\be\label{BS8}
a^{\rm eff}_2=\left(\bar C_2(\mu)+\frac{1}{N} \bar C_1(\mu)\right)
[1+\varepsilon_1^{(B\pi,D)}(\mu)]+
\bar C_1(\mu)\varepsilon_8^{(B\pi,D)}(\mu).
\ee
$\varepsilon_1^{(B\pi,D)}(\mu)$ and $\varepsilon_8^{(B\pi,D)}(\mu)$
are two hadronic parameters defined by
\be\label{BS9}
\varepsilon_1^{(B\pi,D)}(\mu)\equiv
\frac{\langle \pi^0D^0|(\bar c u)_{V-A}(\bar d b)_{V-A}|\bar B^0\rangle}
{\langle O_2 \rangle_F}-1
\ee
and
\be\label{BS10}
\varepsilon_8^{(B\pi,D)}(\mu)\equiv 2
\frac{\langle \pi^0D^0|(\bar c T^a u)_{V-A}
(\bar d T^a b)_{V-A}|\bar B^0\rangle}
{\langle O_2 \rangle_F}~.
\ee
Again the $\mu$ and scheme dependences of $\varepsilon_i$ in
(\ref{BS9}) and (\ref{BS10}) cancel the corresponding
dependences in $\bar C_i(\mu)$ so that the effective coefficient
$a^{\rm eff}_2$ is $\mu$ and scheme independent.
Similarly one can derive the formula (\ref{3a}) by using
$B^-\to D^0 K^-$ or other decay belonging to class III.

Following section 5.1 of \cite{BJLW} and using the experience
accumulated in previous sections it is straightforward to
find the explicit $\mu$ and scheme dependences of the hadronic
parameters $\varepsilon_i(\mu)$. To this end we note
that the $\mu$ dependence of the matrix elements of the
operators $O_\pm$ is given by  
\be\label{BS110}
\langle O_\pm(\mu)\rangle = U_\pm(\mb,\mu) \langle O_\pm(\mb)\rangle~,
\ee
where the evolution function $U_\pm(\mb,\mu)$ 
including NLO QCD corrections is given as in (\ref{B9P}) by 
\begin{equation}\label{BS11}
U_\pm(\mb,\mu)=\left[1+\frac{\alpha_s(\mb)}{4\pi}J_\pm\right]
      \left[\frac{\alpha_s(\mu)}{\alpha_s(\mb)}\right]^{d_\pm}
\left[1-\frac{\alpha_s(\mu)}{4\pi}J_\pm\right]
\end{equation}
with $J_\pm$ and $d_\pm$ in (\ref{B10}). Note the different ordering
of scales in (\ref{BS110}) from the one in the evolution of
Wilson coefficients in (\ref{EVOLC}).

Having these formulae at hand it is straightforward to show
that the $\mu$-dependence of $\varepsilon_1(\mu)$ and
$\varepsilon_8(\mu)$ is governed by the following equations:
\begin{eqnarray}\label{BS20}
1+\varepsilon_1(\mu)&=&
\frac{1}{2}
\left[\left(1+\frac{1}{N}\right)[1+\R1(\mb)]+\E8(\mb)\right]
U_+(\mb,\mu)\\
&+&
\frac{1}{2}
\left[\left(1-\frac{1}{N}\right)[1+\R1(\mb)]-\E8(\mb)\right]
U_-(\mb,\mu),   \nonumber
\end{eqnarray}

\begin{eqnarray}\label{BS21}
\varepsilon_8(\mu)&=&
\frac{1}{2}
\left[\left(1-\frac{1}{N}\right)\E8(\mb)+
\left(1-\frac{1}{N^2}\right)[1+\R1(\mb)]\right]
U_+(\mb,\mu)\\
&+&
\frac{1}{2}
\left[\left(1+\frac{1}{N}\right)\E8(\mb)-
\left(1-\frac{1}{N^2}\right)[1+\R1(\mb)]\right]
U_-(\mb,\mu).
 \nonumber
\end{eqnarray}
It is a very good exercise to derive these formulae and any student
who wants to test her (his) skills in this field should try it.

These formulae reduce to the ones given in \cite{NS97} when $J_\pm$ in
(\ref{BS11}) are set to zero. They give both the $\mu$-dependence
and renormalization scheme dependence of $\varepsilon_i$. The
latter dependence has not been considered in \cite{NS97}.
We will return to these expressions in a moment. First, however,
we would like to formulate the generalized factorization in a
more transparent manner.
\subsection{A Different Formulation}
In order to be able to discuss the relation of our presentation
\cite{BUSI} to the
one of \cite{NS97} we have used until now, as in \cite{NS97}, the
hadronic parameters $\R1(\mu)$ and $\E8(\mu)$ to describe
non-factorizable contributions. 
It appears to us that it is more convenient to work instead
with two other parameters defined simply by \cite{BUSI}
\be\label{BS22}
a^{\rm eff}_1=a_1(\mu)+\xi^{\rm NF}_1(\mu),
\quad\quad
a^{\rm eff}_2=a_2(\mu)+\xi^{\rm NF}_2(\mu),
\ee
where $a_i(\mu)$ are defined in (\ref{BS23}).
Comparison with (\ref{BS4}) and (\ref{BS8}) gives
\be\label{xi1}
\xi^{\rm NF}_1(\mu)=\R1(\mu) a_1(\mu)+\E8(\mu) \bar C_2(\mu),
\ee
\be\label{xi2}
\xi^{\rm NF}_2(\mu)=\bar\R1(\mu) a_2(\mu)+\bar\E8(\mu) \bar C_1(\mu),
\ee
where
\be\label{E18}
\R1(\mu)=\R1^{(BD,\pi)}, \qquad \E8(\mu)=\E8^{(BD,\pi)},
\ee
\be\label{E19}
\bar\R1(\mu)=\R1^{(B\pi,D)}, \qquad \bar\E8(\mu)=\E8^{(B\pi,D)}. 
\ee

In the framework of the
strict factorization hypothesis $\xi^{\rm NF}_i(\mu)$
are set to zero. Their
$\mu$ and scheme dependences can in principle
be found by using the dependences of $\bar C_i(\mu)$ given in
section 7
and of $\varepsilon_i(\mu)$ in (\ref{BS20}) and (\ref{BS21}).
To this end, however, one needs the determination of the
non-perturbative parameters $\varepsilon_i(\mu)$ and
$\bar\varepsilon_i(\mu)$ at a single
value of $\mu$. If, as done in \cite{NS97}, $a_i^{\rm eff}$
are universal parameters, the determination of $\varepsilon_i(\mu)$
and $\bar\varepsilon_i(\mu)$ is only possible if one also
makes the following {\it universality} assumptions:
\be\label{FACTU}
\R1(\mu)=\bar\R1(\mu), \qquad \E8(\mu)=\bar\E8(\mu).
\ee
In \cite{NS97} such an assumption was unnecessary as $\R1(\mu)$
has been set to zero and only $\E8(\mu)$ has been extracted
from the data.

With the assumptions in (\ref{FACTU}), $\R1(\mu)$ and $\E8(\mu)$
can indeed be found once the effective parameters $a_i^{\rm eff}$
have been determined experimentally. Using (\ref{BS4}) and
(\ref{BS8}) together with (\ref{FACTU}) we find
\be\label{E1MU}
\R1(\mu)=\frac{\bar C_1(\mu) a_1^{\rm eff}-\bar C_2(\mu)a_2^{\rm eff}}
          {\bar C^2_1(\mu)-\bar C^2_2(\mu)}-1~,
\ee
\be\label{E2MU}
\E8(\mu)=\frac{a_2^{\rm eff}}{\bar C_1(\mu)}-
\left(\frac{\bar C_2(\mu)}{\bar C_1(\mu)}+\frac{1}{N}\right) 
[1+\R1(\mu)]~.
\ee
On the other hand  $\xi_i^{\rm NF}(\mu)$ can be determined without
the universality assumption (\ref{FACTU}) from two decays simply
as follows
\be\label{BS24}
 \xi^{\rm NF}_1(\mu)=a^{\rm eff}_1-a_1(\mu)~,
\quad\quad
 \xi^{\rm NF}_2(\mu)=a^{\rm eff}_2-a_2(\mu)~.
\ee

Formulae in (\ref{BS24}) make it clear that the strict factorization
in which $\xi_i^{\rm NF}(\mu)$ vanish can be at best correct at a single
value of $\mu$, the so-called factorization scale $\mu_f$. In the
first studies of factorization $\mu_f=\mb$ 
has been assumed. It has been concluded that such a choice is
not in accord with the data.
The idea of the generalized factorization as formulated in
\cite{Cheng,Soares,NS97}
is to allow $\mu_f$ to be different from $\mb$ and to extract
first non-factorizable parameters $\varepsilon_i(\mb)$ from the
data. Subsequently factorization scale $\mu_f$ can be found
by requiring these parameters to vanish.

In the numerical analysis of this procedure done in \cite{NS97}
one additional assumption has been made. Using large $N$ arguments
it has been argued that $\R1(\mu)$ can be set to zero while
$\E8(\mu)$ can be sizable. The resulting expressions
for $a_i^{\rm eff}$ are then
\be\label{NSF}
a_1^{\rm eff}=\bar C_1(\mb), 
\qquad a_2^{\rm eff}=a_2(\mb)+\bar C_1(\mb)\E8(\mb)~,
\ee
where additional small terms have been dropped in order to obtain
the formula for $a_1^{\rm eff}$. Using subsequently
the extracted value $a_2^{\rm eff}=0.21\pm0.05$ together with
the  coefficients $\bar C_i(\mb)$ from \cite{WEISZ} one finds
$\E8(\mb)=0.12\pm0.05$ \cite{NS97}. Next assuming $\E8(\mu_f)=0$ one
can find the factorization scale $\mu_f$
by inverting the formula \cite{NS97}
\be\label{UF}
\E8(\mb)=-\frac{4\as(\mb)}{3\pi}\ln\frac{\mb}{\mu_f}~,
\ee
which follows from (\ref{BS21}) with $\E8(\mu_f)=0$ and $\R1(\mb)=0.$
Thus
\be\label{UF1}
 \mu_f=\mb \exp\left[\frac{3\pi\E8(\mb)}{4\as(\mb)}\right]~.
\ee
Taking $\mb=4.8~\gev$ and $\as(m_b)=0.21$ 
(corresponding to $\as(\mz)=0.118$) we find using $\E8(\mb)=0.12\pm0.05$
a rather large factorization scale $\mu_f=(15.9+11.3-6.6)~\gev$,
by roughly a factor of 3-4 higher than $\mb$. This implies that
non-factorizable contributions in hadronic matrix elements at scales
close to $\mb$ are sizable. This is also signalled by the 
value of $\E8(\mb)\approx0.12$ which is larger than the factorizable
contribution $a_2(\mb)=0.09$ to the effective parameter
$a_2^{\rm eff}=0.21\pm0.05$.

We would like to emphasize that such an interpretation of the
analysis of Neubert and Stech \cite{NS97} would be 
misleading. As stressed in \cite{AJB94a} the coefficient $a_2(\mu)$
is very strongly dependent on the renormalization scheme.
Consequently for a given value of $a_2^{\rm eff}$ also
$\xi_2^{NF}(\mb)$ and $\E8(\mb)$ are strongly scheme dependent.
This shows, that a meaningful analysis of the
$\mu$-dependences in non-leptonic decays, such as the search for the
factorization scale $\mu_f$, cannot be be made without simultaneously
considering the scheme dependence. This is evident if one recalls that
any variation of $\mu_f$ in the leading logarithm is equivalent to
a shift in constant non-logarithmic terms. The latter represent
NLO contributions in the renormalization group improved
perturbation theory and must be included for a meaningful extraction
of $\mu_f$ or any other scale like $\Lms$. However, once  the NLO
contributions are taken into account, the renormalization scheme
dependence enters the analysis and consequently the factorization
scale $\mu_f$ at which the non-factorizable hadronic parameters
$\xi_i^{NF}(\mu_f)$ or $\varepsilon_i(\mu_f)$ vanish is renormalization
scheme dependent. Formula (\ref{21}) exhibits all these statements
very clearly.
  
From this discussion it becomes clear that for any chosen scale
$\mu_f=\ord(\mb)$, it is always possible to find a renormalization
scheme for which
\be\label{xifac}
\xi_1^{NF}(\mu_f)=\xi_2^{NF}(\mu_f)=0~.
\ee
Indeed as seen in (\ref{BS24}) $\xi_i^{NF}(\mu)$ depend through
$a_i(\mu)$ on $\kappa_\pm$ (see section 7)
which characterize a given renormalization scheme. The choice
of $\kappa_\pm$
corresponds to a particular finite renormalization of the operators
$O_\pm$ in addition to the renormalization in the NDR scheme. It
is then straightforward to find the values of $\kappa_\pm$ which
assure that for a chosen scale $\mu_f$ the conditions in (\ref{xifac})
are satisfied. We find \cite{BUSI}
\be\label{kappa+}
\kappa_+=3
\left[\frac{3}{4}\frac{a_1^{\rm eff}+a_2^{\rm eff}}{W_+(\mu_f)}-1\right] 
\frac{4\pi}{\as(\mu_f)}-3 (J_+)_{\rm NDR}~,
\ee
\be\label{kappa-}
\kappa_-=\frac{3}{2}
\left[\frac{3}{2}\frac{a_1^{\rm eff}-a_2^{\rm eff}}{W_-(\mu_f)}-1\right] 
\frac{4\pi}{\as(\mu_f)}-\frac{3}{2} (J_-)_{\rm NDR}~,
\ee
where
\be\label{W+-}
W_\pm(\mu_f)=\left[\frac{\alpha_s(M_W)}{\alpha_s(\mu_f)}\right]^{d_\pm}
\left[1+\frac{\alpha_s(M_W)}{4\pi}(B_\pm-J_\pm)\right]
\ee
with $(J_\pm)_{\rm NDR}$ being the values of $J_\pm$ in the NDR scheme.
$W_\pm(\mu_f)$ are clearly renormalization scheme independent as
$B_\pm-J_\pm$ and $d_\pm$ are scheme independent.
\subsection{Numerical Analysis}
Before presenting the numerical analysis of the formulae derived in
the preceding subsections, it is important  to clarify the difference
between the Wilson coefficients in (\ref{10N}) used by
us and the ones employed in \cite{NS97}. In \cite{NS97} scheme
independent coefficients $\tilde z_\pm(\mu)$ of \cite{WEISZ}  
instead of $z_\pm(\mu)$ have been used. These are obtained by
multiplying $z_\pm(\mu)$ by $(1-B_\pm \alpha_s(\mu)/4\pi)$ 
so that
\begin{equation}\label{T11}
\tilde z_\pm(\mu)=
      \left[\frac{\alpha_s(M_W)}{\alpha_s(\mu)}\right]^{d_\pm}
\left[1+\frac{\alpha_s(M_W)-\alpha_s(\mu)}{4\pi}(B_\pm-J_\pm)\right].
\end{equation}
These coefficients are clearly not the coefficients of the operators
$O_\pm$. 
In order to be consistent, the matrix elements $\langle O_\pm \rangle$
should then be replaced by
\be\label{TOPM}
\langle \tilde O_\pm \rangle=
(1+B_\pm \alpha_s(\mu)/4\pi) \langle O_\pm \rangle.
\ee
This, however, has not been done in \cite{NS97}. This explains, to a large
extent, why our results for $\varepsilon_8(\mb)$  differ considerably from 
the ones quoted in \cite{NS97}.
We strongly advice the practitioners of
non-leptonic decays not to use the scheme independent coefficients
of \cite{WEISZ} in phenomenological applications. These coefficients
have been introduced to test the compatibility of different 
renormalization schemes and can only be used for phenomenology
together with $\langle \tilde O_\pm \rangle$. This would however
unnecessarily complicate the analysis and it is therefore advisable
to  work with the true coefficients $\bar C_i(\mu)$ of the operators $O_i$ 
as given in (\ref{10N}).

In \cite{NS97} the values of $a_i^{\rm eff}$ given in (\ref{8N}) 
have been
extracted from existing data on two-body B--decays.
In order to illustrate various points made until now,
we take the central values of $a_i^{\rm eff}$ in (\ref{8N})
and calculate $\varepsilon_i(\mu)$
and $\xi_i^{NF}(\mu)$ as  functions of $\mu$ in the range
$2.5~\gev \le \mu \le 10~\gev$ for the NDR and HV schemes. The
results are shown in fig.~\ref{SILV1}  and fig.~\ref{SILV2}.
We observe that $\varepsilon_1(\mu)$ and 
$\xi^{\rm NF}_1(\mu)$ are only weakly $\mu$ and scheme
dependent in accordance with the findings in \cite{AJB94a},
where these dependences have been studied for $a_i(\mu)$
defined in (\ref{BS23}). The strong $\mu$ and scheme dependences
of $a_2(\mu)$ found there translate into similar strong dependences
of $\varepsilon_8(\mu)$ and $\xi_2^{\rm NF}(\mu)$.

\begin{figure}   
    \begin{center}
\setlength{\unitlength}{0.1bp}
\begin{picture}(3600,2160)(0,0)
\special{psfile=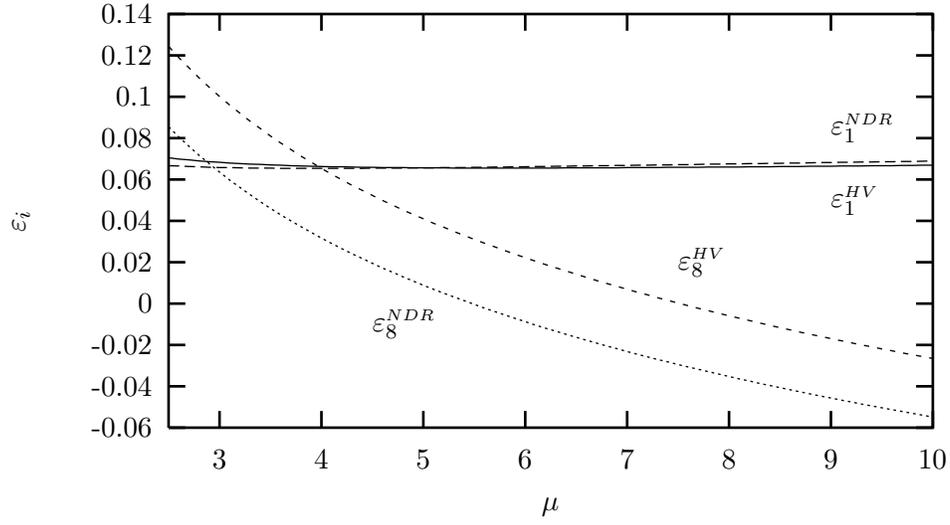 llx=0 lly=0 urx=720 ury=504 rwi=7200}
\put(1428,870){\makebox(0,0)[l]{$\varepsilon_8^{\scriptscriptstyle NDR}$}}
\put(2580,1104){\makebox(0,0)[l]{$\varepsilon_8^{\scriptscriptstyle HV}$}}
\put(3156,1611){\makebox(0,0)[l]{$\varepsilon_1^{\scriptscriptstyle NDR}$}}
\put(3156,1338){\makebox(0,0)[l]{$\varepsilon_1^{\scriptscriptstyle HV}$}}
\put(2100,180){\makebox(0,0){$\mu$}}
\put(120,1260){%
\special{ps: gsave currentpoint currentpoint translate
270 rotate neg exch neg exch translate}%
\makebox(0,0)[b]{\shortstack{$\varepsilon_i$}}%
\special{ps: currentpoint grestore moveto}%
}
\put(3540,360){\makebox(0,0){10}}
\put(3156,360){\makebox(0,0){9}}
\put(2772,360){\makebox(0,0){8}}
\put(2388,360){\makebox(0,0){7}}
\put(2004,360){\makebox(0,0){6}}
\put(1620,360){\makebox(0,0){5}}
\put(1236,360){\makebox(0,0){4}}
\put(852,360){\makebox(0,0){3}}
\put(600,2040){\makebox(0,0)[r]{0.14}}
\put(600,1884){\makebox(0,0)[r]{0.12}}
\put(600,1728){\makebox(0,0)[r]{0.1}}
\put(600,1572){\makebox(0,0)[r]{0.08}}
\put(600,1416){\makebox(0,0)[r]{0.06}}
\put(600,1260){\makebox(0,0)[r]{0.04}}
\put(600,1104){\makebox(0,0)[r]{0.02}}
\put(600,948){\makebox(0,0)[r]{0}}
\put(600,792){\makebox(0,0)[r]{-0.02}}
\put(600,636){\makebox(0,0)[r]{-0.04}}
\put(600,480){\makebox(0,0)[r]{-0.06}}
\end{picture}
    \end{center}
    \caption[]{$\varepsilon_{1,8}(\mu)$ in the NDR and HV schemes.}
    \label{SILV1}
\end{figure}

\begin{figure}   
    \begin{center}
\setlength{\unitlength}{0.1bp}
\begin{picture}(3600,2160)(0,0)
\special{psfile=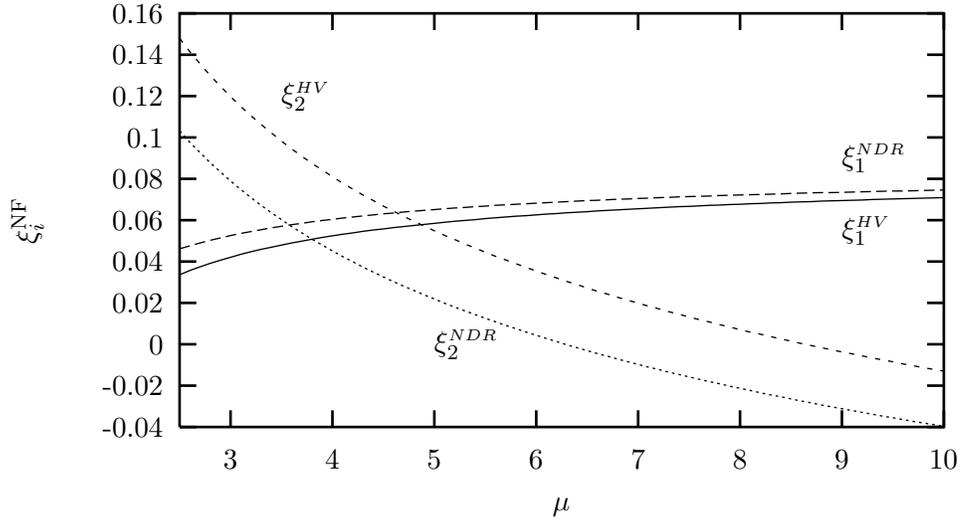 llx=0 lly=0 urx=720 ury=504 rwi=7200}
\put(1620,792){\makebox(0,0)[l]{$\xi_2^{\scriptscriptstyle NDR}$}}
\put(1044,1728){\makebox(0,0)[l]{$\xi_2^{\scriptscriptstyle HV}$}}
\put(3156,1494){\makebox(0,0)[l]{$\xi_1^{\scriptscriptstyle NDR}$}}
\put(3156,1221){\makebox(0,0)[l]{$\xi_1^{\scriptscriptstyle HV}$}}
\put(2100,180){\makebox(0,0){$\mu$}}
\put(120,1260){%
\special{ps: gsave currentpoint currentpoint translate
270 rotate neg exch neg exch translate}%
\makebox(0,0)[b]{\shortstack{$\xi^{\rm NF}_i$}}%
\special{ps: currentpoint grestore moveto}%
}
\put(3540,360){\makebox(0,0){10}}
\put(3156,360){\makebox(0,0){9}}
\put(2772,360){\makebox(0,0){8}}
\put(2388,360){\makebox(0,0){7}}
\put(2004,360){\makebox(0,0){6}}
\put(1620,360){\makebox(0,0){5}}
\put(1236,360){\makebox(0,0){4}}
\put(852,360){\makebox(0,0){3}}
\put(600,2040){\makebox(0,0)[r]{0.16}}
\put(600,1884){\makebox(0,0)[r]{0.14}}
\put(600,1728){\makebox(0,0)[r]{0.12}}
\put(600,1572){\makebox(0,0)[r]{0.1}}
\put(600,1416){\makebox(0,0)[r]{0.08}}
\put(600,1260){\makebox(0,0)[r]{0.06}}
\put(600,1104){\makebox(0,0)[r]{0.04}}
\put(600,948){\makebox(0,0)[r]{0.02}}
\put(600,792){\makebox(0,0)[r]{0}}
\put(600,636){\makebox(0,0)[r]{-0.02}}
\put(600,480){\makebox(0,0)[r]{-0.04}}
\end{picture}
    \end{center}
    \caption[]{$\xi^{\rm NF}_{1,2}(\mu)$ in the NDR and HV schemes.}
    \label{SILV2}
\end{figure}

We make the following observations:
\bi
\item
$\varepsilon_1(\mu)$ and $\xi_1^{\rm NF}(\mu)$ are non-zero in the
full range of $\mu$ considered.
\item
$\varepsilon_8(\mu)$ and $\xi_2^{\rm NF}(\mu)$ vary strongly with
$\mu$ and vanish in the NDR scheme for $\mu=5.5~\gev$ and 
$\mu=6.3~\gev$
respectively. The corresponding values in the HV scheme are
$\mu=7.5~\gev$ and $\mu=8.6~\gev$.
\item
There is no value of $\mu=\mu_f$ in the full range considered for
which $\varepsilon_1(\mu)$ and $\varepsilon_8(\mu)$ or equivalently
$\xi_1^{\rm NF}(\mu)$ and $\xi_2^{\rm NF}(\mu)$ simultaneously
vanish. We also observe contrary to expectations in \cite{NS97}
that $\varepsilon_1(\mu)$ is not necessarily smaller than
$\varepsilon_8(\mu)$. In fact the large $N$ arguments presented
in \cite{NS97} that $\varepsilon_1(\mu)=\ord(1/N^2)$ and
$\varepsilon_8(\mu)=\ord(1/N)$, imply strictly speaking only
that the $\mu$-dependence of $\varepsilon_8(\mu)$
is much stronger than that of $\varepsilon_1(\mu)$,
which we indeed see in figs.~\ref{SILV1}  and \ref{SILV2}.
 The hierarchy of their actual
values is a dynamical question. Even if the large $N$-counting-rules
$\varepsilon_1(\mu)=\ord(1/N^2)$ and
$\varepsilon_8(\mu)=\ord(1/N)$ are true independently of the
factorization hypothesis \cite{EW,BGR}, it follows from our analysis
that once the generalized factorization hypothesis is made, the
extracted values of $\varepsilon_i$ violate for some range of $\mu$
the large-N rule $\varepsilon_1\ll\varepsilon_8$. 
\ei

We can next investigate for which renormalization scheme characterized
by $\kappa_\pm$ the factorization is exact at $\mu_f=\mb=4.8~\gev$.
We call this choice the ``factorization scheme" (FS).
Using the central values in (\ref{8N}) and $\Lms^{(5)}=225\mev$
we find by means of
(\ref{kappa+}) and (\ref{kappa-})
\be\label{KPKM}
\kappa_+=13.5~, \qquad  \kappa_-=3.9~~~~~~~({\rm FS}).
\ee
These values deviate considerably from the NDR values $\kappa_\pm=0$
and the HV values $\kappa_\pm=\mp 4 $. Yet one can verify that for these
values $J_+=6.13 $ and $J_-=1.17$ and consequently in this scheme the NLO
corrections at $\mu=\mb$ remain perturbative.
In table \ref{tabf} we give the values of 
$\xi_i^{\rm NF}(\mu)$ for the NDR,HV and FS schemes. 

The discussion of this subsection casts some doubts on the
usefulness of the formulation in \cite{NS97} with respect to the study of
non-factorizable contributions to non-leptonic decays.

\begin{table}[htb]
\caption[]{$\xi^{\rm NF}_{1,2}(\mu)$ as functions of $\mu$
for different schemes and $\Lms^{(5)}=225\mev$.}
\label{tabf}
\begin{center}
\begin{tabular}{|c|c|c|c||c|c|c|}
\hline
& \multicolumn{3}{c||}{$\xi^{\rm NF}_1(\mu)$} &
  \multicolumn{3}{c| }{$\xi^{\rm NF}_2(\mu)$} \\
\hline
$\mu [{\rm GeV}]$ &NDR & HV & FS & NDR & HV & FS  \\
\hline
\hline
2.5 & 0.046 & 0.035 & --0.033 & 0.102 & 0.144 & 0.075 \\
\hline
5.0 & 0.065 & 0.059 &   0.001 & 0.022 & 0.055 &--0.004 \\
\hline
7.5 & 0.071 & 0.067 & 0.014 & --0.016 & 0.013 &--0.041 \\
\hline
10.0 & 0.074 & 0.071 & 0.021 &--0.039 & -0.013 &--0.064 \\
\hline
\end{tabular}
\end{center}
\end{table}

\subsection{Generalized Factorization and $N^{\rm eff}$}
The generalized factorization  presented in \cite{Cheng,GNF,AKL98} is
similar in spirit but includes more dynamics than the formulation
in \cite{NS97}. Unfortunately, as we will demonstrate below, also this 
approach has its weak points. Let us then briefly describe the basic
idea.

As pointed sometime ago in \cite{BJLW1,rome2} and recently
discussed in \cite{Cheng,GNF,AKL98},
it is always possible
to calculate the  scale and scheme dependence of the hadronic matrix 
elements in perturbation theory by simply calculating the matrix elements
of the relevant operators between the quark states. 
Combining these scheme and scale dependent contributions with the
Wilson coefficients $C_i(\mu)$ one obtaines the effective coefficients
$C_i^{\rm eff}$ which are free from these dependences. If one neglects
in addition final state interactions and other possible non-factorizable
contributions the decay amplitudes can be generally written as follows
\begin{equation}\label{ALI}
A=\langle H_{eff}\rangle =\frac{G_F}{\sqrt{2}} V_{CKM}
\lbrack C_1^{\rm eff}\langle O_1\rangle^{\rm tree} +C_2^{\rm eff}
\langle O_2\rangle^{\rm tree}  \rbrack~,
\end{equation}
where $\langle O_i\rangle^{\rm tree}$ denote tree level matrix elements.
The proposal in \cite{Cheng,GNF,AKL98} is to use (\ref{ALI}) and to apply 
the idea of the factorization to the tree level matrix elements.
In this approach then the effective parameters $a_{1,2}^{\rm eff}$
are given by 
\begin{equation}\label{BS23F}
a_1^{\rm eff}=C^{\rm eff}_1+\frac{1}{N^{\rm eff}} C^{\rm eff}_2 \qquad
a_2^{\rm eff}=C^{\rm eff}_2+\frac{1}{N^{\rm eff}} C^{\rm eff}_1
\end{equation}
with analogous expressions for $a_{i}^{\rm eff}$ ($i=3-10$) parametrizing
penguin contributions.
Here
$N^{\rm eff}$  is treated as a phenomenological parameter which
models those non-factorizable contributions to the hadronic matrix elements
which have not been included in $C_i^{\rm eff}$.
In particular it has been suggested in \cite{Cheng,GNF,AKL98} that
the values for $N^{\rm eff}$ extracted from the data on two-body
non-leptonic decays should teach us about the pattern of 
non-factorizable contributions.

In particular when calculating the effective coefficients $C_i^{\rm eff}$, 
the authors of  \cite{GNF,AKL98}  have included a
subset of contributions to the perturbative matrix elements, which is
sufficient to cancel the scale and scheme dependence of the Wilson
coefficients.
Unfortunately the results of such calculations are generally gauge
dependent and suffer from the dependence on the infrared regulator
and generally on the assumptions about the external momenta. We
have discussed this already in detail in section 6 but it is instructive
to discuss this briefly once more in the context of the analyses 
in \cite{Cheng,GNF,AKL98}.

The Green function of the renormalized operator $O$,
for a given choice of the ultraviolet regularization (NDR or HV for example), 
a choice of the external momenta $p$ and of the gauge parameter $\lambda$, 
is given by 
\begin{equation}
  \label{eq:matel}
  \Gamma_O^\lambda (p) = 1 + \frac{\alpha_s}{4 \pi} 
\left(-\frac{\gamma^{(0)}}{2}
  \ln(\frac{-p^2}{\mu^2}) + \hat{r} \right),
\end{equation}
with
\begin{equation}
\hat{r}= \hat r^{NDR,HV} + \lambda \hat r^\lambda.
\label{eq:rral}
\end{equation}
The matrices $\hat r^{NDR,HV}$ depend on the choice of the external momenta and
on the ultraviolet regularization, while $\hat r^\lambda$ is 
regularization- and
gauge-independent, but depends on the external momenta. 
It is clearly possible to define a renormalization scheme in which, 
for given external momenta and gauge parameter, $\Gamma_O^\lambda (p)
= 1$, or in other words $\langle O \rangle_{p,\lambda}=\langle O
\rangle^{\rm tree}$ (this corresponds to the RI scheme discussed in
\cite{rome2}). However, the definition of the renormalized 
operators will now depend on the choice of the gauge and of the
external momenta. If one were able, for example by means of lattice
QCD, to compute the matrix element of the operator using the same
renormalization prescription, the dependences on the gauge and on the
external momenta would cancel between the Wilson coefficient and the
matrix element. If, on the contrary, the matrix elements are estimated
using factorization, no trace is kept of the renormalization
prescription and the final result is gauge and infrared dependent. 

In  \cite{GNF,AKL98} scale- and scheme-independent effective
Wilson coefficients $C_i^{\rm eff}$ 
have been obtained by adding to $C_i(\mu)$ the
contributions coming from vertex-type quark matrix elements, denoted
by $\hat r_V$ and $\hat\gamma_V$. In particular
\begin{eqnarray}
  \label{eq:cali}
  C_1^{\rm eff}&=& 
C_1(\mu) + \frac{\alpha_s}{4 \pi}\left( r_V^T + \gamma_V^T
  \log \frac{m_b}{\mu}\right)_{1j} C_j(\mu),\nonumber \\
  C_2^{\rm eff}&=& 
C_2(\mu)  + \frac{\alpha_s}{4 \pi}\left( r_V^T + \gamma_V^T
  \log \frac{m_b}{\mu}\right)_{2j} C_j(\mu)
\end{eqnarray}
where the index $j$ runs through all contributing operators, also
penguin operators considered in \cite{Cheng,GNF,AKL98}.

 It is evident from the above discussion that $\hat r_V$  
depends not only on the
external momenta, but also on the gauge chosen.   
For example, in \cite{GNF,AKL98} the following result for $\hat r_V$ is
quoted:
\begin{equation}
  \label{eq:rvali}
  \hat r_V=\left(
    \begin{array}{cccccc}
      \frac{7}{3} & - 7 & 0 & 0 & 0 & 0\\
      - 7 & \frac{7}{3} & 0 & 0 & 0 & 0\\ 
      0 & 0 & \frac{7}{3} & - 7 & 0 & 0\\ 
      0 & 0 & - 7 & \frac{7}{3} & 0 & 0\\ 
      0 & 0 & 0 & 0 & - \frac{1}{3} & 1\\ 
      0 & 0 & 0 & 0 & - 3 & \frac{35}{3} 
    \end{array}
    \right).
\end{equation}
This result is valid in the Landau gauge ($\lambda=0$); 
in an arbitrary gauge, with the same choice of external momenta used
to obtain (\ref{eq:rvali}) one would get
\begin{equation}
  \label{eq:rvtot}
  \hat r_V = \hat r_V (\lambda=0) + \lambda r_V^\lambda,
\end{equation}
with $\hat r_V (\lambda=0)$ given in (\ref{eq:rvali}) and
\begin{equation}
  \label{eq:rvluca}
  r_V^\lambda=\left(
    \begin{array}{cccccc}
      - \frac{5}{6} & - \frac{3}{2} & 0 & 0 & 0 & 0\\
      - \frac{3}{2} & - \frac{5}{6} & 0 & 0 & 0 & 0\\
      0 & 0 & - \frac{5}{6} & - \frac{3}{2} & 0 & 0\\ 
      0 & 0 & - \frac{3}{2} & - \frac{5}{6} & 0 & 0\\ 
      0 & 0 & 0 & 0 & - \frac{11}{6} & \frac{3}{2}\\
      0 & 0 & 0 & 0 & 0 & \frac{8}{3}
    \end{array}
    \right).
\end{equation}
The expressions for the full 
$10 \times 10$ $\hat r$ matrices in the
NDR and HV schemes and in the Feynman and Landau gauges are given in
\cite{rome2}, for a different choice of the external momenta.
The results for the Landau gauge are given in \cite{BJLW1}, where
also penguin diagrams have been included.

Equation (\ref{eq:rvtot}) shows that the definition of the effective
coefficients advocated in \cite{Cheng,GNF,AKL98}
is gauge-dependent. In addition, it also depends on the choice of the
external momenta.
This implies that the effective number of colors extracted in
\cite{Cheng,GNF,AKL98} is also gauge-dependent, and therefore it
cannot have any physical meaning. 
This finding casts some doubts on the
usefulness of the formulation in \cite{Cheng,GNF,AKL98} with respect
to the study of non-factorizable contributions to non-leptonic decays.

The gauge dependences and infrared dependences discussed here 
appear in any calculation of matrix elements of operators
between quark states necessary in the process of matching of the
full theory onto an effective theory as we have seen in section 6.
 Another example can be
found in \cite{BJW90} where the full gauge dependence of the quark
matrix element of the operator $(\bar s d)_{V-A}(\bar s d)_{V-A}$
has been calculated. However, in the process of matching such
unphysical dependences in the effective theory are cancelled by
the corresponding contributions in the full theory so that the
Wilson coefficients are free of such dependences. Similarly
in the case of inclusive decays of heavy quarks, where the spectator
model can be used, they are cancelled by gluon
bremsstrahlung. In exclusive hadron decays there is no meaningful way to
include such effects in a perturbative framework and one is left
with the gauge and infrared dependences in question.

\subsection{Summary}
In this section we have critically analyzed the hypothesis of
the generalized factorization. While the parametrization of the data
in terms of a set of effective parameters discussed in
\cite{NS97,Cheng,GNF,AKL98},
 may appear to be useful,
we do not think that this approach offers convincing means to
analyze the physics of non-factorizable contributions to
non-leptonic decays. In particular:
\begin{itemize}
\item
The renormalization scheme dependence of the non-factorizable
contributions to hadronic matrix elements precludes the
determination of the factorization scale $\mu_f$.
\item
Consequently for any chosen value of $\mu=\ord(m_b)$ 
it is possible to find a renormalization
scheme for which the non-perturbative parameters $\varepsilon_{1,8}$
used in \cite{NS97} to characterize the size of non-factorizable
contributions vanish. The same applies to 
$\xi^{\rm NF}_{1,2}(\mu)$ introduced in (\ref{BS24}).
\item
We point out that the recent extractions of the effective number
of colours $N^{\rm eff}$ from two-body non-leptonic B-decays,
presented in \cite{Cheng,GNF,AKL98},
while $\mu$ and renormalization scheme independent suffer from
gauge dependences and infrared regulator dependences.
\end{itemize}

Our analysis \cite{BUSI} demonstrates clearly the need for
an approach to non-leptonic decays which goes
beyond the generalized factorization discussed recently in
the literature. Some possibilities are offered by dynamical approaches
like QCD sum rules as recently reviewed in \cite{KR98}. 
However, even a phenomenological approach which does not suffer
from the weak points of factorization discussed here, would
be a step forward. Some ideas in this direction will be presented in
\cite{BUSI2}.

\section{$\eps_K$, $B^0$-$\bar B^0$ Mixing and the Unitarity Triangle}
        \label{sec:epsBBUT}
\setcounter{equation}{0}
\subsection{Preliminaries}
Let us next discuss particle--antiparticle mixing which in the past
 has been of fundamental
importance in testing the Standard Model and often has proven to be an
undefeatable challenge for suggested extensions of this model.
Particle--antiparticle mixing is responsible
for the small mass differences between the mass eigenstates of neutral
mesons. Being an FCNC process it involves heavy quarks in loops
and consequently it is a perfect 
testing ground for heavy flavour physics. Let us just recall that
 from the calculation of the
$K_{\rm L}-K_{\rm S}$ mass difference, Gaillard and Lee \cite{GALE} 
were able to estimate the
value of the charm quark mass before charm discovery. On the
other hand $B_d^0-\bar B_d^0$ mixing \cite{ARGUS} gave the first 
indication of a large top quark mass. 
Finally, particle--antiparticle mixing in the $K^0-\bar K^0$ system
offers within the Standard Model a plausible description of
CP violation in $K_L\to\pi\pi$ discovered in 1964 \cite{CRONIN}. 

In this section we will predominantly discuss  the parameter 
$\varepsilon$ 
describing the {\it indirect} CP violation in the $K$ system and  
the mass differences $\Delta M_{d,s}$  which
describe the size of $B_{d,s}^0-\bar B^0_{d,s}$ mixings. 
In the Standard Model all these phenomena
appear first at the one--loop level and as such they are
sensitive measures of the top quark couplings $V_{ti}(i=d,s,b)$ and 
of the top quark mass. 

We have seen in section 2 that tree level 
decays and the unitarity of the CKM
matrix give us already a good information about $V_{tb}$ and $V_{ts}$:
$V_{tb}\approx 1$ and $\mid V_{ts}\mid\;\approx\;\mid V_{cb}\mid$.
Similarly the value of the top quark mass measured by CDF and D0 (see below)
is known within $\pm 4\%$.
Consequently the
main new information to be gained from the quantities discussed here are 
the values of $|V_{td}|$ and of the phase $\delta=\gamma$ in the CKM matrix.
This will allow us to construct the unitarity triangle which has been 
introduced in subsection 2.3.

\begin{figure}[hbt]
\vspace{0.10in}
\centerline{
\epsfysize=1.5in
\epsffile{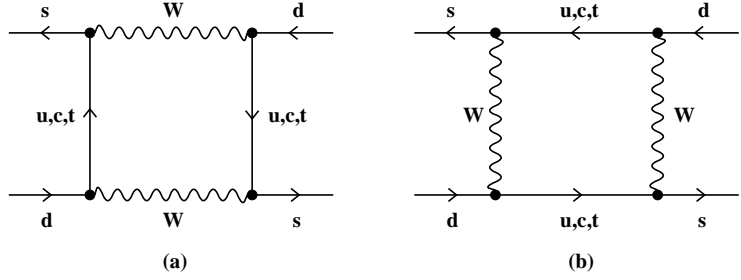}
}
\vspace{0.08in}
\caption[]{Box diagrams contributing to $K^0-\bar K^0$ mixing
in the Standard Model.
\label{L:9}}
\end{figure}

First, however, let us briefly recall the formalism of
particle--antiparticle mixing. We will begin with the $K$--system.
Subsequently we will give
some formulae for $B_{d,s}^0-\bar B^0_{d,s}$ mixings, necessary for the
analysis of the unitarity triangle.
A very detailed discussion of $B_{d,s}^0-\bar B^0_{d,s}$ mixings can
be found in section 8 of a review by Robert Fleischer and myself
\cite{BF97} and in his review \cite{RF97}. The following subsection
borrows a lot from \cite{CHAU83} and \cite{BSSII}.
\subsection{Express Review of $K^0-\bar K^0$ Mixing}
$K^0=(\bar s d)$ and $\bar K^0=(s\bar d)$ are flavour eigenstates which 
in the Standard Model
may mix via weak interactions through the box diagrams in fig.
\ref{L:9}.
We will choose the phase conventions so that 
\be
CP|K^0\rangle=-|\bar K^0\rangle, \qquad   CP|\bar K^0\rangle=-|K^0\rangle.
\ee

In the absence of mixing the time evolution of $|K^0(t)\rangle$ is
given by
\be
|K^0(t)\rangle=|K^0(0)\rangle \exp(-iHt)~, 
\qquad H=M-i\frac{\Gamma}{2}~,
\ee
where $M$ is the mass and $\Gamma$ the width of $K^0$. Similar formula
for $\bar K^0$ exists.

On the other hand, in the presence of flavour mixing the time evolution 
of the $K^0-\bar K^0$ system is described by
\be
i\frac{d\psi(t)}{dt}=\hat H \psi(t) \qquad  
\psi(t)=
\left(\begin{array}{c}
|K^0(t)\rangle\\
|\bar K^0(t)\rangle
\end{array}\right)
\ee
where
\be
\hat H=\hat M-i\frac{\hat\Gamma}{2}
= \left(\begin{array}{cc} 
M_{11}-i\frac{\Gamma_{11}}{2} & M_{12}-i\frac{\Gamma_{12}}{2} \\
M_{21}-i\frac{\Gamma_{21}}{2}  & M_{22}-i\frac{\Gamma_{22}}{2}
    \end{array}\right)
\ee
with $\hat M$ and $\hat\Gamma$ being hermitian matrices having positive
(real) eigenvalues in analogy with $M$ and $\Gamma$. $M_{ij}$ and
$\Gamma_{ij}$ are the transition matrix elements from virtual and physical
intermediate states respectively.
Using
\be
M_{21}=M^*_{12}~, \qquad 
\Gamma_{21}=\Gamma_{12}^*~,\quad\quad {\rm (hermiticity)}
\ee
\be
M_{11}=M_{22}\equiv M~, \qquad \Gamma_{11}=\Gamma_{22}\equiv\Gamma~,
\quad {\rm (CPT)}
\ee
we have
\be\label{MM12}
\hat H=
 \left(\begin{array}{cc} 
M-i\frac{\Gamma}{2} & M_{12}-i\frac{\Gamma_{12}}{2} \\
M^*_{12}-i\frac{\Gamma^*_{12}}{2}  & M-i\frac{\Gamma}{2}
    \end{array}\right)~.
\ee

We can next diagonalize the system to find:

{\bf Eigenstates:}
\be\label{KLS}
K_{L,S}=\frac{(1+\bar\varepsilon)K^0\pm (1-\bar\varepsilon)\bar K^0}
        {\sqrt{2(1+\mid\bar\varepsilon\mid^2)}}
\ee
where $\bar\varepsilon$ is a small complex parameter given by
\be\label{bare3}
\frac{1-\bar\varepsilon}{1+\bar\varepsilon}=
\sqrt{\frac{M^*_{12}-i\frac{1}{2}\Gamma^*_{12}}
{M_{12}-i\frac{1}{2}\Gamma_{12}}}~.
\ee

{\bf Eigenvalues:}
\be
M_{L,S}=M\pm \RE Q  \qquad \Gamma_{L,S}=\Gamma\mp 2 \IM Q
\ee
where
\be
Q=\sqrt{(M_{12}-i\frac{1}{2}\Gamma_{12})(M^*_{12}-i\frac{1}{2}\Gamma^*_{12})}.
\ee
Consequently we have
\be\label{deltak}
\Delta M= M_L-M_S = 2\RE Q
\quad\quad
\Delta\Gamma=\Gamma_L-\Gamma_S=-4 \IM Q.
\ee

It should be noted that the mass eigenstates $K_S$ and $K_L$ differ from 
CP eigenstates
\begin{equation}
K_1={1\over{\sqrt 2}}(K^0-\bar K^0),
  \qquad\qquad CP|K_1\rangle=|K_1\rangle~,
\end{equation}
\begin{equation}
K_2={1\over{\sqrt 2}}(K^0+\bar K^0),
  \qquad\qquad CP|K_2\rangle=-|K_2\rangle~,
\end{equation}
by 
a small admixture of the
other CP eigenstate:
\begin{equation}
K_{\rm S}={{K_1+\bar\varepsilon K_2}
\over{\sqrt{1+\mid\bar\varepsilon\mid^2}}},
\qquad
K_{\rm L}={{K_2+\bar\varepsilon K_1}
\over{\sqrt{1+\mid\bar\varepsilon\mid^2}}}\,
\end{equation}
with $\bar\varepsilon$ defined in (\ref{bare3}).
$\bar\varepsilon$ can also be written as
\be\label{bare}
\frac{1-\bar\varepsilon}{1+\bar\varepsilon}
=\frac{\Delta M-i\frac{1}{2}\Delta\Gamma}
{2 M_{12}-i\Gamma_{12}}\equiv r\exp(i\kappa)~.
\ee

It should be stressed that
the small parameter $\bar\varepsilon$  depends on the 
phase convention
chosen for $K^0$ and $\bar K^0$. Therefore it may not 
be taken as a physical measure of CP violation.
On the other hand $\RE\bar\varepsilon$ and $r$ are independent of
phase conventions. In particular the departure of $r$ from 1
measures CP violation in the $K^0-\bar K^0$ mixing:
\be
r=1+\frac{2 |\Gamma_{12}|^2}{4 |M_{12}|^2+|\Gamma_{12}|^2}
    \IM\left(\frac{M_{12}}{\Gamma_{12}}\right)~.
\ee
Since $\bar\varepsilon$ is $\ord(10^{-3})$, we find, using (\ref{bare3}),
that
\be
\IM M_{12}\ll\RE M_{12}, \quad\quad 
\IM \Gamma_{12}\ll\RE \Gamma_{12}~.
\ee
Consequently to a very good approximation:
\be\label{deltak1}
\Delta M_K = 2 \RE M_{12}, \qquad \Delta\Gamma_K=2 \RE \Gamma_{12}~,
\ee
where we have introduced the subscript K to stress that these formulae apply
only to the $K^0-\bar K^0$ system.

The 
$K_{\rm L}-K_{\rm S}$
mass difference is experimentally measured to be 
\begin{equation}\label{DMEXP}
\Delta M_K=M(K_{\rm L})-M(K_{\rm S}) = 
(3.491\pm 0.009) \cdot 10^{-15} \gev\,.
\end{equation}
In the Standard Model roughly $70\%$ of the measured $\Delta M_K$
is described by the real parts of the box diagrams with charm quark
and top quark exchanges, wherby the contribution of the charm exchanges
is by far dominant. This is related to the smallness of the real parts
of the CKM top quark couplings compared with the corresponding charm
quark couplings. Thus even if the function $S_0(x_t)$ is by a factor
of 1600 larger than $S_0(x_c)$, it cannot compensate for the smallness
of the real top quark couplings. 
Some non-negligible contribution comes from the box diagrams with
simultaneous charm and top exchanges.
The $u$-quark contribution is needed only
for GIM mechanism but otherwise can be neglected.
The remaining $30 \%$ of the measured $\Delta M_K$ is attributed to long 
distance contributions which are difficult to estimate \cite{GERAR}.
It is a useful exercise to check these statements by using 
$\Delta M_K$ in (\ref{deltak}) and the expression for $M_{12}$ given 
in (\ref{eq:M12K}).
Further information with the relevant references can be found in 
\cite{HNa}.

The situation with $\Delta \Gamma_K$ is rather different.
It is fully dominated by long distance effects. Experimentally
one has
\begin{equation}
\Delta \Gamma_K=\Gamma(K_{\rm L})-\Gamma(K_{\rm S}) = 
-7.4 \cdot 10^{-15} \gev\,
\end{equation}
and consequently $\Delta\Gamma_K\approx-2 \Delta M_K$.

With all this information at hand and using  the experimentally 
observed dominance of $\Delta I=1/2$ transitions in $K\to \pi\pi$,
it is possible to derive an important formula for $\bar\varepsilon$
\be\label{basice}
\bar\varepsilon=\frac{i}{1+i}\frac{\IM M_{12}}{\Delta M_K}+
\frac{\xi}{1+i}~, \quad\qquad \xi = \frac{\IM A_0}{\RE A_0}~,
\ee
with the isospin amplitude $A_0$ defined below.
An explicit derivation of (\ref{basice}) can be found in a review
by Chau \cite{CHAU83}. A recent review by Belusevic \cite{BELU} is
also useful in this respect. 
Finally I recommend strongly  excellent lectures
by Yossi Nir \cite{NIRSLAC}, where the issues of phase conventions
are discussed in detail.
\subsection{The First Look at $\varepsilon$ and $\varepsilon'$}
Let us next move to two important CP violating  parameters which 
can be measured experimentally. The route to them proceeds as follows.
It involves the decays $K\to\pi\pi$.

Since a two pion final state is CP even while a three pion final state is CP
odd, $K_{\rm S}$ and $K_{\rm L}$ preferably decay to $2\pi$ and $3\pi$, 
respectively
via the following CP conserving decay modes:
\begin{equation}
K_{\rm L}\to 3\pi {\rm ~~(via~K_2),}\qquad K_{\rm S}\to 2 
\pi {\rm ~~(via~K_1).}
\end{equation}
This difference is responsible for the large disparity in their
life-times. A factor of 579.
However, $K_{\rm L}$ and $K_{\rm S}$ are not CP eigenstates and 
may decay with small branching fractions as follows:
\begin{equation}
K_{\rm L}\to 2\pi {\rm ~~(via~K_1),}\qquad K_{\rm S}\to 3 
\pi {\rm ~~(via~K_2).}
\end{equation}
This violation of CP is called {\it indirect} as it
proceeds not via explicit breaking of the CP symmetry in 
the decay itself but via the admixture of the CP state with opposite 
CP parity to the dominant one.
 The measure for this
indirect CP violation is defined as
\begin{equation}\label{ek}
\varepsilon
={{A(K_{\rm L}\rightarrow(\pi\pi)_{I=0}})\over{A(K_{\rm 
S}\rightarrow(\pi\pi)_{I=0})}},
\end{equation}
where $\varepsilon$ is, contrary to $\bar\varepsilon$ in (\ref{basice}),
independent of the phase conventions.
Following the derivation in \cite{CHAU83} one finds
\be\label{basic2}
\varepsilon=\bar\varepsilon+i \xi~.
\ee
The phase convention dependence of the term $i \xi$ cancells
the convention dependence of $\bar\varepsilon$. We will 
write down a nicer formula for $\varepsilon$ below.

\begin{figure}[hbt]
\centerline{
\epsfysize=1.5in
\epsffile{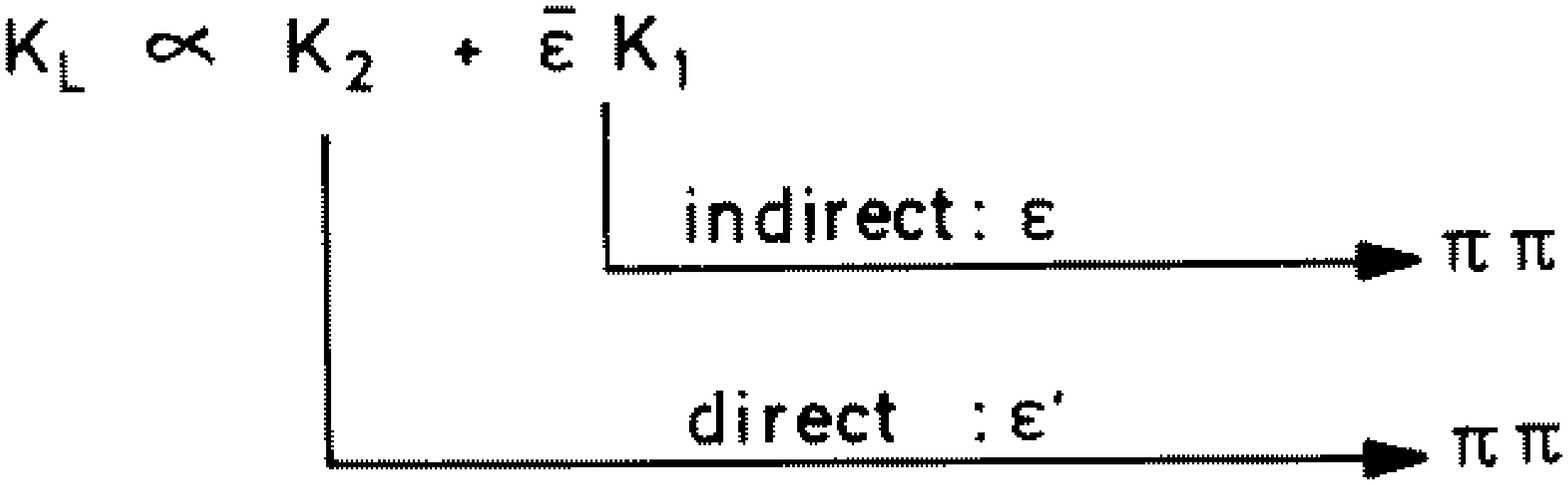}
}
\caption[]{
Indirect versus direct CP violation in $K_L \to \pi\pi$.
\label{fig:14}}
\end{figure}

While {\it indirect} CP violation reflects the fact that the mass
eigenstates are not CP eigenstates, so-called {\it direct}
CP violation is realized via a 
direct transition of a CP odd to a CP even state or vice versa (see
fig.~\ref{fig:14}). 
A measure of such a direct CP violation in $K_L\to \pi\pi$ is characterized
by a complex parameter $\varepsilon'$  defined as
\begin{equation}\label{eprime}
\varepsilon'={1\over {\sqrt 2}}\IM
\left({{A_2}\over{A_0}}\right) e^{i\Phi},\quad\quad
\Phi=\pi/2+\delta_2-\delta_0,
\end{equation}
where
the isospin amplitudes $A_I$ in $K\to\pi\pi$
decays are introduced through
\begin{equation} 
A(K^+\rightarrow\pi^+\pi^0)=\sqrt{3\over 2} A_2 e^{i\delta_2}
\end{equation}
\begin{equation} 
A(K^0\rightarrow\pi^+\pi^-)=\sqrt{2\over 3} A_0 e^{i\delta_0}+ \sqrt{1\over
3} A_2 e^{i\delta_2}
\end{equation}
\begin{equation}
A(K^0\rightarrow\pi^0\pi^0)=\sqrt{2\over 3} A_0 e^{i\delta_0}-2\sqrt{1\over
3} A_2 e^{i\delta_2}\,.
\end{equation} 
Here the subscript $I=0,2$ denotes states with isospin $0,2$
equivalent to $\Delta I=1/2$ and $\Delta I = 3/2$ transitions,
respectively, and $\delta_{0,2}$ are the corresponding strong phases. 
The weak CKM phases are contained in $A_0$ and $A_2$.
The strong phases $\delta_{0,2}$ cannot be calculated, at least, at present.
They can be extracted from $\pi\pi$ scattering. Then
$\Phi\approx \pi/4$.

The isospin amplitudes $A_I$ are complex quantities which depend on
phase conventions. On the other hand, $\varepsilon'$ measures the 
difference between the phases of $A_2$ and $A_0$ and is a physical
quantity.

Experimentally $\varepsilon$ and $\varepsilon'$
can be found by measuring the ratios
\begin{equation}
\eta_{00}={{A(K_{\rm L}\to\pi^0\pi^0)}\over{A(K_{\rm S}\to\pi^0\pi^0)}},
            \qquad
  \eta_{+-}={{A(K_{\rm L}\to\pi^+\pi^-)}\over{A(K_{\rm S}\to\pi^+\pi^-)}}.
\end{equation}
Indeed, assuming $\varepsilon$ and $\varepsilon'$ to be small numbers one
finds
\be
\eta_{00}=\varepsilon-{{2\varepsilon'}\over{1-\sqrt{\omega}}}
            \simeq \varepsilon-2\varepsilon',~~~~
  \eta_{+-}=\varepsilon+{{\varepsilon'}\over{1+\omega/\sqrt{2}}}
            \simeq \varepsilon+ \varepsilon'
\end{equation}
where experimentally $\omega=\RE A_2/\RE A_0=0.045$.

In the absence of direct CP violation $\eta_{00}=\eta_{+-}$.
The ratio ${\varepsilon'}/{\varepsilon}$  can then be measured through
\begin{equation}
\left|{{\eta_{00}}\over{\eta_{+-}}}\right|^2\simeq 1 -6\; 
\RE(\frac{\varepsilon'}{\varepsilon})\,.
\end{equation}
\subsection{Basic Formula for $\eps$}
            \label{subsec:epsformula}
With all this information at hand let us derive a formula for $\varepsilon$
which can be efficiently used in pheneomenological applications.
Using (\ref{basice}) and (\ref{basic2}) we
first find the general formula 
\begin{equation}
\eps = \frac{\exp(i \pi/4)}{\sqrt{2} \Delta M_K} \,
\left( \IM M_{12} + 2 \xi \RE M_{12} \right),
\quad\quad
\xi = \frac{\IM A_0}{\RE A_0}.
\label{eq:epsdef}
\end{equation}
The two terms in (\ref{eq:epsdef}) are separately phase convention
dependent but there sum is free from this dependence.
The off-diagonal 
element $M_{12}$ in
the neutral $K$-meson mass matrix represents $K^0$-$\bar K^0$
mixing. It is given by
\begin{equation}
2 m_K M^*_{12} = \langle \bar K^0| \Heff(\Delta S=2) |K^0\rangle\,,
\label{eq:M12Kdef}
\end{equation}
where $\Heff(\Delta S=2)$ is the effective Hamiltonian for the 
$\Delta S=2$ transitions.
That $ M^*_{12}$ and not $ M_{12}$ stands on the l.h.s of this formula,
is evident from (\ref{MM12}). The factor $2 m_K$ reflects our normalization
of external states.

To lowest order in electroweak interactions $\Delta S=2$ transitions 
are induced
through the box diagrams of fig. \ref{L:9}. Including
 QCD corrections in the manner analogous to the one already discussed for
$\Delta B=2 $ transitions in Section 8.3 one has \cite{BJW90}
\begin{eqnarray}\label{hds2}
{\cal H}^{\Delta S=2}_{\rm eff}&=&\frac{G^2_{\rm F}}{16\pi^2}M^2_W
 \left[\lambda^2_c\eta_1 S_0(x_c)+\lambda^2_t \eta_2 S_0(x_t)+
 2\lambda_c\lambda_t \eta_3 S_0(x_c, x_t)\right] \times
\nonumber\\
& & \times \left[\as^{(3)}(\mu)\right]^{-2/9}\left[
  1 + \frac{\as^{(3)}(\mu)}{4\pi} J_3\right]  Q(\Delta S=2) + h. c.
\end{eqnarray}
where
$\lambda_i = V_{is}^* V_{id}^{}$. Here
$\mu<\mu_c=\ord(m_c)$.
In (\ref{hds2}),
the relevant operator
\begin{equation}\label{qsdsd}
Q(\Delta S=2)=(\bar sd)_{V-A}(\bar sd)_{V-A},
\end{equation}
is multiplied by the corresponding coefficient function.
This function is decomposed into a
charm-, a top- and a mixed charm-top contribution.
This form is obtained upon eliminating $\lambda_u$
by means of the unitarity of the CKM matrix and setting $x_u=0$. 
The functions $S_0$  are given in (\ref{S0})--(\ref{BFF}).

Short-distance QCD effects are described through the correction
factors $\eta_1$, $\eta_2$, $\eta_3$ and the explicitly
$\alpha_s$-dependent terms in (\ref{hds2}). 
$\eta_2$ is the analogue of $\eta_B$ discussed in Section 8.3.
The calculation of $\eta_1$ and $\eta_3$ is more involved and
is discussed in \cite{HNa,HNb}.
$\eta_{1-3}$ are defined in analogy to (\ref{ETANLO}). This means that
in $\ord(\as)$ they are independent of the renormalization
scales and the renormalization scheme for the operator $Q(\Delta S)$.
The NLO values of $\eta_i$ are given as follows \cite{HNa,BJW90,HNb}:
\begin{equation}
\eta_1=1.38\pm 0.20,\qquad
\eta_2=0.57\pm 0.01,\qquad
  \eta_3=0.47\pm0.04~.
\end{equation}
The quoted errors reflect the remaining theoretical uncertainties due to
leftover $\mu$-dependences at $\ord(\as^2)$ and $\Lambda_{\overline{MS}}$.
The factor $\eta_1$ plays only a minor role in the analysis of
$\varepsilon$ but its enhanced value through NLO corrections 
is essential for the $K_{\rm L}-K_{\rm S}$ mass difference.
We refer to \cite{HNa} for the discussion of $\Delta M_K$.

Defining, in analogy to (\ref{Def-Bpar0}),  the renormalization group 
invariant parameter $\hat B_K$ by
\begin{equation}
\hat B_K = B_K(\mu) \left[ \alpha_s^{(3)}(\mu) \right]^{-2/9} \,
\left[ 1 + \frac{\alpha_s^{(3)}(\mu)}{4\pi} J_3 \right]
\label{eq:BKrenorm}
\end{equation}
\begin{equation}
\langle \bar K^0| (\bar s d)_{V-A} (\bar s d)_{V-A} |K^0\rangle
\equiv \frac{8}{3} B_K(\mu) F_K^2 m_K^2
\label{eq:KbarK}
\end{equation}
and using (\eqn{hds2}) one finds
\begin{equation}
M_{12} = \frac{G_{\rm F}^2}{12 \pi^2} F_K^2 \hat B_K m_K \mw^2
\left[ {\lambda_c^*}^2 \eta_1 S_0(x_c) + {\lambda_t^*}^2 \eta_2 S_0(x_t) +
2 {\lambda_c^*} {\lambda_t^*} \eta_3 S_0(x_c, x_t) \right],
\label{eq:M12K}
\end{equation}
where $F_K$ is the $K$-meson decay constant and $m_K$
the $K$-meson mass. 

To proceed further we neglect the last term in (\eqn{eq:epsdef}) as it
 constitutes at most a 2\,\% correction to $\eps$. This is justified
in view of other uncertainties, in particular those connected with
$B_K$.
Inserting (\eqn{eq:M12K}) into (\eqn{eq:epsdef}) we find
\begin{equation}
\eps=C_{\eps} \hat B_K \IM\lambda_t \left\{
\RE\lambda_c \left[ \eta_1 S_0(x_c) - \eta_3 S_0(x_c, x_t) \right] -
\RE\lambda_t \eta_2 S_0(x_t) \right\} \exp(i \pi/4)\,,
\label{eq:epsformula}
\end{equation}
where we have used the unitarity relation $\IM\lambda_c^* = {\rm
Im}\lambda_t$ and  have neglected $\RE\lambda_t/\RE\lambda_c
 = \ord(\lambda^4)$ in evaluating $\IM(\lambda_c^* \lambda_t^*)$.
The numerical constant $C_\eps$ is given by
\begin{equation}
C_\eps = \frac{G_{\rm F}^2 F_K^2 m_K \mw^2}{6 \sqrt{2} \pi^2 \Delta M_K}
       = 3.78 \cdot 10^4 \, .
\label{eq:Ceps}
\end{equation}
To this end we have used the experimental value of $\Delta M_K$ 
in (\ref{DMEXP}). In principle we could use the theoretical
value for $\Delta M_K$ but in view of the presence of long
distance contributions it is safer to use the experimental
value. In this context it should be stressed that 
the parameter $\varepsilon$ 
being related to CP violation and top quark physics should
be dominated by short distance contributions and well approximated
by the imaginary parts of the box diagrams.
Consequently the only non-perturbative uncertainty in 
(\ref{eq:epsformula}) resides in $\hat B_K$.

Using the standard parametrization of (\eqn{2.72}) to evaluate ${\rm
Im}\lambda_i$ and $\RE\lambda_i$, setting the values for $s_{12}$,
$s_{13}$, $s_{23}$ and $\mt$ in accordance with experiment
 and taking a value for $\hat B_K$ (see below), one can
determine the phase $\delta$ by comparing (\eqn{eq:epsformula}) with the
experimental value for $\eps$
\begin{equation}\label{eexp}
\varepsilon^{exp}
=(2.280\pm0.013)\cdot10^{-3}\;e^{i{\pi\over 4}}\,.
\end{equation}

Once $\delta$ has been determined in this manner one can find the
apex $(\bar\varrho,\bar\eta)$ of the unitarity triangle
in fig. \ref{fig:utriangle}   by using 
\begin{equation}\label{2.84a} 
\varrho=\frac{s_{13}}{s_{12}s_{23}}\cos\delta,
\qquad
\eta=\frac{s_{13}}{s_{12}s_{23}}\sin\delta
\end{equation}
and
\begin{equation}\label{2.88da}
\bar\varrho=\varrho (1-\frac{\lambda^2}{2}),
\qquad
\bar\eta=\eta (1-\frac{\lambda^2}{2}).
\end{equation}

For a given set ($s_{12}$, $s_{13}$, $s_{23}$,
$\mt$, $\hat B_K$) there are two solutions for $\delta$ and consequently two
solutions for $(\bar\varrho,\bar\eta)$. 
This will be evident from the analysis of the unitarity triangle discussed
in detail below.

Finally we have to say a few words about the non-perturbative
parameter $\hat B_K$. There is a long history of evaluating  this parameter
in various non-perturbative approaches. A short review of older results
can be found in \cite{BF97}.
The present status of quenched lattice calculations has been recently
reviewed by Gupta \cite{GUPTA98}.
The most accurate result for $B_K(2~\gev)$
using lattice method is obtained by JLQCD collaboration  \cite{JLQCD}:
$B_K(2~\gev)=0.628\pm0.042$. 
A similar result has been published by Gupta, Kilcup and
Sharpe \cite{GKS} last year.
The APE collaboration \cite{APE} 
finds $B_K(2~\gev)=0.66\pm0.11$ which is
consistent with JLQCD and GKS. In order to convert these values into
$\hat B_K$ by means of (\ref{eq:BKrenorm}) one has to face the issue
of the choice of the number of flavours $f$. Fortunately the values for
$\hat B_K$ for $f=0$ and $f=3$ corresponding to the JLQCD result,
turn out to be very similar: $\hat B_K=0.87\pm0.06$ and
$\hat B_K=0.84\pm0.06$, respectively. The final present lattice value
given by Gupta is then
\be
(\hat B_K)_{\rm Lattice}=0.86\pm0.06\pm0.06
\ee
where the second error is attributed to quenching. The
corresponding result from APE  is $\hat B_K=0.93\pm0.16$.
On the other hand a recent analysis in
the chiral quark model gives surprisingly a value as high as 
$\hat B_K=1.1\pm 0.2$ \cite{BERT97}. In our numerical analysis presented 
below we will use 
\begin{equation}\label{BKT}
\hat B_K=0.75\pm 0.15 \,.
\end{equation}
which is in the ball park of various lattice estimates and
$\hat B_K=0.70\pm 0.10$ from
the $1/N$ approach \cite{BBG0,Bijnens}.
These values are higher than those found using QCD Hadronic Duality
approach   ($\hat B_K=0.39\pm0.10$) \cite{Prades} 
or using the SU(3) symmetry and
PCAC ($\hat B_K=1/3$) \cite{Donoghue}.
  
As we will see below, $\hat B_K\le 0.75 $ requires simultaneously high
values of $|V_{ub}/V_{cb}|$ and $\vcb$ in order to be able to fit
the experimental value of $\varepsilon$.

\subsection{Basic Formula for $B^0$-$\bar B^0$ Mixing}
            \label{subsec:BBformula}
The strength of the $B^0_{d,s}-\bar B^0_{d,s}$ mixings
is described by the mass differences
\begin{equation}
\Delta M_{d,s}= M_H^{d,s}-M_L^{d,s}
\end{equation}
with ``H'' and ``L'' denoting {\it Heavy} and {\it Light} respectively. 
In contrast to $\Delta M_K$ , in this case the long distance contributions
are estimated to be very small and $\Delta M_{d,s}$ is very well
approximated by the relevant box diagrams. 
Moreover, due $m_{u,c}\ll m_t$ 
only the top sector can contribute significantly to 
$B_{d,s}^0-\bar B_{d,s}^0$ mixings.
The charm sector and the mixed top-charm contributions are
entirely negligible. This can be easily verified and is left as an useful
exercise.

 $\Delta M_{d,s}$ can be expressed
in terms of the off-diagonal element in the neutral B-meson mass matrix
by using the formulae developed previously for the K-meson system.
One finds
\begin{equation}
\Delta M_q= 2 |M_{12}^{(q)}|, \qquad q=d,s.
\label{eq:xdsdef}
\end{equation}
This formula differs from $\Delta M_K=2 \RE M_{12}$ because in the
B-system $\Gamma_{12}\ll M_{12}$.

Equivalently, the mixing can be described by
\begin{equation}
x_q \equiv \frac{\Delta M_q}{\Gamma_{B_q}},
\end{equation}
where  $\Gamma_{B_q} = 1/\tau_{B_q}$ with
$\tau_{B_q}$ being the corresponding lifetimes.
However, working with $\Delta M_q$ instead of $x_q$
avoids the experimental errors in lifetimes. 

The off-diagonal
term $M_{12}$ in the neutral $B$-meson mass matrix is then given by
a formula analogous to (\ref{eq:M12Kdef})
\begin{equation}
2 m_{B_q} |M_{12}^{(q)}| = 
|\langle \bar B^0_q| \Heff(\Delta B=2) |B^0_q\rangle|,
\label{eq:M12Bdef}
\end{equation}
where 
in the case of $B_d^0-\bar B_d^0$
mixing 
\begin{eqnarray}\label{hdb2}
{\cal H}^{\Delta B=2}_{\rm eff}&=&\frac{G^2_{\rm F}}{16\pi^2}M^2_W
 \left(V^\ast_{tb}V_{td}\right)^2 \eta_{B}
 S_0(x_t)\times
\nonumber\\
& &\times \left[\alpha^{(5)}_s(\mu_b)\right]^{-6/23}\left[
  1 + \frac{\alpha^{(5)}_s(\mu_b)}{4\pi} J_5\right]  Q(\Delta B=2) + h. c.
\end{eqnarray}
Here $\mu_b=\ord(m_b)$,
\begin{equation}\label{qbdbd}
Q(\Delta B=2)=(\bar bd)_{V-A}(\bar bd)_{V-A}
\end{equation}
and \cite{BJW90}
\begin{equation}
\eta_B=0.55\pm0.01.
\end{equation}
Finally $J_5=1.627$ in the NDR scheme. 
In the case of  $B_s^0-\bar B_s^0$ mixing one should simply replace
$d\to s$ in (\ref{hdb2}) and (\ref{qbdbd}) with all other quantities
unchanged.

We next reapeat what we have done already in Section 8.3.
Defining the renormalization group invariant parameters $\hat B_q$
by
\begin{equation}\label{Def-Bpar1}
\hat B_{B_q} = B_{B_q}(\mu) \left[ \as^{(5)}(\mu) \right]^{-6/23} \,
\left[ 1 + \frac{\as^{(5)}(\mu)}{4\pi} J_5 \right]
\label{eq:BBrenorm}
\end{equation}
\begin{equation}
\langle \bar B^0_q| (\bar b q)_{V-A} (\bar b q)_{V-A} |B^0_q\rangle
\equiv \frac{8}{3} B_{B_q}(\mu) F_{B_q}^2 m_{B_q}^2\,,
\label{eq:BbarB}
\end{equation}
where
$F_{B_q}$ is the $B_q$-meson decay constant
and using (\ref{hdb2}) one finds
\begin{equation}
\Delta M_q = \frac{G_{\rm F}^2}{6 \pi^2} \eta_B m_{B_q} 
(\hat B_{B_q} F_{B_q}^2 ) \mw^2 S_0(x_t) |V_{tq}|^2,
\label{eq:xds}
\end{equation}
which implies two useful formulae
\begin{equation}\label{DMD}
\Delta M_d=
0.50/{\rm ps}\cdot\left[ 
\frac{\sqrt{\hat B_{B_d}}F_{B_d}}{200\mev}\right]^2
\left[\frac{\mtb(\mt)}{170\gev}\right]^{1.52} 
\left[\frac{\vtd}{8.8\cdot10^{-3}} \right]^2 
\left[\frac{\eta_B}{0.55}\right]  
\end{equation}
and
\begin{equation}\label{DMS}
\Delta M_{s}=
15.1/{\rm ps}\cdot\left[ 
\frac{\sqrt{\hat B_{B_s}}F_{B_s}}{240\mev}\right]^2
\left[\frac{\mtb(\mt)}{170\gev}\right]^{1.52} 
\left[\frac{\vts}{0.040} \right]^2
\left[\frac{\eta_B}{0.55}\right] \,.
\end{equation}

There is a vast literature on the calculations of $F_{B_d}$ and
$\hat B_d$.
The most recent world averages from lattice are \cite{Flynn,Bernard}
\begin{equation}
F_{B_d}=(175\pm 25)\mev\,, \qquad
\hat B_{B_d}=1.31\pm 0.03\,.
\end{equation}
This result for $F_{B_d}$ is compatible with the results obtained 
with the help of QCD sum rules   \cite{QCDSF}.
In our numerical analysis we will use
\be
F_{B_d}\sqrt{\hat B_{B_d}}=(200\pm 40)\mev.
\ee
The experimental situation on
$\Delta M_d$ taken from Gibbons \cite{Gibbons}
 is given in table \ref{tab:inputparams}. 
\subsection{Standard Analysis of the Unitarity Triangle}\label{UT-Det}
With all these formulae at hand we can now summarize the standard
analysis of the unitarity triangle in fig. \ref{fig:utriangle}. 
It proceeds in five steps.

{\bf Step 1:}

{}From  $b\to c$ transition in inclusive and exclusive $B$ meson decays
one finds $\vcb$ and consequently the scale of the unitarity triangle:
\begin{equation}
\vcb\quad \Longrightarrow\quad\lambda \vcb= \lambda^3 A
\end{equation}

{\bf Step 2:}

{}From  $b\to u$ transition in inclusive and exclusive $B$ meson decays
one finds $\vub$ and consequently the side $CA=R_b$ of the unitarity
triangle:
\begin{equation}\label{rb}
\left| \frac{V_{ub}}{V_{cb}} \right|
 \quad\Longrightarrow \quad R_b=\sqrt{\bar\varrho^2+\bar\eta^2}=
4.44 \cdot \left| \frac{V_{ub}}{V_{cb}} \right|
\end{equation}

{\bf Step 3:}

{}From experimental value of $\varepsilon$ (\ref{eexp}) 
and the formula (\ref{eq:epsformula}) one 
derives, using the approximations (\ref{2.51})--(\ref{2.53}), 
the constraint
\begin{equation}\label{100}
\bar\eta \left[(1-\bar\varrho) A^2 \eta_2 S_0(x_t)
+ P_0(\varepsilon) \right] A^2 \hat B_K = 0.226,
\end{equation}
where
\begin{equation}\label{102}
P_0(\varepsilon) = 
\left[ \eta_3 S_0(x_c,x_t) - \eta_1 x_c \right] \frac{1}{\lambda^4},
\qquad
x_t=\frac{\mt^2}{\mw^2}.
\end{equation}
 $P_0(\varepsilon)=0.31\pm0.05$ summarizes the contributions
of box diagrams with two charm quark exchanges and the mixed 
charm-top exchanges. The error in $P_0(\varepsilon)$ is dominated by the
uncertainties in $\eta_3$ and $m_c$.
However, the $P_0(\varepsilon)$
term contributes only $25\%$ to (\ref{100}) and these uncertainties
constitute only  a few percent uncertainty in the constraint
(\ref{100}). 
Recalling that
$\mt$ and the relevant QCD factors $\eta_2$ and $\eta_3$ 
are rather precisely known, we conclude that
the main uncertainties in the constraint (\ref{100}) reside in
$\hat B_K$ and to some extent in $A^4$ which multiplies the leading term.

\begin{figure}[hbt]
\vspace{0.010in}
\centerline{
\epsfysize=4in
\rotate[r]{
\epsffile{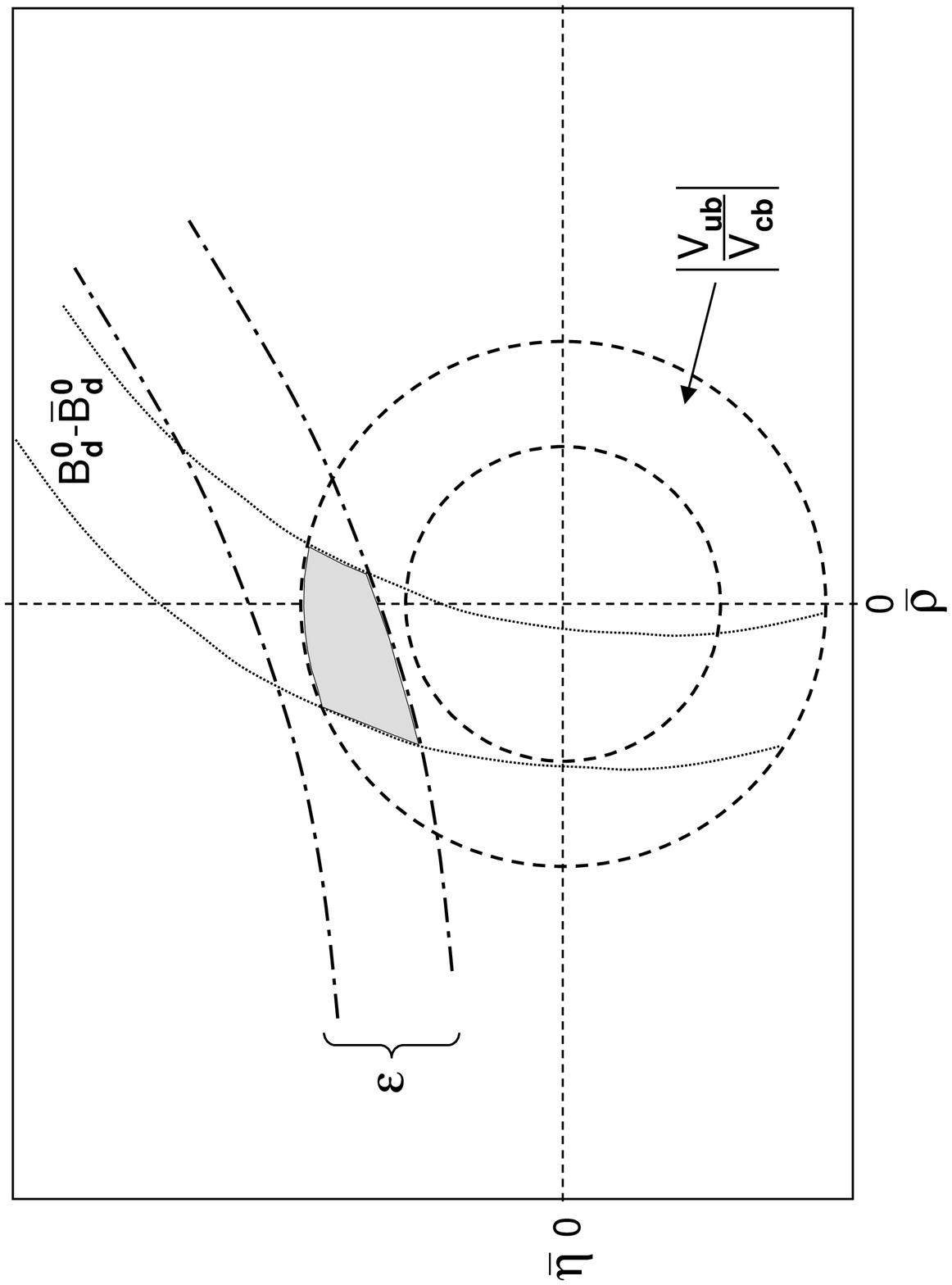}
}}
\vspace{0.0108in}
\caption[]{Schematic determination of Unitarity Triangle.
\label{L:10}}
\end{figure}
Equation (\ref{100}) specifies 
a hyperbola in the $(\bar \varrho, \bar\eta)$
plane.
This hyperbola intersects the circle found in step 2
in two points which correspond to the two solutions for
$\delta$ mentioned earlier. This is illustrated in fig. \ref{L:10}.
The position of the hyperbola (\ref{100}) in the
$(\bar\varrho,\bar\eta)$ plane depends on $\mt$, $|V_{cb}|=A \lambda^2$
and $\hat B_K$. With decreasing $\mt$, $|V_{cb}|$ and $\hat B_K$ the
$\eps$-hyperbola moves away from the origin of the
$(\bar\varrho,\bar\eta)$ plane. When the hyperbola and the circle
(\ref{rb}) touch each other lower bounds consistent with $\eps_K^{\rm
exp}$  can be found \cite{Buras}:
\begin{eqnarray}
(\mt)_{\rm min} &=& \mw \left[ \frac{1}{2 A^2} \left( \frac{1}
{A^2 \hat B_K R_b} - 1.4 \right) \right]^{0.658}
\label{eq:mtmin} \\
\left| \frac{V_{ub}}{V_{cb}} \right|_{\rm min} &=&
\frac{\lambda}{1-\lambda^2/2} \,
\left[ A^2 \hat B_K \left( 2 x_t^{0.76} A^2 + 1.4 \right) \right]^{-1}
\label{eq:Vubcbmin} \\
(\hat B_K)_{\rm min} &=& 
\left[ A^2 R_b \left( 2 x_t^{0.76} A^2 + 1.4 \right)
                    \right]^{-1}.
\label{eq:BKmin}
\end{eqnarray}

{\bf Step 4:}
{}From the observed $B^0_d-\bar B^0_d$ mixing parametrized by $\Delta M_d$ 
the side $BA=R_t$ of the unitarity triangle can be determined:
\begin{equation}\label{106}
 R_t= \frac{1}{\lambda}\frac{|V_{td}|}{\vcb} = 1.0 \cdot
\left[\frac{|V_{td}|}{8.8\cdot 10^{-3}} \right] 
\left[ \frac{0.040}{\vcb} \right]
\end{equation}
with
\begin{equation}\label{VT}
\vtd=
8.8\cdot 10^{-3}\left[ 
\frac{200\mev}{\sqrt{\hat B_{B_d}}F_{B_d}}\right]
\left[\frac{170~GeV}{\mtb(\mt)} \right]^{0.76} 
\left[\frac{\Delta M_d}{0.50/{\rm ps}} \right ]^{0.5} 
\sqrt{\frac{0.55}{\eta_B}}.
\end{equation}

Since $\mt$, $\Delta M_d$ and $\eta_B$ are already rather precisely
known, the main uncertainty in the determination of $\vtd$ from
$B_d^0-\bar B_d^0$ mixing comes from $F_{B_d}\sqrt{B_{B_d}}$.
Note that $R_t$ suffers from additional uncertainty in $\vcb$,
which is absent in the determination of $\vtd$ this way. 
The constraint in the $(\bar\varrho,\bar\eta)$ plane coming from
this step is illustrated in fig.~\ref{L:10}.

{\bf Step 5:}

{}The measurement of $B^0_s-\bar B^0_s$ mixing parametrized by $\Delta M_s$
together with $\Delta M_d$  allows to determine $R_t$ in a different
way. Using (\ref{eq:xds}) and setting $\Delta M^{{\rm max}}_d= 0.482/
\mbox{ps}$ and 
$|V_{ts}/V_{cb}|^{{\rm max}}=0.993$  one finds a useful formula
\cite{ABWAR}:
\begin{equation}\label{107b}
(R_t)_{\rm max} = 1.0 \cdot \xi \sqrt{\frac{10.2/ps}{\Delta M_s}},
\qquad
\xi = 
\frac{F_{B_s} \sqrt{\hat B_{B_s}}}{F_{B_d} \sqrt{\hat B_{B_d}}},
\end{equation}
where $\xi=1$ in the  $SU(3)$--flavour limit.
One should 
note that $\mt$ and $|V_{cb}|$ dependences have been eliminated this way
 and that $\xi$ should in principle 
contain much smaller theoretical
uncertainties than the hadronic matrix elements in $\Delta M_d$ and 
$\Delta M_s$ separately.  
The most recent values relevant for (\ref{107b}) are:
\begin{equation}\label{107c}
\Delta M_s > 10.2/ ps ~(95\%~{\rm  C.L.})
\qquad\quad
\xi=1.15\pm 0.05
\end{equation}
The first number is the improved lower bound from ALEPH \cite{Drell}.
The second number comes from quenched lattice calculations summarized
in \cite{Flynn} and \cite{Bernard}.
A similar result has been obtained using QCD sum rules \cite{NAR}.

The fate of the usefulness of the bound (\ref{107b}) depends clearly
on both $\Delta M_s$ and $\xi$ as well as on the type of the error
analysis. We will return to this point soon.
For $\xi=1.2$ 
the lower bound on $\Delta M_s$ in (\ref{107c}) implies $R_t\le 1.20$
which, as we will see, has a moderate impact on the unitarity triangle
obtained using the scanning method and 
the first four steps alone. 

Finally, I would like to point out that whereas step 5 can give, 
in contrast
to step 4, the value for $R_t$ free of the $\vcb$ uncertainty, it does
not provide at present a more accurate value of $\vtd$ if the scanning
method, discussed below, is used. The point is, that 
having $R_t$, one determines $\vtd$ by means of the relation (\ref{106})
which, in contrast to (\ref{VT}), depends on $\vcb$. 
In fact as we will see below, the inclusion
of step 5 has, with $\xi=1.2$, a visible impact on $R_t$ without
essentially any impact on the range of $\vtd$ obtained using the scanning
method and the first four steps alone.
\subsection{Numerical Results}\label{sec:standard}
\subsubsection{Input Parameters}
 The input parameters needed to perform the
standard analysis using the first four steps alone
are given in table \ref{tab:inputparams}.
We list here the "present" errors based on what we have discussed
above, as well as the "future" errors. The latter are a mere guess,
but as we will see in sections 13 and 14, these are the errors
one should aim at, in order that the standard analysis could be
competitive in the CKM determination with the cleanest rare decays and 
the CP asymmetries in B-decays. 

 $\mt$ in table~\ref{tab:inputparams} 
 refers
to the running current top quark mass normalized at $\mu=\mt$:
$\mtb(\mt)$ and is obtained from the value 
$\mt^{Pole}=175\pm 6\gev$ measured by CDF and D0 by means of the
relation.
 \begin{equation}\label{POLE}
\mtb(\mt)=\mt^{{\rm Pole}}
\left[ 1-\frac{4}{3}\frac{\alpha_s(m_t)}{\pi}\right].
\end{equation}
Thus for $\mt={\cal O}(170\gev)$, $\mtb(\mt)$ is typically
by $8\gev$ smaller than $m_t^{\rm Pole}$. 
In principle known $\ord(\as^2)$ corrections to the relation
(\ref{POLE}) could also be included which would decrease the value
of $\mtb(\mt)$ by roughly $1~\gev$.
Yet this would not be really consistent with the rest of the
analysis which does not include the next--to--NLO corrections.

\begin{table}[thb]
\caption[]{Collection of input parameters.\label{tab:inputparams}}
\vspace{0.4cm}
\begin{center}
\begin{tabular}{|c|c|c|c|}
\hline
{\bf Quantity} & {\bf Central} & {\bf Present} & {\bf Future} \\
\hline
$|V_{cb}|$ & 0.040 & $\pm 0.003$ & $\pm 0.001 $\\
$|V_{ub}/V_{cb}|$ & 0.080 & $\pm 0.020$ & $\pm 0.005 $ \\
$\hat B_K$ & 0.75 & $\pm 0.15$ & $\pm 0.05$ \\
$\sqrt{\hat B_d} F_{B_{d}}$ & $200\mev$ & $\pm 40\mev$ &$\pm 10\mev$ \\
$\mt$ & $167\gev$ & $\pm 6\gev$ & $\pm 3\gev $\\
$\Delta M_d$ & $0.464~\mbox{ps}^{-1}$ & $\pm 0.018~\mbox{ps}^{-1}$ 
& $\pm 0.006~\mbox{ps}^{-1}$\\ 
\hline
\end{tabular}
\end{center}
\end{table}
\subsubsection{$\left| V_{ub}/V_{cb} \right|$,
$\left| V_{cb} \right|$ and $\varepsilon_K$}

The values for $\left| V_{ub}/V_{cb} \right|$ 
and $\left| V_{cb} \right|$ in table \ref{tab:inputparams}
are not correlated with
each other. On the other hand such a correlation is present in
the analysis of the CP violating parameter $\varepsilon$ which
is roughly proportional to the fourth power of $\left| V_{cb}\right|$
and linear in $\left|V_{ub}/V_{cb} \right|$. It follows
that not all values in table \ref{tab:inputparams} are simultaneously
consistent with the observed value of $\varepsilon$.
This 
has been emphasized in particular by
Herrlich and Nierste \cite{HNb} and in \cite{BBL}. 
Explicitly one has using (\ref{eq:Vubcbmin}):

\begin{equation}
\left| \frac{V_{ub}}{V_{cb}} \right|_{\rm min}=
\frac{0.225}{\hat B_K A^2(2 x_t^{0.76}A^2+1.4)}.
\end{equation}

\begin{figure}[hbt]
\vspace{0.10in}
\centerline{
\epsfysize=4.5in
\rotate[r]{\epsffile{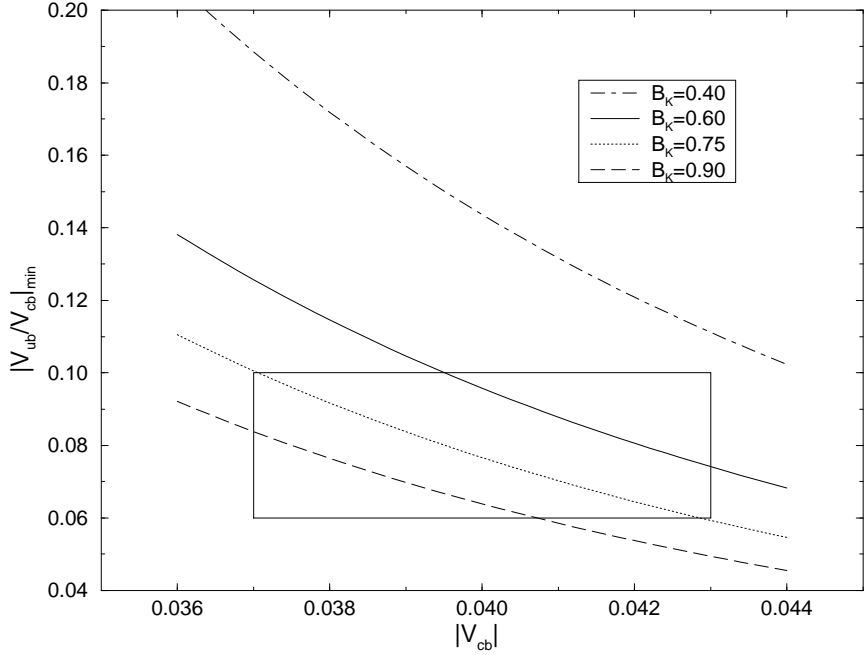}}
}
\vspace{0.08in}
\caption{Lower bound on $\vub$ from $\varepsilon_K$.}\label{fig:bound}
\end{figure}

This bound is shown as a function of $\vcb$ for different
values of $\hat B_K$ and $\mt=173\gev$ in fig.\ \ref{fig:bound}. 
We observe that simultaneously
small values of $\left| V_{ub}/V_{cb} \right|$ and $\left| V_{cb} \right|$,
although still consistent with the ones given in 
table \ref{tab:inputparams}, are not allowed
by the size of indirect CP violation observed in $K \to \pi\pi$.

\begin{table}[thb]
\caption[]{Present output of the Standard Analysis. 
 $\lambda_t=V^*_{ts} V_{td}$.\label{TAB2}}
\vspace{0.4cm}
\begin{center}
\begin{tabular}{|c||c||c|}\hline
{\bf Quantity} & {\bf Scanning} & {\bf Gaussian} \\ \hline
$\mid V_{td}\mid/10^{-3}$ &$6.9 - 11.3$ &$ 8.6\pm 1.1$ \\ \hline
$\mid V_{ts}/V_{cb}\mid$ &$0.959 - 0.993$ &$0.976\pm 0.010$  \\ \hline
$\mid V_{td}/V_{ts}\mid$ &$0.16 - 0.31$ &$0.213\pm 0.034$  \\ \hline
$\sin(2\beta)$ &$0.36 - 0.80$ &$ 0.66\pm0.13 $ \\ \hline
$\sin(2\alpha)$ &$-0.76 - 1.0$ &$ 0.11\pm 0.55 $ \\ \hline
$\sin(\gamma)$ &$0.66 - 1.0 $ &$ 0.88\pm0.10 $ \\ \hline
$\IM \lambda_t/10^{-4}$ &$0.86 - 1.71 $ &$ 1.29\pm 0.22 $ \\ \hline
\end{tabular}
\end{center}
\end{table}

\subsubsection{Output of the Standard Analysis}
The output of the standard analysis depends to some extent on the
error analysis. This should be always remembered in view of the fact
that different authors use different procedures. In order to illustrate
this  I show in tables \ref{TAB2} ("present") and \ref{TAB3} ("future") 
the results for various quantities of interest
using two types of error analyses:

\begin{itemize}
\item
Scanning: Both the experimentally measured numbers and the theoretical input
parameters are scanned independently within the errors given in
table~\ref{tab:inputparams}. 
\item
Gaussian: The experimentally measured numbers and the theoretical input 
parameters are used with Gaussian errors.
\end{itemize}
Clearly the "scanning" method is a bit conservative. On the other
hand using Gaussian distributions for theoretical input parameters
can be certainly questioned. 
I think that
at present the conservative "scanning" method should be preferred,
although one certainly would like to have a better method. Interesting
new methods have been presented in \cite{FRENCH,PAGA}.
They provide more stringent bounds on the apex of the unitarity triangle
than presented here. I must admitt that I did not find time yet
to analyze these papers to the extend that I could say anything profound
about them here. I hope to do it soon.
The analysis discussed here has been done in collaboration with Matthias 
Jamin and Markus Lautenbacher \cite{BJL96b}.

In figs.~\ref{fig:utdata} and  \ref{fig:utdataf}  we show the ranges
 for the upper
corner A of the UT in the case of the "present" input and "future" input
respectively. The circles correspond to $R_t^{max}$ from 
(\ref{107b})
using $\xi=1.20$ and $(\Delta M)_s=10/ps,~15/ps$ and $25/ps$, respectively.
The present bound (\ref{107c}) is represented by the first
of these circles. 
This bound has not 
been used in
obtaining the results in tables \ref{TAB2} and \ref{TAB3}. 
Its impact will be analysed separately
below.
The circles from $B^0_d-\bar B^0_d$ mixing are not shown explicitly
for reasons to be explained below. The impact of $\Delta M_d$ can however
be easily seen by comparing the shaded area with the area one would find
by using the lower $\varepsilon$-hyperbola and the $R_b$-circles alone.
\begin{figure}[thb]
\vspace{0.10in}
\centerline{
\epsfysize=3.4in
\rotate[r]{\epsffile{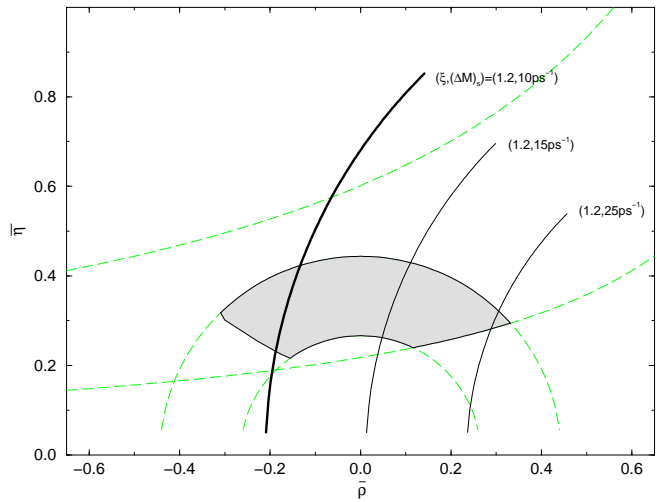}}
}
\vspace{0.08in}
\caption[]{
Unitarity Triangle 1998.
\label{fig:utdata}}
\end{figure}
The allowed region has a typical "banana" shape which can be found
in many other analyses \cite{BLO,ciuchini:95,HNb,ALUT,FRENCH,PAGA}. 
The size of
the banana and its position depends on the assumed input
parameters and on the error analysis which varies from paper
to paper. The results in figs. \ref{fig:utdata} and  \ref{fig:utdataf}
correspond to a simple independent 
scanning of all parameters within one standard deviation.
I should remark that the plots in \cite{PAGA} give substantially smaller
allowed ranges in the $(\bar\varrho,\bar\eta)$ plane and look more
like potatoes than bananas.

\begin{figure}[thb]
\vspace{0.10in}
\centerline{
\epsfysize=3.6in
\rotate[r]{\epsffile{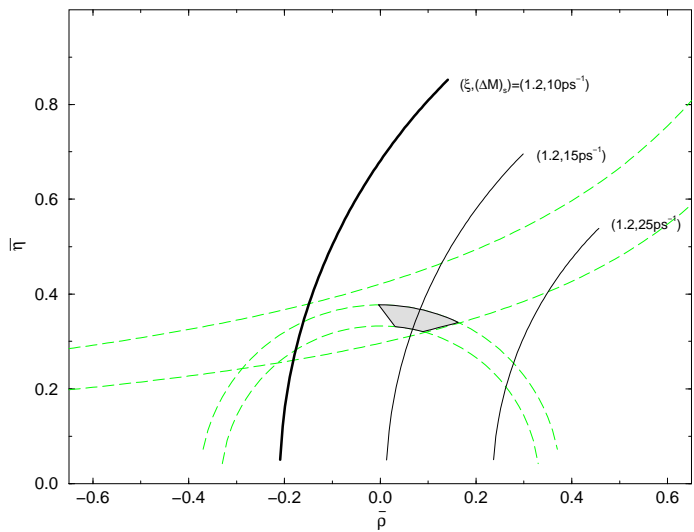}}
}
\vspace{0.08in}
\caption[]{
Unitarity Triangle 2008.
\label{fig:utdataf}}
\end{figure}

As seen in fig.~\ref{fig:utdata} our present knowledge of
the unitarity triangle is still rather poor. Fig.~\ref{fig:utdataf}
demonstrates very clearly that this situation may change dramatically
in the future provided the errors in the input parameters will be decreased
as anticipated in our "future" scenario. 

Comparing the results for $\vtd$ given in table \ref{TAB2} 
with the ones obtained on the basis of unitarity alone (\ref{uni1}) 
we observe that
the inclusion of the constraints from $\varepsilon$ and $\Delta M_d$
had a considerable impact on the allowed range for this CKM matrix
element. This impact will be amplified in the future as seen in
table \ref{TAB3}. An inspection shows that with our input parameters
the lower bound on $\vtd$ is governed by  $\varepsilon_K$, whereas
the upper bound by $\Delta M_d$.

Next we observe that whereas the angle $\beta$ is rather
constrained, the uncertainties in 
$\alpha$ and $\gamma$ are  huge: 
\be\label{ap}
35^\circ\le \alpha \le 115^\circ~,
\quad
11^\circ\le \beta \le 27^\circ~,
\quad
41^\circ\le \gamma \le 134^\circ~.
\ee
The situation will improve when the "future" scenario
will be realized:
\be\label{af}
70^\circ\le \alpha \le 93^\circ~,
\quad
19^\circ\le \beta \le 22^\circ~,
\quad
65^\circ\le \gamma \le 90^\circ~.
\ee
Finally we would like to comment on the impact of the bound on 
$\Delta M_s$
given in (\ref{107c}) if the scanning method is used.
This impact is still
rather small except for the upper limits for $\vtd/\vts$ and $\gamma$
which are lowered in the "scanning'' version to $0.27$ and $129^\circ$
respectively. Larger impact of the bound on $\Delta M_s$ on various
parameters is
found by using the methods in \cite{FRENCH,PAGA}.
\subsubsection{Correlation between $\varepsilon_K$ and $\Delta M_d$}
Now, why did we omitt the explicit circles from $B^0_d-\bar B^0_d$ mixing 
in the plots of unitarity triangles above ? I have to answer this
question because some of my colleagues suspected that a plot similar
to the one in fig.~\ref{fig:utdata} and shown already at the Rochester
conference in Warsaw was wrong. At first  one would expect that the
left border of the allowed area coming from $B^0_d-\bar B^0_d$ mixing
should have a shape similar to the circles coming from 
$\Delta M_d/\Delta M_s$ and shown in the figures above. This expectation
is correct at fixed values of $m_t$ and $\vcb$. Yet once these
two parameters are varied in the allowed ranges, this is no longer
true. In fact one can easily convince oneself that the uncertainties
coming from $\mt$ and $\vcb$ in the analyses of $\varepsilon_K$ and
$\Delta M_d$ cannot be represented simultaneously in the 
$(\bar\varrho,\bar\eta)$ plane in terms of nice hyperbolas
and nice circles. This is simply related to the correlation between
$\varepsilon_K$ and $\Delta M_d$ due to $m_t$ and $\vcb$. Neglecting
this correlation one finds for instance that the most negative value of 
$\bar\varrho$ corresponds to the maximal values of $(m_t,\vcb)$ in the
case of $\varepsilon_K$ and to the minimal values of $(m_t,\vcb)$ in the
case of $B^0_d-\bar B^0_d$ which is of course inconsistent. In 
figs. \ref{fig:utdata} and  \ref{fig:utdataf} we have decided
to show the $\varepsilon_K$-hyperbolas. Consequently the impact
of $B^0_d-\bar B^0_d$ mixing had to be found numerically and as
seen it is not described by a circle. Since $m_t$ is already
very well known, this discussion mainly applies to the $\vcb$ dependence.
Finally it should be stressed that similar correlations have to
be taken into account in the future when various rare decays discussed in
subsequent sections will enter the game of the determination of the
unitarity triangle. Needless to say, the radius $R^{max}_t$ 
determined through
(\ref{107b}) and shown in the UT plots, being independent of
$(\mt,\vcb)$, is not subject to the correlation in question.

\begin{table}[thb]
\caption[]{Future output of the Standard Analysis. 
 $\lambda_t=V^*_{ts} V_{td}$.\label{TAB3}}
\vspace{0.4cm}
\begin{center}
\begin{tabular}{|c||c||c|}\hline
{\bf Quantity} & {\bf Scanning} & {\bf Gaussian} \\ \hline
$\mid V_{td}\mid/10^{-3}$ &$8.1 - 9.2$ &$ 8.6\pm 0.3$ \\ \hline
$\mid V_{ts}/V_{cb}\mid$ &$0.969 - 0.983$ &$0.976\pm 0.004$  \\ \hline
$\mid V_{td}/V_{ts}\mid$ &$0.20 - 0.24$ &$0.215\pm 0.010$  \\ \hline
$\sin(2\beta)$ &$0.61 - 0.70$ &$ 0.67\pm0.03 $ \\ \hline
$\sin(2\alpha)$ &$-0.11 - 0.66.0$ &$ 0.21\pm 0.21 $ \\ \hline
$\sin(\gamma)$ &$0.90 - 1.0 $ &$ 0.96\pm0.03 $ \\ \hline
$\IM \lambda_t/10^{-4}$ &$1.21 - 1.41 $ &$ 1.29\pm 0.06 $ \\ \hline
\end{tabular}
\end{center}
\end{table}
\subsection{Final Remarks}
In this section we have completed the determination of the CKM matrix.
It is given by the values of $|V_{us}|$, $\vcb$ and $|V_{ub}|$ in
(\ref{vcb}) and (\ref{v13}), the results in table~\ref{TAB2} and
the unitarity triangle shown in fig.~\ref{fig:utdata}. Clearly
the accuracy of this determination is not impressive. We have
stressed, however, that in ten years from now the standard analysis
may give the results shown in table~\ref{TAB3} and fig.~\ref{fig:utdataf}.
Moreover a single precise measurement of $\Delta M_s$ in the future
will have a very important impact on the allowed area in the 
$(\bar\varrho,\bar\eta)$ plane. Such a measurement should come from
SLD and later from LHC.

Having the values of CKM parameters at hand, we can use them to predict
various branching ratios of radiative, rare and CP-violating decays.
This we will do in the subsequent three sections. We will see there,
that the poor knowledge of CKM parameters precludes precise predictions of
a number of interesting branching ratios at present. This may change in
the next decade as stressed above.

\clearpage
\section{$\epe$ in the Standard Model}\label{EpsilonPrime}
\setcounter{equation}{0}
\subsection{Preliminaries}
Direct CP violation remains one of the important targets 
of contemporary particle physics. In this respect the search 
for direct CP violation in $K\to\pi\pi$ decays plays a special
role as already sixteen years have been devoted to this enterprise.
In this case,
a non-vanishing value of the ratio Re($\epe$) defined in (\ref{eprime}) 
would give the first
signal for direct CP violation ruling out superweak models
\cite{wolfenstein:64}.
The experimental situation of Re($\varepsilon'/\varepsilon$) is,
however, unclear
at present:
\begin{equation}\label{eprime1}
\RE(\varepsilon'/\varepsilon) =\left\{ \begin{array}{ll}
(23 \pm 7)\cdot 10^{-4} & \cite{barr:93} \\
(7.4 \pm 5.9)\cdot 10^{-4} & \cite{gibbons:93}.\end{array} \right.
\end{equation}

While the result of the NA31 collaboration at CERN  \cite{barr:93}
clearly indicates direct CP violation, the value of E731 at Fermilab
\cite{gibbons:93} is compatible with superweak models
 in which $\varepsilon'/\varepsilon = 0$.
 Hopefully, during the next two years the experimental situation concerning
$\varepsilon'/\varepsilon$ will be clarified through the improved
measurements by the two collaborations at the $10^{-4}$ level and by
the KLOE experiment at  DA$\Phi$NE. A recent discussion of superweak
models can be found in \cite{HALL}. I will not consider them here.

There is no question about that direct CP violation is present in
the Standard Model. Yet accidentally it could turn out that it will be
difficult to see it in $K \to \pi\pi$ decays.  Indeed as we will
discuss in detail below, in the Standard
Model $\varepsilon'/\varepsilon $ is governed by QCD penguins and
electroweak (EW) penguins. We have met them already in connection
with B-decays in Section 8. In spite of being suppressed by
$\alpha/\alpha_s$ relative to QCD penguin contributions, 
electroweak penguin contributions have to be included because of the
additional enhancement factor ${\rm Re}A_0/{\rm Re}A_2=22$ 
(see (\ref{eq:epsprim})--(\ref{eq:ReA0data})) relative
to QCD penguins. With increasing $\mt$ the EW penguins become
increasingly important \cite{flynn:89,buchallaetal:90} and, entering
$\varepsilon'/\varepsilon$ with the opposite sign to QCD penguins,
suppress this ratio for large $\mt$. For $\mt\approx 200\,\gev$ the ratio
can even be zero \cite{buchallaetal:90}.  Because of this strong
cancellation between two dominant contributions and due to uncertainties
related to hadronic matrix elements of the relevant local operators, a
precise prediction of $\varepsilon'/\varepsilon$ is not possible at
present. We will discuss this in detail below.
\subsection{History of $\epe$}
The first calculations of $\epe$ for $\mt \ll \mw$ and in the leading
order approximation can be found in \cite{GW79}. 
For $\mt \ll \mw$ only QCD
penguins play a substantial role. Over the eighties these calculations
were refined through the inclusion of isospin braking in the
quark masses \cite{donoghueetal:86,burasgerard:87,lusignoli:89},
the inclusion of QED penguin effects for $\mt \ll \mw$
\cite{BW84,donoghueetal:86,burasgerard:87}, 
and through improved estimates of hadronic matrix elements in
the framework of the $1/N$ approach \cite{bardeen:87}. 
This era of $\epe$ culminated
in the analyses in \cite{flynn:89,buchallaetal:90}, where QCD
penguins, electroweak penguins ($\gamma$ and $Z^0$ penguins)
and the relevant box diagrams were included for arbitrary
top quark masses. The strong cancellation between QCD penguins
and electroweak penguins for $m_t > 150~\gev$ found in these
papers was confirmed by other authors \cite{PW91}.

All these calculations were done in the leading logarithmic
approximation (e.g.\ one-loop anomalous dimensions of the relevant
operators) with the exception of the $\mt$-dependence which in 
the analyses \cite{flynn:89,buchallaetal:90,PW91} has been already
included at the NLO level. While such a procedure is not fully
consistent, it allowed for the first time to exhibit the strong
$\mt$-dependence of the electroweak penguin contributions,
which is not seen in a strict leading logarithmic approximation.

During the nineties considerable progrees has been made by
calculating complete NLO corrections to $\varepsilon'$
\cite{BJLW1,BJLW2,BJLW,ROMA1,ROMA2}. Together with the NLO
corrections to $\varepsilon$ and $B^0-\bar B^0$ mixing
discussed in the previous section, this allows
a complete NLO analysis of $\varepsilon'/\varepsilon$ including
constraints from the observed indirect CP violation ($\varepsilon$)
and  $B_{d,s}^0-\bar B_{d,s}^0$ mixings ($\Delta M_{d,s}$). The improved
determination of the $V_{ub}$ and $V_{cb}$ elements of the CKM matrix,
the improved estimates of hadronic matrix elements using the lattice 
approach as well as other non-perturbative approaches 
and in particular the determination of the top quark mass
$\mt$ had of course also an important impact on
$\varepsilon'/\varepsilon$. 

After these general remarks let us discuss 
$\epe$ in explicit terms. Other reviews of $\epe$ can be found
in \cite{WW,BERT98}.
\subsection{Basic Formulae}
           \label{subsec:epeformulae}
The direct CP violation in $K \to \pi\pi$ is described by the parameter
$\varepsilon'$ defined in (\ref{eprime}).
The latter formula  can be rewritten 
in terms of the real and imaginary parts of 
the amplitudes $A_0 \equiv
A(K \to (\pi\pi)_{I=0})$ and $A_2 \equiv
A(K \to (\pi\pi)_{I=2})$ as follows:
\begin{equation}
\eps' = -\frac{\omega}{\sqrt{2}} \xi (1 - \Omega) \exp(i \Phi) \, ,
\label{eq:epsprim}
\end{equation}
where
\begin{equation}
\xi = \frac{\IM A_0}{\RE A_0} \, , \quad
\omega = \frac{\RE A_2}{\RE A_0} \, , \quad
\Omega = \frac{1}{\omega} \frac{\IM A_2}{\IM A_0}
\label{eq:xiomega}
\end{equation}
and $\Phi \approx \pi/4$. Let us immediately emphasize the most 
important features
of various terms in (\ref{eq:epsprim}):
\bi
\item
$\IM A_0$ is dominated by QCD penguins and is very weakly dependent
on $\mt$. 
\item
$\IM A_2$ increases strongly with $\mt$
and for large $\mt$ is dominated by electroweak penguins. It receives
also a sizable contribution from isospin braking $(m_u\not=m_d)$ which
conspires with electroweak penguins to cancel substantially the
QCD penguin contribution in $\IM A_0$. 
The factor $1/\omega\approx 22$
in $\Omega$ giving a large enhancement is to a large extend responsible
for this cancellation.
\ei

When using (\ref{eq:epsprim}) and (\ref{eq:xiomega}) in phenomenological
applications one usually takes $\RE A_0$ and $\omega$ from
experiment, i.e.
\begin{equation}
\RE A_0 = 3.33 \cdot 10^{-7}\gev,
\qquad
\RE A_2 = 1.50 \cdot 10^{-8}\gev,
\qquad
\omega = 0.045,
\label{eq:ReA0data}
\end{equation}
where the last relation reflects the so-called $\Delta I=1/2$ rule. The
main reason for this strategy is the unpleasant fact that until today
nobody succeded in fully explaining this rule which to a large extent is
believed to originate in the long-distance QCD contributions
\cite{DI12}. 
On the other hand the
imaginary parts of the amplitudes in (\ref{eq:xiomega}) being related to
CP violation and the top quark physics should be dominated by
short-distance contributions. Therefore $\IM A_0$ and $\IM A_2$ are
usually calculated using the effective Hamiltonian for $\Delta S=1$
transitions:
\begin{equation}
\Heff(\Delta S=1) = 
\frac{G_{\rm F}}{\sqrt{2}} V_{us}^* V_{ud}^{} \sum_{i=1}^{10}
\left( z_i(\mu) + \tau \; y_i(\mu) \right) Q_i(\mu) 
\label{eq:HeffdF1:1010}
\end{equation}
with $\tau=-V_{ts}^* V_{td}^{}/(V_{us}^* V_{ud}^{})$.

The operators $Q_i$ are the analogues of the ones given in 
(\ref{O1})-(\ref{O3}) and (\ref{O4})-(\ref{O6}).
They are given explicitly  as follows:

{\bf Current--Current :}
\begin{equation}\label{OS1} 
Q_1 = (\bar s_{\alpha} u_{\beta})_{V-A}\;(\bar u_{\beta} d_{\alpha})_{V-A}
~~~~~~Q_2 = (\bar s u)_{V-A}\;(\bar u d)_{V-A} 
\end{equation}

{\bf QCD--Penguins :}
\begin{equation}\label{OS2}
Q_3 = (\bar s d)_{V-A}\sum_{q=u,d,s}(\bar qq)_{V-A}~~~~~~   
 Q_4 = (\bar s_{\alpha} d_{\beta})_{V-A}\sum_{q=u,d,s}(\bar q_{\beta} 
       q_{\alpha})_{V-A} 
\end{equation}
\begin{equation}\label{OS3}
 Q_5 = (\bar s d)_{V-A} \sum_{q=u,d,s}(\bar qq)_{V+A}~~~~~  
 Q_6 = (\bar s_{\alpha} d_{\beta})_{V-A}\sum_{q=u,d,s}
       (\bar q_{\beta} q_{\alpha})_{V+A} 
\end{equation}

{\bf Electroweak--Penguins :}
\begin{equation}\label{OS4} 
Q_7 = {3\over 2}\;(\bar s d)_{V-A}\sum_{q=u,d,s}e_q\;(\bar qq)_{V+A} 
~~~~~ Q_8 = {3\over2}\;(\bar s_{\alpha} d_{\beta})_{V-A}\sum_{q=u,d,s}e_q
        (\bar q_{\beta} q_{\alpha})_{V+A}
\end{equation}
\begin{equation}\label{OS5} 
 Q_9 = {3\over 2}\;(\bar s d)_{V-A}\sum_{q=u,d,s}e_q(\bar q q)_{V-A}
~~~~~Q_{10} ={3\over 2}\;
(\bar s_{\alpha} d_{\beta})_{V-A}\sum_{q=u,d,s}e_q\;
       (\bar q_{\beta}q_{\alpha})_{V-A} \,.
\end{equation}
Here, $e_q$ denotes the electrical quark charges reflecting the
electroweak origin of $Q_7,\ldots,Q_{10}$. 

The Wilson coefficient functions $z_i(\mu)$ and $ y_i(\mu)$
were calculated including
the complete next-to-leading order (NLO) corrections in
\cite{BJLW1,BJLW2,BJLW,ROMA1,ROMA2}. The details
of these calculations can be found there and in the review
\cite{BBL}. Only the coefficients $ y_i(\mu)$ enter the evaluation
of $\epe$. Examples of their numerical values are given in table 
\ref{tab:wc10smu13}.
Extensive tables for $ y_i(\mu)$ can be found in \cite{BBL}.

\begin{table}[htb]
\caption[]{$\Delta S=1 $ Wilson coefficients at $\mu=\mc=1.3\gev$ for
$\mt=170\gev$ and $f=3$ effective flavours.
$|z_3|,\ldots,|z_{10}|$ are numerically irrelevant relative to
$|z_{1,2}|$. $y_1 = y_2 \equiv 0$.
\label{tab:wc10smu13}}
\begin{center}
\begin{tabular}{|c|c|c|c||c|c|c||c|c|c|}
\hline
& \multicolumn{3}{c||}{$\Lms^{(4)}=245\mev$} &
  \multicolumn{3}{c||}{$\Lms^{(4)}=325\mev$} &
  \multicolumn{3}{c| }{$\Lms^{(4)}=405\mev$} \\
\hline
Scheme & LO & NDR & HV & LO & 
NDR & HV & LO & NDR & HV \\
\hline
$z_1$ & -0.550 & -0.364 & -0.438 & -0.625 & 
-0.415 & -0.507 & -0.702 & -0.469 & -0.585 \\
$z_2$ & 1.294 & 1.184 & 1.230 & 1.345 & 
1.216 & 1.276 & 1.399 & 1.251 & 1.331 \\
\hline
$y_3$ & 0.029 & 0.024 & 0.027 & 0.034 & 
0.029 & 0.033 & 0.039 & 0.034 & 0.039 \\
$y_4$ & -0.054 & -0.050 & -0.052 & -0.061 & 
-0.057 & -0.060 & -0.068 & -0.065 & -0.068 \\
$y_5$ & 0.014 & 0.007 & 0.014 & 0.015 & 
0.005 & 0.016 & 0.016 & 0.002 & 0.018 \\
$y_6$ & -0.081 & -0.073 & -0.067 & -0.096 & 
-0.089 & -0.081 & -0.113 & -0.109 & -0.097 \\
\hline
$y_7/\aem$ & 0.032 & -0.031 & -0.030 & 0.039 & 
-0.030 & -0.028 & 0.045 & -0.029 & -0.026 \\
$y_8/\aem$ & 0.100 & 0.111 & 0.120 & 0.121 & 
0.136 & 0.145 & 0.145 & 0.166 & 0.176 \\
$y_9/\aem$ & -1.445 & -1.437 & -1.437 & -1.490 & 
-1.479 & -1.479 & -1.539 & -1.528 & -1.528 \\
$y_{10}/\aem$ & 0.588 & 0.477 & 0.482 & 0.668 & 
0.547 & 0.553 & 0.749 & 0.624 & 0.632 \\
\hline
\end{tabular}
\end{center}
\end{table}

Using the Hamiltonian in (\ref{eq:HeffdF1:1010}) and the experimental
values for $\varepsilon$, $\RE A_0$ and $\omega$ the ratio $\epe$ can be
written as follows:
\begin{equation}
\frac{\varepsilon'}{\varepsilon} = 
\IM \lambda_t\cdot \left[ P^{(1/2)} - P^{(3/2)} \right],
\label{eq:epe}
\end{equation}
where
\begin{eqnarray}
P^{(1/2)} & = & r \sum y_i \langle Q_i\rangle_0
(1-\Omega_{\eta+\eta'})~,
\label{eq:P12} \\
P^{(3/2)} & = &\frac{r}{\omega}
\sum y_i \langle Q_i\rangle_2~,~~~~~~
\label{eq:P32}
\end{eqnarray}
with
\begin{equation}
r = \frac{G_{\rm F} \omega}{2 |\eps| \RE A_0}~, 
\qquad
\langle Q_i\rangle_I \equiv \langle (\pi\pi)_I | Q_i | K \rangle .
\label{eq:repe}
\end{equation}
One should note that the overall strong phases in $\varepsilon'$ and 
$\varepsilon$ cancel
in the ratio to an excellent approximation.
The sum in (\ref{eq:P12}) and (\ref{eq:P32}) runs over all contributing
operators. $P^{(3/2)}$ is fully dominated by electroweak penguin
contributions. $P^{(1/2)}$ on the other hand is governed by QCD penguin
contributions which are suppressed by isospin breaking in the quark
masses ($m_u \not= m_d$). The latter effect is described by

\begin{equation}
\Omega_{\eta+\eta'} = \frac{1}{\omega} \frac{(\IM A_2)_{\rm
I.B.}}{\IM A_0}\,.
\label{eq:Omegaeta}
\end{equation}
For $\Omega_{\eta+\eta'}$ we will take
\begin{equation}
\Omega_{\eta+\eta'} = 0.25 \pm 0.05\,,
\label{eq:Omegaetadata}
\end{equation}
which is in the ball park of the values obtained in the $1/N$ approach
\cite{burasgerard:87} and in chiral perturbation theory
\cite{donoghueetal:86,lusignoli:89}. $\Omega_{\eta+\eta'}$ is
independent of $\mt$.

The main source of uncertainty in the calculation of
$\epe$ are the hadronic matrix elements $\langle Q_i \rangle_I$.
They depend generally
on the renormalization scale $\mu$ and on the scheme used to
renormalize the operators $Q_i$. These two dependences are canceled by
those present in the Wilson coefficients $y_i(\mu)$ so that the
resulting physical $\epe$ does not (in principle) depend on $\mu$ and on the
renormalization scheme of the operators.  Unfortunately the accuracy of
the present non-perturbative methods used to evalutate $\langle Q_i
\rangle_I$, like lattice methods, the $1/N$ expansion, chiral
quark models and
chiral effective lagrangians, is not
sufficient to obtain the required $\mu$ and scheme dependences of
$\langle Q_i \rangle_I$. A brief review of the existing methods 
including most recent developments will be given below.

In view of this situation it has been suggested in \cite{BJLW} to
determine as many matrix elements $\langle Q_i \rangle_I$ as possible
from the leading CP conserving $K \to \pi\pi$ decays, for which the
experimental data are summarized in (\ref{eq:ReA0data}). To this end it
turned out to be very convenient to determine $\langle Q_i \rangle_I$
at the scale $\mu = \mc$.  Using the renormalization group evolution one
can then find $\langle Q_i \rangle_I$ at any other scale $\mu \not=
\mc$. The details of this procedure can be found in
\cite{BJLW}. We will briefly summarize the most important results of this
work below.

\subsection{Hadronic Matrix Elements}
\subsubsection{Preliminaries}
It is customary to express the matrix elements
$\langle Q_i \rangle_I$ in terms of non-perturbative parameters
$B_i^{(1/2)}$ and $B_i^{(3/2)}$ as follows:
\begin{equation}
\langle Q_i \rangle_0 \equiv B_i^{(1/2)} \, \langle Q_i
\rangle_0^{\rm (vac)}\,,
\qquad
\langle Q_i\rangle_2 \equiv B_i^{(3/2)} \, \langle Q_i
\rangle_2^{\rm (vac)} \,.
\label{eq:1}
\end{equation}
The label ``vac'' stands for the vacuum
insertion estimate of the hadronic matrix elements in question. 
The full list of $\langle Q_i\rangle_I$ is given in \cite{BJLW}.
 It suffices to give here only a few examples:
\begin{eqnarray}
\langle Q_1 \rangle_0 &=& -\,\frac{1}{9} X B_1^{(1/2)} \, ,
\label{eq:Q10} \\
\langle Q_2 \rangle_0 &=&  \frac{5}{9} X B_2^{(1/2)} \, ,
\label{eq:Q20} \\
\langle Q_6 \rangle_0 &=&  -\,4 \sqrt{\frac{3}{2}} 
\left[ \frac{m_{\rm K}^2}{\ms(\mu) + \md(\mu)}\right]^2
\frac{F_\pi}{\kappa} \,B_6^{(1/2)} \, ,
\label{eq:Q60}\\ 
\langle Q_1 \rangle_2 &=& 
\langle Q_2 \rangle_2 = \frac{4 \sqrt{2}}{9} X B_1^{(3/2)} \, ,
\label{eq:Q122} \\
\langle Q_i \rangle_2 &=&  0 \, , \qquad i=3,\ldots,6 \, ,
\label{eq:Q362} \\
\langle Q_8 \rangle_2 &=& 
  -\left[ \frac{\kappa}{2 \sqrt{2}} \langle \overline{Q_6} \rangle_0
          + \frac{\sqrt{2}}{6} X
   \right] B_8^{(3/2)} \, ,
\label{eq:Q82} \\
\langle Q_9 \rangle_2 &=& 
   \langle Q_{10} \rangle_2 = \frac{3}{2} \langle Q_1 \rangle_2 \, ,
\label{eq:Q9102}
\end{eqnarray}
where
\begin{equation}
\kappa = 
         \frac{F_\pi}{F_{\rm K} - F_\pi} \, ,
\qquad
X = \sqrt{\frac{3}{2}} F_\pi \left( m_{\rm K}^2 - m_\pi^2 \right) \, ,
\label{eq:XQi}
\end{equation}
and
\begin{equation}
\langle \overline{Q_6} \rangle_0 =
   \frac{\langle Q_6 \rangle_0}{B_6^{(1/2)}} \, .
\label{eq:Q60bar}
\end{equation}
In the vacuum insertion method $B_i=1$ independent of $\mu$. In QCD,
however, the hadronic parameters $B_i$ generally depend on the
renormalization scale $\mu$ and the renormalization scheme considered.
\subsubsection{$(V-A) \otimes (V-A)$ Operators}
Let us now extract some matrix from the data on $\RE A_0$ and $\RE A_2$
in (\ref{eq:ReA0data}). To this end we follow \cite{BJLW}.
One notes first that in view of the smallness of 
$\tau=\ord(10^{-4})$ entering
(\ref{eq:HeffdF1:1010}), the
real amplitudes in (\ref{eq:ReA0data}) are governed by the coefficients
$z_i(\mu)$. The method of extracting some of the matrix elements
from the data as proposed in \cite{BJLW} relies then on the
fact that due to the GIM mechanism for $ \mu\ge m_c$ 
the coefficients $z_i(\mu)$ of
the penguin operators (i=3....10) vanish at the matching scale $\mu_c$
(between the four-quark and three-quark effective theories) in the HV
scheme and are negligible in the NDR scheme. 
However, it should be remembered that the smallness or even vanishing
of $z_i(\mu)$ for $ \mu\ge m_c$ is characteristic for mass independent
renormalization schemes. In other schemes, in which the disparity of
$m_u$ and $m_c$ is felt well above $\mu=m_c$, the GIM cancellation is
incomplete and $z_i(m_c)$ for penguin operators are larger than in the
HV and NDR schemes. Examples of the leading order calculations
of this type can be found in \cite{ANII}.   

Staying within the NDR and HV schemes, we can however set $z_i(m_c)=0$
for $i\not=1,2$ 
to find
\begin{equation}
\langle Q_1(m_c) \rangle_2 = \langle Q_2(m_c) \rangle_2 =
\frac{10^6\gev^2}{1.77} \frac{\RE A_2}{z_+(m_c)} =
\frac{8.47 \cdot 10^{-3}\gev^3}{z_+(m_c)}
\label{eq:Q122data}
\end{equation}
with $z_+=z_1+z_2$ and
\begin{equation}
\langle Q_1(\mc) \rangle_0 = \frac{10^6\gev^2}{1.77} \frac{\RE
A_0}{z_1(\mc)} - \frac{z_2(\mc)}{z_1(\mc)} \langle Q_2(\mc) \rangle_0\,.
\label{eq:Q10mc}
\end{equation}
These formulae are easy to derive and are left as a useful homework
problem.

Comparing next (\ref{eq:Q122data})  with (\ref{eq:Q122}) one 
finds immediately
\begin{equation}
B_1^{(3/2)}(m_c) = \frac{0.363}{z_+(m_c)}\,,
\label{eq:B321}
\end{equation}
which using table \ref{tab:wc10smu13} gives for $\mc=1.3\gev$ and 
$\Lms^{(4)}=325\mev$
\begin{equation}
B_{1,NDR}^{(3/2)}(\mc) =  0.453\,,
\qquad
B_{1,HV}^{(3/2)}(\mc) =  0.472 \, .
\label{eq:B321mc}
\end{equation}
The extracted values for $B_1^{(3/2)}$ are by more than a factor of two
smaller than the vacuum insertion estimate.
They are compatible with the $1/N_c$ value $B_1^{(3/2)}(1\gev) \approx
0.55$ \cite{bardeen:87} and are somewhat smaller than the lattice result
$B_1^{(3/2)}(2\gev) \approx 0.6$ \cite{ciuchini:95}.
As analyzed in \cite{BJLW},
$B_1^{(3/2)}(\mu)$ decreases slowly with increasing $\mu$.
As seen in (\ref{eq:Q9102}), this analysis gives also
$\langle Q_9(\mc) \rangle_2$ and $\langle Q_{10}(\mc) \rangle_2$.

In order to extract $B_1^{(1/2)}(\mc)$ and $B_2^{(1/2)}(\mc)$ from
(\ref{eq:Q10mc})
one can make the very plausible assumption,
valid in known non-perturbative approaches, that
 $\langle Q_-(\mc) \rangle_0 \ge
\langle Q_+(\mc) \rangle_0 \ge 0$, where $Q_\pm=(Q_2\pm Q_1)/2$.
This gives  for $\Lms^{(4)}=325\mev$
\begin{equation}
B_{2,NDR}^{(1/2)}(\mc) =  6.6 \pm 1.0,
\qquad
B_{2,HV}^{(1/2)}(\mc) =  6.2 \pm 1.0 \, .
\label{eq:B122mc}
\end{equation}
The extraction of $B_1^{(1/2)}(\mc)$ and of analogous parameters
$B_{3,4}^{(1/2)}(\mc)$ are presented in detail in \cite{BJLW}.
$B_1^{(1/2)}(\mc)$ depends very sensitively on $B_2^{(1/2)}(\mc)$ and
its central value is as high as 15. $B_4^{(1/2)}(\mc)$ is 
typically by (10--15)\,\% lower than $B_2^{(1/2)}(\mc)$. In any case
this analysis shows very large deviations from the results of the
vacuum insertion method.

\subsubsection{$(V-A) \otimes (V+A)$ Operators}
The matrix elements of the $(V-A) \otimes (V+A)$ operators $Q_5$--$Q_8$
cannot be constrained by CP conserving data and one has to rely on
existing non-perturbative methods to calculate them. 
This is rather unfortunate because the QCD penguin operator $Q_6$
and the electroweak penguin operator $Q_8$, having large Wilson
coefficients and large hadronic matrix elements, play the dominant
role in $\epe$. 

We will now review the present status of $B_i$ factors describing
the matrix elements of $Q_5-Q_8$ operators as obtained in various
non-perturbative approaches. We will pay particular attention to
the parameters $B^{(1/2)}_{6}$ and $B^{(3/2)}_{8}$ which are most
important for the evaluation of $\epe$. We recall that $B_i=1$
in the vacuum insertion method.

\subsubsection{$B^{(1/2)}_{6}$ and $B^{(3/2)}_{8}$ from Lattice}
We begin with lattice calculations. These have been reviewed
recently by Gupta \cite{GUPTA98} and the APE collaboration \cite{APE}. 
The most reliable
results are found for $B^{(3/2)}_{7,8}$. The ``modern" quenched
estimates for these parameters, which supercede all previously
reported values are collected in table \ref{tab:317}, which has been
taken from Gupta and quenched a bit. The first three calculations
use perturbative matching between lattice and continuum, the last
one uses non-perturbative matching. Since all three groups agree
within perturbative matching and the non-perturbative matching
should be preferred, I conclude (probably naively) that the best
quenched lattice values are
\be\label{LAT}
(B^{(3/2)}_{7})_{\rm lattice}(2~\gev)=0.72\pm 0.05,
\quad\quad
(B^{(3/2)}_{8})_{\rm lattice}(2~\gev)=1.03\pm 0.03
\ee
where the errors are purely statistical. Concerning the
lattice results for $B^{(1/2)}_{5,6}$ the situation is
 worse. The old results read
$B^{(1/2)}_{5,6}(2~\gev)=1.0 \pm 0.2$ \cite{kilcup:91,sharpe:91}.
More accurate estimates for $B^{(1/2)}_{6}$ have been recently
obtained in \cite{kilcup:98}: 
$B^{(1/2)}_{6}(2~\gev)=0.67 \pm 0.04\pm 0.05$
(quenched) and $B^{(1/2)}_{6}(2~\gev)=0.76 \pm 0.03\pm0.05$
($f=2$). However, as stressed by Gupta, the systematic
errors in this analysis are not really under control.
We have to conclude, that there are no solid predictions for
$B^{(1/2)}_{5,6}$ from the lattice at present.
\begin{table}[thb]
\caption[]{ Lattice results for $B^{(3/2)}_{7,8} (2~\gev)$ obtained
by various groups. 
\label{tab:317}}
\begin{center}
\begin{tabular}{|c|c|c|c|}\hline
  { Fermion type}& $B^{(3/2)}_7$& $B^{(3/2)}_8$ & Matching \\
 \hline
Staggered\cite{GKS}& $0.62(3)(6)$ &$0.77(4)(4)$ & 1-loop \\
Wilson\cite{G67}& $0.58(2)(7)$ &$0.81(3)(3)$ & 1-loop \\
Clover\cite{APE}& $0.58(2)$ &$0.83(2)$ & 1-loop \\
Clover\cite{APE}& $0.72(5)$ &$1.03(3)$ &  Non-pert. \\
\hline
\end{tabular}
\end{center}
\end{table}
\subsubsection{$B^{(1/2)}_{6}$ and $B^{(3/2)}_{8}$ from the 1/N Approach}
The 1/N approach to weak hadronic matrix elements was introduced
in \cite{bardeen:87}. 
In this approach the 1/N expansion becomes a loop expansion
in an effective meson theory. In the strict large N limit only
the tree level matrix elements of $Q_6$ and $Q_8$ contribute
and one finds (\ref{eq:Q60}) and (\ref{eq:Q82}) with
\be\label{LN}
B^{(1/2)}_{6}=B^{(3/2)}_{8}=1, \quad\quad{\rm (Large-N~Limit)} 
\ee
while $B^{(1/2)}_{5}=B^{(3/2)}_{8}=0$. The latter fact is not disturbing,
however, as the operators $Q_5$ and $Q_7$, having small Wilson
coefficients are immaterial for $\epe$.

Now, $B^{(1/2)}_{6}$ and $B^{(3/2)}_{8}$ as given in (\ref{LN}) are
clearly $\mu$-independent. At first sight this appears as a problem.
But in fact it is not! The point is that $Q_{6,8}$ are
density$\times$density operators as one can see by writing them
with the help of the Fierz reordering as follows
\be\label{DENSITY}
Q_6=-2 \sum_{q=u,d,s} \bar s(1+\gamma_5)q\bar q (1-\gamma_5)d,
\quad\quad
Q_8=-3 \sum_{q=u,d,s}e_q \bar s(1+\gamma_5)q\bar q (1-\gamma_5)d.
\ee
Consequently their $\mu$-dependences are related to 
the $\mu$-dependence of the quark masses and the tree level 
factorizable contributions to
$\langle Q_6\rangle_0$ and $\langle Q_8\rangle_2$ are
$\mu$-dependent through the factor $1/\ms^2(\mu)$ as seen in
(\ref{eq:Q60}) and (\ref{eq:Q82}). This should be contrasted with
the matrix elements of $(V-A)\otimes(V-A)$ operators, which are
$\mu$-independent in the large-N limit. The $\mu$-dependence
of $1/\ms^2(\mu)$ in $\langle Q_6\rangle_0$ and $\langle Q_8\rangle_2$
is exactly cancelled in the decay amplitude by the diagonal
evolution (no mixing) of the Wilson coefficients $y_6(\mu)$ and
$y_8(\mu)$ taken in the large-N limit.

Indeed, the $\mu$-dependence of $1/\ms^2(\mu)$ is governed in LO
by $2\gamma_m^{(0)}=12 C_F$. On the other hand, the one-loop
anomalous dimensions of $Q_{6,8}$, which govern the diagonal
evolution of $y_{6,8}(\mu)$ are given by
\be
\gamma_{66}^{(0)}=-2 \gamma_m^{(0)}+\frac{2f}{3},
\quad\quad
\gamma_{88}^{(0)}=-2 \gamma_m^{(0)}.
\ee
Since for large N, $\gamma_m^{(0)}\sim\ord(N)$, we find indeed
$\gamma_{66}^{(0)}= \gamma_{88}^{(0)}=-2 \gamma_m^{(0)}$
in the large-N limit \cite{burasgerard:87}. 
Going back to the respective evolutions
of $\ms(\mu)$ and $y_{7,8}(\mu)$ we indeed confirm the cancellation
of the $\mu$-dependence in question. This feature is preserved
at the two-loop level as discussed in \cite{BJLW1}. 
One can go even further
and demonstrate numerically for $N=3$ that the parameters
$B^{1/2}_{6}$ and $B^{3/2}_{8}$ depend only very weakly on $\mu$,
when $\mu\ge 1~\gev$. In such a numerical
renormalization study in \cite{BJLW} the
factors $B_{6}^{(1/2)}$ and $B_{8}^{(3/2)}$  have been set to unity 
at $\mu=\mc$.
Subsequently the evolution of the matrix elements in the range $1\gev
\le \mu \le 4\gev$ has been calculated showing that for the NDR scheme
$B_{5,6}^{(1/2)}$ and $B_{7,8}^{(3/2)}$ were $\mu$ independent within
an accuracy of (2--3)\,\%. The $\mu$ dependence in the HV scheme has
been found to be stronger but still below 10\,\%.

In view of the fact that for $B^{(1/2)}_{6}=B^{(3/2)}_{8}=1$ and the
known value of $\mt$, there is a strong cancellation between
gluon and electroweak penguin contributions to $\epe$, it is
important to investigate whether the $1/N$ corrections significantly
affect this cancellation. First attempt in this direction has been
made by the Dortmund group \cite{heinrichetal:92, paschos:96}, 
which incorporating in part chiral
loops found an enhancement of $B_{6}^{(1/2)}$ and a suppression
of $B_{8}^{(3/2)}$. From \cite{paschos:96} $B^{(1/2)}_6=1.3$ and 
$B^{(3/2)}_8=0.7$ can be extracted.
 
Recently another Dortmund team \cite{DORT98}, in collaboration with Bill
Bardeen, performed this time a complete investigation
of $\langle Q_6\rangle_0$ and $\langle Q_8\rangle_2$ in the
twofold expansion in powers of external momenta $p$, and in 
$1/N$. Their final result gives $\langle Q_6\rangle_0$ and 
$\langle Q_8\rangle_2$ including the orders $p^2$ and $p^0/N$.
For $\langle Q_8\rangle_2$ also the term $p^0$ contributes.
Of particular interest are the $\ord(p^0/N)$ contributions
resulting from non-factorizable chiral loops which are
important for the matching between long- and short-distance
contributions. The cut-off scale $\Lambda_c$ in these
non-factorizable diagrams is identified with the QCD renormalization
scale $\mu$ which enters the Wilson coefficients. In contrast
to the matrix elements of $Q_{1,2}$ in which the $\Lambda_c$
dependence was quadratic \cite{bardeen:87}, 
the $\Lambda_c$ dependence in the present
case is logarithmic which improves the matching considerably.
There are several technical and conceptual improvements in \cite{DORT98}
over the first attempt in \cite{heinrichetal:92, paschos:96} 
and also over the original approach \cite{bardeen:87}.
Therefore I strongly recommend to read \cite{DORT98}, which is clearly
written.

\begin{table}[thb]
\caption[]{ Results for $B^{(3/2)}_{7,8}$ obtained
in the $1/N$ approach. 
\label{tab:318}}
\begin{center}
\begin{tabular}{|c||c|c|c|c|}\hline
  & $\Lambda_c=0.6~\gev$ & $\Lambda_c=0.7~\gev$ & 
$\Lambda_c=0.8~\gev$ & $\Lambda_c=0.9~\gev$ \\
 \hline\hline
$B_{6}^{(1/2)}$ & $1.10$ &$0.96$ & $0.84$ & $ 0.72 $ \\
                & $(1.30)$ &$(1.19)$ & $(1.09)$ & $(0.99) $ \\
$B_{8}^{(3/2)}$ & $0.66$ &$0.59$ & $0.52$ & $ 0.46 $ \\
                & $(0.71)$ &$(0.65)$ & $(0.60)$ & $(0.54) $ \\
\hline
\end{tabular}
\end{center}
\end{table}
In table \ref{tab:318}, taken from \cite{DORT98}, we show the values of
$B_{6}^{(1/2)}$ and $B_{8}^{(3/2)}$ as functions of the cut-off
scale $\Lambda_c$. The results depend on whether $F_\pi$ or
$F_K$ is used in the calculation, the difference being of higher
order. The results using $F_K$ are shown in the parentheses.
The decrease of both B--factors with $\Lambda_c=\mu$ is qualitatively
consistent with their $\mu$-dependence found for $\mu\ge 1$ in
\cite{BJLW}, but it is much stronger. Clearly one could also
expect stronger $\mu$-dependence in the analysis of \cite{BJLW}
for $\mu\le 1~\gev$, but in view of large perturbative corrections
for such small scales a meaningfull test of the dependence in
table \ref{tab:318} cannot be made.
We note that for $\Lambda_c=0.7~\gev$  the value of $B_{6}^{(1/2)}$
is close to unity as in the large-N limit. However, $B_{8}^{(3/2)}$
is considerably suppressed.

It is difficult to decide which value should be used in phenomenology
of $\epe$. On the one hand, for $\Lambda_c\ge 0.6~\gev$ neglected 
contributions from vector mesons in the loops should be included.
On the other hand for $\Lambda_c=\mu= 0.6~\gev$ it is difficult
to make contact with short distance calculations and with the
lattice results which are obtained for $\mu=2~\gev$. As for
$\mu \ge 1~\gev$ the parameters in table \ref{tab:318} are expected to be
almost $\mu$-independent, let us take the values at $\Lambda_c=0.9~\gev$
as the main result of \cite{DORT98}. 
Averaging over the $F_\pi$- and $F_K$-choices
we find
\be\label{LNN}
B^{(1/2)}_{6}=0.85\pm0.13~, \quad\quad B^{(3/2)}_{8}=0.50\pm 0.04~,
\quad\quad(\Lambda_c=0.9~\gev)
\ee
where the errors should not be taken too seriously. The value
of $B^{(1/2)}_{6}$ is compatible with the corresponding lattice 
results, whereas $ B^{(3/2)}_{8}$ is found to be substantially
smaller in the $1/N$ approach. On the other hand, it will be
tempting later on to calculate $\epe$ for the choice 
$\Lambda_c=0.6~\gev$ which gives instead:
\be\label{LNN1}
B^{(1/2)}_{6}=1.2\pm0.1~, \quad\quad B^{(3/2)}_{8}=0.68\pm 0.03~,
\quad\quad (\Lambda_c=0.6~\gev).
\ee
In view of the large correction to
$B^{(3/2)}_{8}$ one might question the convergence of the $1/N$
expansion. However, the non-factorizable contributions
considered in \cite{DORT98} represent the first term in a new type 
of a series
absent in the large-N limit and consequently there are no strong
reasons for questioning the convergence of the $1/N$ expansion on
the basis of these results. In this context one should also
remark that the lattice studies discussed previously use tree level
chiral perturbation theory to relate the matrix elements
$\langle \pi\pi| Q_i|K\rangle $ to $\langle\pi| Q_i|K\rangle $
which are calculated on the lattice. It is conceivable that
including chiral loops in this relation would decrease the
value of $ B^{(3/2)}_{8}$.

Finally I would like to express one criticism of the approach
in \cite{DORT98} as well as in 
\cite{bardeen:87}. It is the lack of any reference to the
renormalization scheme dependence which is necessary for
a complete matching at the NLO level.

\subsubsection{$B^{(1/2)}_{6}$ and $B^{(3/2)}_{8}$ from 
the Chiral Quark Model}
Effective Quark Models of QCD can be derived in the framework of 
the extended Nambu-Jona-Lasinio model of chiral symmetry 
breaking \cite{NJL}.
For kaon decays and in particular for $\epe$, an extensive analysis
of this model inclusive chiral loops, gluon and $\ord(p^4)$ corrections
has been performed over the last years by the Trieste group
\cite{TR96,TR97}. The
crucial parameters in this approach is a mass parameter $M$ and
the condensates $\langle\bar q q\rangle $ and $\langle\as GG\rangle$.
They can be constrained by imposing the $\Delta I=1/2$ rule.

Since there exists a recent nice review \cite{BERT98} by the Trieste 
group of their
approach, I will only quote here their estimate of the relevant 
$B_i$ parameters. They are
\be\label{TRIESTE}
B^{(1/2)}_{6}=1.6\pm0.3 \quad\quad B^{(3/2)}_{8}=0.92\pm 0.02~.
\quad\quad{(\rm Chiral~QM)}
\ee
We observe a rather large enhancement of $B^{(1/2)}_{6}$, not observed
by other groups, and a
moderate suppression of $B^{(3/2)}_{8}$. These parameters correspond
roughly to the scale $\mu=0.8~\gev$. Looking at the table
\ref{tab:318} we may expect a $10\%$ reduction of these values,
had the scale $\mu=0.9~\gev$ been used.

\subsubsection{Strategy for $(V-A) \otimes (V+A)$ Operators}
We have seen that various approaches differ in their estimates
of the most important parameters $B^{(1/2)}_{6}$ and $B^{(3/2)}_{8}$.
In table \ref{tab:31739} we collect the central values from various
approaches discussed above. In the case of lattice we have chosen
various possible scenarios in view of different results obtained
by various groups. Similarly in the case of the $1/N$ approach
we have chosen two sets of B-values corresponding to two values
of $\Lambda_c$. 
Even if
the $B_i$ factors given in this table are all within say $50\%$ from the
vacuum insertion estimate, they give rather different results
for central values of $\epe$ as illustrated in the last two columns
of this table. How these values have been obtained will be discussed 
a few pages below.
\begin{table}[thb]
\caption[]{ Results for $\epe$ in units of $10^{-4}$ 
for three choices of $\ms(\mc)$
and the central values of  $B^{(1/2)}_6$ and $B^{(3/2)}_8$ 
obtained in various approaches. $\IM\lambda_t=1.29\cdot 10^{-4}$
and $\mt=167\gev$ have been used.
\label{tab:31739}}
\begin{center}
\begin{tabular}{|c|c|c||c|c|c|}\hline
  Approach & $B^{(1/2)}_6$& $B^{(3/2)}_8$ & $150~\mev$& 
 $125~\mev$ & $100~\mev$ \\ \hline
  VIA    & $1.0$ &$1.0$ & $3.2$ & $5.2$ & $8.8$ \\
\hline
Lattice 1    & $1.0$ &$0.81$ & $4.2$ & $6.6$ & $10.9$  \\
Lattice 2    & $1.0$ &$1.03$ & $3.0$ & $5.0$ & $8.4$ \\
Lattice 3    & $0.76$ &$0.81$ & $1.7$ & $3.1$ & $5.7$ \\
Lattice 4    & $0.76$ &$1.03$ & $0.6$ & $1.5$ & $3.2$  \\
\hline
1/N (I) & $0.85$ &$0.50$ & $4.3$ & $6.7$ & $11.1$  \\
1/N (II) & $1.2$ &$0.68$ & $6.9$ & $10.4$ &$ 16.7$\\
\hline
Chiral QM & $1.6$ &$0.92$ & $9.7$ & $14.4$ & $22.7$  \\
\hline
\end{tabular}
\end{center}
\end{table}

Concerning $B_{7,8}^{(1/2)}$ one can simply set $B_{7,8}^{(1/2)}=1$ as
the matrix elementes $\langle Q_{7,8} \rangle_0$ play only a minor role
in the $\epe$ analysis. I should however stress that whereas lattice
results are consistent with this choice, this is not the case for
the chiral quark model \cite{TR97} in which values as high 
as 2.5 are found.

Concerning $B_5^{(1/2)}$ and $B_7^{(3/2)}$ we will simply
set them equal to $B^{(1/2)}_{6}$ and $B^{(3/2)}_{8}$ respectively.
This is consistent with the lattice results and the chiral
quark model. There are no results for these parameters from
the $1/N$ approach beyond the large-N limit.

In summary the treatment of $\langle Q_i \rangle_{0,2}$, $i=5,\ldots 8$
in \cite{BJLW,BBL,BJL96a} is to set
\begin{equation}
B_{7,8}^{(1/2)}(\mc) = 1,
\qquad
B_5^{(1/2)}(\mc) = B_6^{(1/2)}(\mc),
\qquad
B_7^{(3/2)}(\mc) = B_8^{(3/2)}(\mc)
\label{eq:B1278mc}
\end{equation}
and to treat $B_6^{(1/2)}(\mc)$ and $B_8^{(3/2)}(\mc)$ as free
parameters. In particular, in addition to estimates obtained
by other groups, we will
show below the results for $\epe$ when these parameters are varied 
in the ranges
\begin{equation}
B_6^{(1/2)}(\mc)=1.0 \pm 0.2,
\qquad
B_8^{(3/2)}(\mc)=1.0\pm 0.2.
\label{eq:B78mc}
\end{equation}
and
\begin{equation}
B_6^{(1/2)}(\mc)=1.0 \pm 0.2,
\qquad
B_8^{(3/2)}(\mc)=0.7\pm 0.2.
\label{eq:B87mc}
\end{equation}
The range (\ref{eq:B78mc}) corresponds to the variation of the $B_i$ parameters
in the neighbourhood of the large--N limit. The range (\ref{eq:B87mc})
gives a rough description of the fact that in recent analyses 
most approaches find $B_8^{(3/2)}$ to be smaller than $B_6^{(1/2)}$.
This range will be analyzed at the end of this section.

After this long exposition of $B_i$ parameters let us then
incorporate the collected information in the formula for
$\epe$ in a manner useful for phenomenological applications.

\subsection{An Analytic Formula for $\epe$}
           \label{subsec:epeanalytic}
As shown in \cite{buraslauten:93}, it is possible to cast the formal
expression for $\epe$ in (\ref{eq:epe})
into an analytic formula which exhibits the $\mt$ dependence
together with the dependence on $\ms$, $\Lms^{(4)}$,
 $B_6^{(1/2)}$ and $B_8^{(3/2)}$.
Such an analytic formula should be useful for those phenomenologists
and experimentalists who are not interested in getting involved with
the technicalities discussed above.

In order to find an analytic expression for $\epe$, which exactly
reproduces the numerical results based on the formal OPE method,
one uses the PBE presented in Section 3.3. The updated  
analytic formula for $\epe$ of \cite{buraslauten:93} 
presented in \cite{BJL96a}
is given as follows:
\begin{equation}
\frac{\varepsilon'}{\varepsilon} = {\rm Im}\lambda_t \cdot F(x_t) \, ,
\label{eq:3}
\end{equation}
where
\begin{equation}
F(x_t) =
P_0 + P_X \, X_0(x_t) + P_Y \, Y_0(x_t) + P_Z \, Z_0(x_t) 
+ P_E \, E_0(x_t) 
\label{eq:3b}
\end{equation}
and
\begin{equation}
{\rm Im}\lambda_t = {\rm Im} V_{ts}^*V_{td} = |V_{\rm ub}| \, 
|V_{\rm cb}| \, \sin \delta = \eta \, \lambda^5 \, A^2
\label{eq:4}
\end{equation}
in the standard parameterization of the CKM matrix
(\ref{2.72}) and in the Wolfenstein parameterization
(\ref{2.75}), respectively. 

The $\mt$-dependent functions in (\ref{eq:3b}) are given in 
(\ref{E0}) and (\ref{X0})--(\ref{Z0}).
The coefficients $P_i$ are given in terms of $B_6^{(1/2)} \equiv
B_6^{(1/2)}(\mc)$, $B_8^{(3/2)} \equiv B_8^{(3/2)}(\mc)$ and $\ms(\mc)$
as follows:
\begin{equation}
P_i = r_i^{(0)} + R_s 
\left(r_i^{(6)} B_6^{(1/2)} + r_i^{(8)} B_8^{(3/2)} \right) \, ,
\label{eq:pbePi}
\end{equation}
where
\be\label{RS}
R_s=\left[ \frac{158\mev}{\ms(\mc)+\md(\mc)} \right]^2.
\ee
The $P_i$ are renormalization scale and scheme independent. They depend,
however, on $\Lms^{(4)}$. In table~\ref{tab:pbendr} we give the numerical
values of $r_i^{(0)}$, $r_i^{(6)}$ and $r_i^{(8)}$ for different values
of $\Lms^{(4)}$ at $\mu=\mc$ in the NDR renormalization scheme. 
The
coefficients $r_i^{(0)}$, $r_i^{(6)}$ and $r_i^{(8)}$ depend only very
weakly on
$\ms(\mc)$ as the dominant $\ms$ dependence has been factored out. The
numbers given in table~\ref{tab:pbendr} correspond to $\ms(\mc)=150\,\mev$.
However, even for $\ms(\mc)\approx100\mev$, the analytic expressions given
here reproduce the numerical calculations of $\epe$ given below 
to better than $4\%$.
For different scales $\mu$ the numerical values in the tables change
without modifying the values of the $P_i$'s as it should be. The values
of $B_6^{(1/2)}$ and $B_8^{(3/2)}$ should also be  modified, 
in principle, but in view of the comments made previously it
is a good approximation to keep them $\mu$-independent
for $\mu\ge 1~\gev$.

Concerning the scheme dependence only the $r_0$ coefficients
are scheme dependent at the NLO level. Their values in the HV
scheme are given in the last row of table~\ref{tab:pbendr}.
The coefficients $r_i$, 
$i=X, Y, Z, E$ are on the other hand scheme independent at NLO. 
This is related to the fact that the $\mt$
dependence in $\epe$ enters first at the NLO level and consequently all
coefficients $r_i$ in front of the $\mt$ dependent functions must be
scheme independent. 
Consequently, when changing the renormalization scheme, one is only
obliged to change appropriately $B_6^{(1/2)}$ and $B_8^{(3/2)}$ in the
formula for $P_0$ in order to obtain a scheme independence of $\epe$.
In calculating $P_i$ where $i \not= 0$, $B_6^{(1/2)}$ and $B_8^{(3/2)}$
can in fact remain unchanged, because their variation in this part
corresponds to higher order contributions to $\epe$ which would have to
be taken into account in the next order of perturbation theory.

For similar reasons the NLO analysis of $\epe$ is still insensitive to
the precise definition of $\mt$. In view of the fact that the NLO
calculations needed to extract $\IM \lambda_t$ (see previous section) 
have been done with $\mt=\overline{m}_t(\mt)$ we will also use  this 
definition in calculating $F(x_t)$. 

\begin{table}[thb]
\caption[]{PBE coefficients for $\epe$ for various $\Lms^{(4)}$ in 
the NDR scheme.
The last row gives the $r_0$ coefficients in the HV scheme.
\label{tab:pbendr}}
\begin{center}
\begin{tabular}{|c||c|c|c||c|c|c||c|c|c|}
\hline
& \multicolumn{3}{c||}{$\Lms^{(4)}=245\mev$} &
  \multicolumn{3}{c||}{$\Lms^{(4)}=325\mev$} &
  \multicolumn{3}{c| }{$\Lms^{(4)}=405\mev$} \\
\hline
$i$ & $r_i^{(0)}$ & $r_i^{(6)}$ & $r_i^{(8)}$ &
      $r_i^{(0)}$ & $r_i^{(6)}$ & $r_i^{(8)}$ &
      $r_i^{(0)}$ & $r_i^{(6)}$ & $r_i^{(8)}$ \\
\hline
0 &
   --2.674 &   6.537 &   1.111 &
   --2.747 &   8.043 &   0.933 &
   --2.814 &   9.929 &   0.710 \\
$X$ &
    0.541 &   0.011 &       0 &
    0.517 &   0.015 &       0 &
    0.498 &   0.019 &       0 \\
$Y$ &
    0.408 &   0.049 &       0 &
    0.383 &   0.058 &       0 &
    0.361 &   0.068 &       0 \\
$Z$ &
    0.178 &  --0.009 &  --6.468 &
    0.244 &  --0.011 &  --7.402 &
    0.320 &  --0.013 &  --8.525 \\
$E$ &
    0.197 &  --0.790 &   0.278 &
    0.176 &  --0.917 &   0.335 &
    0.154 &  --1.063 &   0.402 \\
\hline
0 &
   --2.658 &   5.818 &   0.839 &
   --2.729 &   6.998 &   0.639 &
   --2.795 &   8.415 &   0.398 \\
\hline
\end{tabular}
\end{center}
\end{table}

The analytic formulae given above are useful for numerical
calculations, but in order to identify the dominant terms in an
elegant manner,
we follow Gupta \cite{GUPTA98} and rewrite it as
\be\label{GUPT}
\frac{\varepsilon'}{\varepsilon} = {\rm Im}\lambda_t \cdot 
\left[c_0+(c_6 B_6^{(1/2)} + c_8 B_8^{(3/2)}) R_s \right].
\end{equation}
For $\mt=167~\gev$ the values of the coefficients $c_i$ are given in
table \ref{cen}.

\begin{table}[thb]
\caption[]{The coefficients $c_i$ for various $\Lms^{(4)}$ in 
the NDR and HV schemes and $\mt=167~\gev$.
\label{cen}}
\begin{center}
\begin{tabular}{|c||c|c||c|c||c|c|}
\hline
& \multicolumn{2}{c||}{$\Lms^{(4)}=245\mev$} &
  \multicolumn{2}{c||}{$\Lms^{(4)}=325\mev$} &
  \multicolumn{2}{c| }{$\Lms^{(4)}=405\mev$} \\
\hline
${\rm Scheme}$ & ${\rm NDR}$ & ${\rm HV}$ &
${\rm NDR}$ & ${\rm HV}$ & ${\rm NDR}$ & ${\rm HV}$ \\
\hline
$c_0$ &
   --1.264 &   --1.248 &
   --1.359 &   --1.341 &
   --1.430 &   --1.411   \\
$c_6$ &
    6.387 &   5.668 &
    7.873 &   6.828 &
    9.735 &   8.221  \\
$c_8$ &
    --3.259 & --3.531 &
    --4.063 & --4.357 &
    --5.041 & --5.353  \\
\hline
\end{tabular}
\end{center}
\end{table}

The inspection of tables~\ref{tab:pbendr} and \ref{cen} shows
that within a few percent
\be
c_6=r_0^{(6)},\quad\quad c_8=r_0^{(8)}+ r_Z^{(8)} Z_0(x_t),
\ee 
whereby $c_8$ is dominated by the second term.
Thus we conclude 
that the terms involving $r_0^{(6)}$ and $r_Z^{(8)}$ dominate the ratio
$\epe$. Moreover, the function $Z_0(x_t)$ representing a gauge invariant
combination of $Z^0$- and $\gamma$-penguins grows rapidly with $\mt$
and due to $r_Z^{(8)} < 0$ these contributions suppress $\epe$ strongly
for large $\mt$ \cite{flynn:89,buchallaetal:90} as stressed at the
beginning of this section. 

\subsection{The Status of the Strange Quark Mass}
It seems appropriate to summarize now the present status of 
the value of the strange quark mass.
In the case of quenched lattice QCD this has been recently done
by Gupta \cite{GUPTA98}. His final result based on 1997 world data is
\be
\ms(2\gev)=(110\pm25)~\mev.
\ee
It is expected that unquenching will lower this value but it is difficult
to tell by how much. 

Gupta summarized also the most recent values
for $\ms(2\gev)$ obtained using QCD sum rules. The older values (in $\mev$)
are $144\pm 21$ \cite{narison:95}, $137\pm23$ \cite{jaminmuenz:95}
$148\pm 15$ \cite{chetyrkinetal:95}, whereas the most recent ones are
found to be $91-116$ \cite{Paver} and $115\pm 22$ \cite{Jamin97}.
On the other hand the following {\it lower bounds} on $\ms(2\gev)$ have
been derived: $118-189$ \cite{Yndurain}, $88\pm 9$ \cite{Dosch},
$104-116$ \cite{DERAF}. We observe that the QCD sum rule results are
consistent with quenched lattice values although generally they are
somewhat higher.

We conclude that the error on $\ms$ is still rather large.
Therefore it will be 
useful to present, few pages below,
 the results for $\epe$ for two values of $\ms(\mc)$:
\begin{equation}\label{msvalues}
\ms(\mc)=(150\pm20)\,\mev
\quad
{\rm and}
\quad
\ms(\mc)=(125\pm20)\,\mev
\end{equation}
corresponding (see table \ref{tab:ms}) roughly to
$\ms(2~\gev)=(129\pm17)\,\mev$ and $\ms(2~\gev)=(107\pm17)\,\mev$,
respectively.

Finally one should remark that the decomposition of the relevant hadronic
matrix elements of penguin operators into a product of $B_i$ factors times
$1/m_s^2$, although useful in the $1/N_c$ approach, is in principle 
unnecessary in a brute
force method like the lattice approach and in certain methods using
effective lagrangians. It is to be expected that the
future lattice calculations will directly give the relevant hadronic 
matrix elements and the issue of $\ms$ in connection with $\epe$ will
effectively disappear.

\subsection{Numerical Results for $\epe$}
In order to complete the analysis of $\epe$ one needs the value of ${\rm
Im}\lambda_t$. Since this value has been already determined in 
section \ref{sec:standard} (see table \ref{TAB2}), 
we are ready to present the results for $\epe$. 
In order to gain some insight in what is going on, let us take
the formula (\ref{GUPT}) and insert the central value 
$\IM\lambda_t=1.29\cdot 10^{-4}$ together with the NDR-values in 
table \ref{cen} for  $\Lms^{(4)}=325~\mev$. We find then
\be\label{GUPT1}
\frac{\varepsilon'}{\varepsilon} =  
\left[-1.75+(10.15\cdot B_6^{(1/2)} -5.24\cdot B_8^{(3/2)}) R_s \right]
\cdot 10^{-4}
\end{equation}
with $R_s$ defined in (\ref{RS}).

Our ``central" formula (\ref{GUPT1}) gives then the values of $\epe$
collected in table \ref{tab:31739}. We observe that for higher values
of $\ms$ the lattice and the 1/N approach (I) give values of $\epe$ in
the ball park of a few $10^{-4}$. Higher values are obtained for
the 1/N approach (II) 
and in particular in the chiral quark
model which even in the first scenario for $m_s$ gives value of $\epe$
close to $\ord(10^{-3})$. For smaller values of $\ms$ all approaches
give higher values of $\epe$ although only the last two give results
consistent with the NA31 value.
The results for the $1/N$ approach (II) are only shown for illustration.
A proper analysis of this case would require the calculation of Wilson
coefficients for $\mu$ well below $1~\gev$, which we do not want to do.

When analyzing these numbers some caution is needed. 
Our ``central" formula (\ref{GUPT1}) includes certain inputs which
are not necessarily the same in all approaches. For instance our
value of $\hat B_K$ is lower than the values obtained in the
  lattice and chiral model approaches. Similarly the value $c_0$
is very much constrained by the incorporation of 
the $\Delta I=1/2$ rule which
cannot be obtained using VIA. In addition in a given approach
$c_6$ and $c_8$ may differ somewhat from the ones used. But since
they are dominated by the short distance Wilson coefficients these
changes cannot be large and
we belive that our formula is not too bad and gives some
insight in what is going on.

On the other hand, once one begins to vary all input parameters
the differences between various approaches wash out to some extend.
We note for instance that the coefficients in tables \ref{tab:pbendr}
and \ref{cen}
exhibit a sizable $\Lms^{(4)}$-dependence leading to almost linear
dependence of $\epe$ on this parameter as pointed out in \cite{BJLW}.  

Let me than present results of the Munich group based on the input
parameters of section 10 and the choice of $B_i$ parameters
summarized in (\ref{eq:B78mc}). To this end exact expressions for
$\epe$ have been used.

For $m_s(\mc)=150\pm20\mev$ one finds \cite{BJL96a}
\begin{equation}
-1.2 \cdot 10^{-4} \le \epe \le 16.0 \cdot 10^{-4}
\label{eq:eperangenew}
\end{equation}
and
\begin{equation}
\epe= ( 3.6\pm 3.4) \cdot 10^{-4}
\label{eq:eperangefinal}
\end{equation}
for the ``scanning'' method and the ``gaussian'' method discussed
in section \ref{sec:standard}, respectively.
Using on the other hand $\ms(\mc)=(125\pm20)\mev$ one finds 
respectively \cite{BJL96b}:
\begin{equation}
-0.5\cdot 10^{-4} \le \epe \le 25.2 \cdot 10^{-4}
\label{eq:eperangenewa}
\end{equation}
and
\begin{equation}
\epe= ( 6.1\pm 5.2) \cdot 10^{-4}
\label{eq:eperangefinala}
\end{equation}
In \cite{BJL96a} the choice $\ms(\mc)=(100\pm20)\mev$ has been
considered giving $0 \le \epe \le 43.0 \cdot 10^{-4}$ and
$\epe= ( 10.4\pm 8.3) \cdot 10^{-4}$ respectively, but such low
values of $\ms(\mc)$ seem now rather improbable.

In table \ref{tab:31738} we compare these results with the existing
results obtained by various groups. There exists no recent 
phenomenological analysis from the Dortmund group based on the $B_i$
parameters obtained in \cite{DORT98}. The older result 
$\epe=(9.9\pm4.1)\cdot 10^{-4}$ from this group will certainly be
superceded by a new analysis which hopefully will be available soon.

We observe that
the result for $m_s( \mc)=150\pm20\mev$ 
in (\ref{eq:eperangefinal}) agrees rather well with
the 1996 analysis of the Rome group \cite{ciuchini:96}.
On the other hand the range in (\ref{eq:eperangenew}) shows that for
particular choices of the input parameters, values for $\epe$ as high as
$16\cdot 10^{-4}$ cannot be excluded. Such high values are
found if simultaneously  $\vub=0.10$, $B_6^{(1/2)}=1.2$, 
$B_8^{(3/2)}=0.8$, $B_K=0.6$,
$\ms(\mc)=130$ MeV, $\Lms^{(4)}=405\mev$ and low values of $\mt$ still
consistent with $\varepsilon_K$ and the observed $B_d^0-\bar B_d^0$ 
mixing
are chosen. It is, however, evident from  the comparision of
(\ref{eq:eperangenew}) and (\ref{eq:eperangefinal})  that such 
high values of $\epe$ and generally values above $10^{-3}$ 
are very improbable for $\ms(\mc)={\cal O}(150\mev)$.

\begin{table}[thb]
\caption[]{ Results for $\epe$ in units of $10^{-4}$ obtained
by various groups. The labels (S) and (G) in the last column
stand for ``Scanning'' and ``Gaussian'' respectively, as discussed
in the text. 
\label{tab:31738}}
\begin{center}
\begin{tabular}{|c|c|c|c||c|}\hline
  {\bf Reference}& $B^{(1/2)}_6$& $B^{(3/2)}_8$ & $\ms(\mc)[\mev]$ &
 $\epe[10^{-4}]$ \\ \hline
Munich
\cite{BJL96a}& $1.0\pm 0.2$ &$1.0\pm0.2$ & $150\pm20$ & $-1.2\to 16.0$ (S) \\
Munich
\cite{BJL96a}& $1.0\pm 0.2$ &$1.0\pm0.2$ & $150\pm20$ & $3.6\pm 3.4$ (G) \\
Munich
\cite{BJL96b}& $1.0\pm 0.2$ &$1.0\pm0.2$ & $125\pm20$ & $-0.5\to 25.2$ (S) \\
Munich
\cite{BJL96b}& $1.0\pm 0.2$ &$1.0\pm0.2$ & $125\pm20$ & $6.1\pm 5.2$ (G) \\
\hline
Rome
\cite{ciuchini:96}& $1.0\pm 0.2$ &$1.0\pm0.2$ & $150\pm20$ & 
$4.6\pm 3.0$ (G) \\
\hline
Trieste
\cite{BERT98}& $1.6\pm 0.3$ &$0.92\pm0.02$ & $-$ & 
$7\to 31$ (S) \\
\hline
Dubna-DESY
\cite{BELKOV} 
& $1.0$ &$1.0$ & $-$ & $-3.0 \to 3.6$ (S) \\
\hline
\end{tabular}
\end{center}
\end{table}

We observe that our ``gaussian'' result for $\ms(\mc)=(125\pm20)\mev$
agrees well with the E731
value and,
as stressed in \cite{BJL96a}, the decrease of $\ms$
 even below $\ms(\mc)= 100$ MeV is insufficient to bring 
the Standard Model in agreement with
the NA31 result provided $B_6=B_8=1.$
However, for $B_6>B_8$, sufficiently large values of
$\IM\lambda_t$ and $\Lms^{(4)}$, and small values of $\ms$, the values
of $\epe$ in the Standard Model can be as large as $(1-2)\cdot 10^{-3}$
and consistent with the NA31 result.
In order to see this explicitly we present in table \ref{tab:31731} the
values of $\epe$ for three choices of $\ms(\mc)$ and for selective
sets of other input parameters keeping  $\mt=167\,\gev$
fixed. 

The Trieste group finds generally higher values of
$\epe$, with the central value around $17\cdot 10^{-4}$ and consequently
consistent with the NA31 result. On the basis of table \ref{tab:31739}
we expect the $\epe$ from the Dortmund group to be below the one from
Trieste but generally higher than the results from Munich and Rome
for the same value of $\ms$.  

Finally I should comment on the results of \cite{BELKOV} where
$\epe$ has been investigated in the framework of an effective chiral
lagrangian approach. In this approach the values of
$B_6^{(1/2)}$ and $B_8^{(3/2)}$ cannot be calculated and the authors
set them to unity in order to obtain the values quoted in table
\ref{tab:31738}. In spite of joined efforts with Bill Bardeen to
understand this work
and discussions with these authors I failed to appreciate fully this
approach. These authors find $\epe$ consistent
with zero.

\begin{table}[thb]
\caption[]{ Values of $\epe$ in units of $10^{-4}$ 
for specific values of various input parameters at $\mt=167\,\gev$. 
\label{tab:31731}}
\begin{center}
\begin{tabular}{|c|c|c|c|c||c|}\hline
  $\IM\lambda_t[10^{-4}]$& 
$\Lms^{(4)}[MeV]$& $B^{(1/2)}_6$& $B^{(3/2)}_8$ & 
$\ms(\mc)[\mev]$ &
 $\epe[10^{-4}]$ \\ \hline
       &       &      &     & $100$ & $~8.8$ \\
$1.3$ & $325$ & $1.0$&$1.0$& $125$ & $~5.2$ \\
       &       &      &     & $150$ & $~3.2$ \\
 \hline
       &       &      &     & $100$ & $11.2$ \\
$1.3$ & $405$ & $1.0$&$1.0$ & $125$ & $~6.8$ \\
       &       &      &     & $150$ & $~4.2$ \\
 \hline
       &       &      &     & $100$ & $13.8$ \\
$1.6$ & $405$ & $1.0$&$1.0$ & $125$ & $~8.3$ \\
       &       &      &     & $150$ & $~5.2$ \\
 \hline\hline
       &       &      &     & $100$ & $12.2$ \\
$1.3$ & $325$ & $1.0$&$0.7$ & $125$ & $~7.5$ \\
       &       &      &     & $150$ & $~4.8$ \\
 \hline
       &       &      &     & $100$ & $15.4$ \\
$1.3$ & $405$ &$1.0$&$0.7$ & $125$ & $~9.5$ \\
       &       &      &     & $150$ & $~6.2$ \\
 \hline
       &       &      &     & $100$ & $19.0$ \\
$1.6$ & $405$ &$1.0$&$0.7$ & $125$ & $11.7$ \\
       &       &      &     & $150$ & $~7.6$ \\
 \hline
\end{tabular}
\end{center}
\end{table}

\subsection{Summary}
The fate of $\epe$ in the Standard Model after the
improved measurement of $\mt$ and complete NLO calculations of
short distance coefficients, depends sensitively on the values of
$|V_{ub}/V_{cb}|$, $\Lms^{(4)}$ and in particular on 
$B_6^{(1/2)}$, $B_8^{(3/2)}$ and $\ms$.
The predictions for $\epe$ obtained by
various groups are summarized in table \ref{tab:31738}. 
This table and the table \ref{tab:31731} show very clearly that any 
value for $\epe$ in
the range
\begin{equation}
0 \le \epe \le 3 \cdot 10^{-3}
\label{eq:epera}
\end{equation}
is still possible within the Standard Model at present, although most
estimates lie below $10^{-3}$ and in the range of E731 result.
Time will show which of the groups came closest to the true prediction.
It appears that most calculations give values of $B_6^{(1/2)}$ rather
close to unity and $B_8^{(3/2)}$ below one so that the inequality
$B_6^{(1/2)}\ge B_8^{(3/2)}$ should be expected to be true. If this
feature will survive more precise calculations and $\ms(\mc)$
will be eventually found in the range 
$125~\mev\le\ms(\mc)\le 150~\mev$ then $\epe$ within the Standard Model
should be somewhere between $5\cdot 10^{-4}$ and $1\cdot 10^{-3}$.
As an example let us then finally take the range (\ref{eq:B87mc}):
$B_6^{(1/2)}=1.0\pm 0.2$ and $B_8^{(3/2)}=0.7\pm 0.2$. Then the gaussian
analysis gives \cite{BJL96b}
\begin{equation}\label{eprimef}
\varepsilon'/\varepsilon =\left\{ \begin{array}{ll}
(5.3 \pm 3.8)\cdot 10^{-4}~, &~~\ms(\mc)=150\pm20\mev \\
(8.5 \pm 5.9)\cdot 10^{-4}~, & ~~\ms(\mc)=125\pm20\mev .\end{array} \right.
\end{equation}
In my opinion these results give the best representation of the
present status of $\epe$ in the Standard Model. 

One prominent physicist once told me that a person who spent
fifteen years in a given field should have enough insight into the
matters to be able to make predictions even if this is impossible
from a scientific point of view. In 1983 I made the first encounter
with $\epe$ and if the above was true I should have by now in my head
a precise prediction for $\epe$ within the Standard Model.
Clearly I do not have it, but I like to bet. Here is my bet for
the $\epe$ in the Standard Model
\begin{equation}
\epe= ( 7\pm 1) \cdot 10^{-4}.
\label{eq:bet}
\end{equation}
It is rather close to the central value of the Fermilab result in
(\ref{eprime}).
The value in (\ref{eq:bet}) corresponds to the average of the values in
(\ref{eprimef}) and the error is the one expected from new experiments.
Whether the new data will find this value is not really important as there
could be new physics invalidating my expectations.

On a more scientific level,
let us hope that the future experimental and theoretical results will
be sufficiently accurate to be able to see whether $\epe\not=0$,
whether the Standard Model agrees with the data or
whether some new physics can be discovered in this ratio. In any case the
coming years should be very exciting. 

\section{$B\to X_s\gamma$} 
\setcounter{equation}{0}
\subsection{General Remarks}
The rare decay $B\to X_s\gamma$ plays an important role in 
present day phenomenology. 
The effective Hamiltonian for $B\to X_s\gamma$ at scales 
$\mu_b={\cal O}(m_b)$
is given by
\begin{equation} \label{Heff_at_mu}
{\cal H}_{\rm eff}(b\to s\gamma) = - \frac{G_{\rm F}}{\sqrt{2}} V_{ts}^* V_{tb}
\left[ \sum_{i=1}^6 C_i(\mu_b) Q_i + C_{7\gamma}(\mu_b) Q_{7\gamma}
+C_{8G}(\mu_b) Q_{8G} \right]\,,
\end{equation}
where in view of $\mid V_{us}^*V_{ub} / V_{ts}^* V_{tb}\mid < 0.02$
we have neglected the term proportional to $V_{us}^* V_{ub}$.
Here $Q_1....Q_6$ are the usual four-fermion operators whose
explicit form is given in (\ref{O1})--(\ref{O3}). 
The remaining two operators,
characteristic for this decay, are the {\it magnetic--penguins}
\begin{equation}\label{O6B}
Q_{7\gamma}  =  \frac{e}{8\pi^2} m_b \bar{s}_\alpha \sigma^{\mu\nu}
          (1+\gamma_5) b_\alpha F_{\mu\nu},\qquad            
Q_{8G}     =  \frac{g}{8\pi^2} m_b \bar{s}_\alpha \sigma^{\mu\nu}
   (1+\gamma_5)T^a_{\alpha\beta} b_\beta G^a_{\mu\nu}  
\end{equation}
originating in the diagrams of fig.~\ref{L:12}.
In order to derive the contribution of $Q_{7\gamma}$ to the
Hamiltonian in (\ref{Heff_at_mu}), in the absence of QCD corrections,
one multiplies the vertex in (\ref{MGP})
by ``i'' and makes the replacement 
\begin{equation}
2i\sigma_{\mu\nu}q^\nu\to-\sigma^{\mu\nu}F_{\mu\nu}.
\end{equation}
Analogous
procedure gives the contribution of $Q_{8G}$.

\begin{figure}[hbt]
\vspace{0.10in}
\centerline{
\epsfysize=1.5in
\epsffile{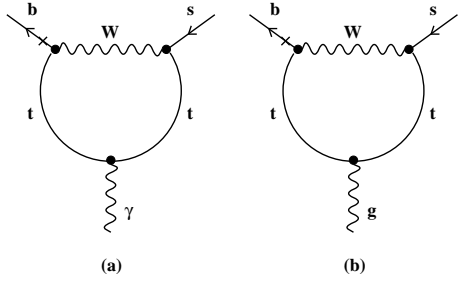}
}
\vspace{0.08in}
\caption[]{Magnetic Photon (a) and Gluon (b) Penguins.
\label{L:12}}
\end{figure}
It is the magnetic $\gamma$-penguin which plays the crucial role in
this decay. However, the role of the dominant current-current
operator $Q_2$ should not be underestimated.
Indeed the short distance QCD effects involving in particular the
mixing between $Q_2$  and $Q_{7\gamma}$ are very important in this decay.
They are known
\cite{Bert,Desh} to enhance $C_{7\gamma}(\mu_b)$ 
substantially, so that the resulting branching ratio
$Br(B\to X_s\gamma)$ turns out to be by a factor 
of 3 higher than it would be without QCD effects.
Since the first analyses
in \cite{Bert,Desh} a lot of progress has been made in calculating
these important  QCD effects beginning with the work in \cite{Grin,Odon}. 
We will briefly summarize this progress.

A peculiar feature of the renormalization group analysis 
in $B\to X_s\gamma$ is that the mixing under infinite renormalization 
between
the set $(Q_1...Q_6)$ and the operators $(Q_{7\gamma},Q_{8G})$ vanishes
at the one-loop level. Consequently in order to calculate 
the coefficients
$C_{7\gamma}(\mu_b)$ and $C_{8G}(\mu_b)$ in the leading logarithmic
approximation, two-loop calculations of ${\cal{O}}(e g^2_s)$ 
and ${\cal{O}}(g^3_s)$
are necessary. The corresponding NLO analysis requires the evaluation
of the mixing in question at the three-loop level. 
This peculiar feature caused
that the first fully correct calculation of the leading  anomalous
dimension matrix relevant for this decay
has been obtained only in 1993 \cite{CFMRS:93,CFRS:94}.
It has been
confirmed subsequently in \cite{CCRV:94a,CCRV:94b,Mis:94}.

As of 1998 also the NLO corrections to $B\to X_s\gamma$ 
have been completed.
It was a joint effort of many groups. Let us summarize this progress: 
\bi
\item
The $O(\alpha_s)$
corrections to $C_{7\gamma}(\mu_W)$ and $C_{8G}(\mu_W)$ have been first
calculated in \cite{Yao1} and recently confirmed by several groups
\cite{GH97,BKP2,GAMB}.
\item
The two-loop
mixing involving the operators
$Q_1.....Q_6$ and the two-loop mixing
in the sector $(Q_{7\gamma},Q_{8G})$ has been calculated in 
\cite{ACMP,WEISZ,BJLW1,BJLW,ROMA1,ROMA2} 
and \cite{MisMu:94}, respectively.  
Finally after a heroic effort  the three loop mixing between
the set $(Q_1...Q_6)$ and the operators $(Q_{7\gamma},Q_{8G})$
 has been completed at the end of 1996 \cite{CZMM}.
As a byproduct the authors of \cite{CZMM} confirmed the existing
two-loop anomalous dimension matrix in the $Q_1...Q_6$ sector.
\item
One-loop matrix elements 
$\langle s\gamma {\rm gluon}|Q_i| b\rangle$ have been calculated in 
\cite{AG2,Pott} and the very difficult two-loop corrections to 
$\langle s\gamma |Q_i| b\rangle$ have been presented in \cite{GREUB}.
\ei

We will now discuss all these achievements in explicit terms.
In order to appreciate the importance of NLO calculations for this
decay it is instructive to discuss first the leading logarithmic
approximation.
\subsection{The Decay $B\to X_s\gamma$ in the Leading Log Approximation}
         \label{sec:Heff:Bsgamma:lo}
\subsubsection{Anomalous Dimension Matrix}
It is instructive to to discuss first the mixing between 
the sets $Q_1,\ldots,Q_6$ and $Q_{7\gamma},Q_{8G}$ in $\hat\gamma_s^{(0)}$.
To this end I use the work done in colaboration with Misiak, M\"unz
and Pokorski \cite{BMMP:94}.
The point is that this mixing resulting from two-loop
diagrams is generally regularization scheme dependent. This is
certainly disturbing because the matrix $\hat\gamma_s^{(0)}$, being the
first term in the expansion for $\hat\gamma_s$, is usually scheme
independent.  As we will show below, there is a simple way to circumvent 
this difficulty \cite{BMMP:94}.

As noticed in \cite{CFMRS:93,CFRS:94} the regularization scheme
dependence of $\hat\gamma_s^{(0)}$ in the case of $b\to s\gamma$ and
$b\to s g$ is signaled in the finite parts of the one-loop matrix 
elements of $Q_1,\ldots,Q_6$
for on-shell photons or gluons.  They vanish in any 4-dimensional
regularization scheme and in the HV scheme but some of them are
non-zero in the NDR scheme.  One has
\begin{equation}
\langle Q_i \rangle_{\rm one-loop}^\gamma =
y_i \, \langle Q_{7\gamma} \rangle_{\rm tree},
\qquad i=1,\ldots,6
\label{eq:defy}
\end{equation}
and
\begin{equation}
\langle Q_i\rangle_{\rm one-loop}^G =
z_i \, \langle Q_{8G} \rangle_{\rm tree},
\qquad i=1,\ldots,6.
\end{equation}

In the HV scheme all the $y_i$'s and $z_i$'s vanish, while in the NDR
scheme $\vec{y} = (0,0,0,0,-\frac{1}{3},-1)$ and $\vec{z} =
(0,0,0,0,1,0)$.  This regularization scheme dependence is canceled by a
corresponding regularization scheme dependence in $\hat\gamma_s^{(0)}$
as first demonstrated in \cite{CFMRS:93,CFRS:94}. It should be
stressed that the numbers $y_i$ and $z_i$ come from divergent, i.e.
purely short-distance parts of the one-loop integrals. So no reference
to the spectator-model or to any other model for the matrix elements is
necessary here.

In view of all this  it is convenient in the leading order to introduce
the so-called ``effective coefficients'' \cite{BMMP:94} for the
operators $Q_{7\gamma}$ and $Q_{8G}$ which are regularization scheme
independent. They are given as follows:
\begin{equation} \label{eq:defc7eff}
C^{(0)eff}_{7\gamma}(\mu_b) =
C^{(0)}_{7\gamma}(\mu_b) + \sum_{i=1}^6 y_i C^{(0)}_i(\mu_b)
\end{equation}
and 
\begin{equation}\label{eq:defc8eff}
C^{(0)eff}_{8G}(\mu_b) = C^{(0)}_{8G}(\mu_b) 
+ \sum_{i=1}^6 z_i C^{(0)}_i(\mu_b).
\end{equation}
One can then introduce a scheme-independent vector
\begin{equation} 
\vec{C}^{(0)eff}(\mu_b) = \left( C^{(0)}_1(\mu_b),\ldots, C^{(0)}_6(\mu_b), 
C^{(0)eff}_{7\gamma}(\mu_b),C^{(0)eff}_{8G}(\mu_b) \right) \, .
\end{equation}
From the RGE for $\vec{C}^{(0)}(\mu)$ it is straightforward
to derive the RGE for $\vec{C}^{(0)eff}(\mu)$. It has the form
\begin{equation} \label{RGEeff}
\mu \frac{d}{d \mu} C^{(0)eff}_i(\mu) = 
\frac{\as}{4\pi} \gamma^{(0)eff}_{ji} C^{(0)eff}_j(\mu)
\end{equation}
where
\begin{equation} \label{def.geff}
\gamma^{(0)eff}_{ji} = \left\{ \begin{array}{ccl}
\gamma^{(0)}_{j7} +
\sum_{k=1}^6 y_k\gamma^{(0)}_{jk} -y_j\gamma^{(0)}_{77} -z_j\gamma^{(0)}_{87}
&\quad& $i=7$,\ $j=1,\ldots,6$ \\
\gamma^{(0)}_{j8} +
\sum_{k=1}^6 z_k\gamma^{(0)}_{jk} -z_j\gamma^{(0)}_{88}
&\quad& $i=8$,\ $j=1,\ldots,6$ \\
\gamma^{(0)}_{ji} &\quad& \mbox{otherwise.}
\end{array}
\right.
\end{equation}
The matrix $\hat\gamma^{(0)eff}$ is a scheme-independent quantity.
It equals the matrix which one would directly obtain from two-loop
diagrams in the HV scheme.  In order to simplify the notation we will
omit the label ``eff'' in the expressions for the elements of this
effective one loop anomalous dimension matrix given below and keep it
only in the Wilson coefficients of the operators $Q_{7\gamma}$ and
$Q_{8G}$.

We are now ready to give the leading anomalous dimension matrix 
relevant for the calculation of the $B\to X_s\gamma $ rate in the
LO approximation.
The
$6 \times 6$ submatrix of $\hat\gamma^{(0)}$ involving the operators
$Q_1,\ldots,Q_6$ is given in (\eqn{eq:gs0Kpp}). Here we only give the
remaining non-vanishing entries of $\hat\gamma^{(0)}$
\cite{CFMRS:93,CFRS:94}.

The elements $\gamma^{(0)}_{i7}$ with $i=1,\ldots,6$ are:
\begin{eqnarray}
\gamma^{(0)}_{17} = 0, &\qquad&  \gamma^{(0)}_{27} =
\frac{104}{27} C_F
\label{eq:g0127} \\
\gamma^{(0)}_{37} = -\frac{116}{27} C_F
 &\qquad&  \gamma^{(0)}_{47}  = \left(\frac{104}{27} u -\frac{58}{27}d
\right) C_F
\label{eq:g0347} \\
\gamma^{(0)}_{57} = \frac{8}{3} C_F &\qquad&
\gamma^{(0)}_{67} = \left( \frac{50}{27}d -\frac{112}{27}u \right) C_F
\label{eq:g0567}
\end{eqnarray}
The elements $\gamma^{(0)}_{i8}$ with $i=1,\ldots,6$ are:
\begin{eqnarray}
\gamma^{(0)}_{18} = 3, &\quad& \gamma^{(0)}_{28} =
\frac{11}{9} N-\frac{29}{9}\frac{1}{N}
\label{eq:g0128} \\
\gamma^{(0)}_{38} = \frac{22}{9} N-\frac{58}{9}\frac{1}{N}+3 f
 &\quad& \gamma^{(0)}_{48}  = 
6+\left(\frac{11}{9} N -\frac{29}{9}\frac{1}{N}\right) f
\label{eq:g0348} \\
\gamma^{(0)}_{58} = -2 N+\frac{4}{N} -3 f  &\quad&
\gamma^{(0)}_{68} = -4-\left( \frac{16}{9} N -
\frac{25}{9}\frac{1}{N}\right) f
\label{eq:g0568}
\end{eqnarray}

Finally the $2\times 2$ one-loop anomalous dimension matrix in the
sector $Q_{7\gamma},Q_{8G}$ is given by \cite{Grin}
\begin{eqnarray}
\gamma^{(0)}_{77} = 8 C_F
&\qquad&
\gamma^{(0)}_{78} = 0
\label{gammaB0} \\
\gamma^{(0)}_{87} = -\frac{8}{3} C_F
&\qquad&
\gamma^{(0)}_{88} = 16 C_F - 4 N
\nn
\end{eqnarray}
\subsubsection{Renormalization Group Evolution}
         \label{sec:Heff:BXsgamma:RGE}
The coefficients $C_i(\mu_b)$ in (\ref{Heff_at_mu}) can be calculated
by using
\begin{equation}
\vec C(\mu_b)= \hat U_5(\mu_b,\mu_W)\vec C(\mu_W)
\end{equation}
Here $ \hat U_5(\mu_b,\mu_W)$ is the $8\times 8$ evolution matrix which is
given in general terms in (\eqn{u0jj}) with $\hat\gamma$ being this
time an $8\times 8$ anomalous dimension matrix. In the leading order
$\hat U_5(\mu_b,\mu_W)$ is to be replaced by $\hat U_5^{(0)}(\mu_b,\mu_W)$ 
and
the initial conditions by $\vec C^{(0)}(\mu_W)$ with \cite{Grin}
\begin{equation}\label{c2}
C^{(0)}_2(\mu_W) = 1                               
\end{equation}
\begin{equation}\label{c7}
C^{(0)}_{7\gamma} (\mu_W) = \frac{3 x_t^3-2 x_t^2}{4(x_t-1)^4}\ln x_t + 
   \frac{-8 x_t^3 - 5 x_t^2 + 7 x_t}{24(x_t-1)^3}
   \equiv -\frac{1}{2} D'_0(x_t)
\end{equation}
\begin{equation}\label{c8}
C^{(0)}_{8G}(\mu_W) = \frac{-3 x_t^2}{4(x_t-1)^4}\ln x_t +
   \frac{-x_t^3 + 5 x_t^2 + 2 x_t}{8(x_t-1)^3}                               
   \equiv -\frac{1}{2} E'_0(x_t)
\end{equation}
In LO
all remaining coefficients are set to zero at $\mu=\mu_W$. 

Using the techniques developed in section 5, the leading order results for 
the Wilson coefficients of all operators
entering the effective Hamiltonian in (\ref{Heff_at_mu}) can be written
in an analytic form. They are \cite{BMMP:94}
\begin{eqnarray}
\label{coeffs}
C_j^{(0)}(\mu_b)    & = & \sum_{i=1}^8 k_{ji} \eta^{a_i}
  \qquad (j=1,\ldots,6)  \\
\label{C7eff}
C_{7\gamma}^{(0)eff}(\mu_b) & = & 
\eta^\frac{16}{23} C_{7\gamma}^{(0)}(\mu_W) + \frac{8}{3}
\left(\eta^\frac{14}{23} - \eta^\frac{16}{23}\right) C_{8G}^{(0)}(\mu_W) +
 C_2^{(0)}(\mu_W)\sum_{i=1}^8 h_i \eta^{a_i},
\\
\label{C7Geff}
C_{8G}^{(0)eff}(\mu_b) & = & 
\eta^\frac{14}{23} C_{8G}^{(0)}(\mu_W) 
   + C_2^{(0)}(\mu_W) \sum_{i=1}^8 \bar h_i \eta^{a_i},
\end{eqnarray}
with
\begin{eqnarray}
\eta & = & \frac{\as(\mu_W)}{\as(\mu_b)}, 
\end{eqnarray}
and $C_{7\gamma}^{(0)}(\mu_W)$
and $ C_{8G}^{(0)}(\mu_W)$ given in (\ref{c7}) and (\ref{c8}),
respectively. The numbers $a_i$ and $k_{ji}$ have been already given
in section 8.4. For convenience we give again the values 
of $a_i$ together with $h_i$ and $\bar h_i$ 
in table \ref{tab:akh}.

\begin{table}[htb]
\caption[]{Magic Numbers.
\label{tab:akh}}
\begin{center}
\begin{tabular}{|r|r|r|r|r|r|r|r|r|}
\hline
$i$ & 1 & 2 & 3 & 4 & 5 & 6 & 7 & 8 \\
\hline
$a_i $&$ \frac{14}{23} $&$ \frac{16}{23} $&$ \frac{6}{23} $&$
-\frac{12}{23} $&$
0.4086 $&$ -0.4230 $&$ -0.8994 $&$ 0.1456 $\\
$h_i $&$ 2.2996 $&$ - 1.0880 $&$ - \frac{3}{7} $&$ -
\frac{1}{14} $&$ -0.6494 $&$ -0.0380 $&$ -0.0185 $&$ -0.0057 $\\
$\bar h_i $&$ 0.8623 $&$ 0 $&$ 0 $&$ 0
 $&$ -0.9135 $&$ 0.0873 $&$ -0.0571 $&$ 0.0209 $\\
\hline
\end{tabular}
\end{center}
\end{table}

Let us perform a quick numerical analysis of (\ref{C7eff}) and
(\ref{C7Geff}).
Using the leading $\mu_b$-dependence of $\as$:
\begin{equation} 
\as(\mu_b) = \frac{\as(\mz)}{1 
- \beta_0 \frac{\as(\mz)}{2\pi} \, \ln(\mz/\mu_b)} 
\label{eq:asmumz}
\end{equation} 
one finds the results in table \ref{tab:c78effnum}.

\begin{table}[htb]
\caption[]{Wilson coefficients $C^{(0){\rm eff}}_{7\gamma}$ and 
$C^{(0){\rm eff}}_{8G}$
for $\mt = 170 \gev$ and various values of $\as^{(5)}(\mz)$ and $\mu$.
\label{tab:c78effnum}}
\begin{center}
\begin{tabular}{|c||c|c||c|c||c|c|}
\hline
& \multicolumn{2}{c||}{$\as^{(5)}(\mz) = 0.113$} &
  \multicolumn{2}{c||}{$\as^{(5)}(\mz) = 0.118$} &
  \multicolumn{2}{c| }{$\as^{(5)}(\mz) = 0.123$} \\
\hline
$\mu [\gev]$ & 
$C^{(0){\rm eff}}_{7\gamma}$ & $C^{(0){\rm eff}}_{8G}$ &
$C^{(0){\rm eff}}_{7\gamma}$ & $C^{(0){\rm eff}}_{8G}$ &
$C^{(0){\rm eff}}_{7\gamma}$ & $C^{(0){\rm eff}}_{8G}$ \\
\hline
 2.5 & --0.328 & --0.155 & --0.336 & --0.158 & --0.344 & --0.161 \\
 5.0 & --0.295 & --0.142 & --0.300 & --0.144 & --0.306 & --0.146 \\
 7.5 & --0.277 & --0.134 & --0.282 & --0.136 & --0.286 & --0.138 \\
10.0 & --0.265 & --0.130 & --0.269 & --0.131 & --0.273 & --0.133 \\
\hline
\end{tabular}
\end{center}
\end{table}

Two features of these results should be emphasised:
\begin{itemize}
\item
The strong enhancement of the
coefficient $C^{(0){\rm eff}}_{7\gamma}$ by short distance QCD effects which 
we illustrate by the relative numerical importance of the three terms in
expression (\ref{C7eff}).
For instance, for $\mt = 170\gev$, $\mu_b = 5\gev$ and $\as^{(5)}(\mz)
=0.118$ one obtains
\begin{eqnarray}
C^{(0){\rm eff}}_{7\gamma}(\mu_b) &=&
0.695 \; C^{(0)}_{7\gamma}(\mu_W) +
0.085 \; C^{(0)}_{8G}(\mu_W) - 0.158 \; C^{(0)}_2(\mu_W)
\nn\\
 &=& 0.695 \; (-0.193) + 0.085 \; (-0.096) - 0.158 = -0.300 \, .
\label{eq:C7geffnum}
\end{eqnarray}
In the absence of QCD we would have $C^{(0){\rm eff}}_{7\gamma}(\mu_b) =
C^{(0)}_{7\gamma}(\mu_W)$ (in that case one has $\eta = 1$). Therefore, the
dominant term in the above expression (the one proportional to
$C^{(0)}_2(\mu_W)$) is the additive QCD correction that causes the
enormous QCD enhancement of the \Bsg rate \cite{Bert,Desh}.
It originates solely from the two-loop diagrams. On the other hand, the
multiplicative QCD correction (the factor 0.695 above) tends to
suppress the rate, but fails in the competition with the additive
contributions.

In the case of $C^{(0){\rm eff}}_{8G}$ a similar enhancement is observed
\begin{eqnarray}
C^{(0){\rm eff}}_{8G}(\mu_b) &=&
0.727 \; C^{(0)}_{8G}(\mu_W) - 0.074 \; C^{(0)}_2(\mu_W)
\nn \\
 &=& 0.727 \; (-0.096) - 0.074 = -0.144 \, .
\label{eq:C8Geffnum}
\end{eqnarray}
\item
A strong $\mu_b$-dependence of both coefficients   
as first stressed by Ali and Greub \cite{AG1} and confirmed
in \cite{BMMP:94}. 
Since \Bsg is dominated by QCD effects, it is not 
surprising 
that this scale-uncertainty in the leading order 
is particularly large. We will investigate this scale
uncertainty in a moment.
\end{itemize}
\subsubsection{Scale Uncertainties at LO}
In calculating $Br(B\to X_s\gamma)$ it is customary to use the
spectator model in which the inclusive decay $B\to X_s\gamma$
is approximated by the partonic decay $b\to s\gamma$. That is
one uses the following approximate equality: 

\begin{equation}\label{ratios}
\frac{\Gamma(B \to X_s \gamma)}
     {\Gamma(B \to X_c e \bar{\nu}_e)}
 \simeq                                                     
\frac{\Gamma(b \to s \gamma)}
     {\Gamma(b \to c e \bar{\nu}_e)} \equiv R_{{\rm quark}},
\end{equation}
where the quantities on the r.h.s are calculated in the spectator model
corrected for short-distance QCD effects. The normalization to the
semileptonic rate is usually introduced in order to reduce the
uncertainties due to the CKM matrix
elements and factors of $\mb^5$ in the r.h.s. of (\ref{ratios}).
Additional support for the approximation given above comes from the
heavy quark expansions.  Indeed the spectator model has been shown to
correspond to the leading order approximation of an expansion in
$1/\mb$.  The first corrections appear at the ${\cal O}(1/\mb^2)$
level and will be discussed at the end of this section. 

The leading
logarithmic calculations 
\cite{Grin,CFRS:94,CCRV:94a,Mis:94,AG1,BMMP:94} 
can be summarized in a compact form
as follows:
\begin{equation}\label{main}
R_{{\rm quark}} =\frac{Br(B \to X_s \gamma)}
     {Br(B \to X_c e \bar{\nu}_e)}=
 \frac{|V_{ts}^* V_{tb}^{}|^2}{|V_{cb}|^2} 
\frac{6 \alpha}{\pi f(z)} |C^{(0){\rm eff}}_{7}(\mu_b)|^2\,,
\end{equation}
where
\begin{equation}\label{g}
f(z) = 1 - 8z + 8z^3 - z^4 - 12z^2 \ln z           
\quad\mbox{with}\quad
z =
\frac{m^2_{c,pole}}{m^2_{b,pole}}
\end{equation}
is the phase space factor in $Br(B \to X_c e \bar{\nu}_e)$ and
$\alpha=e^2/4\pi$. In order to find (\ref{main}) only the tree level
matrix element $<s\gamma|Q_{7\gamma}|B>$ has to be computed. 

\noindent
There are three scale uncertainties present in (\ref{main}):
\begin{itemize}
\item
The low energy scale $\mu_b=\ord(m_b)$ at which the Wilson
Coefficient $C_{7}^{(0){\rm eff}}(\mu_b)$ is evaluated.
\item
The high energy scale $\mu_W=\ord(\mw)$ at which 
the full theory is matched with the effective five-quark theory.
In LO this scale enters only $\eta$.
$C_{7}^{(0)}(\mu_W)$ and  $C_{8}^{(0)}(\mu_W)$
serve in LO as initial
conditions to the renormalization group evolution from $\mu_W$ down
to $\mu_b$. As seen explicitly in (\ref{c7}) and (\ref{c8}) they do
not depend on $\mu_W$.
\item
The scale $\mu_t=\ord(m_t)$ at which the running top quark mass is
defined. In LO it enters only $x_t$: 
\be\label{xt}
x_t=\f{\mtb^2(\mu_t)}{\mw^2}.
\ee
As we stressed in connection with $B^0-\bar B^0$ mixing
in section 8.3, $\mu_W$ and $\mu_t$ do not have to be
equal. Initially when the top quark and the W-boson are integrated
out, it is convenient in the process of matching to keep
$\mu_t=\mu_W$. Yet one has always the freedom to redefine the top
quark mass and to work with $\mtb(\mu_t)$ where $\mu_t\not=\mu_W$.
\end{itemize}

It is evident from the formulae above that in LO the variations of
$\mu_b$, $\mu_W$ and $\mu_t$ remain uncompensated which results
in potential theoretical uncertainties in the predicted branching
ratio.
  In the context of phenomenological analyses of $B \to X_s\gamma$,
the uncertainty due to $\mu_b$ has been discussed
\cite{AG1,BMMP:94,CZMM,GREUB,BKP1}. The
uncertainties due to $\mu_W$
and $\mu_t$ have been analyzed first in \cite{BKP1} and 
recently in \cite{BG98}. I will follow here my own work with
Axel Kwiatkowski and Nicolas Pott \cite{BKP1}.

\noindent
It is customary to estimate the uncertainties due to $\mu_b$ by
varying it in the range $\mb/2\le\mu_b\le 2\mb$. Similarly one
can vary $\mu_W$ and $\mu_t$ in the ranges $\mw/2\le\mu_W\le 2\mw$
and $\mt/2\le\mu_t\le 2\mt$ respectively. Specifically in our
numerical analysis we will consider the ranges
\be\label{ranges1}
2.5~\gev\le\mu_b\le 10~\gev
\ee
and
\be\label{ranges}
40~\gev\le\mu_W\le 160~\gev\qquad 80~\gev\le \mu_t\le 320\gev
\ee
In the LO analysis we use the leading order formula for
$\as(\mu_b)$ in (\ref{eq:asmumz})
with $\alpha_s(\mz)=0.118$ and
\be\label{mbar}
\mtb(\mu_t)=\mtb(\mt)
\left[\f{\as(\mu_t)}{\as(\mt)}\right]^{\f{4}{\beta_0}}.
\ee
Here $\beta_0=23/3$.
We set $\mtb(\mt)=168~\gev$
and $\mt\equiv m_{t,pole}=176~\gev$.

\noindent
Varying $\mu_b$, $\mu_W$ and $\mu_t$ in the ranges (\ref{ranges1})
and (\ref{ranges})  we
find the following  uncertainties in the branching
ratio \cite{BKP1}:
\begin{equation}\label{LOmu1}
\Delta Br(B\to X_s \gamma)=\left\{ \begin{array}{ll}
\pm 22\% & (\mu_b) \\
\pm 13\% & (\mu_W) \\
\pm 3 \% & (\mu_t) \end{array} \right.
\end{equation}
The fact that the $\mu_W$-uncertainty is smaller than
the $\mu_b$ uncertainty is entirely due to $\as(\mu_W)<\as(\mu_b)$. 
Still this uncertainty is rather disturbing as it introduces an error of
approximately $\pm 0.40\cdot 10^{-4}$ in the branching ratio.
The
smallness of the $\mu_t$-uncertainty is related to the weak $x_t$
dependence of $C_{7}^{(0)}(\mu_W)$ and  $C_{8}^{(0)}(\mu_W)$
which in the range of interest can be well approximated by
\be
C_{7}^{(0)}(\mu_W)=-0.122~ x_t^{0.30}
\qquad  C_{8}^{(0)}(\mu_W)=-0.072~ x_t^{0.19}.
\ee
Thus even if $161\gev\le\mtb(\mu_t)\le 178\gev$ for $\mu_t$ in 
(\ref{ranges}),
the $\mu_t$ uncertainty in  $Br(B\to X_s \gamma)$ is small.
This should be contrasted with  $B_s\to\mu\bar\mu$,
$K_L\to\pi^0\nu\bar\nu$ and $ B_{d,s}^0-\bar B_{d,s}^0$ mixings, 
where $\mu_t$ uncertainties in LO have been
found \cite{BB2,BJW90} to be $\pm 13\%$, $\pm 10\%$ and $\pm 9\%$ 
respectively.

A critical analysis of theoretical and
experimental
uncertainties present in the prediction for Br(\Bsg) based on the
formula (\ref{main}) has been made in \cite{BMMP:94} 
with the result that the error in the Standard Model prediction
in the LO approximation is dominated by
the scale ambiguities. 
The final result of the LO analysis in \cite{BMMP:94} which 
omitted the $\mu_t$ and $\mu_W$ uncertainties was
\be\label{LORES}
Br(B{\to}X_s \gamma)_{\rm LO} =
 (2.8 \pm 0.8)  \times 10^{-4}
\ee
Similar result has been found in \cite{AG1}.

These finding made 
it  clear already in 1993 that  a  complete
NLO analysis of \Bsg was very desirable. 
Such a complete next-to-leading
calculation of \Bsg was described in \cite{BMMP:94} in general terms. 
As demonstrated formally there, the cancellation of the dominant 
$\mu_b$-dependence in the leading
order can then be  achieved. While this formal NLO analysis was
very encouraging with respect to the reduction of the $\mu_b$-dependence,
it could obviously not provide the actual size of Br(\Bsg) after
the inclusion of NLO corrections. Fortunately four years later such
a complete NLO analysis exists and the impact of NLO corrections on
Br(\Bsg) can be analysed in explicit terms. This is precisely 
what we will do now.

\subsection{\Bsg Beyond Leading Logarithms}
         \label{sec:Heff:Bsgamma:nlo}
\subsubsection{Master Formulae}
The formula (\ref{main}) modifies after the inclusion of NLO
corrections as follows \cite{CZMM}:
\be \label{ration}
R_{{\rm quark}} = 
\frac{|V_{ts}^* V_{tb}|^2}{|V_{cb}|^2} 
\frac{6 \alpha}{\pi f(z)} F \left( |D|^2 + A \right)\,,
\ee
where
\be \label{factor}
F = \f{1}{\kappa(z)} 
    \left( \f{\overline{m}_b(\mu=m_b)}{m_{b,{\rm pole}}} \right)^2
    = 
    \f{1}{\kappa(z)} \left( 1 - \f{8}{3} \f{\as(m_b)}{\pi} \right),
\ee
 \be \label{Dvirt}
D = C_{7\gamma}^{(0){\rm eff}}(\mu_b) + \frac{\as(\mu_b)}{4 \pi} \left\{ 
C_{7\gamma}^{(1){\rm eff}}(\mu_b) + \sum_{i=1}^8 C_i^{(0){\rm eff}}(\mu_b) 
\left[ r_i + \gamma_{i7}^{(0){\rm eff}} \ln \frac{m_b}{\mu_b} 
\right] \right\}
\ee
and $A$ is discussed below. 

Let us explain the origin of various new contributions:
\begin{itemize}
\item
First $\kappa(z)$
is the QCD correction to the semileptonic decay
\cite{CM78}. To a good approximation it is given by \cite{KIMM}
\be \label{kap}
\kappa(z) = 1 - \frac{2 \as (\bar\mu_b)}{3 \pi}
\left[(\pi^2-\frac{31}{4})(1-z)^2+\frac{3}{2}\right] \,.
\ee
An exact analytic formula for $\kappa(z)$ can be found in \cite{N89}.
Here $\bar\mu_b=\ord(m_b)$ is a scale in the calculation of QCD corrections
to the semi-leptonic rate which is generally different from the one used
in the $b\to s\gamma$ transition. In this respect we differ from Greub et al.
\cite{GREUB} who set $\bar\mu_b=\mu_b$. 
\item
The second factor in (\ref{factor}) originates as follows.
The \Bsg rate is proportional to $m_{b,{\rm pole}}^3$ present in the
two body phase space and to $\overline{m}_b(\mu=m_b)^2$ present in 
$<s\gamma|Q_{7\gamma}|B>^2$. On the other hand the semileptonic
rate is is proportional to $m_{b,{\rm pole}}^5$ present in the
three body phase space. Thus the $m_b^5$ factors present in
both rates differ by a ${\cal O}(\as)$ correction which has
been consistently omitted in the leading logarithmic approximation
but has to be included now.
\item
For similar reason the variable $z$ entering $f(z)$ and $\kappa(z)$
can be more precisely specified at the NLO level to be 
\cite{GREUB,CZMM}:
\be \label{g(z)}
 z = \frac{m_{c,{\rm pole}}}{m_{b,{\rm pole}}}=0.29\pm0.02 
\ee
which is obtained from $m_{b,{\rm pole}} = 4.8 \pm 0.15$~GeV and
$m_{b,{\rm pole}}-m_{c,{\rm pole}}=3.40$~GeV. 
This gives
\be \label{kf}
\kappa(z)=0.879\pm0.002\approx 0.88\,,
\qquad
f(z)=0.54\pm 0.04\,.
\end{equation}
\item
The amplitude $D$ in (\ref{Dvirt}) includes two types of new
contributions. The first $\as$-correction originates in
the NLO correction to the Wilson coefficients of $Q_{7\gamma}$:
\be \label{C.expanded}
C^{\rm eff}_{7\gamma}(\mu_b) = C^{(0){\rm eff}}_{7\gamma}(\mu_b) + 
\frac{\as(\mu_b)}{4 \pi} C^{(1){\rm eff}}_{7\gamma}(\mu_b)\,. 
\ee
It is this correction which requires the calculation of the
three-loop anomalous dimensions \cite{CZMM}. An explicit formula for
$C^{(1){\rm eff}}_{7\gamma}(\mu_b)$ has been given for the first time
in \cite{CZMM}. We will give a generalization of this formula 
in a moment.

The two remaining corrections in (\ref{Dvirt}) come from one-loop
matrix elements $<s\gamma|Q_{7\gamma}|B>$ and 
$<s\gamma|Q_{8G}|B>$ and from two-loop matrix elements
$<s\gamma|Q_i|B>$ of the remaining operators. These two-loop
matrix elements have been calculated in \cite{GREUB}. The coefficients
of the logarithm are the relevant elements in the leading
anomalous dimension matrix. The explicit logarithmic 
$\mu_b$ dependence in the last term in $D$ will play an important role
few pages below.

Now $C^{(1){\rm eff}}_{7\gamma}(\mu_b)$ is renormalization scheme dependent.
This scheme dependence is cancelled by the one present in the
constant terms $r_i$. Actually ref. \cite{GREUB} does not provide
the matrix elements of the QCD-penguin operators and consequently
$r_i~(i=3-6)$ are unknown. However, the Wilson coefficients
of QCD-penguin operators are very small and this omission is 
most probably immaterial.
\item
The term $A$ in (\ref{ration}) originates from the bremsstrahlung
corrections and the necessary virtual corrections needed for the
cancellation of the infrared divergences. These have been
calculated in \cite{AG2,Pott} and are also considered in 
\cite{CZMM,GREUB} in the
context of the full analysis.  Since the virtual corrections
are also present in the terms $r_i$ in $D$, care must be taken
in order to avoid double counting. This is discussed in detail
in \cite{CZMM} where an explicit formula for $A$ can be found.
It is the equation (32) of \cite{CZMM}.

Actually $A$ depends on  an explicit lower cut on the
photon energy 
\be\label{phs}
E_{\gamma} > 
( 1 - \delta ) E_{\gamma}^{{\rm max}} \equiv ( 1 - \delta ) \frac{m_b}{2}.
\ee
Moreover $A$ is divergent in the limit $\delta \to 1$.
In order to cancel this divergence one would have to consider
the sum of \Bsg and $b{\to}X_s$ decay rates.
However, the divergence at $\delta{\to}1$ is very slow. 
In order to
allow an easy comparison with previous experimental and theoretical
publications the authors in \cite{CZMM} choose $\delta = 0.99$.
Further details on the $\delta$-dependence can be found in this paper.
\item
Finally the values of $\as(\mu_b)$ in all the
above formulae are calculated with the use of the NLO expression 
for the strong coupling constant:
\be \label{alphaNLL1}
\as(\mu) = \frac{\as(M_Z)}{v(\mu)} \left[1 - \f{\beta_1}{\beta_0} 
           \frac{\as(M_Z)}{4 \pi}    \f{\ln v(\mu)}{v(\mu)} \right],
\ee
where 
\be \label{v(mu1)}
v(\mu) = 1 - \beta_0 \frac{\as(M_Z)}{2 \pi} 
\ln \left( \frac{M_Z}{\mu} \right),
\ee
$\beta_0 = \frac{23}{3}$ and $\beta_1 = \frac{116}{3}$.
\end{itemize}

Generalizing the formula (21) of \cite{CZMM} to include $\mu_t$ and $\mu_W$
dependences one finds \cite{BKP1}
\bea     \label{c7eff1}
C^{(1)eff}_7(\mu_b) &=& 
\eta^{\f{39}{23}} C^{(1)eff}_7(\mu_W) + \f{8}{3} \left( \eta^{\f{37}{23}} 
- \eta^{\f{39}{23}} \right) C^{(1)eff}_8(\mu_W) 
\nonumber \\ &&
+\left( \f{297664}{14283} \eta^{\f{16}{23}}-\f{7164416}{357075} 
\eta^{\f{14}{23}} 
       +\f{256868}{14283} \eta^{\f{37}{23}} -\f{6698884}{357075} 
\eta^{\f{39}{23}} \right) C_8^{(0)}(\mu_W) 
\nonumber \\ &&
+\f{37208}{4761} \left( \eta^{\f{39}{23}} - 
\eta^{\f{16}{23}} \right) C_7^{(0)}(\mu_W) 
+ \sum_{i=1}^8 (e_i \eta E_0(x_t) + f_i + g_i \eta) \eta^{a_i}
\nonumber \\ &&
+\Delta C^{(1)eff}_7(\mu_b),
\eea
where
in the $\overline{MS}$ scheme 
\bea\label{GENC7}
C_7^{(1)eff}(\mu_W)&=& C_7^{(1)eff}(M_W)+
8 x_t \f{\partial C_7^{(0)}(\mu_W)}{\partial x_t}\ln\f{\mu_t^2}{\mw^2}
\nonumber \\ &&
+\left(\f{16}{3}C_7^{(0)}(\mu_W)-\f{16}{9} C_8^{(0)}(\mu_W)
+\f{\gamma_{27}^{(0){\rm eff}}}{2}\right) \ln \frac{\mu_W^2}{\mw^2} 
\eea
\bea\label{GENC8}
C_8^{(1)eff}(\mu_W)&=& C_8^{(1)eff}(M_W)+
8 x_t \f{\partial C_8^{(0)}(\mu_W)}{\partial x_t}\ln\f{\mu_t^2}{\mw^2} 
\nonumber \\ &&
+\left(\f{14}{3}C_8^{(0)}(\mu_W)
+\f{\gamma_{28}^{(0){\rm eff}}}{2}\right) \ln\frac{\mu_W^2}{\mw^2} 
\eea
\be\label{GB981}
\Delta C^{(1)eff}_7(\mu_b)=
\sum_{i=1}^8 \left(\frac{2}{3}e_i  + 6 l_i \right) \eta^{a_i+1}
\ln\frac{\mu_W^2}{\mw^2} 
\ee
Here ($x=x_t$)
\bea
C_7^{(1)eff}(M_W) &=& \f{-16 x^4 -122 x^3 + 80 x^2 -  8 x}{9 (x-1)^4} 
{\rm Li}_2 \left( 1 - \f{1}{x} \right)
                  +\f{6 x^4 + 46 x^3 - 28 x^2}{3 (x-1)^5} \ln^2 x 
\nonumber \\ &&
                  +\f{-102 x^5 - 588 x^4 - 2262 x^3 + 3244 x^2 - 1364 x +
208} {81 (x-1)^5} \ln x
\nonumber \\ &&
                  +\f{1646 x^4 + 12205 x^3 - 10740 x^2 + 2509 x - 436}
{486 (x-1)^4} 
\vspace{0.2cm} \\
C_8^{(1)eff}(M_W) &=& \f{-4 x^4 +40 x^3 + 41 x^2 + x}{6 (x-1)^4} 
{\rm Li}_2 \left( 1 - \f{1}{x} \right)
                  +\f{ -17 x^3 - 31 x^2}{2 (x-1)^5} \ln^2 x 
\nonumber \\ &&
                  +\f{ -210 x^5 + 1086 x^4 +4893 x^3 + 2857 x^2 - 1994 x
+280} {216 (x-1)^5} \ln x
\nonumber \\ &&
        +\f{737 x^4 -14102 x^3 - 28209 x^2 + 610 x - 508}{1296 (x-1)^4}
\eea
and
\be
E_0(x) = \frac{x (18 -11
x - x^2)}{12 (1-x)^3} + \frac{x^2 (15 - 16 x + 4 x^2)}{6 (1-x)^4} \ln
x-\frac{2}{3} \ln x.
\ee

The formulae for $C_{7,8}^{(1)eff}(M_W)$ given above and presented in 
\cite{CZMM} are obtained from
the results in \cite{Yao1,GH97,BKP2,GAMB} by using the general formulae 
for the effective coefficient functions in (\ref{eq:defc7eff}) and
(\ref{eq:defc8eff}). For $\mu_W=\mu_t=\mw$ the formulae above 
reduce to the ones given
in \cite{CZMM}. We have put back the superscript "eff" in
(\ref{GENC7}) and (\ref{GENC8}) to emphasize that the effective
anomalous dimensions should be used here.

\begin{table}[htb]
\caption[]{Magic Numbers.
\label{tab:akh1}}
\begin{center}
\begin{tabular}{|r|r|r|r|r|r|r|r|r|}
\hline
$i$ & 1 & 2 & 3 & 4 & 5 & 6 & 7 & 8 \\
\hline
$a_i $&$ \frac{14}{23} $&$ \frac{16}{23} $&$ \frac{6}{23} $&$
-\frac{12}{23} $&$
0.4086 $&$ -0.4230 $&$ -0.8994 $&$ 0.1456 $\\
$e_i$ &$\frac{4661194}{816831}$&$ -\frac{8516}{2217}$ &$  0$ &$  0$ 
        & $ -1.9043$  & $  -0.1008$ & $ 0.1216$  &$ 0.0183$\\
$f_i$ & $-17.3023$ & $8.5027 $ & $ 4.5508$  & $ 0.7519$
        & $  2.0040 $ & $  0.7476$  &$ -0.5385$  & $ 0.0914$\\
$g_i$ & $14.8088$ &  $ -10.8090$  &$ -0.8740$  & $ 0.4218$ 
        & $  -2.9347$   & $ 0.3971$  & $ 0.1600$  & $ 0.0225$ \\
$l_i$ & $0.5784$ &  $ -0.3921$  &$ -0.1429$  & $ 0.0476$ 
        & $  -0.1275$   & $ 0.0317$  & $ 0.0078$  & $ -0.0031$ \\
\hline
\end{tabular}
\end{center}
\end{table}

The numbers $e_i$--$g_i$ and $l_i$ are given in table \ref{tab:akh1}. 
These  numbers as well as the numerical coefficients
in (\ref{c7eff1}) can be confirmed easily by using the anomalous dimension 
matrices  in \cite{CZMM} and the techniques developed in section 5. 

For completeness we give here some information on the relevant
NLO anomalous dimension matrix $\gamma^{(1)}_s$.
The $6\times 6$ two-loop submatrix of $\gamma^{(1)}_s$ involving
the operators $Q_1,\ldots,Q_6$ is given in (\ref{eq:gs1ndrN3Kpp}).
The two-loop generalization of (\ref{gammaB0}) has been calculated 
in \cite{MisMu:94}. It is given for both NDR and HV
schemes as follows
\begin{eqnarray}
\gamma^{(1)}_{77} &=& 
   C_F \left(\frac{548}{9} N - 16 C_F - \frac{56}{9} f \right)
\nn \\
\gamma^{(1)}_{78} &=& 0
\label{gammaB1} \\
\gamma^{(1)}_{87} &=& 
   C_F \left(-\frac{404}{27} N +\frac{32}{3} C_F +\frac{56}{27} f \right)
\nn \\
\gamma^{(1)}_{88} &=& -\frac{458}{9} -\frac{12}{N^2}+ \frac{214}{9} N^2 +
   \frac{56}{9} \frac{f}{N} - \frac{13}{9} f N
\nn
\end{eqnarray}

The generalization of (\ref{eq:g0127})--(\ref{eq:g0568}) to next-to-leading
order requires three loop calculations . The result can be found in
\cite{CZMM}.

The constants $r_i$ resulting from the calculations of NLO corrections
to decay matrix elements \cite{GREUB} are collected in \cite{CZMM}.
It should be stressed that the basis of the operators with $i=1-6$ used
in \cite{CZMM} differs from the standard basis used in the literature
\cite{BBL,GREUB} and here. The basis used in \cite{CZMM} has been chosen in 
order to avoid $\gamma_5$ problems in the three-loop calculations
peformed in the NDR scheme. This has to be remembered when using
formulae of this paper.
In particular the constants $r_i$ calculated in \cite{GREUB} have
to be transformed to the basis of \cite{CZMM}. As pointed out
this year in \cite{GAMB} and in particular by Kagan and Neubert 
\cite{KN98}
this tranformation made originally in \cite{CZMM} contained some errors.
The corrected values of $r_i$ can be found in the hep--version of
\cite{CZMM} and in \cite{KN98}. 
The numerical analysis given below is based on these
new values.

For the discussion below it will be useful to have \cite{CFMRS:93}
\be
\gamma_{27}^{(0){\rm eff}}=\f{416}{81} \qquad
\gamma_{28}^{(0){\rm eff}}=\f{70}{27}
\ee
which enter (\ref{GENC7}) and (\ref{GENC8}) respectively.
They can be obtained from (\ref{eq:g0127}) and (\ref{eq:g0128}).

\subsubsection{Going Beyond the Spectator Model}
In order to calculate the final rate we
have to pass from the calculated $b$-quark decay rates to 
the $B$-meson decay rates. Relying on the
Heavy Quark Expansion (HQE) calculations one finds \cite{CZMM}
\be \label{BR} 
Br(B{\to}X_s \gamma) = Br(B{\to}X_c e \bar{\nu_e})
\cdot R_{{\rm quark}} 
\left( 1 - \frac{\delta^{NP}_{sl}}{m_b^2}
         + \frac{\delta^{NP}_{rad}}{m_b^2} \right),
\ee
where $\delta^{{\rm NP}}_{{\rm sl}}$ and $\delta^{{\rm NP}}_{{\rm rad}} $
parametrize nonperturbative corrections to the semileptonic and
radiative $B$-meson decay rates, respectively. 

Following \cite{FLS96}, one can express
$\delta^{{\rm NP}}_{{\rm sl}}$ and $\delta^{{\rm NP}}_{{\rm rad}} $
in terms of the HQET parameters $\lambda_1$ and $\lambda_2$

\be
\delta^{{\rm NP}}_{{\rm sl}}  = \f{1}{2} \lambda_1 
+\left(\frac{3}{2}- \frac{6 (1-z)^4}{f(z)}\right) \lambda_2.
\ee

\be
\delta^{{\rm NP}}_{{\rm rad}}  = \f{1}{2} \lambda_1 - \f{9}{2} \lambda_2.
\ee
where $f(z)$ is given (\ref{g}).

The value of $\lambda_2$ is known from $B^*$--$B$ mass splitting
\be
\lambda_2 = \f{1}{4} ( m_{B^*}^2 - m_B^2 ) \simeq 0.12\;{\rm GeV}^2.
\ee
The value of $\lambda_1$ is controversial. Fortunately it cancels out
in the r.h.s. of (\ref{BR}).

The two nonperturbative corrections in (\ref{BR}) are
both around $4\%$ in magnitude and tend to cancel each other. In
effect, they sum up to only around $1\%$. As stressed in \cite{CZMM},
such a small number has
to be taken with caution. 
Indeed, one has to remember that the four-quark operators 
$Q_1...... Q_6$ have not been included in the calculation of
$\delta^{{\rm NP}}_{{\rm rad}}$. Contributions from these operators could
potentially give one- or two-percent effects. Nevertheless, it seems
reasonable to conclude that the total nonperturbative $1/m_b^2$ 
correction to
(\ref{BR}) is well below 10\%, i.e.\ it is smaller than the
inaccuracy of the perturbative calculation of $R_{{\rm quark}}$.

In additions to the $1/m_b^2$ corrections one has to consider
long distance contributions to $B\to X_s\gamma$. These are not easy
to calculate and until recently most estimates were based on phenomenological
models. In these model estimates long distance contributions 
are expected to arise dominantly 
from transitions $B \to \sum_i V_i+X_s \to \gamma X_s$ where
$V_i=J/\psi,\psi^\prime,...$ and are found  to be below
$10\%$ \cite{LDGAMMA}. 

A more modern way to estimate these long distance corrections
is to use heavy quark expansions treating the charm quark as
a heavy quark. As pointed out
by Voloshin \cite{VOL96} and also discussed by other
authors \cite{LRW97}, 
these non-perturbative corrections
originate in the photon coupling to a virtual $c\bar c$ loop and
their general structure is given by 
$$
(\Lambda_{\rm QCD}^2/\mc^2)(\Lambda_{\rm QCD}\mb/\mc^2)^n
$$
with $(n=0,1..)$. The term $n=0$ can be estimated reliably.
Originally a 3\% suppression of the decay rate by this term has been
found  in \cite{VOL96} 
Subsequently, however, an overall sign error 
in this estimate has been pointed out in \cite{BUC97} so that this
$1/m_c^2$ correction  is positive. 

Since
$\Lambda_{\rm QCD}\mb/\mc^2\approx 0.6 $, the terms with $n>0$
are not necessarily much smaller. Although the presence of
unknown matrix elements in these contributions does not
allow a definite estimate of their actual size, the analyses in
\cite{VOL96,LRW97} find that these contributions are weighted by
very small calculable coefficients. Consequently these higher order 
contributions
are expected to be  substantially smaller than the $n=0$ term 
and  the $3\%$ {\it enhancement} from $1/m_c^2$ corrections found in
\cite{BUC97} appears to be a good estimate of the long distance 
contributions to the $B\rightarrow X_s\gamma$ decay rate. 
We will include this enhancement in the numerical analysis below.

\subsubsection{Numerical Analysis at NLO}
Let us investigate how much the uncertainties in
(\ref{LOmu1}) are reduced after including NLO corrections.
We begin this discussion by demonstrating analytically
that the $\mu_b$, $\mu_W$ and $\mu_t$ dependences present in 
$C^{(0){\rm eff}}_{7}(\mu_b)$ are indeed cancelled at $\ord(\as)$
by the explicit scale dependent terms in (\ref{Dvirt}) and (\ref{GENC7}). 
The scale dependent terms in (\ref{GENC8}) do not enter this cancellation
at this order in $\as$ in $B\to X_s \gamma$. On the other hand
they are responsible for the cancellation of the scale dependences
in $C^{(0){\rm eff}}_{8}(\mu_b)$ relevant for the $b \to s~{\rm gluon}$
transition. 

\noindent
Expanding the three terms in (\ref{C7eff}) in $\as$ and
keeping the leading logarithms we find:
\be\label{E1}
\eta^\frac{16}{23} C_{7}^{(0)}(\mu_W)=
\left (1+\f{\as}{4\pi}\f{16}{3}\ln\f{\mu_b^2}{\mu^2_W}\right)
C_{7}^{(0)}(\mu_W)
\ee

\be\label{E2}
 \frac{8}{3}
   \left(\eta^\frac{14}{23} - \eta^\frac{16}{23}\right) 
C_{8}^{(0)}(\mu_W)= 
-\f{\as}{4\pi}\f{16}{9}\ln\f{\mu_b^2}{\mu^2_W}
C_{8}^{(0)}(\mu_W)
\ee

\be\label{E3}
\sum_{i=1}^8 h_i \eta^{a_i}= \f{\as}{4\pi}\f{23}{3}\ln\f{\mu_b^2}{\mu^2_W}
\sum_{i=1}^8 h_i a_i =\f{208}{81} \f{\as}{4\pi}\ln\f{\mu_b^2}{\mu^2_W}
\ee
respectively. In (\ref{E3}) we have used $\sum h_i=0$. 
Inserting these expansions into (\ref{Dvirt}),
we observe that the $\mu_W$ dependences
in (\ref{E1}), (\ref{E2}) and (\ref{E3})
are precisely cancelled  by the three  explicit logarithms in
(\ref{GENC7}) involving $\mu_W$, respectively. Similarly one can convince
oneself that the $\mu_t$-dependence of $C^{(0){\rm eff}}_{7}(\mu_b)$
is cancelled at $\ord(\as)$ by 
the $\ln \mu_t^2/\mw^2$ term in (\ref{GENC7}).
Finally and most importantly the $\mu_b$ dependences in
(\ref{E1}), (\ref{E2}) and (\ref{E3}) are cancelled by the explicit
logarithms in (\ref{Dvirt}) which result from the calculation of
the one-loop
matrix elements $<s\gamma|Q_{7\gamma}|B>$ and 
$<s\gamma|Q_{8G}|B>$ and the two-loop matrix element
$<s\gamma|Q_2|B>$ as discussed previously. Interestingly
the scale dependent term in (\ref{GB981}) does not contribute to
any cancellation of the $\mu_W$ dependence
at this order in $\as$ due to the relation
\be
\sum_{i=1}^8 \left(\frac{2}{3}e_i  + 6 l_i \right)=0.
\ee
which can be verified by using the table \ref{tab:akh1}.

Clearly there remain small $\mu_b$, $\mu_W$ and $\mu_t$ dependences in
(\ref{ration}) which can only be reduced by going beyond the NLO
approximation. They constitute the theoretical uncertainty which
should be taken into account in estimating the error in the
prediction for $Br(B\to X_s\gamma)$. For this reason also the term
$\Delta C^{(1)eff}_7(\mu_b)$ in (\ref{GENC7}), originally omitted in
\cite{BKP1}), has to be kept as pointed out in \cite{BG98}.

\noindent
Using the two-loop generalization of $(\ref{mbar})$ from Section 4.7
and varying $\mu_b$, $\mu_W$ and $\mu_t$ in the ranges (\ref{ranges1})
and (\ref{ranges}) one
finds \cite{BKP1} the following respective uncertainties in the branching
ratio after the inclusion of NLO corrections:
\begin{equation}\label{NLOm}
\Delta Br(B\to X_s \gamma)=\left\{ \begin{array}{ll}
\pm 4.3\% & (\mu_b) \\
\pm 1.1\% & (\mu_W) \\
\pm 0.4\% & (\mu_t) \end{array} \right.
\end{equation}

This reduction of the $\mu_b$-uncertainty by roughly a factor of seven 
 relative to $\pm 22\%$ in LO is impressive.
The remaining $\mu_W$ and $\mu_t$ uncertainties are negligible.

Next we would like to comment on the uncertainty due to variation of
$\bar\mu_b$ in $\kappa(z)$ given in (\ref{kap}). In \cite{GREUB}
the choice $\bar\mu_b=\mu_b$ has been made. Yet in my opinion
such a treatment is not really correct, since the scale $\bar\mu_b$ in
the semi-leptonic decay has nothing to do with the scale $\mu_b$
in the renormalization group evolution in the $B\to X_s\gamma$
decay. 
Varying $\bar\mu_b$ in the range $2.5\gev\le\mu_b\le 10\gev$ we find
\begin{equation}\label{NLOm1}
\Delta Br(B\to X_s \gamma)=\pm 1.7\% \quad (\bar\mu_b)
\end{equation}
Since the $\mu_b$ and $\bar\mu_b$ uncertainties are uncorrelated we
can add them in quadrature finding $\pm 4.6\%$ for the total
scale uncertainty due to $\mu_b$ and $\bar\mu_b$. 
The addition of the uncertainties in $\mu_t$ and $\mu_W$ in
(\ref{NLOm}) modifies this result slightly and the total scale 
uncertainty in $Br(B \to X_s\gamma)$ amounts then to
\be\label{stheon}
\Delta Br(B{\to}X_s \gamma) = \pm 4.8\% \quad({\rm scale})
 \ee

It should be stressed that this pure theoretical 
uncertainty related to the truncation of the perturbative series
should be distinguished from parametric uncertainties related
to $\alpha_s$, the quark masses etc. discussed below.

In our numerical calculations we have included all corrections
in the NLO approximation. To work  consistently
in this order, we have in particular
expanded the various factors in (\ref{ration}) in  $\alpha_s$ and discarded
all NNLO terms of order $\alpha_s^2$ which resulted in the process
of multiplication. This treatment is different
from \cite{CZMM,GREUB}, where the $\alpha_s$ corrections in (\ref{factor}) 
have not been expanded in the evaluation of 
(\ref{ration}) and therefore some higher order corrections have been kept.
Different scenarios of partly incorporating higher order corrections
by expanding or not expanding various factors in (\ref{ration})
affect the branching ratio by $\Delta Br(B\rightarrow X_s\gamma)\approx
\pm 3.0 \%$. This number indicates that indeed the scale uncertainty
in (\ref{stheon}) realistically  estimates the magnitude of yet
unknown higher order corrections.   
The remaining uncertainties are due to the values of the various 
input parameters.
In order to obtain the final result for the branching ratio
we have used  the  parameters given in table \ref{tabbsg}.

\begin{table}[htb]
\caption[]{Input parameter values and their uncertainties.
The masses are given in GeV.
\label{tabbsg}}
\begin{center}
\begin{tabular}{|c|c|c|c|c|c|c|c|}
\hline
 & $\as(M_Z)$ & $m_{t,pole}$   & $m_{c,pole}/m_{b,pole}$ 
& $m_{b,pole}$ & $\alpha_{em}^{-1}$ 
& $|V_{ts}^{\star}V_{tb}|/V_{cb}$ & $Br(B\to X_c e\bar\nu_e)$ \\
\hline
{\rm Central} & $0.118$ & $176$  & $0.29$ & $4.8$  &  
  $130.3$  & $0.976$ &$0.104$ \\
\hline
{\rm Error} & $\pm 0.003$   & $\pm 6.0$  & $\pm 0.02$ & $\pm 0.15$  &  
  $\pm 2.3$  & $\pm 0.010$ &$\pm 0.004$ \\
\hline
\end{tabular} 
\end{center}
\end{table}

\begin{table}[htb]
\caption[]{Uncertainties in $Br(B \to X_s\gamma)$ due to various 
sources.\label{tab:akh2}}
\begin{center}
\begin{tabular}{|c|c|c|c|c|c|c|c|}
\hline
{\rm Scales} & $\as(M_Z)$ & $m_{t,pole}$  & $m_{c,pole}/m_{b,pole}$ 
& $m_{b,pole}$ & $\alpha_{em}$ & CKM angles & $B\to X_c e\bar\nu_e$ \\
\hline
$\pm 4.8\%$ & $\pm 2.9\%$   & $\pm 1.7\%$  & $\pm 5.4\%$ & $\pm 0.7\%$  &  
  $\pm 1.8\%$  & $\pm 2.0\%$ &$\pm  3.8\%$ \\
\hline
\end{tabular} 
\end{center}
\end{table}

Adding all the uncertainties 
in quadrature we find  
\be\label{sfin}
Br(B{\to}X_s \gamma) =(3.60 \pm 0.17~({\rm scale})~\pm 0.28~({\rm par})) 
  \times 10^{-4}
= (3.60 \pm 0.33)  \times 10^{-4}
\ee
where we show separately scale and parametric uncertainties.
The relative importance of various
uncertainties is shown in table \ref{tab:akh2}. Similar results
can be found in \cite{CZMM,BG98}. 
We observe that inclusion of NLO corrections
increased the value of the LO prediction in (\ref{LORES}) by 
roughly $25\%$. Simultaneously the total error has been decreased
by more than a factor of two. The shift upwards is mainly caused by the
$\ord(\alpha_s)$ corrections to the matrix elements of the
contributing operators calculated in \cite{GREUB}, 
not to the Wilson coefficients.
One has to remember, however, that this feature is valid in
the NDR scheme considered here and may not be true in another
renormalization scheme without changing the total result
for the decay rate.

We also observe that 
the parametric uncertainties dominate the theoretical
error at present. Once these parametric uncertainties will be reduced
in the future the smallness of the scale uncertainties achieved
through very involved QCD calculations, in particular in 
\cite{CZMM,GREUB,AG2,Pott,Yao1,GH97,BKP2}, can be better appreciated.
This reduction of the theoretical error in the Standard Model
prediction for $Br(B{\to}X_s \gamma)$ could turn out to be very
important in the searches for new physics when the experimental
data improve. 
\subsection{Recent Developments}
Very recently electroweak $\ord(\alpha)$ corrections to $R_{\rm quark}$
have been investigated in an interesting paper by a student of 
this school, Andrzej Czarnecki, and Bill Marciano \cite{CZMA}.
A study of $\ord(\alpha)$ corrections to $R_{\rm quark}$ must
entail two-loop electroweak contributions to $b \to s\gamma$
as well as one loop corrections to $b\to ce\nu$. A complete
calculation of all $\ord(\alpha)$ contributions would be a very
heroic task, but it is already valuable to identify potentially
dominant contributions.

One obvious question is the scale $\mu$ in 
$\alpha_{\rm em}\equiv e^2(\mu)/4\pi$ which is rather arbitrary
if corrections $\ord(\alpha)$ are not considered. In all recent
calculations $\mb\le \mu\le\mw$ has been used, giving
$1/\alpha_{\rm em}=130.3\pm 2.3$. The inclusion of fermion
loop contributions in the photon propagator indicates \cite{CZMA},
however, that $\alpha$ renormalized at $q^2=0$, i.e
$\alpha=1/137.036$ is more appropriate. This reduces the branching
ratio by roughly $5\%$. The fermion loops in the W-propagator
bring a reduction of $2\%$. Two other reductions, 
each of roughly $1\%$,
come from short distance photonic corrections to $b\to s\gamma$
and $b\to ce\nu$. The total reduction of $R_{\rm quark}$ 
quoted in \cite{CZMA} amounts then to $(9\pm2)\%$ where the
error is a guess-estimate of unknown corrections.
With this reduction
the branching ratio in (\ref{sfin}) becomes
\be\label{sfincm}
Br(B{\to}X_s \gamma) 
= (3.28 \pm 0.30)  \times 10^{-4}~.
\ee
The $\pm 2\%$ error in the estimate of $\ord(\alpha)$ corrections
is compensated by the fact that $\alpha$ has a negligible error
compared to $\alpha_{em}$ in table~\ref{tabbsg}.
Personally, I am not yet convinced that the $\ord(\alpha)$ 
reduction is as high as $9\%$. The reduction of $5\%$ through
the change $\alpha_{\rm em}\to \alpha$ appears rather plausible.
On the other hand the same sign of three smaller corrections
could turn out to be accidental and other corrections, not considered
yet, could well cancel them. 
Some indication for this is given by a very recent analysis of
Strumia \cite{STRUMIA}, who performed a complete calculation of
two--loop electroweak contributions to $B\to X_s\gamma$ in the
large $\mt$ limit, finding them below $1\%$.
In spite of this reservation, the
calculation of Czarnecki and Marciano certainly indicates that
a reduction of $Br(B{\to}X_s \gamma)$ through $\ord(\alpha)$ corrections
by $\ord(5\%)$ is certainly possible. A more detailed investigation
of this issue would be desirable at some stage in the future. 

Finally I would like to mention here a very recent paper of Kagan and
Neubert \cite{KN98} who also made an extensive analysis of $B\to X_s\gamma$.
Reanalyzing in detail the issue of the 
photon-spectrum and of $\delta$ in (\ref{phs}) and including also
QED corrections, Kagan and Neubert arrive
at the estimate of $Br(B{\to}X_s \gamma)$, 
which in spite of some differences at intermediate stages agrees very
well with (\ref{sfincm}). Since the analysis in \cite{KN98} is very
recent, I am not in a position to make any useful comments on it.
Certainly of interest is their reanalysis of the extraction of
the total decay rate $Br(B{\to}X_s \gamma)$ from the experimental 
photon spectrum, which I will briefly mention below.

\subsection{Experimental Status}
After all this theoretical exposition it is really time to summarize
the present data.
The branching ratio for $B \to X_s \gamma$  found 
by the CLEO collaboration already in 1994 \cite{CLEO2} is given by
\begin{equation}\label{EXP}
Br(B \to X_s\gamma) = (2.32 \pm 0.57 \pm 0.35) \times 10^{-4}
\end{equation}
and the very recent preliminary update from CLEO reads \cite{CLEO98}
\begin{equation}\label{EXP98}
Br(B \to X_s\gamma) = (2.50 \pm 0.47 \pm 0.39) \times 10^{-4}\,.
\end{equation}
On the other hand the recent ALEPH measurement of the corresponding
branching ratio for b--hadrons (mesons and baryons) produced
in $Z^0$ decays
reads \cite{ALEPH}
\begin{equation}\label{EXP2}
Br(H_b \to X_s\gamma) = (3.11 \pm 0.80 \pm 0.72) \times 10^{-4}.
\end{equation}
which is compatible with the CLEO result.
In (\ref{EXP})-(\ref{EXP2})
 the first error is statistical and the second is systematic.
As stressed already by several authors in the literature the
measurements in \cite{CLEO2} and \cite{ALEPH} are quite different
and should not be expected to give identical results.

Now, the experimental results given above, are obtained by
measuring the high-energy part of the photon spectrum and the
extrapolation to the total rate. This requires theoretical
understanding of the photon spectrum. Improving recently the
analysis of the photon spectrum, Kagan and Neubert \cite{KN98}
find that the result in (\ref{EXP}) should actually read
\begin{equation}\label{EXPKN}
Br(B \to X_s\gamma) = (2.66 \pm 0.56_{\rm exp} \pm 0.45_{\rm th})
\times 10^{-4}\,,
\end{equation}
and that the central value in (\ref{EXP98}) should be increased
to 2.8. It will be interesting to watch the further development
and to have a new official CLEO value including this new insight.

The theoretical estimates in (\ref{sfin}) and (\ref{sfincm}) 
are somewhat higher than experimental data.
However, within
 the remaining theoretical
and in particular experimental uncertainties, the Standard Model value
is compatible with experiment. 
\subsection{A Look Beyond the Standard Model}
The
inclusive  radiative \Bsg decay  plays an important role in the 
indirect searches 
for physics beyond the Standard Model and 
places already now rather strong constraints on some new physics
scenarios. 
The possible non-standard contributions can indeed be of the same order of
magnitude of the Standard Model loop effects discussed above.
This is well illustrated by the simplest of these extensions,
where the Higgs sector of the Standard Model is enlarged to include two
doublets (Two Higgs Doublet Models, or 2HDM), 
leading to three new  physical fields, two  neutral  scalars 
(CP even and odd) and one charged scalar.
 In this context, 
only the charged Higgs $H^\pm$ contributes to the Wilson 
coefficients $C_{7\gamma}$ and  $C_{8G}$.
 Its interaction with  quarks  is described by the Lagrangian
\be
{\cal L}= (2\sqrt{2}G_F)^{1/2}\sum_{i,j=1}^3\bar u_i
\left( A_u m_{u_i}V_{ij}\frac{1-\gamma_5}{2}-
A_d V_{ij}m_{d_i}\frac{1+\gamma_5}{2}\right) d_j H^++{\rm h.c.}
\ee
Here $i,j$ are generation indices, $m_{u,d}$ are quark masses, and $V$ is the
CKM matrix. 
The fermions may then acquire their masses in two ways:
the first possibility, referred to as Model I, is that both up and down quarks
get their masses from the same Higgs doublet $H_2$, and
\be
A_u=A_d=1/\tan \beta ~,
\ee
where $\tan \beta$ is the ratio of the v.e.v. of $H_2$ and $H_1$.
In the case of
Model II, up quarks get their masses from Yukawa couplings to $H_2$,
while down quarks get masses from couplings to $H_1$, and
\be
A_u=-1/A_d=1/\tan \beta ~.
\ee 
Model II is particularly interesting because it is realized in the minimal
supersymmetric extension of the SM.
The charged-Higgs contributions to $C_{7\gamma}$ and  $C_{8G}$
are functions of the top and charged Higgs masses and of $\tan\beta$
whose LO expressions are given in \cite{chwil}.
Recently, the complete NLO corrections to $Br(B\to X_s\gamma)$ 
in the 2HDM have been computed
\cite{GAMB,BG98}. Partial results can also be found in \cite{strum}.

With respect to the Standard Model, in Model II the branching ratio is
strongly enhanced for a light charged Higgs and the decoupling 
at large $M_H$ takes place very slowly.
 This leads to very stringent bounds on
${\rm M_H}$ for any 
particular value of $\tan\beta$.
Actually, for $\tan\beta>1$, the dependence on $\tan\beta$ is very mild
and practically saturates for $\tan \beta \ge 2$.
Using the current  CLEO 95\% CL upper bound 
$Br(B\to X_s\gamma) < 4.2 \times 10^{-4}$
and adopting a conservative approach to evaluate the theoretical uncertainty
(scanning), 
one obtains lower bounds on ${\rm M_H}$ of $\approx 250$GeV,
independent of $\tan\beta$ \cite{GAMB,BG98}.
On the other hand, 
adding different theoretical errors in quadrature leads to 
${\rm M_H}>370$GeV. 
Indeed, these bounds are quite sensitive to
the errors of the theoretical prediction and to the details of the
calculation \cite{GAMB}.
For instance including Czarnecki-Marciano $\ord({\alpha})$ corrections
would weaken them significantly.
 Improving the calculation to
the NLO has also had 
important effects on these bounds, since the theoretical error
is significantly reduced \cite{GAMB}.
Finally, it is  clear that
 one of the reasons we are  able to obtain such strong bounds on
${\rm M_H}$ is the poor agreement between the Standard Model prediction 
and CLEO measurement, and
that the situation may drastically change with new experimental results.
In the case of a heavy Higgs, 
a resummation of the leading logarithms of ${\rm M_H}/\mw$ has been 
performed in \cite{anl}.

For what concerns Model I, in that case 
the charged-Higgs contribution reduces the 
value of $Br(B\to X_s\gamma)$ and therefore no significant bound can be 
obtained. On
the other hand, it is interesting that new physics effects can
bring the prediction for \Bsg closer to the CLEO value. 
A significant effect can only be expected for small $\tan \beta$, since
in Model I all charged-Higgs contributions vanish in the limit of
large $\tan \beta$, as $\tan^{-2} \beta$. However, in that case the top Yukawa
coupling grows and strong limitations come from high energy measurements, in
particular of $R_b$. It can be concluded \cite{GAMB} that 
the reduction of $Br(B\to X_s\gamma)$  can be at best about 20{\%}. 

A more general class of multi-Higgs models, where only one charged Higgs does
not decouple and its couplings are arbitrary and may violate CP, 
 has been studied at LO in \cite{multiH}
and more recently at NLO in \cite{BG98}.

In the MSSM, the charged Higgs loops are accompanied by chargino-squark 
contributions which can partly compensate the
effect of the charged Higgs. Therefore the above bounds do not apply to the
MSSM, except in some scenarios, like gauge-mediated supersymmetric models 
\cite{rattazzi}, where the Higgs contribution is known to dominate over the
chargino loops, because the squarks are generally heavy.
 Indeed, in the supersymmetric limit, there
is an exact cancellation of different contributions \cite{io}. In the
realistic case of broken supersymmetry, this cancellation is spoiled but,
if charginos and stops are light, it may still be partially effective.
A complete analysis at LO in the MSSM can be found in 
\cite{berto}. Although no direct limit on ${\rm M_H}$ 
can be set, $b\to s\gamma$ has
helped in ruling out very large portions of the SUSY parameter space.
It can be expected that a NLO analysis would further enhance  this 
exclusion potential.
\subsection{Summary and Outlook}
The rare decay $B\to X_s\gamma$ plays at present together with
$B^0_{d,s}-\bar B^0_{d,s}$ mixing the central role
in loop induced transitions in the $B$-system. On the theoretical
side considerable progress has been made recently by calculating
NLO corrections, thereby reducing the large $\mu_b$ uncertainties
present in the leading order. This way the error in the
prediction for $Br(B\to X_s\gamma)$ as
given in (\ref{sfin}), and in  (\ref{sfincm}) 
after including QED corrections, 
has been decreased down to roughly
$\pm 10\%$ compared with $\pm (25-30)\%$ in the leading order.
Since during last two years the central value for 
$Br(B\to X_s\gamma)$ was changing constantly due to inclusion
of various small corrections and different error analyses, 
it is hard to imagine that the
result in  (\ref{sfincm}) is the final word
on this subject. It will be interesting to see how this value
will look like in five years from now. 

On the experimental side considerable progress has been made
by CLEO \cite{CLEO96} in the case of $Br (B^0_d\to K^*\gamma)$, which
we left out due to space limitations.
 It is very
desirable to obtain now an improved measurement of 
$Br(B\to X_s\gamma)$.
Indeed,
in the forthcoming years much more
precise measurements of $Br(B{\to}X_s \gamma)$ are expected from the
upgraded CLEO detector, as well as from the B-factories at SLAC and KEK.

Confrontation of these new
improved experimental values with the already rather precise theoretical
Standard Model estimate may shed some light on whether some
physics beyond this model is required to fit the improved data.

More
on $B\to X_s\gamma$, in particular on the photon spectrum and 
the determination
of $\vtd/\vts$ from $B\to X_{s,d}\gamma$, can be found in
\cite{ALUT,ALIB,Photon,KN98}. 
CP violation in $B \to K^* \gamma$ and $B \to \varrho \gamma$ 
is discussed in \cite{GSW95}.

\section{ Rare $K$- and $B$-Decays}
         \label{sec:HeffRareKB}
\setcounter{equation}{0}
\subsection{General Remarks}
            \label{sec:HeffRareKB:overview}
We will now move to discuss
the semileptonic rare FCNC
transitions $\kpn$,  $K_{\rm L}\to\pi^0\nu\bar\nu$, $B\to X_{s,
d}\nu\bar\nu$ and $B_{s,d}\to l^+l^-$ paying particular attantion
to the first two decays.
The presentation given here overlaps considerably with the ones given
in the reviews \cite{BBL,BF97}, although there are some differences.
In particular the  decay $\klm$ will not be considered here in view of 
space limitations.
Some details on this decay, which is not as theoretically clean as the ones 
discussed here, can be found  in the latter reviews and in \cite{CPRARE}.
On the other hand we will provide certain derivations which cannot
be found in \cite{BBL,BF97}. 
Moreover we discuss briefly two--loop electroweak contributions and
make a few remarks on the physics beyond the Standard Model.

The decay modes considered here are very
similar in their structure which differs considerably from the one
encountered in
the decays $K \to \pi\pi$ and
 $B \to X_s \gamma$  discussed in previous
sections. In particular: 

\begin{itemize}
\item
Within the Standard Model all the decays listed above are loop-induced
semileptonic FCNC processes determined only 
by $Z^0$-penguin and box diagrams which we encountered already
on many occasions in these lectures.
Thus, a distinguishing feature of the present class of decays
is the absence of a photon penguin contribution. For the decay modes
with neutrinos in the final state this is obvious, since the photon
does not couple to neutrinos. For the mesons decaying into a charged
lepton pair the photon penguin amplitude vanishes due to vector current
conservation. Consequently the decays in question are governed by the
functions $X_0(x_t)$ and $Y_0(x_t)$ (see (\ref{X0}) and (\ref{Y0}))
which as seen in (\ref{PBE1}) and (\ref{PBE2}) exhibit strong 
$\mt$-dependences.
\item
A particular and very important advantage of these decays
is their clean theoretical character.
This is related to the fact that
the low energy hadronic
matrix elements required are just the matrix elements of quark currents
between hadron states, which can be extracted from the leading
(non-rare) semileptonic decays. Other long-distance contributions
are negligibly small. As a consequence of these features,
the scale ambiguities, inherent to perturbative QCD, essentially
constitute  the only theoretical uncertainties 
present in the analysis of these decays.
These theoretical uncertainties have been considerably reduced
through the inclusion of
the next-to-leading QCD corrections 
 \cite{BB1,BB2,BB3} as we will demonstrate below. 
\item
The investigation of these low energy rare decay processes in
conjunction with their theoretical cleanliness, allows to probe,
albeit indirectly, high energy scales of the theory and in particular
to measure the top quark couplings $V_{ts}$ and $V_{td}$.
Moreover $\klpn$  offers
a clean determination of the Wolfenstein parameter $\eta$ and 
as we will stress below offers the cleanest measurement
of $\IM\lambda_t= \IM V^*_{ts} V_{td}$ which governs all  
CP violating  $K$-decays. 
However, the very fact
that these processes are based on higher order electroweak effects
implies
that their branching ratios are expected to be very small and not easy to
access experimentally.
\end{itemize}

\begin{table}[htb]
\caption[]{
Order of magnitude of CKM parameters relevant for the various decays,
expressed in powers of the Wolfenstein parameter $\lambda=0.22$. In the
case of $K_{\rm L}\to\pi^0\nu\bar\nu$, which is CP-violating, only the
imaginary parts of $\lambda_{c, t}$ contribute.
\label{tab:lambdaexp}}
\begin{center}
\begin{tabular}{|r|c|c|c|c|}
\hline
&$\kpn$&$K_{\rm L}\to\pi^0\nu\bar\nu$&$B\to X_s\nu\bar\nu$&
$B\to X_d\nu\bar\nu$\\
&~&~&$B_s\to l^+l^-$&$B_d\to l^+l^-$\\  \hline
$\lambda_c$&$\sim\lambda$&(${\rm Im}\lambda_c\sim\lambda^5$)&
$\sim\lambda^2$&$\sim\lambda^3$\\  \hline
$\lambda_t$&$\sim\lambda^5$&(${\rm Im}\lambda_t\sim\lambda^5$)&
$\sim\lambda^2$&$\sim\lambda^3$ \\
\hline
\end{tabular}
\end{center}
\end{table}

The effective Hamiltonians governing the decays
$\kpn$, $K_{\rm L}\to\pi^0\nu\bar\nu$,
$B\to X_{s, d}\nu\bar\nu$ and $B\to l^+l^-$
resulting from the $Z^0$-penguin and box-type contributions can all be
written in the following general form:
\begin{equation}\label{hnr} 
{\cal H}_{\rm eff}={G_{\rm F} \over{\sqrt 2}}{\alpha\over 2\pi 
\sin^2\Theta_{\rm W}}
 \left( \lambda_c F(x_c) + \lambda_t F(x_t)\right)
 (\bar nn^\prime)_{V-A}(\bar rr)_{V-A}\,,  \end{equation}
where $n$, $n^\prime$ denote down-type quarks
($n, n^\prime=d, s, b$ but $n\not= n^\prime$) and $r$ leptons,
$r=l, \nu_l$ ($l=e, \mu, \tau$). The $\lambda_i$ are products of CKM elements,
in the general case $\lambda_i=V^*_{in}V_{in^\prime}^{}$. Furthermore
$x_i=m^2_i/M^2_W$.
The functions $F(x_i)$ describe the dependence on the internal
up-type quark masses $m_i$ (and on lepton masses if necessary)
and are understood to include QCD corrections.
They are increasing functions of the quark masses, a property that is
particularly important for the top contribution.
Since $F(x_c)/F(x_t)\approx
\ord(10^{-3})\ll 1$ the top contributions are by far dominant unless there
is a partial compensation through the CKM factors $\lambda_i$. 
 As seen in
table~\ref{tab:lambdaexp} such a partial compensation takes place in
$\kpn$  and consequently in this decay internal
charm contribution, albeit smaller than the top contribution,
has to be kept. On the other hand in the remaining decays the
charm contributions can be safely neglected. Since the charm contributions
involve QCD corrections with $\as(m_c)$, the scale uncertainties in 
$\kpn$  are found to be larger 
than in the remaining decays in which the QCD effects enter only
through $\as(m_t) < \as(m_c)$.
After these general remarks let us enter some details. Other reviews
of rare decays can be found in \cite{CPRARE,BF97}.
\subsection{The Decay \kpnn}
            \label{sec:HeffRareKB:kpnn}
\subsubsection{The effective Hamiltonian}
The effective Hamiltonian for $\kpn$  can
be written as
\begin{equation}\label{hkpn} 
{\cal H}_{\rm eff}={G_{\rm F} \over{\sqrt 2}}{\alpha\over 2\pi 
\sin^2\Theta_{\rm W}}
 \sum_{l=e,\mu,\tau}\left( V^{\ast}_{cs}V_{cd} X^l_{\rm NL}+
V^{\ast}_{ts}V_{td} X(x_t)\right)
 (\bar sd)_{V-A}(\bar\nu_l\nu_l)_{V-A} \, .
\end{equation}
The index $l$=$e$, $\mu$, $\tau$ denotes the lepton flavour.
The dependence on the charged lepton mass resulting from the box-graph
is negligible for the top contribution. In the charm sector this is the
case only for the electron and the muon but not for the $\tau$-lepton.

We have discussed the top quark contribution already in section 8.2
but there is no harm when we repeat certain things in order to
have the most important information about this decay in one place.

The function $X(x_t)$ relevant for the top part is given by
\begin{equation}\label{xx9} 
X(x_t)=X_0(x_t)+\aspi X_1(x_t) 
\end{equation}
with the leading contribution $X_0(x)$ given in (\ref{X0})
and the QCD correction \cite{BB2}
\begin{eqnarray}\label{xx1}
X_1(x_t)=&-&{23x_t+5x_t^2-4x_t^3\over 3(1-x_t)^2}
+{x_t-11x_t^2+x_t^3+x_t^4\over (1-x_t)^3}\ln x_t
\nonumber\\
&+&{8x_t+4x_t^2+x_t^3-x_t^4\over 2(1-x_t)^3}\ln^2 x_t
-{4x_t-x_t^3\over (1-x_t)^2}L_2(1-x_t)
\nonumber\\
&+&8x_t{\partial X_0(x_t)\over\partial x_t}\ln x_\mu\,,
\end{eqnarray}
where $x_\mu=\mu_t^2/M^2_W$ with $\mu_t=\ord(m_t)$ and
\begin{equation}\label{l2UU} 
L_2(1-x)=\int^x_1 dt {\ln t\over 1-t}   \,.
\end{equation}
The $\mu_t$-dependence of the last term in (\ref{xx1}) cancels to the
considered order the $\mu_t$-dependence of the leading term 
$X_0(x_t(\mu))$.
The leftover $\mu_t$-dependence in $X(x_t)$ is tiny and will be given
in connection with the discussion of the branching ratio below.

The function $X$ in (\ref{xx9})
can also be written as
\begin{equation}\label{xeta9}
X(x_t)=\eta_X\cdot X_0(x_t), \qquad\quad \eta_X=0.985,
\end{equation}
where $\eta_X$ summarizes the NLO corrections represented by the second
term in (\ref{xx9}).
With $\mt\equiv \mtb(\mt)$ the QCD factor $\eta_X$
is practically independent of $m_t$ and $\Lambda_{\overline{MS}}$.

The expression corresponding to $X(x_t)$ in the charm sector is the function
$X^l_{\rm NL}$. It results from the NLO calculation \cite{BB3} and is given
explicitly in \cite{BB3,BBL}.
The inclusion of NLO corrections reduced considerably the large
$\mu_c$ dependence
(with $\mu_c={\cal O}(m_c)$) present in the leading order expressions
for the charm contribution
 \cite{novikovetal:77,ellishagelin:83,dibetal:91,PBE0}.
Varying $\mu_c$ in the range $1\gev\le\mu_c\le 3\gev$ changes $X_{\rm NL}$
by roughly $24\%$ after the inclusion of NLO corrections to be compared
with $56\%$ in the leading order. Further details can be found in
\cite{BB3,BBL}. The impact of the $\mu_c$ uncertainties on the
resulting branching ratio $Br(\kpn)$ is discussed below.

The
numerical values for $X_{\rm NL}$ for $\mu = \mc$ and several values of
$\Lms^{(4)}$ and $\mc(\mc)$ are given in table \ref{tab:xnlnum}. 
The net effect of QCD corrections is to suppress the charm contribution
by roughly $30\%$.

\begin{table}[htb]
\caption[]{The functions $X^e_{\rm NL}$ and $X^\tau_{\rm NL}$
for various $\Lms^{(4)}$ and $\mc$.
\label{tab:xnlnum}}
\begin{center}
\begin{tabular}{|c|c|c|c|c|c|c|}
\hline
& \multicolumn{3}{c|}{$X^e_{\rm NL}/10^{-4}$} &
  \multicolumn{3}{c|}{$X^\tau_{\rm NL}/10^{-4}$} \\
\hline
$\Lms^{(4)}\ [\mev]\;\backslash\;\mc\ [\gev]$ &
1.25 & 1.30 & 1.35 & 1.25 & 1.30 & 1.35 \\
\hline
245 & 10.32  & 11.17  & 12.04 & 6.94 & 7.63 & 8.36 \\
285 & 10.02  & 10.86  & 11.73 & 6.64 & 7.32 & 8.04 \\
325 &  9.71  & 10.55  & 11.41 & 6.32 & 7.01 & 7.72 \\
365 &  9.38  & 10.22  & 11.08 & 6.00 & 6.68 & 7.39 \\
405 &  9.03  &  9.87  & 10.72 & 5.65 & 6.33 & 7.04 \\
\hline
\end{tabular}
\end{center}
\end{table}

\subsubsection{Deriving the Branching Ratio}
The relevant hadronic
matrix element of the weak current $(\bar sd)_{V-A}$ can be extracted,
with the help of isospin symmetry from
the leading decay $K^+\to\pi^0e^+\nu$.
Consequently the resulting theoretical
expression for  the branching fraction $Br(K^+\to\pi^+\nu\bar\nu)$ can
be related to the experimentally well known quantity
$Br(K^+\to\pi^0e^+\nu)$. Let us demonstrate this.

The effective Hamiltonian for the tree level decay $K^+\to\pi^0 e^+\nu$
is given by
\begin{equation}\label{kp0} 
{\cal H}_{\rm eff}(K^+\to\pi^0 e^+\nu)
={G_{\rm F} \over{\sqrt 2}}
 V^{\ast}_{us}
 (\bar su)_{V-A}(\bar\nu_e e)_{V-A} \, .
\end{equation}
Using isospin symmetry we have
\be\label{iso1}
\langle \pi^+|(\bar sd)_{V-A}|K^+\rangle=\sqrt{2}
\langle \pi^0|(\bar su)_{V-A}|K^+\rangle.
\ee
Consequently neglecting differences in the phase space of these two decays,
due to $m_{\pi^+}\not=m_{\pi^0}$ and $m_e\not=0$, we find 
\be\label{br1}
\frac{Br(\kpn)}{Br(K^+\to\pi^0 e^+\nu)}=
{\alpha^2\over |V_{us}|^2 2\pi^2 
\sin^4\Theta_{\rm W}}
 \sum_{l=e,\mu,\tau}\left| V^{\ast}_{cs}V_{cd} X^l_{\rm NL}+
V^{\ast}_{ts}V_{td} X(x_t)\right|^2~.
\end{equation}
\subsubsection{Basic Phenomenology}
We are now ready to present the expression for the branching fraction
$Br(\kpn)$ and to collect various formulae relevant for phenomenological
applications.
Using (\ref{br1}) 
and including isospin breaking corrections one finds
\begin{equation}\label{bkpn}
Br(\kpn)=\kappa_+\cdot\left[\left({\imlt\over\lambda^5}X(x_t)\right)^2+
\left({\relc\over\lambda}P_0(X)+{\relt\over\lambda^5}X(x_t)\right)^2
\right]~,
\end{equation}
\begin{equation}\label{kapp}
\kappa_+=r_{K^+}{3\alpha^2 Br(K^+\to\pi^0e^+\nu)\over 2\pi^2
\sin^4\Theta_{\rm W}}
 \lambda^8=4.11\cdot 10^{-11}\,,
\end{equation}
where we have used
\begin{equation}\label{alsinbr}
\alpha=\frac{1}{129},\qquad \sin^2\Theta_{\rm W}=0.23, \qquad
Br(K^+\to\pi^0e^+\nu)=4.82\cdot 10^{-2}\,.
\end{equation}
Here $\lambda_i=V^\ast_{is}V_{id}$ with $\lambda_c$ being
real to a very high accuracy. $r_{K^+}=0.901$ summarizes isospin
breaking corrections in relating $\kpn$ to $K^+\to\pi^0e^+\nu$.
These isospin breaking corrections are due to quark mass effects and 
electroweak radiative corrections and have been calculated in
\cite{MP}. Next
\begin{equation}\label{p0k}
P_0(X)=\frac{1}{\lambda^4}\left[\frac{2}{3} X^e_{\rm NL}+\frac{1}{3}
 X^\tau_{\rm NL}\right]
\end{equation}
with the numerical values for $X_{\rm NL}^l$ given in table \ref{tab:xnlnum}.
The corresponding values for $P_0(X)$ as a function of
$\Lambda^{(4)}_{\overline{MS}}$ and $m_c\equiv m_c(m_c)$ 
are collected in
table \ref{tab:P0Kplus}. We remark that a negligibly small term
$\sim (X^e_{\rm NL}-X^{\tau}_{\rm NL})^2 $ has been discarded in
(\ref{bkpn}).

\begin{table}[htb]
\caption[]{The function $P_0(X)$ for various $\Lms^{(4)}$ and $m_c$.
\label{tab:P0Kplus}}
\begin{center}
\begin{tabular}{|c|c|c|c|}
\hline
&\multicolumn{3}{c|}{$P_0(X)$}\\
\hline
$\Lms^{(4)}$ $\backslash$ $m_c$ & $1.25\gev$ & $1.30\gev$ & $1.35\gev$  \\
\hline
$245\mev$ & 0.393 & 0.426 & 0.462 \\
$285\mev$ & 0.380 & 0.413 & 0.448 \\
$325\mev$ & 0.366 & 0.400 & 0.435 \\
$365\mev$ & 0.352 & 0.386 & 0.420 \\
$405\mev$ & 0.337 & 0.371 & 0.405 \\ 
\hline
\end{tabular}
\end{center}
\end{table}

Using the improved Wolfenstein parametrization and the approximate
formulae (\ref{2.51}) -- (\ref{2.53}) we can next put 
(\ref{bkpn}) into a more transparent form \cite{BLO}:
\begin{equation}\label{108}
Br(K^{+} \to \pi^{+} \nu \bar\nu) = 4.11 \cdot 10^{-11} A^4 X^2(x_t)
\frac{1}{\sigma} \left[ (\sigma \bar\eta)^2 +
\left(\varrho_0 - \bar\varrho \right)^2 \right]\,,
\end{equation}
where
\begin{equation}\label{109}
\sigma = \left( \frac{1}{1- \frac{\lambda^2}{2}} \right)^2\,.
\end{equation}

The measured value of $Br(K^{+} \to \pi^{+} \nu \bar\nu)$ then
determines  an ellipse in the $(\bar\varrho,\bar\eta)$ plane  centered at
$(\varrho_0,0)$ with 
\begin{equation}\label{110}
\varrho_0 = 1 + \frac{P_0(X)}{A^2 X(x_t)}
\end{equation}
and having the squared axes
\begin{equation}\label{110a}
\bar\varrho_1^2 = r^2_0, \qquad \bar\eta_1^2 = \left( \frac{r_0}{\sigma}
\right)^2
\end{equation}
where
\begin{equation}\label{111}
r^2_0 = \frac{1}{A^4 X^2(x_t)} \left[
\frac{\sigma \cdot Br(K^{+} \to \pi^{+} \nu \bar\nu)}
{4.11 \cdot 10^{-11}} \right]\,.
\end{equation}
Note that $r_0$ depends only on the top contribution.
The departure of $\varrho_0$ from unity measures the relative importance
of the internal charm contributions.

The ellipse defined by $r_0$, $\varrho_0$ and $\sigma$ given above
intersects with the circle (\ref{2.94}).  This allows to determine
$\bar\varrho$ and $\bar\eta$  with 
\begin{equation}\label{113}
\bar\varrho = \frac{1}{1-\sigma^2} \left( \varrho_0 - \sqrt{\sigma^2
\varrho_0^2 +(1-\sigma^2)(r_0^2-\sigma^2 R_b^2)} \right), \qquad
\bar\eta = \sqrt{R_b^2 -\bar\varrho^2}
\end{equation}
and consequently
\begin{equation}\label{113aa}
R_t^2 = 1+R_b^2 - 2 \bar\varrho,
\end{equation}
where $\bar\eta$ is assumed to be positive.

In the leading order of the Wolfenstein parametrization
\begin{equation}\label{113ab}
\sigma \to 1, \qquad \bar\eta \to \eta, \qquad \bar\varrho \to \varrho
\end{equation}
and $Br(K^+ \to \pi^+ \nu \bar\nu)$ determines a circle in the
$(\varrho,\eta)$ plane centered at $(\varrho_0,0)$ and having the radius
$r_0$ of (\ref{111}) with $\sigma =1$. Formulae (\ref{113}) and
(\ref{113aa}) then simplify to \cite{BB3}
\begin{equation}\label{113a}
R_t^2 = 1 + R_b^2 + \frac{r_0^2 - R_b^2}{\varrho_0} - \varrho_0, \qquad
\varrho = \frac{1}{2} \left( \varrho_0 + \frac{R_b^2 - r_0^2}{\varrho_0}
\right).
\end{equation}
Given $\bar\varrho$ and $\bar\eta$ one can determine $V_{td}$:
\begin{equation}\label{vtdrhoeta}
V_{td}=A \lambda^3(1-\bar\varrho-i\bar\eta),\qquad
|V_{td}|=A \lambda^3 R_t.
\end{equation}
At this point a few remarks are in
order:
\begin{itemize}
\item
The long-distance contributions to $\kpn$ have been studied in
\cite{RS} and found to be
very small: a few percent of the charm contribution to the amplitude at
most, which is savely negligible.
\item
The determination of $|V_{td}|$ and of the unitarity triangle requires
the knowledge of $V_{cb}$ (or $A$) and of $|V_{ub}/V_{cb}|$. Both
values are subject to theoretical uncertainties present in the existing
analyses of tree level decays. Whereas the dependence on
$|V_{ub}/V_{cb}|$ is rather weak, the very strong dependence of
$Br(\kpn)$ on $A$ or $V_{cb}$ makes a precise prediction for this
branching ratio difficult at present. We will return to this below.
\item
The dependence of $Br(\kpn)$ on $\mt$ is also strong. However $\mt$
is known already  within $\pm 4\%$ and
consequently the related uncertainty in 
$Br(\kpn)$ is substantialy smaller than the corresponding uncertainty 
due to $V_{cb}$.
\item
Once $\varrho$ and $\eta$ are known precisely from CP asymmetries in
$B$ decays, some of the uncertainties present in (\ref{108}) related
to $|V_{ub}/V_{cb}|$ (but not to $V_{cb}$) will be removed.
\item
A very clean determination of $\sin 2\beta$ without essentially
any dependence on $m_t$ and $V_{cb}$ can be made by combining
$Br(\kpn)$ with $Br(\klpn)$ discussed below. 
\end{itemize}

\subsubsection{Numerical Analysis of \kpnn}
\label{sec:Kpnn:NumericalKp}
Let us begin the numerical analysis by  investigating the uncertainties 
in the prediction for $Br(\kpn)$ and in the determination of  $|V_{td}|$
related to the choice of the renormalization scales $\mu_t$
and $\mu_c$ in the top part and the charm part, respectively. To this end we
will fix the remaining parameters as follows:
\begin{equation}\label{mcmtnum}
\mc\equiv \mcb(\mc)=1.3\gev, \qquad \mt\equiv \mtb(\mt)=170\gev
\end{equation}
\begin{equation}\label{vcbubnum}
V_{cb}=0.040, \qquad |V_{ub}/V_{cb}|=0.08\,.
\end{equation}
In the case of $Br(\kpn)$ we need the values of both $\bar\varrho$
and $\bar\eta$. Therefore in this case we will work with
\begin{equation}\label{rhetnum}
\bar\varrho=0, \qquad\quad  \bar\eta=0.36
\end{equation}
rather than with $|V_{ub}/V_{cb}|$. Finally we will set
$\Lambda_{\overline{MS}}^{(4)}=0.325\gev$ and
$\Lambda_{\overline{MS}}^{(5)}=0.225\gev$ for the charm part and top
part, respectively.
We then vary the scales $\mu_c$ and $\mu_t$ entering $m_c(\mu_c)$
and $m_t(\mu_t)$, respectively, in the ranges
\begin{equation}\label{muctnum}
1\gev\leq\mu_c\leq 3\gev, \qquad 100\gev\leq\mu_t\leq 300\gev\,.
\end{equation}

The results of such an analysis are as follows \cite{BBL}:
The uncertainty in $Br(\kpn)$
\begin{equation}\label{varbkpnLO}
0.68\cdot 10^{-10}\leq Br(\kpn)\leq 1.08\cdot 10^{-10}
\end{equation}
present in the leading order is reduced to
\begin{equation}\label{varbkpnNLO}
0.79\cdot 10^{-10}\leq Br(\kpn)\leq 0.92\cdot 10^{-10}
\end{equation}
after including NLO corrections. 
The difference in the numerics compared to \cite{BBL} results
from $r_{K^+}=1$ used there.
Similarly one finds
\begin{equation}\label{varvtdLO}
8.24\cdot 10^{-3}\leq |V_{td}|\leq 10.97\cdot 10^{-3} \qquad {\rm LO}
\end{equation}
\begin{equation}\label{varvtdNLO}
9.23\cdot 10^{-3}\leq |V_{td}|\leq 10.10\cdot 10^{-3}  \qquad {\rm NLO}\,,
\end{equation}
where $Br(\kpn)=0.9\cdot 10^{-10}$ has been set. We observe that including
the full next-to-leading corrections reduces the uncertainty in the
determination of $|V_{td}|$ from $\pm 14\%$ (LO) to $\pm 4.6\%$ (NLO)
in the present example. The main bulk of this theoretical error stems
from the charm sector. Indeed, keeping $\mu_c=m_c$ fixed and varying
only $\mu_t$, the uncertainties in the determination of $|V_{td}|$
would shrink to $\pm 4.7\%$ (LO) and $\pm 0.6\%$ (NLO).
Similar comments apply to $Br(\kpn)$ where, as seen in
(\ref{varbkpnLO}) and (\ref{varbkpnNLO}), the theoretical uncertainty
due to $\mu_{c,t}$ is reduced from $\pm 22\%$ (LO) to $\pm 7\%$ (NLO).

Finally using the input parameters of table \ref{tab:inputparams}
(``present") and performing two
types of error analysis one finds \cite{BJL96b}
\begin{equation}\label{kpnr}
Br(\kpn)=\left\{ \begin{array}{ll}
(9.1 \pm 3.8)\cdot 10^{-11} & {\rm Scanning} \\
(8.0 \pm 1.6) \cdot 10^{-11} & {\rm Gaussian}\,, \end{array} \right.
\end{equation}
where the error comes dominantly from the uncertainties in the CKM
parameters.
The corresponding analysis with the ``future" input parameters gives
\begin{equation}\label{kpnr0}
Br(\kpn)=\left\{ \begin{array}{ll}
(8.0 \pm 1.6)\cdot 10^{-11} & {\rm Scanning} \\
(7.8 \pm 0.7) \cdot 10^{-11} & {\rm Gaussian}\,, \end{array} \right.
\end{equation}
 
\subsubsection{$\vtd$ from $K^+\to\pi^+\nu\bar\nu$}
Once $Br(K^+\to\pi^+\nu\bar\nu)\equiv Br(K^+)$ is measured, $\vtd$ can be
extracted subject to various uncertainties:
\be\label{vtda}
\frac{\sigma(\vtd)}{\vtd}=\pm 0.04_{scale}\pm \frac{\sigma(\vcb)}{\vcb}
\pm 0.7 \frac{\sigma(\bar\mc)}{\bar\mc}
\pm 0.65 \frac{\sigma( Br(K^+))}{Br(K^+)}~.
\ee
Taking $\sigma(\vcb)=0.002$, $\sigma(\bar\mc)=100\mev$ and
$\sigma( Br(K^+))=10\%$ and adding the errors in quadrature we find that
$\vtd$ can be determined with an accuracy of $\pm 10\%$ in the present
example. This number
is increased to $\pm 11\%$ once the uncertainties due to $\mt$,
$\alpha_s$ and $|V_{ub}|/\vcb$ are taken into account. Clearly this
determination can be improved although a determination of $\vtd$ with
an accuracy better than $\pm 5\%$ seems rather unrealistic.
\subsubsection{Summary and Outlook}
The accuracy of the Standard Model prediction for $Br(\kpn)$ has
improved considerably during the last five years. Indeed in 1992
 ranges like $(5-80)\cdot 10^{-11}$
could be found in the literature. This progress can be traced back to the
improved values of $\mt$ and $\vcb$ and to the inclusion of NLO
QCD corrections which considerably reduced the scale uncertainties
in the charm sector. I expect that further progress
in the determination of CKM parameters via the standard analysis of
section \ref{sec:standard} could reduce 
the errors in (\ref{kpnr}) by at least a
factor of two during the next five years.
A numerical example is given in (\ref{kpnr0}).

Now, what about the experimental status of this decay ?
Until August 97 the experimental lower bound on $Br(K^+\to \pi^+\nu\bar\nu)$
was \cite{Adler95}: $Br(\kpn)<2.4 \cdot 10^{-9}$.
One of the high-lights of August 97 was the observation by BNL787
collaboration at Brookhaven \cite{Adler97} 
of one event consistent with the signature expected for this decay.
The branching ratio:
\be\label{kp97}
Br(K^+ \rightarrow \pi^+ \nu \bar{\nu})=
(4.2+9.7-3.5)\cdot 10^{-10}
\end{equation}
has the central value  by a factor of 4 above the Standard Model
expectation but in view of large errors the result is compatible with the
Standard Model. This new result implies that $\vtd$ lies in the range
$0.006<\vtd< 0.06 $ which is substantially larger than the range
from the standard analysis of section 10. The analysis of additional
data on $K^+\to \pi^+\nu\bar\nu$ present on tape at BNL787 should narrow
this range in the near future considerably.
In view of the clean character of this decay a measurement of its
branching ratio at the level of $ 2 \cdot 10^{-10}$ 
would signal the presence of physics
beyond the Standard Model. The Standard Model sensitivity is
expected to be reached at AGS around the year 2000 \cite{AGS2}.
Also Fermilab with the Main Injector 
could measure this decay \cite{Cooper}.
\subsection{The Decay $K_{\rm L}\to\pi^0\nu\bar\nu$}
            \label{sec:HeffRareKB:klpinn1}
\subsubsection{The effective Hamiltonian}
The effective
Hamiltonian for $K_{\rm L}\to\pi^0\nu\bar\nu$
is given as follows:
\begin{equation}\label{hxnu}
{\cal H}_{\rm eff} = {G_{\rm F}\over \sqrt 2} {\alpha \over
2\pi \sin^2 \Theta_{\rm W}} V^\ast_{ts} V_{td}
X (x_t) (\bar sd)_{V-A} (\bar\nu\nu)_{V-A} + h.c.\,,   
\end{equation}
where the function $X(x_t)$, present already in $\kpn$,
includes NLO corrections and is given in (\ref{xx}). 

As we will demonstrate shortly, $\klpn$  proceeds in the Standard Model 
almost
entirely through direct CP violation \cite{littenberg:89}. Consequently it
is completely dominated by short-distance loop diagrams with top quark
exchanges. The charm contribution can be fully
neglected and the theoretical uncertainties present in $\kpn$ due to
$m_c$, $\mu_c$ and $\Lambda_{\overline{MS}}$ are absent here. 
Consequently the rare decay $\klpn$ is even cleaner than $\kpn$
and is very well suited for the determination of 
the Wolfenstein parameter $\eta$ and $\imlt$.

Before going into the details it is appropriate to clarify one point
\cite{NIR96,BUCH96}. It is usually stated in the literature that the
decay $\klpn$ is dominated by {\it direct} CP violation. Now
the standard definition of the direct CP violation (see section 8
of \cite{BF97}) requires the presence of strong phases which are
completely negligible in $\klpn$. Consequently the violation of
CP symmetry in $\klpn$ arises through the interference between
$K^0-\bar K^0$ mixing and the decay amplitude. This type of CP
violation is often called {\it mixing-induced} CP violation.
However, as already pointed out by Littenberg \cite{littenberg:89}
and demonstrated explictly in a moment,
the contribution of CP violation to $\klpn$ via $K^0-\bar K^0$ mixing 
alone is tiny. It gives $Br(\klpn) \approx 5\cdot 10^{-15}$.
Consequently, in this sence,  CP violation in $\klpn$ with
$Br(\klpn) = {\cal O}(10^{-11})$ is a manifestation of CP violation
in the decay and as such deserves the name of {\it direct} CP violation.
In other words the difference in the magnitude of CP violation in
$K_{\rm L}\to\pi\pi~(\varepsilon)$ and $\klpn$ is a signal of direct
CP violation and measuring $\klpn$ at the expected level would
rule out superweak scenarios. More details on this
issue can be found in \cite{NIR96,BUCH96,BB96}.
\subsubsection{Deriving the Branching Ratio}
Let us derive the basic formula for $Br(\klpn)$ in a manner analogous
to the one for  $Br(K^+ \to \pi^+ \nu \bar\nu)$. To this end we
consider one neutrino flavour and define the complex function:
\begin{equation}\label{hxnu1}
F = {G_{\rm F}\over \sqrt 2} {\alpha \over
2\pi \sin^2 \Theta_{\rm W}} V^\ast_{ts} V_{td}
X (x_t).   
\end{equation}
Then the effective Hamiltonian in (\ref{hxnu}) can be written as
\begin{equation}\label{hxnu2}
{\cal H}_{\rm eff} =  F (\bar sd)_{V-A} (\bar\nu\nu)_{V-A}+
F^\ast (\bar ds)_{V-A} (\bar\nu\nu)_{V-A}~.
\end{equation}
Now, from (\ref{KLS}) we have
\be\label{KLS1}
K_L=\frac{1}{\sqrt{2}}
[(1+\bar\varepsilon)K^0+ (1-\bar\varepsilon)\bar K^0]
\ee
where we have neglected
$\mid\bar\varepsilon\mid^2\ll 1$. Thus the amplitude
for $K_L\to\pi^0\nu\bar\nu$ is given by
\be\label{ampkl0}
A(K_L\to\pi^0\nu\bar\nu)=
\frac{1}{\sqrt{2}}
\left[F(1+\bar\varepsilon) \langle \pi^0|(\bar sd)_{V-A}|K^0\rangle
+ 
F^\ast (1-\bar\varepsilon) \langle \pi^0|(\bar ds)_{V-A}|\bar K^0\rangle
 \right] (\bar\nu\nu)_{V-A}.
\ee
Recalling
\be\label{DEF}
CP|K^0\rangle = - |\bar K^0\rangle, \quad\quad
C|K^0\rangle =  |\bar K^0\rangle
\ee
we have
\be
\langle \pi^0|(\bar ds)_{V-A}|\bar K^0\rangle=-
\langle \pi^0|(\bar sd)_{V-A}|K^0\rangle,
\ee
where the minus sign is crucial for the subsequent steps.

Thus we can write
\be\label{bmpkl0}
A(K_L\to\pi^0\nu\bar\nu)=
\frac{1}{\sqrt{2}}
\left[F(1+\bar\varepsilon) -F^\ast (1-\bar\varepsilon)\right]
 \langle \pi^0|(\bar sd)_{V-A}| K^0\rangle
 (\bar\nu\nu)_{V-A}.
\ee
Now the terms $\bar\varepsilon$ can be safely neglected in comparision
with unity, which implies that the indirect CP violation
(CP violation in the $K^0-\bar K^0$ mixing) is negligible in this decay.
We have then
\be
F(1+\bar\varepsilon) -F^\ast (1-\bar\varepsilon)=
{G_{\rm F}\over \sqrt 2} {\alpha \over
\pi \sin^2 \Theta_{\rm W}} \IM (V^\ast_{ts} V_{td})
\cdot X(x_t).   
\end{equation}
Consequently using isospin relation
\be
\langle \pi^0|(\bar ds)_{V-A}|\bar K^0\rangle=
\langle \pi^0|(\bar su)_{V-A}|K^+\rangle
\ee
together with (\ref{kp0}) and taking into account the difference
in the lifetimes of $K_L$ and $K^+$ we have after summation over three
neutrino flavours
\be\label{br2}
\frac{Br(K_L\to\pi^0\nu\bar\nu)}{Br(K^+\to\pi^0 e^+\nu)}=
3\frac{\tau(K_L)}{\tau(K^+)}
{\alpha^2\over |V_{us}|^2 2\pi^2 
\sin^4\Theta_{\rm W}}
 \left(\IM \lambda_t \cdot X(x_t)\right)^2
\end{equation}
where $\lambda_t=V^{\ast}_{ts}V_{td}$.
\subsubsection{Master Formulae for $Br(\klpn)$}
\label{sec:Kpnn:MasterKL}
Using (\ref{br2}) we can write $Br(\klpn)$ simply as
follows
\begin{equation}\label{bklpn}
Br(K_{\rm L}\to\pi^0\nu\bar\nu)=\kappa_{\rm L}\cdot
\left({\imlt\over\lambda^5}X(x_t)\right)^2
\end{equation}
\begin{equation}\label{kapl}
\kappa_{\rm L}=\frac{r_{K_{\rm L}}}{r_{K^+}}
 {\tau(K_{\rm L})\over\tau(K^+)}\kappa_+ =1.80\cdot 10^{-10}
\end{equation}
with $\kappa_+$ given in (\ref{kapp}) and
$r_{K_{\rm L}}=0.944$ summarizing isospin
breaking corrections in relating $\klpn$ to $K^+\to\pi^0e^+\nu$
\cite{MP}.

Using the Wolfenstein
parametrization we can rewrite (\ref{bklpn}) as
\begin{equation}\label{bklpnwol1}
Br(\klpn)=1.80\cdot 10^{-10} \eta^2 A^4 X^2(x_t)
\end{equation}
or
\begin{equation}\label{bklpnwol2}
Br(\klpn)=3.29\cdot 10^{-5} \eta^2 |V_{cb}|^4 X^2(x_t)
\end{equation}
or using 
\begin{equation}\label{xxappr}
X(x_t)=0.65\cdot x_t^{0.575}
\end{equation}
as
\begin{equation}
Br(K_{\rm L}\to\pi^0\nu\bar\nu)=
3.0\cdot 10^{-11}
\left [ \frac{\eta}{0.39}\right ]^2
\left [\frac{\mtb(\mt)}{170~GeV} \right ]^{2.3} 
\left [\frac{\mid V_{cb}\mid}{0.040} \right ]^4 \,.
\label{bklpn1}
\end{equation}

The determination of $\eta$ using $Br(\klpn)$ requires the knowledge
of $V_{cb}$ and $\mt$. The very strong dependence on $V_{cb}$ or $A$
makes a precise prediction for this branching ratio difficult at
present.

\subsubsection{$\vcb$ and $\IM \lambda_t$  from $K_L\to\pi^0\nu\bar\nu$}
It was pointed out in \cite{AJB94} that the strong
dependence of $Br(\klpn)$ on $V_{cb}$, together with the clean nature of
this decay, can be used to determine this element without any hadronic
uncertainties. To this end $\eta$ and $m_t$ have to be known with
sufficient precision in addition to $Br(\klpn)$. 
Inverting (\ref{bklpn1})
one finds
\begin{equation}\label{vcbklpn}
|V_{cb}|=40.0\cdot 10^{-3} \sqrt{\frac{0.39}{\eta}}
\left[\frac{170\gev}{\mtb(\mt)}\right]^{0.575}
\left[\frac{Br(\klpn)}{3\cdot 10^{-11}}\right]^{1/4}\,.
\end{equation}
We note that the weak dependence of $V_{cb}$ on $Br(\klpn)$ allows
to achieve a high precision for this CKM element even when $Br(\klpn)$
is known with only relatively moderate accuracy, e.g.\ 10--15\%.

With $\eta$ 
determined one day from CP asymmetries in B-decays
and $\mt$ measured very precisely at LHC and NLC,
a measurement of $Br(\klpn)$ with an accuracy of $10\%$
would determine $\vcb$ with an error of $\pm 0.001$.
A comparision of
this determination of $|V_{cb}|$ with the usual one in tree level
B-decays would offer an excellent test of the Standard Model
and in the case of discrepancy would signal physics beyond 
it.

On the other hand inverting (\ref{bklpn}) and using (\ref{xxappr})
 one finds \cite{BB96}:
\begin{equation}\label{imlta}
\IM\lambda_t=1.36\cdot 10^{-4} 
\left[\frac{170\gev}{\mtb(\mt)}\right]^{1.15}
\left[\frac{Br(\klpn)}{3\cdot 10^{-11}}\right]^{1/2}\,.
\end{equation}
(\ref{imlta}) offers
 the cleanest method to measure $\IM\lambda_t$;
even better than the CP asymmetries
in $B$ decays discussed briefly in the next section.
\subsubsection{Numerical Analysis of \klpnn}
\label{sec:Kpnn:NumericalKL}
The $\mu_t$-uncertainties present in the function $X(x_t)$ have 
already been
discussed in connection with $\kpn$. After the inclusion of NLO
corrections they are so small that they can be neglected for all
practical purposes. 
At the level of $Br(\klpn)$ the ambiguity in the choice of $\mu_t$ is
reduced from $\pm 10\%$ (LO) down to $\pm 1\%$ (NLO), which
considerably increases the predictive power of the theory. Varying
$\mu_t$ according to (\ref{muctnum}) and using the input parameters
as in the case of $\kpn$ we find that the uncertainty
in $Br(\klpn)$
\begin{equation}\label{varbklpnLO}
2.53\cdot 10^{-11}\leq Br(\klpn)\leq 3.08\cdot 10^{-11}
\end{equation}
present in the leading order is reduced to
\begin{equation}\label{varbklpnNLO}
2.64\cdot 10^{-11}\leq Br(\klpn)\leq 2.72\cdot 10^{-11}
\end{equation}
after including NLO corrections. This means that the theoretical
uncertainty in the determination of $\eta$ amounts to only $\pm 0.7\%$
which is safely negligible.

Using the input parameters of table \ref{tab:inputparams}
one finds \cite{BJL96b}
\begin{equation}\label{klpnr4}
Br(\klpn)=\left\{ \begin{array}{ll}
(2.8 \pm 1.7)\cdot 10^{-11} & {\rm Scanning} \\
(2.6 \pm 0.9) \cdot 10^{-11} & {\rm Gaussian} \end{array} \right.
\end{equation}
where the error comes dominantly from the uncertainties in the CKM
parameters. The corresponding analysis with the ``future" input
parameters gives
\begin{equation}\label{klpnr5}
Br(\klpn)=\left\{ \begin{array}{ll}
(2.7 \pm 0.5)\cdot 10^{-11} & {\rm Scanning} \\
(2.6 \pm 0.3) \cdot 10^{-11} & {\rm Gaussian} \end{array} \right.
\end{equation}
\subsubsection{Summary and Outlook}
The accuracy of the Standard Model prediction for $Br(\klpn)$ has
improved considerably during the last five years. Indeed in 1992
values as high as $15\cdot 10^{-11}$ could be found in the
literature. This progress can be traced back mainly to the
improved values of $\mt$ and $\vcb$ and to some extent to 
the inclusion of NLO QCD corrections.
I expect that further progress
in the determination of CKM parameters via the standard analysis of
section 9 could reduce the errors in (\ref{klpnr4}) by at least a
factor of two during the next five years.
A numerical example is given in (\ref{klpnr5}).

The present upper bound on $Br(K_{\rm L}\to \pi^0\nu\bar\nu)$ from
FNAL experiment E799 \cite{XX97} is 
\begin{equation}\label{KLD}
Br(\klpn)<1.8 \cdot 10^{-6}\,.
\end{equation}
This is about five orders of magnitude above the Standard Model expectation
(\ref{klpnr4}).

How large could $Br(\klpn)$ really be? As shown  in \cite{NIR96}
one can easily derive by means of isospin symmetry the following 
{\it model independent} bound:
\begin{equation}
Br(\klpn) < 4.4 \cdot Br(\kpn)
\end{equation}
which through (\ref{kp97})  gives
\begin{equation}\label{B108}
Br(\klpn) < 6.1 \cdot 10^{-9}
\end{equation}
This bound is much stronger than the direct experimental bound in
(\ref{KLD}).

Now FNAL-E799 expects to reach
the accuracy ${\cal O}(10^{-8})$ and
a very interesting new experiment
at Brookhaven (BNL E926) \cite{AGS2} 
expects to reach the single event sensitivity $2\cdot 10^{-12}$
allowing a $10\%$ measurement of the expected branching ratio. 
There are furthermore plans
to measure this gold-plated  decay with comparable sensitivity
at Fermilab \cite{FNALKL} and KEK \cite{KEKKL}.
\begin{figure}[hbt]
\vspace{0.10in}
\centerline{
\epsfysize=2.7in
\epsffile{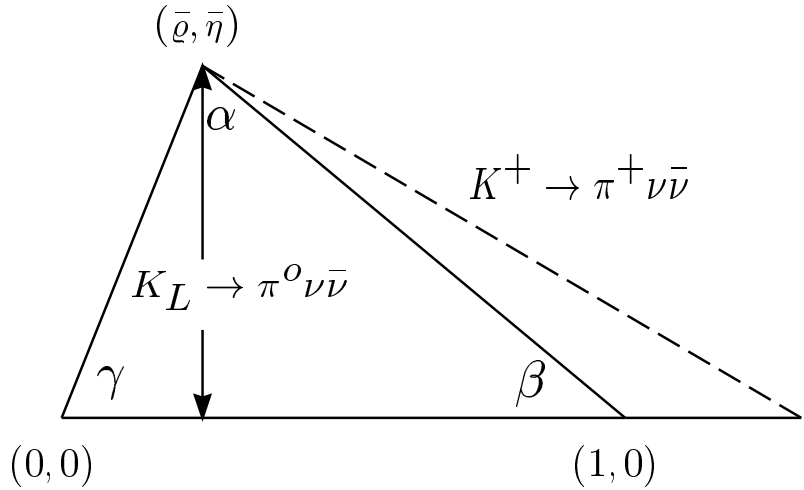}
}
\vspace{0.08in}
\caption{Unitarity triangle from $K\to\pi\nu\bar\nu$.}\label{fig:KPKL}
\end{figure}

\subsection{Unitarity Triangle and $\sin 2\beta$ from $K\to\pi\nu\bar\nu$}
\label{sec:Kpnn:Triangle}
The measurement of $Br(\kpn)$ and $Br(\klpn)$ can determine the
unitarity triangle completely, (see fig.~\ref{fig:KPKL}), 
provided $\mt$ and $V_{cb}$ are known \cite{BH}.
Using these two branching ratios simultaneously allows to eliminate
$|V_{ub}/V_{cb}|$ from the analysis which removes a considerable
uncertainty. Indeed it is evident from (\ref{bkpn}) and
(\ref{bklpn}) that, given $Br(\kpn)$ and $Br(\klpn)$, one can extract
both $\imlt$ and $\relt$. One finds \cite{BB4,BBL}
\begin{equation}\label{imre}
\imlt=\lambda^5{\sqrt{B_2}\over X(x_t)}\qquad
\relt=-\lambda^5{{\relc\over\lambda}P_0(X)+\sqrt{B_1-B_2}\over X(x_t)}\,,
\end{equation}
where we have defined the ``reduced'' branching ratios
\begin{equation}\label{b1b2}
B_1={Br(\kpn)\over 4.11\cdot 10^{-11}}\qquad
B_2={Br(\klpn)\over 1.80\cdot 10^{-10}}\,.
\end{equation}
Using next the expressions for $\imlt$, $\relt$ and $\relc$ given
in (\ref{2.51})--(\ref{2.53}) we find
\begin{equation}\label{rhetb}
\bar\varrho=1+{P_0(X)-\sqrt{\sigma(B_1-B_2)}\over A^2 X(x_t)}\,,\qquad
\bar\eta={\sqrt{B_2}\over\sqrt{\sigma} A^2 X(x_t)}
\end{equation}
with $\sigma$ defined in (\ref{109}). An exact treatment of the CKM
matrix shows that the formulae (\ref{rhetb}) are rather precise
\cite{BB4}. The error in $\bar\eta$ is below 0.1\% and
$\bar\varrho$ may deviate from the exact expression by at most
$\Delta\bar\varrho=0.02$ with essentially negligible error for
$0\leq\bar\varrho\leq 0.25$.

Using (\ref{rhetb}) one finds subsequently \cite{BB4}
\begin{equation}\label{sin}
r_s=r_s(B_1, B_2)\equiv{1-\bar\varrho\over\bar\eta}=\cot\beta\,, \qquad
\sin 2\beta=\frac{2 r_s}{1+r^2_s}
\end{equation}
with
\begin{equation}\label{cbb}
r_s(B_1, B_2)=\sqrt{\sigma}{\sqrt{\sigma(B_1-B_2)}-P_0(X)\over\sqrt{B_2}}\,.
\end{equation}
Thus within the approximation of (\ref{rhetb}) $\sin 2\beta$ is
independent of $V_{cb}$ (or $A$) and $m_t$. An exact treatment of
the CKM matrix confirms this finding to a high accuracy. The
dependence on $V_{cb}$ and $m_t$ enters only at order
$\ord(\lambda^2)$ and as a numerical analysis shows this
dependence can be fully neglected.

It should be stressed that $\sin 2\beta$ determined this way depends
only on two measurable branching ratios and on the function
$P_0(X)$ which is completely calculable in perturbation theory.
Consequently this determination is free from any hadronic
uncertainties and its accuracy can be estimated with a high degree
of confidence. 

An extensive numerical analysis of the formulae above has been presented
in \cite{BB4,BB96}. We summarize the results of the latter paper.
Assuming that the branching ratios are known to within $\pm 10\%$
\begin{equation}\label{bkpkl}
Br(\kpn)=(1.0\pm 0.1)\cdot 10^{-10}\,,\qquad
Br(\klpn)=(3.0\pm 0.30)\cdot 10^{-11}
\end{equation}
and choosing 
\begin{equation}\label{mtcv}
\mt=(170\pm 3)\gev,\quad P_0(X)=0.40\pm0.06,\quad
|V_{cb}|=0.040\pm 0.002,
\end{equation}
one finds the results given in the second column of table 
\ref{tabkb1}.
In the third column the results for the choice
$|V_{cb}|=0.040\pm 0.001$ are shown.
It should be remarked that the quoted errors for the input parameter
are quite reasonable if one keeps in mind
that it will take  five years to achieve the accuracy
assumed in (\ref{bkpkl}). The error in $P_0(X)$ in (\ref{mtcv}) 
results from the errors (see table \ref{tab:P0Kplus} and (\ref{muctnum})) 
in $\Lms^{(4)}$, $m_c$ and $\mu_c$ added quadratically.
Doubling the error in $m_c$ would give $P_0(X)=0.40\pm 0.09$ and
an increase of the errors in $|V_{td}|/10^{-3}$, $\bar\varrho$ and
$\sin 2\beta$ by at most $\pm 0.2$, $\pm 0.02$ and $\pm 0.01$
respectively, without any changes in $\bar\eta$ and 
${\rm Im}\lambda_t$.

\begin{table}
\caption[]{Illustrative example of the determination of CKM
parameters from $K\to\pi\nu\bar\nu$ for two choices of
$V_{cb}$ and other parameters given in the text.
\label{tabkb1}}
\begin{center}
\begin{tabular}{|c||c|c|}\hline
&$|V_{cb}|=0.040\pm 0.002$&$|V_{cb}|=0.040\pm 0.001$.\\ 
\hline
\hline
$|V_{td}|/10^{-3}$&$10.3\pm 1.1$&$10.3\pm 0.9$\\ 
\hline
$|V_{ub}/V_{cb}|$&$0.089\pm 0.017$
&$0.089\pm 0.011$ \\
\hline 
$\bar\varrho$&$-0.10\pm 0.16$ &$-0.10\pm 0.12$\\
\hline
$\bar\eta$&$0.38\pm 0.04$&$0.38\pm 0.03$\\
\hline
$\sin 2\beta$&$0.62\pm 0.05$&$0.62\pm 0.05$\\
\hline
${\rm Im}\lambda_t/10^{-4}$&$1.37\pm 0.07$
&$1.37\pm 0.07$ \\
\hline
\end{tabular}
\end{center}
\end{table}

We observe that respectable determinations of all considered 
quantities except for 
$\bar\varrho$ can be obtained.
Of particular interest are the accurate determinations of
$\sin 2\beta$ and of ${\rm Im}\lambda_t$.
The latter quantity as seen in (\ref{imre}) 
can be obtained from
$K_{\rm L}\to\pi^0\nu\bar\nu$ alone and does not require knowledge
of $V_{cb}$.

As pointed out in \cite{BB96},
$K_{\rm L}\to\pi^0\nu\bar\nu$ appears to be the best decay to 
measure ${\rm Im}\lambda_t$; even better than the CP asymmetries
in $B$ decays discussed in the next section.
The importance of measuring accurately  ${\rm Im}\lambda_t$ is evident.
It plays a central role in the phenomenology of CP violation
in $K$ decays and is furthermore equivalent to the 
Jarlskog parameter $J_{\rm CP}$ \cite{CJ}, 
the invariant measure of CP violation in the Standard Model, 
$J_{\rm CP}=\lambda(1-\lambda^2/2){\rm Im}\lambda_t$.

The accuracy to which $\sin 2\beta$ can be obtained from
$K\to\pi\nu\bar\nu$ is, in the  example discussed above, 
comparable to the one expected
in determining $\sin 2\beta$ from CP asymmetries in $B$ decays prior to
LHC experiments.  In this case $\sin 2\beta$ is determined best by
measuring CP violation in $B_d\to J/\psi K_{\rm S}$.
Using the formula  for the corresponding time-integrated 
CP asymmetry one finds an
interesting connection between rare $K$ decays and $B$ physics \cite{BB4}
\begin{equation}\label{kbcon}
{2 r_s(B_1,B_2)\over 1+r^2_s(B_1,B_2)}=
-a_{\mbox{{\scriptsize CP}}}(B_d\to J/\psi K_{\mbox{{\scriptsize S}}})
{1+x^2_d\over x_d}
\end{equation}
which must be satisfied in the Standard Model. We stress that except
for $P_0(X)$ given in table \ref{tab:P0Kplus} all quantities in
(\ref{kbcon}) can be directly measured in experiment and that this
relationship is essentially independent of $m_t$ and $V_{cb}$.
Due to very small theoretical uncertainties in (\ref{kbcon}), this
relation is particularly suited for tests of CP violation in the
Standard Model and offers a powerful tool to probe the physics
beyond it.
Further comparision between the potential of $K \to \pi \nu\bar\nu$ and
CP asymmetries in $B$ decays will be given in section 14.

Finally we  compare the determination of the unitarity triangle by
means of $K\to\pi\nu\bar\nu$ with
the one by means of the standard analysis of the unitarity triangle.
The results obtained from $K\to\pi\nu\bar\nu$ corresponding to table
\ref{tabkb1}
are given in the second and the third column of table 
\ref{tabkb2}. In the fourth and fifth column the corresponding results
of the standard analysis of the unitarity triangle are shown.
We observe that a considerable progress,
when compared with the present analysis of the unitarity triangle,
can be achieved through the measurements of $K\to\pi\nu\bar\nu$
decays.
\begin{table}
\caption[]{Illustrative example of the determination of CKM
parameters from $K\to\pi\nu\bar\nu$ and from the standard
analysis of the unitarity triangle.
\label{tabkb2}}
\vspace{0.4cm}
\begin{center}
\begin{tabular}{|c||c|c||c|c|}\hline
&$\sigma(|V_{cb}|)=\pm 0.002$ & $\sigma(|V_{cb}|)=\pm 0.001$
& {\rm Present} & {\rm Future}
\\ 
\hline
\hline
$\sigma(|V_{td}|) $& $\pm 10\% $ & $ \pm 9\% $
& $\pm 24\%$ & $\pm 7\%$\\ 
\hline 
$\sigma(\bar\varrho) $ & $\pm 0.16$ &$\pm 0.12$
& $\pm 0.32$  & $\pm 0.08$\\
\hline
$\sigma(\bar\eta)$ & $\pm 0.04$&$\pm 0.03$
&$\pm 0.12 $ & $\pm 0.03 $\\
\hline
$\sigma(\sin 2\beta)$ & $\pm 0.05$&$\pm 0.05$
& $\pm 0.22 $ & $\pm 0.05$\\
\hline
$\sigma({\rm Im}\lambda_t)$&$\pm 5\%$ &$\pm 5\%$ 
& $\pm 33\%$ & $\pm 8\%$\\
\hline
\end{tabular}
\end{center}
\end{table}
\subsection{$K\to\pi\nu\bar\nu$ Beyond the Standard Model}
In view of the very clean character of $K\to\pi\nu\bar\nu$,
these decays are very suitable for the study of new physics
effects. One example is the relation (\ref{kbcon}). Recently
several extensive analyses of supersymmetry effects in general
supersymmetric models have been presented in \cite{NIR96,GN1,BRS}
where further references can be found. In the MSSM these
effects are found to be very small but in certain more 
general scenarios of supersymmetry enhancements or
suppressions of $Br(K^+\to\pi^+\nu\bar\nu)$ and
$Br(K_L\to\pi^0\nu\bar\nu)$ by factors 2-3 cannot be
excluded. 
Model independent studies of these decays
can be found in \cite{NIR96,BRS}. The corresponding analyses in
various no--supersymmetric extensions of the Standard Model are 
listed in \cite{KLBSM}.
In particular, enhancement of $Br(K_L\to\pi^0\nu\bar\nu)$
by 1--2 orders of magnitude above the Standard Model
expectations is according to \cite{HHW98} 
still possible in four-generation models.
\subsection{The Decays $B\to X_{s,d}\nu\bar\nu$}
            \label{sec:HeffRareKB:klpinn2}
\subsubsection{Effective Hamiltonian}
The decays $B\to X_{s,d}\nu\bar\nu$ are the theoretically
cleanest decays in the field of rare $B$-decays.
They are dominated by the same $Z^0$-penguin and box diagrams
involving top quark exchanges which we encountered already
in the case of $\kpn$ and $\klpn$ except for the appropriate
change of the external quark flavours. Since the change of external
quark flavours has no impact on the $m_t$ dependence,
the latter is fully described by the function $X(x_t)$ in
(\ref{xx}) which includes
the NLO corrections \cite{BB2}. The charm contribution as
discussed at the beginning of this section is fully neglegible
here and the resulting effective Hamiltonian is very similar to
the one for $\klpn$ given in (\ref{hxnu}). 
For the decay $B\to X_s\nu\bar\nu$ it reads
\begin{equation}\label{bxnu}
{\cal H}_{\rm eff} = {G_{\rm F}\over \sqrt 2} {\alpha \over
2\pi \sin^2 \Theta_{\rm W}} V^\ast_{tb} V_{ts}
X (x_t) (\bar bs)_{V-A} (\bar\nu\nu)_{V-A} + h.c.   
\end{equation}
with $s$ replaced by $d$ in the
case of $B\to X_d\nu\bar\nu$.
 
The theoretical uncertainties related to the renormalization
scale dependence are as in $\klpn$ and 
can be essentially neglected. The same applies to long distance
contributions considered in \cite{BUC97}.
On the other hand $B\to X_{s,d}\nu\bar\nu$ are CP conserving and
consequently the relevant branching ratios are sensitive to 
$\vtd$ and $\vts$ as opposed to $Br(\klpn)$ in which
$\IM (V^\ast_{ts} V_{td})$ enters. As we will stress below the
measurement of both
$B\to X_{s}\nu\bar\nu$ and $B\to X_{d}\nu\bar\nu$ offers the
cleanest determination of the ratio $\vtd/\vts$.

\subsubsection{The Branching Ratios}
The calculation of the branching fractions for $B\to X_{s,d}\nu\bar\nu$ 
can be done similarly to $B\to X_s \gamma$ 
in the spectator model corrected for short distance QCD effects.
Normalizing as in these latter decays 
to $Br(B\to X_c e\bar\nu)$ and summing over three neutrino 
flavours one finds

\begin{equation}\label{bbxnn}
\frac{Br(B\to X_s\nu\bar\nu)}{Br(B\to X_c e\bar\nu)}=
\frac{3 \alpha^2}{4\pi^2\sin^4\Theta_{\rm W}}
\frac{|V_{ts}|^2}{|V_{cb}|^2}\frac{X^2(x_t)}{f(z)}
\frac{\bar\eta}{\kappa(z)}\,.
\end{equation}
Here $f(z)$ is the phase-space factor for $B\to X_c
e\bar\nu$ defined already in (\eqn{g}) and $\kappa(z)$ is the
corresponding QCD correction given in (\eqn{kap}). The
factor $\bar\eta$ represents the QCD correction to the matrix element
of the $b\to s\nu\bar\nu$ transition due to virtual and bremsstrahlung
contributions and is given by the well known expression
\begin{equation}\label{etabar}
\bar\eta=\kappa(0)=
1+\frac{2\alpha_s(m_b)}{3\pi}\left(\frac{25}{4}-\pi^2\right)
\approx 0.83\,.
\end{equation}
In the case of $B\to X_d\nu\bar\nu$ one has to replace $V_{ts}$ by
$V_{td}$ which results in a decrease of the branching ratio by
roughly an order of magnitude.

It should be noted that $Br(B \to X_s \nu\bar\nu)$ as given in
(\eqn{bbxnn}) is in view of $|V_{ts}/V_{cb}|^2 \approx 0.95 \pm 0.03$
essentially independent of the CKM parameters and the main uncertainty
resides in the value of $\mt$ which is already rather precisely
known. Setting $Br(B\to X_ce\bar\nu)=10.4\%$, $f(z)=0.54$,
$\kappa(z)=0.88$ and using the values in (\ref{alsinbr})
 we have
\begin{equation}
Br(B \to X_s \nu\bar\nu) = 3.7 \cdot 10^{-5} \,
\frac{|V_{ts}|^2}{|V_{cb}|^2} \,
\left[ \frac{\mtb(\mt)}{170\gev} \right]^{2.30} \, .
\label{eq:bxsnnnum}
\end{equation}

Taking next, in accordance with (\ref{kf}), $\kappa(z)=0.88$,
$f(z)=0.54\pm 0.04$ and 
$Br(B\to X_ce\bar\nu)=(10.4\pm 0.4)\%$
and using the input parameters of table \ref{tab:inputparams}
one finds \cite{BJL96b}
\begin{equation}\label{klpnr3}
Br(B \to X_s \nu\bar\nu)=\left\{ \begin{array}{ll}
(3.4 \pm 0.7)\cdot 10^{-5} & {\rm Scanning} \\
(3.2 \pm 0.4) \cdot 10^{-5} & {\rm Gaussian}\,. \end{array} \right.
\end{equation}
 
What about the data? 
One of the high-lights of FCNC-1996 was the upper bound:
\begin{equation}\label{124}
Br(B\to X_s \nu\bar\nu) < 7.7\cdot 10^{-4} 
\quad
(90\%\,\,\mbox{C.L.})
\end{equation}
obtained for the first time by ALEPH \cite{Aleph96}.
This is only a factor of 20 above the Standard Model expectation.
Even if the actual measurement of this decay is extremly difficult,
all efforts should be made to measure it. One should also 
make attempts to measure $Br(B\to X_d \nu\bar\nu)$. Indeed 

\begin{equation}\label{bratio}
\frac{Br(B\to X_d\nu\bar\nu)}{Br(B\to X_s\nu\bar\nu)}=
\frac{|V_{td}|^2}{|V_{ts}|^2}
\end{equation} 
offers the
cleanest direct determination of $\vtd/\vts$ as all uncertainties related
to $\mt$, $f(z)$ and $Br(B\to X_ce\bar\nu)$ cancel out.
\subsection{The Decays $B_{s,d}\to l^+l^-$}
\subsubsection{The Effective Hamiltonian}
The decays $B_{s,d}\to l^+l^-$ are after $B\to X_{s,d}\nu\bar\nu$ 
the theoretically cleanest decays in the field of rare $B$-decays.
They are dominated by the $Z^0$-penguin and box diagrams
involving top quark exchanges which we encountered already
in the case of $B\to X_{s,d}\nu\bar\nu$   except that due to
charged leptons in the final state the charge flow in the
internal lepton line present in the box diagram is reversed.
This results in a different $\mt$ dependence summarized
by the function  $Y(x_t)$, the NLO generalization \cite{BB2}
of the function $Y_0(x_t)$ given in (\ref{Y0}).
The charm contributions as
discussed at the beginning of this section are fully negligible
here and the resulting effective Hamiltonian is given 
for $B_s\to l^+l^-$ as follows:

\begin{equation}\label{hyll}
{\cal H}_{\rm eff} = -{G_{\rm F}\over \sqrt 2} {\alpha \over
2\pi \sin^2 \Theta_{\rm W}} V^\ast_{tb} V_{ts}
Y (x_t) (\bar bs)_{V-A} (\bar ll)_{V-A} + h.c.   \end{equation}
with $s$ replaced by $d$ in the
case of $B_d\to l^+l^-$.

The function $Y(x)$ is given by
\begin{equation}\label{yyx}
Y(x_t) = Y_0(x_t) + \aspi Y_1(x_t)\,,
\end{equation}
where $Y_0(x_t)$ can be found in (\ref{Y0})
and $Y_1(x_t)$ in (\ref{yy1}).
The leftover $\mu_t$-dependence in $Y(x_t)$ is tiny and amounts to
an uncertainty of $\pm 1\%$ at the level of the branching ratio.
We recall that $Y(x_t)$ can also be written as
\begin{equation}\label{yeta2}
Y(x_t)=\eta_Y\cdot Y_0(x_t)\,, \qquad\quad \eta_Y=1.026\pm 0.006\,,
\end{equation}
where $\eta_Y$ summarizes the NLO corrections.
With $\mt\equiv \mtb(\mt)$ this QCD factor
depends only very weakly on $m_t$. The range in (\ref{yeta2})
corresponds to $150\gev\leq m_t\leq 190\gev$. The dependence on
$\Lambda_{\overline{MS}}$ can be neglected. 

\subsubsection{The Branching Ratios}
The branching ratio for $B_s\to l^+l^-$ is given by \cite{BB2}
\begin{equation}\label{bbll}
Br(B_s\to l^+l^-)=\tau(B_s)\frac{G^2_{\rm F}}{\pi}
\left(\frac{\alpha}{4\pi\sin^2\Theta_{\rm W}}\right)^2 F^2_{B_s}m^2_l m_{B_s}
\sqrt{1-4\frac{m^2_l}{m^2_{B_s}}} |V^\ast_{tb}V_{ts}|^2 Y^2(x_t)
\end{equation}
where $B_s$ denotes the flavour eigenstate $(\bar bs)$ and $F_{B_s}$ is
the corresponding decay constant. Using
(\ref{alsinbr}), (\ref{yeta2}) and (\ref{PBE2}) we find in the
case of $B_s\to\mu^+\mu^-$
\begin{equation}\label{bbmmnum}
Br(B_s\to\mu^+\mu^-)=3.5\cdot 10^{-9}\left[\frac{\tau(B_s)}{1.6
\mbox{ps}}\right]
\left[\frac{F_{B_s}}{210\mev}\right]^2 
\left[\frac{|V_{ts}|}{0.040}\right]^2 
\left[\frac{\mtb(\mt)}{170\gev}\right]^{3.12}.
\end{equation}

The main uncertainty in this branching ratio results from
the uncertainty in $F_{B_s}$.
Using the input parameters of table \ref{tab:inputparams}
together with $\tau(B_s)=1.6$ ps and $F_{B_s}=(210\pm 30)\mev$ 
one finds \cite{BJL96b}
\begin{equation}\label{klpnr1}
Br(B_s\to\mu^+\mu^-)=\left\{ \begin{array}{ll}
(3.6 \pm 1.9)\cdot 10^{-9} & {\rm Scanning} \\
(3.4 \pm 1.2) \cdot 10^{-9} & {\rm Gaussian.} \end{array} \right.
\end{equation}

For $B_d\to\mu^+\mu^-$ a similar formula holds with obvious
replacements of labels $(s\to d)$. Provided the decay constants
$F_{B_s}$ and $F_{B_d}$ will have been calculated reliably by
non-perturbative methods or measured in leading leptonic decays one
day, the rare processes $B_{s}\to\mu^+\mu^-$ and $B_{d}\to\mu^+\mu^-$
should offer clean determinations of $|V_{ts}|$ and $|V_{td}|$. 
In particular the ratio
\begin{equation}
\frac{Br(B_d\to\mu^+\mu^-)}{Br(B_s\to\mu^+\mu^-)}
=\frac{\tau(B_d)}{\tau(B_s)}
\frac{m_{B_d}}{m_{B_s}}
\frac{F^2_{B_d}}{F^2_{B_s}}
\frac{|V_{td}|^2}{|V_{ts}|^2}
\end{equation}
having smaller theoretical uncertainties than the separate
branching ratios should offer a useful measurement of
$\vtd/\vts$. Since $Br(B_d\to\mu^+\mu^-)= {\cal O}(10^{-10})$
this is, however, a very difficult task. For $B_s \to \tau^+\tau^-$
and $B_s\to e^+e^-$ one expects branching ratios ${\cal O}(10^{-6})$
and ${\cal O}(10^{-13})$, respectively, with the corresponding branching 
ratios for $B_d$-decays by one order of magnitude smaller.

We should also remark that in conjunction with a future measurement of 
$x_s$, the branching
ratio $Br(B_s\to \mu\bar\mu)$ could help to determine 
the non-perturbative parameter $B_{B_s}$ and consequently allow
a test of existing non-perturbative methods \cite{B95}:
\begin{equation}
 B_{B_s}=
\left[\frac{x_s}{22.1}\right]
\left[\frac{\mtb(\mt)}{170~\mbox{GeV}} \right]^{1.6} 
\left[\frac{4.2\cdot 10^{-9}}{Br(B_s\to \mu\bar\mu)} \right] \,.
\end{equation}
This test could be of course affected by new physics contributions.
\subsubsection{Outlook}
What about the data?

The bounds on $B_{s,d}\to l\bar l$ are still
many orders of magnitude away from Standard Model expectations.
The best bounds come from CDF \cite{CDFMU}. One has:
\begin{equation}\label{MUBOUND}
Br(B_s\to\mu^+\mu^-)\le 
2.6\cdot 10^{-6}~~~~~(95\% C.L.)
\end{equation}
and $Br(B_d\to\mu^+\mu^-)\le 8.6\cdot 10^{-7}$.
CDF should reach in Run II the
sensitivity of $1\cdot 10^{-8}$ and $4\cdot 10^{-8}$ for
$B_d\to \mu\bar\mu$ and $B_s\to \mu\bar\mu$, respectively.
It is hoped that these decays will be observed at
LHC-B. The experimental status of $B\to\tau^+\tau^-$ and its
usefulness in tests of the physics beyond the Standard Model
is discussed in \cite{GLN96}.
\subsection{Higher Order Electroweak Effects in Rare Decays}
Until now we have considered various penguin and box diagrams
contributing to rare decays together with QCD corrections.
In none of these contributions the role of the neutral Higgs boson 
$H^0$ has been felt. Since the couplings of $H^0$ to fermions
are proportional to fermion masses, contributions of internal
$H^0$ are very strongly suppressed unless $H^0$ couples at both
ends of its propagator to the top. This situation appears first
at two-loop level in electroweak interactions. 
Examples of such diagrams can be constructed from diagrams
(a)--(c) in fig. \ref{L:7} by replacing there the gluon propagator
by the $H^0$-propagator. Even more important diagrams are obtained
by replacing $W^\pm$ and the gluon by the fictitious $\phi^\pm$
Higgs exchanges with the appropriate change in internal fermion
propagators.

Once the higher order electroweak contributions are considered and
one  recalls the extensive precision electroweak studies at 
$Z^0$-factories, an obvious question arises. What about the
ambiguities in rare meson decays stemming from various possible 
definitions of electroweak parameters? We have seen in this section
that the branching ratios $Br(K_L\to\pi^0\nu\bar\nu)$,
$Br(K^+\to\pi^+\nu\bar\nu)$, $Br(B\to X_{d,s}\nu\bar\nu)$ and
$Br(B\to l^-l^+)$ all had the following generic structure
\be\label{HO1}
Br \sim \frac{G^2_F \alpha^2(\mz)}{\sin^4\Theta_W}
\lbrack F(x_t)\rbrack^2,
\ee
where we have suppressed the charm contribution to 
$Br(K^+\to\pi^+\nu\bar\nu)$.

Now, there are several definitions of $\sin^2\Theta_W$. For
instance, $\sin^2\Theta_W=0.224$ in the on-shell scheme,
whereas the effective $\sin^2\hat\Theta_W|_{\rm eff}=0.230$.
These two choices result in branching ratios which differ by
$5.6\%$ to be compared with uncertainties of $1-2\%$ from
QCD after NLO corrections have been taken into account.
There is of course also the question of the scale in $\alpha$.
This is analogous to the recent discussion of two--loop
electroweak effects in $B\to X_s\gamma$ presented in section
12.4 and the related issue of $\alpha(\mu)$ there.

Clearly, in order to reduce such uncertainties,
one has to consider two-loop electroweak contributions
to the rare decays in question. Such an analysis has been performed
in \cite{BB97} in the large $\mt$-limit. Schematically the formula
(\ref{HO1}) reads now
\be\label{HO2}
Br \sim \frac{G^2_F \alpha^2(\mz)}{\sin^4\Theta_W}
\left[ F(x_t)+c G_F \mt^2 \frac{\mt^2}{\mw^2}\right]
\ee
where the second term represents  two-loop electroweak corrections
for large $\mt$. The scheme dependence of this term cancels in the 
large $\mt$ limit, the scheme dependence of $\sin^2\Theta_W$. 
Moreover the proper scale in $\alpha$ turns out to be $\mz$ as
anticipated (\ref{HO1}) and in all our calculations before.
Evidently the decays in question being governed by short distance
penguin and box contributions involve $\alpha(\mz)$, as opposed
to $B\to X_s\gamma$, where due to the on-shell photons $\alpha(m_e)$
matters.

The
large $\mt$ estimate of the full two-loop electroweak corrections
can be only trusted within a factor of two. Yet the residual parameter
uncertainties after the inclusion of these corrections turns out
to be less than $2\%$, which is well below the experimental
sensitivity in the forseeable future. Similarly for
$\sin^2\hat\Theta_W|_{\rm eff}=0.230$, used previously in our numerical
estimates, there is an enhancement of various branching ratios by
$1-2\%$ which can also be neglected. It should be stressed that all
these effects cancel in the determination of $\sin 2\beta$ from
$K\to\pi\nu\bar\nu$. Further details can be found in \cite{BB97}. 
\section{Future Visions}
\setcounter{equation}{0}
\subsection{Preliminaries}
Let us next have a look in the future and ask the question how well various
parameters of the Standard Model can be determined provided the
cleanest decays  have been measured
to some respectable precision. We have made already such an exercise in
section \ref{sec:Kpnn:Triangle}
using the decays $\klpn$ and $\kpn$. Now we want to make
an analogous analysis using CP-asymmetries in $B$-decays. This way we
will be able to compare the potentials of the CP asymmetries in
determining the parameters of the Standard Model with those
of the cleanest rare $K$-decays: $K_{\rm L}\to\pi^0\nu\bar\nu$ and
$K^+\to\pi^+\nu\bar\nu$. This section is based on 
\cite{BLO,AJB94,BB96,B95}.
\subsection{CP-Asymmetries in B-Decays}
CP violation in B-decays is certainly one of the most important 
targets of B-factories and of dedicated B-experiments at hadron 
facilities. It is well known that CP violating effects are expected
to occur in a large number of channels at a level attainable at 
forthcoming experiments. Moreover there exist channels which
offer the determination of CKM phases essentially without any hadronic
uncertainties. Since extensive reviews on CP violation in B decays can 
be found in the literature \cite{BF97,NQ,RF97}, 
let me concentrate only on the most important points.

The classic determination of $\alpha$ by means of the
time dependent CP  asymmetry in the decay
$B_d^0 \rightarrow \pi^+ \pi^-$ 
is affected by the "QCD penguin pollution" which has to be
taken care of in order to extract $\alpha$. 
The recent CLEO results for penguin dominated decays indicate that
this pollution could be substantial as stressed recently in particular
in \cite{ITAL}.
The most popular strategy to deal with this "penguin problem''
is the isospin analysis of Gronau and London \cite{CPASYM}. It
requires however the measurement of $Br(B^0\to \pi^0\pi^0)$ which is
expected to be below $10^{-6}$: a very difficult experimental task.
For this reason several, rather involved, strategies \cite{SNYD} 
have been proposed which
avoid the use of $B_d \to \pi^0\pi^0$ in conjunction with
$a_{CP}(\pi^+\pi^-,t)$. They are reviewed in \cite{BF97}. 
 It is to be seen which of these methods
will eventually allow us to measure $\alpha$ with a respectable precision.
It is however clear that the determination of this angle is a real
challenge for both theorists and experimentalists.

The CP-asymmetry in the decay $B_d \rightarrow \psi K_S$ allows
 in the Standard Model
a direct measurement of the angle $\beta$ in the unitarity triangle
without any theoretical uncertainties \cite {BSANDA}.
Of considerable interest \cite{RF97,PHI} is also the pure penguin decay
$B_d \rightarrow \phi K_S$, which is expected to be sensitive
to physics beyond the Standard Model. Comparision of $\beta$
extracted from $B_d \rightarrow \phi K_S$ with the one from
$B_d \rightarrow \psi K_S$ should be important in this
respect. An analogue of $B_d \rightarrow \psi K_S$ in $B_s$-decays
is $B_s \rightarrow \psi \phi$. The CP asymmetry measures here
$\eta$ \cite{B95} in the Wolfenstein parametrization. It is very
small, however, and this fact makes it a good place to look for the 
physics beyond the Standard Model. In particular the CP violation
in $B^0_s-\bar B^0_s$ mixing from new sources beyond the Standard
Model should be probed in this decay.

The two theoretically cleanest methods for the determination of $\gamma$
are: i) the full time dependent analysis of 
$B_s\to D^+_s K^{-}$ and $\bar B_s\to D^-_s K^{+}$  \cite{adk}
and ii) the well known triangle construction due to Gronau and Wyler 
\cite{Wyler}
which uses six decay rates $B^{\pm}\to D^0_{CP} K^{\pm}$,
$B^+ \to D^0 K^+,~ \bar D^0 K^+$ and  $B^- \to D^0 K^-,~ \bar D^0 K^-$.
Both methods are  unaffected by penguin contributions. 
The first method is experimentally very
challenging because of the
expected large $B^0_s-\bar B^0_s$ mixing. The second method is problematic
because of the small
branching ratios of the colour supressed channel $B^{+}\to D^0 K^{+}$
and its charge conjugate,
giving a rather squashed triangle and thereby
making
the extraction of $\gamma$ very difficult. Variants of the latter method
which could be more promising have been proposed in \cite{DUN2,V97}.
It appears that these methods will give useful results at later stages
of CP-B investigations. In particular the first method will be feasible
only at LHC-B.

All this has been known already for some time and is well documented
in the literature \cite{BF97,RF97}. Let us now be more explicit on
the most recent developments which deal with the extraction of
the angle $\gamma$ from the decays $B^0_d\to\pi^-K^+$, 
$B^+\to\pi^+K^0$ and their charge conjugates  
\cite{PAPIII}--\cite{defan}. These modes, which have recently been observed 
by the CLEO collaboration \cite{cleo}, should allow us to obtain direct 
information on $\gamma$ at future $B$-factories (BaBar, 
BELLE, CLEO III) (for interesting feasibility studies, see 
\cite{groro,wuegai,babar}). At present, there are only experimental results 
available for the combined branching ratios of these modes, i.e.\ averaged 
over decay and its charge conjugate, suffering from large hadronic
uncertainties. 

In order to determine the CKM angle $\gamma$ by using the strategy proposed
in \cite{PAPIII} (see also \cite{groro}), the separate branching ratios 
for $B^0_d\to\pi^-K^+$, $B^+\to\pi^+K^0$ and their charge conjugates are 
needed, i.e.\ the combined branching ratios are not sufficient, and an 
additional input is required to fix the magnitude of a certain decay 
amplitude $T$, which is usually referred to as a ``tree'' amplitude. Using 
arguments based on the factorization discussed in section 9, one expects 
that a future theoretical 
uncertainty of $|T|$ as small as ${\cal O}(10\%)$ may be achievable 
\cite{groro,wuegai}. Unfortunately detailed studies show, that the properly 
defined 
amplitude $T$ is actually not just a colour-allowed ``tree'' amplitude, 
where factorization may work reasonably well \cite{bjorken}. 
It receives also contributions from penguin and annihilation topologies 
due to certain rescattering effects \cite{defan,bfm} and consequently
the expectations in \cite{groro,wuegai}
appear too optimistic. In any case, some model dependence enters in the 
extracted value of $\gamma$ by means of these decays.

In this context
an interesting method for constraining $\gamma$,
which  
does not suffer from a model 
dependence related to $|T|$, is the method of 
Fleischer and Mannel \cite{fm2}. 
This method
uses only the combined rates for $B^{\pm}\to\pi^{\pm}K$ and 
$B_d\to\pi^{\mp}K^{\pm}$.
Assuming that the final state interactions and electroweak penguin 
contributions are small,
one finds  the bound:
\be\label{FMBOUND}
\sin^2\gamma \le 
\frac{Br(B_d\to\pi^{\mp}K^{\pm})}{Br(B^{\pm}\to\pi^{\pm}K)}\equiv R~.
\ee
The Fleischer-Mannel bound is of particular interest because the most 
recent CLEO data give $R=0.65\pm 0.40$ \cite{cleo}.
If true,
the FM--bound with $R<1$
would exclude the region around $\bar\varrho=0$ in the 
$(\bar\varrho,\bar\eta)$ space  putting 
the "$\gamma=90^\circ$ club" \cite{BjSt} into serious difficulties. It
should be stressed that excluding the region around $\bar\varrho=0$
would have a profound impact on the unitarity triangle dividing the
allowed region for its apex into well separated regions with
$\bar\varrho<0$ and $\bar\varrho>0$. The former could then probably
be eliminated by improving the lower bound on $\Delta M_s$ leaving
only a small allowed area with $\bar\varrho>0$. 
More details on
the implications of the FM--bound can be found in 
\cite{fm2,GNF,FRENCH}.

The crucial questions
then are, whether R is indeed smaller than unity  and whether
the assumptions used to obtain the FM bound can be justified.
The first question will hopefully be answered by CLEO and future
B factories. Here we concentrate on the second question.
Indeed,
the theoretical accuracy of the FM bound on $\gamma$ is limited by 
rescattering processes of the kind $B^+\to\{\pi^0K^+,\,\pi^0K^{\ast +},\,
\rho^0K^{\ast +},\,\ldots\,\}\to\pi^+K^0$ \cite{gewe}--\cite{atso} (for
earlier references, see \cite{FSI}), and by contributions from electroweak 
penguins \cite{groro,neubert,fm3}, which led to considerable interest in 
the recent literature. 

In order to gain some insight into this issue,
a completely general 
parametrization of the $B^+\to\pi^+K^0$ and $B^0_d\to\pi^-K^+$ decay 
amplitudes was presented in \cite{defan}, 
relying only on the isospin symmetry of strong 
interactions and the phase structure of the Standard Model. This 
parametrization leads to the following transparent expression for the 
minimal value of $R$:
\begin{equation}\label{Rmin}
R_{\rm min}=\kappa\,\sin^2\gamma\,+\,
\frac{1}{\kappa}\left(\frac{A_0}{2\,\sin\gamma}\right)^2,
\end{equation}
where the ``pseudo-asymmetry'' $A_0$ is defined by
\begin{equation}
A_0\equiv\frac{Br(B^0_d\to\pi^-K^+)-
Br(\overline{B^0_d}\to\pi^+K^-)}{Br(B^+\to\pi^+K^0)+
Br(B^-\to\pi^-\overline{K^0})}=
A_{\rm CP}(B_d\to\pi^\mp K^\pm)\,R~.
\end{equation}
Rescattering and electroweak penguin effects are included through the 
parameter $\kappa$, which is given by 
\begin{equation}
\kappa=\frac{1}{w^2}\left[\,1+2\,(\epsilon\,w)\cos\Delta+
(\epsilon\,w)^2\,\right]
\end{equation}
with
\begin{equation}\label{w-def}
w\equiv\sqrt{1+2\,\rho\,\cos\theta\cos\gamma+\rho^2}\,.
\end{equation}
The parameters $\rho$ and $\epsilon$ measure the ``strengths'' of the
rescattering processes and electroweak penguin contributions, respectively,
and $\theta$ and $\Delta$ are CP-conserving strong phases. Simple model 
estimates typically give values of $\rho$ and $\epsilon$ at the level of 
$1\%$. However, in a recent attempt to evaluate rescattering processes such 
as $B^+\to\{\pi^0K^+\}\to\pi^+K^0$, it is found that $\rho$ may be as large 
as ${\cal O}(10\%)$ \cite{fknp}. A similar feature arises also in a simple 
model to describe final-state interactions, which assumes elastic 
rescattering processes and has been proposed in \cite{gewe,neubert}. 
Also electroweak penguins may play a more important role than naively 
expected \cite{groro,neubert,fm3}, so that $\epsilon$ may actually be of 
${\cal O}(10\%)$. 

A detailed study of the impact of these effects on the generalized
bound on $\gamma$ 
related to (\ref{Rmin}) was performed in \cite{defan}. The ``original''
bound derived in \cite{fm2} corresponds to $\kappa=1$ and sets effectively
the asymmetry $A_0$ to zero. As soon as a non-vanishing experimental 
result for 
$A_0$ has been established, also an interval around $\gamma=0^\circ$ and 
$180^\circ$ can be ruled out, while the impact on the excluded region around 
$90^\circ$ is rather small \cite{defan}. 

An interesting feature of the rescattering effects is that they may lead to 
sizeable CP violation in the decay $B^+\to\pi^+K^0$ 
\cite{gewe}--\cite{atso}, 
in contrast to simple 
quark-level estimates, from which at most a few percent for this CP 
asymmetry~\cite{pert-pens} could be expected. 
This CP asymmetry provides a first step towards 
the experimental control of rescattering processes \cite{defan}. The 
rescattering effects can be included in the generalized bounds on 
$\gamma$ completely 
by using additional experimental information on the decay $B^+\to K^+
\overline{K^0}$ and its charge conjugate \cite{defan,rf-FSI}. 
Different strategies to constrain rescattering effects have also been
considered in \cite{fknp}.

At first sight, an experimental study of $B^+\to K^+ \overline{K^0}$ appears 
to be challenging, since model estimates performed at the perturbative quark 
level give a combined branching ratio 
$Br(B^\pm\to K^\pm K)={\cal O}(10^{-6})$, which is one order of 
magnitude below the present upper limit $2.1\times10^{-5}$ obtained by the 
CLEO collaboration. However, as was pointed out in \cite{defan,rf-FSI}, 
rescattering processes may well enhance this branching ratio by 
${\cal O}(10)$, so that it may be possible to study this 
mode to obtain insights into final state interactions at future $B$-factories.
Also electroweak penguins can be constrained by using additional information
\cite{defan}, and certainly experiment will tell us one day how important
rescattering processes and electroweak penguins in $B\to\pi K$ decays really
are. An interesting probe of $\gamma$ is also provided by $B_s\to K
\overline{K}$ decays, which can be combined with their $B_{u,d}
\to\pi K$ counterparts through the $SU(3)$ flavour symmetry \cite{bskk}. 

Finally I would like to mention a recent interesting paper of Lenz,
Nierste and Ostermaier \cite{LNO}, 
where inclusive direct CP-asymmetries in 
charmless $B^{\pm}$-decays including QCD effects have been studied.
These asymmetries should offer useful means to constrain the unitarity
triangle.

\subsection{CP-Asymmetries in $B$-Decays versus $K \to \pi \nu\bar\nu$}
Let us next compare the potentials of the CP asymmetries in
determining the parameters of the Standard Model with those
of the cleanest rare $K$-decays: $K_{\rm L}\to\pi^0\nu\bar\nu$ and
$K^+\to\pi^+\nu\bar\nu$.

To this end let us assume that the problems with the determination
of $\alpha$ will be solved somehow. Since in the usual rescaled 
unitarity triangle  one side is known, it suffices to measure
two angles to determine the triangle completely. This means that
the measurements of $\sin 2\alpha$ and $\sin 2\beta$ can determine
the parameters $\varrho$ and $\eta$.
As the standard analysis of the unitarity triangle of section 10
shows, $\sin 2\beta$ is expected to be large: $\sin 2\beta=0.58\pm 0.22$
implying the time-integrated CP asymmetry  
$a_{\rm CP}(B_d\to J/\psi K_{\rm S})$
as high as $(30 \pm 10)\%$.
The prediction for $\sin 2\alpha$ is very
uncertain on the other hand $(0.1\pm0.9)$ and even a rough measurement
of $\alpha$ would have a considerable impact on our knowledge of
the unitarity triangle as stressed in \cite{BLO,BB96}.

Measuring then $\sin 2\alpha$ and $\sin 2\beta$ from CP asymmetries in
$B$ decays allows, in principle, to fix the 
parameters $\bar\eta$ and $\bar\varrho$, which can be expressed as
\cite{AJB94}
\begin{equation}\label{ersab}
\bar\eta=\frac{r_-(\sin 2\alpha)+r_+(\sin 2\beta)}{1+
r^2_+(\sin 2\beta)}\,,\qquad
\bar\varrho=1-\bar\eta r_+(\sin 2\beta)\,,
\end{equation}
where $r_\pm(z)=(1\pm\sqrt{1-z^2})/z$.
In general the calculation of $\bar\varrho$ and $\bar\eta$ from
$\sin 2\alpha$ and $\sin 2\beta$ involves discrete ambiguities.
As described in \cite{AJB94}
they can be resolved by using further information, e.g.\ bounds on
$|V_{ub}/V_{cb}|$, so that eventually the solution (\ref{ersab})
is singled out.

Let us then consider two scenarios of the measurements of CP asymmetries 
in $B_d\to\pi^+\pi^-$ and $B_d\to J/\psi K_{\rm S}$, expressed in terms 
of $\sin 2\alpha$ and
$\sin 2\beta$:
\begin{equation}\label{sin2a2bI}
\sin 2\alpha=0.40\pm 0.10\,, \qquad \sin 2\beta=0.70\pm 0.06
\qquad ({\rm scenario\ I})
\end{equation}
\begin{equation}\label{sin2a2bII}
\sin 2\alpha=0.40\pm 0.04\,, \qquad \sin 2\beta=0.70\pm 0.02
\qquad ({\rm scenario\ II})\,.
\end{equation}
Scenario I corresponds to the accuracy being aimed for at $B$-factories
and HERA-B prior to the LHC era. An improved precision can be anticipated from
LHC experiments, which we illustrate with the scenario II.

In table \ref{tabkb} this way of the determination of
the Standard Model parameters is compared with the analogous analysis
using $\klpn$ and $\kpn$ which has been presented in section 13. We
recall that in the latter analysis
the following input has been used:
\begin{equation}\label{vcbmt}
|V_{cb}|=0.040\pm 0.002(0.001)\,, \qquad m_t=(170\pm 3) \mbox{GeV}
\end{equation}
\begin{equation}\label{bklkp}
Br(K_{\rm L}\to\pi^0\nu\bar\nu)=(3.0\pm 0.3)\cdot 10^{-11}\,,\qquad
Br(K^+\to\pi^+\nu\bar\nu)=(1.0\pm 0.1)\cdot 10^{-10}\,.
\end{equation}
The value  $|V_{cb}|=0.040\pm 0.002(0.001)$ is also used in B physics
scenarios I and II respectively.

\begin{table}
\caption[]{Illustrative example of the determination of CKM
parameters from $K\to\pi\nu\bar\nu$ and B-decays.
\label{tabkb}}
\vspace{0.4cm}
\begin{center}
\begin{tabular}{|c||c||c|c|}\hline
&$K\to\pi\nu\bar\nu$ 
& {\rm Scenario I} & {\rm Scenario II}
\\ 
\hline
\hline
$\sigma(|V_{td}|) $& $\pm 10\% (9\% )$
& $\pm 5.5\% (3.5\%)$ & $\pm 5.0\% (2.5\%)$\\ 
\hline 
$\sigma(\bar\varrho) $ & $\pm 0.16 (0.12)$
& $\pm 0.03$  & $\pm 0.01$\\
\hline
$\sigma(\bar\eta)$ & $\pm 0.04(0.03)$
&$\pm 0.04 $ & $\pm 0.01 $\\
\hline
$\sigma(\sin 2\beta)$ & $\pm 0.05$
& $\pm 0.06 $ & $\pm 0.02$\\
\hline
$\sigma({\rm Im}\lambda_t)$&$\pm 5\%$ 
& $\pm 14\%(11\%)$ & $\pm 10\%(6\%)$\\
\hline
\end{tabular}
\end{center}
\end{table}
As can be seen in table \ref{tabkb}, the CKM determination
using $K\to\pi\nu\bar\nu$ is competitive with the one based
on CP violation in $B$ decays in scenario I, except for $\bar\varrho$ which
is less constrained by the rare kaon processes.
On the other hand as advertised previously ${\rm Im}\lambda_t$ 
is better determined
in $K\to\pi\nu\bar\nu$ even if scenario II is considered.
The virtue of the comparision of the determinations
of various parameters using CP-B asymmetries with the determinations
in very clean decays $K\to\pi\nu\bar\nu$ is that any substantial deviations
from these two determinations would signal new physics beyond the
Standard Model.
 Formula (\ref{kbcon}) is an example of such a comparison.

\subsection{Unitarity Triangle from $\klpn$ and $\sin 2\alpha$}
Next, results from CP asymmetries in $B$ decays could also be
combined with measurements of $K\to\pi\nu\bar\nu$.
As an illustration we would like to present a scenario \cite{BB96}
where
the unitarity triangle is determined by $\lambda$, $V_{cb}$,
$\sin 2\alpha$ and $Br(K_{\rm L}\to\pi^0\nu\bar\nu)$.
In this case $\bar\eta$ follows directly from 
$Br(K_{\rm L}\to\pi^0\nu\bar\nu)$ (\ref{bklpn1}) and $\bar\varrho$ is
obtained using \cite{AJB94}
\begin{equation}\label{rhoalpha}
\bar\varrho=\frac{1}{2}-\sqrt{\frac{1}{4}-\bar\eta^2+
\bar\eta r_-(\sin 2\alpha)}\,,
\end{equation}
where $r_-(z)$ is defined after (\ref{ersab}).
The advantage of this strategy is that most CKM quantities are
not very sensitive to the precise value of $\sin 2\alpha$.
Moreover a high accuracy in 
${\rm Im}\lambda_t$ is automatically guaranteed. As shown in
table \ref{tabkl2a}, very respectable results can be expected
for other quantities as well with only modest requirements
on the accuracy of $\sin 2\alpha$. 
It is conceivable that theoretical uncertainties due to penguin
contributions could eventually be brought under control at least
to the level assumed in table \ref{tabkl2a}. 
\begin{table}
\caption[]{Determination of the CKM matrix from $\lambda$, $V_{cb}$,
$K_{\rm L}\to\pi^0\nu\bar\nu$ and $\sin 2\alpha$ from the CP asymmetry
in $B_d\to\pi^+\pi^-$ \cite{BB96}. Scenario A (B) assumes
$V_{cb}=0.040\pm 0.002 (\pm 0.001)$
and $\sin 2\alpha=0.4\pm 0.2 (\pm 0.1)$. In both cases we take
$Br(K_{\rm L}\to\pi^0\nu\bar\nu)\cdot 10^{11}=3.0\pm 0.3$ and
$\mt=(170\pm 3)\gev$. 
\label{tabkl2a}}
\begin{center}
\begin{tabular}{|c||c|c|c|}\hline
&&A&B \\
\hline
\hline
$\bar\eta$&$0.380$&$\pm 0.043$&$\pm 0.028$ \\
\hline
$\bar\varrho$&$0.070$&$\pm 0.058$&$\pm 0.031$ \\
\hline
$\sin 2\beta$&$0.700$&$\pm 0.077$&$\pm 0.049$ \\
\hline
$|V_{td}|/10^{-3}$&$8.84$&$\pm 0.67$&$\pm 0.34$ \\
\hline
$|V_{ub}/V_{cb}|$&$0.087$&$\pm 0.012$&$\pm 0.007$ \\
\hline 
\end{tabular}
\end{center}
\end{table}
As an alternative, $\sin 2\beta$ from $B_d\to J/\psi K_{\rm S}$ 
could be used as an independent input instead of $\sin 2\alpha$.
Unfortunately the combination of $K_{\rm L}\to\pi^0\nu\bar\nu$ and
$\sin 2\beta$ tends to yield somewhat less restrictive constraints
on the unitarity triangle \cite{BB96}. 
On the other hand it has of course the
advantage of being practically free of any theoretical uncertainties.   

\subsection{Unitarity Triangle and $\vcb$ from $\sin 2\alpha$,
$\sin 2\beta$ and $\klpn$}
As proposed in \cite{AJB94},
unprecedented precision for all basic CKM
parameters could be achieved by combining the cleanest $K$ and 
$B$ decays. 
While $\lambda$ is obtained as usual from
$K\to\pi e\nu$, $\bar\varrho$ and $\bar\eta$ could be determined
from $\sin 2\alpha$ and $\sin 2\beta$ as measured in CP
violating asymmetries in $B$ decays. Given $\eta$, one could
take advantage of the very clean nature of $K_{\rm L}\to\pi^0\nu\bar\nu$
to extract $A$ or, equivalently $|V_{cb}|$. As seen in (\ref{vcbklpn}),
this determination
benefits further from the very weak dependence of $|V_{cb}|$ on
the $K_{\rm L}\to\pi^0\nu\bar\nu$ branching ratio, which is only with
a power of $0.25$. Moderate accuracy in $Br(K_{\rm L}\to\pi^0\nu\bar\nu)$
would thus still give a high precision in $|V_{cb}|$.
As an example we take $\sin 2\alpha=0.40\pm 0.04$,
$\sin 2\beta=0.70\pm 0.02$ and 
$Br(K_{\rm L}\to\pi^0\nu\bar\nu)=(3.0\pm 0.3)\cdot 10^{-11}$,
$m_t=(170\pm 3)$ GeV. 
This yields \cite{BB96}:
\begin{equation}\label{rhetvcb}
\bar\varrho=0.07\pm 0.01\,,\qquad
\bar\eta=0.38\pm 0.01\,,\qquad
|V_{cb}|=0.0400\pm 0.0013\,,
\end{equation}
which would be a truly remarkable result. Again the comparision of
this determination of $|V_{cb}|$ with the usual one in tree level
$B$-decays would offer an excellent test of the Standard Model
and in the case of discrepancy would signal physics beyond 
it.  

\subsection{Unitarity Triangle from $R_t$ and $\sin 2\beta$} 
Another strategy is to use the measured value of $R_t$ together with
$\sin 2\beta$. Useful measurements of $R_t$ can be achieved
using the ratios $Br(B\to X_d \nu\bar\nu)/Br(B\to X_s \nu\bar\nu)$,
$\Delta M_d/\Delta M_s$,
$Br(B_d\to l^+l^-)/Br(B_s \to l^+l^-)$
 and $Br(\kpn)$. Then (\ref{ersab})
is replaced by \cite{B95}
\begin{equation}\label{5a}
\bar\eta=\frac{R_t}{\sqrt{2}}\sqrt{\sin 2\beta \cdot r_{-}(\sin 2\beta)}\,,
\quad\quad
\bar\varrho = 1-\bar\eta r_{+}(\sin 2\beta)\,.
\end{equation}
The numerical results of this exercise can be found in \cite{B95}.
Additional strategies involving the angle $\gamma$ 
can be found in \cite{BLO}.
\section{Summary and Outlook}
\setcounter{equation}{0}
We are approaching the end of our tour. I hope that some of you
enjoyed reading these  lectures as much as I did preparing,
delivering and finally writing them.
The collection of many techniques and formulae should be useful
in various phenomenological applications and constitutes hopefully
a good introduction to future research.
I hope that I have convinced the students that the field of weak decays
plays an important 
role in the deeper understanding of the Standard Model 
and particle physics in general.
Indeed the field of weak decays and of CP violation is one of the least
understood sectors of the Standard Model.
Even if the Standard Model is fully consistent with the existing data for
weak decay processes, the near future could change 
this picture
dramatically through the advances in experiment and theory.
In particular the experimental work
done in the next ten
years at BNL, CERN, CORNELL, DA$\Phi$NE, DESY, 
FNAL, KEK, SLAC and eventually LHC will certainly 
have considerable impact on this field.

Before closing these lectures with a few final messages, I would
like to make a list of things we could expect in the next ten years.
This list is certainly very biased by my own interests but could
be useful anyway. Here we go:

\begin{itemize}
\item
The error on the CKM elements $\vcb$ and $\vub$ could be decreased 
below 0.002 and 0.01, respectively. This progress should come mainly from
Cornell, $B$-factories and new theoretical efforts. It would have
considerable impact on the unitarity triangle and would improve
theoretical predictions for rare and CP-violating decays sensitive
to these elements.
\item
The error on $\mt$ should be decreased down to $\pm 3\gev$
at Tevatron in the Main Injector era and to $\pm 1\gev$ at LHC.
\item
The improved measurements of $\epe$ with the accuraccy of
 $\pm (1-2) \cdot 10^{-4}$ 
from CERN, FNAL and DA$\Phi$NE should give some insight into the 
physics of 
direct CP violation inspite of large theoretical uncertainties. 
Excluding confidently the superweak models would be an important result. 
In this respect measurements of CP-violating asymmetries in charged $B$
decays will also play an outstanding role. These experiments can be
performed e.g.\ at CLEO since no time-dependences
are needed. The situation concerning hadronic uncertainties is quite similar
to $\epe$. Although these CP asymmetries cannot be calculated
reliably, any measured non-vanishing values would unambiguously rule out 
superweak scenarios. Simultaneously one should hope 
that some definite progress in calculating relevant hadronic matrix elements 
will be made. 
\item
More events for $K^+\to\pi^+\nu\bar\nu$ could in principle
be seen at BNL already this or next year. In view of the theoretical 
cleanliness of this decay an observation of events at the $2\cdot 10^{-10}$
level would signal physics beyond the Standard Model.
A detailed study of this very
important decay requires, however, new experimental ideas and
new efforts. The new efforts \cite{AGS2,Cooper} in this direction allow 
to hope that
a measurement of $Br(\kpn)$ with an accuracy of $\pm 10 \%$ should
be possible before 2005. This would have a very important impact
on the unitarity triangle and would constitute an important test of
the Standard Model.
\item
The future improved inclusive $B \to X_{s,d} \gamma$ measurements
confronted with improved Standard Model predictions could
give the first signals of new physics. It appears that the errors
on the input parameters could be lowered further and the
theoretical error on $Br(B\to X_s\gamma)$ could be decreased
confidently down to $\pm 8 \%$ in the next years. The same
accuracy in the experimental branching ratio will hopefully
come soon from CLEO II and later from KEK and SLAC. 
This may, however, be insufficient to
disentangle new physics contributions although such an accuracy
should put important constraints on the physics beyond the Standard
Model. It would also be desirable to look for $B \to X_d \gamma$,
but this is clearly a much harder task.
\item
Similar comments apply to transitions $B \to X_s l^+l^-$ (not discussed
here)
which appear to be even  more sensitive to new physics contributions
than $ B \to X_{s,d} \gamma$. An observation of
$B \to X_s \mu\bar\mu$ is expected from D0 and $B$-physics dedicated
experiments at the beginning of the next 
decade. The distributions of various kind when measured should
be very useful in the tests of the Standard Model and its extensions.
\item
The theoretical status of $K_{\rm L}\to \pi^0 e^+ e^-$ and of 
$K_{\rm L}\to \mu\bar\mu$, which we did not cover here, 
should be improved to confront future
data. Experiments at DA$\Phi$NE should be very helpful in this
respect. The first events of $K_{\rm L}\to \pi^0 e^+ e^-$ should
come in the first years of the next decade from KAMI at FNAL.
The experimental status of $K_{\rm L}\to \mu\bar\mu$, with the 
experimental error of $\pm 7\%$ to be decreased soon down to $\pm 1\%$,
is truly impressive.
\item
The newly approved experiment at BNL to
measure $Br(\klpn)$ at the $\pm 10\%$ level before 2005 may make a decisive
impact on the field of CP violation. 
In particular $\klpn$ seems to allow the
cleanest determination of $\imlt$. Taken together with $\kpn$
a very clean determination of $\sin 2 \beta$ can be obtained.
\item
The measurement of the $B^0_s-\bar B^0_s$ mixing and in particular of
$B \to X_{s,d}\nu\bar\nu$ and 
$B_{s,d}\to \mu\bar\mu$ will take most probably longer time but
as stressed in these lectures all efforts should be made to measure
these transitions. Considerable progress on $B^0_s-\bar B^0_s$ mixing
should be expected from HERA-B, SLAC and TEVATRON in the first years
of the next decade. LHC-B should measure it to a high precision.
With the improved calculations of $\xi$ in (\ref{107b}) this will have
important impact on the determination of $\vtd$ and on the
unitarity triangle. 
\item
Clearly future precise studies of CP violation at SLAC-B, KEK-B, 
HERA-B, CORNELL, FNAL and  LHC-B providing first
direct measurements of $\alpha$, $\beta$ and $\gamma$ may totally
revolutionize our field. In particular the first signals
of new physics could be found in the $(\bar\varrho,\bar\eta)$ plane.
During the recent years several, in some cases quite sophisticated and
involved, strategies have been developed to extract these angles with
small or even no hadronic uncertainties. Certainly the future will bring
additional methods to determine $\alpha$, $\beta$ and $\gamma$. 
Obviously it is very desirable to have as many such strategies as possible
available in order to overconstrain the unitarity triangle and to resolve 
certain discrete ambiguities which are a characteristic feature of these 
methods.
\item
The forbidden or strongly suppressed transitions such as
$D^0-\bar D^0$ mixing and $K_{\rm L}\to \mu e$ are also very
important in this respect. Considerable progress in this area
should come from the experiments at BNL, FNAL and KEK.
\item
On the theoretical side,
one should hope that the non-perturbative
methods will be considerably improved so that various $B_i$ parameters
will be calculated with sufficient precision. It is very important
that simultaneously with advances in lattice QCD, further efforts
are being made in finding efficient analytical tools for calculating
QCD effects in the long distance regime. This is, in particular very
important in the field of non-leptonic decays, where one should
not expect too much from our lattice friends in the coming ten years
unless somebody will get a brilliant idea which will revolutionize
lattice calculations. The accumulation of data for non-leptonic $B$
and $D$
decays at Cornell, SLAC, KEK and FNAL should teach us more 
about the role of non-factorizable contributions and in particular
about the final state interactions. 
In this context, in the case of K-decays, important
lessons will come from DA$\Phi$NE which is an excellent machine
for testing chiral perturbation theory and other non-perturbative
methods. 
\end{itemize}

In any case the field of weak decays and in particular of the FCNC 
transitions and of CP violation have a great future and
one should expect that they could dominate particle physics in the first 
part of the next decade. 
Clearly the next ten years should be very exciting in this field
and it is advisable
to buy shares before it is too late.

\section{Final Messages}

The two weeks I have spent in Les Houches in August 1997 will remain in my
memory for ever. Therefore I would like to close these lectures by thanking
those who contributed most to this happening.

First of all I would like to thank Rajan Gupta and Francois David for
inviting me to this school and keeping me busy. In particular I would like to
thank Rajan for creating such a pleasent atmosphere and his persistent
e-mails reminding me that it is time to finish writing up these lectures.

However my warmest thanks go to the students of this school 
who made the sixteen hours of my presence in front of the blackboard
and the remaining time a real joy. In particular:
\bi
\item
 Many thanks to the magnificant seven: Fabien Motsch, Markus Peter, Solveig
Skadhauge, Thomas Teubner, Anja Werthenbach, Joerg Westphalen and Stefan
Wienzerl for keeping me alive during a two day mountain expedition.
Champagne offered after this tour by a very special
student of this school, Leung Ka Chun, will never be forgotten.
\item
 The results of our expedition appeared in hep-ph/9708777 under the title
``No Loops beyond the Trees in the Splittorff Renormalization Scheme", where
further details can be found. Splittorff, the youngest student of the school
was the only one of this Les Houches session to climb Mont Blanc. There is
nothing exciting in hep-ph/9708777 except one thing: 
this work will go down in history as yet another 
Buras et al. paper.
\item
 Many thanks to Luca Girlanda, Nicos Irges and Leszek Motyka 
for arranging table tennis
championships and to Andrzej Czarnecki for giving me Polysporin which allowed
me to reach quarter finals where I was slaughtered by a spanish
matador (Francisco Guerrero).
\item
 From all these remarks it is clear that I had rather close contacts with
the students of this school. 
Yet my closest companions, day and night, were
the washing machine and the dryer both placed next to my room.
The lively discussions, in particular at night, in front of my door
forced me to work hard on my lectures, except for the last night of my stay
when following the advice of the sole experimentalist
of the school (Fabien Motsch) I switched off these two important
inventions of this century.
\ei

I hope that these final comments made it clear why I have enjoyed this school
so much. Many thanks to all of you.    

Particular thanks go to Markus Lautenbacher for creating many figures
and a number of numerical calculations. 
I would also like to thank Robert Fleischer, Paolo Gambino, 
Axel Kwiatkowski, Mikolaj Misiak, Nicolas Pott
and Luca Silvestrini for helpful discussions during  the preparation of
these lectures. 

This work has been supported by the
German Bundesministerium f{\"u}r Bildung and Forschung under contract 
06 TM 874  and DFG Project Li 519/2-2.

\vfill\eject


\begin{thebibliography}{999}
\bibitem{CAB}
N. Cabibbo, Phys. Rev. Lett. {\bf 10} (1963) 531.
\bibitem{KM}
{ M. Kobayashi and K. Maskawa},
 { Prog. Theor. Phys.} {\bf 49} (1973) 652.
\bibitem{OPE}
K.G. Wilson, { Phys. Rev.} {\bf 179} (1969) 1499;
K.G. Wilson and W. Zimmermann, { Comm. Math. Phys.} 
{\bf 24} (1972) 87.
\bibitem{ZIMM}
W. Zimmermann, in Proc. 1970 Brandeis Summer Institute in
Theor. Phys, (eds. S. Deser, M. Grisaru and H. Pendleton),
MIT Press, 1971, p.396; 
{ Ann. Phys.} {\bf 77} (1973) 570.
\bibitem{SUMA}
{ E.C.G. Sudarshan and R.E. Marshak}, Proc. Padua-Venice Conf. on 
Mesons and Recently Discovered Particles (1957).
\bibitem{GF}
{ R.P. Feynman and M. Gell-Mann,}
 { Phys. Rev.} {\bf 109} (1958) 193.
\bibitem{WIT}
E. Witten, { Nucl. Phys.} {\bf B 120} (1977) 189.
\bibitem{REGM}
E.C.G. Stueckelberg and A. Petermann, Helv. Phys. Acta {\bf 26} (1953)
499;
M. Gell--Mann and F.E. Low, { Phys. Rev.} {\bf 95} (1954) 1300;
L.V. Ovsyannikov, Dokl. Acad. Nauk SSSR {\bf 109} (1956) 1112;
K. Symanzik, Comm. Math. Phys. {\bf 18} (1970) 227;
C.G. Callan Jr, { Phys. Rev.} {\bf D 2} (1970) 1541.
\bibitem{HV1}
G. 't Hooft,
{ Nucl. Phys.} {\bf B 61} (1973) 455.
\bibitem{Weinberg}
S. Weinberg, { Phys. Rev.} {\bf D 8} (1973) 3497.
\bibitem{DIAG}
D. Zeppenfeld, Z. Phys. {\bf C 8} (1981) 77;
L.L. Chau, { Phys. Rev.} {\bf D 43} (1991) 2176;
M. Gronau, J.L. Rosner and D. London, Phys. Rev. Lett. {\bf 73} (1994) 21;
O.F. Hernandez, M. Gronau, J.L. Rosner and D. London,
{ Phys. Lett.} {\bf B 333} (1994) 500, { Phys. Rev.} {\bf D 50} (1994) 4529.
\bibitem{HQE1}
J. Chay, H. Georgi and B. Grinstein,
{ Phys. Lett.} {\bf B 247} (1990) 399.
\bibitem{HQE2}
I.I. Bigi, N.G. Uraltsev and A.I. Vainshtein,
{ Phys. Lett.} {\bf B 293} (1992) 430
[E: {\bf B 297} (1993) 477].
I.I. Bigi, M.A. Shifman, N.G. Uraltsev and A.I. Vainshtein,
Phys. Rev. Lett. {\bf 71} (1993) 496;
B. Blok, L. Koyrakh, M.A. Shifman and A.I. Vainshtein,
{ Phys. Rev.} {\bf D 49} (1994) 3356 [E: {\bf D 50} (1994) 3572].
\bibitem{HQE3}
A.V. Manohar and M.B. Wise,
{ Phys. Rev.} {\bf D 49} (1994) 1310.
\bibitem{GIM1}
{ S.L. Glashow, J. Iliopoulos and L. Maiani}
{ Phys. Rev.} {\bf D 2} (1970) 1285.
\bibitem{PBE0}
{ G. Buchalla, A.J. Buras and M.K. Harlander,} { Nucl. Phys.}
 {\bf B 349} (1991) 1.
\bibitem{BBL}
{ G. Buchalla, A.J. Buras and M. Lautenbacher,} 
{ Rev. Mod. Phys} {\bf 68} (1996) 1125.
\bibitem{BF97}
{ A.J. Buras and R. Fleischer,} hep-ph/9704376, to appear in \cite{HFII}.
\bibitem{HFII}
A.J. Buras and M. Lindner, Heavy Flavours II, World Scientific,
1998.
\bibitem{MUTA}
T. Muta, Foundations of Chromodynamics, World Scientific, 1987.
\bibitem{BOOK1}
M.E. Peskin and D.V. Schroeder, An Introduction to Quantum Field
Theory, Addison-Wesley Publishing Company.
\bibitem{BOOK1b}
F. Mandl and G. Shaw, Quantum Field Theory, John Wiley $\&$ Sons.
\bibitem{BOOK1c}
T.-P. Cheng and L.-F. Li, Gauge Theory of Elementary Particle
Physics, Clarendon Press, Oxford. 
\bibitem{BOOK1a}
L.H. Ryder, Quantum Field Theory, Cambridge University Press.
\bibitem{BOOK2}
J.F. Donoghue, E. Golowich and B.R. Holstein, Dynamics of the
Standard Model, Cambridge Monographs.
\bibitem{BOOK3}
D. Bailin and A. Love, Introduction to Gauge Field Theory,
Adam Hilger, Bristol and Boston.
\bibitem{BOOK4}
S. Pokorski, Gauge Field Theory, Cambridge Monographs.
\bibitem{BOOK5}
S. Weinberg, The Quantum Theory of Fields, Cambridge University
Press.
\bibitem{Collins}
J.C. Collins, Renormalization, Cambridge University Press.
\bibitem{CHAU}
{ L.L. Chau and W.-Y. Keung}, 
{ Phys. Rev. Lett.} {\bf 53} (1984) 1802.
\bibitem{PDG}
{ Particle Data Group,} { Phys. Rev.} {\bf D 54} (1996) 1.
\bibitem{WO}
{ L. Wolfenstein}, { Phys. Rev. Lett.} {\bf 51} (1983) 1945.
\bibitem{HALE}
{ H. Harari and M. Leurer,} { Phys. Lett.} {\bf B 181} (1986) 123.
\bibitem{BLO}
{ A.J. Buras, M.E. Lautenbacher and G. Ostermaier,}
{ Phys. Rev.} {\bf D 50} (1994) 3433.
\bibitem{schubert}
{ M. Schmidtler and K.R. Schubert}, { Z. Phys.} {\bf C 53}
(1992) 347.
\bibitem{js}
{ C. Jarlskog and R. Stora},
{ Phys. Lett.} {\bf B 208} (1988) 268.
\bibitem{LER1}
{ H. Leutwyler and M. Roos}, { Z. Physik} {\bf C25} (1984) 91.
\bibitem{DHK}
{ J.F. Donoghue, B.R. Holstein and S.W. Klimt,}
{ Phys. Rev.} {\bf D35} (1987) 934.
\bibitem{Gibbons}
{ L. Gibbons}, in proceedings of the 28th International Conference
on High Energy Physics, July 1996, Warsaw, Poland, page 183.
\bibitem{SUV}
{ M. Shifman, N.G. Uraltsev and A. Vainshtein,}
{ Phys. Rev.} {\bf D51} (1995) 2217;
I. Bigi, M. Shifman and N. Uraltsev, Ann. Rev. Nucl. Part. Sci.
47 (1997) 591.
\bibitem{Neubert}
{  M. Neubert,} { Phys. Lett.} {\bf B338} (1994) 84;
{ Int. J. Mod. Phys.} {\bf A11} (1996) 4173.
\bibitem{Braun}
{P. Ball, M. Beneke and V.M. Braun,} 
{ Phys. Rev.} {\bf D52} (1995) 3929.
\bibitem{CZMI}
A. Czarnecki, { Phys. Rev. Lett.} {\bf 76} (1996) 4124.
A. Czarnecki and K. Melnikov, 
{ Phys. Rev. Lett.} {\bf 78} (1997) 3630.
\bibitem{CLEOU}
{J.P. Alexander} et al. (CLEO), CLNS 96/1419, CLEO 96-9 (1996).
\bibitem{IL}
{ T. Inami and C.S. Lim,}
{ Progr. Theor. Phys.} {\bf 65} (1981) 297.
\bibitem{DDD}
G. Burdman, hep-ph/9407378, hep-ph/9508349.
\bibitem{AB80}
A.J. Buras,
{ Rev. Mod. Phys} {\bf 52} (1980) 199.
\bibitem{WEISZ}
{ A.J. Buras and P.H. Weisz,}
{ Nucl. Phys.} {\bf B 333} (1990) 66.
\bibitem{HV}
{ G. 't Hooft and M. Veltman},
{ Nucl. Phys.} {\bf B 44} (1972) 189.
\bibitem{BM}
{ P. Breitenlohner and D. Maison,} { Comm. Math. Phys.}
{\bf 52} (1977) 11, 39, 55.
\bibitem{Bo}
G. Bonneau, { Phys. Lett.} {\bf B 94} (1980) 147;
{ Nucl. Phys.} {\bf B 177} (1981) 523.
\bibitem{Si}
W. Siegel, { Phys. Lett.} {\bf B 84} (1979) 193.
\bibitem{ACMP}
{ G. Altarelli, G. Curci, G. Martinelli and S. Petrarca,}
{ Nucl. Phys.} {\bf B 187} (1981) 461.
\bibitem{Ma}
D. Maison, { Phys. Lett.} {\bf B 150} (1985) 39.
\bibitem{NT}
H. Nicolai and P.K. Townsend, { Phys. Lett.} {\bf B 93} (1980) 111;
P. Majumdar, E.C. Poggio and H.J. Schnitzer, 
{ Phys. Rev.} {\bf D 21} (1980) 2203.
\bibitem{CAND}
{ R.~Grigjanis, P.J.~O'Donnell, M.~Sutherland and H.~Navelet,} 
{ Phys.~Lett.} {\bf B213} (1988) 355;
{ Phys.~Lett.} {\bf B286} (1992) 413 E.
\bibitem{MISD}
M. Misiak, { Phys.~Lett.} {\bf B321} (1994) 113.
\bibitem{AD}
{ D.A. Akyeampong and R. Delbourgo}, { Nuovo Cim.}
{\bf 17A} (1973) 578, {\bf 18A} (1973) 94, {\bf 19A} (1974) 219.
\bibitem{KNS}
J.G. K\"orner, N. Nasrallah and K. Schilcher, 
{ Phys. Rev.} {\bf D 41} (1990) 888.
\bibitem{JaLau}
M. Jamin and M. Lautenbacher, Comput. Phys. Commun. 74 (1993) 265.
\bibitem{BBDM}
{ W.A. Bardeen, A.J. Buras, D.W. Duke and T. Muta},
{ Phys. Rev.} {\bf D 18} (1978) 3998.
\bibitem{Gross}
D.J. Gross,  Methods in Fleld Theory, (eds. R. Balian and J. Zinn-Justin),
North-Holland, 1976, p. 141.
\bibitem{Schmelling}
{ M. Schmelling}, in proceedings of the 28th International Conference
on High Energy Physics, July 1996, Warsaw, Poland, page 91.
\bibitem{MAR80}
W.J. Marciano, { Phys. Rev.} {\bf D 12} (1975) 3861.
\bibitem{BB1}
{ G. Buchalla and A.J. Buras,}
{ Nucl. Phys.} {\bf B 398} (1993) 285.
\bibitem{GH97}
{ C. Greub and T. Hurth,} { Phys. Rev.} {\bf D 56} (1997) 2934.
\bibitem{BKP2}
A.J. Buras, A. Kwiatkowski and N. Pott,
{ Nucl. Phys.} {\bf B 517} (1998) 353. 
\bibitem{BJLW1}
{ A.J. Buras, M. Jamin, M.E. Lautenbacher and P.H. Weisz,}
{ Nucl. Phys.} {\bf B 370} (1992) 69;
{ Nucl. Phys.} {\bf B 400} (1993) 37.
\bibitem{BJLW2}
{ A.J. Buras, M. Jamin and M.E. Lautenbacher,}
{ Nucl. Phys.} {\bf B 400} (1993) 75.
\bibitem{CURCI}
C. Curci and G. Ricciardi, { Phys. Rev.} {\bf D 47} (1993) 2991.
\bibitem{MISTRIK}
K. Chetyrkin, M. Misiak and M{\"u}nz, hep-ph/9711280; hep-ph/9711266.
\bibitem{MAIANI}
G. Altarelli and L. Maiani, { Phys. Lett.} {\bf B 52} (1974) 351;
M.K. Gaillard and B.W. Lee, { Phys. Rev. Lett.} {\bf 33} (1974) 108.  
\bibitem{BJLW}
{ A.J. Buras, M. Jamin and M.E. Lautenbacher,}
{ Nucl. Phys.} {\bf B 408} (1993) 209.
\bibitem{ROMA1}
{ M. Ciuchini, E. Franco, G. Martinelli and L. Reina,}
{ Phys. Lett.} {\bf B 301} (1993) 263.
\bibitem{ROMA2}
{ M. Ciuchini, E. Franco, G. Martinelli and L. Reina,}
{ Nucl. Phys.} {\bf B 415} (1994) 403.
\bibitem{GREEK}
N. Tracas and N. Vlachos, { Phys. Lett.} {\bf B 115} (1982) 419.
\bibitem{BKP1}
A.J. Buras, A. Kwiatkowski and N. Pott, 
{ Phys. Lett.} {\bf B 414} (1997) 157.
\bibitem{BuMu:94}
{ A.J. Buras and M. M{\"u}nz,}
{ Phys. Rev.} {\bf D 52} (1995) 186.
\bibitem{DuGr}
M.J. Dugan and B. Grinstein,
 { Phys. Lett.} {\bf B 256} (1991) 239.
\bibitem{HNE}
S. Herrlich and U. Nierste, { Nucl. Phys.} {\bf B 445} (1995) 39.
\bibitem{SH94}
S. Herrlich, Technical Univesity, PhD Thesis 1994 (in German).
\bibitem{UN95}
U. Nierste, Technical Univesity, PhD Thesis 1995.
\bibitem{FLEISCHP}
R. Fleischer, { Zeit. Phys.} {\bf C 62} (1994) 81.
\bibitem{PENGUIN}
A.I. Vainshtein, V.I. Zakharov and M.A. Shifman, JEPT {\bf 45} (1977) 670.
\bibitem{BH}
{ A.J. Buras and M.K. Harlander,} {\it A Top Quark Story, in
Heavy Flavours,} eds. A.J. Buras and M. Lindner, World Scientific,
1992, p.58.
\bibitem{BBHLS}
{ G. Buchalla, A.J. Buras, M.K. Harlander, M.E. Lautenbacher and C. Salazar,}
{ Nucl. Phys.} {\bf B 355} (1991) 305.
\bibitem{MW96}
P. Cho, M. Misiak and D. Wyler, { Phys. Rev.} {\bf D 54} (1996) 3329.
\bibitem{AAA}
A. Ali, Th. Mannel and Ch. Greub, { Zeit. Phys.} {\bf C 67} (1995) 417.
\bibitem{AJB94a}
A.J. Buras, { Nucl. Phys.} {\bf B 434} (1995) 606.
\bibitem{BJW90}
{ A.J. Buras, M. Jamin, and P.H. Weisz,}
{ Nucl. Phys.} {\bf B347} (1990) 491;\\
J. Urban, F. Krauss, U.Jentschura and G. Soff, 
{ Nucl. Phys.} {\bf B523} (1998) 40. 
\bibitem{BB3}
{ G. Buchalla and A.J. Buras,}
{ Nucl. Phys.} {\bf B 412} (1994) 106.
\bibitem{HNa}
{ S. Herrlich and U. Nierste,}
{ Nucl. Phys.} {\bf B419} (1994) 292. 
\bibitem{HNb}
{ S.~Herrlich and  U.~Nierste},
{ Phys. Rev.} {\bf D52} (1995) 6505; 
{ Nucl. Phys.} {\bf B476} (1996) 27. 
\bibitem{MisMu:94}
{ M.Misiak and M. M{\"u}nz,}
{ Phys. Lett.} {\bf B344} (1995) 308.
\bibitem{Buch:93}
{ G. Buchalla,} { Nucl. Phys.} {\bf B 391} (1993) 501.
\bibitem{Bagan}
{ E. Bagan, P.Ball, V.M. Braun and P.Gosdzinsky,}
{ Nucl. Phys.} {\bf B 432} (1994) 3;
{ E. Bagan} { et al.,} { Phys. Lett.} {\bf B 342} (1995) 362;
{\bf B 351} (1995) 546.
\bibitem{JP}
{ M. Jamin and A. Pich,}
{ Nucl.~Phys.} {\bf B425} (1994) 15.
\bibitem{BB2}
{ G. Buchalla and A.J. Buras,}
{ Nucl. Phys.} {\bf B 400} (1993) 225.
\bibitem{BB5}
{ G. Buchalla and A.J. Buras,}
{ Phys. Lett.} {\bf B 336} (1994) 263.
\bibitem{BLMM}
{ A. J. Buras, M. E. Lautenbacher, M. Misiak and M. M{\"u}nz,}
{ Nucl.~Phys.} {\bf B423} (1994) 349.
\bibitem{Mis:94}
{ M. Misiak,}
{ Nucl.~Phys.} {\bf B393} (1993) 23;
{ Erratum}, { Nucl.~Phys.} {\bf B439} (1995) 461.
\bibitem{Potte}
N. Pott, hep-ph/9710503.
\bibitem{Krakauer}
J. Krakauer, "Into Thin Air", Villard Books, New York, 1997.
\bibitem{BF95}
A.J. Buras and R. Fleischer, { Phys. Lett.} {\bf B 341} (1995) 379.
\bibitem{ITAL}
M. Ciuchini, E. Franco, G. Martinelli, and  L. Silvestrini,
{ Nucl. Phys.} {\bf B501} (1997) 271;
M. Ciuchini, R. Contino, E. Franco, G. Martinelli, and  L. Silvestrini,
{ Nucl. Phys.} {\bf B512} (1998) 3.
\bibitem{AG2} 
{  A.~Ali, and  C.~Greub,} { Z.Phys.} {\bf C49} (1991) 431;  
{ Phys.~Lett.} {\bf B259} (1991) 182;
{ Phys.~Lett.} {\bf B361} (1995) 146.
\bibitem{Yao1} {  K.~Adel and Y.P.~Yao,} 
{ Modern Physics Letters} {\bf A8} (1993) 1679;
{ Phys. Rev.} {\bf D 49} (1994) 4945.
\bibitem{Pott} 
{ N. Pott,} { Phys. Rev.} {\bf D 54} (1996) 938.
\bibitem{GREUB}
{ C. Greub, T. Hurth and D. Wyler,} { Phys.~Lett.} {\bf B380} 
(1996) 385; { Phys. Rev.} {\bf D 54} (1996) 3350; 
{ C. Greub and T. Hurth,} hep-ph/9608449.
\bibitem{CZMM}
{ K.G. Chetyrkin, M. Misiak and M. M{\"u}nz,} 
{ Phys. Lett.} {\bf B400} (1997) 206; hep-ph/9612313. 
\bibitem{GAMB}
M. Ciuchini, G. Degrassi, P. Gambino and G.F. Giudice, 
hep-ph/9710335.
\bibitem{strum} 
P. Ciafaloni, A. Romanino, and A. Strumia, 
hep-ph/9710312.
\bibitem{BUSI}
A.J. Buras and L. Silvestrini, TUM-HEP-315/98, hep-ph/9806278.
\bibitem{KR98}
A. Khodjamirian and R. R\"uckl, hep-ph/9801443, to appear in \cite{HFII}.
\bibitem{FEYNMAN}
J. Schwinger, { Phys. Rev. Lett.} {\bf 12} (1964) 630; 
R.P. Feynman, in {\it Symmetries in Particle Physics}, ed. A. Zichichi,
Acad. Press 1965, p.167; O. Haan and B. Stech, 
{ Nucl. Phys.} {\bf B 22}  (1970) 448.  
\bibitem{STECHF}
D. Fakirov and B. Stech, { Nucl. Phys.} {\bf B 133}  (1978) 315;
L.L. Chau, Phys. Rep. {\bf 95} (1983) 1.  
\bibitem{BAUER}
M. Wirbel, B. Stech and M. Bauer, { Z. Phys.}{\bf C 29} (1985) 637.
M. Bauer, B. Stech and M. Wirbel, { Z. Phys.}{\bf C 34} (1987) 103.
\bibitem{NEUBERT}
M. Neubert, V. Rieckert, B. Stech and Q.P. Xu, in ``Heavy Flavours",
 eds. A.J. Buras and M. Lindner (World Scientific, Singapore, 1992),
p. 286.
\bibitem{BJORKEN}
J.D. Bjorken,
{ Nucl. Phys.} {\bf B } (Proc. Suppl.) 11 (1989) 325;
SLAC-PUB-5389.
\bibitem{DUGAN}
M.J. Dugan and B. Grinstein,
{ Phys. Lett.} {\bf B 255} (1991) 583.
\bibitem{NS97}
M. Neubert and B. Stech,
[hep-ph/9705292], to appear in \cite{HFII};
B. Stech [hep-ph/9706384];
M.Neubert, Nucl. Phys. {\bf B } (Proc. Suppl.) 64 (1998) 474, 
[hep-ph/9801269].
\bibitem{ISGUR}
C. Reader and N. Isgur,
{ Phys. Rev.} {\bf D 47} (1993) 1007.
\bibitem{ITALY}
M. Ciuchini, R. Contino, E. Franco, G. Martinelli, L. Silvestrini,
hep-ph/9801420.
\bibitem{LNF}
D. Du and Z. Xing, { Phys. Lett.} {\bf B 312} (1993) 199;
A. Deandrea et al., { Phys. Lett.} {\bf B 318} (1993) 549,
{ Phys. Lett.} {\bf B 320} (1994) 170;
N.G. Deshpande, B. Dutta, S. Oh, { Phys. Rev.} {\bf D 57} (1998) 5723,
hep-ph/9712445.
\bibitem{Cheng}
H.-Y. Cheng, { Phys. Lett.} {\bf B 335} (1994) 428,
{ Phys. Lett.} {\bf B 395} (1997) 345;
H.-Y. Cheng and B. Tseng, [hep-ph/9708211], [hep-ph/9803457].
\bibitem{Soares}
J.M. Soares, { Phys. Rev.} {\bf D 51} (1995) 3518.
\bibitem{GNF}
A. Ali and C. Greub, { Phys. Rev.} {\bf D57} (1998) 2996;
A. Ali, J. Chay, C. Greub and P. Ko, 
{ Phys. Lett.} {\bf B 424} (1998) 161.
\bibitem{AKL98}
A. Ali, G. Kramer and C.-D. L\"u, hep-ph/9804363.
\bibitem{EW}
 E. Witten,
{ Nucl. Phys.} {\bf B 160} (1979) 57.
\bibitem{BGR}
 A.J. Buras, J.M. G{\'e}rard and R. R\"uckl,
{ Nucl. Phys.} {\bf B 268} (1986) 16.
\bibitem{rome2}
M. Ciuchini, E. Franco, G. Martinelli, L. Reina and L. Silvestrini, 
Z.Phys. {\bf C68} (1995) 239.
\bibitem{BUSI2}
A.J. Buras and L. Silvestrini, work in progress.
\bibitem{GALE}
{ M.K. Gaillard and B.W. Lee,} 
{ Phys. Rev.} {\bf D10} (1974) 897.
\bibitem{ARGUS}
{ H. Albrecht et al. (ARGUS)}, { Phys. Lett.} {\bf B192} (1987) 245;
{ M. Artuso et al. (CLEO)}, { Phys. Rev. Lett.} {\bf 62} (1989) 2233.
\bibitem{CRONIN}
{ J.H. Christenson, J.W. Cronin, V.L. Fitch and R. Turlay},
{ Phys. Rev. Lett.} {\bf 13} (1964) 128. 
\bibitem{RF97}
{ R. Fleischer}, { Int. J. of Mod. Phys.}
 {\bf A12} (1997) 2459.
\bibitem{CHAU83}
L.L. Chau, { Physics Reports}, {\bf 95} (1983) 1.
\bibitem{BSSII}
A.J. Buras, W. Slominski and H. Steger,
{ Nucl. Phys.} {\bf B245} (1984) 369.
\bibitem{GERAR}
J. Bijnens, J.-M. G{\'e}rard and G. Klein, 
{ Phys. Lett.} {\bf B257} (1991) 191.
\bibitem{BELU}
R. Belusevic, KEK preprint 97--264 (1998).
\bibitem{NIRSLAC}
Y. Nir, SLAC-PUB-5874 (1992).
\bibitem{GUPTA98}
R. Gupta, hep-ph/9801412.
\bibitem{JLQCD}
S. Aoki et al., JLQCD collaboration, 
{ Phys. Rev. Lett.} {\bf 80} (1998) 5271. 
\bibitem{GKS}
G. Kilcup, R. Gupta and S.R. Sharpe, 
{ Phys. Rev.} {\bf D57} (1998) 1654.
\bibitem{G67}
R. Gupta, T. Bhattacharaya, and S.R. Sharpe, 
{ Phys. Rev.} {\bf D55} (1997) 4036.
\bibitem{APE}
L. Conti, A. Donini, V. Gimenez, G.Martinelli, M. Talevi and
A. Vladikas, hep-lat/9711053.
\bibitem{BERT97}
S. Bertolini, J.O. Eeg, M. Fabbrichesi and E.I. Lashin,
{ Nucl. Phys.} {\bf B514} (1998) 63.
\bibitem{BBG0}
{W.A. Bardeen, A.J. Buras and J.-M. G\'erard,}
{ Phys. Lett.} {\bf B211} (1988) 343;
 {J-M. G\'erard,} { Acta Physica Polonica} {\bf B21} (1990) 257. 
\bibitem{Bijnens}
{ J. Bijnens and J. Prades,} { Nucl. Phys.} {\bf B444} (1995) 523. 
\bibitem{Prades}
{ A. Pich and E. de Rafael,} { Phys. Lett.} {\bf B158} (1985) 477;
{ J. Prades} {  et al,} { Z. Phys.}  {\bf C51} (1991) 287.
\bibitem{Donoghue}
{ J.F. Donoghue, E. Golowich and B.R. Holstein,}
{ Phys. Lett.} {\bf B119} (1982) 412.
\bibitem{Flynn}
{ J. Flynn}, in proceedings of the 28th International Conference
on High Energy Physics, July 1996, Warsaw, Poland, page 335; 
 J.M. Flynn and C.T. Sachrajda, hep-lat/9710057, 
to appear in \cite{HFII}.
\bibitem{Bernard}
C. Bernard, hep-ph/9709460.
\bibitem{QCDSF}
{ E. Bagan, P. Ball, V.M. Braun and H.G. Dosch},
{ Phys. Lett.} {\bf B278} (1992) 457;
{ M. Neubert}, { Phys. Rev.} {\bf D45} (1992) 2451 and references
therein.
\bibitem{Buras}
A.J. Buras, { Phys. Lett.} {\bf B317} (1993) 449.
\bibitem{ABWAR}
A.J. Buras, hep-ph/9610461.
\bibitem{Drell}
The LEP B Oscillations Working Group, LEPBOSC 97/002.3 (August 14, 1997).
\bibitem{NAR}
{S. Narison,}
{ Phys. Lett.} {\bf B322} (1994) 247.
\bibitem{FRENCH}
Y. Grossman, Y. Nir, S. Plaszczynski and M. Schune,
{ Nucl.~Phys.} {\bf B511} (1998) 69.
\bibitem{PAGA}
P. Paganini, F. Parodi, P. Roudeau and A. Stocchi, hep-ph/9711261;
F. Parodi, P. Roudeau and A. Stocchi, hep-ph/9802289.
\bibitem{BJL96b}
{ A.J. Buras, M.Jamin and M.E. Lautenbacher,} 1997, unpublished;
A.J. Buras, hep-ph/9711217.
\bibitem{ciuchini:95}
{ M.~Ciuchini}, { E.~Franco}, { G.~Martinelli}, {L.~Reina
 and   L.~Silvestrini},
 { Z. Phys.} {\bf C68} (1995) 239.
\bibitem{ALUT}
{ A. Ali and D. London,}
{ Z. Phys.} {\bf C65} (1995) 431; 
{ Nucl. Phys. B} (proc. Suppl.) {\bf 54A} (1997) 297;
A. Ali, hep-ph/9801270.
\bibitem{barr:93}
{ G.~D. Barr} { et~al.},
{ Phys. Lett.} {\bf B317} (1993) 233.
\bibitem{gibbons:93}
{ L.~K. Gibbons} { et~al.},
{ Phys. Rev. Lett.} {\bf 70} (1993) 1203.
\bibitem{wolfenstein:64}
{ L.~Wolfenstein},
 { Phys. Rev. Lett.} {\bf 13} (1964) 562.
\bibitem{HALL}
R. Barbieri, L. Hall, A. Stocchi, and  N. Weiner, hep-ph/9712252.
\bibitem{flynn:89}
{ J.~M. Flynn} and { L.~Randall},
{ Phys. Lett.} {\bf B224} (1989) 221; erratum ibid.\ { Phys.
  Lett.} {\bf B235} (1990) 412.
\bibitem{buchallaetal:90}
{ G.~Buchalla}, { A.~J. Buras}, and { M.~K. Harlander},
{ Nucl. Phys.} {\bf B337} (1990) 313.
\bibitem{GW79}
{ F.J. Gilman and M.B. Wise,} { Phys. Lett.} {\bf B83} (1979) 83;
{ B. Guberina and R.D. Peccei,} { Nucl. Phys.} {\bf B163} (1980) 289.
\bibitem{donoghueetal:86} 
{ J.F. Donoghue, E. Golowich, B.R. Holstein and J. Trampetic,}
{ Phys. Lett.} {\bf B179} (1986) 361. 
\bibitem{burasgerard:87}
{ A.~J. Buras} and { J.-M. G{\'e}rard},
{ Phys. Lett.} {\bf B192} (1987) 156.
\bibitem{lusignoli:89}
{ M. Lusignoli,} { Nucl. Phys.} {\bf B325} (1989) 33. 
\bibitem{ANII}
M. Ciuchini, E. Franco and R. Onforio,
{ Mod. Phys. Lett.} {\bf A5} (1990) 2173;
W.A. Bardeen,  A.~J. Buras and  J.-M. G{\'e}rard,
{ Phys. Lett.} {\bf B180} (1986) 133;
{ Nucl. Phys.} {\bf B293} (1987) 787;
H.-Y. Cheng, { Phys. Rev.} {\bf D37} (1988) 1908.
\bibitem{BW84}
{ J. Bijnens and M.B. Wise,} { Phys. Lett.} {\bf B137} (1984) 245.
\bibitem{bardeen:87}
{ W. A. Bardeen}, { A. J. Buras} and { J.-M. G{\'e}rard},
 { Phys. Lett.} {\bf B180} (1986) 133;
{ Nucl. Phys.} {\bf B293} (1987) 787;
{ Phys. Lett.} {\bf B192} (1987) 138.
\bibitem{PW91}
{ E.A. Paschos and Y.L. Wu,} { Mod. Phys. Lett.} {\bf A6} (1991) 93;
{ M. Lusignoli, L. Maiani, G. Martinelli and L. Reina,} 
{ Nucl. Phys.} {\bf B369} (1992) 139.
\bibitem{WW}
{ B. Winstein and L. Wolfenstein,} { Rev. Mod. Phys.} {\bf 65} (1993)
1113.
\bibitem{BERT98}
S. Bertolini, M. Fabbrichesi and J.O. Eeg, hep-ph/9802405.
\bibitem{DI12}
{ W. A. Bardeen}, { A. J. Buras} and { J.-M. G{\'e}rard},
{ Phys. Lett.} {\bf B192} (1987) 138;
{ A. Pich and E. de Rafael}, { Nucl. Phys.} {\bf B358} (1991) 311;
{ M. Neubert and B. Stech}, { Phys. Rev.} {\bf D 44} (1991) 775;
{ M. Jamin and A. Pich}, { Nucl. Phys.} {\bf B425} (1994) 15; 
{ J. Kambor, J. Missimer and D. Wyler},
{ Nucl. Phys.} {\bf B346} (1990) 17;
{ Phys. Lett.} {\bf B261} (1991) 496;
{ V. Antonelli, S. Bertolini, M. Fabrichesi, and E.I. Lashin},
{ Nucl. Phys.} {\bf B469} (1996) 181.
\bibitem{kilcup:91}
{ G.~W. Kilcup},
 { Nucl. Phys. (Proc. Suppl.)} {\bf B20} (1991) 417.
\bibitem{sharpe:91}
{ S.~R. Sharpe},
 { Nucl. Phys. (Proc. Suppl.)} {\bf B20} (1991) 429.
\bibitem{kilcup:98}
D. Pekurovsky and G. Kilcup, hep-lat/9709146.
\bibitem{heinrichetal:92}
{ J.~Heinrich}, { E.~A. Paschos}, { J.-M. Schwarz}, and { Y.~L.
  Wu},
{ Phys. Lett.} {\bf B279} (1992) 140.
\bibitem{paschos:96}
{ E.~A. Paschos},
 review presented at the 27th Lepton-Photon Symposium,
  Beijing, China (August 1995).
  \bibitem{DORT98}
T. Hambye, G.O. K\"ohler, E.A. Paschos, P.H. Soldan and W.A. Bardeen,
hep-ph/9802300.
\bibitem{NJL}
D. Espriu, E. de Rafael and J. Taron, { Nucl. Phys.} {\bf B345} (1990) 22;
J. Bijnens, Phys. Rept. {\bf 265} (1996) 369.
\bibitem{TR96}
S. Bertolini, J.O. Eeg and  M. Fabbrichesi,
{ Nucl. Phys.} {\bf B449} (1995) 197;
{ Nucl. Phys.} {\bf B476} (1996) 225.
\bibitem{TR97}
S. Bertolini, J.O. Eeg, M. Fabbrichesi and E.I. Lashin,
{ Nucl. Phys.} {\bf B514} (1998) 93.
\bibitem{BJL96a}
{ A.~J. Buras}, { M.~Jamin}, and { M.~E. Lautenbacher},
{ Phys. Lett.} {\bf B389} (1996) 749.
\bibitem{buraslauten:93}
{ A.~J. Buras} and { M.~E. Lautenbacher},
{ Phys. Lett.} {\bf B318} (1993) 212.
\bibitem{narison:95}
{ S.~Narison},
{ Phys. Lett.} {\bf B358} (1995) 113.
\bibitem{jaminmuenz:95}
{ M.~Jamin} and { M.~M{\"u}nz},
 { Z. Phys.} {\bf C66} (1995) 633.
\bibitem{chetyrkinetal:95}
K.~G. Chetyrkin, C.~A. Dominguez,  D.~Pirjol, and 
  K.~Schilcher,
{ Phys. Rev.} {\bf D51} (1995) 5090;
K.~G. Chetyrkin, D.~Pirjol, and 
  K.~Schilcher,
{ Phys. Lett.} {\bf B404} (1997) 337.
\bibitem{Paver}
P. Colangelo, F. De Fazio, G. Nardulli, and N. Paver,
{ Phys. Lett.} {\bf B408} (1997) 340.
\bibitem{Jamin97}
M. Jamin, Nucl. Phys. B. Proc. Suppl. {\bf 64} (1998) 250.
\bibitem{Yndurain}
F.J. Yndurain, hep-ph/9708300.
\bibitem{Dosch}
H.G. Dosch and S. Narison, { Phys. Lett.} {\bf B417} (1998) 173.
\bibitem{DERAF}
L. Lellouch, E. de Rafael, and J. Taron, 
{ Phys. Lett.} {\bf B414} (1997) 195.
\bibitem{ciuchini:96}
{ M. Ciuchini}, Nucl. Phys. B. Proc. Suppl. {\bf 59} (1997) 149.
\bibitem{BELKOV}
A.A. Belkov, G. Bohm, A.V. Lanyov, A.A. Moshkin, hep-ph/9704354.
\bibitem{Bert} 
{ S.~Bertolini, F.~Borzumati and A.~Masiero,} 
{ Phys. Rev. Lett.} {\bf 59} (1987) 180.
\bibitem{Desh} 
{ N.~G.~Deshpande, P.~Lo, J.~Trampetic, G.~Eilam and P. Singer}
{ Phys. Rev. Lett.} {\bf 59} (1987) 183.
\bibitem{Grin} 
{ B.~Grinstein, R.~Springer and M.B.~Wise,} 
{ Nucl.~Phys.} {\bf B339} (1990) 269.
\bibitem{Odon} 
{ R.~Grigjanis, P.J.~O'Donnell, M.~Sutherland and H.~Navelet,} 
{ Phys.~Lett.} {\bf B213} (1988) 355;
{ Phys.~Lett.} {\bf B286} (1992) 413 E.
\bibitem{CFMRS:93}
{ M. Ciuchini, E. Franco, G. Martinelli, L. Reina and L. Silvestrini,}
 { Phys.~Lett.} {\bf B316} (1993) 127.
\bibitem{CFRS:94}
{ M. Ciuchini, E. Franco, L. Reina and L. Silvestrini,}
{ Nucl.~Phys.} {\bf B421} (1994) 41.
\bibitem{CCRV:94a}
{ G.~Cella, G.~Curci, G.~Ricciardi and  A.~Vicer{\'e},}
{ Phys.~Lett.} {\bf B325} (1994) 227.
\bibitem{CCRV:94b}
{ G.~Cella, G.~Curci, G.~Ricciardi and A.~Vicer{\'e},}
{ Nucl.~Phys.} {\bf B431} (1994) 417.
\bibitem{AG1} 
{ A.~Ali, and  C.~Greub,} { Z.Phys.} {\bf C60} (1993) 433.  
\bibitem{BMMP:94}
{ A. J. Buras, M. Misiak, M. M{\"u}nz and S. Pokorski,}
{ Nucl.~Phys.} {\bf B424} (1994) 374.
\bibitem{BG98}
F.M. Borzumati and Ch. Greub, hep-ph/9802391.
\bibitem{CM78} 
{ N. Cabibbo and L. Maiani}, 
{ Phys.~Lett.} {\bf B79} (1978) 109.
\bibitem{KIMM}
{ C.S. Kim and A.D. Martin},
{ Phys.~Lett.} {\bf B225} (1989) 186.
\bibitem{N89} 
{ Y. Nir,}
{ Phys.~Lett.} {\bf B221} (1989) 184.
\bibitem{KN98}
A.L. Kagan and M. Neubert, hep-ph/9805303.
\bibitem{FLS96} 
{ A.F.~Falk, M.~Luke and M.~Savage,}
{ Phys. Rev.} {\bf D53} (1996) 2491.
\bibitem{LDGAMMA}
{ D. Atwood, B. Blok, and A. Soni}, 
{ Int. J. Mod. Phys.} {\bf A11} (1996) 3743;
{ H.-Y. Cheng,} { Phys. Rev.} {\bf D51} (1995) 6228;
{ E. Golowich and S. Pakvasa,} { Phys. Rev.} {\bf D51} (1995) 1215;
{ G. Ricciardi}, { Phys.~Lett.} {\bf B355} (1995) 313;
{ A. Khodjamirian, G. Stoll and D. Wyler},
{ Phys.~Lett.} {\bf B358} (1995) 129;
{ G. Eilam, A. Ioannissian and R.R. Mendel}, 
{ Z. Phys.} {\bf C71} (1996) 95;
{ G. Eilam, A. Ioannissian, R.R. Mendel and P. Singer},
 { Phys. Rev.} {\bf D53} (1996) 3629;
{ J.M. Soares,} { Phys. Rev.} {\bf D53} (1996) 241;
{ J. Milana,} { Phys. Rev.} {\bf D53} (1996) 1403;
{ N.G. Deshpande, X.-G. He and J. Trampetic,}
{ Phys.~Lett.} {\bf B367} (1996) 362.
\bibitem{VOL96}
{ M.B. Voloshin}, { Phys.~Lett.} {\bf B397} (1997) 275.
\bibitem{LRW97}
{ A. Khodjamirian, R. R\"uckl, G. Stoll and D. Wyler},
{ Phys.~Lett.} {\bf B402} (1997) 167;
{ Z. Ligeti, L. Randall and M.B. Wise}, 
{ Phys.~Lett.} {\bf B402} (1997) 178;
{ A.K. Grant, A.G. Morgan, S. Nussinov and R.D. Peccei},
 { Phys. Rev.} {\bf D56} (1997) 3151.
\bibitem{BUC97}
G. Buchalla, G. Isidori and S.-J. Rey, 
{ Nucl.~Phys.} {\bf B511} (1998) 594.
\bibitem{CZMA}
A. Czarnecki and W.J. Marciano, hep-ph/9804252.
\bibitem{STRUMIA}
A. Strumia, hep-ph/9804274. 
\bibitem{CLEO2} { M.S. Alam} { et. al} (CLEO), 
{ Phys. Rev. Lett.} {\bf 74} (1995) 2885.
\bibitem{CLEO98}
S. Glenn (CLEO), talk presented at the Meeting of the American Physics
Society, Columbus, Ohio, 18-21 March 1998.
\bibitem{ALEPH}
R. Barate et al., (ALEPH), CERN-EP/98-044.
\bibitem{chwil} 
W.S. Hou and R.S. Willey, { Phys.~Lett.} {\bf B202} (1988) 591; 
B. Grinstein, R. Springer, and M. Wise, 
{ Nucl.~Phys.} {\bf B339} (1990) 269. 
\bibitem{anl}
H. Anlauf, { Nucl.~Phys.} {\bf B430} (1994) 245.
\bibitem{multiH} 
P. Krawczyk and S. Pokorski, 
{ Nucl.~Phys.} {\bf B364} (1991) 10;
  Y. Grossmann, Y. Nir, R. Rattazzi, in \cite{HFII}.
\bibitem{rattazzi} 
See for instance G.F. Giudice, R. Rattazzi, hep-ph/9801271.
\bibitem{io} R. Barbieri and G.F. Giudice, 
{ Phys.~Lett.} {\bf B309} (1993) 86.
\bibitem{berto} S. Bertolini, F. Borzumati, A. Masiero, and G. Ridolfi, 
{ Nucl. Phys.} {\bf B353} (1991) 591;
N. Oshimo, { Nucl.~Phys.} {\bf B404} (1993) 20; 
R. Garisto, J.N. Ng, { Phys.~Lett.} {\bf B315} (1993) 372; 
M.A. Diaz, { Phys.~Lett.} {\bf B304} (1993) 278; 
Y. Okada, { Phys.~Lett.} {\bf B315} (1993) 119; 
F. Borzumati, { Z. Phys.} {\bf C63} (1994) 291; 
P. Nath and R. Arnowitt, { Phys.~Lett.} {\bf B336} (1994) 395; 
S. Bertolini and F. Vissani, { Z. Phys.} {\bf C67} (1995) 513; 
J. Lopez et al., { Phys. Rev.} {\bf D51} (1995) 147. 
\bibitem{CLEO96}
CLEO II, Contribution (PA05-093) to the 28th International Conference
on High Energy Physics, July 1996, Warsaw, Poland.
\bibitem{ALIB}
{ A. Ali}, hep-ph/9606324, hep-ph/9610333.
\bibitem{Photon}
{ I. Bigi et al.,} { Phys.~Rev.~Lett.} {\bf 71} (1993) 496;
{ Int. J. Mod. Phys.} {\bf A9} (1994) 2467;
{ G. Korchemsky and G. Sterman,} { Phys.~Lett.} {\bf B340} (1994) 96;
{ M. Neubert}, { Phys. Rev.} {\bf D49} (1994) 4623;
{ A. Ali and C. Greub}, { Phys.~Lett.} {\bf B361} (1995) 146;
{ A. Kapustin, Z. Ligeti and H.D. Politzer,}
{ Phys.~Lett.} {\bf B357} (1995) 653;
{ R.D. Dikeman, M. Shifman and R.G. Uraltsev},
{ Int. J. Mod. Phys.} {\bf A11} (1996) 571;
{ N. Pott,} { Phys. Rev.} {\bf D 54} (1996) 938.
\bibitem{GSW95}
{ C. Greub, H. Simma and D. Wyler},
{ Nucl.~Phys.} {\bf B434} (1995) 39; Erratum-ibid,
{\bf B444} (1995) 447.
\bibitem{CPRARE}
{ L. Littenberg and G. Valencia,}
{ Ann. Rev. Nucl. Part. Sci.} {\bf 43} (1993) 729;
{ J.L. Ritchie and S.G. Wojcicki,} { Rev. Mod. Phys.} {\bf 65} (1993)
1149; A. Pich, hep-ph/9610243; G. D'Ambrosio and G. Isidori,
hep-ph/9611284.
\bibitem{novikovetal:77}
{ V.A. Novikov, A.I. Vainshtein, V.I. Zakharov and M.A. Shifman,}
Phys. Rev. {\bf D16}, (1977) 223.
\bibitem{ellishagelin:83}
{ J. Ellis and J.S. Hagelin,} { Nucl.~Phys.} {\bf B217} (1983) 189.
\bibitem{dibetal:91}
{ C.O. Dib, I. Dunietz and F.J. Gilman,} { Mod. Phys. Lett.}
{\bf A6} (1991) 3573.
\bibitem{MP}
{ W. Marciano and Z. Parsa}, Phys. Rev. {\bf D53}, R1 (1996).
\bibitem{RS}
{ D. Rein and L.M. Sehgal,} { Phys. Rev.} {\bf D39} (1989) 3325;
{ J.S. Hagelin and L.S. Littenberg,} { Prog. Part. Nucl. Phys.}
{\bf 23} (1989) 1;
{ M. Lu and M.B. Wise,} { Phys. Lett.} {\bf B324} (1994) 461;
{ S. Fajfer}, [hep-ph/9602322]; { C.Q. Geng, I.J. Hsu and Y.C. Lin},
{ Phys. Rev.} {\bf D54} (1996) 877.
\bibitem{Adler95}
{ S. Adler} et al., { Phys. Rev. Lett.} {\bf 76} (1996) 1421.
\bibitem{AGS2}
{ L. Littenberg and J. Sandweiss}, eds., AGS2000, Experiments for the
21st Century, BNL 52512.
\bibitem{Cooper}
{ P. Cooper, M. Crisler, B. Tschirhart and J. Ritchie}
(CKM collaboration), 
EOI for measuring $Br(K^+\to\pi^+\nu\bar\nu)$ at the Main Injector,
Fermilab EOI 14, 1996.
\bibitem{littenberg:89}
{ L. Littenberg,} { Phys. Rev.} {\bf D39} (1989) 3322.
\bibitem{NIR96}
{ Y. Grossman, Y. Nir and R. Rattazzi}, [hep-ph/9701231] in \cite{HFII}.
\bibitem{BUCH96}
{ G. Buchalla}, hep-ph/9612307.
\bibitem{BB96}
{ G. Buchalla} and { A.J. Buras},
 { Phys. Rev.} {\bf D54} (1996) 6782.
\bibitem{AJB94}
{ A.J. Buras}, { Phys. Lett.} {\bf B333} (1994) 476.
\bibitem{XX97}
T. Nakaya, in proceedings of FCNC97, page 105.
\bibitem{Adler97}
S. Adler et al., { Phys. Rev. Lett.} {\bf 79}, (1997) 2204.
\bibitem{AGS2000}
L. Littenberg and J. Sandweiss, eds., AGS2000, Experiments for the 21st 
Century, BNL 52512.
\bibitem{FNALKL}
{ K. Arisaka et al.,} KAMI conceptual design report, FNAL, June 1991.
\bibitem{KEKKL}
{ T. Inagaki, T. Sato and T. Shinkawa,} Experiment to search for the
decay \klpnn\, at KEK 12 GeV proton synchrotron, 30 Nov. 1991.
\bibitem{KLBSM}
Y. Grossman and Y. Nir, { Phys. Lett.} {\bf B398} (1997) 163;
C.E. Carlson, G.D. Dorada and M. Sher,
{ Phys. Rev.} {\bf D54} (1996) 4393; G. Burdman, hep-ph/9705400;
A. Berera, T.W. Kephart and M. Sher, 
{ Phys. Rev.} {\bf D56} (1997) 7457;
Gi-Chol Cho, hep-ph/9804327.
\bibitem{HHW98}
T. Hattori, T. Hasuike and S. Wakaizumi, hep-ph/9804412.
\bibitem{BRS}
A.J. Buras, A. Romanino and L. Silvestrini, 
{ Nucl. Phys.} {\bf B520} (1998) 3.
\bibitem{GN1}
Y. Nir and M.P. Worah, hep-ph/9711215.
\bibitem{BB4}
{ G. Buchalla and A.J. Buras}, 
{ Phys. Lett.} {\bf B333} (1994) 221.
\bibitem{CJ}
{ C. Jarlskog,} { Phys. Rev. Lett.} {\bf 55}, (1985) 1039;
{ Z. Phys.} {\bf C29} (1985) 491.
\bibitem{Aleph96}
ALEPH Collaboration, Contribution (PA10-019) to 
the 28th International Conference
on High Energy Physics, July 1996, Warsaw, Poland.
\bibitem{B95}
{ A.J. Buras,} { Nucl. Instr. Meth.} {\bf A368} (1995) 1.
\bibitem{CDFMU}
F. Abe et al. (CDF), { Phys. Rev.} {\bf D57} (1998) R3811.
\bibitem{GLN96}
Y. Grossman, Z. Ligeti and E. Nardi,
{ Phys. Rev.} {\bf D55} (1997) 2768.
\bibitem{BB97}
{ G. Buchalla} and { A.J. Buras}, 
{ Phys. Rev.} {\bf D57} (1998) 216.
\bibitem{NQ}
{Y. Nir and H.R. Quinn}
{ Ann. Rev. Nucl. Part. Sci.} {\bf 42}
(1992) 211 and
 in " B Decays ", ed S. Stone
 (World Scientific, 1994),
p. 520; {I. Dunietz,} ibid p.550 and refs. therein.
\bibitem{CPASYM}
{ M. Gronau and D. London,} { Phys. Rev. Lett.}
 {\bf 65} (1990) 3381.
\bibitem{SNYD}
A. Snyder and H.R. Quinn, { Phys. Rev.} {\bf D48} (1993) 2139;
{ A.J. Buras and R. Fleischer,}
{ Phys. Lett.} {\bf B360} (1995) 138;
J.P. Silva and L. Wolfenstein, 
{ Phys. Rev.} {\bf D49} (1995) R1151; 
{ A.S. Dighe, M. Gronau and J. Rosner}, 
{ Phys. Rev.} {\bf D54} (1996) 3309; 
R. Fleischer and T. Mannel, 
{ Phys. Lett.} {\bf B397} (1997) 269;
C.S. Kim, D. London and T. Yoshikawa, 
{ Phys. Rev.} {\bf D57} (1998) 4010. 
\bibitem{BSANDA}
{ I.I.Y. Bigi and A.I. Sanda,}
{ Nucl. Phys.} {\bf B193} (1981) 85.
\bibitem{PHI}
D. London and A. Soni, { Phys. Lett.} {\bf B407} (1997) 61;
Y. Grossman and M.P. Worah, { Phys. Lett.} {\bf B395} (1997) 241;
M. Ciuchini et al., { Phys. Rev. Lett.} {\bf B79} (1997) 978;
R. Barbieri and and A. Strumia, { Nucl. Phys.} {\bf B508} (1997) 3.
\bibitem{adk}
{ R. Aleksan, I. Dunietz and B. Kayser,}
 { Z.Phys.} {\bf C54} (1992) 653;\\
R. Fleischer and I. Dunietz, { Phys. Lett.} {\bf B387} (1996) 361.
\bibitem{Wyler}
{ M. Gronau and D. Wyler,} { Phys. Lett.} {\bf B265} (1991) 172.
\bibitem{DUN2}
M. Gronau and D. London, { Phys. Lett.} {\bf B253} (1991) 483.
{ I. Dunietz}, { Phys. Lett.} {\bf B270} (1991) 75.
\bibitem{V97}
D. Atwood, I. Dunietz and A. Soni, 
{ Phys. Rev. Lett.} {\bf B78} (1997) 3257.
\bibitem{PAPIII}
R. Fleischer, { Phys.\ Lett.} {\bf B365} (1996) 399.
 \bibitem{fm2}
R. Fleischer and T. Mannel, {Phys.\ Rev.} {\bf D57}
(1998) 2752.
\bibitem{groro}
M. Gronau and J.L. Rosner, hep-ph/9711246, hep-ph/9712287.
\bibitem{wuegai}
F. W{\"u}rthwein and P. Gaidarev, hep-ph/9712531.
\bibitem{defan}
R. Fleischer, hep-ph/9802433.
\bibitem{cleo}
R. Godang  et al., hep-ex/9711010.
\bibitem{babar}
The BaBar Physics Book, preprint SLAC-R-504, in preparation.
\bibitem{bjorken}
J.D. Bjorken, { Nucl.\ Phys.}~{\bf B} (Proc.\ Suppl.)
{\bf 11} (1989) 325; SLAC-PUB-5389 (1990), published in the proceedings
of the SLAC Summer Institute 1990, p.\ 167.
\bibitem{bfm}
A.J. Buras, R. Fleischer and T. Mannel, hep-ph/9711262.
\bibitem{gewe}
J.-M. G{\'e}rard and J. Weyers, hep-ph/9711469; D. Del{\'e}pine,
J.-M. G{\'e}rard, J. Pestieau and J. Weyers, hep-ph/9802361;
J.-M. G{\'e}rard, J. Pestieau and J. Weyers, hep-ph/9803328.
\bibitem{neubert}
M. Neubert, hep-ph/9712224.
\bibitem{fknp}
A.F. Falk, A.L. Kagan, Y. Nir and A.A. Petrov, 
{Phys.\ Rev.} {\bf D57} (1998) 4290.
\bibitem{atso}
D. Atwood and A. Soni (1997), hep-ph/9712287, hep-ph/9712252.
\bibitem{FSI}
L. Wolfenstein, {Phys.\ Rev.} {\bf D52} (1995) 537; 
J. Donoghue, E. Golowich, A.~Petrov and J. Soares, 
{ Phys. Rev. Lett.} {\bf 77} (1996) 2178; 
B. Blok and I. Halperin, { Phys. Lett.} {\bf B385} (1996) 324; 
B. Blok, M. Gronau and J.L.~Rosner,
{ Phys.\ Rev.\ Lett.} {\bf 78} (1997) 3999.
\bibitem{BjSt}
B. Stech, { Phys. Lett.} {\bf B130} (1983) 189; 
J. Bjorken, hep-ph/9706524.
\bibitem{fm3}
R. Fleischer and T. Mannel, hep-ph/9706261.
\bibitem{pert-pens}
R. Fleischer, Z. Phys. {\bf C58} (1993) 483 and 
{\bf C62} (1994) 81; G. Kramer, W.F. Palmer and H. Simma,
 Z. Phys. {\bf C66} (1995) 429.
\bibitem{rf-FSI}
R. Fleischer, hep-ph/9804319.
\bibitem{bskk}
R. Fleischer, hep-ph/9710331.
\bibitem{LNO}
A. Lenz, U. Nierste and G. Ostermaier, hep-ph/9802202; 
U. Nierste, hep-ph/9805388.
\end{thebibliography}
\end{document}